\let\mypdfximage\pdfximage
\def\pdfximage{\immediate\mypdfximage}
\documentclass{emulateapj}
\usepackage{natbib}
\usepackage{graphicx}
\usepackage{times}
\usepackage{xspace} 
\usepackage{booktabs,longtable,tabu}
\usepackage{tabularx}

\newcommand {\apgt} {\ {\raise-.5ex\hbox{$\buildrel>\over\sim$}}\ }
\newcommand {\aple} {\ {\raise-.5ex\hbox{$\buildrel<\over\sim$}}\ }

\newcommand {\asca} {{\it ASCA}\xspace}

\newcommand {\chandra} {{\it Chandra}\xspace}
\newcommand {\xmm} {{\it XMM-Newton}\xspace}
\newcommand {\nustar} {{\it NuSTAR}\xspace}

\newcommand {\integral} {{\it INTEGRAL}\xspace}
\newcommand {\swiftxrt} {{\it Swift}~XRT\xspace}
\newcommand {\swiftbat} {{\it Swift}~BAT\xspace}

\newcommand {\wise} {{\it WISE}\xspace}

\newcommand {\sixum} {$6$~$\mu$m\xspace}
\newcommand {\Lsixum} {$L_{\rm 6\mu m}$\xspace}

\newcommand {\tnet} {$t_{\rm net}$\xspace}
\newcommand {\fprob} {$P_{\rm False}$\xspace}

\newcommand {\brnu} {$\mathrm{BR}_{\mathrm{Nu}}$\xspace}

\newcommand {\lx} {$L_{\mathrm{X}}$\xspace}
\newcommand {\Lsoft} {$L_{\mathrm{2-10\ keV}}$\xspace}

\newcommand {\Lhard} {$L_{\mathrm{10-40\ keV}}$\xspace}

\newcommand {\fx} {$f_{\mathrm{X}}$\xspace}

\newcommand {\hardband} {$8$--$24$~keV\xspace}
\newcommand {\softband} {$3$--$8$~keV\xspace}
\newcommand {\fullband} {$3$--$24$~keV\xspace}

\newcommand {\wone} {$W{\rm 1}$\xspace}
\newcommand {\wtwo} {$W{\rm 2}$\xspace}
\newcommand {\wthree} {$W{\rm 3}$\xspace}
\newcommand {\wfour} {$W{\rm 4}$\xspace}
\newcommand {\wonewtwo} {$W{\rm 1}$--$W{\rm 2}$\xspace}
\newcommand {\wtwowthree} {$W{\rm 2}$--$W{\rm 3}$\xspace}
\newcommand {\flow} {$F_{\mathrm{soft}}$\xspace}
\newcommand {\flowTotal} {$F_{\mathrm{soft}}^{\mathrm{30}}$\xspace}
\newcommand {\fwedge} {$f_{\mathrm{wedge}}$\xspace}

\newcommand{\siiv}{\mbox{\ion{Si}{4}}\xspace}
\newcommand{\mgii}{\mbox{\ion{Mg}{2}}\xspace}
\newcommand{\nev}{\mbox{[\ion{Ne}{5}]}\xspace}
\newcommand{\neiii}{\mbox{[\ion{Ne}{3}]}\xspace}
\newcommand{\nii}{\mbox{[\ion{N}{2}]}\xspace}
\newcommand{\oi}{\mbox{[\ion{O}{1}]}\xspace}
\newcommand{\oii}{\mbox{[\ion{O}{2}]}\xspace}
\newcommand{\oiii}{\mbox{[\ion{O}{3}]}\xspace}
\newcommand{\cii}{\mbox{\ion{C}{2}]}\xspace}
\newcommand{\ciii}{\mbox{\ion{C}{3}]}\xspace}
\newcommand{\civ}{\mbox{\ion{C}{4}}\xspace}
\newcommand{\heii}{\mbox{\ion{He}{2}}\xspace}
\newcommand{\feii}{\mbox{\ion{Fe}{2}}\xspace}
\newcommand{\sii}{\mbox{[\ion{S}{2}]}\xspace}
\newcommand{\lya}{Ly$\mathrm{\alpha}$\xspace}
\newcommand{\halpha}{H$\mathrm{\alpha}$\xspace}
\newcommand{\hbeta}{H$\mathrm{\beta}$\xspace}
\newcommand{\hgamma}{H$\mathrm{\gamma}$\xspace}
\newcommand{\hdelta}{H$\mathrm{\delta}$\xspace}

\newcommand{\typeii}{Type~2\xspace}

\newcommand {\ergpersec} {erg~s$^{-1}$\xspace}
\newcommand {\fluxunit} {erg~s$^{-1}$~cm$^{-2}$\xspace}
\newcommand {\fnuunit} {erg~s$^{-1}$~cm$^{-2}$~Hz$^{-1}$\xspace}
\newcommand {\nhunit} {cm$^{-2}$\xspace}
\newcommand {\degrees} {$^{\circ}$\xspace}
\newcommand {\persqarcmin} {$\mathrm{arcmin}^{-2}$\xspace}

\newcommand {\wavdetect}{$\mathtt{wavdetect}$\xspace}

\newcommand {\mytorus}{{\sc MYTorus}\xspace}

\begin{document}

\title{The {\it NuSTAR} Serendipitous Survey: the 40 Month Catalog and the Properties of the Distant High Energy X-ray Source
  Population}

\author{G.~B.~Lansbury\altaffilmark{1,2,$\dagger$}, 
D.~Stern\altaffilmark{3}, 
J.~Aird\altaffilmark{2,1}, 
D.~M.~Alexander\altaffilmark{1}, 
C.~Fuentes\altaffilmark{4},  
F.~A.~Harrison\altaffilmark{5}, 
E.~Treister\altaffilmark{4,6},  
F.~E.~Bauer\altaffilmark{6,7,8}, 
J.~A.~Tomsick\altaffilmark{9}, 
M.~Balokovi\'c\altaffilmark{5}, 
A.~Del~Moro\altaffilmark{10,1},  
P.~Gandhi\altaffilmark{1,11}, 
M.~Ajello\altaffilmark{12}, 
A.~Annuar\altaffilmark{1}, 
D.~R.~Ballantyne\altaffilmark{13}, 
S.~E.~Boggs\altaffilmark{9}, 
W.~N.~Brandt\altaffilmark{14,15,16}, 
M.~Brightman\altaffilmark{5},
C.-T.~J.~Chen\altaffilmark{14}, 
F.~E.~Christensen\altaffilmark{17},
F.~Civano\altaffilmark{18,19},
A.~Comastri\altaffilmark{20}, 
W.~W.~Craig\altaffilmark{17,21},
K.~Forster\altaffilmark{5}, 
B.~W.~Grefenstette\altaffilmark{5},
C.~J.~Hailey\altaffilmark{22}, 
R.~C.~Hickox\altaffilmark{23},
B.~Jiang\altaffilmark{5},
H.~D.~Jun\altaffilmark{3},
M.~Koss\altaffilmark{24}, 
S.~Marchesi\altaffilmark{12},
A.~D.~Melo\altaffilmark{4},
J.~R.~Mullaney\altaffilmark{25}, 
G.~Noirot\altaffilmark{3,26},
S.~Schulze\altaffilmark{6,7},
D.~J.~Walton\altaffilmark{3,5},
L.~Zappacosta\altaffilmark{27}, 
W.~W.~Zhang\altaffilmark{28}}

\affil{$^1$Centre for Extragalactic Astronomy, Department of Physics,
  Durham University, South Road,
  Durham, DH1 3LE, UK}
\affil{$^{2}$Institute of Astronomy, University of Cambridge,
  Madingley Road, Cambridge, CB3 0HA, UK; $^{\dagger}$gbl23@ast.cam.ac.uk}
\affil{$^{3}$Jet Propulsion Laboratory, California Institute of
  Technology, 4800 Oak Grove Drive, Mail Stop 169-221, Pasadena, CA
  91109, USA}
\affil{$^{4}$Universidad de Concepci\'{o}n, Departamento de
  Astronom\'{\i}a, Casilla 160-C, Concepci\'{o}n, Chile}
\affil{$^{5}$Cahill Center for Astrophysics, 1216 East California
  Boulevard, California Institute of Technology, Pasadena, CA 91125,
  USA}
\affil{$^{6}$Instituto de Astrof\'{\i}sica, Facultad de F\'{i}sica,
  Pontificia Universidad Cat\'{o}lica de Chile, 306, Santiago 22,
  Chile}
\affil{$^{7}$Millennium Institute of Astrophysics, Vicu\~{n}a Mackenna 4860, 7820436 Macul, Santiago, Chile}
\affil{$^{8}$Space Science Institute, 4750 Walnut Street, Suite 205,
  Boulder, Colorado 80301, USA}
\affil{$^{9}$Space Sciences Laboratory, 7 Gauss Way, University of
  California, Berkeley, CA 94720-7450, USA}
\affil{$^{10}$Max-Planck-Institut f\"ur Extraterrestrische Physik
  (MPE), Postfach 1312, D85741, Garching, Germany}
\affil{$^{11}$School of Physics and Astronomy, University of Southampton, Highfield, Southampton SO17 1BJ, UK}
\affil{$^{12}$Department of Physics and Astronomy, Clemson University, Clemson, SC 29634-0978, USA}
\affil{$^{13}$Center for Relativistic Astrophysics, School of Physics,
  Georgia Institute of Technology, Atlanta, GA 30332, USA}
\affil{$^{14}$Department of Astronomy and Astrophysics, 
  The Pennsylvania State University, University Park, PA 16802, USA}
\affil{$^{15}$Institute for Gravitation and the Cosmos, The Pennsylvania
State University, University Park, PA 16802, USA}
\affil{$^{16}$Department of Physics, The Pennsylvania State University,
University Park, PA 16802, USA}
\affil{$^{17}$DTU Space-National Space Institute, Technical University of
  Denmark, Elektrovej 327, DK-2800 Lyngby, Denmark}
\affil{$^{18}$Yale Center for Astronomy and Astrophysics, Physics
  Department, Yale University, New Haven, CT 06520, USA}
\affil{$^{19}$Harvard-Smithsonian Center for Astrophysics, 60 Garden Street,
  Cambridge, MA 02138, USA}
\affil{$^{20}$INAF Osservatorio Astronomico di Bologna, via Ranzani 1, 40127 Bologna, Italy}
\affil{$^{21}$Lawrence Livermore National Laboratory, Livermore, CA
  94550, USA}
\affil{$^{22}$Columbia Astrophysics Laboratory, 550 W 120th Street,
  Columbia University, NY 10027, USA}
\affil{$^{23}$Department of Physics and Astronomy, Dartmouth College,
  6127 Wilder Laboratory, Hanover, NH 03755, USA}
\affil{$^{24}$Institute for Astronomy, Department of Physics, ETH
  Zurich, Wolfgang-Pauli-Strasse 27, CH-8093 Zurich, Switzerland}
\affil{$^{25}$Department of Physics and Astronomy, The University of
  Sheffield, Hounsfield Road, Sheffield, S3 7RH, UK}
\affil{$^{26}$Universit\'e Paris-Diderot Paris VII, Universit\'e de Paris Sorbonne Cit\'e (PSC), 75205 Paris Cedex 13, France}
\affil{$^{27}$INAF Osservatorio Astronomico di Roma, via Frascati 33,
  00040 Monte Porzio Catone (RM), Italy}
\affil{$^{28}$NASA Goddard Space Flight Center, Greenbelt, MD 20771,
  USA}



\begin{abstract}

We present the first full catalog and science results for the
\nustar serendipitous
survey. The catalog incorporates
data taken during the first 40 months of \nustar operation, which
provide $\approx 20$~Ms of effective exposure time over $331$ fields,
with an areal coverage of $13$~deg$^2$, and $497$ sources
detected in total over the $3$--$24$~keV energy range. 
There are $276$ sources with spectroscopic redshifts and
classifications, 
largely resulting from our extensive campaign of ground-based spectroscopic followup.
We characterize the overall sample in terms of the X-ray, optical, and infrared
source properties.
The sample is primarily comprised of active galactic nuclei (AGNs), detected over a large range
in redshift from $z=0.002$ to $3.4$ (median of $\left< z \right> = 0.56$), but
also includes $16$ spectroscopically confirmed Galactic sources. There
is a large range in X-ray flux, from $\log (f_{\rm 3-24keV} / \mathrm{erg\
  s^{-1}\ cm^{-2}})\approx -14$ to $-11$, and in rest-frame
$10$--$40$~keV luminosity, from $\log (L_{\rm 10-40keV} / \mathrm{erg\
  s^{-1}})\approx 39$ to $46$, with a median of $44.1$. Approximately
$79\%$ of the \nustar sources have lower
energy ($<10$~keV) X-ray counterparts from \xmm, \chandra,
and \swiftxrt. 
The mid-infrared (MIR) analysis, using \wise all-sky survey data,
shows that MIR AGN color selections
miss a large fraction of the \nustar-selected AGN population, from
$\approx 15\%$ at the highest luminosities ($L_{\rm X}>10^{44}$~\ergpersec) to $\approx 80\%$ at the
lowest luminosities ($L_{\rm X}<10^{43}$~\ergpersec). 
Our optical spectroscopic analysis finds that the observed fraction of
optically obscured AGNs (i.e., the \typeii fraction) is $F_{\rm
  Type\ 2}=53^{+14}_{-15}\%$, for a well-defined subset of the $8$--$24$~keV
selected sample. This is higher, albeit at a low significance level,
than the \typeii fraction measured for redshift- and
luminosity-matched AGNs selected by $<10$~keV X-ray missions.

\end{abstract}


\section{Introduction}
\label{Introduction}

Since the late 1970s, which saw the advent of focusing X-ray
observatories in space 
 (e.g., \citealt{Giacconi79}), X-ray surveys have provided fundamental advances in our
understanding of growing supermassive black holes (e.g.,
\citealt{Fabian92,Brandt05,Alexander12,Brandt15}).
X-rays provide the most direct and efficient means of identifying
active galactic nuclei (AGNs; the sites of rapid mass accretion onto
supermassive black holes), since the effects of both
line-of-sight absorption and dilution by host-galaxy light are
comparatively low at X-ray energies. 
The collection of X-ray surveys over the last few decades have ranged from
wide-area all-sky surveys to deep pencil-beam surveys, allowing 
the evolution of AGN obscuration and the X-ray luminosity function to
be measured for a wide range in luminosity and redshift (up to
$z\approx 5$; 
e.g., see \citealt{Brandt15} for a review). The deepest surveys with
\chandra and \xmm have directly resolved the majority ($\approx 70$--$90\%$) of the
$\lesssim 8$~keV cosmic X-ray background (CXB) into individual objects (e.g.,
\citealt{Worsley05,Hickox06,Xue12}).

Until very recently, the most sensitive X-ray surveys (e.g., with
\chandra and \xmm) have been limited to photon energies of $<10$~keV, and are therefore
biased against the identification of heavily obscured AGNs (for which the
line-of-sight column density exceeds $N_{\rm H}\sim \mathrm{a\ few}\times 10^{23}$~\nhunit). This bias is especially strong
at $z\lesssim1$, but becomes less so for higher redshifts where the spectral
features of absorption, and the penetrating higher energy X-rays, are shifted into the observed-frame X-ray
energy window. The result is a complicated AGN selection function,
which is challenging to correct for without a full knowledge of
the prevalence of highly absorbed systems.
These photon energies are also low compared to the peak of
the CXB (at $\approx 20$--$30$~keV), meaning that spectral
extrapolations are required to characterize the AGN population
responsible for the CXB peak. High energy ($>10$~keV) X-ray surveys
with non-focusing X-ray
observatories (e.g., \swiftbat and \integral) have {\it directly}
resolved $\approx 1$--$2\%$ of the CXB peak into individual
AGNs \citep{Krivonos07,Ajello08,Bottacini12}. These surveys have
been successful in characterizing the local high-energy emitting AGN
population \citep[e.g., ][]{Tueller08,Burlon11,Vasudevan13,Ricci15}
but, being largely confined to $z\lesssim0.1$,
there is limited scope for evolutionary studies. 

A great breakthrough in studying the high-energy X-ray emitting
population is the {\it Nuclear Spectroscopic Telescope Array} (\nustar;
\citealt{Harrison13}), the first orbiting observatory with the ability
to focus X-ray light at energies $>10$~keV, resulting in a two orders
of magnitude increase in sensitivity over previous non-focusing missions. This
has opened up the possibility to study large, cleanly selected samples
of high-energy emitting AGNs in the distant universe for the first time.
The \nustar extragalactic survey program has provided the
first measurements of the $>10$~keV AGN luminosity functions at
$z>0.1$ \citep{Aird15b}, and has directly resolved a large fraction
($35\pm 5\%$) of the CXB at $8$--$24$~keV \citep{Harrison16}. 
In addition, both the survey program and targetted \nustar campaigns
have demonstrated the importance of high-energy 
coverage for accurately constraining the intrinsic properties of distant AGNs
\citep[e.g.,][]{DelMoro14,Lansbury14,Luo14,Civano15,Lansbury15,LaMassa16}, 
especially in the case of the most
highly absorbed Compton-thick (CT) systems (where $N_{\rm H}>1.5\times
10^{24}$~\nhunit). 

The \nustar extragalactic survey is the largest scientific program,
in terms of time investment, undertaken with \nustar and is one of the
highest priorities of the mission. There are two
main ``blind survey'' components.
Firstly, deep blank-field \nustar surveys have been performed in
the following well-studied fields: the Extended \chandra Deep Field South
(ECDFS; \citealt{Lehmer05}), for which the total areal coverage with \nustar is
$\approx 0.33$~$\mathrm{deg}^{2}$ (\citealt{Mullaney15}, hereafter
M15); the Cosmic Evolution Survey field (COSMOS;
\citealt{Scoville07}), which has $\approx 1.7$~$\mathrm{deg}^{2}$ of \nustar
coverage (\citealt{Civano15}, hereafter C15); the Extended Groth Strip
(EGS; \citealt{Groth94}), with $\approx 0.25$~$\mathrm{deg}^{2}$ of coverage (Aird et al., in
prep.); the northern component of the Great Observatories Origins Deep
Survey North (GOODS-N; \citealt{Dickinson03}),
with $\approx 0.07$~$\mathrm{deg}^{2}$ of coverage (Del Moro et
al., in prep.); and the Ultra Deep Survey field (UDS; \citealt{Lawrence07}), with $\approx 0.4$~$\mathrm{deg}^{2}$ of coverage (Masini et
al., in prep.). Secondly, a wide-area
``serendipitous survey'' has been performed by searching the majority of \nustar
pointings for chance background sources. An initial look
at $10$ serendipitous survey sources was presented in \citet{Alexander13}. 
Serendipitous surveys represent an efficient and economical way to
sample wide sky areas, and provide substantial data sets with which to
examine the X-ray emitting population and search for extreme
sources. They have been undertaken with
many X-ray missions over the last few decades (e.g.,
\citealt{Gioia90,Comastri01,Fiore01,Harrison03,Nandra03,Gandhi04a,Kim04,Ueda05,Watson09,Evans10,Evans14}).

In this paper, we describe the \nustar serendipitous
survey and present the first catalog, compiled from data which
span the first $40$~months of \nustar operation.
The serendipitous survey is a
powerful component of the \nustar survey programme, with the largest overall sample
size, the largest areal coverage ($\approx 13$~$\mathrm{deg}^{2}$), and regions with comparable
sensitivity to the other \nustar surveys in well-studied fields. 
Section \ref{nudata} details the \nustar observations, data reduction, source detection, and
photometry.
We match to counterparts at
lower X-ray energies (from \chandra, \xmm, and \swiftxrt; Section
\ref{cparts_xray}), and at optical and infrared (IR) wavelengths (Section
\ref{cparts_optIR}). We have undertaken an extensive
campaign of ground-based spectroscopic followup, 
crucial for obtaining source redshifts and classifications, which
is described in Section \ref{spec_opt}. Our results for
the X-ray, optical, and IR properties of
the overall sample are presented in Sections \ref{results_xray},
\ref{results_opt}, and \ref{results_ir}, respectively. We summarize the 
main results in Section \ref{summary}.
All uncertainties and limits are quoted at the $90\%$ confidence
level, unless otherwise stated. We assume the flat $\Lambda$CDM cosmology from WMAP7 \citep{Komatsu11}.

\section{The \nustar Data}
\label{nudata}

The \nustar observatory (launched in 2012 June; \citealt{Harrison13}) is comprised of two
independent telescopes (A and B),
identical in design, the focal plane modules of which are hereafter
referred to as FPMA and FPMB. The modules have fields-of-view (FoVs) of
$\approx12\arcmin \times 12\arcmin$, which overlap in sky coverage.
The observatory is sensitive between $3$~keV and $78$~keV.
The main energy band that we focus on here is the $3$--$24$~keV band;
this is the most useful band for the relatively faint sources detected
in the \nustar extragalactic surveys, since the
combination of instrumental background and a decrease in effective
area with increasing energy means that source photons are unlikely to be
detected at higher energies (except for the brightest sources). 
\nustar provides over an order of
  magnitude improvement in angular resolution compared to previous
  generation hard X-ray observatories: the point-spread function (PSF)
  has a full width at half maximum (FWHM) of $18\arcsec$ and a half-power diameter of
  $58\arcsec$, and is relatively constant across the FoV. The
  astrometric accuracy of \nustar is $8\arcsec$
  for the brightest
  targets ($90\%$ confidence; \citealt{Harrison13}). This 
  worsens with decreasing photon counts, reaching a positional
  accuracy of $\approx 20\arcsec$ for
  the faintest sources (as we demonstrate in Section \ref{cparts_xray}).

Here we describe the observations, data
reduction and data-analysis procedures used for the \nustar serendipitous
survey: Section \ref{serendip_obs} describes the \nustar observations
which have been incorporated as part of the survey; Section \ref{data_processing} details the data
reduction procedures used to generate the \nustar science data;
Section \ref{nu_srcdet} provides details of the source detection
approach; Section \ref{nu_phot} outlines the photometric measurements
for source counts, band ratios, fluxes and luminosities; and Section
\ref{nu_cat} describes the final source catalog.

\subsection{The serendipitous survey observations}
\label{serendip_obs}

The serendipitous 
 survey is the largest area
  blind survey undertaken with \nustar. The survey is achieved
  by searching the background regions of almost every non-survey \nustar
  pointing for background sources unassociated with the original
  science target. 
The survey approach is well-suited to \nustar since
  there are generally large regions of uncontaminated background.
We exclude from the survey \nustar fields with bright science targets,
identified as fields with $>10^{6}$ counts within $120''$ of the
on-axis position. We also exclude the dedicated extragalactic (COSMOS,
ECDFS, EGS, GOODS-N, UDS) and Galactic survey fields (the Galactic
center survey; \citealt{Mori15,Hong16}; and the Norma Arm survey; Fornasini et al., in prep.).

\begin{figure}
\centering
\includegraphics[width=0.47\textwidth]{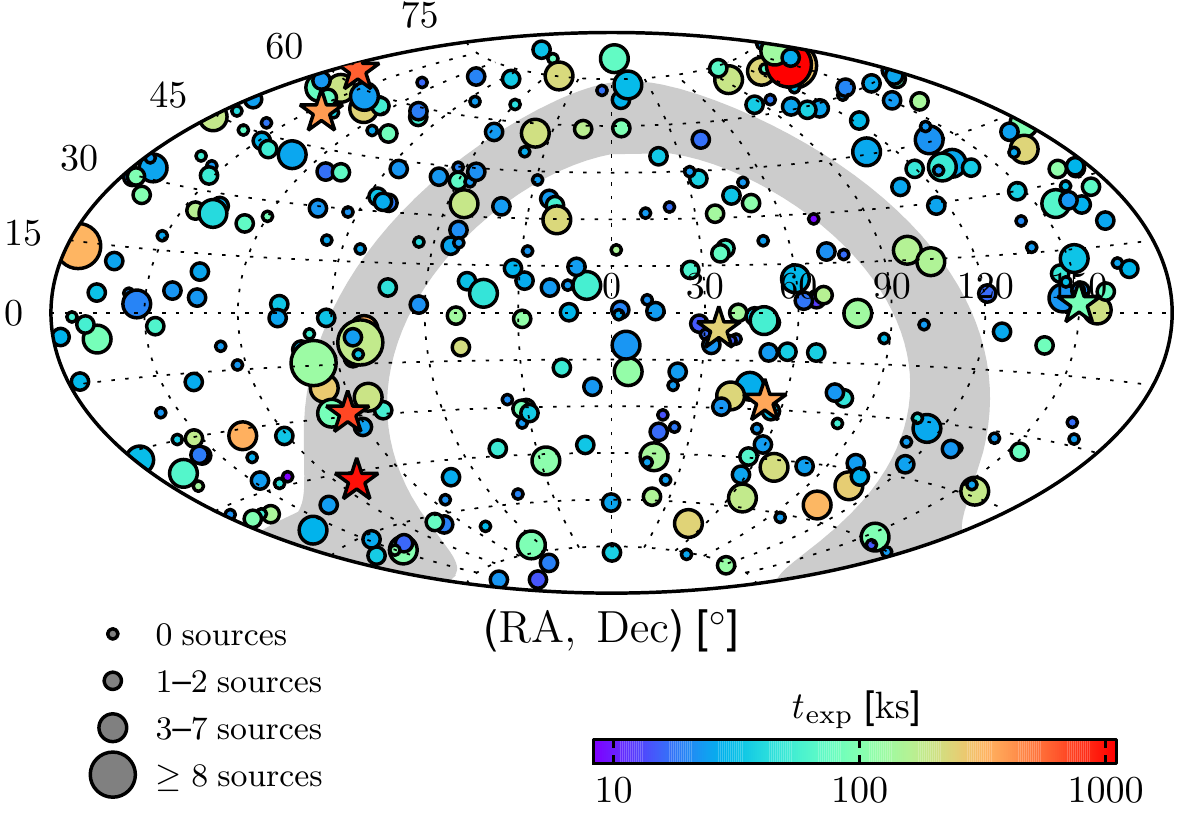}
\caption{Aitoff projection showing the distribution of \nustar
  serendipitous survey fields on the sky, in equatorial coordinates
  (R.A., decl.). The circle sizes correspond to the number of sources
  detected in a given field, and the colors correspond to the cumulative
  exposure time (per FPM) for a given field. 
The locations of the dedicated \nustar surveys in well-studied fields
(COSMOS, ECDFS, EGS, GOODS-N, UDS, the Galactic center, and the Norma Arm), which are not included in the serendipitous
survey, are marked with star symbols (with the colors representing the
maximum unvignetted exposures). Also excluded from the serendipitous
survey are \nustar fields containing bright targets (not shown on this
figure; see Section \ref{serendip_obs}).
The gray area highlights the region $\pm10$\degrees of the Galactic
  plane.}
\label{aitoff}
\end{figure}

Over the period from 2012 July to 2015 November, which is the focus of
the current study, there are $510$ individual \nustar exposures which
have been incorporated into the 
serendipitous survey. These exposures were performed over $331$ unique
fields (i.e., $331$ individual sky regions, each with contiguous
coverage comprised of one or more \nustar exposures), 
yielding a total sky area coverage of $13$~deg$^2$. 
Table~\ref{fields_table} lists the fields chronologically,\footnote{In Table \ref{fields_table} we show the first ten fields
    as an example. The full table, which includes all $331$ fields, is
    available as an electronic table.} 
and provides the following details for each field: the name of the primary \nustar science
target; the number of \nustar exposures; the individual \nustar observation ID(s); the observation date(s); the
  pointing coordinates; the exposure time(s); the number of serendipitous sources
  detected; and flags to indicate the \nustar fields
which were used in the \citet{Aird15b} and \citet{Harrison16}
studies.
\renewcommand*{\arraystretch}{1.1}
\begin{table*}
\centering
\caption{Details of the individual \nustar observations which make up the 
  serendipitous survey}
\begin{tabular}{lcccccccccc} \hline\hline \noalign{\smallskip}
Field ID & Science Target & $N_{\rm obs}$ & Obs.\ ID & Obs.\ Date & R.A.\ (\degrees) & Decl.\ (\degrees) &
   $t_{\rm exp}$ (ks) & $N_{\rm serendips}$ & A15 & H16 \\
(1) & (2) & (3) & (4) & (5) & (6) & (7) & (8) & (9) & (10) & (11) \\
\noalign{\smallskip} \hline \noalign{\smallskip}
1 & 2MASX J05081967+1721483 & 1 & 60006011002 & 2012-07-23 & 77.08 & 17.36 & 16.6 & 0 & 0 & 0 \\
2 & Bkgd BII -11.2 & 1 & 10060003001 & 2012-07-24 & 71.11 & 28.38 & 8.9 & 0 & 0 & 0 \\
3 & 2MASX J04234080+0408017 & 2 & $\cdots$ & $\cdots$ & $\cdots$ & $\cdots$ & 12.3 & 2 & 1 & 1 \\
3a &  & $\cdots$ & 60006005002 & 2012-07-25 & 65.92 & 4.13 & 6.4 & $\cdots$ & $\cdots$ & $\cdots$ \\
3b &  & $\cdots$ & 60006005003 & 2012-07-25 & 65.92 & 4.13 & 5.9 & $\cdots$ & $\cdots$ & $\cdots$ \\
4 & IC 4329A & 1 & 60001045002 & 2012-08-12 & 207.33 & -30.31 & 177.3 & 2 & 0 & 1 \\
5 & Mrk 231 & 2 & $\cdots$ & $\cdots$ & $\cdots$ & $\cdots$ & 74.9 & 4 & 1 & 1 \\
5a &  & $\cdots$ & 60002025002 & 2012-08-26 & 194.06 & 56.87 & 44.3 & $\cdots$ & $\cdots$ & $\cdots$ \\
5b &  & $\cdots$ & 60002025004 & 2013-05-09 & 194.06 & 56.87 & 30.6 & $\cdots$ & $\cdots$ & $\cdots$ \\
6 & NGC 7582 & 2 & $\cdots$ & $\cdots$ & $\cdots$ & $\cdots$ & 33.4 & 2 & 0 & 1 \\
6a &  & $\cdots$ & 60061318002 & 2012-08-31 & 349.60 & -42.37 & 17.7 & $\cdots$ & $\cdots$ & $\cdots$ \\
6b &  & $\cdots$ & 60061318004 & 2012-09-14 & 349.60 & -42.37 & 15.7 & $\cdots$ & $\cdots$ & $\cdots$ \\
7 & AE Aqr & 4 & $\cdots$ & $\cdots$ & $\cdots$ & $\cdots$ & 134.2 & 2 & 1 & 1 \\
7a &  & $\cdots$ & 30001120002 & 2012-09-04 & 310.04 & -0.87 & 7.2 & $\cdots$ & $\cdots$ & $\cdots$ \\
7b &  & $\cdots$ & 30001120003 & 2012-09-05 & 310.04 & -0.87 & 40.5 & $\cdots$ & $\cdots$ & $\cdots$ \\
7c &  & $\cdots$ & 30001120004 & 2012-09-05 & 310.04 & -0.87 & 76.6 & $\cdots$ & $\cdots$ & $\cdots$ \\
7d &  & $\cdots$ & 30001120005 & 2012-09-07 & 310.04 & -0.87 & 9.8 & $\cdots$ & $\cdots$ & $\cdots$ \\
8 & NGC 612 & 1 & 60061014002 & 2012-09-14 & 23.49 & -36.49 & 17.9 & 0 & 0 & 1 \\
9 & 3C 382 & 1 & 60061286002 & 2012-09-18 & 278.76 & 32.70 & 18.0 & 1 & 0 & 0 \\
10 & PBC J1630.5+3924 & 1 & 60061271002 & 2012-09-19 & 247.64 & 39.38 & 17.1 & 1 & 1 & 1 \\
\vdots & \vdots & \vdots & \vdots & \vdots & \vdots & \vdots & \vdots & \vdots & \vdots & \vdots \\
\noalign{\smallskip} \hline \noalign{\smallskip}
\end{tabular}
\begin{minipage}[c]{0.92\textwidth}
\footnotesize
\textbf{Notes.} (1): ID assigned to each field. For fields with
multiple \nustar exposures (i.e., $N_{\rm obs}>1$), each individual component exposure is
listed with a letter suffixed to the field ID (e.g., 3a and 3b). (2): Object name for the primary science target of the \nustar
observation(s). (3): The number of individual \nustar exposures for a
given field ($N_{\rm obs}$). (4): \nustar observation ID. (5): Observation start date.
(6) and (7): Approximate R.A.\ and decl.\ (J2000) coordinates for the aim-point, in decimal degrees.
(8): Exposure time (``ONTIME''; ks), for a single FPM (i.e., averaged over FPMA
and FPMB). (9): The number of serendipitous \nustar sources detected
in a given field. (10) and (11): Binary flags to highlight the serendipitous
survey fields used for the \citet{Aird15b} and \citet{Harrison16}
studies, respectively. This table shows the first ten (out of $331$) fields
only. 
\end{minipage}
\label{fields_table}
\end{table*}
Figure \ref{aitoff} shows an all-sky map of the serendipitous survey fields. 
The fields have a cumulative exposure time of $20.4$~Ms. For
comparison, the \nustar surveys of COSMOS and ECDFS have cumulative
exposure times of $3.1$~Ms and $1.5$~Ms (C15 and M15, respectively).
The serendipitous survey fields cover a
wide range in individual exposure times (from
$\sim 10$~ks to $1$~Ms), and have a median exposure of $28$~ks (these
values correspond to a single \nustar FPM).
For $76\%$ of the fields there is a single \nustar exposure, and for the remainder
there are multiple (from two to $15$) exposures which are combined
together for the science analyses (see Section \ref{data_processing}). 

An important contributor of fields to the \nustar serendipitous survey
is the \nustar ``snapshot survey'' (\citealt{Balokovic14}; Balokovi\'{c} et al. 2017, in prep.), a
dedicated \nustar program targetting \swiftbat-selected AGNs (the
\swiftbat AGNs themselves are not included in the serendipitous survey,
only the background regions of the \nustar observations). For this
work we include $154$ snapshot survey fields observed during the first
$40$ months of \nustar operation. 
These yield $21\%$ of the total serendipitous survey source detections, 
and make up a large fraction of the survey area (accounting for $47\%$
of the fields incorporated, in total).

\subsection{Data processing}
\label{data_processing}

For data reduction, we use HEASoft v.\ 6.15, 
the \nustar Data Analysis Software (NuSTARDAS) v.\ 1.3.0,\footnote{http://heasarc.gsfc.nasa.gov/docs/nustar/analysis} 
and CIAO v.\ 4.8. 
For each of the $510$ obsIDs 
incorporated in the survey, the raw,
unfiltered event files for FPMA and FPMB were processed using the
{\sc nupipeline} program to yield calibrated, cleaned event files.
For source detection and photometry (see Sections \ref{nu_srcdet} and \ref{nu_phot}), we
adopt the observed-frame energy bands which have been 
utilized for the \nustar extragalactic survey programme in general, and
other recent \nustar studies:
$3$--$8$, $3$--$24$, and $8$--$24$~keV \citep[hereafter referred to as
the soft, full, and hard bands; e.g.,
][]{Alexander13, Luo14, Aird15b, Lansbury15, Harrison16}.
To produce individual energy band images from the \nustar event lists
we used the CIAO program {\sc dmcopy} \citep{Fruscione06}.

To produce exposure maps, which account
for the natural dither of the observatory and regions of lower
sensitivity (e.g., chip gaps), we follow
the procedure outlined in detail in Section 2.2.3 of M15. Vignetting
in the optics
results in a decrease in the effective exposure with increasing
distance from the optical axis. We produce both vignetting-corrected and
non-vignetting-corrected exposure maps. The former allow us to
determine the effective exposure at source positions within the FoV and correctly
determine count rates,
while the latter are more appropriate for the scaling of background
counts since the \nustar aperture background component dominates the
background photon counts at $\lesssim30$~keV \citep[e.g.,][]{Wik14}. 

In order to increase sensitivity, we perform source detection (see Section
\ref{nu_srcdet}) and
photometry (see Section \ref{nu_phot}) on the coadded FPMA+FPMB (hereafter ``A+B'') data, 
produced by combining the FPMA and FPMB science data with the HEASoft package {\sc
  ximage}. For fields with multiple obsIDs, we use {\sc ximage} 
to combine the data from individual observations, such that each field
has a single mosaic on which source detection and photometry are
performed.

\subsection{Source detection}
\label{nu_srcdet}

In general, the source-detection procedure follows that adopted in the dedicated
blank-field surveys (e.g., see C15 and M15). A significant difference
with the serendipitous survey, compared to the blank-field surveys, is
the existence of a science target at the FoV aim-point. We account for
the background contribution from such science targets by incorporating
them in the background map generation, as described below. 
We also take two steps
to exclude sources associated with the science target: (1) in cases where
the target has an extended counterpart in the optical or IR bands (e.g., a
low-redshift galaxy or galaxy cluster), we mask out custom-made regions which are
conservatively defined to be larger than the extent of the counterpart in the 
optical imaging coverage (from the SDSS or DSS),
accounting for spatial smearing of emission due to the \nustar PSF; (2) for all point-source detections with
spectroscopic identifications, we assign an ``associated'' flag to
those which have a velocity offset from the science target [$\Delta(c
z)$] smaller than $5\%$ of the total science target velocity. 

Here we summarize the source detection procedure, which is applied
separately for each of the individual \nustar energy bands (soft,
full, and hard) before the individual band source lists are merged to
form the final catalog. For every pixel position across
the \nustar image, a ``false probability'' is calculated to quantify the
chance that the counts measured in a source detection aperture around that position
are solely due to a background fluctuation. 
In this calculation we adopt a circular source detection aperture of radius $20\arcsec$,
which is justified by the tight core of
the \nustar PSF (FWHM$=18\arcsec$), and was also adopted for the
dedicated blank-field surveys (e.g., C15; M15).
To measure the background level at each pixel position, background
counts are first
measured from the \nustar image using an annular aperture of
inner radius $45\arcsec$ and outer radius $90\arcsec$, centered on
that position. These background counts are then 
re-scaled to the $20\arcsec$ source detection aperture according to the ratio of effective
areas (as determined from non-vignetting-corrected exposure maps). This approach allows the
local background to be sampled without significant
contamination from the serendipitous source counts (since the background annulus has a relatively
large inner radius). The background measurement also
accounts for any contaminating photons from the aim-point science target
which, due to the broad wings of the \nustar PSF, can contribute to
the background (if the science target is comparatively bright and offset by
$\lesssim 200''$ from the serendipitous source position).
The Poisson false probability (\fprob) is assessed at each pixel, using the
source and scaled background counts
\citep[e.g.,][]{Lehmer05,Nandra05,Laird09}, to yield a \fprob map. 
From this map we exclude areas 
within $30\arcsec$ of the low-exposure ($<10\%$ of the maximum
exposure) peripheral regions close to the FoV edge,
where there is a steep drop-off in exposure and the
background is poorly characterized. 

 We then perform source detection on
the \fprob map to identify sources. For a full, detailed
description of this source detection procedure we refer the reader to
Section 2.3 of M15. In brief, the SExtractor algorithm
\citep{Bertin96} is used to identify regions of each \fprob map which
fall below a threshold of $\log(P_{\rm False})<-6$ (the approximate average
of the thresholds adopted for the \nustar-COSMOS and \nustar-ECDFS
surveys; C15; M15), producing source lists for each individual energy band. The
coordinates for each detected source are measured at the local minimum in
\fprob. Finally, we merge the sources detected in the different
energy bands to yield a final source list. To achieve this
band-merging, the soft ($3$--$8$~keV) and hard ($8$--$24$~keV) band
detected sources are matched to the full ($3$--$24$~keV) band source list using a matching
radius of $35\arcsec$. The adopted \nustar source coordinates
correspond to the position of the source in the full band,
if there is a detection in this band. Otherwise the coordinates
correspond to the soft band, if there is a detection in this band,
or the hard band if there is no full or soft band
detection. The analyses described below
(e.g., photometry and multiwavelength counterpart matching) are
performed using these adopted source coordinates.
After the above source detection has been performed, we exclude any sources within 
$90\arcsec$ of the central science target position (for comparison, the half-power diameter of the
\nustar PSF is $58\arcsec$). 

\begin{figure}
\centering
\includegraphics[width=0.47\textwidth]{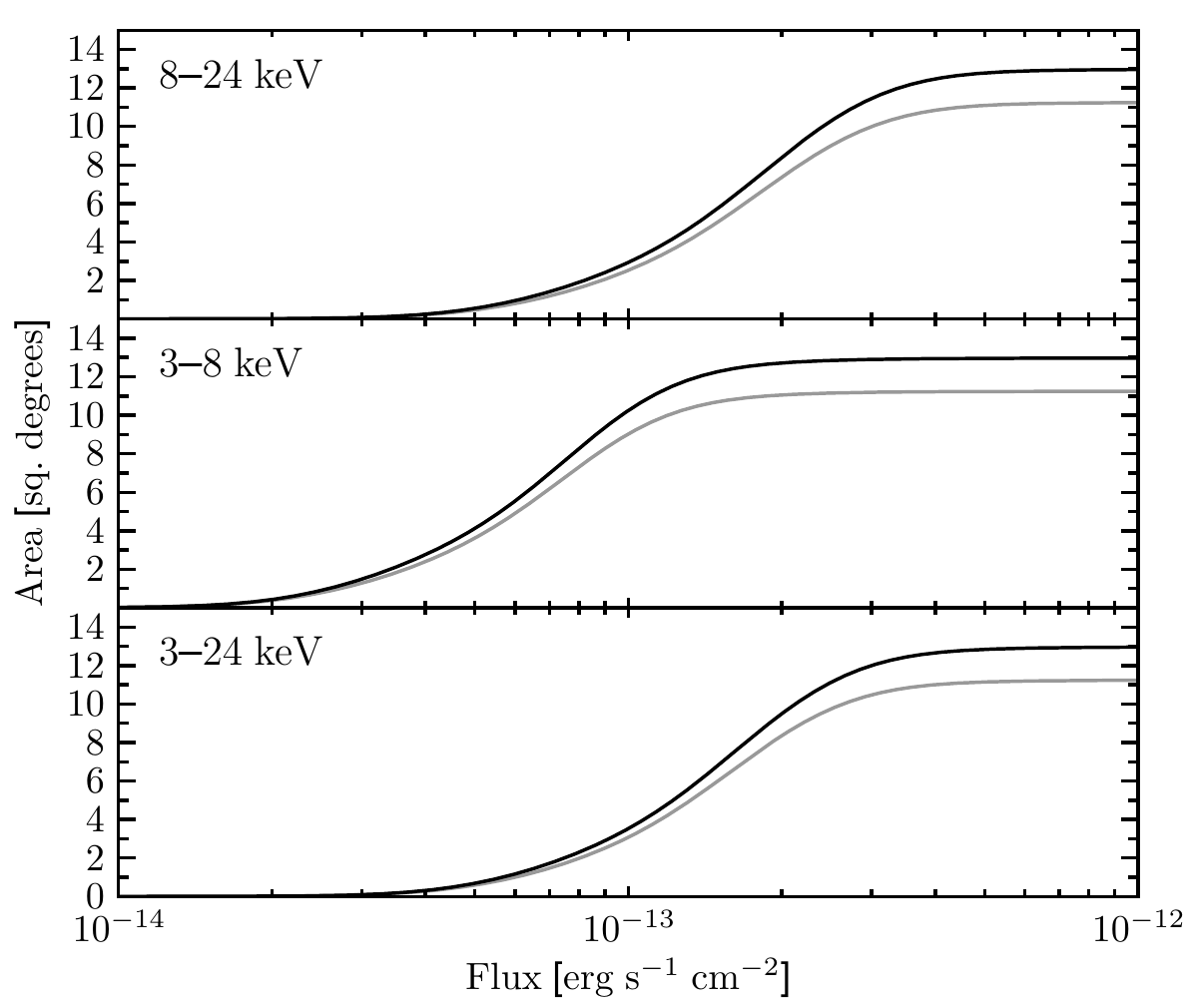}
\caption{Sky coverage (solid angle) of the \nustar serendipitous
  survey as a function of (aperture-corrected) flux sensitivity, for the three main
  energy bands. The black line shows the area curve for the full
  survey, and the gray line shows that for the survey regions outside
  the Galactic plane ($|b|>10$\degrees).}
\label{acurves_ser}
\end{figure}

\begin{figure}
\centering
\includegraphics[width=0.47\textwidth]{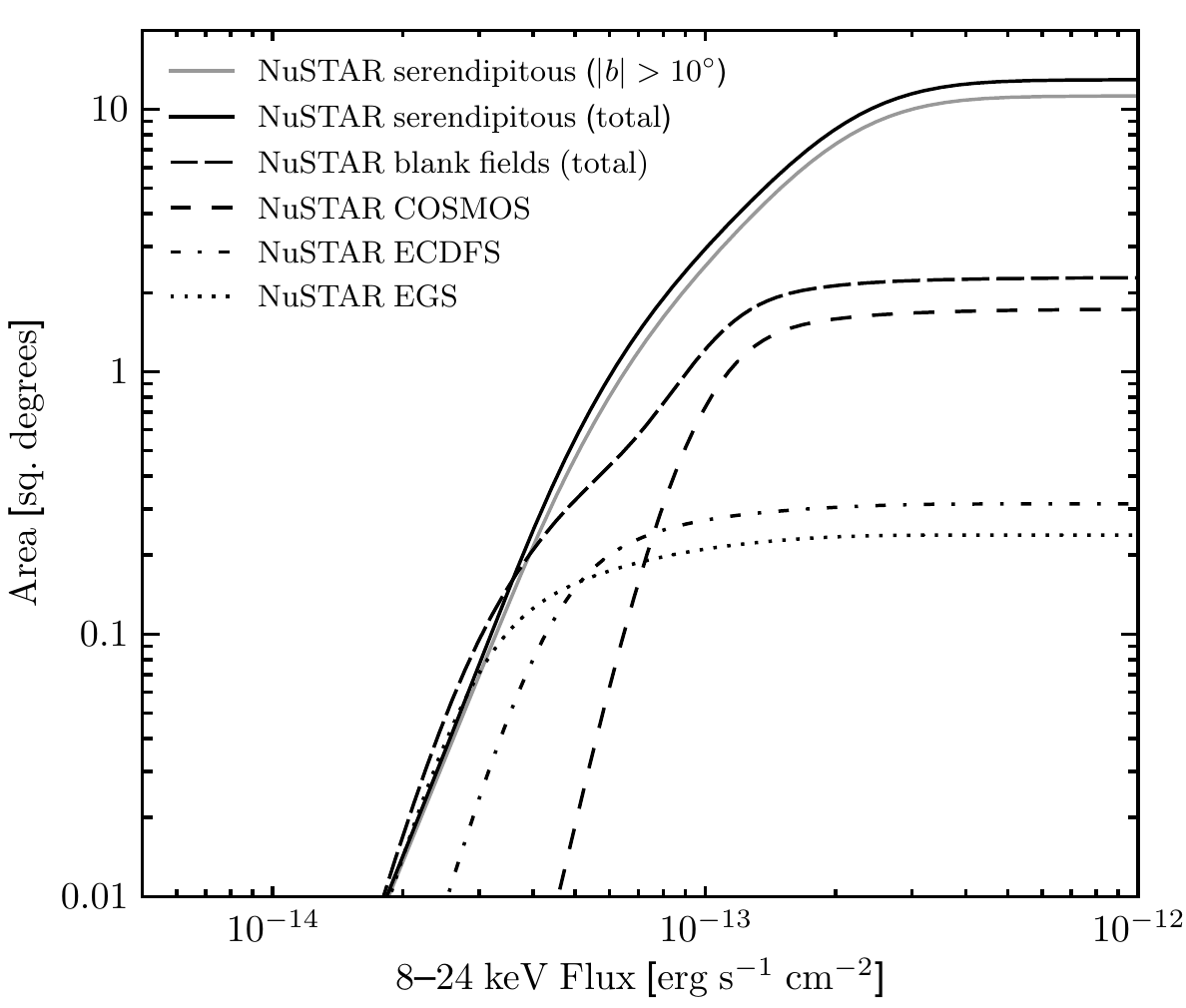}
\caption{Sky coverage (solid angle) of the \nustar serendipitous
  survey as a function of flux sensitivity, for the hard ($8$--$24$~keV)
  energy band, at which \nustar is uniquely sensitive. I.e.,
  the sky coverage for which sources above a given hard
  band flux will be detected in the hard band. The black and
  gray solid lines show the area curves for the overall and the $|b|>10$\degrees
  serendipitous survey, respectively. We compare with the other
  completed components of the \nustar extragalactic surveys program,
  which include the following dedicated blank-field surveys:
  \nustar-COSMOS (dashed line; C15), \nustar-ECDFS (dash-dotted line;
  M15), and \nustar-EGS (dotted line; Aird et al., in prep.). The
  total area for these blank-field surveys (which are not included
  as part of the serendipitous survey) is shown as a long-dashed line.}
\label{acurves_surveys}
\end{figure}

To determine the overall sky coverage of the survey as a function of
flux sensitivity, we sum the sensitivity curves for the $331$
individual fields. For each field the sensitivity curve is
determined by calculating, for every point in the \nustar image
(excluding the low-exposure peripheral regions), the
flux limit corresponding to $\log(P_{\rm False})=-6$ (the detection
threshold), given the background and exposure maps described above
and the count-rate to flux conversion factors listed in Section
\ref{nu_phot}. Figure \ref{acurves_ser} shows the total, summed
sensitivity curves for the serendipitous survey, for the three main
energy bands. 
Figure \ref{acurves_surveys} shows the logarithmic version, compared
to the other components of the \nustar extragalactic surveys
program. The serendipitous survey has the largest solid angle coverage for
most fluxes, and a similar areal coverage to the deepest blank-field
survey (the \nustar-EGS survey)
at the lowest flux limits. In both Figure \ref{acurves_ser} and Figure
\ref{acurves_surveys} we also show the area curves for the subset of
the serendipitous survey which lies outside of the Galactic plane
($|b|>10$\degrees) and is thus relatively free of
Galactic sources. We note that the recent works of
\citet{Aird15b} and \citet{Harrison16}, which presented the luminosity
functions and source number counts for the \nustar extragalactic
survey program, only incorporated serendipitous survey fields at $|b|>20$\degrees (and at
$\mathrm{decl.}>-5$\degrees for \citealt{Aird15b}). 

\subsection{Photometry}
\label{nu_phot}

For each source detected using the above procedure we measure the net
counts, count rates and fluxes, and for sources with spectroscopic
redshifts we calculate rest-frame luminosities. For the aperture photometry, we adopt a circular
aperture of $30\arcsec$ radius to measure the gross (i.e., source plus
background) counts ($S$). The scaled background counts ($B_{\rm src}$) are determined
using the same procedure as for the source detection (Section
\ref{nu_srcdet}), and are subtracted from $S$ to obtain the net
source counts ($S_{\rm net}$). The errors on $S_{\rm net}$ are computed
as $1 + \sqrt{S + 0.75}$ ($84\%$ confidence level; e.g., \citealt{Gehrels86}). 
For sources undetected in a
given band, upper limits for $S_{\rm net}$ are calculated
using the Bayesian approach outlined in \citet{Kraft91}.
To determine the net count rate, we divide $S_{\rm net}$
by the exposure time drawn from the vignetting-corrected exposure
map (mean value within the $30''$ aperture).

Deblending is performed following the procedure outlined in detail in
Section 2.3.2 of M15. In short, for a given source, the contributions from neighboring detections
(within a $90\arcsec$ radius) to the source aperture counts are accounted for
using knowledge of their separation and brightness. The false
probabilities and photometric quantities (e.g., counts, flux) are all
recalculated post-deblending, and included in the catalog in separate columns. Out of the
total $498$ sources in the source catalog, only one is no longer
significant (according to our detection threshold) post-deblending.

\nustar hard-to-soft band ratios (\brnu) are calculated as the
ratio of the $8$--$24$ to $3$--$8$~keV count rates.
For sources with full band counts of $S_{\rm net}>100$, and with
a detection in at least one of the soft or hard bands, 
we derive an effective photon index
($\Gamma_{\rm eff}$); i.e., the spectral slope of a power
law spectrum that is required to produce a given band ratio.

To measure fluxes, we convert from the deblended $30\arcsec$ count rates using
the following conversion factors: $6.7\times 10^{-11}$,
$9.4\times 10^{-11}$ and $13.9 \times 10^{-11}$~$\mathrm{erg\ cm^{-2}\
cts^{-1}}$ for the soft, full
and hard bands, respectively. 
These conversion factors were derived to account for the \nustar response, and assume an
unabsorbed power-law with a photon index of $\Gamma_{\rm eff}=1.8$
(typical of AGN detected by \nustar; e.g., \citealt{Alexander13}). The
conversion factors return aperture-corrected fluxes; i.e., they are corrected to the $100\%$
encircled-energy fraction of the PSF. The general agreement between our
\nustar fluxes and those from \chandra and \xmm (see Section
\ref{cparts_xray}) indicates that the
\nustar flux measurements are reliable. For sources with spectroscopic
redshifts, we determine the rest-frame $10$--$40$~keV luminosity by
extrapolating from a measured observed-frame flux, assuming a photon index
of $\Gamma_{\rm eff}=1.8$. To ensure that the adopted
observed-frame flux energy band corresponds to the rest-frame
$10$--$40$~keV energy band, we use the observed-frame $8$--$24$ and
$3$--$8$~keV bands for sources with redshifts of $z<1.35$ and
$z\ge 1.35$, respectively. For cases with a non-detection
in the relevant band (i.e., $8$--$24$ or $3$--$8$~keV), we
instead extrapolate from the full band ($3$--$24$~keV).

\subsection{The source catalog}
\label{nu_cat}

The serendipitous survey source catalog is available as an electronic
table. In Section A.1 we give a detailed description of
  the $106$ columns that are provided in the catalog. 
In total, the catalog contains $497$ sources which are
significantly detected (according to the definition in Section
\ref{nu_srcdet}) post-deblending, in at least one energy
band. 
Table \ref{stats_table} provides source detection statistics, broken
down for the different combinations of energy bands, and the number of
sources with spectroscopic redshift measurements.
\begin{table}
\centering
\caption{Source statistics for the \nustar serendipitous survey}
\begin{tabularx}{0.47\textwidth}{Xccc} \hline\hline
  \noalign{\smallskip}
Band  & $N$ & $N_{z}$ \\
(1)  & (2) & (3) \\
\noalign{\smallskip} \hline \noalign{\smallskip}
Any band & $497$ & $276$ \\ 
\noalign{\smallskip}
F + S + H & $104$ ($21\%$) & $77$ \\ 
F + S & $116$ ($23\%$) & $82$ \\ 
F + H & $35$ ($7\%$) & $21$ \\ 
S + H & $0$ ($0\%$) & $0$ \\ 
F & $165$ ($33\%$) & $77$ \\ 
S & $53$ ($11\%$) & $16$ \\ 
H & $24$ ($5\%$) & $3$ \\ 
\noalign{\smallskip} \hline \noalign{\smallskip}
\end{tabularx}
\begin{minipage}[c]{0.47\textwidth}
\footnotesize
\textbf{Notes.} (1): F, S, and H refer to sources detected
in the full (\fullband), soft (\softband), and hard
(\hardband) energy bands. E.g.: ``F~+~H'' refers to sources detected
in the full and hard bands only, but not in the soft band; and ``S'' refers to sources detected
exclusively in the soft band. (2): The number of sources
detected post-deblending, 
for a given band or set of bands. 
(3): The number of sources with spectroscopic redshift
measurements.
\end{minipage}
\label{stats_table}
\end{table}

In addition to the primary source detection approach (Section \ref{nu_srcdet}), which has been
used to generate the above main catalog, in Section A.6 we provide a
``secondary catalog'' containing sources that do not appear in the
main catalog (for reasons described therein). However, all results in
this work are limited to the main catalog only (the secondary
catalog is thus briefer in content).

\section{The Multiwavelength Data}
\label{multiwav}

 The positional accuracy of
 \nustar ranges from $\approx 8''$ to $\approx20''$, depending on the source
 brightness (the latter is demonstrated in the following section). For matching to unique counterparts at other
  wavelengths (e.g., optical and IR), a higher astrometric accuracy is required, especially
  toward the Galactic plane where the sky density of sources increases
  dramatically.
  We therefore first match to soft X-ray (\chandra, \xmm, and \swiftxrt)
  counterparts, which have significantly higher positional accuracy
  (Section \ref{cparts_xray}), before proceeding to identify optical and IR
  counterparts (Section \ref{cparts_optIR}), and undertaking optical
  spectroscopy (Section \ref{spec_opt})

\subsection{Soft X-ray counterparts}
\label{cparts_xray}

The \nustar serendipitous survey is mostly composed of fields
  containing well-known extragalactic and Galactic targets. This means
  that the large majority of the serendipitous survey sources also have lower-energy (or ``soft'') X-ray coverage
  from \chandra, \xmm, or \swiftxrt, thanks to the relatively large FoVs of these
  observatories. In addition, short-exposure coordinated \swiftxrt
  observations have been obtained for the majority of the \nustar observations.
Overall, $81\%$ ($401/497$) of the \nustar detections have coverage with \chandra
or \xmm,  
and this increases to $99\%$ ($493/497$) if \swiftxrt coverage is
included.
Only $1\%$ ($4/497$) lack any form of coverage from all of these three soft
X-ray observatories.

We crossmatch with the third \xmm serendipitous source catalog (3XMM;
\citealt{Watson09,Rosen16}) and the \chandra Source Catalog (CSC;
\citealt{Evans10}) using a $30\arcsec$ search radius from each \nustar
source position; the errors in the source matching are dominated by the \nustar
positional uncertainty (as quantified below). Based on the sky
density of X-ray sources with $f_{\rm 2-10keV}\gtrsim 10^{-14}$~\fluxunit
found by \citeauthor{Mateos08} (\citeyear{Mateos08}; for
$|b|>20$\degrees sources in the \xmm serendipitous survey), we
estimate that the $30\arcsec$ radius matching results in a typical spurious match
fraction of $\approx 7\%$ for this flux level and latitude range. 
Overall, we find multiple
matches for $\approx 20\%$ 
of the cases where there is at least one match. In these multiple match cases we assume that the
3XMM or CSC source with the brightest hard-band ($4.5$--$12$~keV and
$2$--$7$~keV, respectively) flux is the correct
counterpart.\footnote{For clarity, throughout the paper we refer to
  the $3$--$8$~keV band as
  the ``soft'' band, since it represents the lower (i.e.,
  ``softer'') end of the energy range for which \nustar is
  sensitive. However, energies of $3$--$8$~keV (and other similar bands;
  e.g., $2$--$7$~keV) are commonly referred to as ``hard'' in the
  context of lower energy X-ray missions such as \chandra and \xmm,
  for which these energies are at the upper end of the telescope sensitivity.}
We provide the positions and \chandra/\xmm $3$--$8$~keV fluxes (\flow)
for these counterparts in the source catalog (see Section A.1).
To assess possible flux contributions from other nearby
  \chandra/\xmm sources, we also
  determine the total combined flux of all 3XMM or CSC sources contained within the
$30\arcsec$ search aperture (\flowTotal). For the $284$ sources which are successfully matched
to 3XMM or CSC, 
$29$ ($10\%$) have \flowTotal values
which exceed \flow by a factor of $> 1.2$, 
and there are only four cases where this factor is $>2$. In other words,
there are few cases where additional nearby X-ray sources
appear to be contributing substantially to the \nustar detected emission.

In addition to the aforementioned catalog matching, we identify
archival \chandra, \xmm and \swiftxrt data that overlap in sky
coverage with the \nustar data. Using these archival data sets, we
manually identify and measure positions for soft X-ray counterparts
which are not already included in the 3XMM and CSC catalogs.
For \chandra we process the archival data using {\sc chandra\_repro},\footnote{http://cxc.harvard.edu/ciao/ahelp/chandra\_repro.html}
for \xmm we analyze data products from the Pipeline Processing System,\footnote{http://www.cosmos.esa.int/web/xmm-newton/pipeline}
and for \swiftxrt we use screened event files (as provided on
HEASARC).\footnote{http://heasarc.gsfc.nasa.gov}
We perform source detection on the archival soft X-ray
($\approx 0.5$--$8$~keV) counts images using the CIAO source detection
algorithm \wavdetect \citep{Freeman02}, which identifies $111$ new soft X-ray
counterparts. 
 $88\%$ of these have high detection significances (false
probabilities of $<10^{-6}$), and $12\%$ have moderate detection
significances (false probabilities of $10^{-6}$--$10^{-4}$). 

In total, soft X-ray counterparts are successfully identified for
$79\%$ ($395/497$) 
of the \nustar detections: $284$ are existing
counterparts in the 3XMM and CSC catalogs, 
with $269$ and $82$ 
counterparts from the individual 3XMM and CSC catalogs, respectively.
Of the remaining $213$ \nustar detections that lack 3XMM and CSC 
counterparts, 
we have manually identified soft X-ray
counterparts in archival data (using \wavdetect
as described above) for $111$ sources, 
of which $27$, $60$, and $24$
are from \chandra, \swiftxrt, and \xmm data, respectively. In addition,
we manually determine new \chandra positions for $12$ sources which
appear in 3XMM and not CSC, but have \chandra coverage, 
thus improving the X-ray position constraints for these sources.
For four of these sources, 
the newly measured \chandra positions were obtained
through our own \chandra observing program aimed at
localizing the X-ray emission for Galactic-candidate \nustar serendipitous sources (Tomsick
et al., in prep.).
For the soft X-ray counterparts which are detected with multiple soft
X-ray observatories, we adopt the position with the highest
accuracy: for $31\%$ ($121/395$) the adopted
position is from \chandra, which has the best positional
accuracy; 
for $54\%$ ($214/395$) the adopted position is from \xmm; 
and for $15\%$ ($60/395$) the adopted counterpart is from \swiftxrt. 

Overall, $21\%$ ($102/497$) 
of the \nustar detections lack soft X-ray
counterparts. 
This can largely be explained as a
result of insufficient-depth soft X-ray coverage, or zero coverage. However, for the
sources with sufficient-depth soft X-ray coverage this may indicate either a spurious \nustar detection, a transient
detection, or the detection of an unidentified contaminating
feature such as stray light (e.g., \citealt{Mori15}). We estimate that
there are $34$ (out of $102$) such sources, 
that lack a soft X-ray counterpart but have sufficiently deep soft
X-ray data (from \chandra or \xmm) that 
we would expect a detection (given the \nustar source flux in the
overlapping $3$--$8$~keV band). 
We retain these
sources in the sample, but note that their inclusion (or exclusion) has a
negligible impact on the results presented herein which are primarily
based on the broader subsample with successful counterpart
identifications and spectroscopic redshift measurements.

\begin{figure}
\centering
\begin{minipage}{0.47\textwidth}
\includegraphics[width=\textwidth]{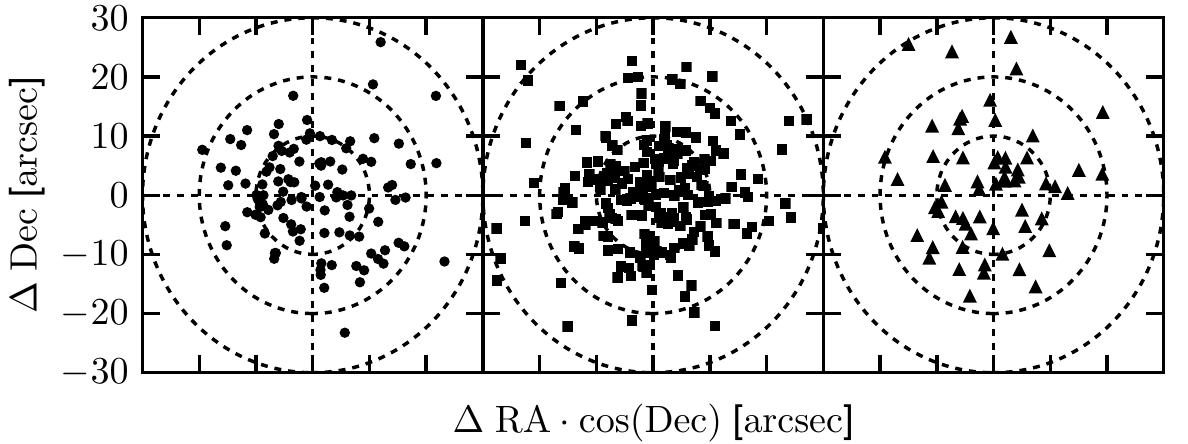}
\end{minipage}
\begin{minipage}{0.47\textwidth}
\includegraphics[width=\textwidth]{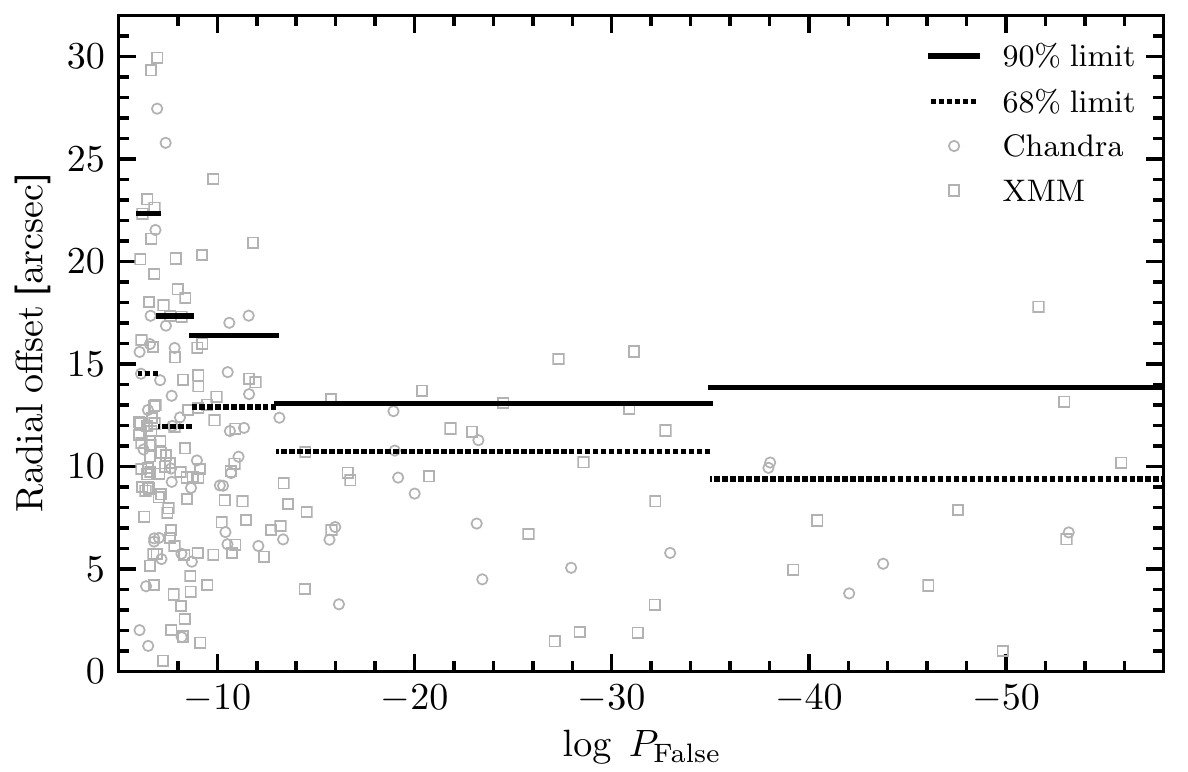}
\end{minipage}
\caption{Upper panel: astrometric offsets between the \nustar coordinates and
  lower-energy X-ray counterpart coordinates as identified with
  \chandra (circles, left panel), \xmm (squares, center panel), and
  \swiftxrt (triangles, right panel). Lower panel: the angular
  separation between \nustar and \chandra/\xmm coordinates, as a
  function of \fprob (source detection significance increases
  towards the right). The solid and dotted lines show the limits in
  angular offset enclosing $90\%$ and $68\%$ of sources, for bins in
  \fprob. Each bin contains $\approx 40$--$50$ sources, except the
  rightmost bin which contains $23$ sources (and extends beyond the
  x-axis upper limit, including all sources with $P_{\rm
    False}<-35$). This figure illustrates the positional accuracy of
  \nustar as a function of source significance.}
\label{delRAdec_nusoft}
\end{figure}

The upper panel of Figure \ref{delRAdec_nusoft} shows the
distribution of positional offsets (in R.A.\ and decl.) for the \nustar sources relative to their
soft X-ray (\chandra, \xmm, and \swiftxrt) counterparts. We find no
evidence for systematic differences in the astrometry between
observatories, since the mean positional offsets are all consistent with
zero: the mean values of $ \Delta \mathrm{RA}\cdot
\cos(\mathrm{Dec})$ and $\Delta \mathrm{Dec}$ are $0.41\pm 1.45''$ and
$0.18 \pm 1.28''$ for \chandra, $-0.19 \pm 1.11''$ and
$0.50 \pm 0.95''$ for \xmm, and $-0.34 \pm 1.97''$ and
$1.70 \pm 2.09''$ for \swiftxrt.

The lower panel of Figure \ref{delRAdec_nusoft} shows the radial
separation (in arcseconds) of \nustar sources from their
well-localized soft X-ray
counterparts (for those sources with \chandra or \xmm counterparts) as
a function of \fprob, thus illustrating the positional accuracy of
  \nustar as a function of source-detection significance. To
  reliably assess the positional accuracy of \nustar, we limit this
  particular analysis to sources with unique matches at soft X-ray
  energies, and thus with higher likelihoods of being correctly
  matched. Assuming zero uncertainty on the \chandra and \xmm positions, the
$90\%$ confidence limit on the 
 \nustar positional uncertainty is $22\arcsec$ for the
 least-significant detections, and $14\arcsec$ for the most-significant detections. If we instead only consider the
  \chandra positions, which are in general more tightly constrained 
  (positional accuracy $\lesssim 1\arcsec$; e.g., see Section
  \ref{cparts_optIR}), then the inferred $90\%$ positional accuracy of
  \nustar improves to $20\arcsec$ and $12\arcsec$ for the
  least-significant and most-significant sources, respectively.

\begin{figure}
\centering
\includegraphics[width=0.47\textwidth]{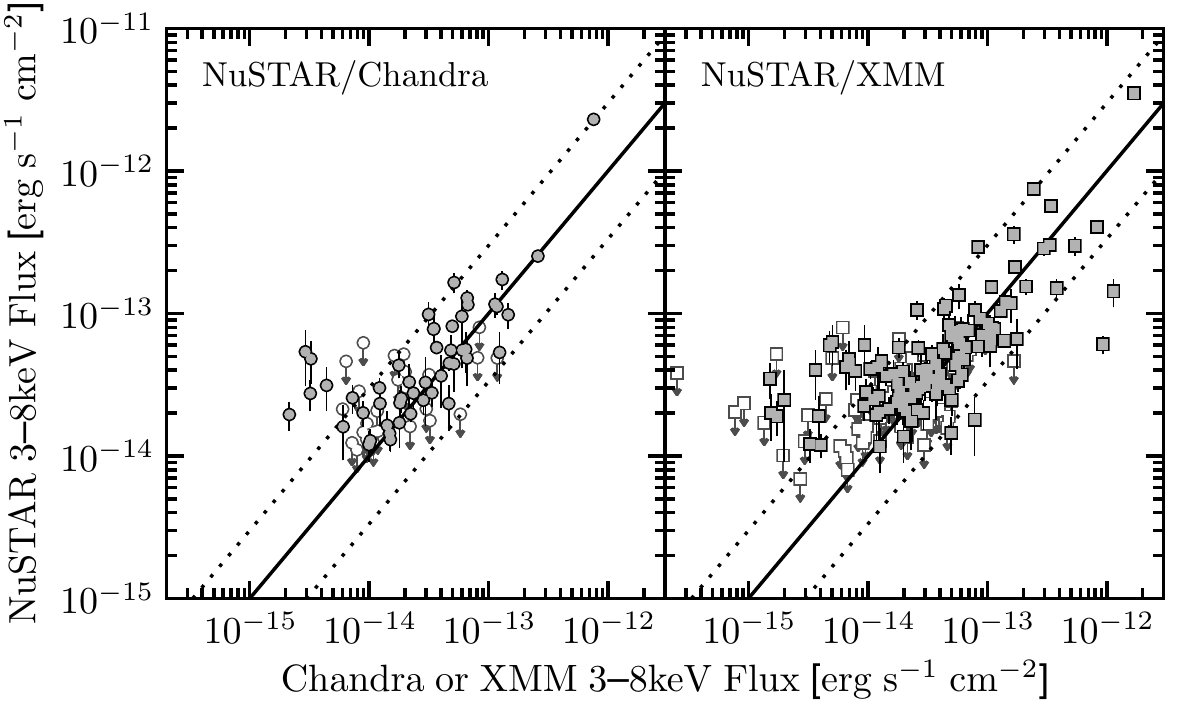}
\caption{Comparison of the \nustar and soft X-ray mission flux (\flow;
  from \chandra or \xmm), at $3$--$8$~keV, for those serendipitous
  survey sources with matched CSC or 3XMM counterparts. The empty symbols show
  upper limits. The 1:1 relation is shown by a solid line, and the dotted
  lines show values a factor of three from this
  relation.}
\label{fnu_fsoft}
\end{figure}

Figure \ref{fnu_fsoft} compares the $3$--$8$~keV fluxes, as measured
by \nustar, with those measured by \chandra and \xmm for the sources
with 3XMM or CSC counterparts. The small flux corrections from
the 3XMM and
CSC energy bands ($4.5$--$12$~keV and
$2$--$7$~keV, respectively) to the $3$--$8$~keV energy band are
described in Section A.1. The majority of sources ($92\%$ and $89\%$ for \chandra and
\xmm, respectively) are consistent with lying within a factor of three
of the 1:1 relation, given the uncertainties, and thus show reasonable
agreement between observatories. Given that the \nustar and the lower-energy
X-ray observations are not contemporaneous, intrinsic source variability is expected
to contribute to the observed scatter. A number of sources at the
lowest X-ray fluxes lie above the relation, due to Eddington
bias. This effect has been observed in the \nustar-ECDFS and \nustar-COSMOS
surveys (M15; C15), and is predicted from simulations (C15). 

Two relatively high-flux 3XMM sources lie significantly below the 
1:1 relation, suggesting that they have
experienced a large decrease in flux (by a factor of $\gtrsim 5$). 
The first,
NuSTAR~J183452-0845.6, 
is a known Galactic magnetar for which the \nustar (2015) flux is lower than
the \xmm (2005 and 2011 combined) flux by a factor of $\approx
15$. This is broadly consistent with the
known properties of the source, which varies in X-ray flux by
orders of magnitude over multi-year
timescales \citep[e.g.,][]{Younes12}. The second
outlying source is extragalactic in origin:
NuSTAR~J133311-3406.8 (hereafter J1333;
$z=0.091$; $L_{\rm 10-40keV}=8\times 10^{42}$~\ergpersec). 
Our NTT (ESO New Technology Telescope) spectrum for
J1333 reveals a NLAGN, with an apparently
asymmetric, blue wing component to the H$\alpha$$+$\nii complex, and
our NTT $R$-band imaging shows a well-resolved, undisturbed
host galaxy. 
Modeling the \xmm ($14$~ks exposure; $\approx 1100$ EPIC source counts
at $0.5$--$10$~keV) and \nustar ($17$~ks exposure; $\approx 75$ source
counts at $3$--$24$~keV) spectra, the former
of which precedes the latter by $\approx9$~years, the 
X-ray spectral flux has decreased by a factor of $\approx 5$ in the
energy band where the observatories overlap in sensitivity
($3$--$10$~keV). 
While this is an outlier, it is not
unexpected to observe one AGN with this level of variability,
given the range of AGN variability observed on decade timescales in
deep $<10$~keV X-ray surveys such as CDFS (e.g., Yang et al.,
submitted). 
The variability could be due to a change in the intrinsic
  X-ray continuum, or the line-of-sight column density (e.g., \citealt{Risaliti05,Markowitz14}).
However, it is not possible to place
informative constraints on spectral shape variability of J1333, since the
\nustar spectral shape is poorly constrained at $3$--$10$~keV
($\Gamma_{\rm eff}=1.2^{+1.3}_{-1.7}$). Deeper,
simultaneous broad-band X-ray coverage would be required to determine
whether a variation in spectral shape accompanies the relatively large
variation in AGN flux. There is
\swiftxrt coverage contemporaneous with the 2014 \nustar data, but
J1333 is undetected by \swiftxrt. The \swiftxrt flux upper limit is consistent
with the \nustar flux, 
and is a factor of $\approx 4.2$ lower than the \xmm flux 
(and thus in agreement with
a factor of $\approx 5$ variation in the X-ray flux).
This source represents the maximum variation in AGN
flux identified in the survey.

\subsection{IR and optical counterparts}
\label{cparts_optIR}

Here we describe the procedure for matching between the $395$ (out of $497$) \nustar
sources with soft X-ray
counterparts (identified in Section \ref{cparts_xray}), and counterparts at
IR and optical wavelengths. The results from this matching are
summarized in Table \ref{optIR_table} (for the sources with
Galactic latitudes of $|b|>10$\degrees). 
We adopt matching radii which are a
compromise between maximizing completeness and minimizing
spurious matches, and take into account the
additional uncertainty (at the level of $1''$) between the X-ray and
the optical/IR positions. 
For \chandra positions we use a
matching radius of $2.5\arcsec$, which is well-motivated based on the known behaviour of the
positional uncertainty as a function of off-axis angle (the majority of \nustar serendipitous sources
lie significantly off-axis) and source counts (e.g.,
\citealt{Alexander03,Lehmer05,Xue11}). 
For \swiftxrt positions we use
a matching radius of $6\arcsec$, justified by the typical
positional uncertainty (statistical plus systematic) which is at the
level of $\approx 5.5\arcsec$ ($90\%$ confidence level; e.g., \citealt{Moretti06,Evans14}).
For \xmm positions we use a matching
radius of $5 \arcsec$, which is motivated by the typical
positional uncertainties of X-ray sources in the \xmm serendipitous
survey (e.g., $\approx 4\arcsec$ at the $90\%$ confidence level
for \xmm bright serendipitous survey sources; \citealt{Caccianiga08}). 

\renewcommand*{\arraystretch}{1.1}
\begin{table}
\centering
\caption{Summary of the optical and IR counterpart matching statistics
  and photometric magnitudes}
\begin{tabular}{lcccccc} \hline\hline \noalign{\smallskip}
Catalog / Band & $N$ & Fraction & $m_{\rm max}$ & $m_{\rm min}$ &
         $\bar{m}$ & $\left< m \right>$ \\
(1) & (2) & (3) & (4) & (5) & (6) & (7) \\
\noalign{\smallskip} \hline \noalign{\smallskip}
Total optical + IR & $290$ & $87.9\%$ & $\cdots$ & $\cdots$ & $\cdots$ & $\cdots$ \\
{\it WISE} (all) & $252$ & $76.4\%$ & $\cdots$ & $\cdots$ & $\cdots$ & $\cdots$ \\
{\it WISE} / {\it W1} & $249$ & $75.5\%$ & $18.4$ & $7.8$ & $15.3$ & $15.5$ \\
{\it WISE} / {\it W2} & $248$ & $75.2\%$ & $17.1$ & $7.9$ & $14.4$ & $14.6$ \\
{\it WISE} / {\it W3} & $194$ & $58.8\%$ & $13.3$ & $4.5$ & $11.2$ & $11.4$ \\
{\it WISE} / {\it W4} & $131$ & $39.7\%$ & $9.9$ & $1.8$ & $8.1$ & $8.4$ \\
Optical (all) & $249$ & $75.5\%$ & $\cdots$ & $\cdots$ & $\cdots$ & $\cdots$ \\
SDSS / $r$ & $121$ & $36.7\%$ & $24.5$ & $11.7$ & $19.6$ & $19.9$ \\
USNOB1 / $R$ & $198$ & $60.0\%$ & $20.9$ & $10.5$ & $18.5$ & $19.1$ \\
\noalign{\smallskip} \hline \noalign{\smallskip}
\end{tabular}
\begin{minipage}[c]{0.44\textwidth}
\footnotesize
\textbf{Notes.} Summary of the optical and IR counterpart
matching for the $330$ \nustar serendipitous survey sources 
with high Galactic latitudes ($|b|>10$\degrees) and soft X-ray 
telescope (\chandra, \swiftxrt, or \xmm) counterpart positions (see
Section \ref{cparts_optIR}). 
(1): The catalog and photometric band (where magnitude statistics are
provided). (2): The number of the \nustar sources successfully matched to
a counterpart in a given catalog. For the \wise all-sky survey
catalog, this is broken down for the four photometric \wise
bands. (3): The fraction of the \nustar sources which are
matched. (4): The maximum (i.e., faintest) 
magnitude for the counterparts in a given catalog and photometric
band. (5): The minimum (i.e., brightest) magnitude. (6): The mean
magnitude. (7): The median magnitude.
\end{minipage}
\label{optIR_table}
\end{table}

\begin{figure}
\centering
\includegraphics[width=0.47\textwidth]{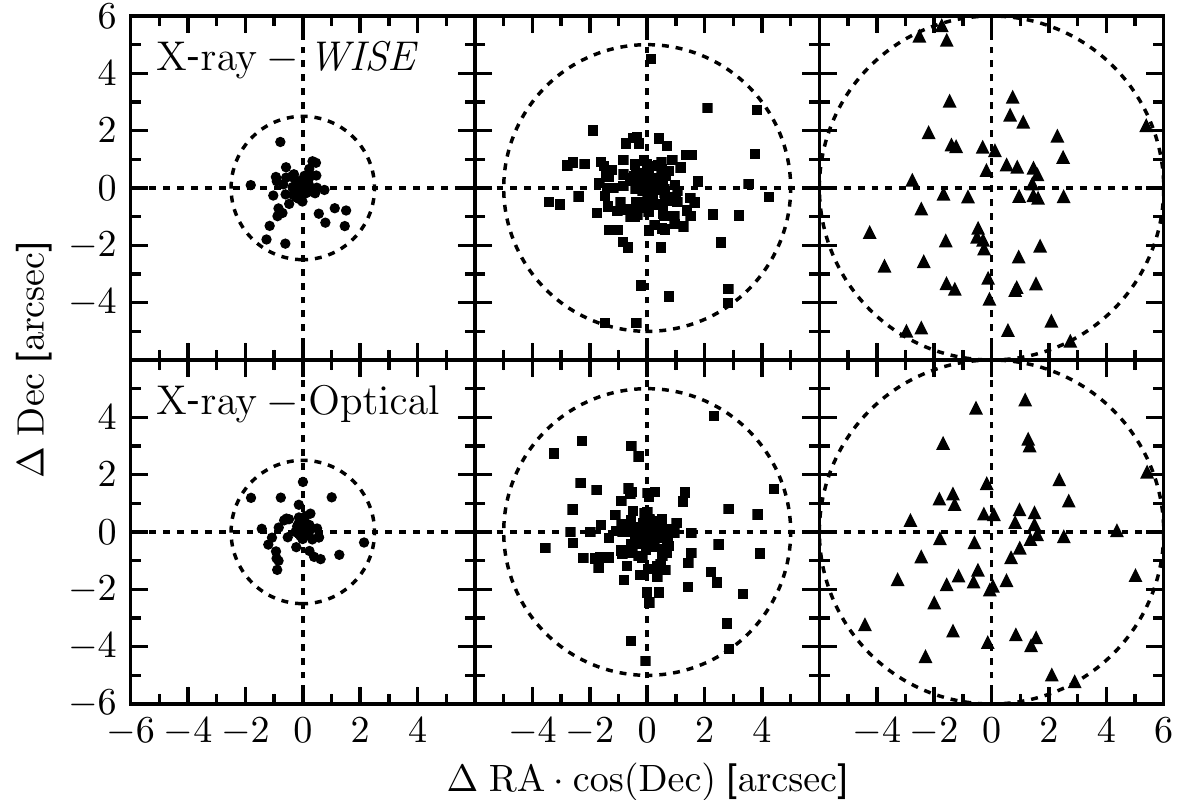}
\caption{Astrometric offsets between the soft X-ray
  counterpart coordinates and the \wise (top row) and optical (bottom
  row) coordinates. The soft X-ray counterparts are from \chandra
  (left column), \xmm (middle column) and
  \swiftxrt (right column). The dashed circles correspond to the search radii for
  each telescope ($2.5\arcsec$, $5\arcsec$ and $6\arcsec$ for \chandra, \xmm and \swiftxrt, respectively).}
\label{delRAdec_optIR}
\end{figure}

To identify IR counterparts, we match to the \wise all-sky
survey catalog \citep{Wright10}. Of the $395$ sources with soft X-ray
counterparts, 
$274$ ($69\%$) have \wise matches. 
In $100\%$ of these cases there is a single unique \wise match
(detected in at least one \wise band).
To identify optical counterparts, we match to the SDSS DR7 catalog
\citep{York00} and the USNOB1 catalog (\citealt{Monet03}). If both
contain matches, we adopt
optical source properties from the former catalog. Of the
$395$ sources with soft X-ray counterparts, $252$ ($64\%$) have a match in at
least one of these optical catalogs: 
$121$ have an SDSS match 
and $131$ without SDSS matches have a USNOB1
match. 
In $77\%$ ($193/252$) of cases there is a single optical
match. 
In the case of multiple matches within the search radius we
adopt the closest source. 
Figure \ref{delRAdec_optIR} shows the
distribution of astrometric offsets between the 
soft X-ray counterparts and the \wise and optical (SDSS and USNOB1)
counterparts.
For Galactic latitudes of $|b|>10$\degrees,
which we focus on for the analysis of \nustar serendipitous survey source properties (Section \ref{Results}),
the spurious matching fractions are low ($\lesssim 10\%$; see Section A.3). 

\begin{figure*}
\centering
\includegraphics[width=1.0\textwidth]{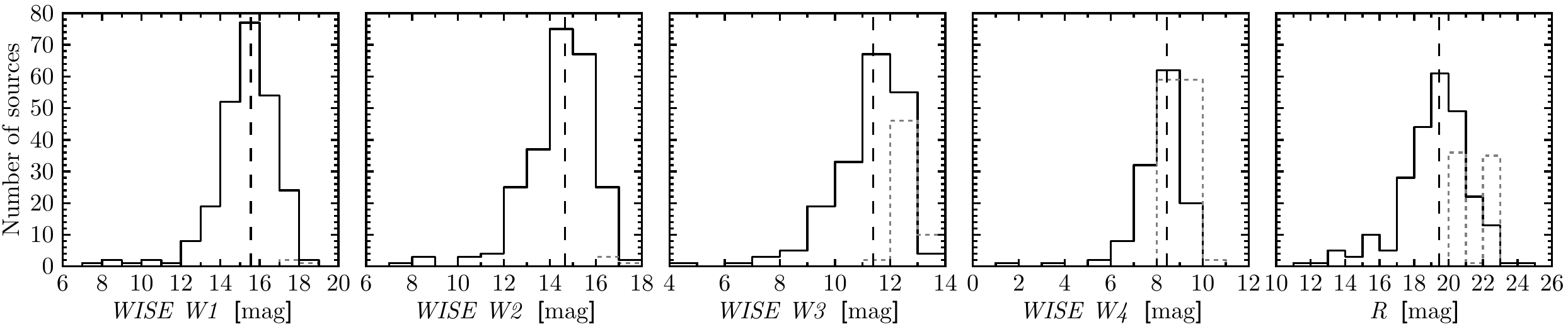}
\caption{Distributions of the MIR and optical magnitudes for the
  \nustar serendipitous survey sources with high
Galactic latitudes ($|b|>10$\degrees) and soft X-ray 
telescope (\chandra, \swiftxrt, or \xmm) counterparts. Left four
panels: magnitude distributions for the four photometric \wise
bands, for the sources with successful matches to the \wise all-sky
survey catalog. For each band, the solid line shows the magnitude
distribution for detected sources, the dashed line marks the median
magnitude of the detections, and the dotted line shows the
distribution of magnitude lower limits for sources undetected in that
band (but detected in other bands). Right panel: the $R$-band
magnitudes (primarily from matching to the SDSS and USNOB1 catalogs)
for all the sources.
}
\label{NoptIR} 
\end{figure*}

For the $143$ (out of $395$) soft X-ray counterparts without SDSS and
USNOB1 matches, 
we determine whether there are detections within the existing
optical coverage, which is primarily photographic plate coverage (obtained
through the DSS) but also includes dedicated $R$-band and multi-band
imaging from our own programs with the ESO-NTT (EFOSC2) and ESO-2.2m (GROND),
respectively. This identifies an additional $33$ optical counterparts. 
For the $110$ non-detections, 
we estimate $R$-band magnitude lower limits from the data (all cases
have coverage, at least from photographic plate observations). These optical non-detections
do not rule out
follow-up spectroscopy; for $21$ of them 
we have successfully performed optical spectroscopy, obtaining
classifications and redshifts, either by identifying an optical
counterpart in pre-imaging
or by positioning the spectroscopic slit on a \wise source within the
X-ray error circle.
In Figure \ref{NoptIR} we show histograms of the \wise and $R$-band
magnitudes for the \nustar sources with soft
X-ray counterparts.

For the $102$ (out of $497$) sources without soft X-ray counterparts, 
the X-ray positional error circle (from \nustar)
is comparatively large (see Section \ref{cparts_xray}), so unique
counterparts cannot be
identified with high confidence. To identify possible counterparts for
these sources, for the purposes of optical spectroscopic followup, we
consider the properties of nearby \wise sources. 
Matching to the \wise all-sky survey, we
identify AGN candidates within a $25''$ radius of the \nustar
position, using the following two criteria: a \wise color of
$W{\rm 1}$--$W{\rm 2}>0.8$ (and $W{\rm 2}<15$; Vega mag; \citealt{Stern12}) or a \wfour band detection.
The \wise \wone, \wtwo, \wthree, and \wfour bands are are centered at
$3.4$~\micron, $4.6$~\micron, $12$~\micron, and $22$~\micron, respectively.
We limit this matching to the $85$ (out of $102$) 
sources at Galactic latitudes above $|b|=10$\degrees.
Given the sky densities of \wise sources which satisfy these criteria, 
($\approx 46$~deg$^{-2}$ and $\approx 730$~deg$^{-2}$, respectively,
for $|b|>10$\degrees), 
the probabilities of chance
matches are $\approx 1\%$ and $\approx 11\%$,
respectively. 
Where multiple such \wise sources are identified, we prioritize those
which satisfy both criteria, then those which satisfy the former criterion.
For $24$ (out of $102$) of these sources 
there is a \wise AGN candidate within the \nustar error circle,
the position of which we match to optical
counterparts. 
The optical and IR counterparts identified in this manner (for \nustar sources without
soft X-ray counterparts) are primarily used for the purposes of undertaking
spectroscopic followup (Section \ref{spec_opt}), and we exclude them
from our analysis of the IR properties of the \nustar serendipitous
survey AGNs (Section \ref{results_ir}), to avoid biasing the results.
For the remaining $78$ (out of $102$) sources at $|b|<10$\degrees or without
matches to \wise AGN candidates, 
we use the available
$R$-band information to obtain magnitude constraints: in cases where there
is at least one optical source within the \nustar error circle, we
adopt the lowest (i.e., brightest) $R$-band magnitude as a lower limit; and
in cases with no optical source within the \nustar error circle, we
adopt the magnitude limit of the imaging data. 

For a large fraction of the sources discussed in this section, the spectroscopic
  followup (Section \ref{spec_opt}) shows evidence for an AGN, which
  provides additional strong support for correct counterpart
  identification (given the low sky density of AGNs). Furthermore, the 
  optical and IR photometric properties of the \nustar serendipitous survey
  counterparts are in agreement with AGNs (see Sections \ref{xray_opt} and
  \ref{wise_colors}).

\subsection{Optical Spectroscopy}
\label{spec_opt}

Optical identifications and source redshifts, obtained through
spectroscopy, are a prerequisite
to the measurement of intrinsic source properties such as luminosity
and the amount of obscuration. A small fraction ($\approx 11\%$; $57/497$) of the \nustar
serendipitous survey sources have pre-existing spectroscopic
coverage, 
primarily from the SDSS. 
However, the majority ($\approx 89\%$) of the
serendipitous survey sources do not have pre-existing 
spectroscopy. 
For that reason, we have undertaken a 
  campaign of dedicated spectroscopic
  followup in the optical--IR bands (Section \ref{ground}), obtaining
  spectroscopic identifications for a large fraction ($56\%$) of the total
  sample. 
  For the high Galactic latitude ($|b|>10$\degrees) samples selected
  in individual bands, this has resulted in a spectroscopic completeness of $\approx 70\%$.
  The analysis of and 
  classifications obtained from these new spectroscopic data, and those from
  pre-existing spectroscopy, are described in Section \ref{opt_analysis}.

\subsubsection{Dedicated followup campaign}
\label{ground}

Since \nustar performs science pointings across the whole sky, a
successful ground-based followup campaign requires the use
  of observatories at a range of geographic latitudes, and preferably
  across a range of dates throughout the sidereal year. This has been
  achieved through observing programmes with, primarily, the following telescopes over a multi-year period:
  the Hale Telescope at Palomar Observatory ($5.1$~m;
  Decl.~$\gtrsim-21$\degrees; PIs F.~A.~Harrison and D.~Stern); Keck I and
  II at the W.~M.~Keck Observatory ($10$~m;
  $-35$\degrees$\lesssim$~Decl.~$\lesssim 75$\degrees; PIs F.~A.~Harrison and
  D.~Stern); the New Technology Telescope
  (NTT) at La Silla Observatory ($3.6$~m; Decl.~$\lesssim 25$\degrees; PI
  G.~B.~Lansbury);\footnote{Program IDs: 093.B-0881, 094.B-0891, 095.B-0951, and 096.B-0947.} 
the Magellan I (Baade) and Magellan II (Clay) Telescopes at Las
  Campanas Observatory ($6.5$~m; Decl.~$\lesssim 25$\degrees; PIs E.~Treister
  and F.~E.~Bauer);\footnote{Program IDs: CN2013B-86, CN2014B-113,
    CN2015A-87, CN2016A-93.} and the Gemini-South observatory
  ($8.1$~m; Decl.~$\lesssim 25$\degrees; PI
  E.~Treister).\footnote{Program ID: GS-2016A-Q-45.}
Table \ref{runsTable} provides a list of the observing runs
undertaken. 
In each case we provide the observing run starting
  date (UT), number of nights, telescope, instrument, and total number
  of spectroscopic redshifts obtained for \nustar serendipitous survey
  sources. 
\begin{table}
\centering
\caption{Chronological list of ground-based observing runs for spectroscopic
  followup of the \nustar serendipitous survey}
\begin{tabular}{llllc} \hline\hline \noalign{\smallskip}
Run ID & UT start date & Telescope & Instrument & Spectra \\
(1) & (2) & (3) & (4) & (5) \\
\noalign{\smallskip} \hline \noalign{\smallskip}
1 & 2012 Oct 10 & Palomar & DBSP & $1$ \\
2 & 2012 Oct 13 & Keck & DEIMOS & $1$ \\
3 & 2012 Nov 09 & Keck & LRIS & $1$ \\
4 & 2012 Nov 20 & Palomar & DBSP & $2$ \\
5 & 2012 Dec 12 & Gemini-South & GMOS & $1$ \\
6 & 2013 Jan 10 & Keck & LRIS & $1$ \\
7 & 2013 Feb 12 & Palomar & DBSP & $2$ \\
8 & 2013 Mar 11 & Palomar & DBSP & $6$ \\
9 & 2013 Jul 07 & Palomar & DBSP & $2$ \\
10 & 2013 Oct 03 & Keck & LRIS & $9$ \\
11 & 2013 Dec 05 & Magellan & MagE & $2$ \\
12 & 2013 Dec 10 & Keck & DEIMOS & $6$ \\
13 & 2014 Feb 22 & Palomar & DBSP & $4$ \\
14 & 2014 Apr 22 & Palomar & DBSP & $6$ \\
15 & 2014 Jun 25 & Keck & LRIS & $12$ \\
16 & 2014 Jun 30 & NTT & EFOSC2 & $8$ \\
17 & 2014 Jul 21 & Palomar & DBSP & $3$ \\
18 & 2014 Sep 25 & Magellan & MagE & $5$ \\
19 & 2014 Oct 20 & Keck & LRIS & $4$ \\
20 & 2014 Dec 23 & Palomar & DBSP & $4$ \\
21 & 2015 Feb 17 & Palomar & DBSP & $5$ \\
22 & 2015 Mar 14 & NTT & EFOSC2 & $17$ \\
23 & 2015 Mar 18 & Magellan & IMACS & $6$ \\
24 & 2015 May 19 & NTT & EFOSC2 & $14$ \\
25 & 2015 Jun 09 & Palomar & DBSP & $1$ \\
26 & 2015 Jun 15 & Palomar & DBSP & $1$ \\
27 & 2015 Jul 17 & Keck & LRIS & $3$ \\
28 & 2015 Jul 21 & Palomar & DBSP & $4$ \\
29 & 2015 Aug 09 & Palomar & DBSP & $6$ \\
30 & 2015 Aug 12 & Keck & LRIS & $6$ \\
31 & 2015 Dec 04 & Keck & LRIS & $28$ \\
32 & 2015 Dec 06 & NTT & EFOSC2 & $25$ \\
33 & 2016 Jan 11 & Palomar & DBSP & $8$ \\
34 & 2016 Feb 05 & Palomar & DBSP & $2$ \\
35 & 2016 Feb 08 & Magellan & MagE & $6$ \\
36 & 2016 Feb 13 & Keck & LRIS & $17$ \\
37 & 2016 Jul 05 & Keck & LRIS & $10$ \\
38 & 2016 Jul 10 & Palomar & DBSP & $6$ \\
\noalign{\smallskip} \hline \noalign{\smallskip}
\end{tabular}
\begin{minipage}[c]{0.45\textwidth}
\footnotesize
\textbf{Notes.} (1): ID assigned to each observing run. (2): Observing run start
date. (3) and (4): The telescope and
instrument used. (5): The number of spectra from a given observing run which
have been adopted, in this work, as the analysed optical spectrum for a
\nustar serendipitous survey source. These correspond
to the individual sources listed in Table \ref{specTable} of Section
A.2, and are primarily ($\approx 93\%$) 
sources with successful redshift measurements and spectroscopic
classifications.
These source numbers exclude the $35$ sources in the secondary catalog
for which we have obtained new spectroscopic identifications (see
Section A.6).
\end{minipage}
\label{runsTable}
\end{table}

The total number of
sources with spectroscopic redshift measurements and classifications is $276$. 
The large majority of
spectroscopic identifications in the northern hemisphere were obtained
using a combination of Palomar and Keck, with the former
being efficient for brighter targets and the latter for fainter
targets. These account for $51\%$ ($141/276$) of the spectroscopically identified
sample. 
Similarly, for the southern hemisphere the majority of
spectroscopic identifications were obtained using the ESO NTT
while complementary Magellan observations were used to
identify the fainter optical sources. These account for $28\%$
($76/276$) of the overall spectroscopically identified
sample. 

Conventional procedures were followed for
  spectroscopic data reduction, using IRAF
  routines. Spectrophotometric standard star observations, from the
  same night as the science observations, were used to flux calibrate the spectra. 

\subsubsection{Spectral Classification and Analysis}
\label{opt_analysis}

All flux-calibrated optical spectra from this work are provided
in Section A.2.
For our instrument setups, the typical observed-frame wavelength range covered
is $\lambda \approx 3500$--$9000$\AA. At lower redshifts, for example
$z=0.3$, 
this results in coverage for the following emission lines common to
AGNs and quasars:
\mgii~$\lambda2800$, \nev~$\lambda3346$\ and $\lambda3426$,
\oii~$\lambda3728$, \neiii~$\lambda3869$,
\hdelta~$\lambda4102$, \hgamma~$\lambda4340$,
\hbeta~$\lambda4861$, \oiii~$\lambda4959$\ and $\lambda5007$,
\oi~$\lambda6300$\ and $\lambda6364$, \nii~$\lambda6548$\ and
$\lambda6584$, \halpha~$\lambda6563$, and \sii~$\lambda6716$\ and $\lambda6731$.
At higher redshifts, for example $z=2$, 
the lines covered are: \lya~$\lambda1216$, \siiv~$\lambda1398$,
\civ~$\lambda1549$, \heii~$\lambda1640$, \ciii~$\lambda1909$,
\cii~$\lambda2326$, and \mgii~$\lambda2800$.

To measure spectroscopic redshifts, we identify emission and
absorption lines, and measure their observed-frame wavelengths
using Gaussian profile fitting. To determine the redshift solution, we crossmatch
the wavelength ratios of the identified lines with a
look-up table of wavelength ratios based on the emission and
absorption lines observed in AGN and galaxy spectra. The final, precise redshift
measurement is then obtained from the Gaussian profile fit to the
strongest line. For the large majority of cases there are multiple lines detected, and
there is only one valid redshift solution. The lines identified for
each individual \nustar source are tabulated in Section A.2. 
There are only five sources where the redshift is based on a single line
identification (marked with ``quality B'' flags in Section A.2).
For four of these, the single emission line detected is identified as
\mgii~$\lambda2800$. In all cases this is well justified: \mgii
is a dominant broad line in quasar spectra, and there is a relatively large
separation in wavelength between the next strong line bluewards of
\mgii (\ciii~$\lambda1909$) and that redwards of \mgii
(\hbeta~$\lambda4861$). This means that \mgii can be observed in isolation
for redshifts of $z\sim 0.8$ in cases where our wavelength coverage is
slightly narrower than usual, or if the other lines (e.g., \ciii and
\hbeta) are below the detection threshold. \mgii can also be clearly
identifiable in higher $S/N$ data due to the shape of the neighboring \feii pseudo-continuum.

We perform optical classifications visually, based on the spectral lines
observed. For the extragalactic sources with available optical spectra and with
identified lines ($253$ sources), 
emission lines are detected
for all but one source (where multiple absorption lines are
identified). Both permitted emission lines (e.g., the Balmer series and \mgii) and
forbidden (e.g., \oiii and \nii) emission lines
are identified for $183$ (out of $253$) sources. 
For these sources, if any permitted line is broader than the forbidden
lines we assign a BLAGN classification, otherwise we assign a NLAGN
classification. There are $58$ (out of $253$) sources where only
permitted (or semi-forbidden) emission lines are identified. 
For the majority of these ($56$ sources) the line profiles are visually broad, 
and we assign a BLAGN classification (these sources predominantly lie at
higher redshifts, with $51$ at $z\gtrsim1$, and have quasar-like continuum-dominated
spectra). For $24$ sources where there is a level of ambiguity as to whether the
permitted lines are broad or not, 
we append the optical classification
(i.e., ``NL'' or ``BL'' in Table \ref{specTable}) with a ``?'' symbol.
For the remaining $11$ sources (out of $253$) with only forbidden line
detections, 
and the single source with absorption line detections
only, 
we assign NLAGN classifications. 

In total we have spectroscopic classifications for $276$ of the \nustar
serendipitous survey sources, 
including the $253$ extragalactic sources mentioned above, an
additional BL~Lac type object, $16$
Galactic ($z=0$) objects, and six
additional (BLAGN and NLAGN) classifications from the
literature.
$222$ of these classifications were assigned using data from the
dedicated observing runs (Table \ref{runsTable}), 
and $54$ using existing data (primarily SDSS) or literature. 
Considering the total classified sample, the majority of the sources ($162$, or $58.7\%$) are
BLAGNs, 
$97$ ($35.1\%$) are NLAGNs, 
one ($0.4\%$) is a BL~Lac type object, 
and the remaining $16$ ($5.8\%$) are Galactic objects 
(e.g., cataclysmic variables and high mass X-ray binaries). Tomsick et
al.\ (in prep.) will present a detailed analysis of the
Galactic subsample. The current spectroscopic completeness (i.e., the
fraction of sources with successful spectroscopic identifications) is
$\approx 70\%$ for the overall serendipitous survey (for
the $|b|>10$\degrees individual band-selected samples), although the completeness is a function of X-ray
flux (see Section \ref{results_opt}). 

In Table \ref{specTable} (see Section A.2) we provide the following for
all \nustar serendipitous survey sources with optical spectra: the spectroscopic
redshift, the optical classification, the identified emission and
absorption lines, individual source notes, and the observing run ID
(linking to Table \ref{runsTable}).

\section{Results and Discussion}
\label{Results}

Here we describe the properties of the \nustar serendipitous survey sources, with
a focus on the high energy X-ray (Section \ref{results_xray}), optical (Section
\ref{results_opt}) and infrared (Section \ref{results_ir}) wavelength
regimes. We compare and contrast with other relevant samples, including: the blank-field \nustar
surveys in well-studied fields (COSMOS and ECDFS); non-survey samples of extreme objects targetted with \nustar; the \swiftbat all-sky
survey, one of the most sensitive high energy X-ray surveys to precede
\nustar; and lower energy ($<10$~keV; e.g., \chandra and \xmm) X-ray surveys. 

\subsection{X-ray properties}
\label{results_xray}

\subsubsection{Basic NuSTAR properties} 
\label{basic_properties}

Overall there are $497$ sources
  with significant detections (post-deblending) in at least one band.
Section \ref{nu_cat} details the source-detection statistics, broken
down by energy band.
 In the $8$--$24$~keV band, which is unique to
\nustar amongst focusing X-ray observatories, there are $163$
detections, i.e. $33\%$ of the sample. 
The \nustar-COSMOS and \nustar-ECDFS
surveys found fractions of \hardband detected sources which are consistent with
this:
$35\%$ ($32/91$ sources; C15) and $39\%$ ($19/49$ sources
post-deblending; M15), respectively. 

The net (cleaned, vignetting-corrected) exposure times per source
(\tnet; for the combined FPMA+B data) have a large
range, from $10$--$1500$~ks, with a median of $60$~ks. For the
$3$--$8$, $8$--$24$, and $3$--$24$~keV bands,
the lowest net source counts ($S_{\rm net}$) for sources with
detections in these bands are $12$, $15$, and $18$,
respectively. The highest $S_{\rm net}$ values are $9880$, $5853$, and $15693$,
respectively, and correspond to one individual source
NuSTAR~J043727--4711.5, a BLAGN at $z=0.051$. The
median $S_{\rm net}$ values are $56$, $62$, and $75$, respectively. 
The count rates range from $0.17$--$52$, $0.11$--$36$, and
$0.13$--$94$~ks$^{-1}$, respectively, and the median count rates are $0.77$,
$0.84$, and $1.1$~ks$^{-1}$, respectively.

\begin{figure}
\centering
\includegraphics[width=0.47\textwidth]{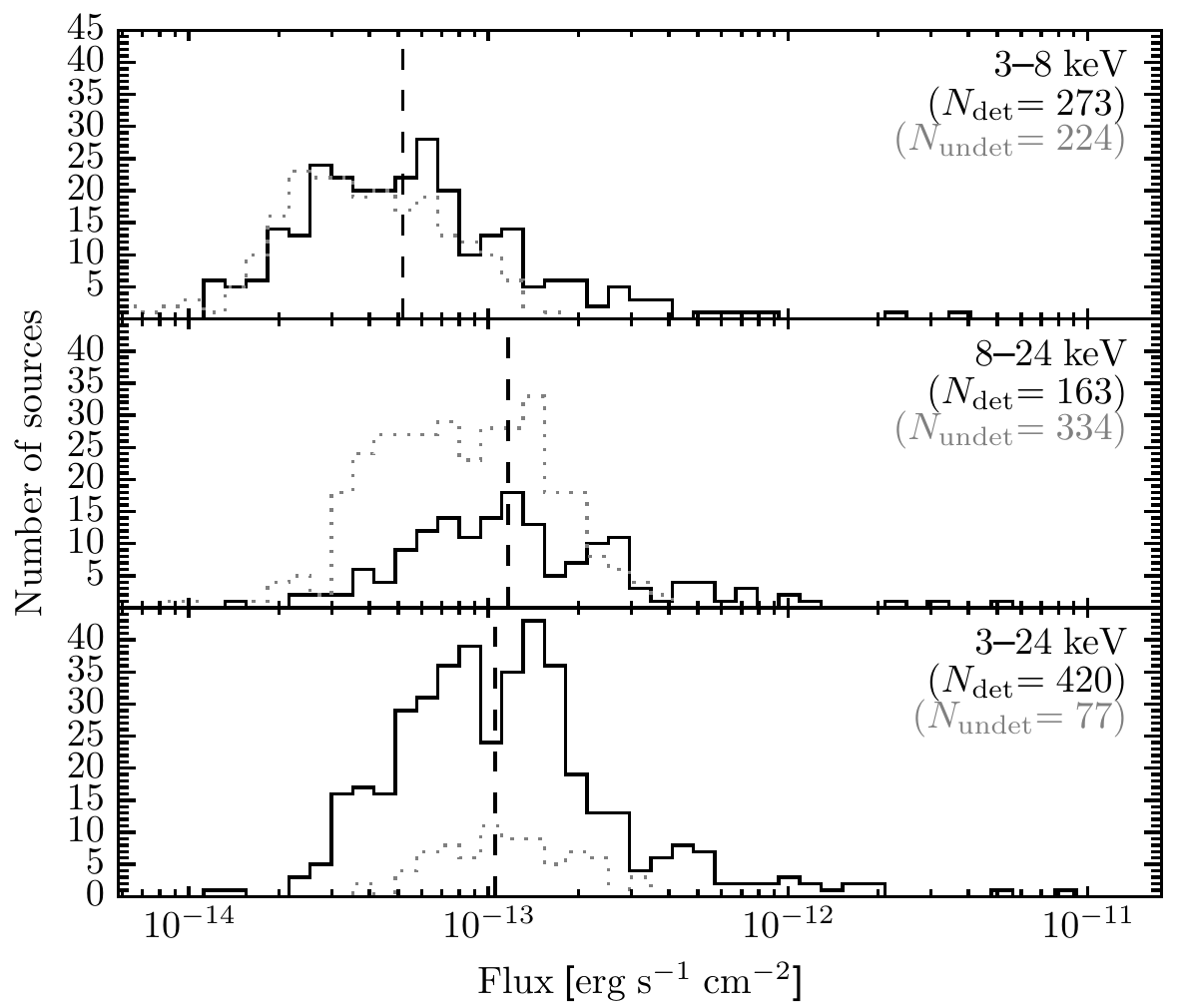}
\caption{Flux distributions in the soft, hard, and full energy bands
  (top, middle, and bottom panels, respectively) for the \nustar serendipitous
  survey sample. 
For each band, the solid line
  shows the flux distribution for sources independently detected in
  that band (the number of these sources, $N_{\rm det}$, is shown
  in black font), and the median flux of the detected sources is marked by
  a dashed line. For each band, the dotted line shows the
  distribution of flux upper limits for sources undetected in that
  band, but independently detected in at least one other band (the
  number of these sources, $N_{\rm undet}$, is shown
  in grey font).}
\label{Nf}
\end{figure}

Figure \ref{Nf} shows the distribution of fluxes for the full sample, for each
energy band. The distributions for detected and undetected
sources (for a given band) are shown separately. For sources which are
detected in the $3$--$8$, $8$--$24$, and $3$--$24$~keV bands,
the faintest fluxes measured are $1.17$, $1.53$, and
$1.22 \times 10^{-14}$~\fluxunit, respectively. The brightest
fluxes are $3.5$, $5.0$, and $8.8 \times 10^{-12}$~\fluxunit,
respectively, and correspond to one individual source
NuSTAR~J075800+3920.4, a BLAGN at $z=0.095$. The median
fluxes are $5.2$, $11.6$, and $10.5 \times 10^{-14}$~\fluxunit, respectively. The dynamic range of the serendipitous survey
exceeds the other \nustar extragalactic survey components. For
comparison, the blank-field ECDFS and COSMOS components span $3$--$24$~keV flux ranges
of $\approx (2$--$10)$ and $(5$--$50) \times 10^{-14}$~\fluxunit, respectively
(C15 and M15). The serendipitous survey pushes to fluxes (both flux
limits and median fluxes) $\sim$~two
orders of magnitude fainter than those achieved by previous-generation
hard X-ray observatories such as \swiftbat
\citep[e.g.,][]{Baumgartner13} and \integral \citep[e.g.,][]{Malizia12}. 

\subsubsection{Band ratios} 
\label{br}

\begin{figure}
\centering
\includegraphics[width=0.47\textwidth]{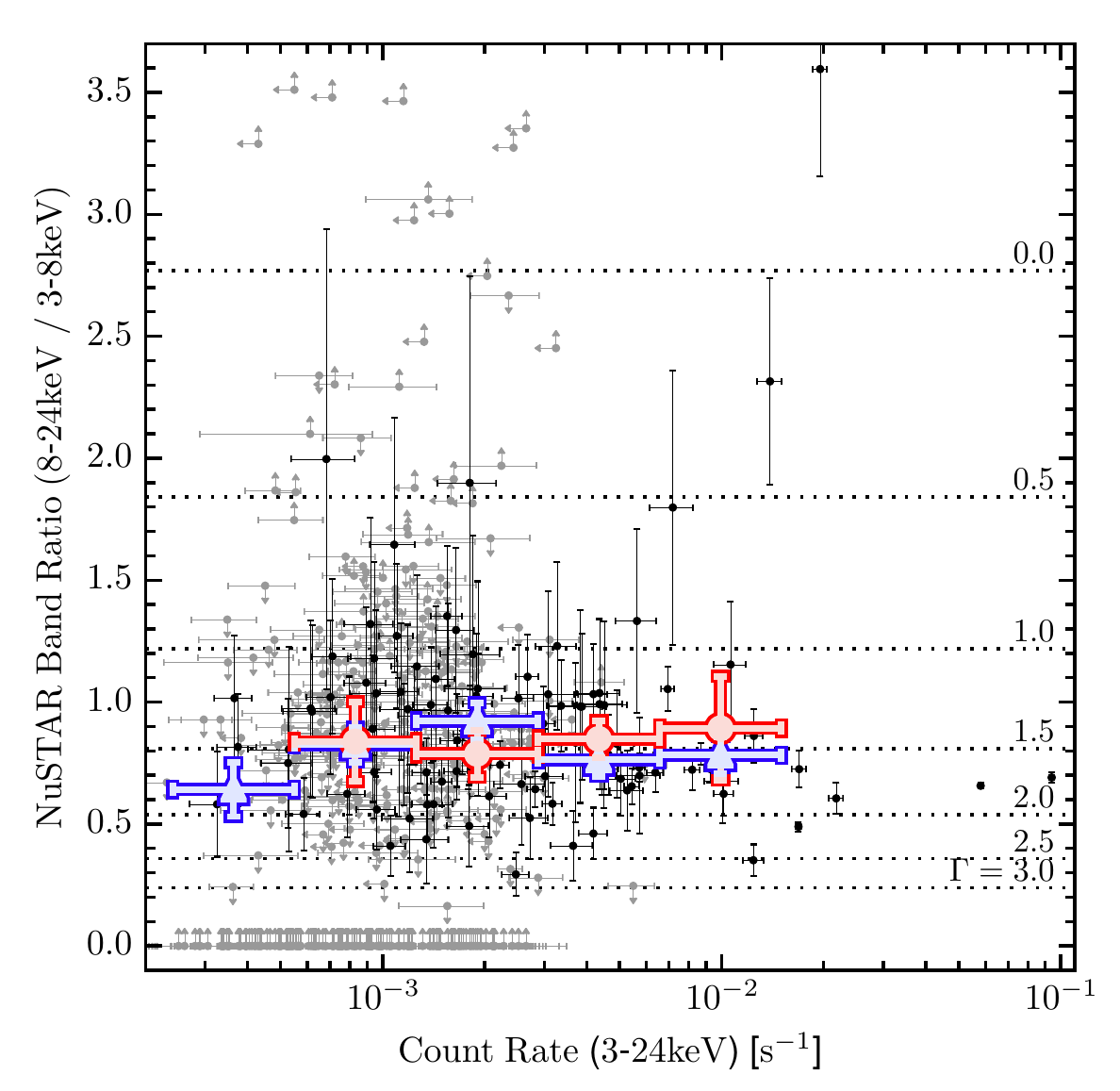}
\caption{The \nustar $8$--$24$ to $3$--$8$~keV band ratio (\brnu) versus
  full-band ($3$--$24$~keV) count rate for the full \nustar
  serendipitous survey sample. Constrained \brnu values are shown in
  black, and those with upper or lower limits are shown in gray. 
  The dotted horizontal lines indicate spectral slopes
  ($\Gamma_{\rm eff}$) to which the band ratios correspond. The overplotted
  red circles show numerical means (binning in full-band count
  rate), for a subset of extragalactic sources with $\log (P_{\rm
    False})<-14$. The overplotted 
  blue triangles show ``stacked'' means obtained from summing the
  net count-rates of all sources, including non-detections, and bootstrapping errors.}
\label{br_cr}
\end{figure}

Figure \ref{br_cr} shows the $8$--$24$ to $3$--$8$~keV band ratios (\brnu) for the full sample of
\nustar serendipitous survey sources, as a function of full-band
($3$--$24$~keV) count rate. 
In order to examine the results for extragalactic sources
only, we remove sources which are spectroscopically confirmed as
having $z=0$ (see Section \ref{spec_opt}) and exclude sources with
Galactic latitudes below $|b|=10$\degrees,
for which there is significant contamination to the
non-spectroscopically identified sample from Galactic sources. 
A large and statistically significant
variation in \brnu is observed across the
sample, with some sources exhibiting extreme spectral slopes
($\Gamma_{\rm eff}\approx3$ at the softest values; $\Gamma_{\rm
  eff}\approx0$ at the hardest values).

In Figure \ref{br_cr}, we overlay mean band ratios and
corresponding errors (in bins of full-band count rate, with an average
of $13$ sources per bin) for a subset of the extragalactic
serendipitous sample with $\log (P_{\rm False})<-14$ in
the full band. This cut in
source significance reduces the fraction of sources with upper or
lower limits in \brnu to only $7\%$, allowing numerical means
to be estimated. The results are consistent with a flat relation in the average band
ratio versus count rate, and a constant average effective photon index of
$\Gamma_{\rm eff}\approx 1.5$.  
This value is consistent with the average effective photon index found
from spectral analyses of sources detected in the 
dedicated \nustar surveys of the ECDFS, EGS and COSMOS fields
($\Gamma_{\rm eff}= 1.59\pm 0.14$; Del Moro et al.\ 2016, in prep).
 This hard average spectral slope suggests numerous obscured
 AGNs within the sample.
The mean band ratios disfavor an increase toward lower
count rates. This is in apparent disagreement with the recent results of
M15 for the \nustar-ECDFS survey, which show an increase towards
lower count rates, albeit for small source numbers with constrained band
ratios. Deep surveys at lower X-ray energies have previously found
an anticorrelation between band ratio and count rate for the
$0.5$--$8$~keV band 
\citep[e.g.,][]{DellaCeca99,Ueda99,Mushotzky00,Tozzi01,Alexander03},
interpreted as being driven by an increase in the number of absorbed
AGNs toward lower count rates. We find no evidence for such an
anticorrelation in the higher energy
$3$--$24$~keV band. This may be understood partly as a result of the
X-ray spectra of AGNs being less strongly affected by absorption in the high
energy \nustar band.

To incorporate the full
  serendipitous sample, including weak and non-detections, we also
  calculate ``stacked'' means in \brnu (also shown in Figure \ref{br_cr}), by summing the
  net count-rates of all sources. 
  The stacked means are also consistent with
  a flat trend in band ratio as function of count-rate. 

\begin{figure}
\centering
\includegraphics[width=0.47\textwidth]{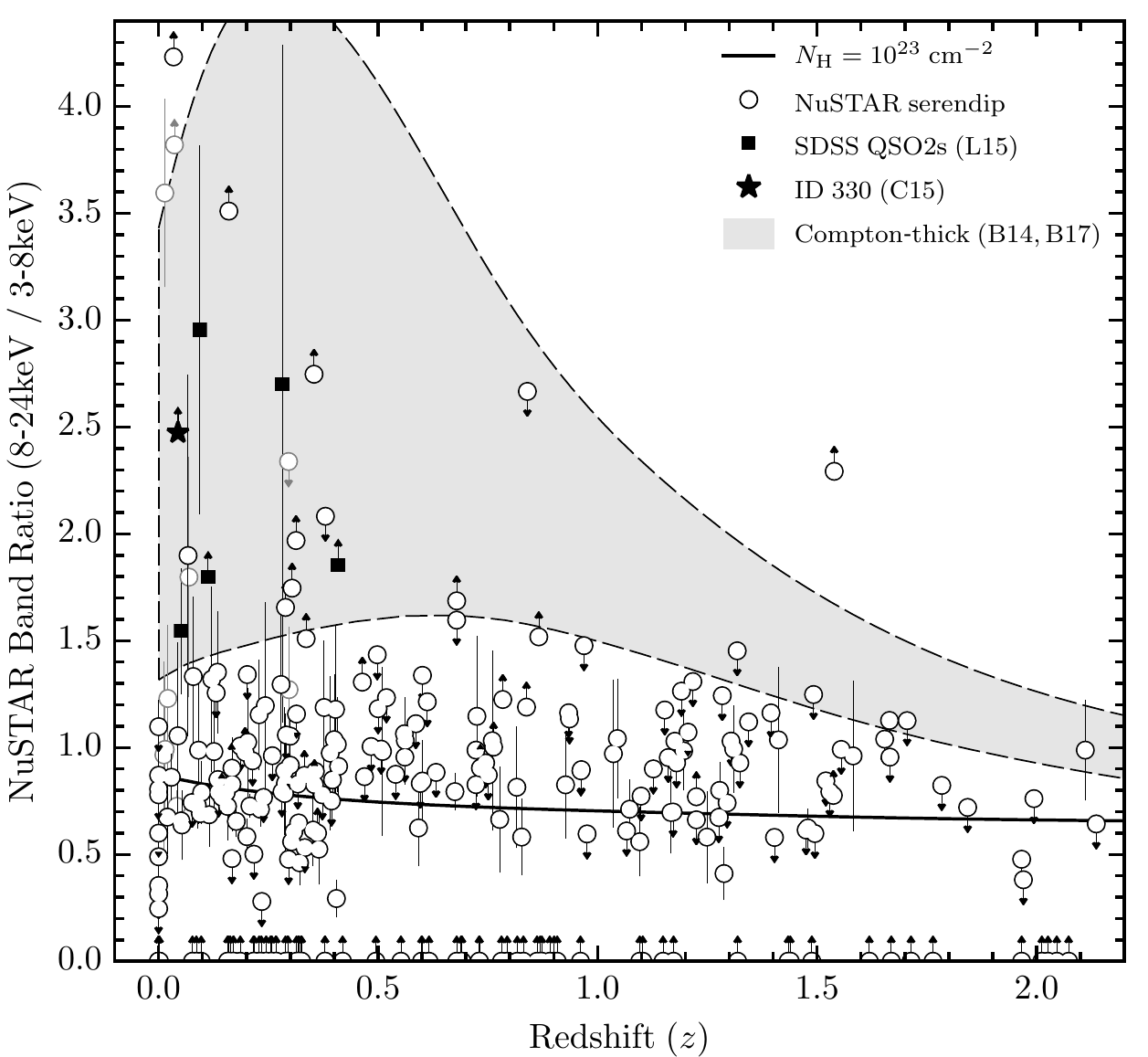}
\caption{\nustar band ratio (\brnu) versus redshift ($z$) for the full \nustar
  serendipitous survey sample (black circles). Sources which are associated with the primary science
  targets of the \nustar observations (according to the $\Delta (cz)$ criterion in Section \ref{nu_srcdet}) are
  shown as lighter gray circles. We compare to other
  \nustar-observed sources targetted for other programs (i.e., not
  part of the serendipitous survey). 
  The black star shows a CT AGN identified in the \nustar-COSMOS
  survey (C15).
  The black squares show heavily obscured
  SDSS-selected \typeii quasars observed with \nustar, for which there is 
  evidence for either CT or close to CT absorption
  \citep{Lansbury14,Gandhi14,Lansbury15}. The gray shaded
  region highlights the 
  parameter space expected for CT (i.e., $N_{\rm H} > 1.5\times
  10^{24}$~\nhunit) AGNs, considering all populations (including
  reflection- and transmission-dominated CT AGNs), based
  on results from the \nustar snapshot survey (\citealt{Balokovic14};
  Balokovi\'{c} et al. 2017, in prep.). This gray region was obtained by
  redshifting the best-fit spectral models of local 
  CT snapshot
  AGNs, for which the X-ray spectra are relatively well constrained.
  The upper and lower extents (dashed lines) represent the $68\%$
  percentiles (i.e., $84\%$ of the CT snapshot AGNs lie above the lower dashed line). 
Serendipitous sources lying at \brnu values within or higher than this
gray region are
good candidates for being CT.
The black track shows a \mytorus model prediction for \brnu as a
function of redshift, for a more moderate column density of $N_{\rm H} = 10^{23}$~\nhunit.}
\label{br_z} 
\end{figure}

While obscured AGNs can be crudely identified using \brnu alone, an estimate
of obscuring columns requires additional knowledge of the source redshifts, which 
shift key spectral features (e.g., the photoelectric absorption cut-off) across the observed energy bands.
Here we use the combination of \brnu and the source redshifts to
identify potentially highly obscured objects. Figure \ref{br_z} shows \brnu versus $z$
for the spectroscopically-identified serendipitous survey sample.
We compare with the band ratios measured for CT, or near-CT, SDSS-selected
\typeii quasars observed with \nustar in a separate targetted program
\citep{Lansbury14,Gandhi14,Lansbury15}, and with tracks (gray region) predicted for
CT absorption based on redshifting the best-fit spectra of local CT AGNs
from the \nustar snapshot survey of \swiftbat AGNs
(\citealt{Balokovic14}; Balokovi\'{c} et al.\ 2017, in prep.).
A number of sources stand out as CT-candidates based on this
analysis. 
While \brnu can only provide a crude estimate of the absorbing
columns, a more detailed investigation of the \nustar spectra and
multiwavelength properties of the CT-candidates can
strengthen the interpretation of these high-\brnu sources as highly
absorbed systems (Lansbury et al., in prep.).

\subsubsection{Redshifts and Luminosities} 
\label{redshifts_luminosities}

\begin{figure}
\centering
\includegraphics[width=0.47\textwidth]{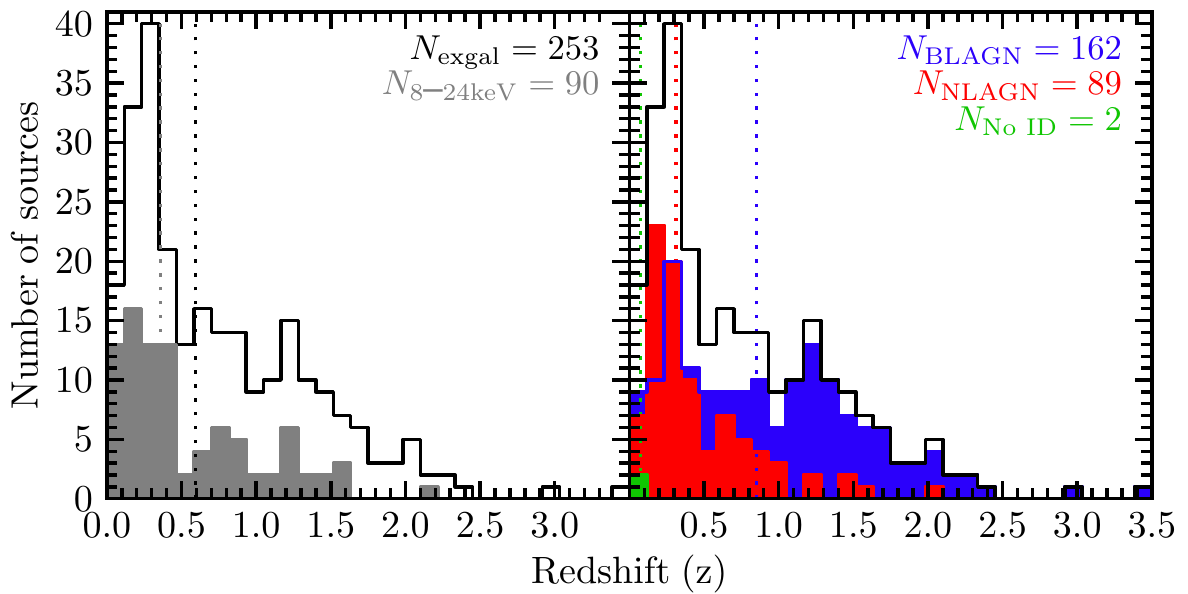}
\caption{The distribution of spectroscopic redshifts for the spectroscopically
  identified \nustar serendipitous survey sample. Galactic ($z=0$)
  sources have been excluded. In addition to the total distribution
  (black line),
  the left panel shows the distribution for the subset which are
  independently detected in the hard band ($8$--$24$~keV; gray
  filled histogram, left panel) and the right panel shows the distribution
  separated by optical classification: BLAGNs are shown in blue; NLAGNs
  are shown in red. The vertical lines mark the median redshifts.}
\label{Nz}
\end{figure}

Of the \nustar serendipitous survey sources with optical spectroscopic coverage
and spectroscopic redshift measurements (described in
Section \ref{spec_opt}), there are $262$ identified as extragalactic. Figure \ref{Nz} shows the redshift
distribution for the extragalactic sources, excluding nine sources
with evidence for being associated to the \nustar targets for their
respective observations (see Section \ref{nu_srcdet}). The redshifts cover a
large range, from $z=0.002$ to $3.43$, with a median of $\left<
z \right> = 0.56$. For the $90$ extragalactic objects with independent
detections in the high-energy band ($8$--$24$~keV), 
to which \nustar is uniquely sensitive,
the median redshift is $\left< z \right> = 0.34$. 
Roughly comparable numbers of NLAGNs and BLAGNs are identified for lower
redshifts ($z\lesssim 1$), but there is a significant bias towards BLAGNs at higher
redshifts. This was also found for the \nustar surveys in well-studied
fields (e.g., C15), and for surveys with sensitive lower energy
($<10$~keV) X-ray observatories such as \chandra and \xmm (e.g.,
\citealt{Barger03,Eckart06,Barcons07}). This effect is largely due to
selection biases against the detection of highly
absorbed AGNs, and against the spectroscopic identification of
the optically fainter NLAGNs (e.g., \citealt{Treister04}).

\begin{figure}
\centering
\includegraphics[width=0.47\textwidth]{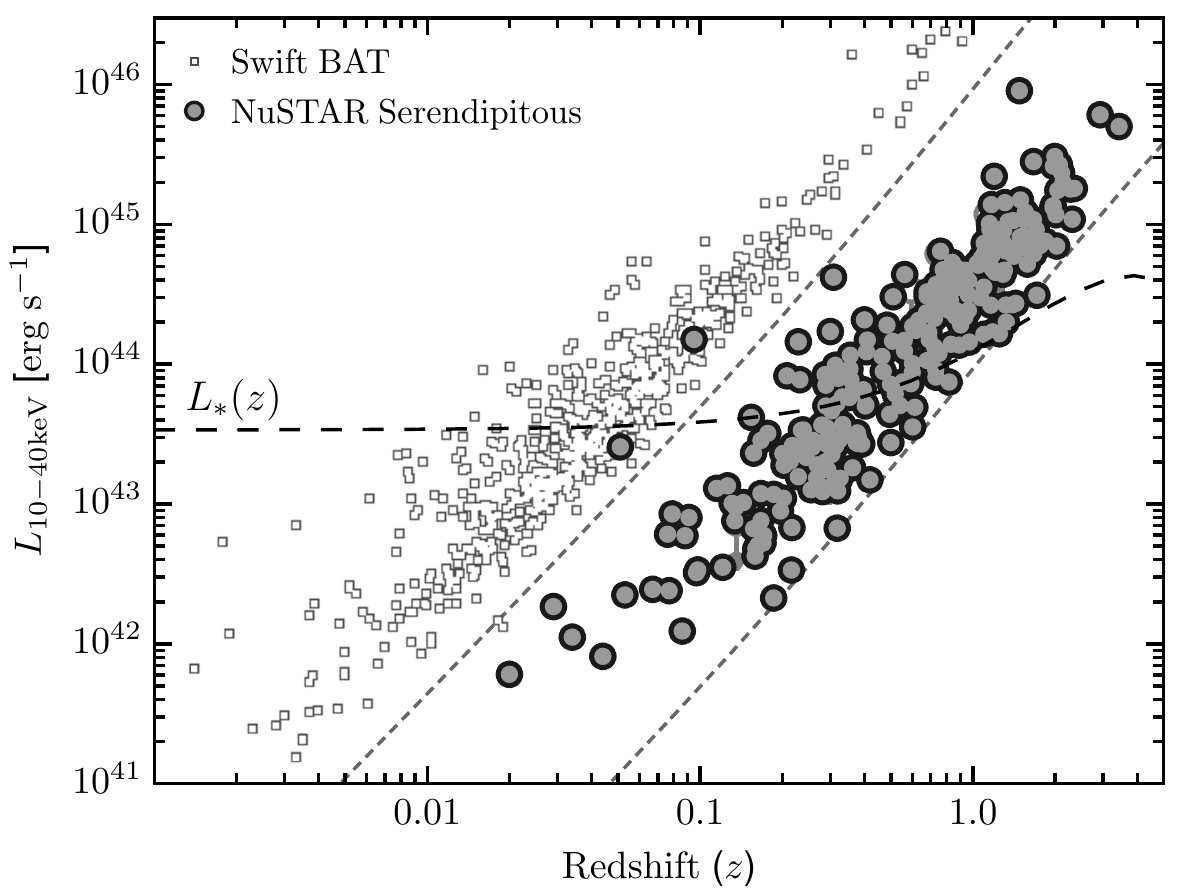}
\caption{Rest-frame $10$--$40$~keV luminosity (\Lhard) versus
  redshift. We compare the \nustar
  serendipitous survey sample
  (circles) with the \swiftbat 70-month all-sky survey
  catalog (squares; \citealt{Baumgartner13}; blazar and BL~Lac types have been excluded). 
\Lhard values for the \swiftbat sample are calculated from the
$14$--$195$~keV values, assuming $\Gamma_{\rm eff}=2.0$ for the $K$-correction
factor (consistent with the median spectral slope for the \swiftbat
sources shown).
The gray short-dashed lines highlight an observed-frame X-ray flux range spanning two orders of magnitude,
  from $2\times 10^{-14}$ to $2\times
  10^{-12}$~\fluxunit. The black long-dashed line shows the
  evolution of the knee of the X-ray luminosity function ($L_{*}$)
  with redshift, as measured by \citet{Aird15a}. The \nustar
  serendipitous survey probes below
  $L_{*}$ at $z\lesssim1$.}
\label{lx_z}
\end{figure}

Figure \ref{lx_z} shows the redshift--luminosity plane for the
rest-frame $10$--$40$~keV band. The luminosities are calculated from the
observed frame \nustar fluxes, assuming an effective photon index of
$\Gamma_{\rm eff}=1.8$
(as detailed in Section \ref{nu_phot}). The \nustar serendipitous survey
covers a large range in $10$--$40$~keV luminosity; the large majority
($99.6\%$; $238/239$) 
of the unassociated sources lie in the range of $L_{\rm 10-40keV} \approx 10^{42}$ to
$10^{46}$~\ergpersec. 
The median luminosity of $1.2\times 10^{44}$~\ergpersec 
is just above
the ``X-ray quasar'' threshold.\footnote{A threshold of $10^{44}$~\ergpersec is often
  adopted to define ``X-ray quasars'', since this roughly agrees with
  the classical optical quasar definition ($M_{B}\leq -23$; \citealt{Schmidt83}) and 
the $L_{\rm X,*}$ value for unobscured AGNs (e.g., \citealt{Hasinger05}).}
There is a single outlying source at very low luminosity and redshift,
NuSTAR~J115851+4243.2 (hereafter J1158; NLAGN; $z = 0.0023$; $L_{\rm 10-40keV} = 1.0 \times
10^{39}$~\ergpersec), hosted by the galaxy IC750. In this case, the
SDSS optical spectrum shows a narrow line AGN superimposed over the
galaxy spectrum. The source is discussed in detail in a
work focusing on the \nustar-selected AGNs with dwarf galaxy hosts (Chen
et al., submitted).
At the other extreme end in luminosity is NuSTAR~J052531-4557.8 
(hereafter J0525; BLAGN; $z=1.479$; $L_{\rm 10-40keV} = 9.0\times
10^{45}$~\ergpersec), also referred to as PKS~0524-460 in the
literature.\footnote{We note that J0525 appears in the
\swiftbat all-sky catalog of \citet{Baumgartner13} as a counterpart to
the source SWIFTJ0525.3-4600. However, this appears to be a mismatch:
an examination of the \swiftbat maps (following
the procedures in \citealt{Koss13}) and the \nustar data shows that
J0525 is undetected by \swiftbat, and a nearby AGN in a foreground
low redshift galaxy ESO~253-G003 ($z=0.042$) instead dominates the
SWIFTJ0525.3-4600 counts.} 
J0525 has an effective \nustar photon index of
$\Gamma_{\rm eff}=1.9^{+0.3}_{-0.2}$, and a \swiftxrt spectrum
which is consistent with zero X-ray absorption. The optical
spectrum of \citet{Stickel93} shows a broad line quasar with strong
\heii, \ciii, and \mgii emission lines. The source is also radio-bright
(e.g., $f_{\rm 1.4GHz}=1.7$~Jy; \citealt{Tingay03}) and has been
classified as a blazar in the literature (e.g.,
\citealt{Massaro09}).

The most distant source detected is an optically unobscured quasar, NuSTAR~J232728+0849.3 (hereafter J2327; BLAGN;
$z=3.430$; $L_{\rm 10-40keV} = 5.0\times 10^{45}$~\ergpersec), 
which represents the highest-redshift AGN
identified in the \nustar survey program to-date. Our Keck
optical spectrum for J2327 shows a quasar spectrum with strong \lya, \civ, and
\ciii emission lines, and a well-detected \lya forest. The source is
consistent with having an observed X-ray spectral slope of
$\Gamma_{\rm eff}=2$ for
both the \nustar spectrum and the \xmm counterpart spectrum,
and is thus in agreement with being unobscured at X-ray energies.
The most distant optically obscured quasar detected is NuSTAR~J125657+5644.6
(hereafter J1256; NLAGN; $z=2.073$; $L_{\rm 10-40keV} = 2.7\times 10^{45}$~\ergpersec). 
Our Keck optical spectrum for J1256
reveals strong narrow \lya, \civ, \heii, and \ciii emission lines.
Analysing the \nustar spectrum in combination with a deep archival
\chandra spectrum ($\approx
360$~ks of exposure in total), we measure a
moderately large line of sight column density of $N_{\rm H}=(1.3\pm 0.4)\times
10^{23}$~\nhunit. This distant quasar is thus obscured in both the optical and X-ray regimes.

In Figure \ref{lx_z} we compare with 
the 70-month \swiftbat all-sky survey \citep{Baumgartner13}. The two
surveys are highly complementary; the \swiftbat all-sky survey
provides a statistical hard X-ray-selected sample of AGNs in the nearby universe
(primarily $z < 0.1$), while the \nustar serendipitous survey provides
an equivalent sample (with comparable source statistics) for the
distant universe.
We compare with the redshift evolution of the knee of the
X-ray luminosity function ($L_{\star}$), as determined by
\citet{Aird15a}.
The \swiftbat all-sky survey samples the population below $L_{\star}$ for
redshifts up to $z\approx 0.05$,
while the \nustar serendipitous survey can achieve this up to
$z\approx 1$.
There is almost no overlap between the two surveys, which sample 
different regions of the parameter space. However, there are two
\nustar sources, outlying in Figure \ref{lx_z}, which have very high
fluxes approaching the detection threshold of \swiftbat:
NuSTAR~J043727-4711.5 ($z=0.051$; $L_{\rm 10-40keV} = 2.6\times
10^{43}$~\ergpersec) and NuSTAR~J075800+3920.4 ($z=0.095$; $L_{\rm
  10-40keV} = 1.5\times 10^{44}$~\ergpersec). 
Both are BLAGNs (based on our Keck and NTT spectra), and are
unobscured at X-ray energies ($\Gamma_{\rm eff}\approx 1.9$). The
former is detected in the 70 month \swiftbat catalog of \citet{Baumgartner13},
and the latter is only detected with \swiftbat at the $\approx
2\sigma$ level, based on the direct examination of the
104 month BAT maps (following the procedures in \citealt{Koss13}).

\begin{figure}
\centering
\includegraphics[width=0.47\textwidth]{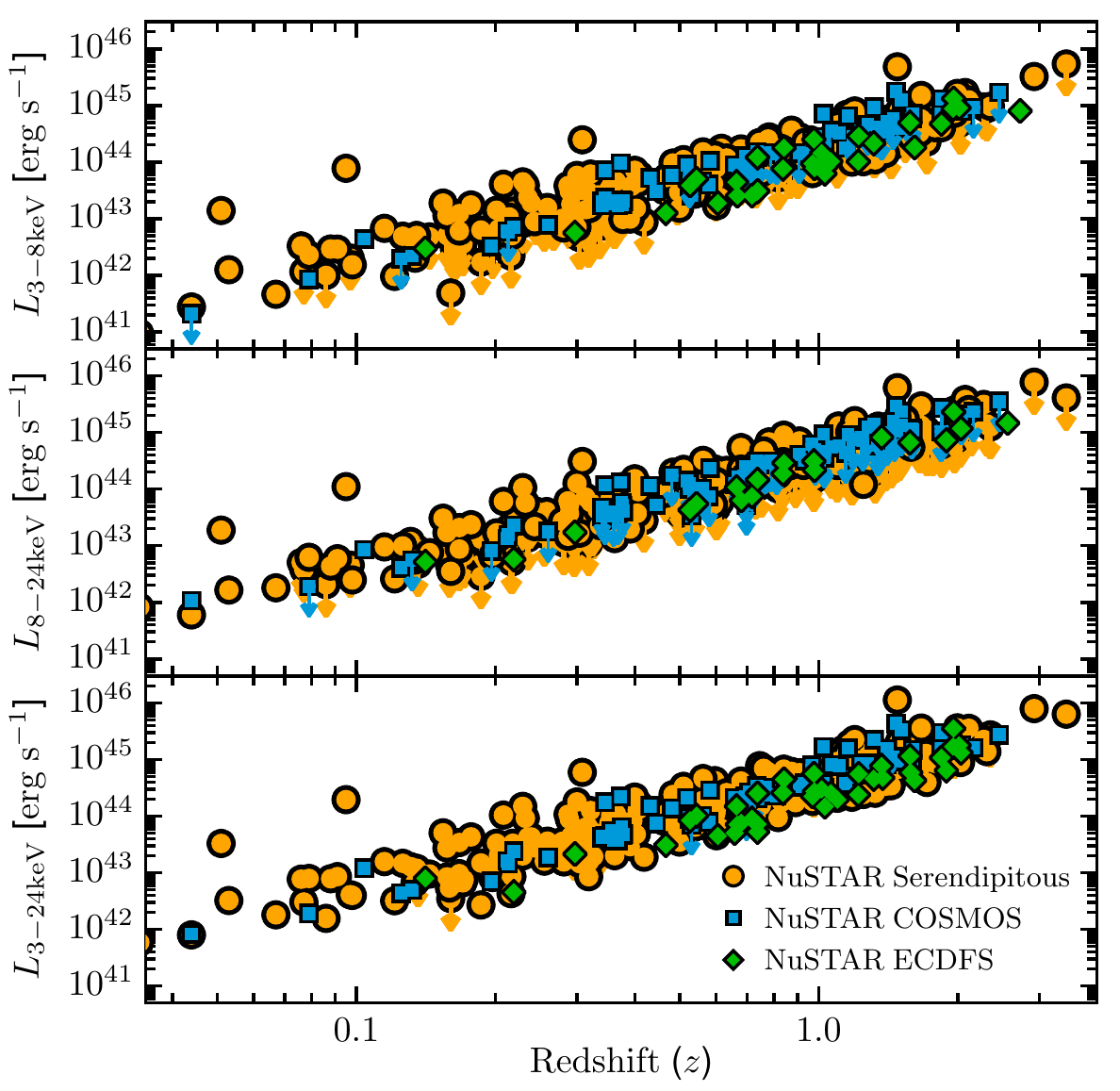}
\caption{Luminosity versus redshift for the three \nustar energy
  bands: $3$--$8$ (top), $8$--$24$ (middle), and $3$--$24$~keV
  (bottom). We compare the \nustar
  serendipitous survey sample
  (orange circles) with the blank-field \nustar surveys of
  COSMOS (blue squares; C15) and ECDFS (green diamonds; M15). 
   }
\label{lx_z_bands}
\end{figure}

In Figure \ref{lx_z_bands} we compare the luminosity--redshift
source distribution with other \nustar extragalactic
survey samples:  the \nustar-ECDFS survey (M15) and
  the \nustar-COSMOS survey (C15). Rest-frame luminosities are shown
  for the standard three \nustar bands (\softband, \fullband, and
  \hardband). The serendipitous survey fills out the broadest range of
  luminosities and redshifts, due to the nature of the coverage
  (a relatively large total area, but with deep sub-areas that push
  to faint flux limits).

\subsection{Optical properties}
\label{results_opt}

\subsubsection{The X-ray--optical flux plane}
\label{xray_opt}

\begin{figure}
\centering
\includegraphics[width=0.47\textwidth]{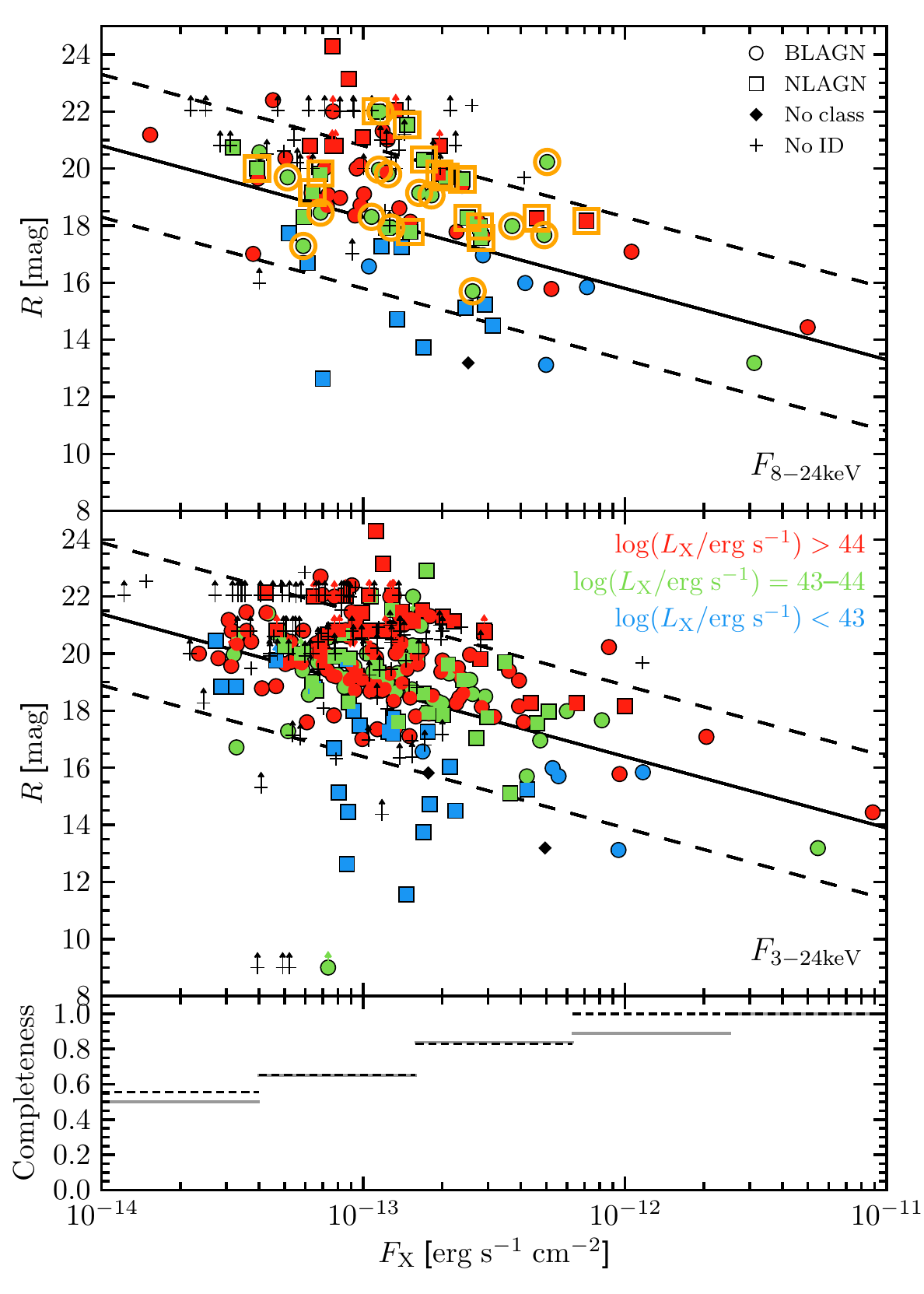}
\caption{$R$-band optical magnitude ($R$) versus X-ray flux (\fx) for the
  hard band ($8$--$24$~keV; top panel) and full band ($3$--$24$~keV;
  middle panel) selected \nustar serendipitous survey samples. 
The blue, green, and red colors highlight
  three X-ray luminosity ranges, from low to high luminosity,
  respectively.
The
  solid and dashed lines indicate constant X-ray-to-optical flux
  ratios of $\log (f_{\rm X}/f_{\rm opt}) = 0$ and $\pm 1$,
  respectively. 
  The hard band subsample for which we calculate a reliable \typeii
  fraction (see Section \ref{type2fraction}) measurement is highlighted with orange outlines.
  In the bottom panel we show the optical spectroscopic completeness of both the
  $8$--$24$ (dashed lines) and $3$--$24$~keV (solid lines) samples as a
  function of \fx, calculated as the number of sources with successful
  optical spectroscopic classifications (see Section \ref{opt_analysis}) divided by the total number of sources
  in a given \fx bin.}
\label{R_fx}
\end{figure}

The X-ray--optical flux plane is a classic diagnostic diagram for
sources detected in X-ray surveys \citep[e.g.,][]{Maccacaro88}.
This plane has recently been explored for the
\nustar-COSMOS sample, using the $i$-band (C15). Here we investigate the
plane using the optical $R$-band for the \nustar serendipitous survey, which
provides a relatively large hard X-ray selected sample spanning a
comparatively wide flux range. The X-ray-to-$R$-band flux ratio ($f_{\rm X}/f_{\rm opt}$)
diagnostic has been widely applied in past
\chandra and \xmm surveys of well-known blank fields \citep[e.g.,][]{Hornschemeier01,Barger03,Fiore03,Xue11}.
Figure \ref{R_fx} shows the optical $R$-band magnitude ($R$) against
X-ray flux (\fx) for the \nustar serendipitous survey sources which
are detected in the hard band ($8$--$24$~keV) and full band
($3$--$24$~keV).
We exclude $|b|<10$\degrees and $z=0$ sources, thus minimizing
contamination from Galactic sources. 
We subdivide the \nustar
sample according to X-ray luminosity and optical spectroscopic classification: objects with
successful identifications as either NLAGNs or BLAGNs;
objects with redshift constraints, but no classification; and objects
with no redshift constraint or classification.
For $R>20$, the sources shown with lower limits in $R$ generally correspond to
a non-detection in the optical coverage, within the X-ray positional error
circle. 
For sources where
it is not possible to obtain an $R$-band constraint (e.g., due to
contamination from a nearby bright star), we plot lower limits at the lower
end of the y-axis.

We compare with the range of X-ray to optical flux ratios typically
observed for AGNs identified in soft X-ray surveys, $-1 < \log (f_{\rm X}/f_{\rm opt}) < 1$
\citep[e.g.,][]{Schmidt98,Akiyama00,Lehmann01}. To illustrate constant 
X-ray-to-optical flux ratios, we adopt the relation of
\citet{McHardy03} and correct to the \nustar energy bands assuming
$\Gamma_{\rm eff}=1.8$. The large majority of 
sources lie at $\log
(f_{\rm X}/f_{\rm opt}) > -1$, in agreement with them being
AGNs. At least $\approx 25\%$ of the hard-band ($8$--$24$~keV) selected sources lie at $\log
(f_{\rm X}/f_{\rm opt}) > 1$, in agreement with the findings for the
lower energy selected X-ray sources detected in the \chandra and \xmm
surveys \citep[e.g.,][]{Comastri02,Fiore03,Brandt05}. Such high $f_{\rm X}/f_{\rm opt}$
values are interpreted as being driven by a combination of relatively
high redshifts and obscuration \citep[e.g.,][]{Alexander01,Hornschemeier01,DelMoro08}. 

\begin{figure}
\centering
\includegraphics[width=0.47\textwidth]{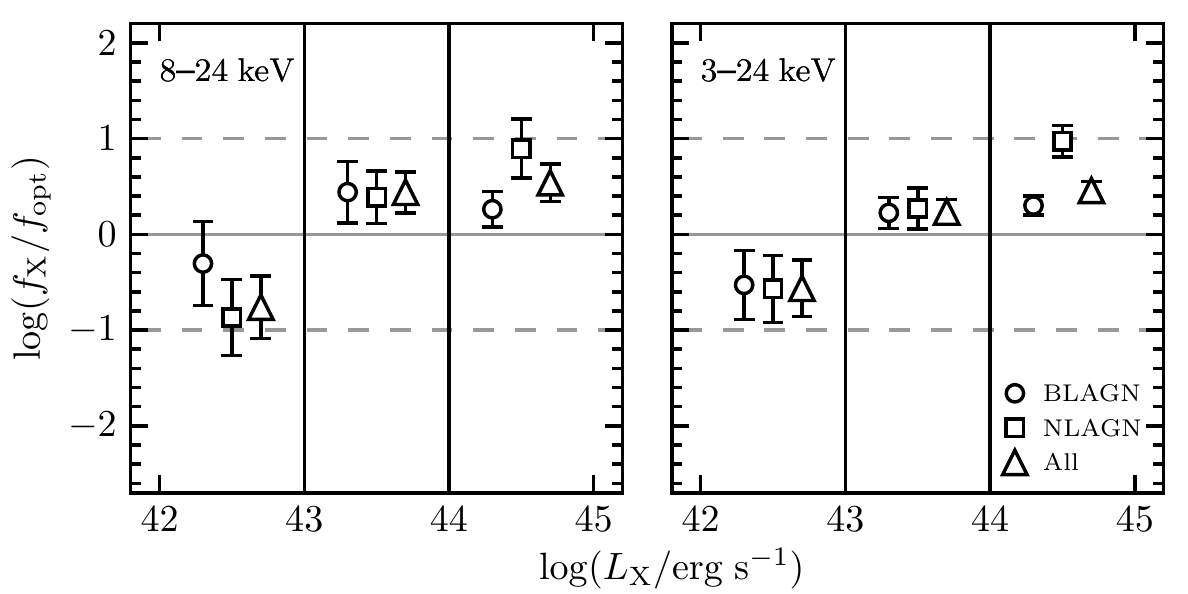}
\caption{
  The X-ray-to-$R$-band flux ratio ($f_{\rm X}/f_{\rm opt}$), as a function of
  luminosity, for hard band ($8$--$24$~keV) selected sources (left);
  and full band ($3$--$24$~keV) selected sources (right). The luminosity
  bins follow those adopted in Figure
  \ref{R_fx}. We show results for the overall spectroscopically
  identified population (triangles), BLAGN only (circles), and NLAGN
  only (squares). 
The solid and dashed horizontal gray lines indicate $\log (f_{\rm X}/f_{\rm opt}) = 0$ and $\pm 1$, respectively. 
The horizontal
offsets of the data points, within each luminosity bin, are arbitrary
and for visualization purposes only. }
\label{fxfopt_lx}
\end{figure}

To demonstrate the dependence on X-ray luminosity and on spectral type,
Figure \ref{fxfopt_lx} shows median $f_{\rm X}/f_{\rm opt}$ values 
for bins of X-ray luminosity, and for the NLAGN and
BLAGN subsamples separately. The low, medium, and high luminosity bins
correspond to $\log(L_{\rm X} / \mathrm{erg\ s^{-1}})<43$,
$43<\log(L_{\rm X} / \mathrm{erg\ s^{-1}})<44$, and $\log(L_{\rm X} /
\mathrm{erg\ s^{-1}})>44$, respectively. 
The observed dependence on luminosity
and on spectral type is consistent between the hard band and the full
band selected samples (left and right panels of Figure
\ref{fxfopt_lx}, respectively). Overall, $f_{\rm X}/f_{\rm opt}$
increases with X-ray luminosity. The increase between the low and
medium luminosity bins is highly significant; for the hard-band
selected sample, the median $\log(f_{\rm X}/f_{\rm opt})$ value increases from $\approx -0.8$
to $\approx 0.5$. There is a marginally significant
overall increase in $f_{\rm X}/f_{\rm opt}$ between the medium and high
luminosity bins, which is driven by a significant
increase in the $f_{\rm X}/f_{\rm opt}$ values of NLAGNs. 
A positive correlation between $f_{\rm X}/f_{\rm opt}$ and \lx
has previously been identified for \chandra and \xmm 
samples of optically obscured AGNs selected at $<10$~keV, over the
same luminosity range (\citealt{Fiore03}). 
Here we have demonstrated a strong positive correlation for high
energy ($\gtrsim 10$~keV) selected AGNs.

In general, the NLAGNs span a wider range in $f_{\rm X}/f_{\rm opt}$
than the BLAGNs, which mostly lie within the range expected for BLAGNs
based on soft X-ray surveys, $-1 < \log (f_{\rm X}/f_{\rm opt}) < 1$. 
The most notable difference between the two classes is in the high-luminosity
bin (which represents the ``X-ray quasar'' regime; $L_{\rm
  X}>10^{44}$~\ergpersec), where the NLAGNs lie significantly
higher than the BLAGNs, with a large fraction at $\log(f_{\rm X}/f_{\rm
  opt})>1$. 
This effect can be understood as a consequence of extinction of the
nuclear AGN emission. For the BLAGNs the nuclear optical--UV emission contributes strongly to
the $R$-band flux, while for the NLAGNs the nuclear optical emission
is strongly suppressed by intervening dust (the corresponding
absorption by gas at X-ray energies is comparatively weak). The effect
is augmented for the high-luminosity bin, where
the higher source redshifts ($\left< z \right> \approx 0.9$) result in the observed-frame optical band
sampling a more heavily extinguished part of the AGN spectrum, while
the observed-frame X-ray band samples a less absorbed part of the
spectrum (e.g., \citealt{DelMoro08}). 
The other main difference between the two classes is seen for the
lowest luminosity bin where,
although the median flux ratios are consistent, the NLAGNs extend to 
lower values of $f_{\rm X}/f_{\rm opt}$ than the BLAGNs, with a handful
of the NLAGNs lying at $\log (f_{\rm X}/f_{\rm opt}) < -1$.

\subsubsection{The type 2 fraction}
\label{type2fraction}

Here we investigate the relative numbers of the optically
obscured (i.e., NLAGN) and optically unobscured (i.e., BLAGN)
populations within the \nustar serendipitous survey sample. 
To provide meaningful constraints on the \typeii fraction (i.e., the observed number
of NLAGNs divided by the total number of NLAGNs+BLAGNs), it is important to understand
the sample completeness.
We therefore investigate a specific subset of the overall sample for
which completeness is well understood: hard
band ($8$--$24$~keV) selected sources with $0.1<z<0.5$,
$2\times 10^{43}<L_{\rm 10-40\ keV}<2\times 10^{44}$~\ergpersec, and $|b|>10$\degrees
(highlighted with orange outlines in the upper panel of Figure
\ref{R_fx}). 
The redshift limit ensures that the subsample has high spectroscopic
completeness (i.e., the majority of sources have redshifts and classifications
from optical spectroscopy; see below), the lower luminosity limit
ensures ``X-ray completeness'' (i.e., the AGN population within this
\lx--$z$ parameter space has fluxes which lie above the \nustar
detection limit; e.g., see Figure \ref{lx_z}), and the upper
luminosity limit is applied to allow comparisons with luminosity-matched
comparison samples (see below). The luminosity range
samples around the knee of the X-ray luminosity function ($L_{*}$) for
these redshifts, $L_{\rm 10-40\ keV}\approx (4$--$7)\times
10^{43}$~\ergpersec (\citealt{Aird15a}). In total, there are
$30$ spectroscopically identified sources (all NLAGNs or BLAGNs)
within this subsample, which have a median redshift of
$\left< z \right>=0.3$. 
Accounting for sources which are not spectroscopically identified, we
estimate an effective spectroscopic completeness of
$97$--$100\%$ for this subsample (details are provided in Section A.4).

\begin{figure}
\centering
\includegraphics[width=0.47\textwidth]{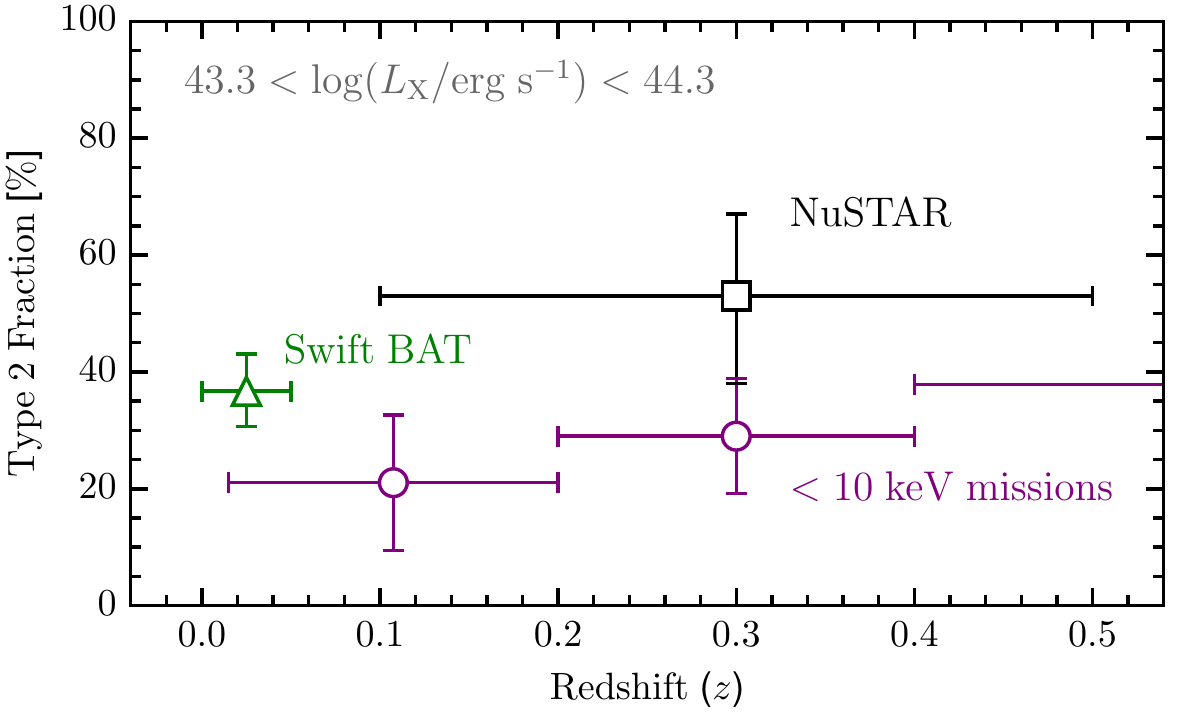}
\caption{Observed \typeii fraction versus redshift for various
  luminosity-matched ($2\times10^{43}<L_{\rm X}<2\times
  10^{44}$~\ergpersec), X-ray selected AGN samples: the black square
  shows a hard band ($8$--$24$~keV) selected subset of the \nustar
  serendipitous survey sample with $0.1<z<0.5$; the green triangle
  shows a subset of the 70-month \swiftbat all-sky survey
  sample ($z<0.05$; \citealt{Baumgartner13}); and the purple circles
  correspond to a $<10$~keV selected AGN sample, compiled from
  multiple X-ray surveys (including \asca, \chandra and \xmm surveys; \citealt{Hasinger08}). The
  horizontal error bars show the redshift limits of each subsample.}
\label{t2_z}
\end{figure}

The observed \typeii fraction for the \nustar hard band-selected
subsample described above is
$F_{\rm Type\ 2}=53^{+14}_{-15}\%$ (binomial uncertainties). 
If we instead use the sources selected in the full band
($3$--$24$~keV), a similar fraction is obtained ($F_{\rm Type\
  2}=48\pm 11\%$).
In Figure \ref{t2_z} we compare with the \typeii fraction for nearby ($z<0.05$)
AGNs similarly selected at high X-ray energies ($>10$~keV). 
To obtain this data point we calculate the observed
\typeii fraction for the 70-month \swiftbat all-sky survey.
Importantly, we use a luminosity-matched subsample of the \swiftbat
survey  ($2\times 10^{43}<L_{\rm
  10-40\ keV}<2\times 10^{44}$~\ergpersec, as for the 
\nustar subsample), since
the \typeii fraction likely varies with luminosity. We apply a redshift cut of $z<0.05$ to
ensure X-ray completeness (redshifts above this threshold push below the flux
limit of \swiftbat for the adopted $L_{\rm 10-40\ keV}$ range; see
Figure \ref{lx_z}). For consistency
with our approach for the \nustar sample, we class
\swiftbat AGNs with intermediate types of $1.9$ or below as BLAGNs, those with
NL~Sy1 type spectra as BLAGNs, and those with galaxy type optical
spectra as NLAGNs. The observed \typeii fraction for this
luminosity-matched \swiftbat sample at $z<0.05$ is $F_{\rm Type\
  2}=37\pm 6\%$, 
slightly lower than our \nustar-measured \typeii fraction at $z\approx 0.3$,
but consistent within the uncertainties.
A caveat to this comparison is that the
spectroscopic completeness of the \swiftbat subsample is unknown;
overall there are $\approx 100$ sources in the \citet{Baumgartner13} catalog
which are consistent with being AGNs but lack an optical spectroscopic redshift and classification, some
of which could potentially lie within the luminosity and redshift
ranges adopted above. 
Making the extreme assumption that these $\approx 100$ sources
all lie in the above luminosity and redshift ranges, and are all
NLAGNs, the maximum possible \swiftbat $F_{\rm Type\ 2}$
value is $66\%$ (which would still be in agreement with the \nustar-measured
fraction). 
Depending on the full duration of the
\nustar mission, the source numbers for the \nustar serendipitous
survey may feasibly increase by a factor of two or more, which
will reduce the uncertainties on the  \typeii
fraction. However, to determine reliably whether there is
evolution in the \typeii fraction of high energy selected
AGNs between $z<0.05$ and $z>0.1$, future studies should
systematically apply the same optical spectroscopic classification
methodologies to both samples. An early indication that the obscured
fraction of AGN might increase with redshift was given by
\citet{LaFranca05}, and this has been further quantified in subsequent
works
\citep[e.g.,][]{Ballantyne06,Treister06,Hasinger08,Merloni14}. The
slope of the increase with redshift is consistent with that found by \citet{Treister06}.

The \typeii fraction has been thoroughly investigated for the AGN population
selected by lower energy ($<10$~keV) X-ray missions such as \chandra and
\xmm. \citet{Hasinger08} presented a relatively complete
$2$--$10$~keV selected sample, compiled from a variety of
surveys with $<10$~keV missions (also see \citealt{Merloni14} for a
more recent study of \xmm-selected sources at $0.3<z<3.5$). We consider the $0.2<z<0.4$
subset of the \citet{Hasinger08} sample,
in order to match to our \nustar subsample in redshift as closely as possible, and we limit to the luminosity range
explored above ($2\times 10^{43}<L_{\rm
  10-40\ keV}<2\times 10^{44}$~\ergpersec; we assume a luminosity band
correction of $L_{\rm
  10-40\ keV}/L_{\rm 2-10\ keV}=1$). The \typeii fraction for this subset of the
\citet{Hasinger08} sample is $F_{\rm Type\ 2}=29\pm 10\%$, 
which is lower
than our \nustar-measured \typeii fraction (see Figure
\ref{t2_z}), but only at a significance level of $\approx 2\sigma$. 
This could be explained as a result of the
different selection functions of different X-ray missions, with the high energy
($>8$~keV) selection of \nustar being less biased against obscured
sources. Another factor to consider is the different classification
methodologies applied. In addition to optical spectroscopic
constraints, \citet{Hasinger08} use additional X-ray
hardness information to classify ambiguous sources as NLAGNs or
BLAGNs. \citet{Hasinger08} do assess the extent to which
the \typeii fraction measurements change if, instead, only the pure optical
spectroscopic classification is adopted (i.e., a similar approach
to our spectroscopic classification for the \nustar sources) and find that, for
the redshift and luminosity ranges explored here, the \typeii fraction
would be somewhat higher but unlikely to
increase by more than a factor of $\approx 1.2$. 

In Figure \ref{t2_z}
we compare with additional luminosity-matched subsamples for the
adjacent redshift bins studied by \citet{Hasinger08}. The
high-energy selected AGN samples (\nustar and \swiftbat) appear to lie
systematically higher in \typeii fraction than the luminosity-matched
lower energy ($<10$~keV) selected AGNs, for the redshift ranges covered.
We note that the Type 2 fraction constraints of \citet{Merloni14} for $<10$~keV
selected AGNs are broadly
consistent with the values shown in Figure \ref{t2_z} (we 
primarily compare with the \citealt{Hasinger08} sample since the
source redshifts and luminosities sampled facilitate a direct comparison
of results).
The apparently small numbers of CT AGNs identified (e.g., see Figure
\ref{br_z}) suggest that the offset in \typeii fraction
is not primarily driven by the uncovering of a new CT population, but
more likely by the selection functions of \nustar and \swiftbat being
generally less biased against significantly obscured AGNs.

\subsection{Infrared properties}
\label{results_ir}

\subsubsection{\wise colors}
\label{wise_colors}

Mid-infrared (MIR; $\gtrsim 5$~\micron) emission from AGNs is primary
emission that has been reprocessed by
circumnuclear dust, and suffers little extinction relative to other
(e.g., optical and soft X-ray) wavelengths.
Color selections using the \wise telescope bands
\citep[e.g.,][]{Assef10,Jarrett11,Stern12,Mateos12,Mateos13,Assef13} can
separate bright AGNs from host-galaxy light (from stars and the interstellar medium) through the identification of a red MIR
spectral slope, and have thus
become widely applied. These selections have the potential to identify large
samples of AGNs with less bias against
heavily obscured systems. However, their effectiveness worsens toward
lower AGN luminosities, where identifying the AGN component of the MIR
spectrum is more problematic.
Here we investigate the MIR properties of our \nustar serendipitous
survey sample, and consider the results with respect to these AGN selection criteria.

\begin{figure}
\centering
\includegraphics[width=0.47\textwidth]{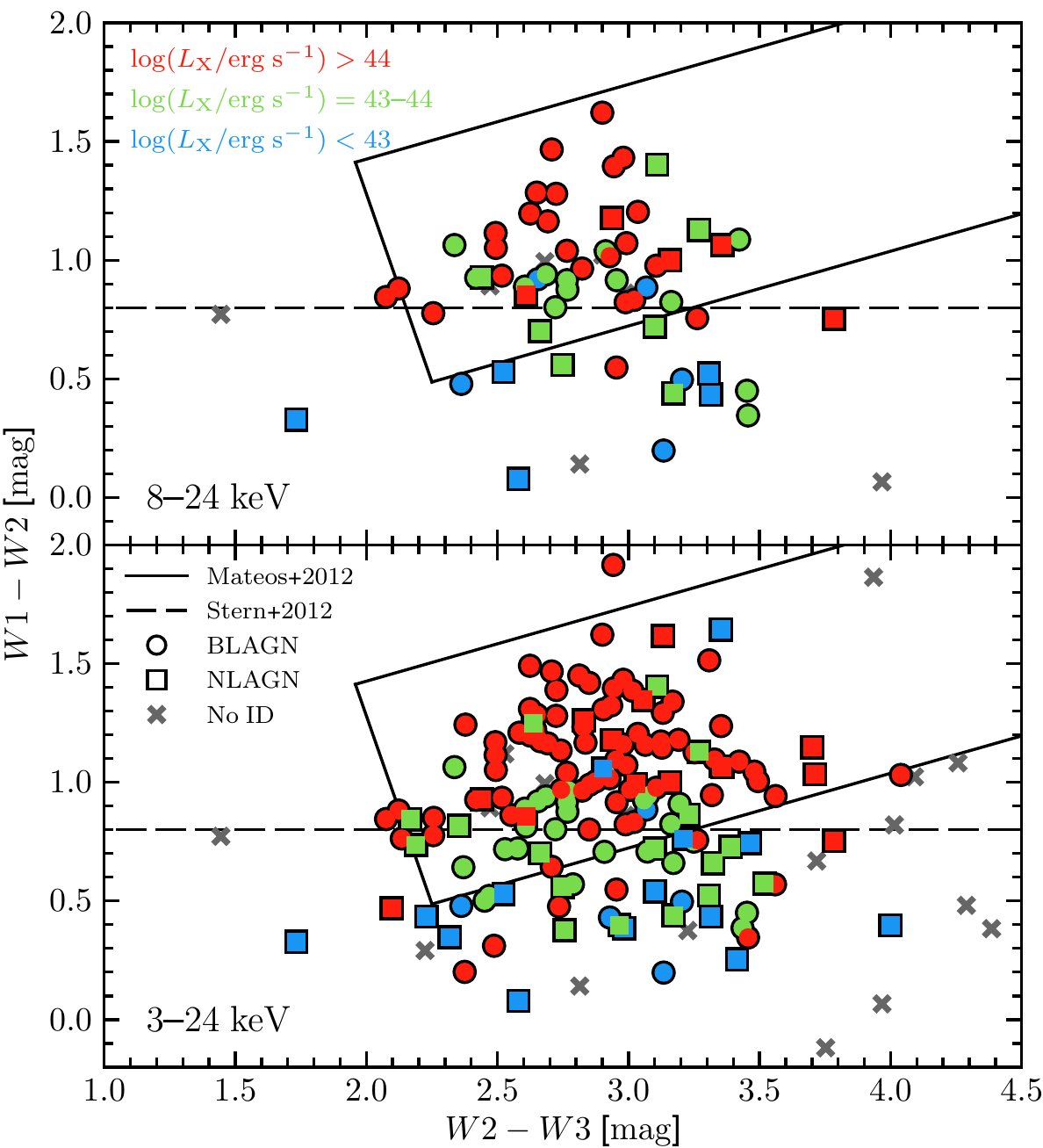}
\caption{\wise color--color diagram of \nustar serendipitous
  survey AGNs as a function of X-ray luminosity (\lx) and source
  classification, for hard-band ($8$--$24$~keV) selected sources (top)
  and full-band ($3$--$24$~keV) selected sources (bottom). 
  BLAGN and NLAGN are shown as circles and squares,
  respectively, while sources without a spectroscopic identification
  are shown as gray crosses. The blue, green, and red colors highlight
  three X-ray luminosity ranges, from low to high luminosity,
  respectively. The luminosities correspond to the selection bands
  used for this analysis (i.e., $L_{\mathrm{8-24\ keV}}$ and
  $L_{\mathrm{3-24\ keV}}$ for the upper and lower panels,
  respectively).
We compare with the AGN `wedge' of \citet{Mateos12} and the AGN
  color cut of \citeauthor[][]{Stern12} (\citeyear{Stern12}; $W{\rm
    1}$--$W{\rm 2}\geq0.8$).}
\label{w2w3_w1w2}
\end{figure}

\begin{figure}
\centering
\includegraphics[width=0.47\textwidth]{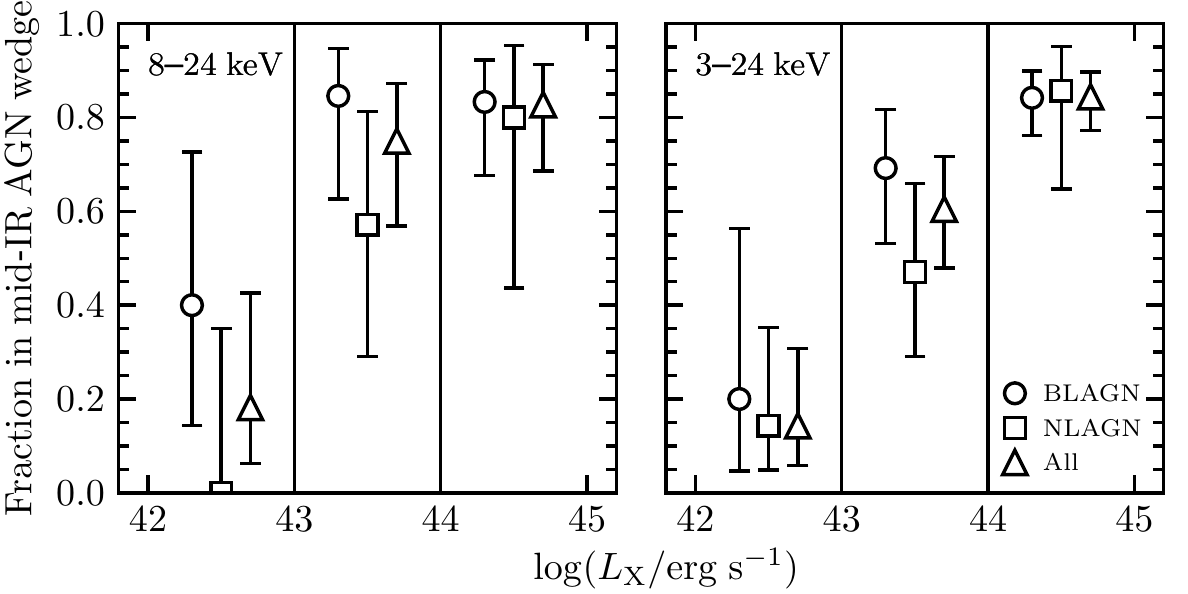}
\caption{The fraction of extragalactic \nustar serendipitous survey sources which are
  selected as AGNs based on MIR colors alone (i.e., they lie in the
  \wise color wedge of \citealt{Mateos12}), as a function of
  luminosity, for hard-band ($8$--$24$~keV) selected sources (left)
  and full-band ($3$--$24$~keV) selected sources (right). The luminosity
  bins follow those adopted in Figure
  \ref{w2w3_w1w2}. We show results for the overall spectroscopically
  identified population (triangles), BLAGN only (circles), and NLAGN
  only (squares). The error bars show binomial uncertainties. The horizontal
offsets of the data points, within each luminosity bin, are arbitrary
and for visualization purposes only. }
\label{inWedge}
\end{figure}

Figure \ref{w2w3_w1w2} shows a \wise color--color diagram (\wonewtwo
versus \wtwowthree) for the \nustar serendipitous survey subsamples
which are selected (i.e., independently detected) in the hard band (\hardband; upper panel) and full
band (\fullband; lower panel). In general, the sources which lie at
higher (i.e., redder) \wonewtwo values have stronger AGN contributions to their MIR
SEDs. We exclude low Galactic latitude sources ($|b|<10$\degrees),
and sources which are spectroscopically confirmed as Galactic. 
In addition, we only consider sources with well constrained X-ray
positions (i.e., with \chandra, \swiftxrt, or \xmm positions), and we limit the
analysis to the fraction of these sources ($70\%$ and $61\%$ for the hard
and full band, respectively) 
with significant detections in all three of the relevant, shorter
wavelength \wise bands (\wone, \wtwo, and \wthree; which are
centered at
$3.4$~\micron, $4.6$~\micron, and $12$~\micron, respectively). 
Figure \ref{w2w3_w1w2} shows the sample subdivided according to X-ray
luminosity and optical spectral classification. In Figure
\ref{inWedge} we show the fraction (\fwedge hereafter) of sources which are selected as
AGNs based on MIR colors alone, according to the selection ``wedge''
of \citet{Mateos12}, as a function of X-ray
luminosity and optical classification.

For the \nustar AGNs selected in the full band (lower panel of Figure
\ref{w2w3_w1w2} and right panel of Figure \ref{inWedge}) the overall fraction
of sources identified as AGNs in the MIR is
$f_{\rm wedge}=64.9^{+5.7}_{-6.2}\%$ ($111/171$). 
Considering sources with optical spectroscopic classifications, the
fractions for the overall BLAGN and NLAGN samples
are $f_{\rm wedge} = 77.6^{+5.9}_{-7.2}\%$ ($83/107$) and
$48.9^{+11.9}_{-11.8}\%$ ($22/45$), respectively. NLAGNs are therefore
significantly less likely to be identified as AGNs based on MIR colors
alone. This
is largely driven by the lower luminosity, on average, of the NLAGNs
(median of $4\times 10^{43}$~\ergpersec)
compared to the BLAGNs (median of $3\times 10^{44}$~\ergpersec), in combination with the fact that \fwedge
decreases toward lower luminosities (see Figure
\ref{inWedge}). Matching the NLAGNs and BLAGNs in luminosity, we do
not find statistically significant differences in $f_{\rm wedge}$ between the
two classes.

For the remainder of our overall sample which lack optical
spectroscopic classifications (gray crosses in Figure
\ref{w2w3_w1w2}), the \wise colors are informative of their likely
properties. A low fraction of these sources lie within the wedge,
 $f_{\rm wedge} = 27.8^{+19.3}_{-13.5}$ ($5/18$). This 
 suggests that, statistically, the
unidentified sources are likely to be less luminous AGNs. In
combination with the poor success rate for optical spectroscopy of
these sources, we expect
that they are likely to be dominated by optically obscured, low
luminosity systems. 

The results in Figure \ref{inWedge} show that MIR selections miss a
significant fraction of the \nustar-selected AGN population, with the missed
fraction increasing from $\approx 15\%$ at high luminosities,
to $\approx 80\%$ at the lower luminosity end. 
The dependence of MIR selections on AGN luminosity has been
identified for lower energy X-ray selected AGN samples (e.g.,
\citealt{Cardamone08,Eckart10}), and is likely primarily driven by a stronger
contribution to the SED from the host galaxy
for lower X-ray luminosities, which results in bluer MIR colors.
The MIR AGN selection wedge of \citet{Mateos12} was defined using the
Bright Ultrahard \xmm survey (BUXS) sample, selected at
$4.5$--$10$~keV, for comparable numbers of spectroscopically identified
AGNs ($114$ BLAGNs and $81$ NLAGNs) as the full-band selected \nustar serendipitous
survey sources incorporated here ($107$ BLAGNs and $45$ NLAGNs), 
and for a similar redshift and luminosity distribution. For the NLAGNs, our results
for $f_{\rm wedge}$ as a function of X-ray luminosity and optical classification are consistent
(given the uncertainties) with those found for the
BUXS sample. Our BLAGNs have marginally lower $f_{\rm
  wedge}$ values than the BUXS BLAGNs. For instance, \citet{Mateos12}
find that the MIR selection is essentially complete for BLAGNs at $L_{\rm
  X}>10^{43}$~\ergpersec (e.g., $f_{\rm wedge} = 100_{-6.6}\%$ and
$96.1^{+3.0}_{-6.3}\%$ for $L_{\rm 2-10keV}=10^{43}$--$10^{44}$ and
$10^{44}$--$10^{45}$~\ergpersec, respectively), while even at the highest luminosities
($L_{\rm 3-24keV}>10^{44}$~\ergpersec) we
find $f_{\rm wedge} = 84.2^{+5.7}_{-8.0}\%$. 

\begin{figure}
\centering
\includegraphics[width=0.47\textwidth]{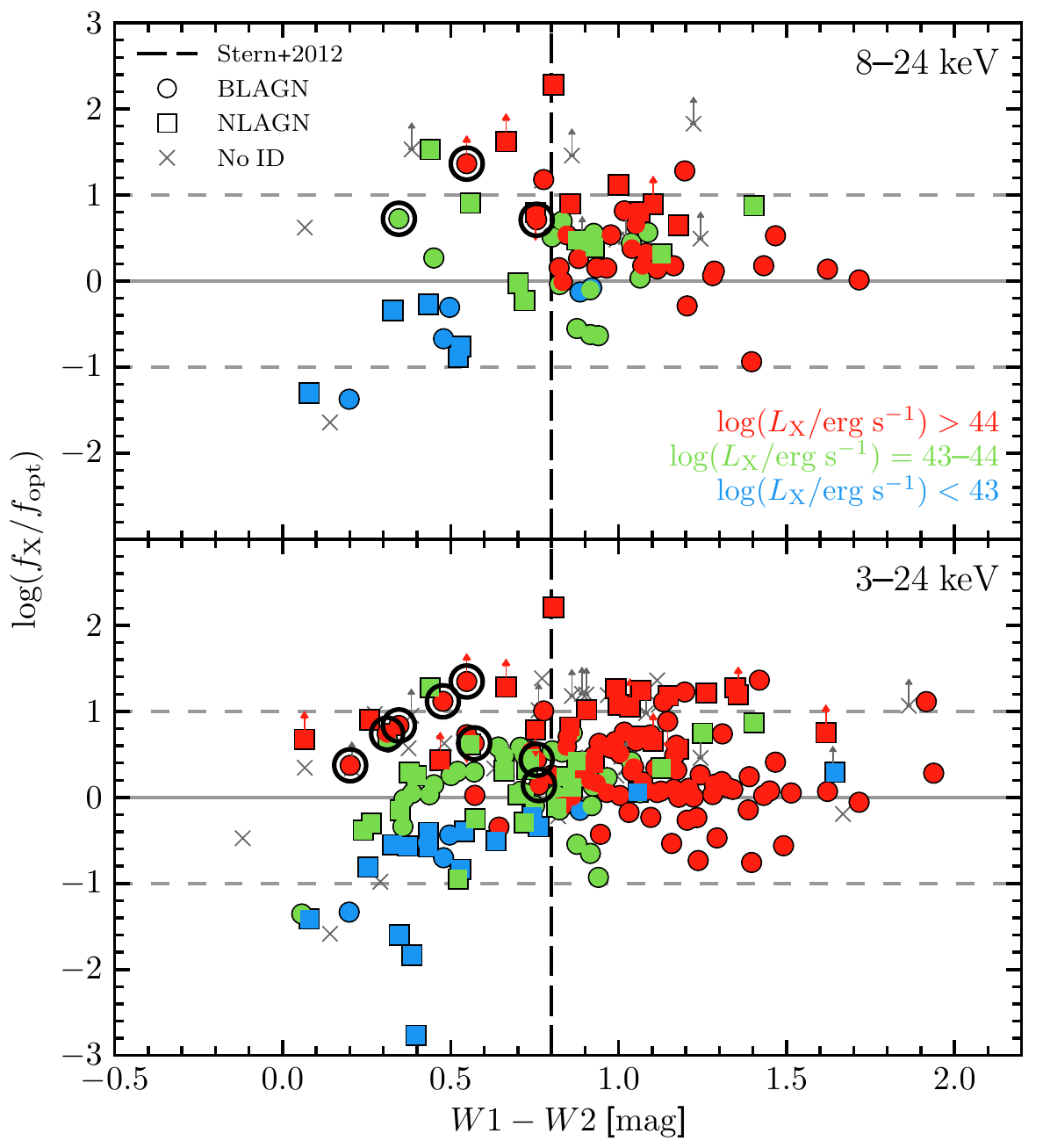}
\caption{
  The X-ray-to-$R$-band flux ratio ($f_{\rm X}/f_{\rm opt}$) versus
  the \wise \wonewtwo color, for hard band ($8$--$24$~keV) selected sources (top);
  and full band ($3$--$24$~keV) selected sources (bottom). The luminosity
  bins and the marker labelling follow those adopted in Figures
  \ref{R_fx} and \ref{w2w3_w1w2}. Eight high luminosity BLAGNs which
are not selected as AGNs in the MIR (see Section \ref{wise_colors})
are highlighted with large black circles.}
\label{fxfopt_w1w2}
\end{figure}

It is notable that the MIR
selection fails for $12$ (i.e., $15.8\%$) of the high
luminosity \nustar-selected BLAGNs, since MIR selections are typically
expected to be close to complete for high luminosity, unobscured
sources. To assess why these sources in particular are not MIR-selected, we compare their 
source properties (e.g., \nustar detection significance, redshift,
$10$--$40$~keV luminosity, $2$--$10$~keV luminosity, brightness,
optical spectra, Galactic latitude) with the $64$
(i.e., the $84.2\%$) high luminosity BLAGNs which are MIR-selected. 
There are no clear statistically significant
differences, with a possible exception: the optical $R$-band magnitude
distributions of the two subsets are different at a moderate
significance level (KS-test p-value of $p=0.037$), with the $12$
MIR-unselected sources skewed to fainter $R$ values (median of
$\left< R\right> =19.9$) than their MIR-selected counterparts ($\left< R\right> =19.4$). This result
increases in significance (to $p=0.0075$) if we limit the comparison to the eight (out
of $12$) MIR-unselected sources which are additionally missed by the
\citet{Stern12} \wonewtwo color AGN selection. 
Comparing the distribution of $f_{\rm X}/f_{\rm opt}$ versus \wonewtwo for
these eight sources with the overall serendipitous sample (see Figure \ref{fxfopt_w1w2}), they overlap
with lower luminosity AGNs where we expect that the relatively
blue \wonewtwo colors are driven by a stronger (relative) contribution
to the MIR SED from the host galaxy. 
The latter could also be true for the
eight MIR-unselected high-\lx BLAGNs if their MIR AGN luminosities are
relatively low compared to the $64$ MIR-selected
counterparts (which are matched in X-ray luminosity). Estimating the
rest-frame \sixum luminosities ($L_{\rm 6\mu m}$) by interpolating
between the relevant observed-frame \wise band magnitudes,\footnote{From the \wise all-sky survey catalog, there
are no indications of bad photometry (e.g., due to blending,
contamination, or confusion) for these eight sources.} we find
that the eight MIR-unselected BLAGNs have a different
$L_{\rm 6\mu m}$ distribution to the MIR-selected counterparts
($p=0.046$), and are indeed skewed to
lower MIR luminosities ($\left< L_{\rm 6\mu m}\right>=3.4\times
10^{44}$~\ergpersec) than
the MIR-selected sources ($\left< L_{\rm 6\mu m}\right>=1.3\times
10^{45}$~\ergpersec). 
In summary, the incompleteness of MIR selections for unobscured high-\lx
AGNs appears to be related to scatter in the intrinsic AGN
properties. The luminous MIR-unselected sources
  could potentially represent AGNs which are lacking in hot dust emission (i.e.,
  ``hot-dust-poor'' AGNs; e.g., \citealt{Hao10}), although the
  inferred hot-dust-poor fraction ($\sim 10$--$15\%$) would be
  unexpectedly high compared to that observed for optically selected quasars ($\sim 1\%$; \citealt{Jun13}).

For the \nustar serendipitous survey sources selected in the hard band (upper panel of Figure
\ref{w2w3_w1w2} and left panel of Figure \ref{inWedge}), for which
\nustar is uniquely sensitive, the results are consistent
with those for the full-band sample, but with greater uncertainties
due to the smaller source numbers. For instance, $f_{\rm wedge} =
67.6^{+8.5}_{-9.8}\%$ ($46/68$) for the
overall hard band selected sample.
We conclude that, while there are some small differences, the MIR
color distribution of the \nustar serendipitous survey
sample is largely consistent with that expected
based on the results for lower-energy ($<10$~keV) selected AGNs.

\subsubsection{X-ray--MIR luminosity plane}
\label{xray_mir}

\begin{figure}
\centering
\includegraphics[width=0.47\textwidth]{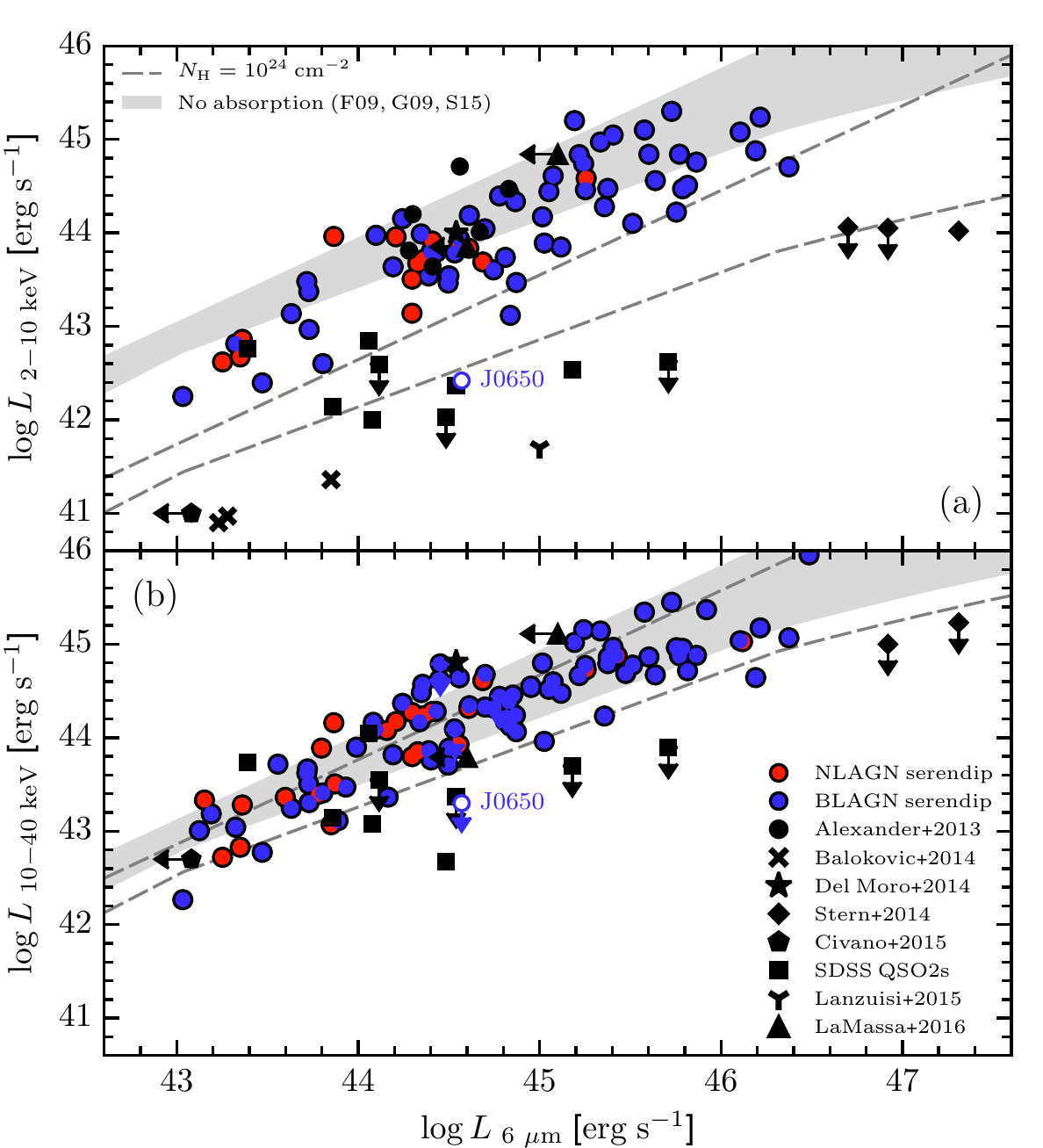}
\caption{Observed (i.e., uncorrected
  for absorption) rest-frame $2$--$10$~keV [\Lsoft; (a)] and
  $10$--$40$~keV X-ray luminosity [\Lhard; (b)] versus
  rest-frame \sixum luminosity (\Lsixum, in $\mathrm{\nu}
  L_{\mathrm{\nu}}$ units). Filled circles show the \nustar serendipitous
  survey sources. We only show sources that
  have detections in the \wise bands necessary to estimate \Lsixum
  (through interpolation),
  and which are AGN-dominated at
  MIR wavelengths according to their \wise colors (based on
  satisfying either the \citealt{Mateos12} or \citealt{Stern12}
  criteria), and thus where we believe \Lsixum to have minimal
  contamination from the host galaxy. The apparently X-ray weak source
  J0650 (see Sections \ref{xray_mir} and A.5) is shown as an empty circle and labelled.
  We compare with other samples:
  \nustar-observed SDSS-selected heavily obscured \typeii quasars (squares; $z=0.05$--$0.49$;
  \citealt{Lansbury14,Gandhi14,Lansbury15}); 
three CT Seyfert 2 AGNs from the \nustar snapshot survey (``$\times$'' symbols; $z\approx0.01$; \citealt{Balokovic14});
luminous and heavily obscured \wise-selected AGNs targetted with \nustar \citep[diamonds;
  $z\approx2$;][]{Stern14}; a heavily obscured quasar identified in
  the \nustar-ECDFS survey \citep[star; $z\approx2$;][]{DelMoro14}; a CT AGN identified in the
  \nustar-COSMOS survey (pentagon; ID~330; $z=0.044$; C15); a
  candidate heavily CT AGN identified in the COSMOS field (triangle;
  $z=0.35$; \citealt{Lanzuisi15a}); and
  \nustar-observed FIRST-2MASS red quasars (triangle;
  $z=0.14$--$0.41$; \citealt{LaMassa16}).
All of the data are compared with the luminosity ratios expected in
the case of zero line-of-sight absorption (gray
region). This region shows the range of intrinsic luminosity ratios
predicted by three different intrinsic relations in the literature:
\citet{Gandhi09}, \citet{Fiore09} and \citet{Stern15}. The dashed lines
illustrate the observed X-ray luminosity suppression expected if the
zero absorption region is absorbed by gas with a column density of $N_{\rm
  H}=10^{24}$~\nhunit. 
}
\label{lx_lmir}
\end{figure}

There is a remarkably tight correlation
between the X-ray luminosities and the MIR luminosities of unobscured AGNs, with
both providing estimates of the intrinsic AGN power
\citep[e.g.,][]{Lutz04,Fiore09,Gandhi09,Lanzuisi09,Ichikawa12,Matsuta12,Asmus15,Mateos15,Stern15}.
Low X-ray to MIR luminosity ratios are interpreted as
being due to either X-ray absorption or intrinsic X-ray weakness.

In Figure \ref{lx_lmir} we show the observed (i.e., uncorrected for
absorption) rest-frame X-ray luminosities
($L_{\rm X}^{\rm obs}$) versus the rest-frame \sixum luminosities
($L_{\rm 6\mu m}$, in $\nu L_{\nu}$ units) for
\nustar serendipitous survey sources. 
We only include sources which
are AGN-dominated at MIR wavelengths according
  to the \wise colors (based on either of the criteria in Section
  \ref{wise_colors}), and thus where we believe the rest-frame \sixum flux to be
  dominated by the AGN rather than host-galaxy light. Additionally, we require
  the sources to be detected in the two observed-frame \wise 
  bands which are interpolated between to estimate $L_{\rm 6\mu m}$ (e.g., \wtwo
  and \wthree for $z<1$).
For the high energy ($10$--$40$~keV) rest-frame X-ray band (bottom panel of Figure
\ref{lx_lmir}), the X-ray luminosities are from \nustar photometry (as
described in Section \ref{nu_phot}). For the low energy
($2$--$10$~keV) rest-frame X-ray band (top panel of Figure
\ref{lx_lmir}), the X-ray luminosities are estimated from CSC or 3XMM
counterpart fluxes (for the top panel, we only show sources with
counterparts in these catalogs).
We compare with other \nustar-observed
samples, including a number of heavily obscured AGNs. To demonstrate
the approximate X-ray to MIR luminosity ratios expected in the cases
of zero absorption and high absorption, we show the intrinsic
X-ray--MIR relation (as measured by multiple studies;
\citealt{Fiore09,Gandhi09,Stern15}) and the same relation after
absorption by $N_{\rm H}=10^{24}$~\nhunit gas, respectively.

At $10$--$40$~keV, the serendipitous survey sources are generally
consistent with both the intrinsic and the highly absorbed
X-ray--MIR relations (which are close together for these
energies). The most outlying source, J0650 (highlighted in Figure \ref{lx_lmir}), has a very low
upper limit in X-ray to MIR luminosity ratio. Notably, for this
source the Keck optical
spectroscopy reveals a narrow line Seyfert 1 (NL~Sy1) spectrum, and we
measure a very
steep $0.5$--$10$~keV X-ray spectrum ($\Gamma_{\rm eff}=3.1$). 
Given these properties, we interpret the low X-ray to MIR ratio as
likely being driven by intrinsic X-ray weakness (in combination with the
steep X-ray spectrum), rather than being driven by extreme absorption
levels. Intrinsic X-ray weakness has previously been identified for
objects in the NL~Sy1 class (e.g.,
\citealt{Miniutti12,Leighly07a,Leighly07b}). A detailed discussion of
J0650 is provided in Section A.5.

At $2$--$10$~keV, the sample shows evidence for significant downwards deviations from
the intrinsic relations, although there is little overlap with the 
known heavily absorbed and CT sources which have been observed in
targetted \nustar programs. 
We note however that this analysis is currently limited to a specific subset of the
serendipitous survey (i.e., sources which are AGN-dominated at MIR
wavelengths, and which are detected in the relevant \wise
bands). Future SED modelling of the broader spectroscopically
identified sample would allow reliable \Lsixum measurements
(disentangling AGN and host galaxy MIR emission) for a more complete
subset of the serendipitous survey sample.

\section{Summary}
\label{summary}

The high sensitivity of \nustar at $\gtrsim10$~keV has provided access to
large samples of high-energy X-ray emitting AGNs in the distant universe, whereas previous observatories
were largely restricted to the local universe ($z\lesssim0.1$). In
this paper we
have presented the first full catalog for the \nustar serendipitous
survey, the largest survey undertaken with \nustar, which incorporates
data from the first 40~months of telescope
operation. The data include $331$ unique fields, with a
total areal coverage of $13\ \mathrm{deg}^{2}$, and a cumulative
exposure time of $\approx 20$~Ms. We have characterized the \nustar
detected AGNs in terms of their X-ray, optical, and IR properties.
Below we summarize the main results:

\begin{itemize}

\item Overall, we detect $497$ sources which are significant
  post-deblending (i.e., after accounting for
  contamination of the photon counts from nearby sources). 
  Of these, $163$ are independently detected in
  the hard ($8$--$24$~keV) energy band; see Section \ref{nu_cat}.

\item The median vignetting-corrected exposure time per source (for the combined
  FPMA+FPMB data) is $\left< t_{\rm net}\right>=60$~ks, and the
  maximum is $1500$~ks. The X-ray fluxes span from $f_{\rm 3-24keV}\approx 10^{-14}$ to
  $10^{-11}$~\fluxunit, with a median value of $\left< f_{\rm 3-24keV}
    \right> = 1.1\times 10^{-13}$~\fluxunit; see Section
    \ref{basic_properties}. The survey reaches flux depths similar to
    the \nustar surveys in well-studied fields (COSMOS, ECDFS, EGS,
    GOODS-N, and UDS)
    over comparable areas (see Section \ref{nu_srcdet}), and is
    $\approx$two orders of magnitude fainter than the \swiftbat surveys; e.g., see
    Section \ref{redshifts_luminosities}.

\item There is a large range in the observed band ratios of AGNs at
  $3$--$24$~keV, which imply a range of effective photon indices going from
  very soft ($\Gamma_{\rm eff}\approx 3$) to very hard ($\Gamma_{\rm eff}\approx 0$); see
  Section \ref{br}. We find no evidence for an
  anticorrelation between band ratio and count rate, as has previously been
  found for lower energy X-ray bands; see Section \ref{br}.

\item A large fraction $79\%$ ($395/497$) of the sources have soft ($<10$~keV)
  X-ray counterparts detected in surveys or archival
data from \xmm, \chandra, and \swiftxrt. 
The \nustar fluxes and the soft X-ray
counterpart fluxes show good agreement for the $3$--$8$~keV energy band, and the maximum
identified variation in AGN flux between the soft X-ray and \nustar observations is a factor of $\approx$five; see Section \ref{cparts_xray}.
The higher positional
accuracies of the soft X-ray observatories relative to \nustar allow us to
reliably match to optical and IR counterparts; see Section \ref{cparts_optIR}.

\item Optical spectroscopic identifications (i.e., redshift
  measurements and source classifications) have been successfully
  obtained for $276$ sources. For the large majority of the sample
  ($222$ sources) this was achieved through
  our extensive campaign of ground-based spectroscopic followup, using a
  range of observatories at multiple geographic latitudes; see Section
  \ref{spec_opt}. $16$ sources are
  spectroscopically confirmed as Galactic. Of the $260$ extragalactic
  sources (AGNs), $162$ ($62.3\%$) 
 are classified as BLAGNs, $97$ ($37.3\%$) 
  are NLAGNs, and one ($0.4\%$) is a BL~Lac; see Section
  \ref{opt_analysis}. While similar numbers of NLAGNs and BLAGNs are
  identified at lower redshifts ($z\lesssim1$) there is a
  bias towards detections of BLAGNs at higher redshifts; this bias has
  been well established for other X-ray missions (e.g., \chandra and
  \xmm); see Section \ref{redshifts_luminosities}.

\item The serendipitous survey AGNs have redshifts covering a wide range, from $z=0.002$ to
  $3.4$, with a median of $\left< z \right> = 0.56$.
 The rest-frame $10$--$40$~keV luminosities also span a wide range, from $L_{\rm
  10-40keV}\approx 10^{39}$ to $10^{46}$~\ergpersec, with a median
value of $\left< L_{\rm 10-40keV} \right> =
10^{44.1}$~\ergpersec. Previous X-ray missions with sensitivity at $>10$~keV
were able to sample the AGN population below the knee of the X-ray
luminosity function ($L_{\star}$) for redshifts up to $z\approx 0.05$, and \nustar extends
this to $z\approx 1$; see Section
  \ref{redshifts_luminosities}. 

\item We present the X-ray--optical flux plane for the optical $R$
  band, and the $8$--$24$~keV and $3$--$24$~keV \nustar bands. 
  The majority of sources have $f_{\rm X}/f_{\rm opt}$ values
  consistent with those expected for AGNs based on the findings of
  previous low energy ($<10$~keV) X-ray observatories. We find a
  strong, positive correlation between $f_{\rm X}/f_{\rm opt}$ and X-ray
  luminosity, in agreement with results at $<10$~keV. We also
  find evidence for significant differences in $f_{\rm X}/f_{\rm opt}$
  between the BLAGNs
  and NLAGNs; see Section \ref{xray_opt}. 

\item We measure a \typeii AGN fraction of $53^{+14}_{-15}\%$ for an effectively spectroscopically
  complete subset of the hard band ($8$--$24$~keV) selected sample at
  $0.1<z<0.5$ and with $2\times 10^{43}<L_{\rm 10-40keV}<2\times
  10^{44}$~\ergpersec. Comparing with luminosity-matched $z<0.05$ AGNs
  selected by the \swiftbat survey, the \nustar-measured \typeii
  fraction for distant AGNs is higher, but consistent within the uncertainties.
  However, the \nustar-measured and \swiftbat-measured \typeii
  fractions appear to be systematically higher than
  those measured for redshift- and luminosity-matched AGNs selected by
  $<10$~keV X-ray missions (e.g., \chandra and \xmm); see Section \ref{type2fraction}.

\item We compare the distribution of \wise \wonewtwo and \wtwowthree colors for \nustar
  AGNs with commonly applied MIR color-selection techniques. The
  fraction of \nustar AGNs which would be selected as AGNs based on
  the MIR colors alone is a strong function of X-ray luminosity, in
  agreement with findings for low energy ($<10$~keV) X-ray selected samples. 
  The fraction of \nustar AGNs missed by MIR
  color-selections is large, ranging between $\approx 15\%$ and $\approx 80\%$
  for the highest luminosities ($L_{\rm X}>10^{44}$~\ergpersec) and
  the lowest luminosities ($L_{\rm X}<10^{43}$~\ergpersec), respectively; see Section
  \ref{wise_colors}. It is notable that a number of luminous \nustar-selected
  BLAGNs are not selected in the MIR, and that this appears to be driven by
  the intrinsic AGN properties; see Section \ref{wise_colors}.

\item We present the X-ray--MIR luminosity plane for sources which are
  AGN-dominated at MIR wavelengths. For both the
  rest-frame $2$--$10$~keV and $10$--$40$~keV bands the large majority of
  the sources are consistent with being scattered around the intrinsic
  $L_{\rm X}$--$L_{\rm 6\mu m}$ relation; see Section
  \ref{xray_mir}. One source is highlighted as having an extremely low
  $L_{\rm 10-40keV}/L_{\rm 6\mu m}$ ratio (J0650; $z=0.32$; $L_{\rm 6\mu m}\approx
  4\times 10^{44}$~\ergpersec; $L_{\rm 10-40keV}<2\times 10^{43}$~\ergpersec). A
  detailed investigation reveals a narrow-line Seyfert 1, likely
  to be intrinsically X-ray weak as opposed to heavily obscured; see
  Section \ref{xray_mir}.

\end{itemize}

The \nustar serendipitous survey presented herein is the largest
sample of distant AGNs selected with a focusing high energy ($\gtrsim
10$~keV) X-ray observatory. 
As the \nustar science operations continue into the future, the
serendipitous survey will continue to grow at a similar rate, and
is likely to eventually achieve a sample size on the order of $\gtrsim 1000$
sources. This will result in improved statistical
constraints on the overall properties of the hard X-ray
emitting source population, and will facilitate the discovery of rare and
extreme sources not sampled as effectively by the smaller-area dedicated
\nustar surveys (e.g., in the COSMOS, ECDFS, EGS, GOODS-N,
and UDS fields). 
A continued program of followup observations will be necessary to
maximize the effectiveness of the serendipitous survey.

\section*{Acknowledgements}

The authors firstly thank the anonymous referee for the constructive comments.
We acknowledge financial support from: the Science and Technology
Facilities Council (STFC) grants ST/K501979/1 (G.B.L.), ST/I001573/1 (D.M.A.), and ST/J003697/2 (P.G.); 
a Herchel Smith Postdoctoral Fellowship of the University of Cambridge (G.B.L.);
the ERC Advanced Grant FEEDBACK 340442 at the University of Cambridge (J.A.); 
a COFUND Junior Research Fellowship from the Institute of Advanced Study, Durham University (J.A.); 
the Leverhulme Trust (D.M.A.);
CONICYT-Chile grants FONDECYT 1120061 and 1160999 (E.T.), 3140534 (S.S.), and Anillo ACT1101 (E.T. and F.E.B.); 
the Center of Excellence in Astrophysics and Associated Technologies (PFB 06; E.T. and F.E.B.);
the NASA Earth and Space Science Fellowship Program, grant NNX14AQ07H (M.B.).
We extend gratitude to Felipe Ardila, Roberto Assef, Eduardo Ba\~nados, Stanislav George Djorgovski, Andrew Drake, Jack Gabel, Audrey Galametz, Daniel Gawerc, David Girou, Marianne Heida, Nikita Kamraj, Peter Kosec, Thomas Kr\"uhler, Ashish Mahabal, Alessandro Rettura, and Aaron Stemo for their support during the ground-based follow-up observations.
We thank John Lucey for unearthing the J1410 spectrum, and Sophie Reed, David Rosario, Mara Salvato and Martin Ward for the informative discussions.
Additional thanks to Eden Stern for lending a hand during the August 2015 Keck run.
This work was supported under NASA Contract No.\ NNG08FD60C, and made use of data from the \nustar mission, a project led by the California Institute of Technology, managed by the Jet Propulsion Laboratory, and funded by the National Aeronautics and Space Administration. We thank the \nustar Operations, Software and Calibration teams for support with the execution and analysis of these observations. This research has made use of the \nustar Data Analysis Software (NuSTARDAS) jointly developed by the ASI Science Data Center (ASDC, Italy) and the California Institute of Technology (USA).

Facilities: \chandra, ESO La Silla, Gemini, Keck, Magellan, \nustar, Palomar, SDSS, {\it Swift}, {\it WISE}, \xmm.

\bibliography{bibliography.bib}{}

\appendix
\label{appendix}

\section{A.1. Description of the {\it NuSTAR} Serendipitous Survey
  Source Catalog}

The \nustar serendipitous survey source catalog, containing $498$ sources in
total, is available as an electronic table. Here we
describe the columns of the catalog, which are summarized in Table \ref{primary_columns_table}.

\begin{table}
\centering
\caption{Column Descriptions for the Primary {\it NuSTAR} Serendipitous Source Catalog}
\begin{tabular}{ll} \hline\hline \noalign{\smallskip}
Column number & Description \\
\noalign{\smallskip} \hline \noalign{\smallskip}
1 & Unique source identification number (ID). \\
2 & Unique \nustar source name. \\
3, 4 & Right ascension (R.A.) and declination (decl.). \\
5--7 & Flags indicating the energy bands for which the source is detected. \\
8--10 & Same as columns 5--7, post-deblending. \\
11--13 & Logarithm of the false probabilities for the three standard
        energy bands. \\
14--16 & Same as columns 11--13, post-deblending. \\
17 & Flag indicating whether the source is significant post-deblending,
  for at least one energy band. \\
18--32 & Total, background, and net source
        counts for the three standard
        energy bands, and associated errors. \\
33--44 & Same as columns 18--32, post-deblending. \\
45--47 & Net vignetting-corrected exposure times at the source
         position, for the combined A+B data. \\
48--62 & Total, background, and net source
        count rates for the three standard
        energy bands, and associated errors. \\
63--68 & Deblended net source
        count rates for the three standard
        energy bands, and associated errors. \\
69--71 & Band ratio and upper and lower errors. \\
72--74 & Effective photon index and upper and lower errors. \\
75--80 & Deblended fluxes in the three standard bands and associated
         errors. \\
81 & Reference for the adopted lower-energy X-ray (\chandra, \xmm or \swiftxrt) counterpart. \\
82, 83 & R.A.\ and decl.\ of the lower-energy X-ray counterpart. \\
84 & Angular separation between the \nustar and lower-energy X-ray
     counterpart positions. \\
85 & $3$--$8$~keV (3XMM or CSC) flux of the lower-energy X-ray counterpart. \\
86 & Total $3$--$8$~keV flux of all (3XMM or CSC) sources within $30\arcsec$ of the \nustar position. \\
87, 88 & R.A.\ and decl.\ of the adopted \wise counterpart. \\
89 & Angular separation between the \nustar and \wise
     counterpart positions. \\
90--97 & \wise magnitudes in the four standard bands and associated errors. \\
98 & Reference for the adopted optical counterpart. \\
99, 100 & R.A.\ and decl.\ of the optical counterpart. \\
101 & Angular separation between the \nustar and optical
     counterpart positions. \\
102 & $R$-band magnitude for the optical counterpart. \\
103 & Spectroscopic redshift. \\
104 & Non-absorption-corrected, rest-frame $10$--$40$~keV luminosity. \\
105 & Binary flag to indicate sources associated with the
      primary science targets of their respective \nustar fields. \\
106 & Binary flag to indicate the sources used for the \citet{Aird15b}
      study. \\
\noalign{\smallskip} \hline \noalign{\smallskip}
\end{tabular}
\begin{minipage}[l]{0.76\textwidth}
\footnotesize
\textbf{Notes.} The full catalog is available as a machine readable
electronic table.
\end{minipage}
\label{primary_columns_table}
\end{table}

{\it Column 1}: the unique source identification numbers (ID), in order of
increasing right ascension (R.A.). 

{\it Column 2}: the unique \nustar source names, following the
IAU-approved format: NuSTAR~JHHMMSS$\pm$DDMM.m, where m is the truncated
fraction of one arcminute for the arcseconds component of the
declination (decl.). 

{\it Columns 3, 4}: the \nustar R.A.\ and decl.\ coordinates (J2000), as
described in Section \ref{nu_srcdet}.

{\it Columns 5--7}: a binary flag indicating whether the source is
detected with a false probability lower than our
threshold of $\log(P_{\rm False})=-6$, for the soft ($3$--$8$~keV),
hard ($8$--$24$~keV), and full ($3$--$24$~keV) bands. These three
bands are abbreviated as SB, HB, and FB, respectively, throughout the
source catalog.

{\it Columns 8--10}: the same as columns 5--7, after deblending has
been performed to account for contamination of the source counts from
very nearby sources (see Section \ref{nu_phot} of this paper, and Section 2.3.2
of M15. Deblending only affects a very small
  fraction of the overall sample (e.g., see Section \ref{nu_phot}).

{\it Columns 11--13}: the logarithm of the false probabilities ($P_{\rm False}$) of the \nustar detected
sources, for the three standard energy bands (see Section \ref{nu_srcdet}).

{\it Columns 14--16}: the same as columns 11--13, after deblending has
been performed.

{\it Column 17}: a binary flag indicating whether the \nustar detected
source remains significant after deblending, in at least one of the
three standard energy bands.

{\it Columns 18--32}: photometric quantities, calculated at the
source coordinates in columns 3 and 4, and using a source aperture of
$30''$ radius (see Section \ref{nu_phot}). The values are
non-aperture-corrected; i.e., they correspond to the $30''$ values,
and have not been corrected to the full PSF values. We provide the total counts
(i.e., all counts within the source aperture) and associated errors
($84\%$ CL), the background counts scaled to the source aperture, and the net
source counts (i.e., total minus background) and associated
errors. For the net source counts, we give $90\%$~CL upper
limits for sources not detected in a given band. Throughout the table, upper
limits are flagged with a $-99$ value in the error column.

{\it Columns 33--44}: the same as columns 18--32, after deblending has
been performed.

{\it Columns 45--47}: the average net, vignetting-corrected exposure
time at the source coordinates (columns 3 and 4), for each energy
band. These correspond to the A+B data, so should be divided by two to
obtain the average exposure per FPM. Units: $\mathrm{s}$.

{\it Columns 48--62}: the non-aperture-corrected total, background, and net count rates (and
associated errors; $84\%$ CL) determined from the photometric values in columns
18--32, and the exposure times in columns 45--47. Units: $\mathrm{s^{-1}}$.

{\it Columns 63--68}: the deblended net count rates, and associated
errors, determined from the photometric values in columns
33--44, and the exposure times in columns 45--47. Units: $\mathrm{s^{-1}}$.

{\it Columns 69--71}: the \nustar band ratio (\brnu) and associated errors, as described in
Section \ref{nu_phot}. Upper limits, lower limits, and sources with no
constraints are flagged with $-99$, $-88$, and $-77$ values, respectively, in the error columns. 

{\it Columns 72--74}: the effective photon index ($\Gamma_{\rm eff}$),
and associated errors, estimated from the band ratio values in columns
69--71 (see Section \ref{nu_phot}).

{\it Columns 75--80}: the observed-frame fluxes and associated errors
($84\%$ CL) for the three
standard energy bands, after deblending has been performed. These are
aperture corrected values (i.e., they correspond to the full \nustar
PSF), and are calculated from the count rates in columns 63--68 using the
conversion factors listed in Section \ref{nu_phot}. Units:
$\mathrm{erg\ s^{-1}\ cm^{-2}}$.

{\it Column 81}: an abbreviated code indicating the origin of the
adopted soft (i.e., low energy; $<10$~keV) X-ray counterpart. CXO\_CSC indicates
counterparts from the \chandra Source Catalog (CSC;
\citealt{Evans10}). XMM\_3XMM indicates counterparts from the third
\xmm serendipitous source catalog (3XMM;
\citealt{Watson09,Rosen16}). CXO\_MAN, XMM\_MAN, and XRT\_MAN indicate
sources manually identified using archival \chandra, \xmm,
and \swiftxrt data, respectively. Section \ref{cparts_xray} details
the counterpart matching.

{\it Columns 82, 83}: the R.A.\ and decl.\ coordinates (J2000) of the
soft X-ray counterpart.

{\it Column 84}: the angular offset between the \nustar position
(columns 3 and 4) and the soft X-ray counterpart position (columns 82
and 83). Units: $\mathrm{arcsec}$.

{\it Column 85}: the observed-frame $3$--$8$~keV flux of the soft
X-ray counterpart, for sources with counterparts in the CSC and 3XMM
catalogs. For CSC sources we convert to the $3$--$8$~keV flux from the $2$--$7$~keV flux using
a conversion factor of $0.83$, and for the 3XMM sources we convert from the
$4.5$--$12$~keV flux using a conversion factor of $0.92$. Units:
$\mathrm{erg\ s^{-1}\ cm^{-2}}$.

{\it Column 86}: the total combined $3$--$8$~keV flux of all (3XMM or CSC)
sources within $30''$ of the \nustar position. Units:
$\mathrm{erg\ s^{-1}\ cm^{-2}}$.

{\it Columns 87, 88}: the R.A.\ and decl.\ coordinates (J2000) of the \wise
counterpart, if there is a match in the \wise all-sky
survey catalog \citep{Wright10}. Section \ref{cparts_optIR} details the \wise counterpart matching.

{\it Column 89}: the angular offset between the \nustar position
(columns 3 and 4) and the \wise counterpart position (columns 87
and 88). Units: $\mathrm{arcsec}$.

{\it Columns 90--97}: the \wise profile-fit magnitudes (and associated errors), for the four standard \wise bands: \wone ($\lambda\approx 3.4$~$\mu$m), \wtwo ($\approx 4.6$~$\mu$m), \wthree
($\approx 12$~$\mu$m), and \wfour ($\approx 22$~$\mu$m). NaN
  error values
indicate \wise upper limits. Units: Vega $\mathrm{mag}$.

{\it Column 98}: an abbreviated code indicating the origin of the
adopted optical counterpart to the \nustar source. 
The code SDSS
indicates sources with soft X-ray counterparts and successful matches
in the SDSS DR7 catalog \citep{York00}. The code USNO indicates
sources with soft X-ray counterparts and successful matches in the
USNOB1 catalog \citep{Monet03}. MAN indicates sources with a soft X-ray
counterpart and a corresponding optical counterpart manually identified in the available optical
coverage. SDSS\_WISE and USNO\_WISE indicate the cases where there is
no soft X-ray counterpart to the \nustar position, but a \wise AGN candidate is
identified within the \nustar error circle and successfully matched to
the SDSS DR7 or USNOB1 catalog (these are mainly used
  as candidates for spectroscopic followup).
We give a detailed description of the
procedure used to identify optical counterparts in Section \ref{cparts_optIR}.

{\it Columns 99, 100}: the R.A.\ and decl.\ coordinates (J2000) of the optical
counterpart, for the sources with SDSS DR7 and USNOB1 matches.

{\it Column 101}: the angular offset between the \nustar position
(columns 3 and 4) and the optical counterpart position (columns 99
and 100). Units: $\mathrm{arcsec}$.

{\it Column 102}: the $R$-band magnitude of the optical
counterpart. For the SDSS DR7 matches, this is calculated as
$R=r-0.16$. 
For the USNOB1 matches, this is taken as the mean of
the two independent photographic plate measurements, R1mag and
R2mag. For the manual identifications, the magnitude is taken from
another optical catalog or manually determined from the imaging data. 
Units: Vega $\mathrm{mag}$.

{\it Column 103}: the spectroscopic redshift of the \nustar
source. The large majority of the redshifts were obtained through our
own campaign of ground-based spectroscopic followup of \nustar serendipitous survey sources
(see Section \ref{ground}).

{\it Column 104}: the rest-frame $10$--$40$~keV luminosity, estimated
from the fluxes in columns 75--80, following the procedure outlined in
Section \ref{nu_phot}. Negative values indicate upper limits. The
luminosities are observed values, uncorrected for any absorption along
the line of sight. The intrinsic luminosities may therefore be higher,
for highly absorbed AGNs. Units: $\mathrm{erg\ s^{-1}}$.

{\it Column 105}: a binary flag indicating the few sources which show
evidence for being associated with the primary science targets of their
respective \nustar observations, according to the definition in Section
\ref{nu_srcdet} [$\Delta (cz) < 0.05cz$].

{\it Column 106}: a binary flag highlighting the sources used in the
\citet{Aird15b} study.

\section{A.2. Optical spectroscopic properties of individual objects}

Here we provide details of the optical spectroscopic properties of
individual sources from the \nustar serendipitous survey. As described
in Section \ref{spec_opt}, these largely result from our dedicated
followup campaign using the Keck, Magellan, NTT, and Palomar
facilities, and also from existing publically available spectroscopy
(primarily SDSS spectroscopy). Individual source spectra
($F_{\mathrm{\nu}}$ versus $\lambda$) are shown in Figure
\ref{spec_all}, and details for individual sources are tabulated in
Table \ref{specTable}, the columns of which are as follows: columns 1
and 2 give the unique source identification number and the unique
\nustar source name, as listed in source catalog; columns
3 and 4 give the source redshift and classification (see Section
\ref{opt_analysis}); column 5 lists the emission or absorption lines
identified (the latter are marked with $\mathrm{\dagger}$ symbols),
which are additionally highlighted in the individual panels of Figure \ref{spec_all};
column 6 gives individual object notes, including references for
literature spectra; and column 7 gives the unique observing run
identification number, as defined in Table \ref{runsTable} (``S'' and
``L'' mark spectra obtained from the SDSS and from elsewhere in the
literature, respectively).

\begin{center}
\renewcommand*{\arraystretch}{1.1}
\begin{longtable}{llccllc} 
\caption[Summary of the optical spectroscopy]{Summary of the optical
  spectroscopy for the \nustar serendipitous survey sources. The
  columns are described in Section A.2.} \label{specTable} \\
\hline \hline \noalign{\smallskip} \multicolumn{1}{c}{ID} & \multicolumn{1}{c}{\nustar Name} & \multicolumn{1}{c}{$z$} &
\multicolumn{1}{c}{Type} & \multicolumn{1}{c}{Lines} &
\multicolumn{1}{c}{Notes} & \multicolumn{1}{c}{Run} \\
\multicolumn{1}{c}{(1)} & \multicolumn{1}{c}{(2)} & \multicolumn{1}{c}{(3)} &
\multicolumn{1}{c}{(4)} & \multicolumn{1}{c}{(5)} &
\multicolumn{1}{c}{(6)} & \multicolumn{1}{c}{(7)} \\
\noalign{\smallskip} \hline 
\endfirsthead
\multicolumn{7}{c}%
{\tablename\ \thetable{} -- continued from previous page} \\
\hline \hline \noalign{\smallskip} \multicolumn{1}{c}{ID} & \multicolumn{1}{c}{\nustar Name} & \multicolumn{1}{c}{$z$} &
\multicolumn{1}{c}{Type} & \multicolumn{1}{c}{Lines} &
\multicolumn{1}{c}{Notes} & \multicolumn{1}{c}{Run} \\ \noalign{\smallskip} \hline \noalign{\smallskip}
\endhead
\noalign{\smallskip}
\hline \multicolumn{7}{r}{{Continued on next page}} \\ 
\endfoot
\noalign{\smallskip}
\hline \hline
\endlastfoot
\noalign{\smallskip}
1 & NuSTARJ000011-7652.5 & 0.053 & $\cdot$$\cdot$$\cdot$ & $\cdot$$\cdot$$\cdot$ & \parbox[t]{4.4cm}{\citet{Jones04,Jones09}} & L \\
2 & NuSTARJ001130+0057.8 & 1.492 & BL & \parbox[t]{6cm}{\mbox{\ion{C}{4}} \mbox{\ion{He}{2}} \mbox{\ion{C}{3}]} \mbox{\ion{Mg}{2}} \mbox{[\ion{O}{2}]} \mbox{[\ion{Ne}{3}]} H$\mathrm{\delta}$} & $\cdot$$\cdot$$\cdot$ & S \\
3 & NuSTARJ001442+8131.9 & 0.365 & BL & \parbox[t]{6cm}{\mbox{\ion{Mg}{2}} \mbox{[\ion{Ne}{5}]} \mbox{[\ion{O}{2}]} H$\mathrm{\beta}$ \mbox{[\ion{O}{3}]} H$\mathrm{\alpha}$ \mbox{[\ion{N}{2}]}} & $\cdot$$\cdot$$\cdot$ & 28 \\
4 & NuSTARJ001542+8134.4 & $\cdot$$\cdot$$\cdot$ & $\cdot$$\cdot$$\cdot$ & $\cdot$$\cdot$$\cdot$ & \parbox[t]{4.4cm}{Continuum detected} & 33 \\
6 & NuSTARJ001852-1026.1 & 0.332 & BL & \parbox[t]{6cm}{\mbox{[\ion{Ne}{5}]} \mbox{[\ion{O}{2}]} \mbox{[\ion{Ne}{3}]} H$\mathrm{\delta}$ H$\mathrm{\gamma}$ H$\mathrm{\beta}$ \mbox{[\ion{O}{3}]} H$\mathrm{\alpha}$ \mbox{[\ion{N}{2}]} \mbox{[\ion{S}{2}]}} & $\cdot$$\cdot$$\cdot$ & 31 \\
7 & NuSTARJ001858-1022.7 & 1.172 & NL & \parbox[t]{6cm}{\mbox{\ion{C}{3}]} \mbox{\ion{Mg}{2}} \mbox{[\ion{Ne}{5}]} \mbox{[\ion{O}{2}]} \mbox{[\ion{Ne}{3}]} H$\mathrm{\delta}$ H$\mathrm{\gamma}$} & $\cdot$$\cdot$$\cdot$ & 31 \\
13 & NuSTARJ002227-1854.7 & 0.287 & NL & \parbox[t]{6cm}{\mbox{\ion{Mg}{2}} \mbox{[\ion{Ne}{5}]} \mbox{[\ion{O}{2}]} \mbox{[\ion{Ne}{3}]} H$\mathrm{\delta}$ H$\mathrm{\beta}$ \mbox{[\ion{O}{3}]} H$\mathrm{\alpha}$ \mbox{[\ion{N}{2}]}} & $\cdot$$\cdot$$\cdot$ & 29 \\
14 & NuSTARJ002544+6818.8 & 0.012 & NL & \parbox[t]{6cm}{\mbox{[\ion{O}{3}]} H$\mathrm{\alpha}$ \mbox{[\ion{N}{2}]} \mbox{[\ion{S}{2}]}} & $\cdot$$\cdot$$\cdot$ & 17 \\
\end{longtable}
\end{center}

\begin{figure*}
\centering
\begin{minipage}[l]{0.325\textwidth}
\includegraphics[width=\textwidth]{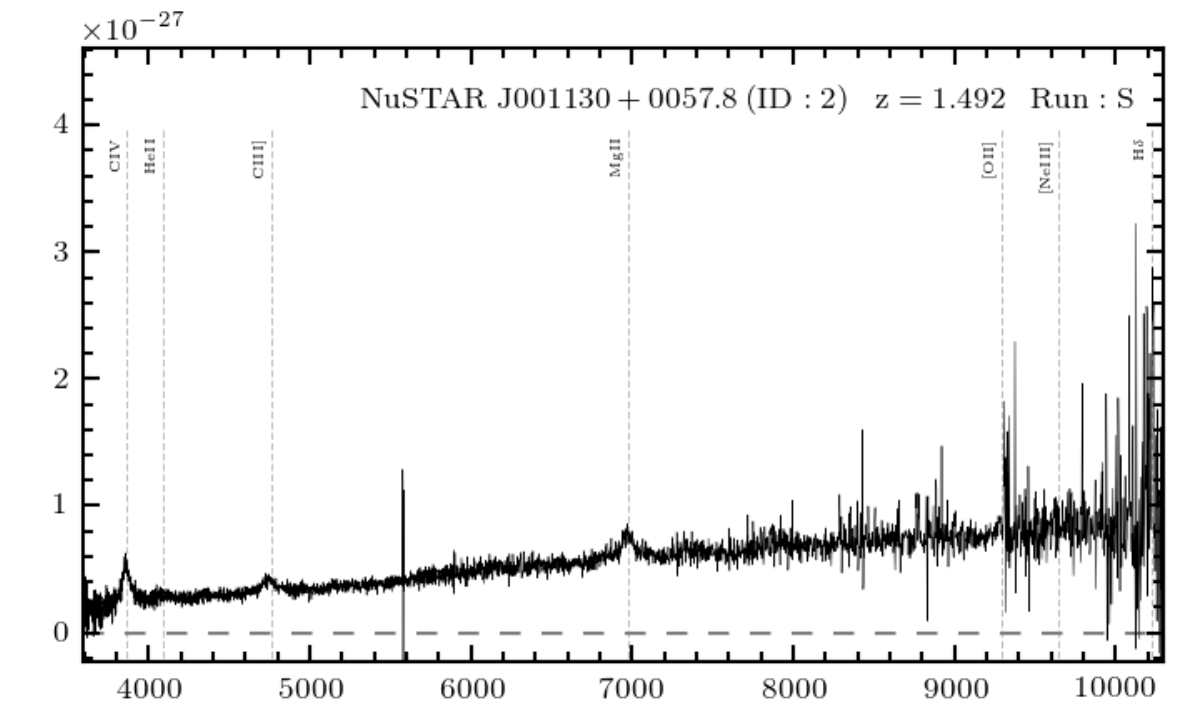}
\end{minipage}
\begin{minipage}[l]{0.325\textwidth}
\includegraphics[width=\textwidth]{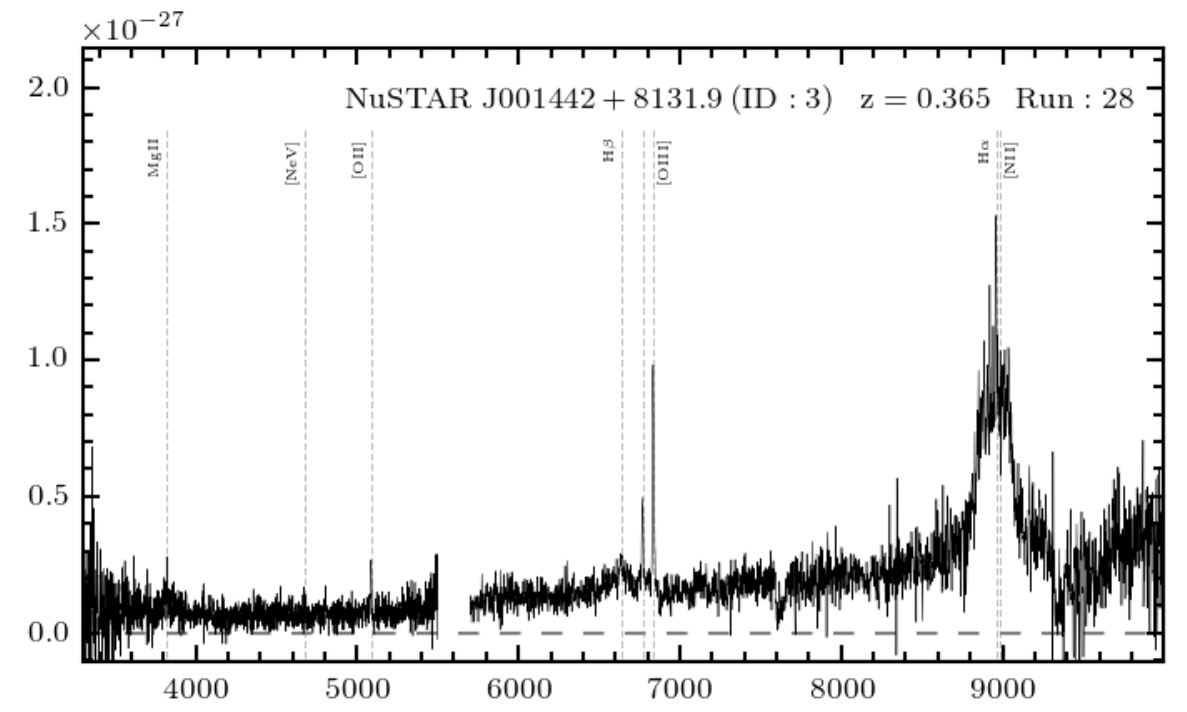}
\end{minipage}
\begin{minipage}[l]{0.325\textwidth}
\includegraphics[width=\textwidth]{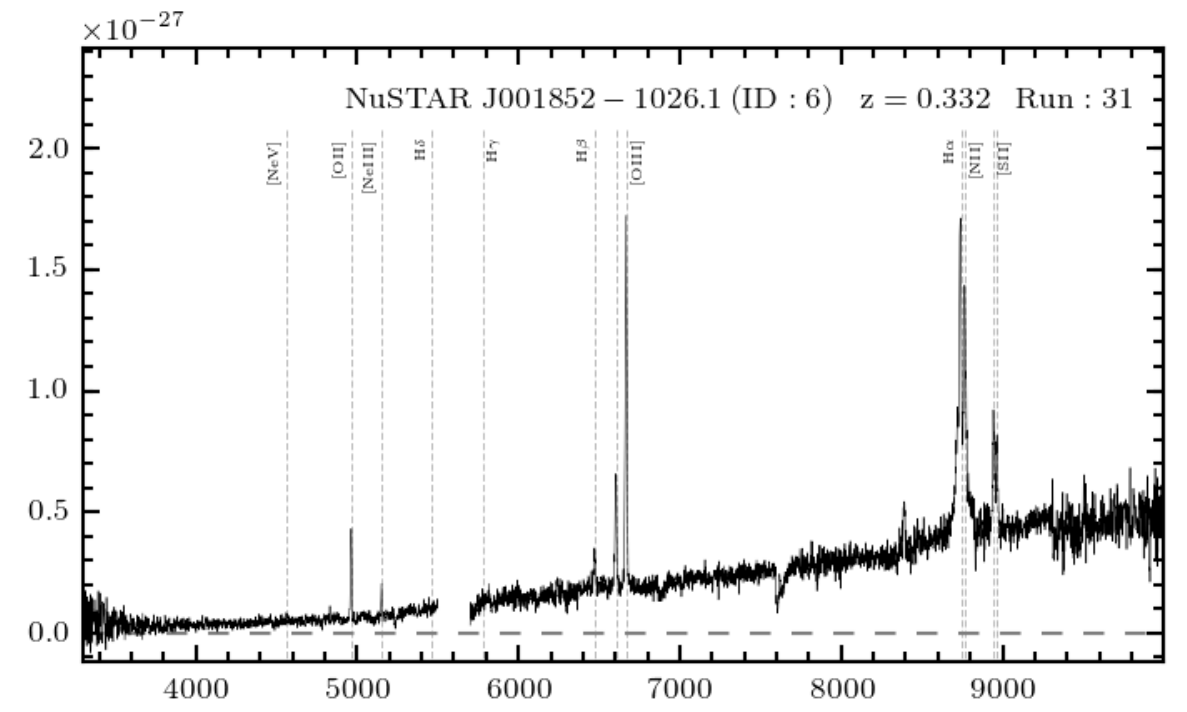}
\end{minipage}
\begin{minipage}[l]{0.325\textwidth}
\includegraphics[width=\textwidth]{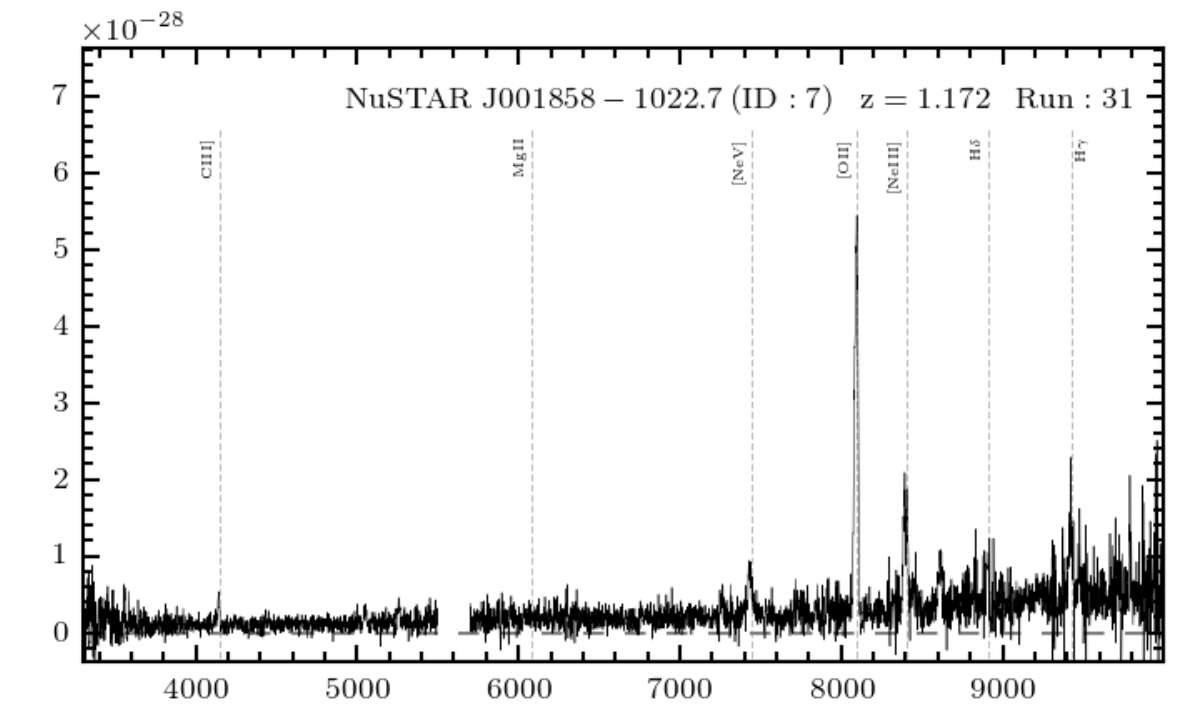}
\end{minipage}
\begin{minipage}[l]{0.325\textwidth}
\includegraphics[width=\textwidth]{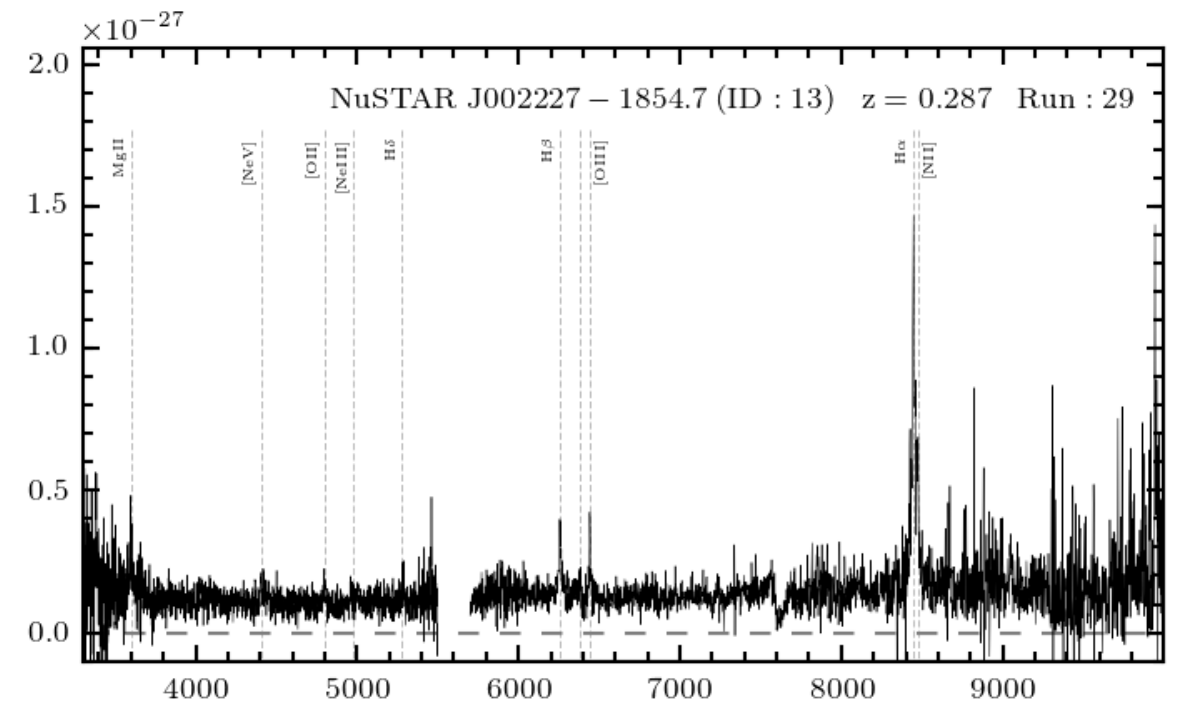}
\end{minipage}
\begin{minipage}[l]{0.325\textwidth}
\includegraphics[width=\textwidth]{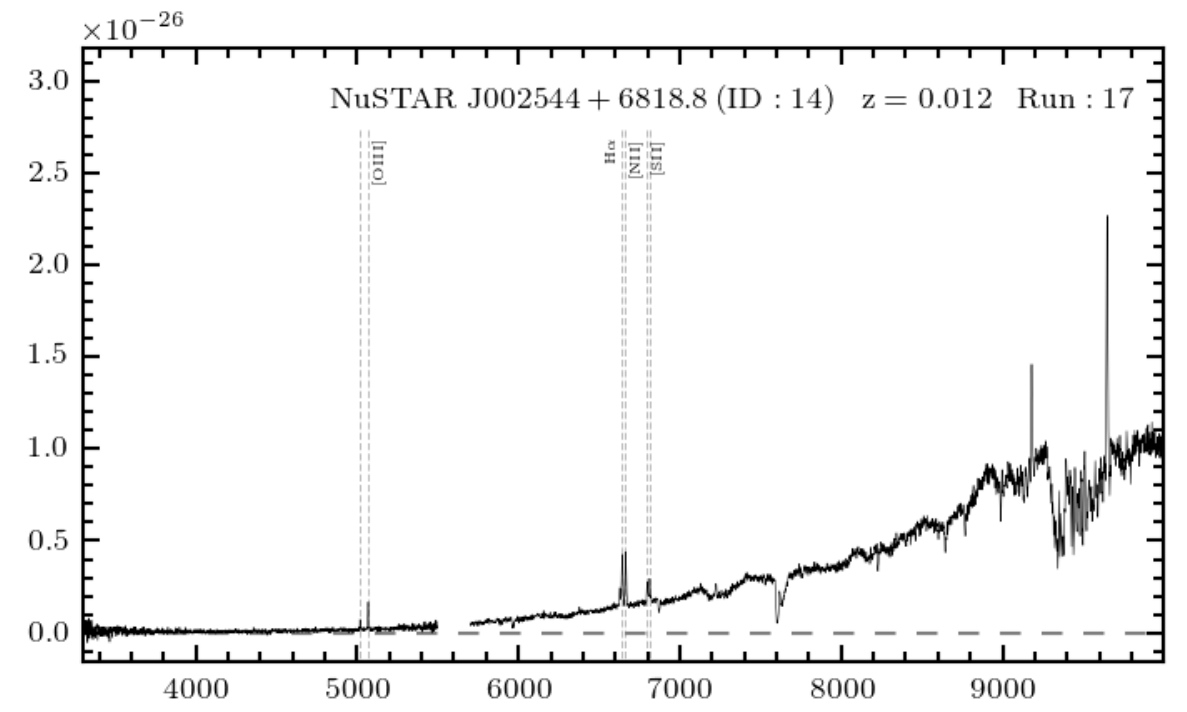}
\end{minipage}
\begin{minipage}[l]{0.325\textwidth}
\includegraphics[width=\textwidth]{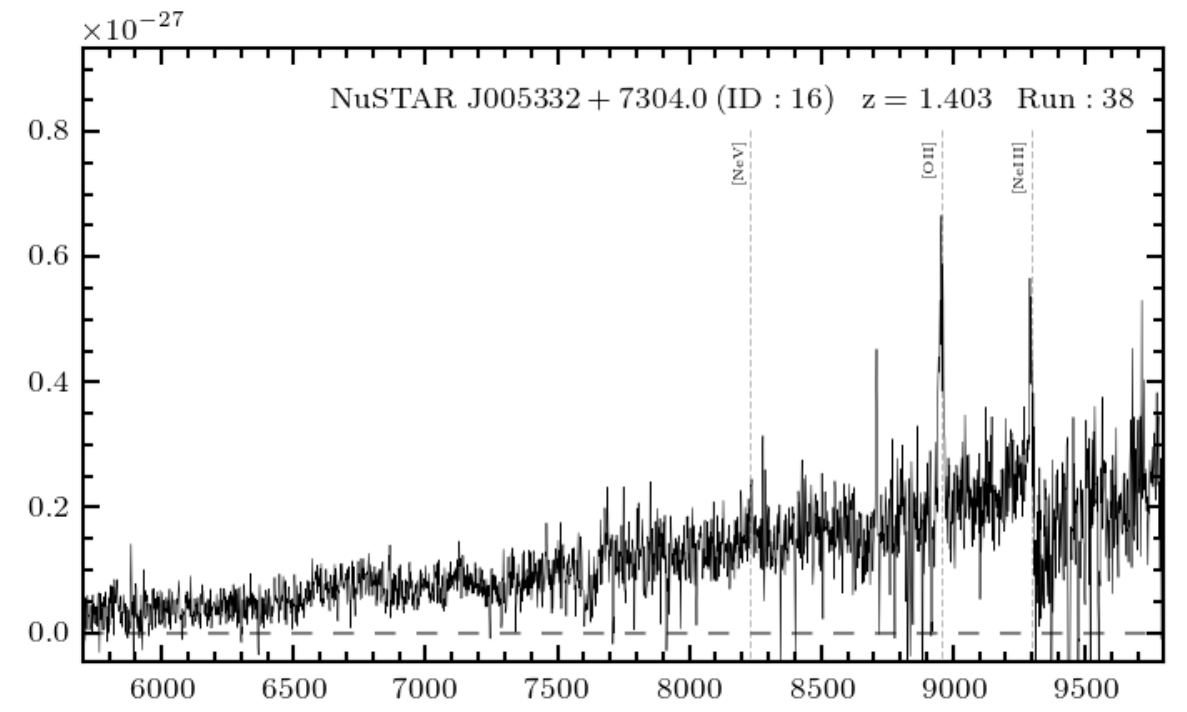}
\end{minipage}
\begin{minipage}[l]{0.325\textwidth}
\includegraphics[width=\textwidth]{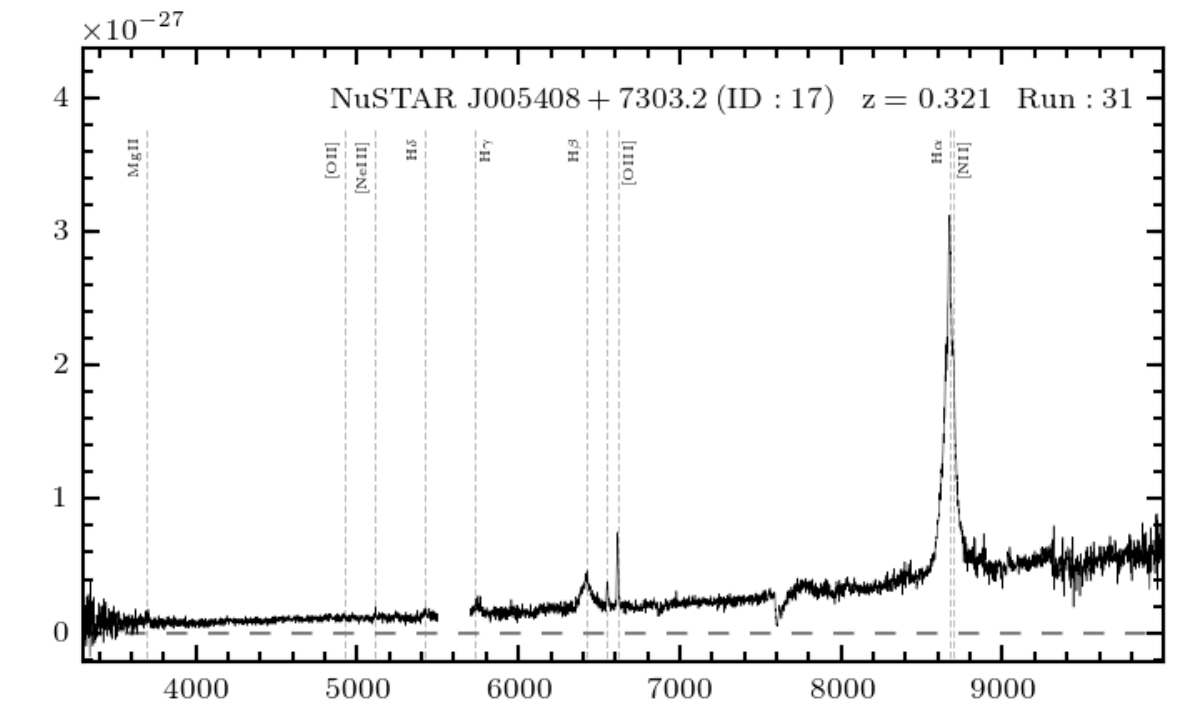}
\end{minipage}
\begin{minipage}[l]{0.325\textwidth}
\includegraphics[width=\textwidth]{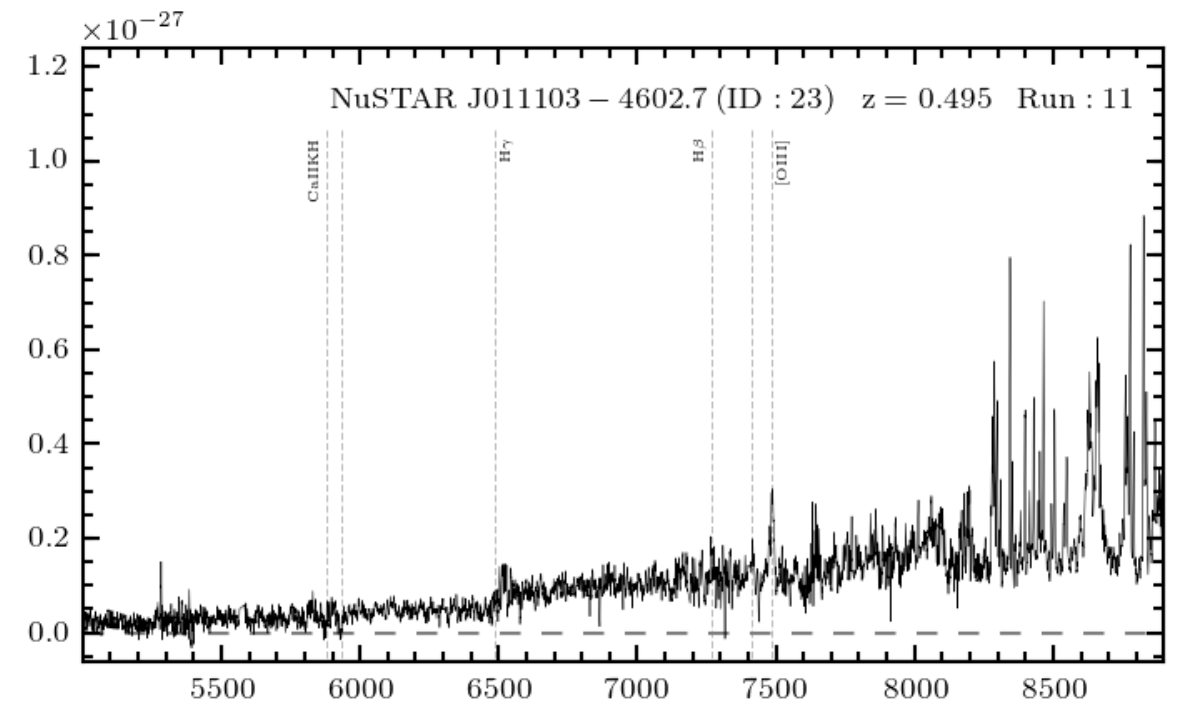}
\end{minipage}
\begin{minipage}[l]{0.325\textwidth}
\includegraphics[width=\textwidth]{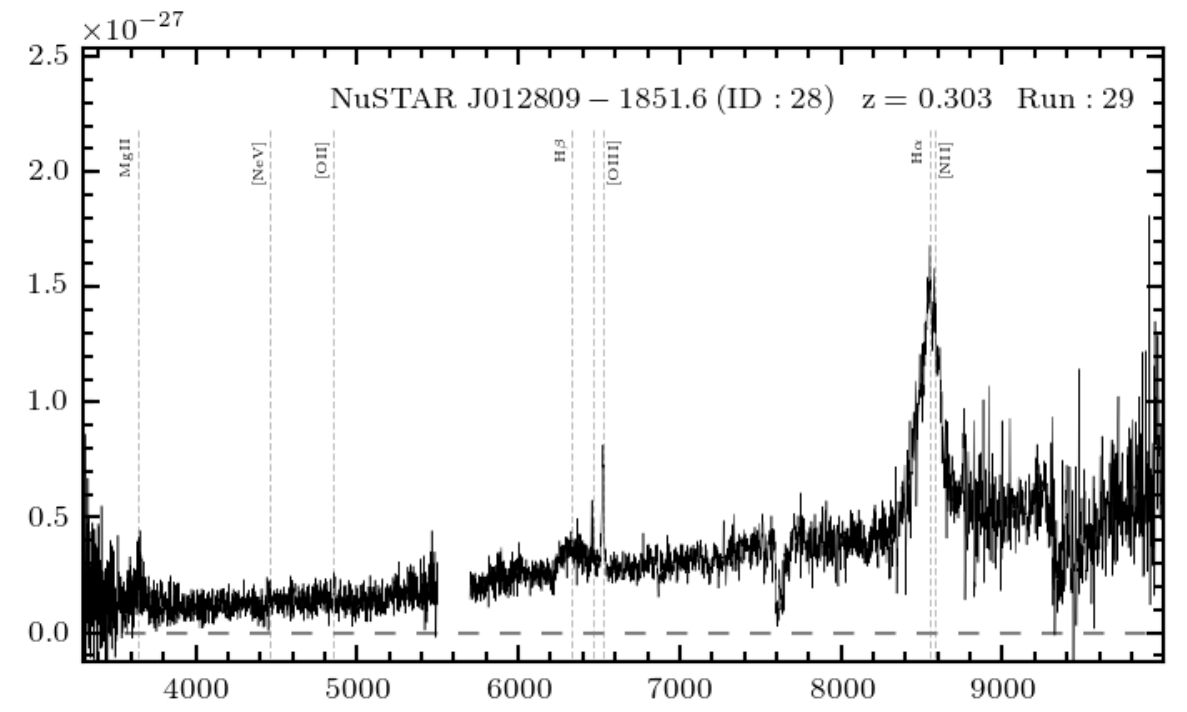}
\end{minipage}
\begin{minipage}[l]{0.325\textwidth}
\includegraphics[width=\textwidth]{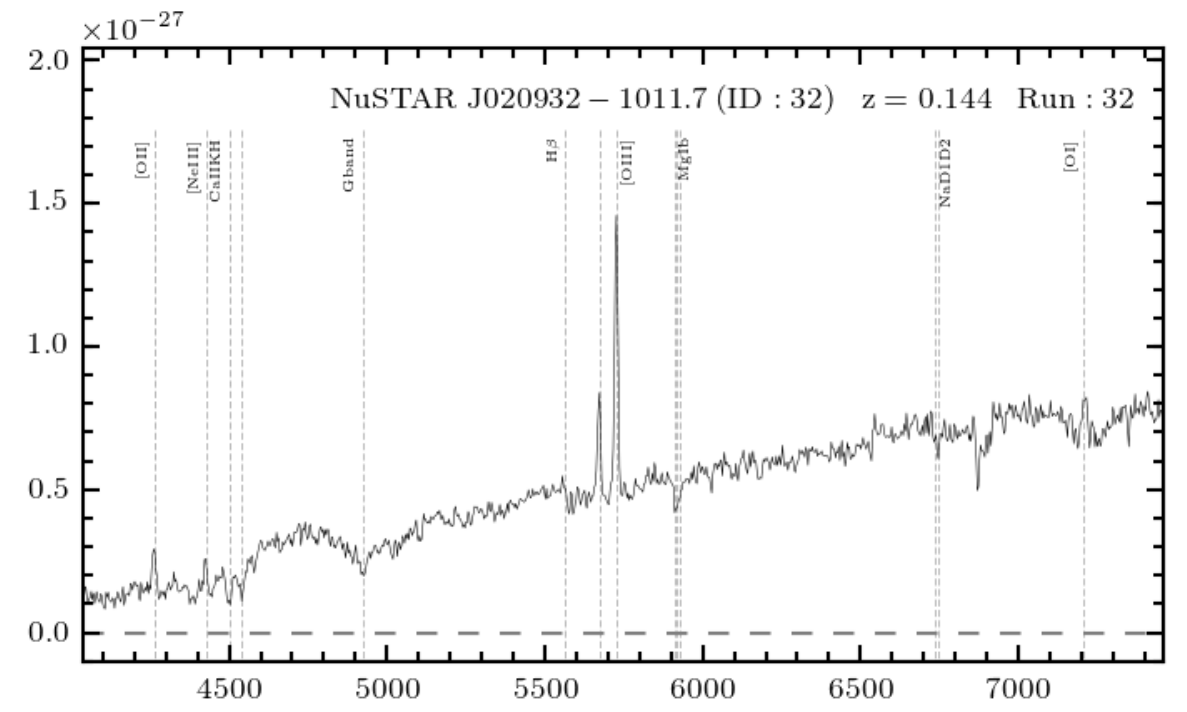}
\end{minipage}
\begin{minipage}[l]{0.325\textwidth}
\includegraphics[width=\textwidth]{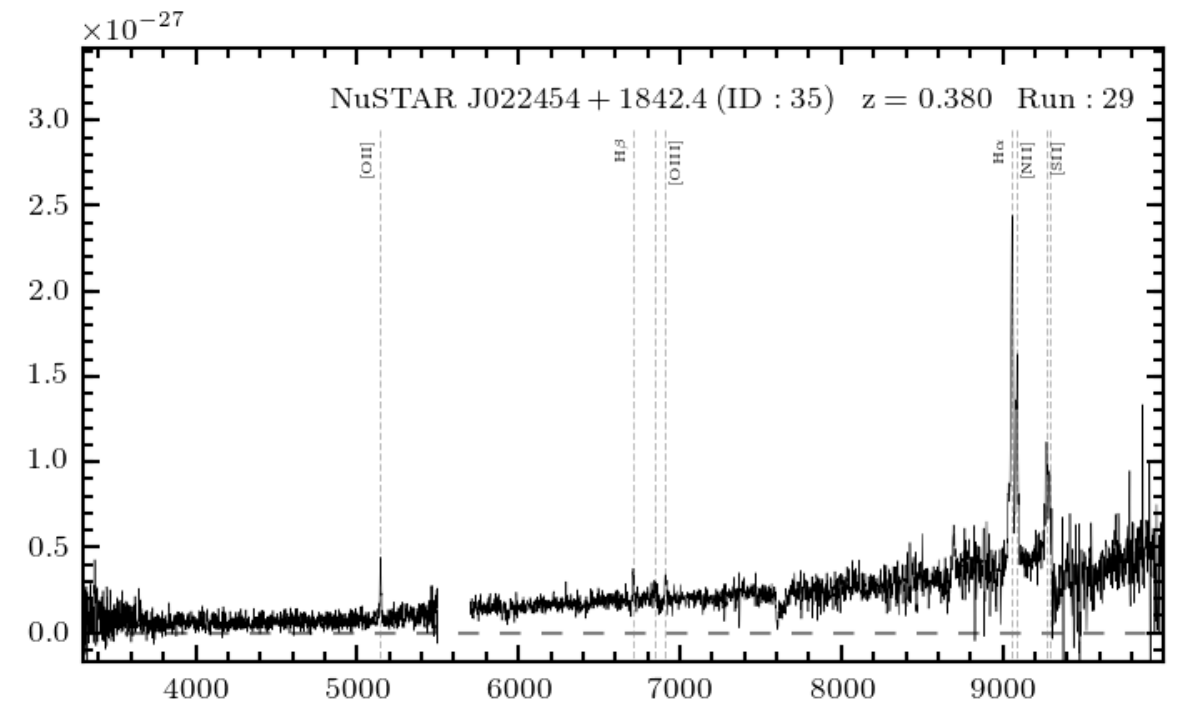}
\end{minipage}
\begin{minipage}[l]{0.325\textwidth}
\includegraphics[width=\textwidth]{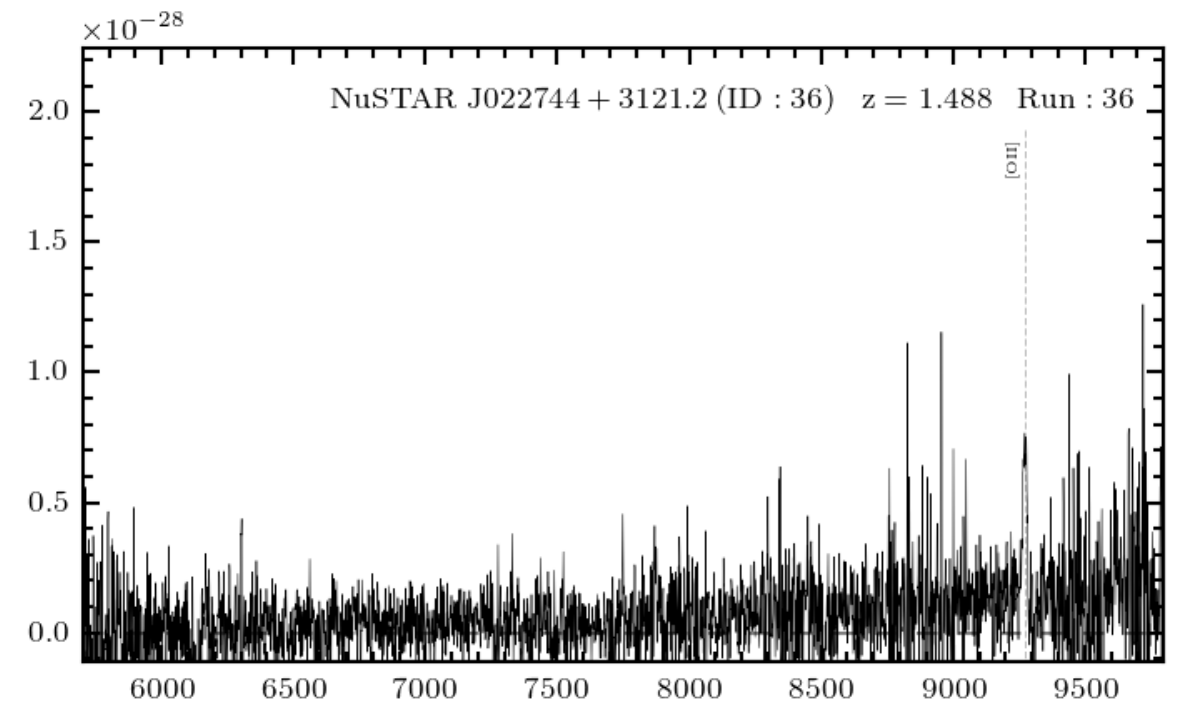}
\end{minipage}
\begin{minipage}[l]{0.325\textwidth}
\includegraphics[width=\textwidth]{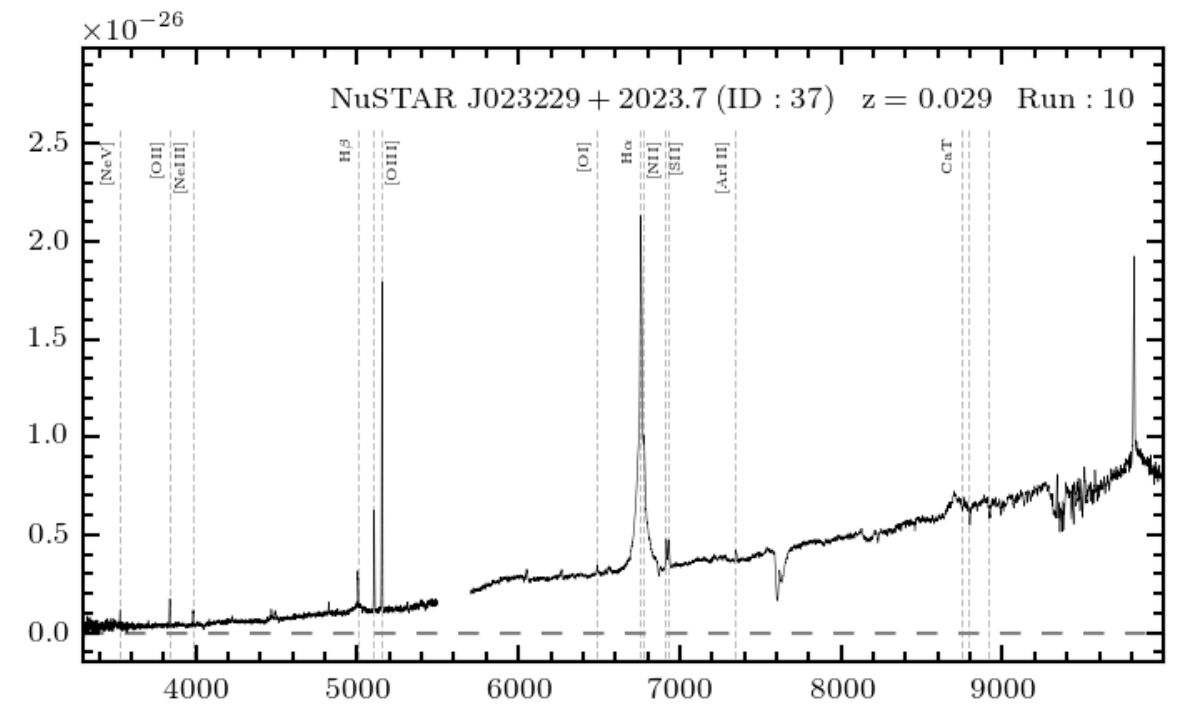}
\end{minipage}
\begin{minipage}[l]{0.325\textwidth}
\includegraphics[width=\textwidth]{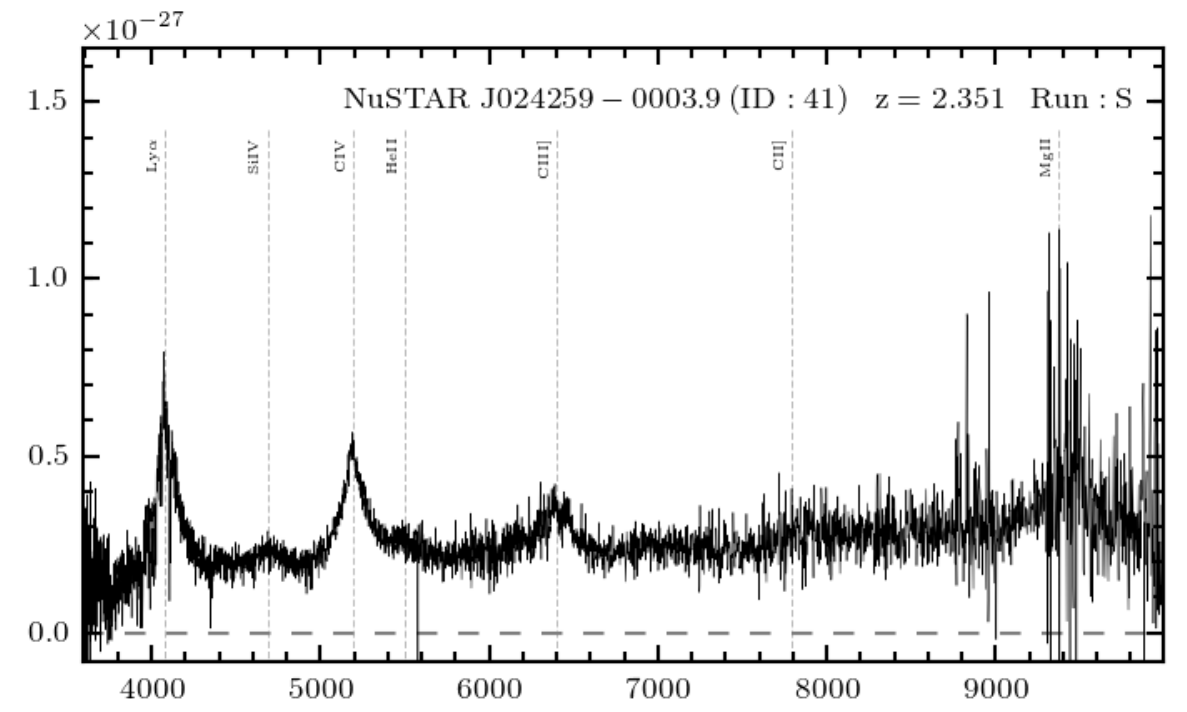}
\end{minipage}
\begin{minipage}[l]{0.325\textwidth}
\includegraphics[width=\textwidth]{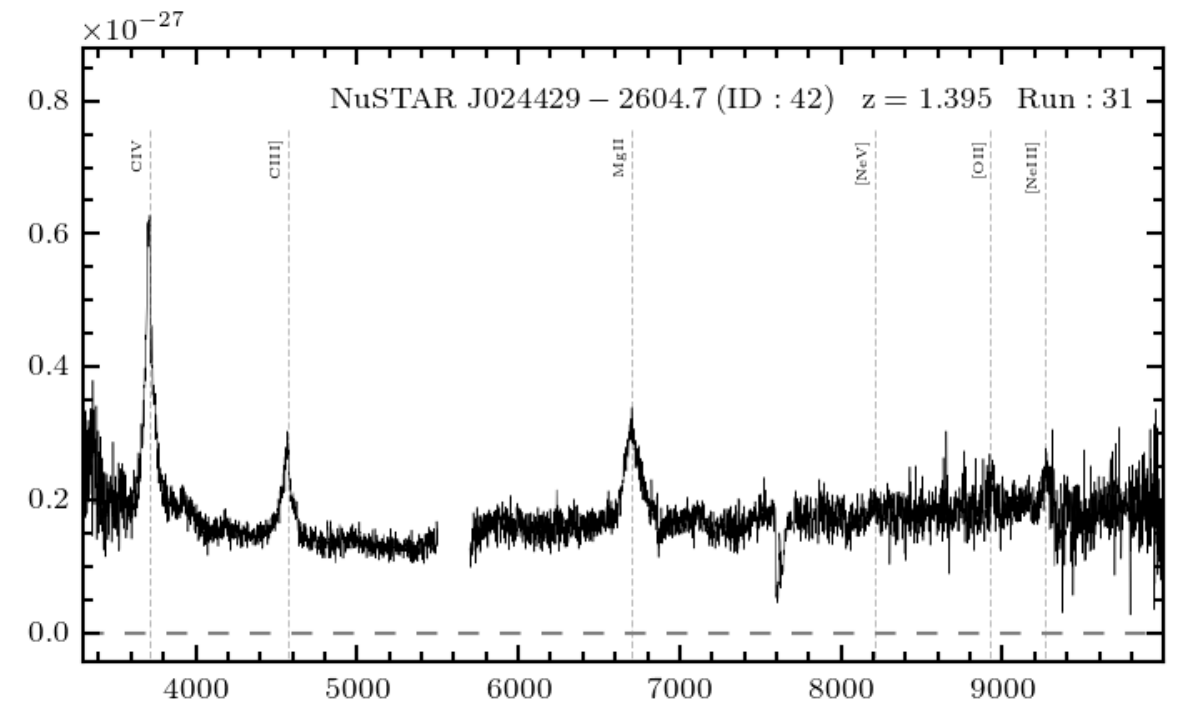}
\end{minipage}
\begin{minipage}[l]{0.325\textwidth}
\includegraphics[width=\textwidth]{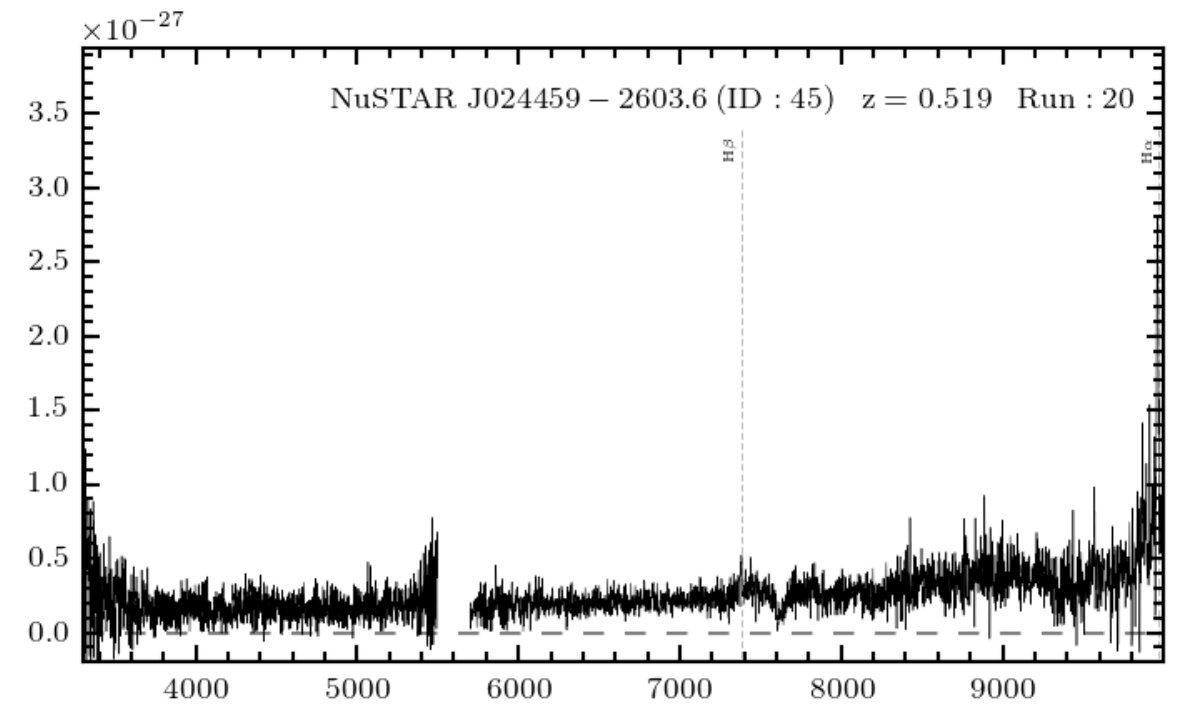}
\end{minipage}
\begin{minipage}[l]{0.325\textwidth}
\includegraphics[width=\textwidth]{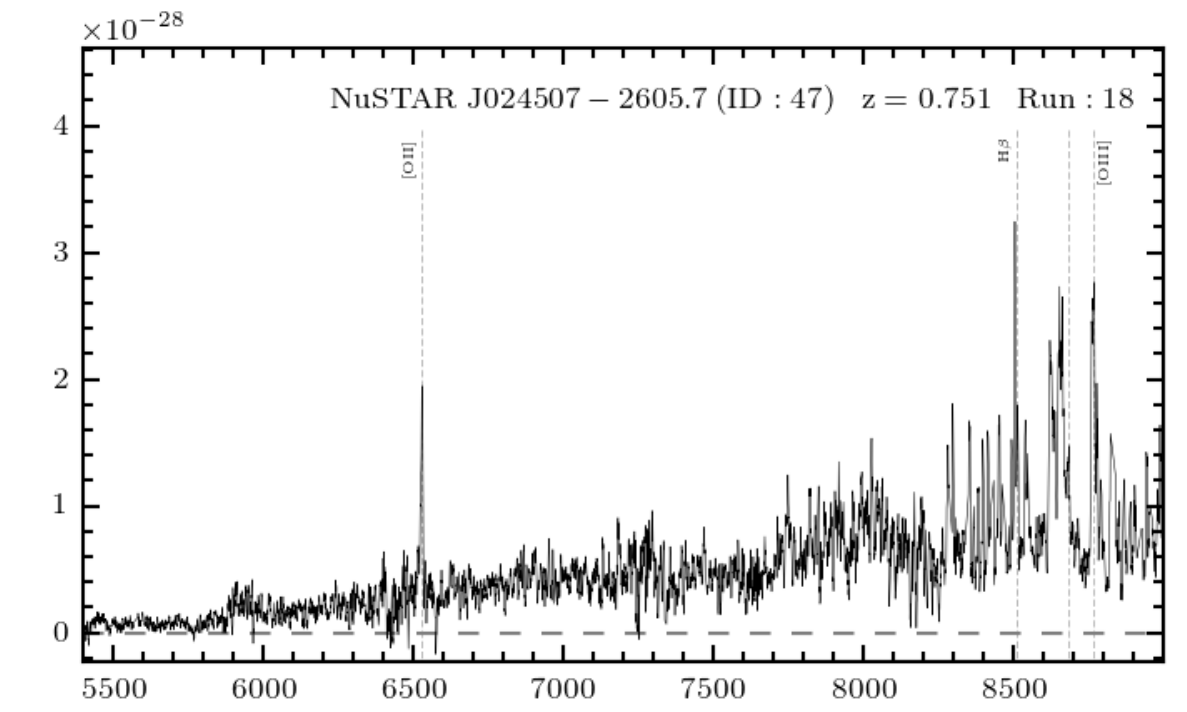}
\end{minipage}
\caption{Optical spectra for the \nustar serendipitous survey sources
  (continued on the following pages). The horizontal axis shows the
  wavelength in units of \AA, and the vertical axis shows the flux
  ($f_{\mathrm{\nu}}$) in units of \fnuunit. Shown in the upper right
  corner are the unique \nustar source name, the unique source ID, the
  source redshift, and the observing run identification number
  (corresponding to Tables \ref{runsTable} and \ref{specTable}; ``S"
  indicates SDSS spectra). The identified emission and absorption
  lines are labelled and marked with vertical dashed gray lines. Full
  resolution figures are available online.}
\end{figure*}
\addtocounter{figure}{-1}
\begin{figure*}
\centering
\begin{minipage}[l]{0.325\textwidth}
\includegraphics[width=\textwidth]{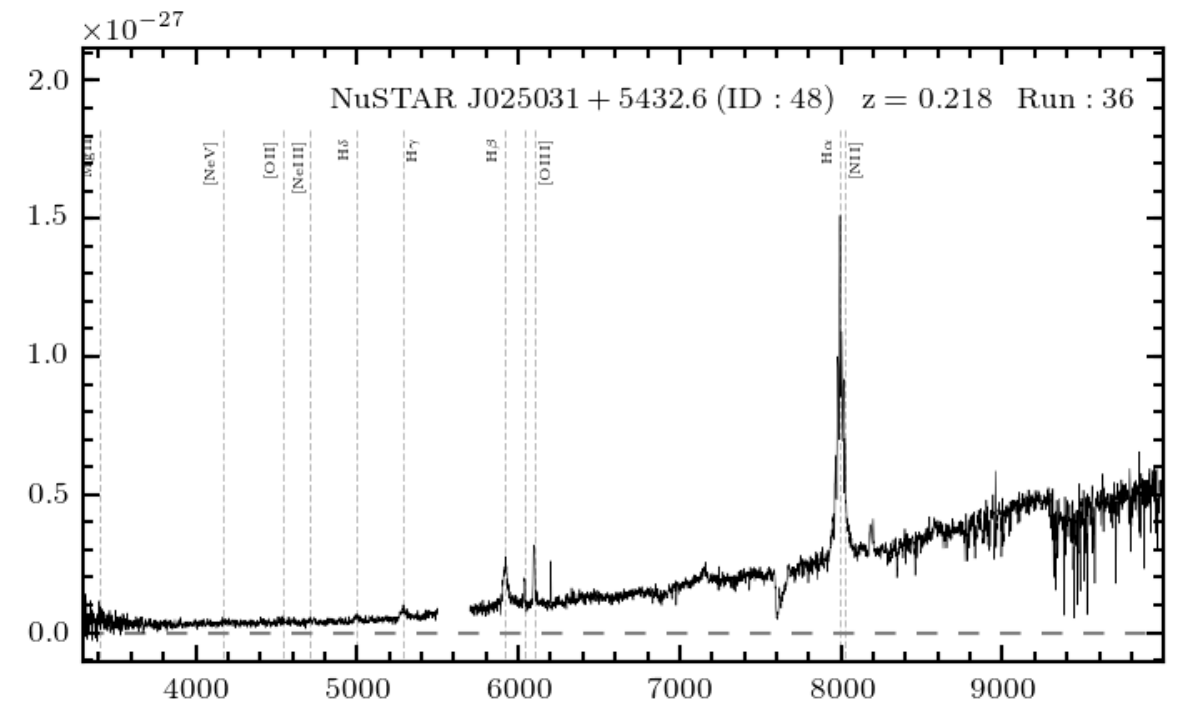}
\end{minipage}
\begin{minipage}[l]{0.325\textwidth}
\includegraphics[width=\textwidth]{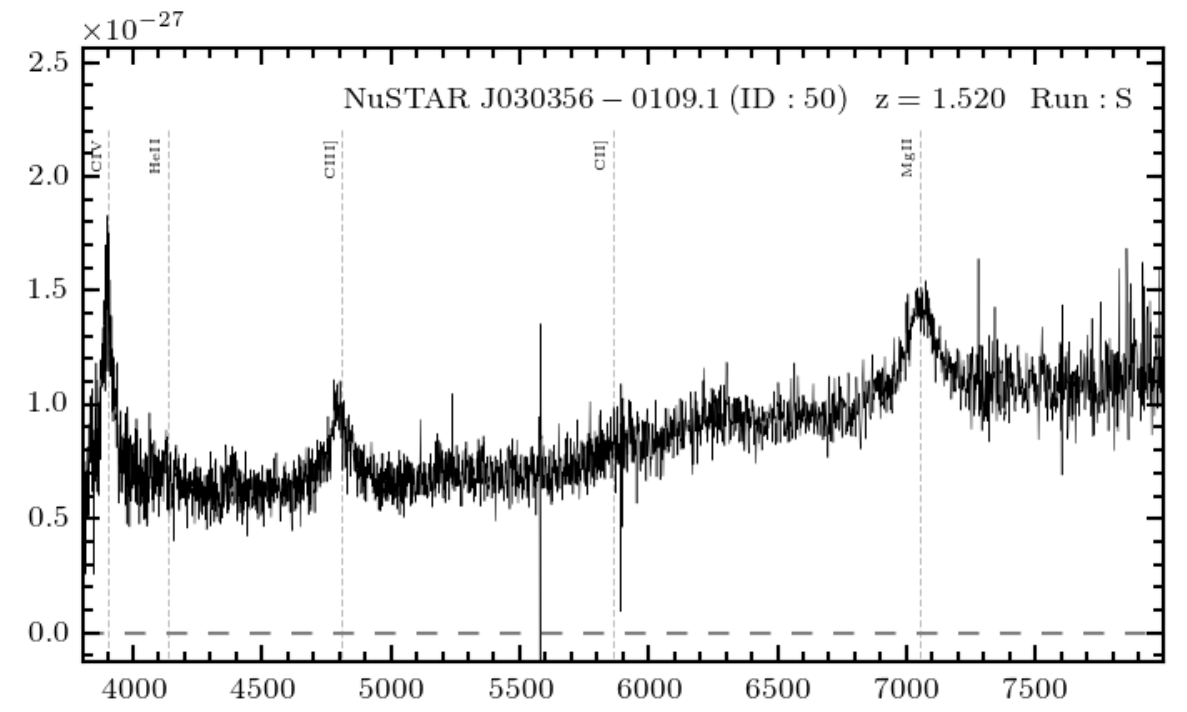}
\end{minipage}
\begin{minipage}[l]{0.325\textwidth}
\includegraphics[width=\textwidth]{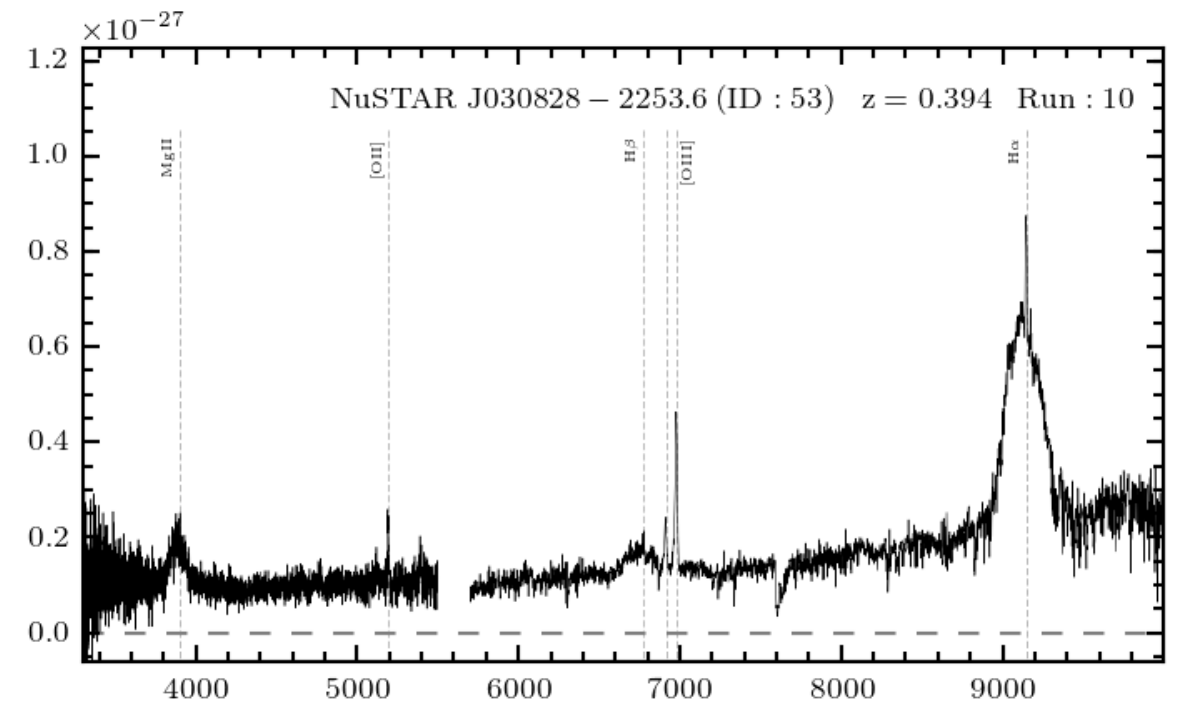}
\end{minipage}
\begin{minipage}[l]{0.325\textwidth}
\includegraphics[width=\textwidth]{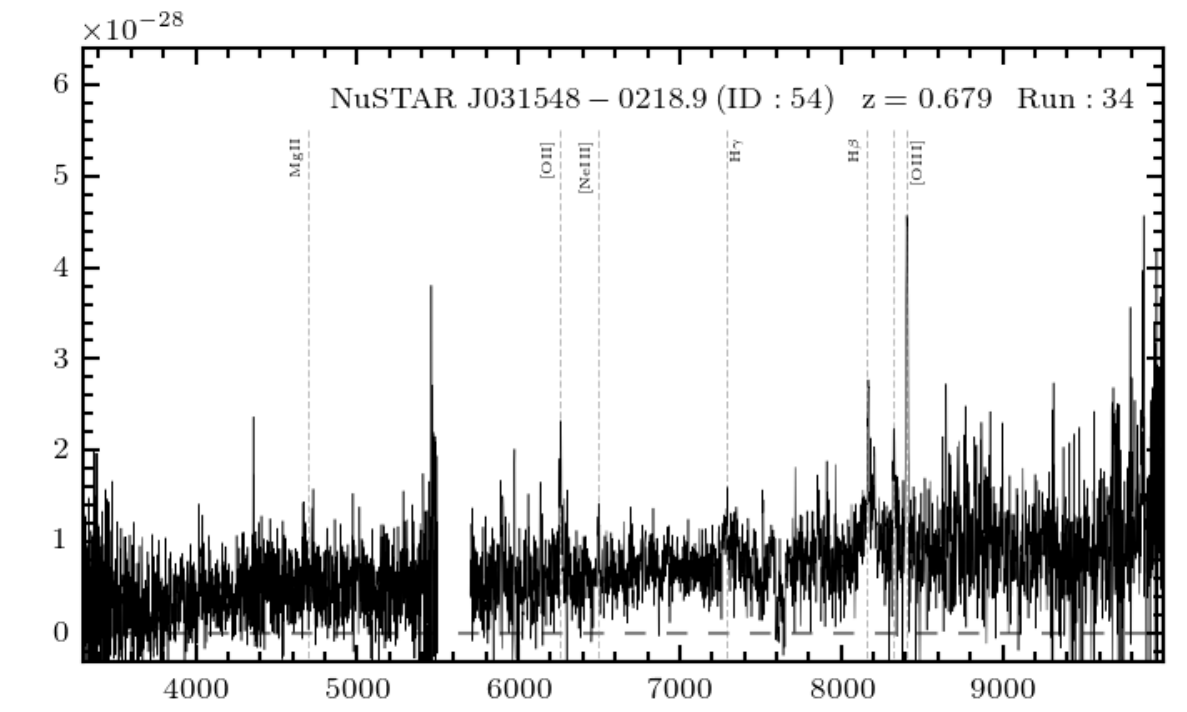}
\end{minipage}
\begin{minipage}[l]{0.325\textwidth}
\includegraphics[width=\textwidth]{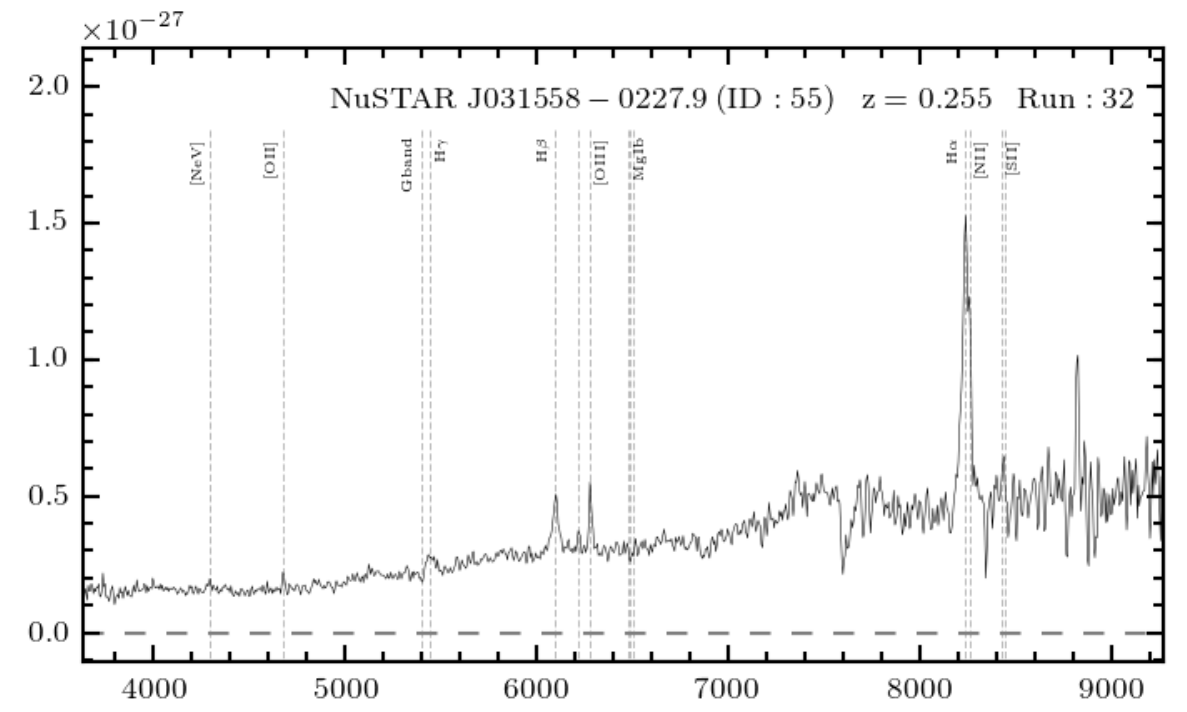}
\end{minipage}
\begin{minipage}[l]{0.325\textwidth}
\includegraphics[width=\textwidth]{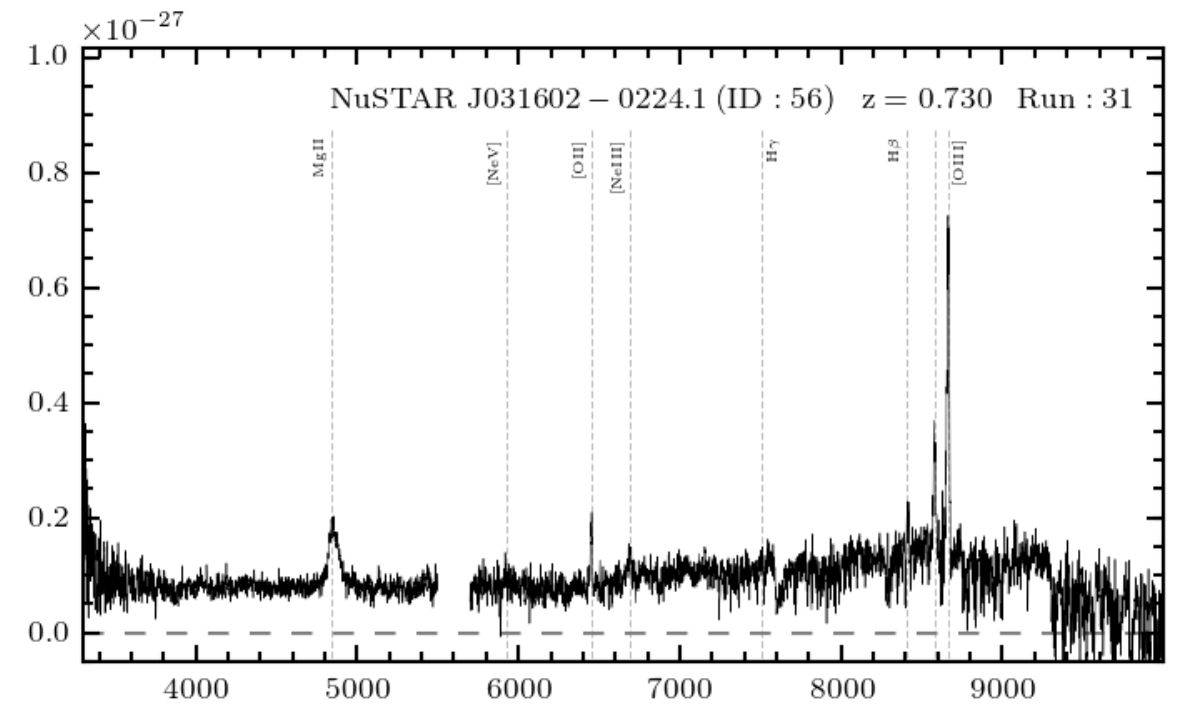}
\end{minipage}
\begin{minipage}[l]{0.325\textwidth}
\includegraphics[width=\textwidth]{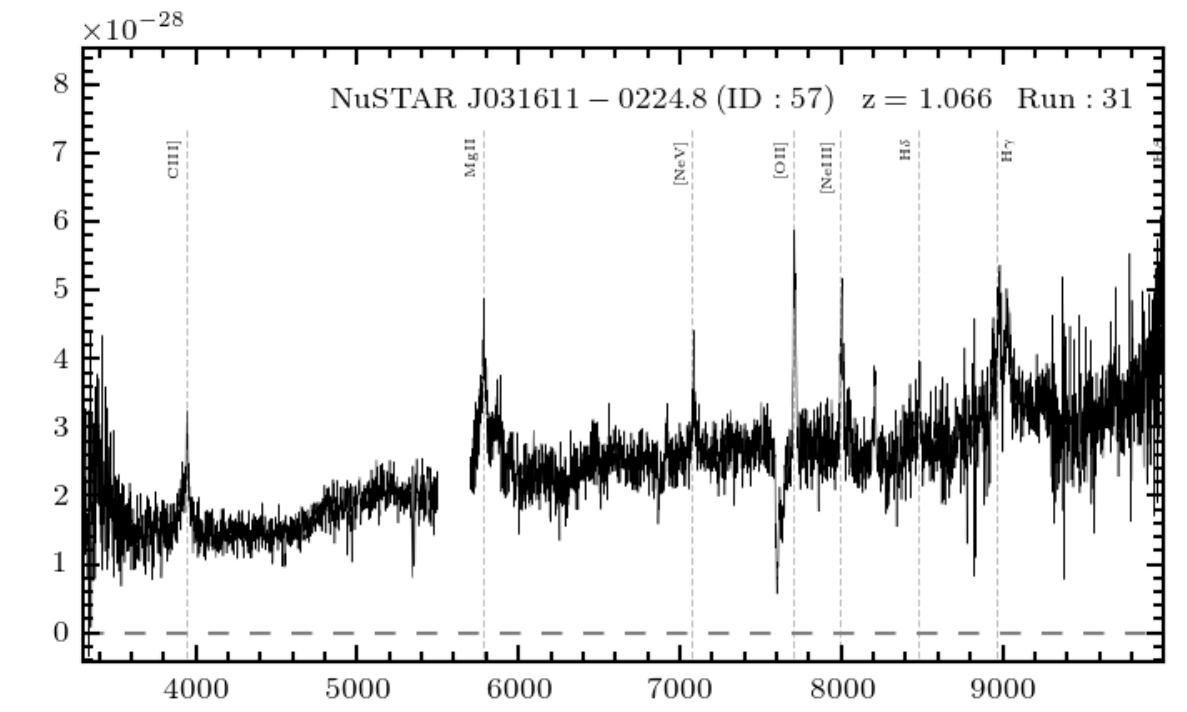}
\end{minipage}
\begin{minipage}[l]{0.325\textwidth}
\includegraphics[width=\textwidth]{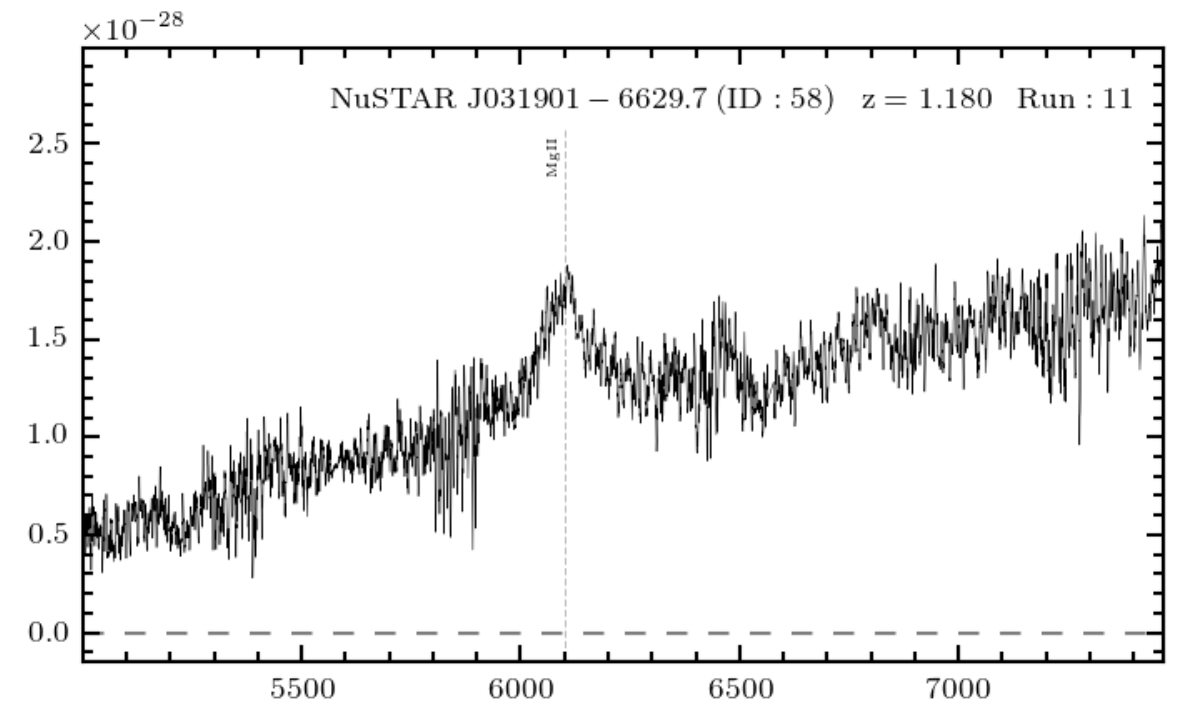}
\end{minipage}
\begin{minipage}[l]{0.325\textwidth}
\includegraphics[width=\textwidth]{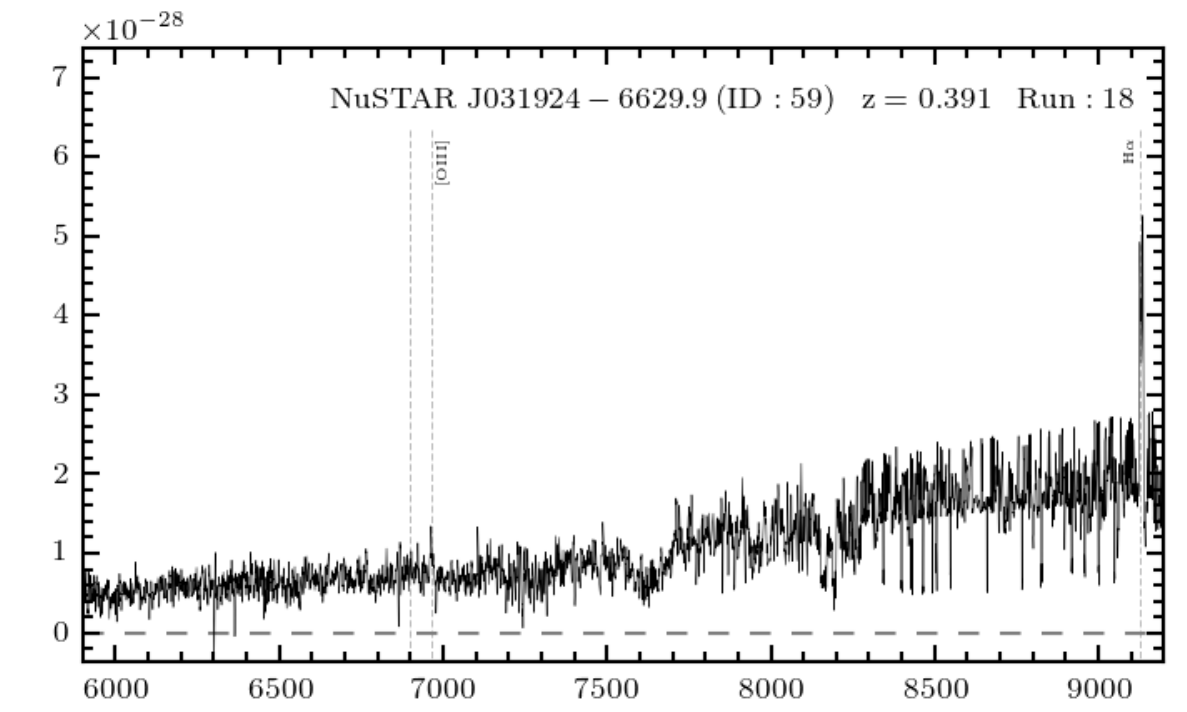}
\end{minipage}
\begin{minipage}[l]{0.325\textwidth}
\includegraphics[width=\textwidth]{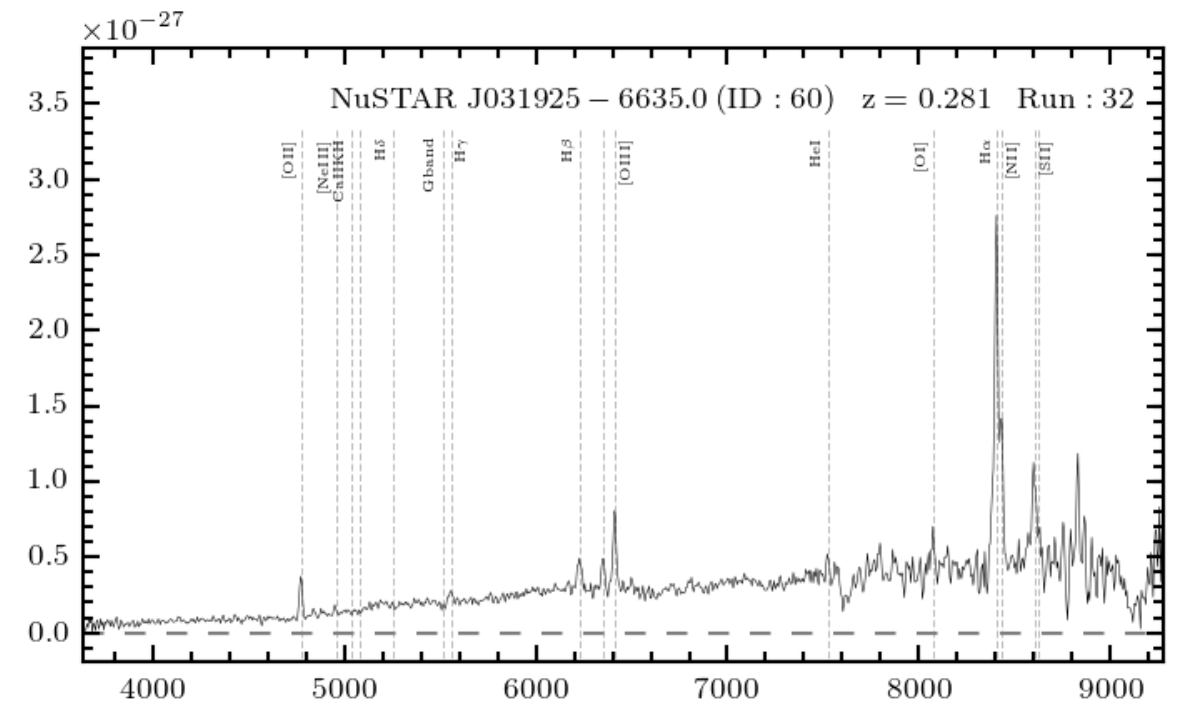}
\end{minipage}
\begin{minipage}[l]{0.325\textwidth}
\includegraphics[width=\textwidth]{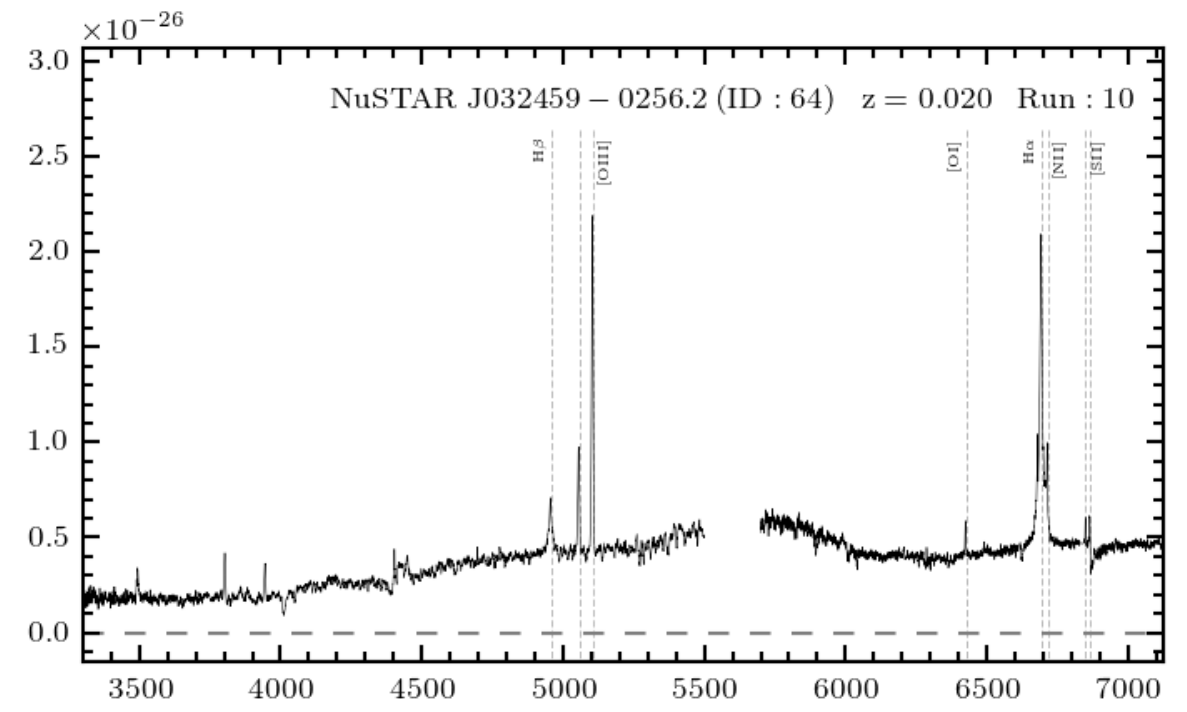}
\end{minipage}
\begin{minipage}[l]{0.325\textwidth}
\includegraphics[width=\textwidth]{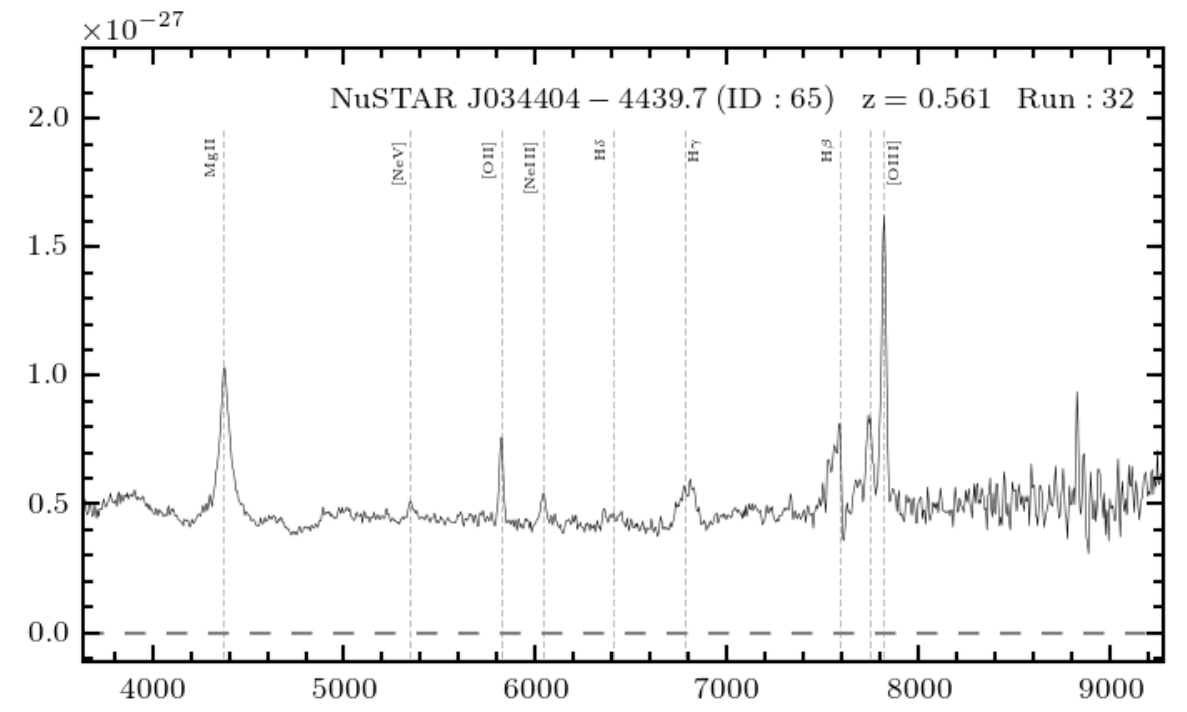}
\end{minipage}
\begin{minipage}[l]{0.325\textwidth}
\includegraphics[width=\textwidth]{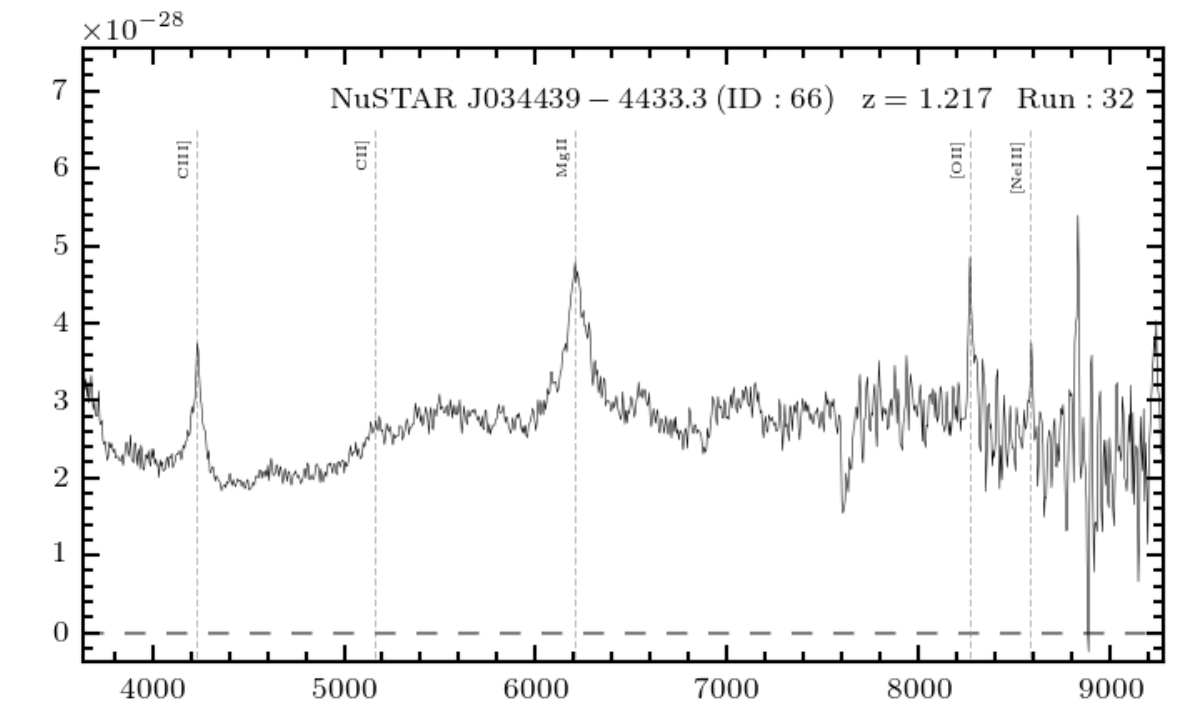}
\end{minipage}
\begin{minipage}[l]{0.325\textwidth}
\includegraphics[width=\textwidth]{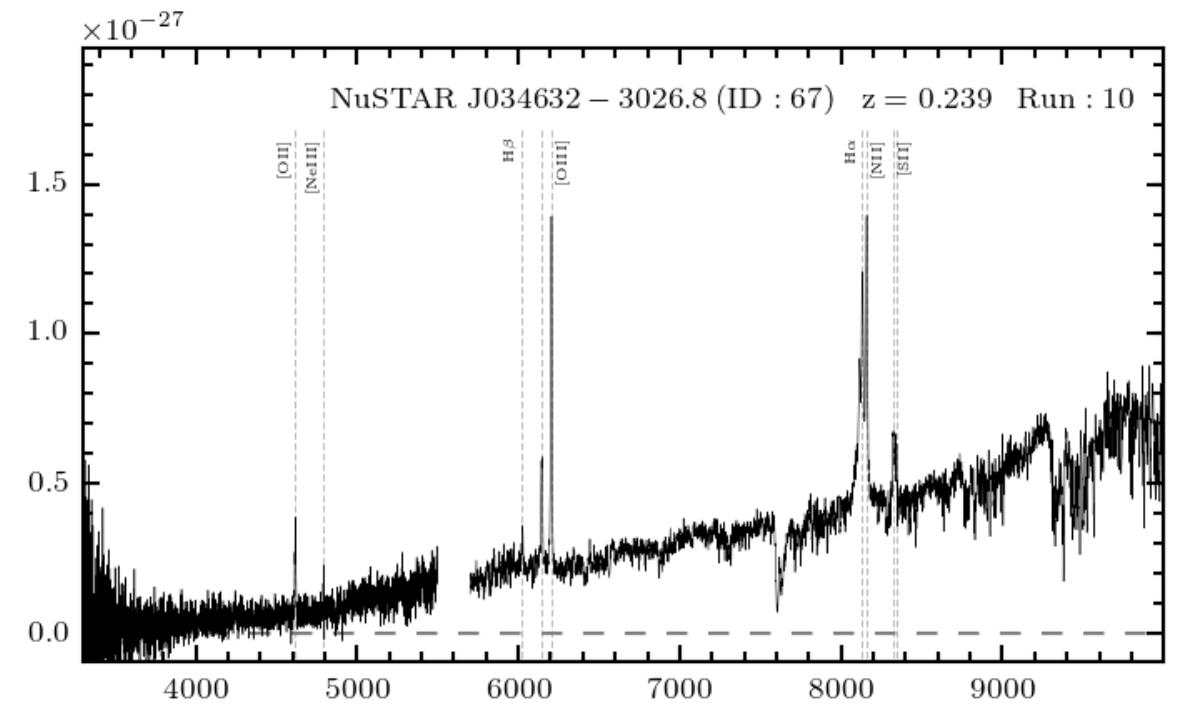}
\end{minipage}
\begin{minipage}[l]{0.325\textwidth}
\includegraphics[width=\textwidth]{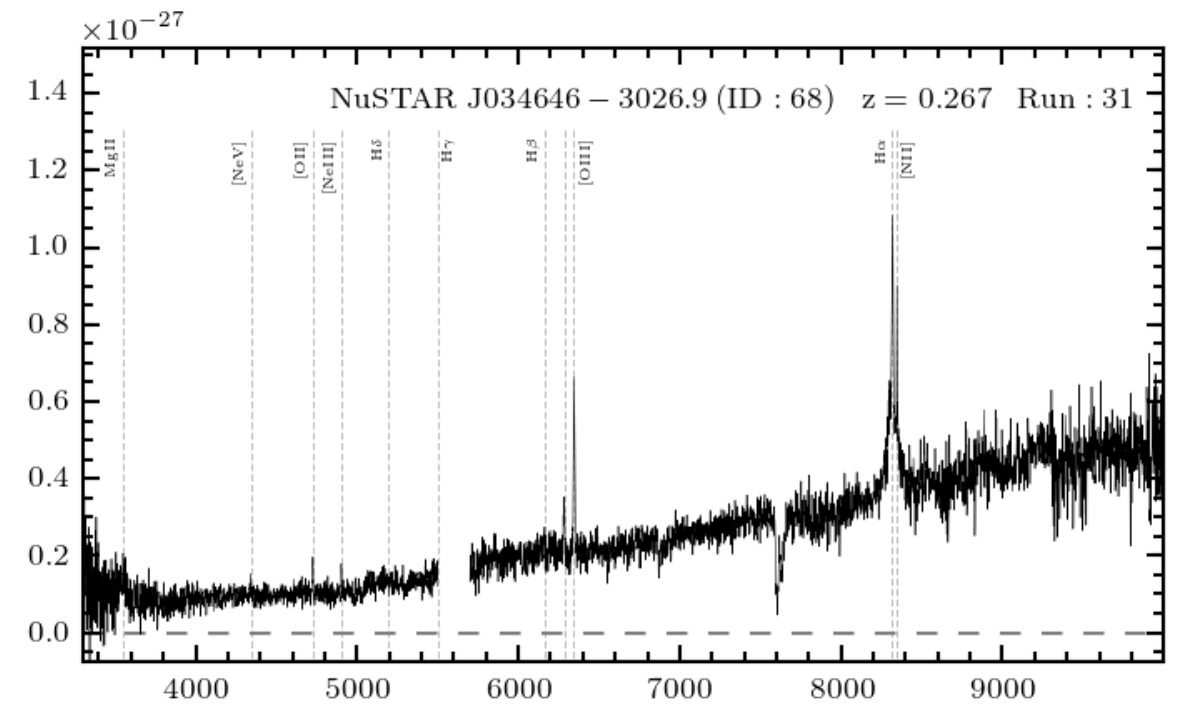}
\end{minipage}
\begin{minipage}[l]{0.325\textwidth}
\includegraphics[width=\textwidth]{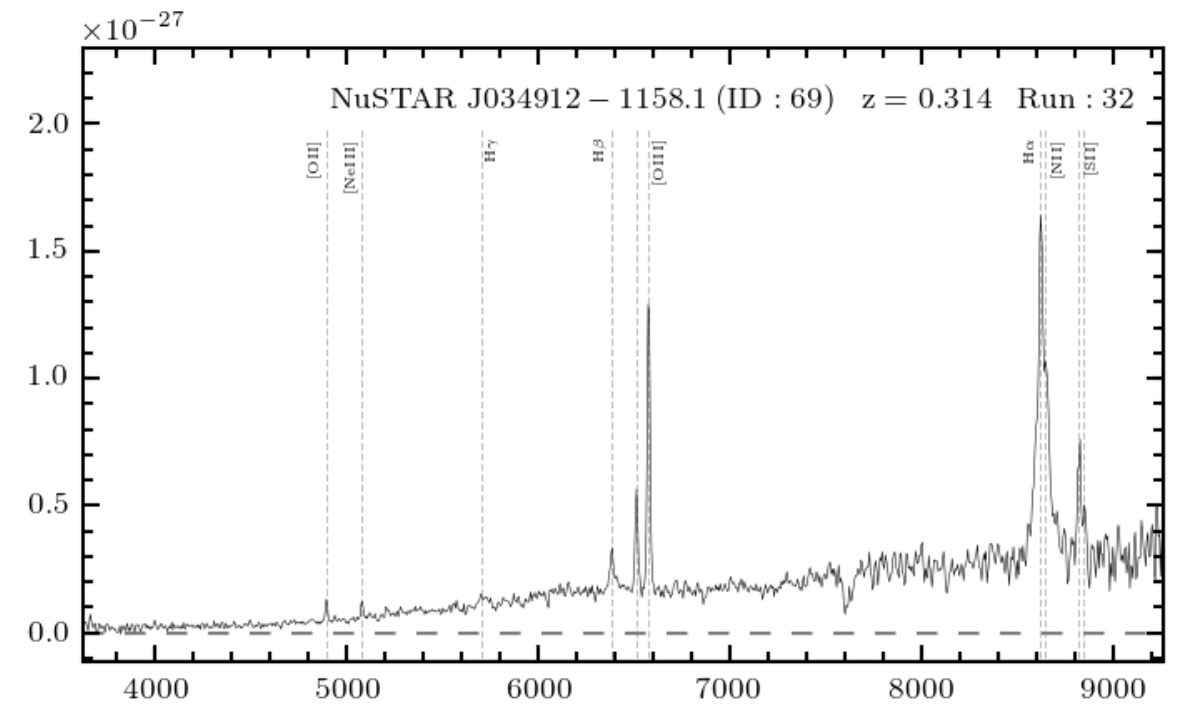}
\end{minipage}
\begin{minipage}[l]{0.325\textwidth}
\includegraphics[width=\textwidth]{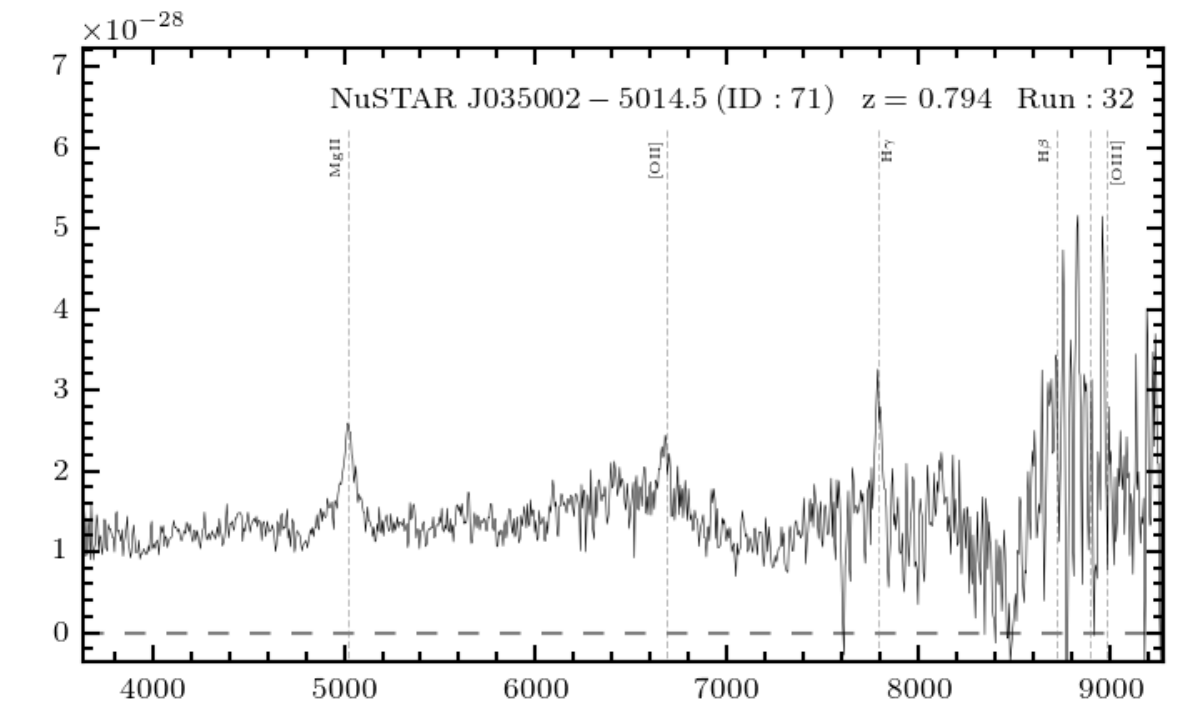}
\end{minipage}
\begin{minipage}[l]{0.325\textwidth}
\includegraphics[width=\textwidth]{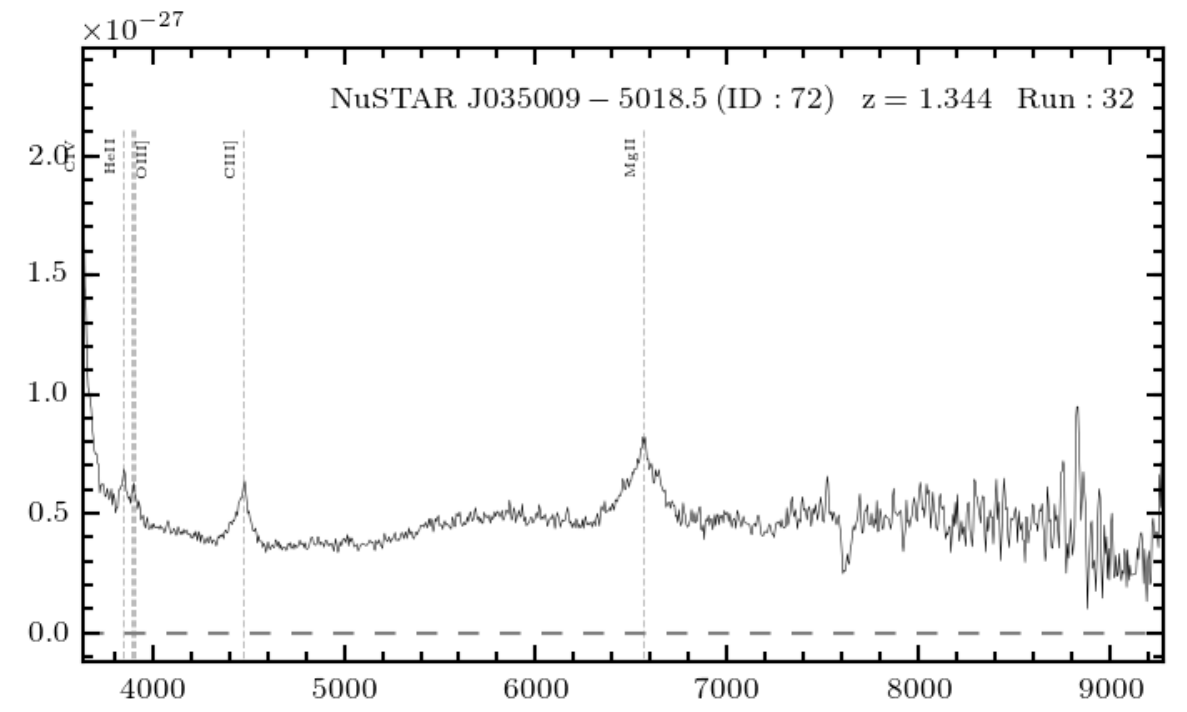}
\end{minipage}
\caption{Continued.}
\end{figure*}
\addtocounter{figure}{-1}
\begin{figure*}
\centering
\begin{minipage}[l]{0.325\textwidth}
\includegraphics[width=\textwidth]{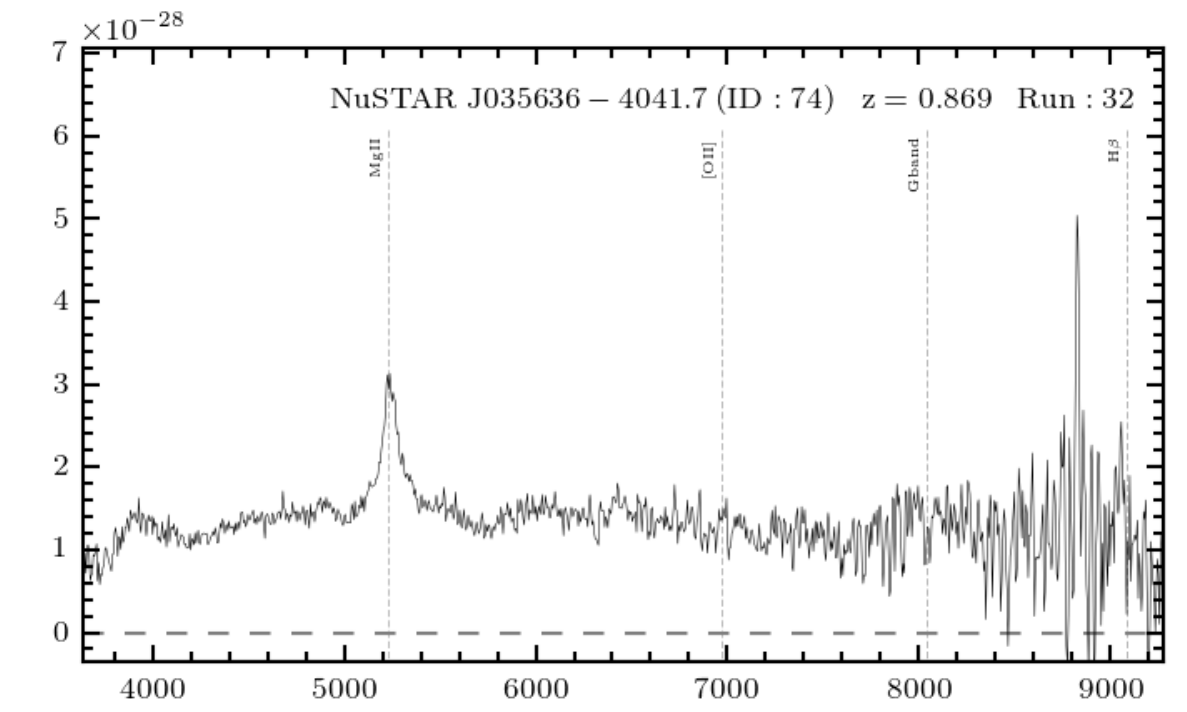}
\end{minipage}
\begin{minipage}[l]{0.325\textwidth}
\includegraphics[width=\textwidth]{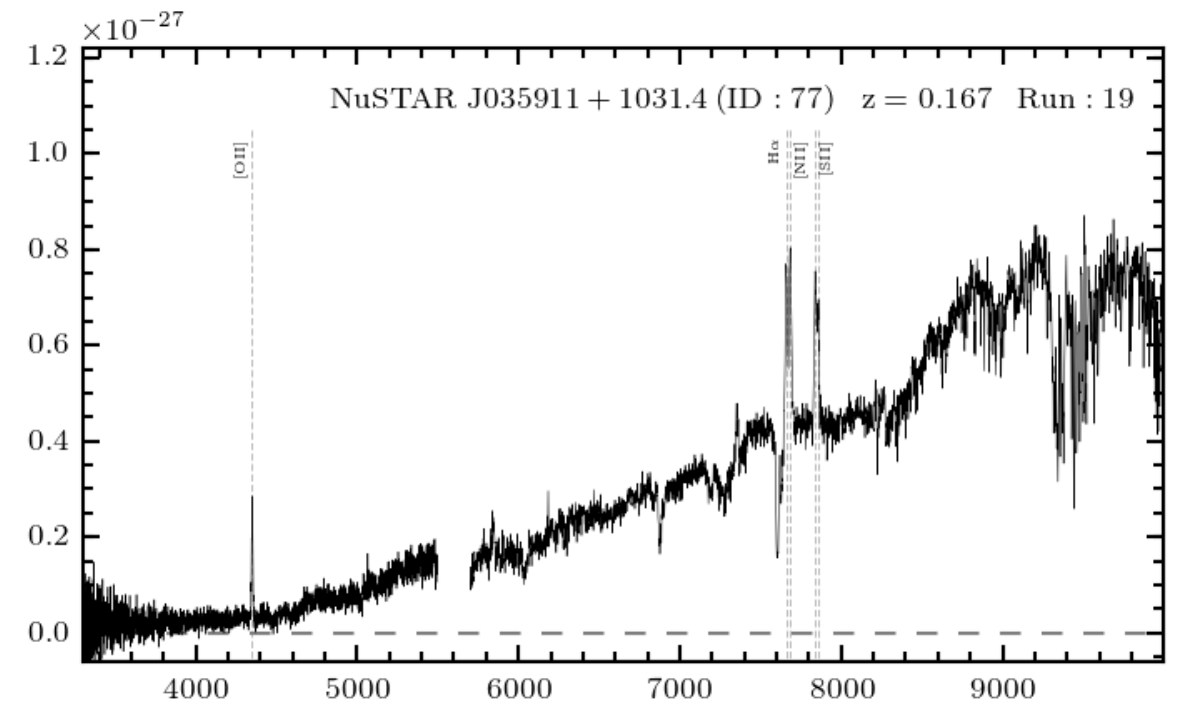}
\end{minipage}
\begin{minipage}[l]{0.325\textwidth}
\includegraphics[width=\textwidth]{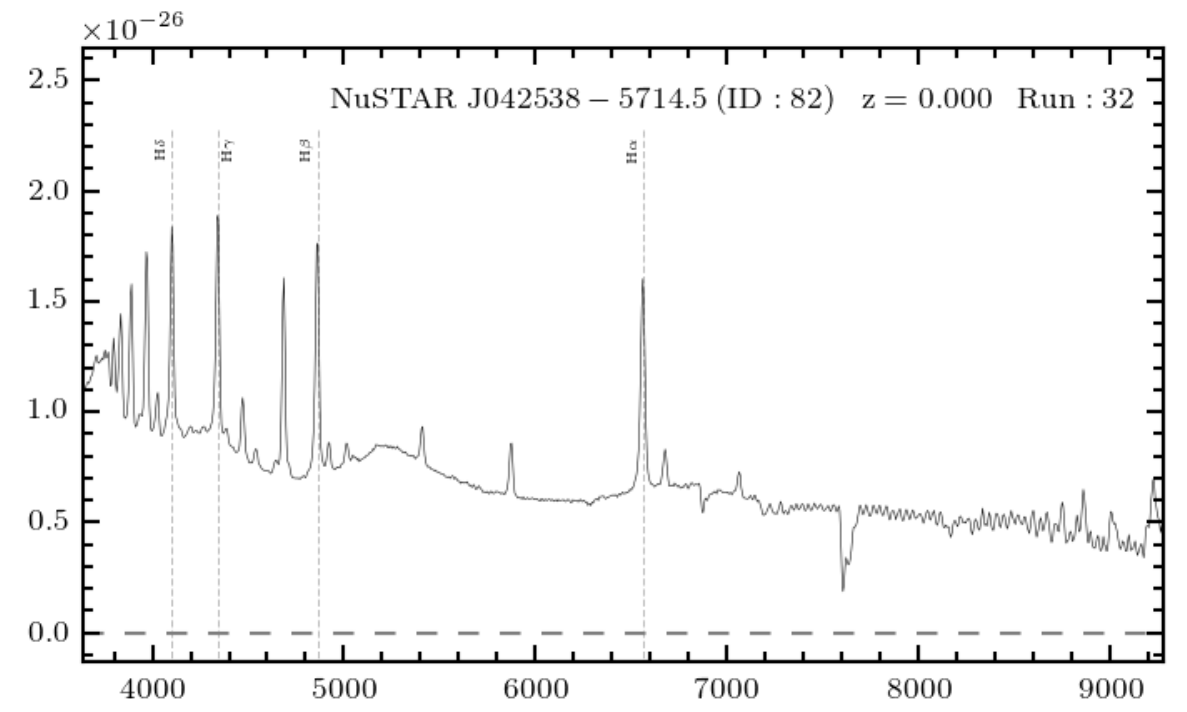}
\end{minipage}
\begin{minipage}[l]{0.325\textwidth}
\includegraphics[width=\textwidth]{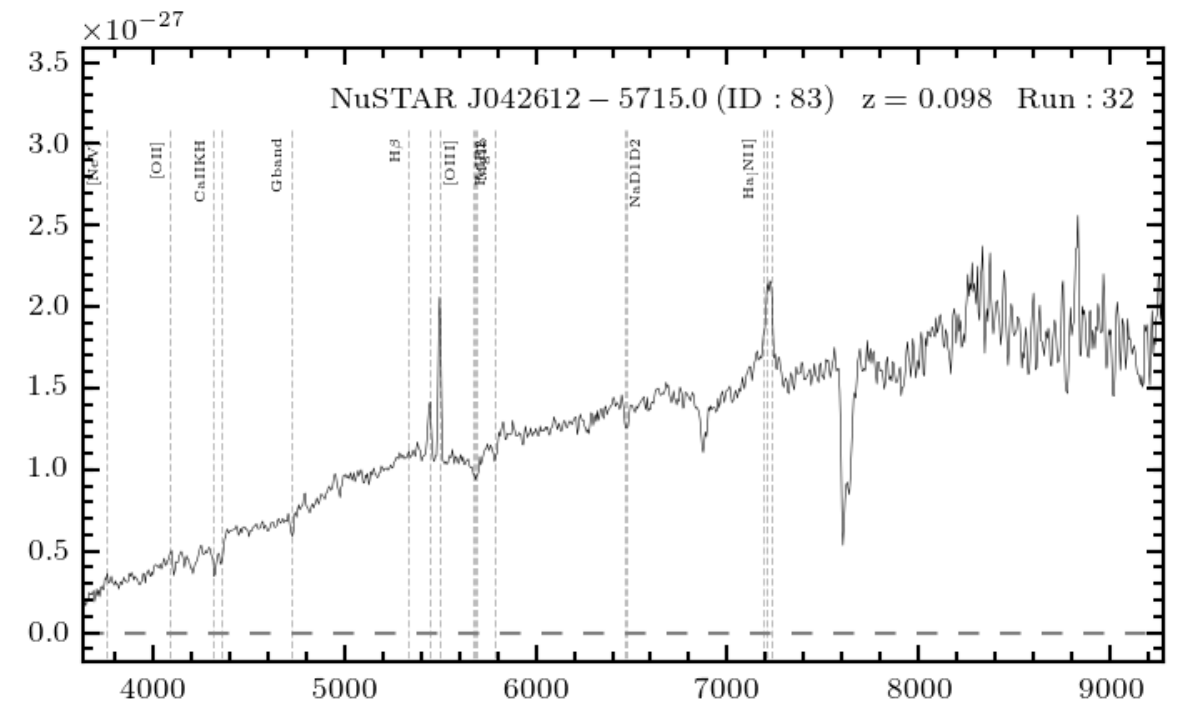}
\end{minipage}
\begin{minipage}[l]{0.325\textwidth}
\includegraphics[width=\textwidth]{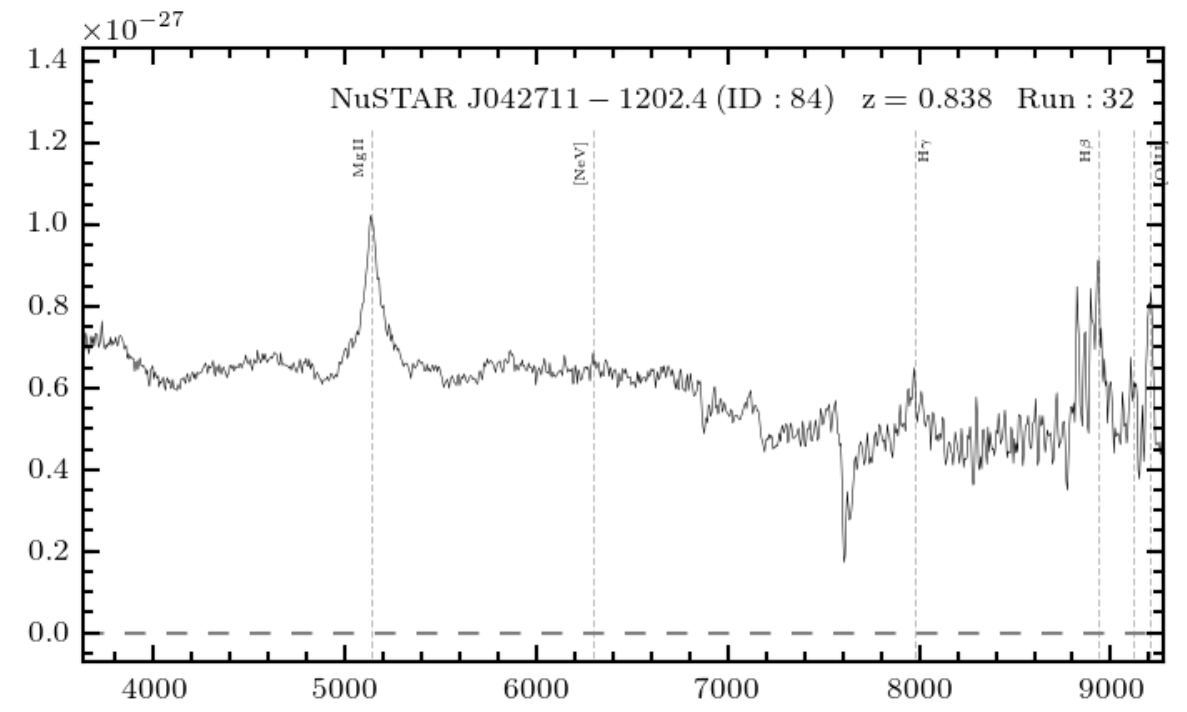}
\end{minipage}
\begin{minipage}[l]{0.325\textwidth}
\includegraphics[width=\textwidth]{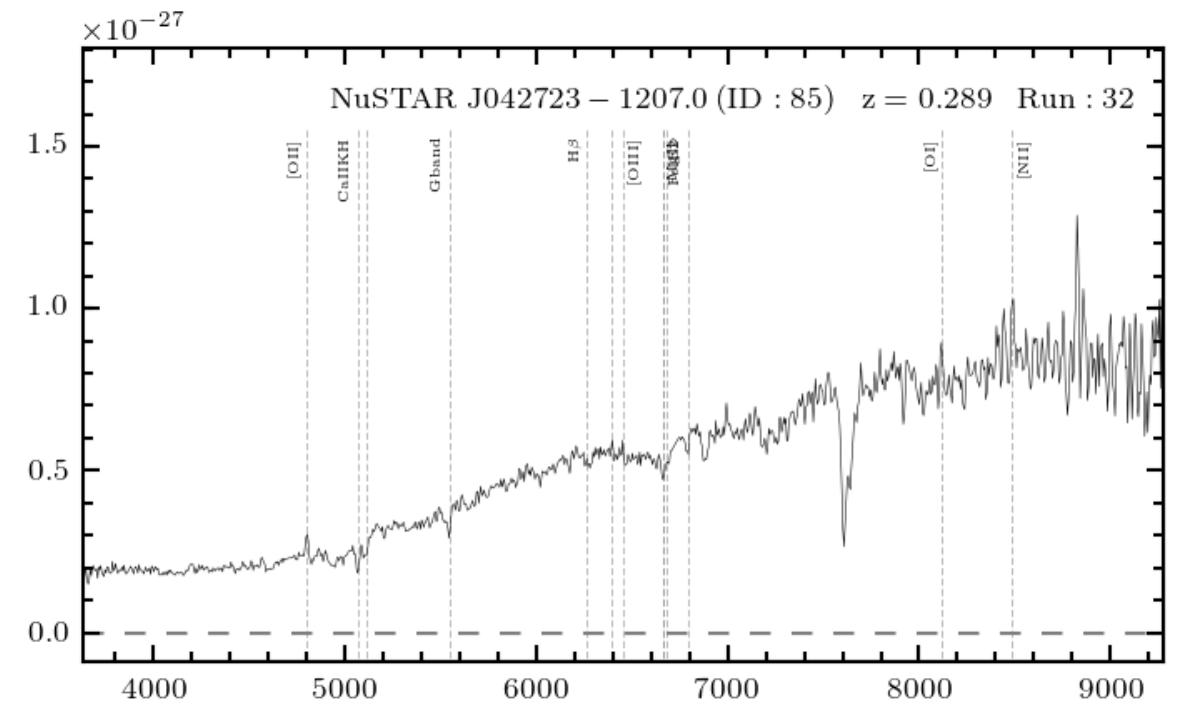}
\end{minipage}
\begin{minipage}[l]{0.325\textwidth}
\includegraphics[width=\textwidth]{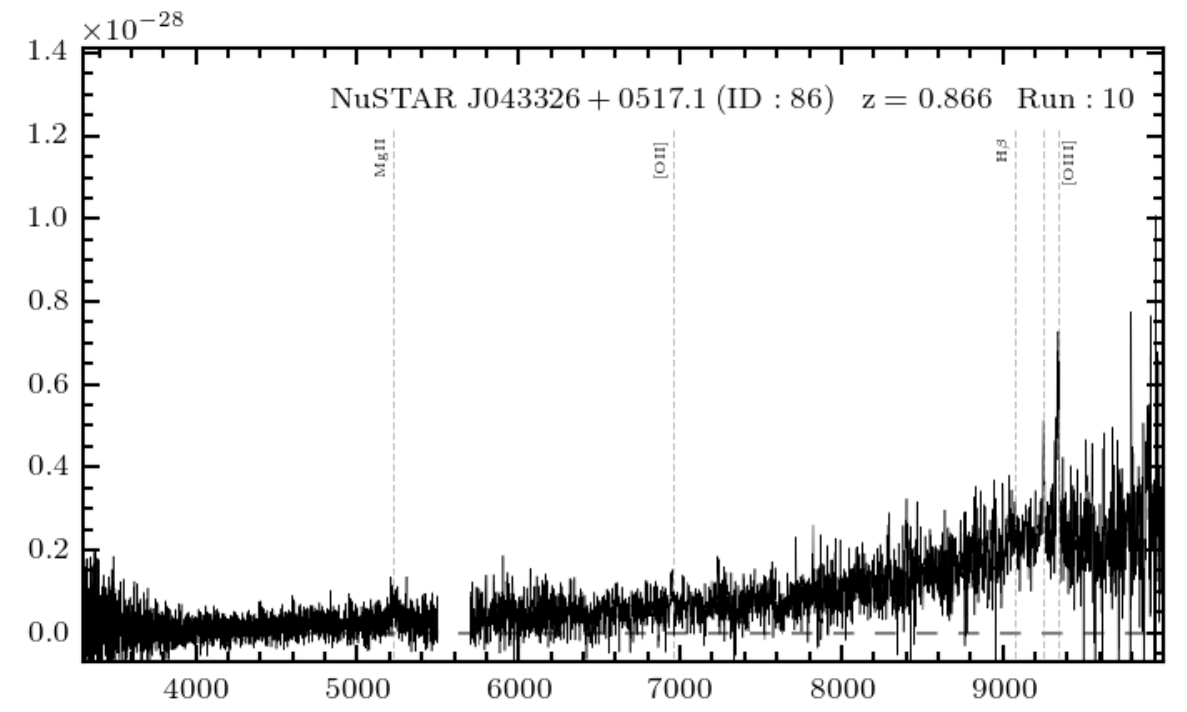}
\end{minipage}
\begin{minipage}[l]{0.325\textwidth}
\includegraphics[width=\textwidth]{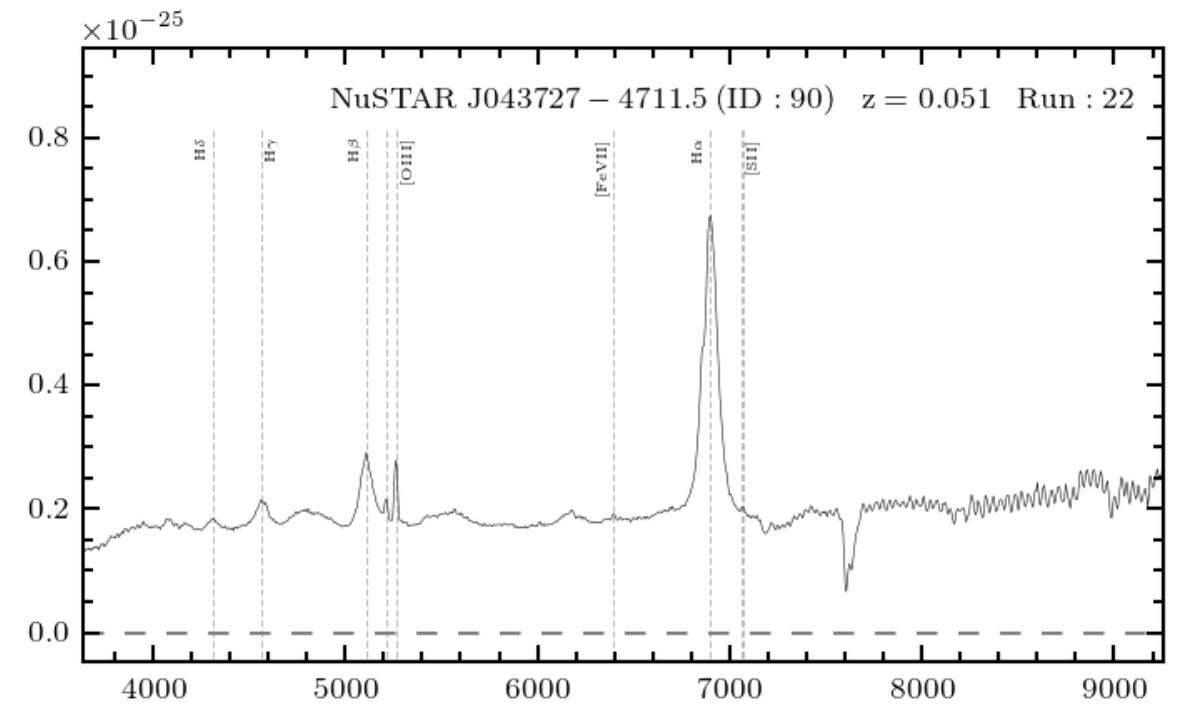}
\end{minipage}
\begin{minipage}[l]{0.325\textwidth}
\includegraphics[width=\textwidth]{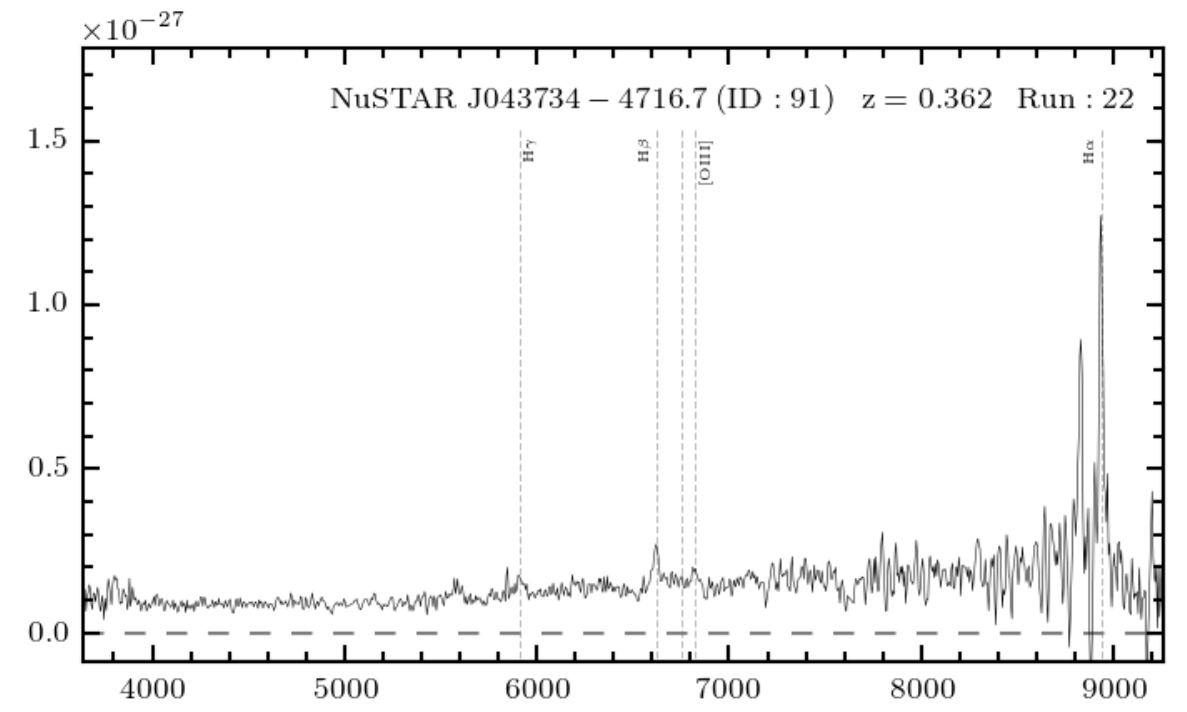}
\end{minipage}
\begin{minipage}[l]{0.325\textwidth}
\includegraphics[width=\textwidth]{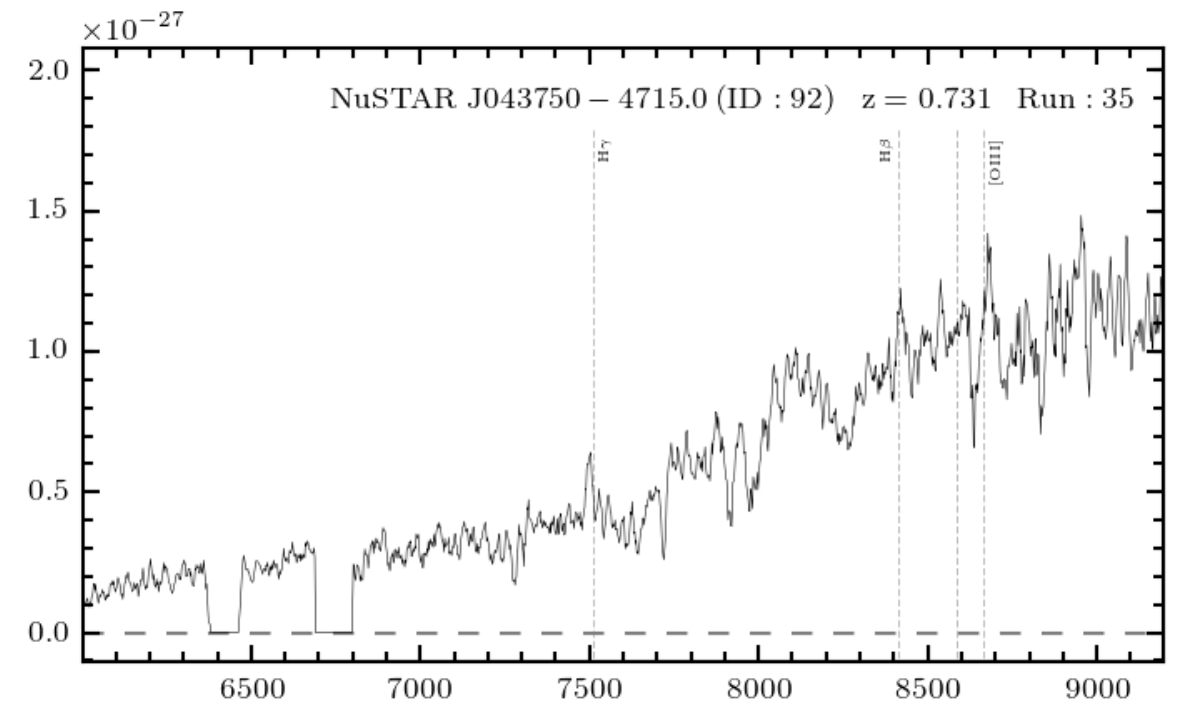}
\end{minipage}
\begin{minipage}[l]{0.325\textwidth}
\includegraphics[width=\textwidth]{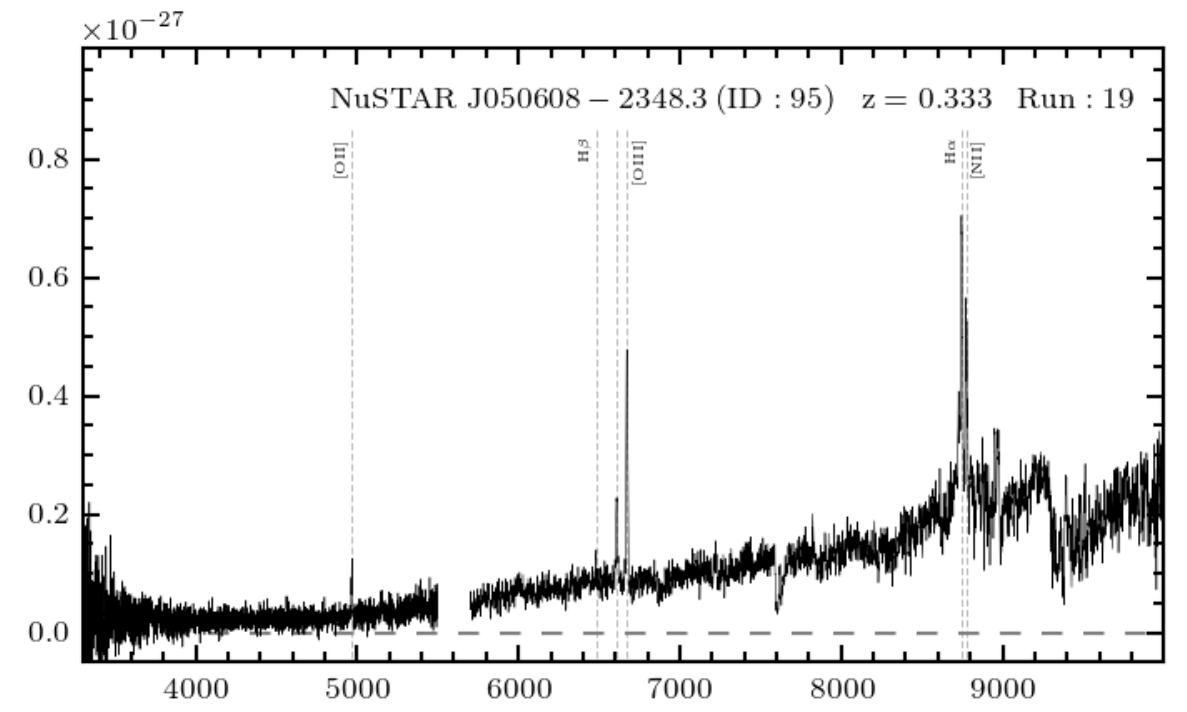}
\end{minipage}
\begin{minipage}[l]{0.325\textwidth}
\includegraphics[width=\textwidth]{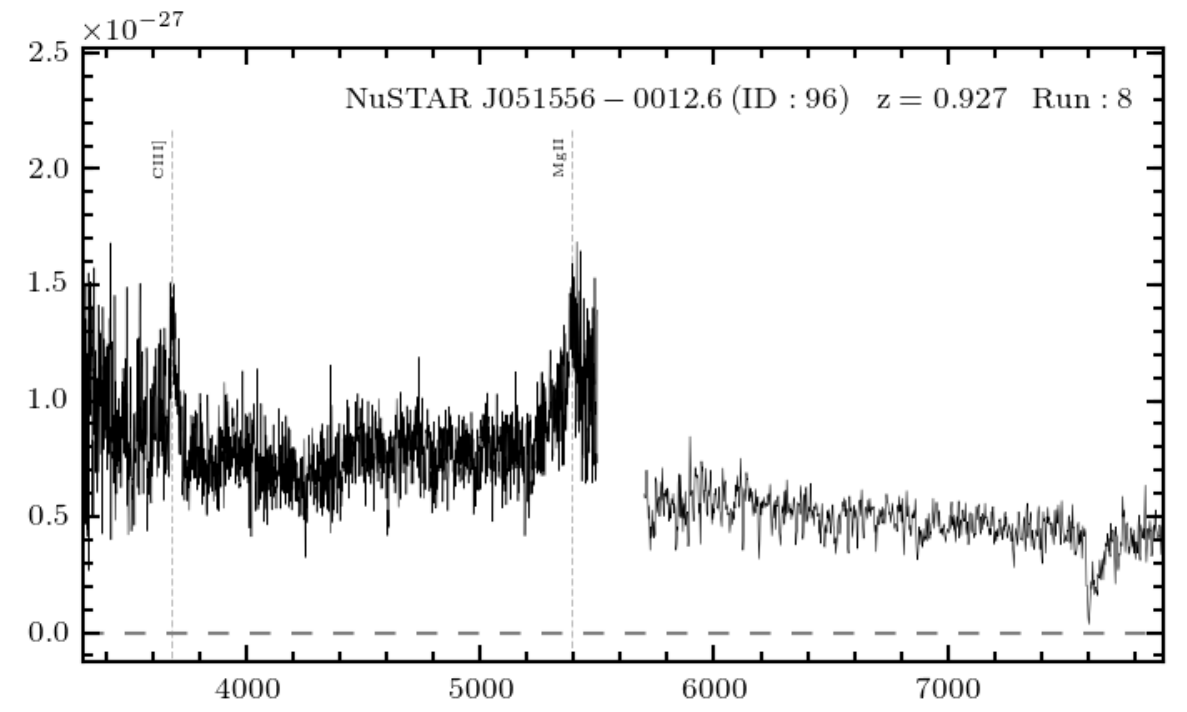}
\end{minipage}
\begin{minipage}[l]{0.325\textwidth}
\includegraphics[width=\textwidth]{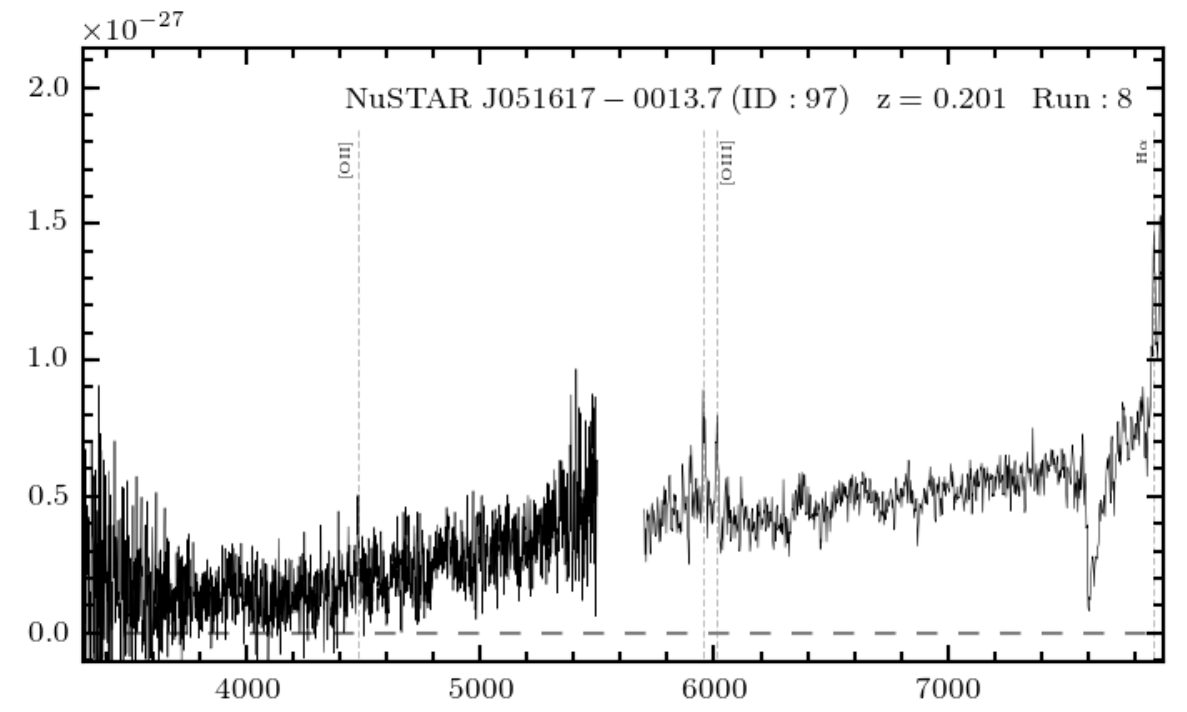}
\end{minipage}
\begin{minipage}[l]{0.325\textwidth}
\includegraphics[width=\textwidth]{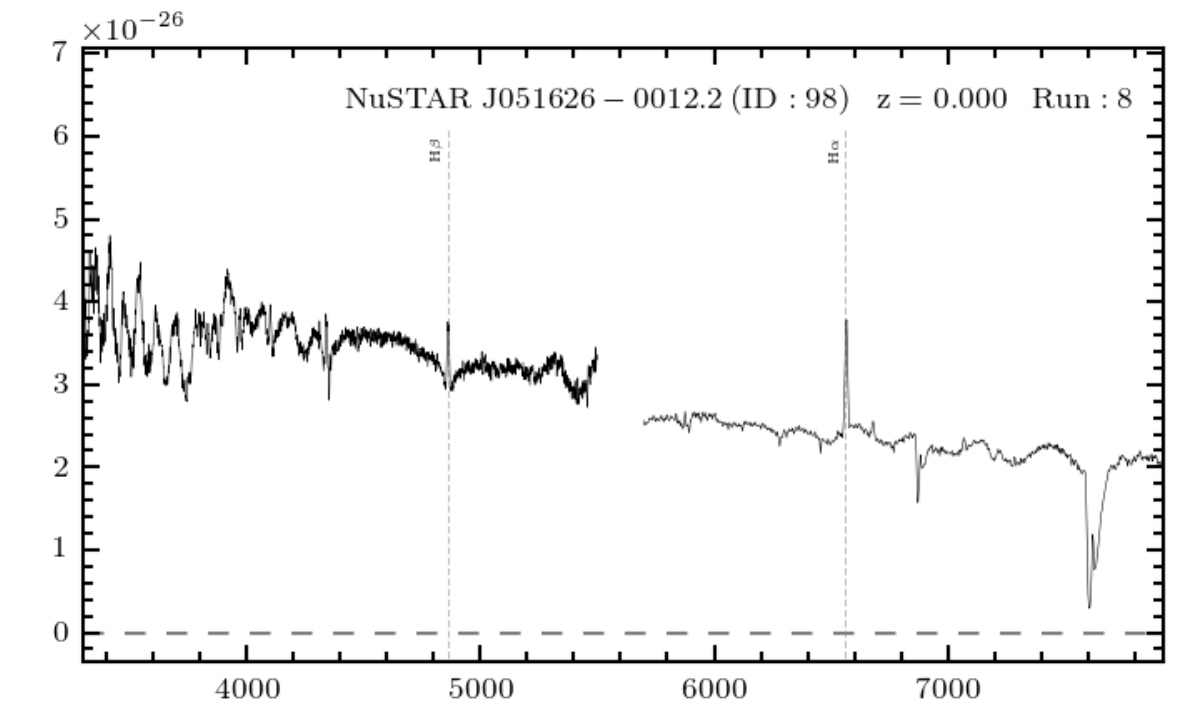}
\end{minipage}
\begin{minipage}[l]{0.325\textwidth}
\includegraphics[width=\textwidth]{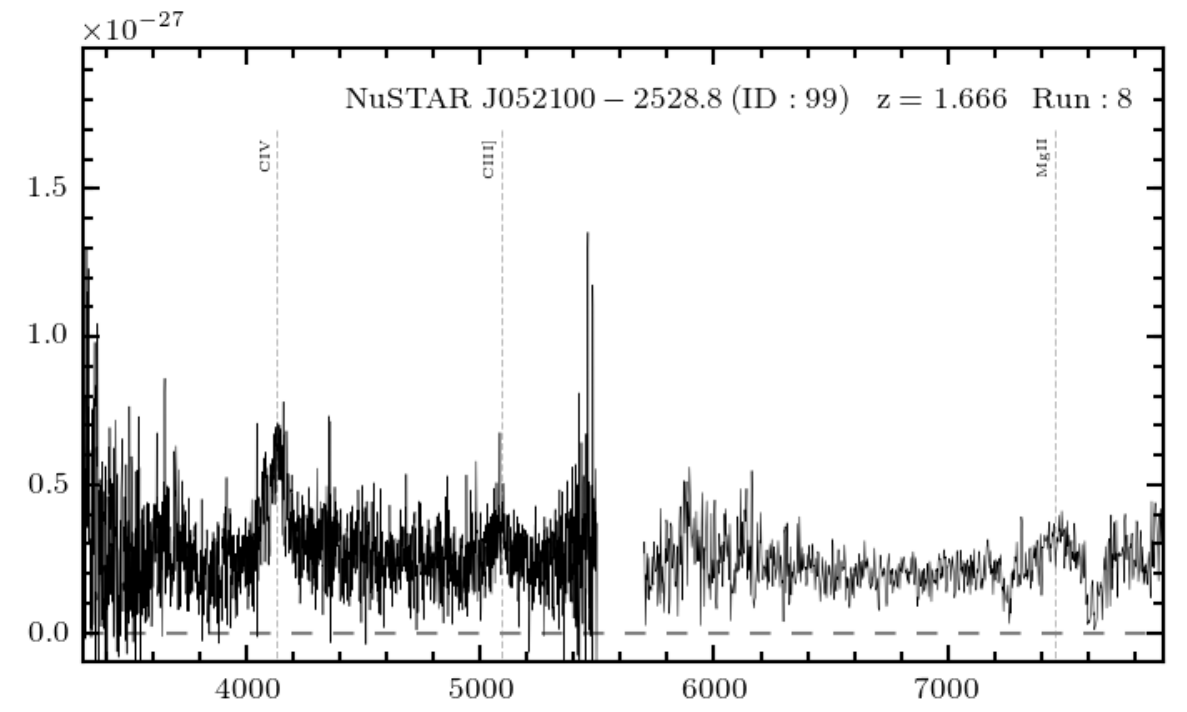}
\end{minipage}
\begin{minipage}[l]{0.325\textwidth}
\includegraphics[width=\textwidth]{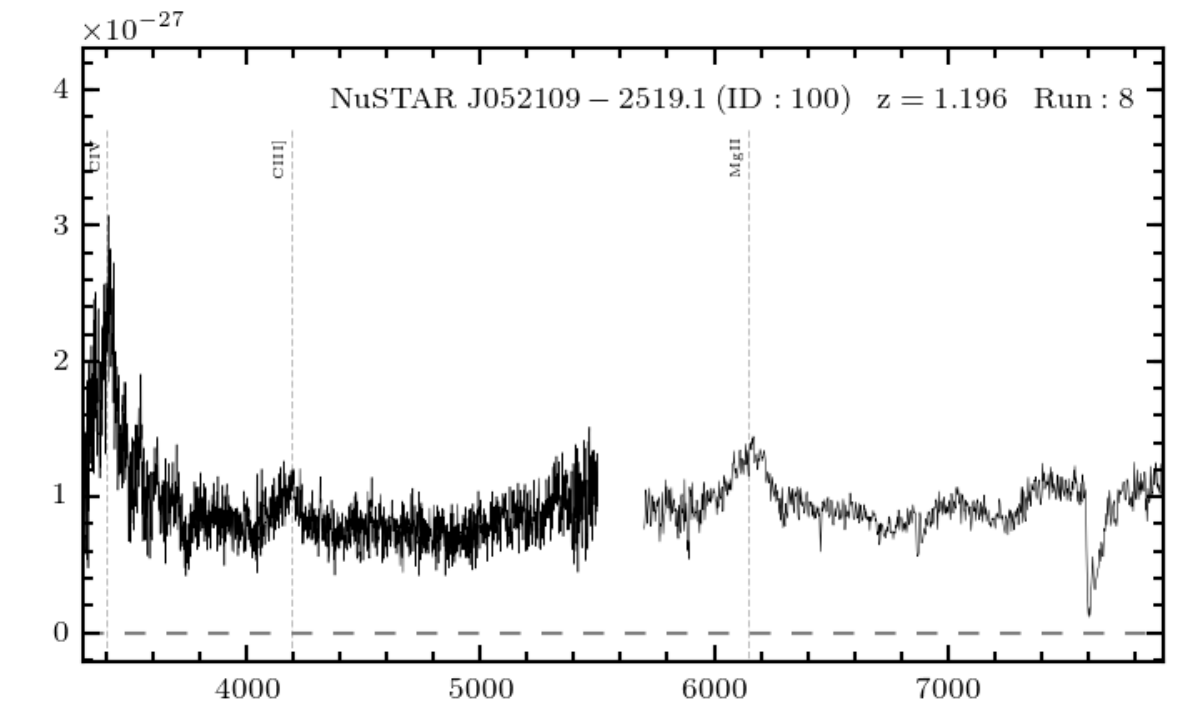}
\end{minipage}
\begin{minipage}[l]{0.325\textwidth}
\includegraphics[width=\textwidth]{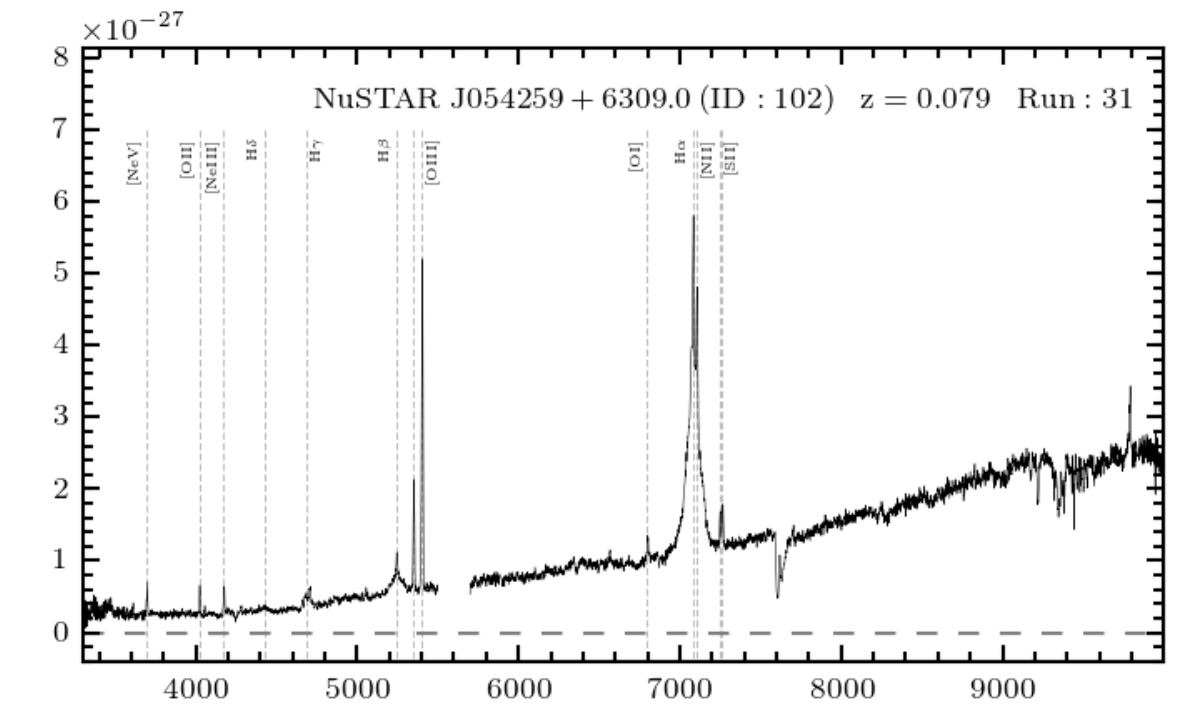}
\end{minipage}
\begin{minipage}[l]{0.325\textwidth}
\includegraphics[width=\textwidth]{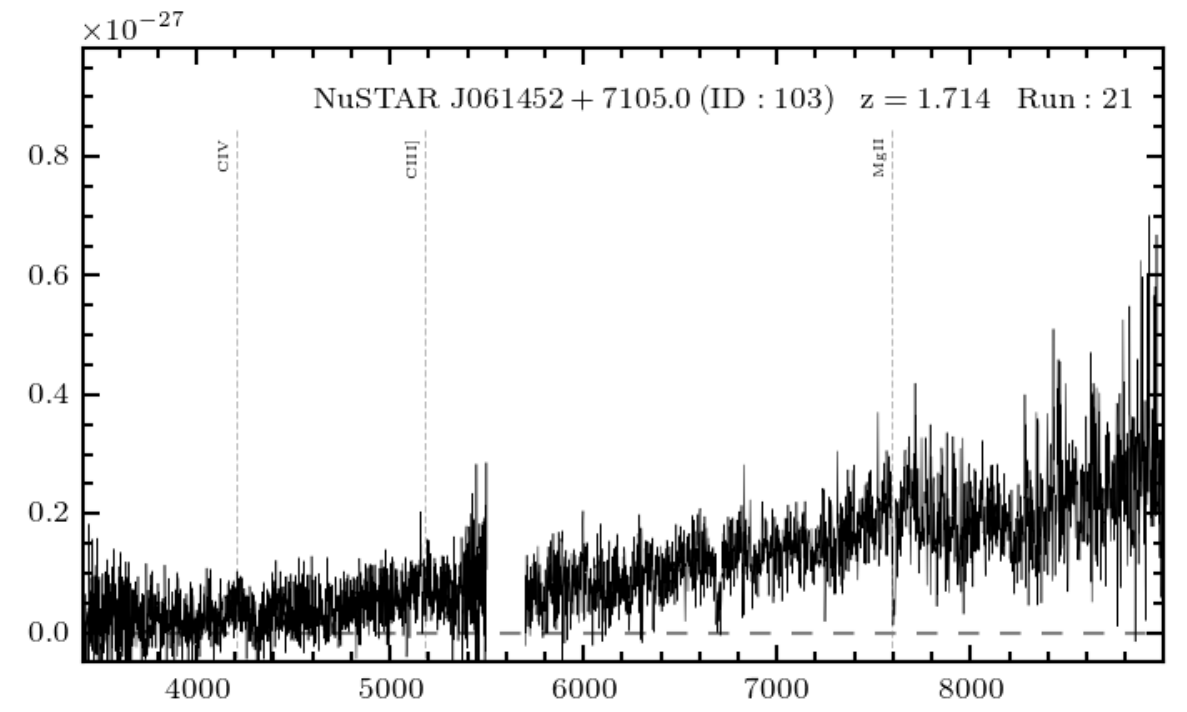}
\end{minipage}
\caption{Continued.}
\end{figure*}
\addtocounter{figure}{-1}
\begin{figure*}
\centering
\begin{minipage}[l]{0.325\textwidth}
\includegraphics[width=\textwidth]{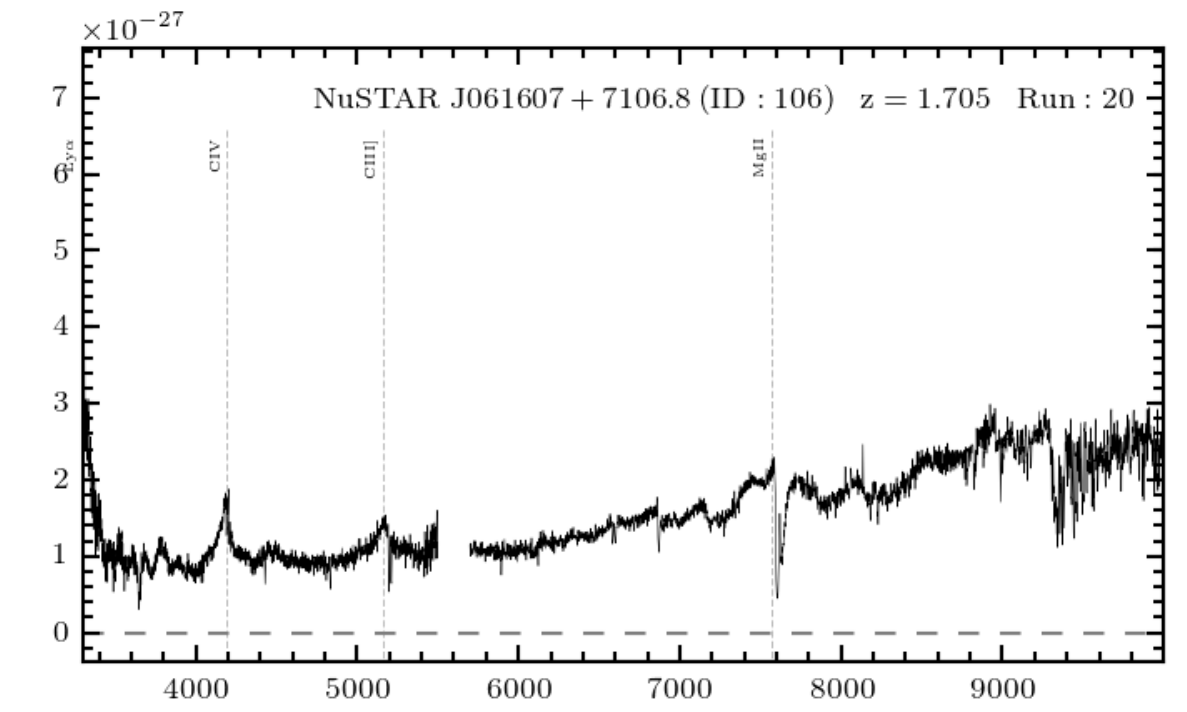}
\end{minipage}
\begin{minipage}[l]{0.325\textwidth}
\includegraphics[width=\textwidth]{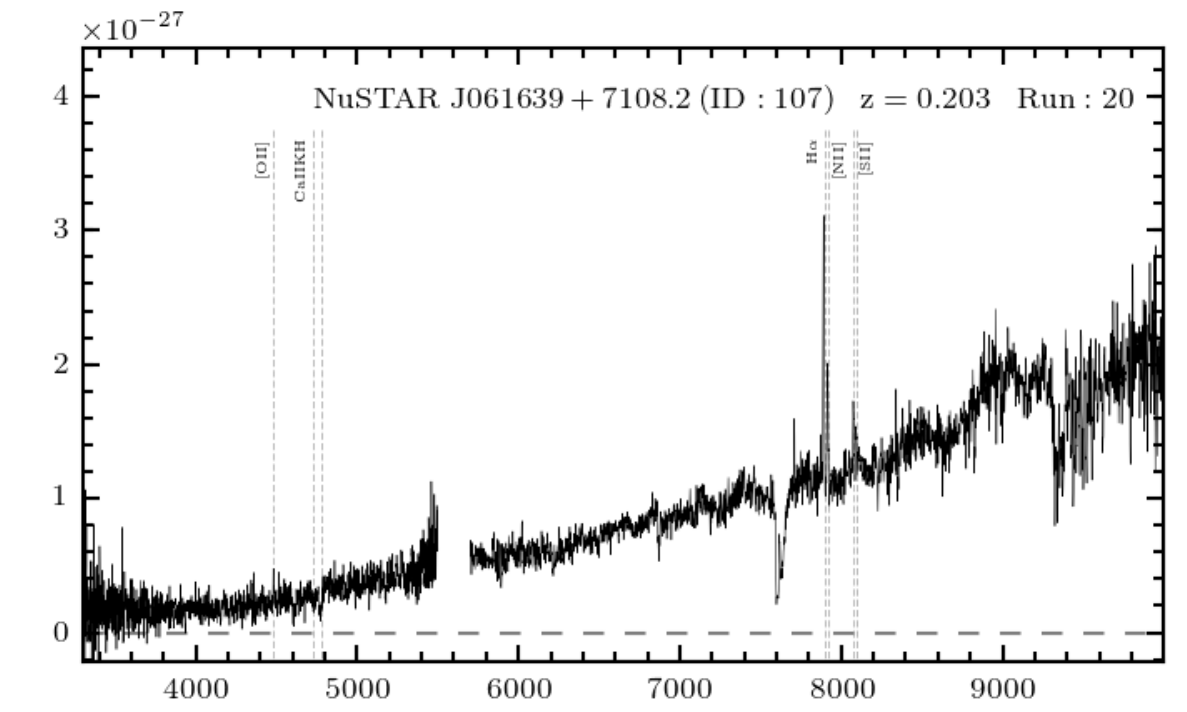}
\end{minipage}
\begin{minipage}[l]{0.325\textwidth}
\includegraphics[width=\textwidth]{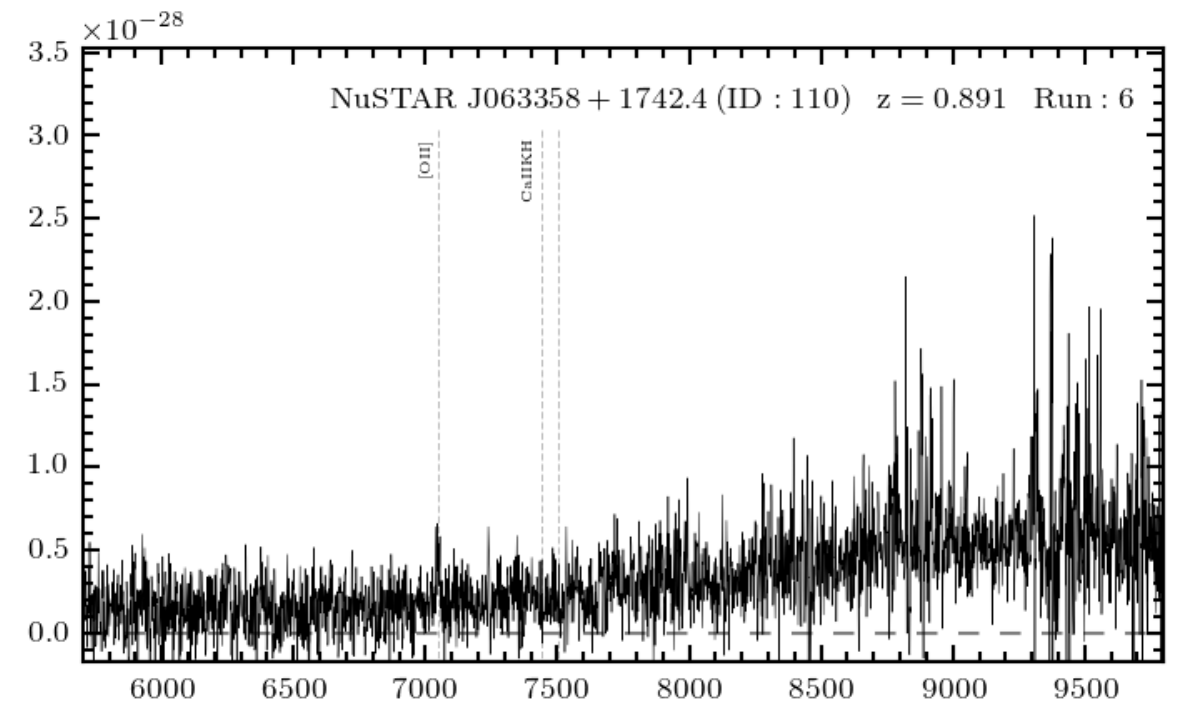}
\end{minipage}
\begin{minipage}[l]{0.325\textwidth}
\includegraphics[width=\textwidth]{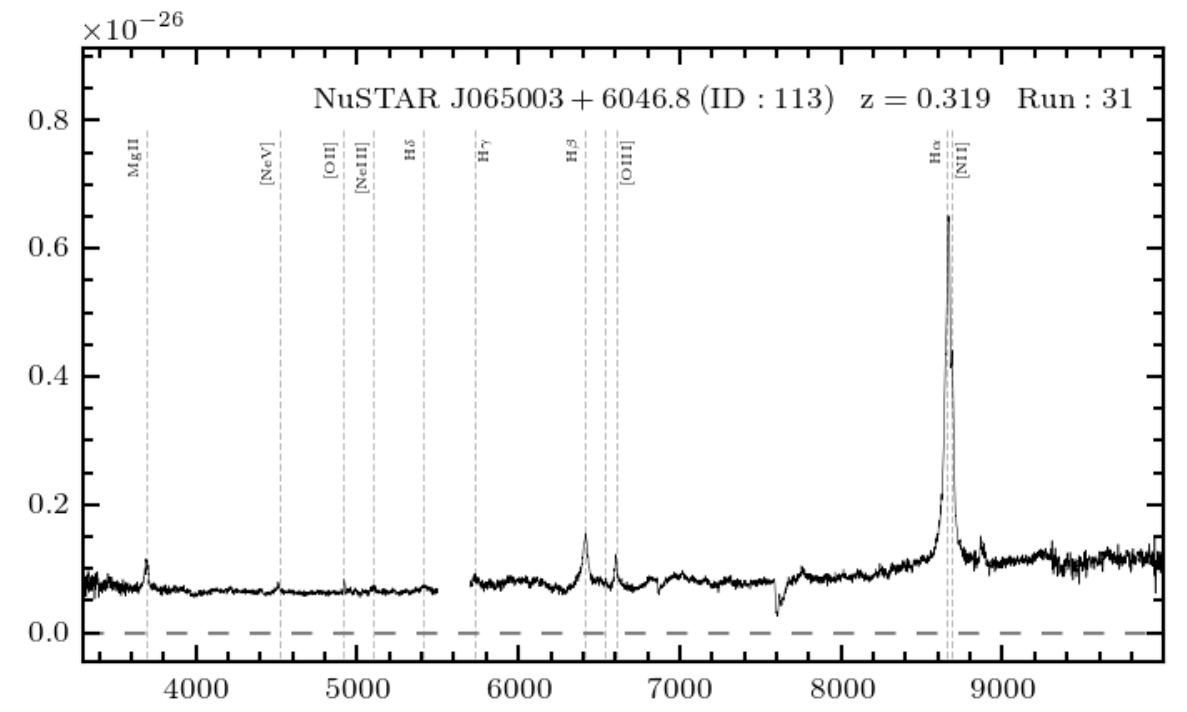}
\end{minipage}
\begin{minipage}[l]{0.325\textwidth}
\includegraphics[width=\textwidth]{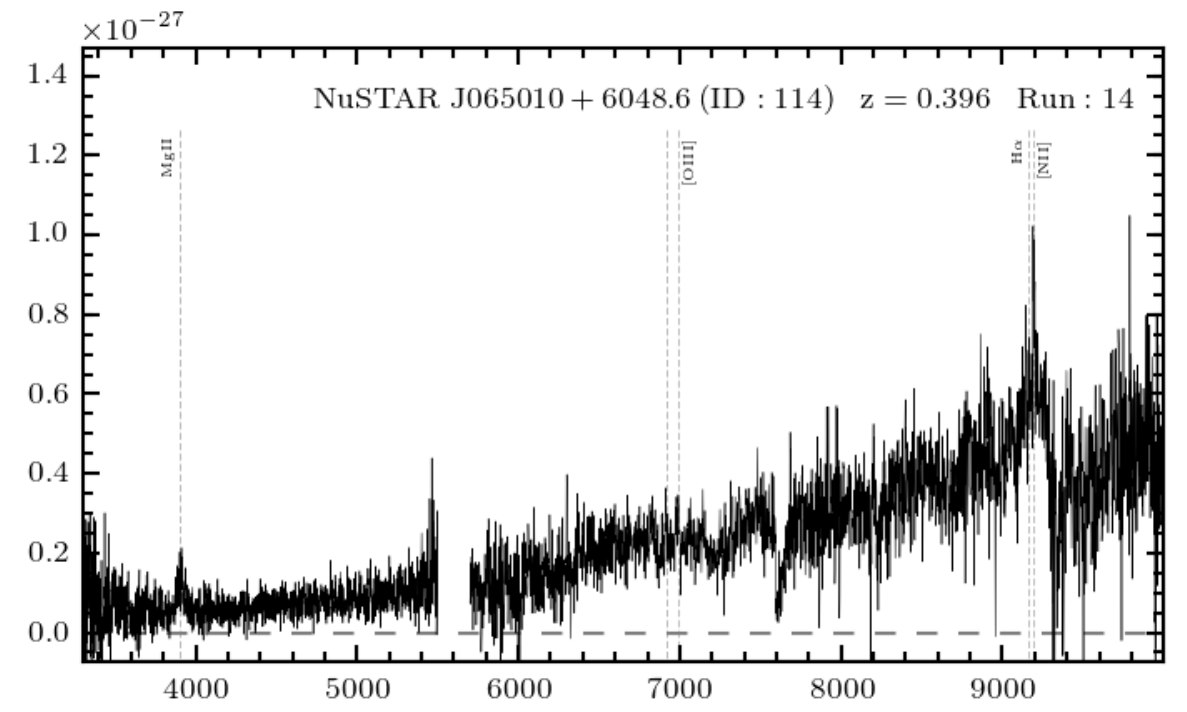}
\end{minipage}
\begin{minipage}[l]{0.325\textwidth}
\includegraphics[width=\textwidth]{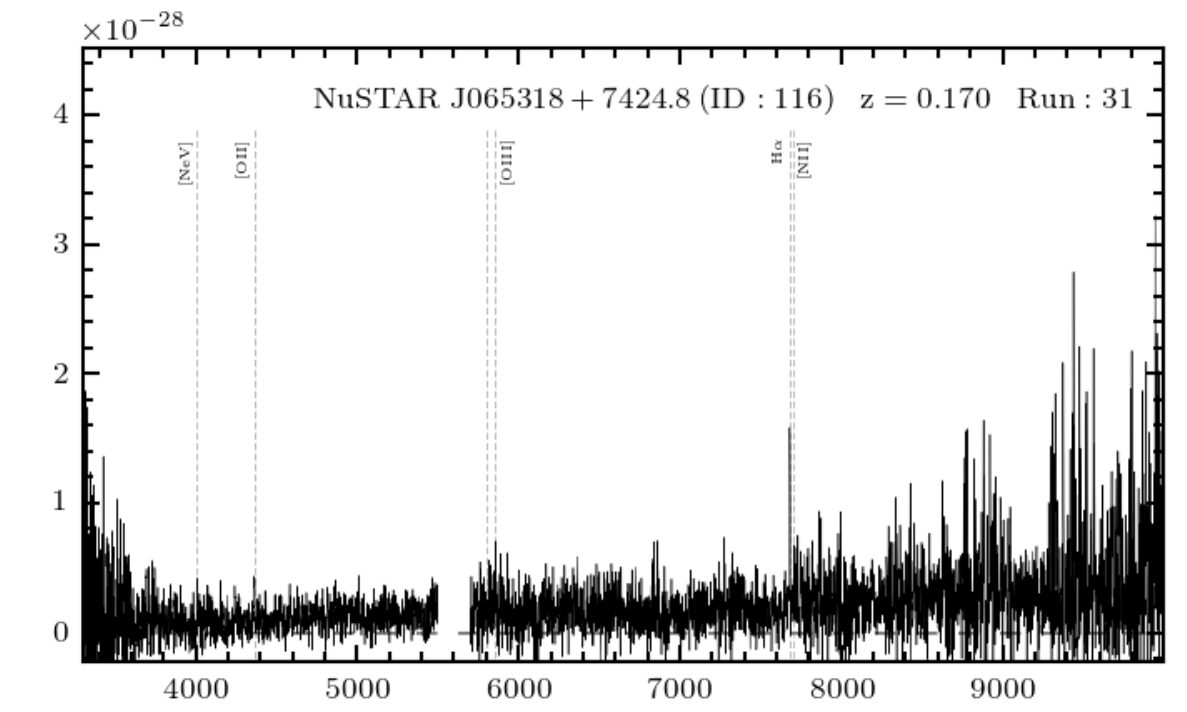}
\end{minipage}
\begin{minipage}[l]{0.325\textwidth}
\includegraphics[width=\textwidth]{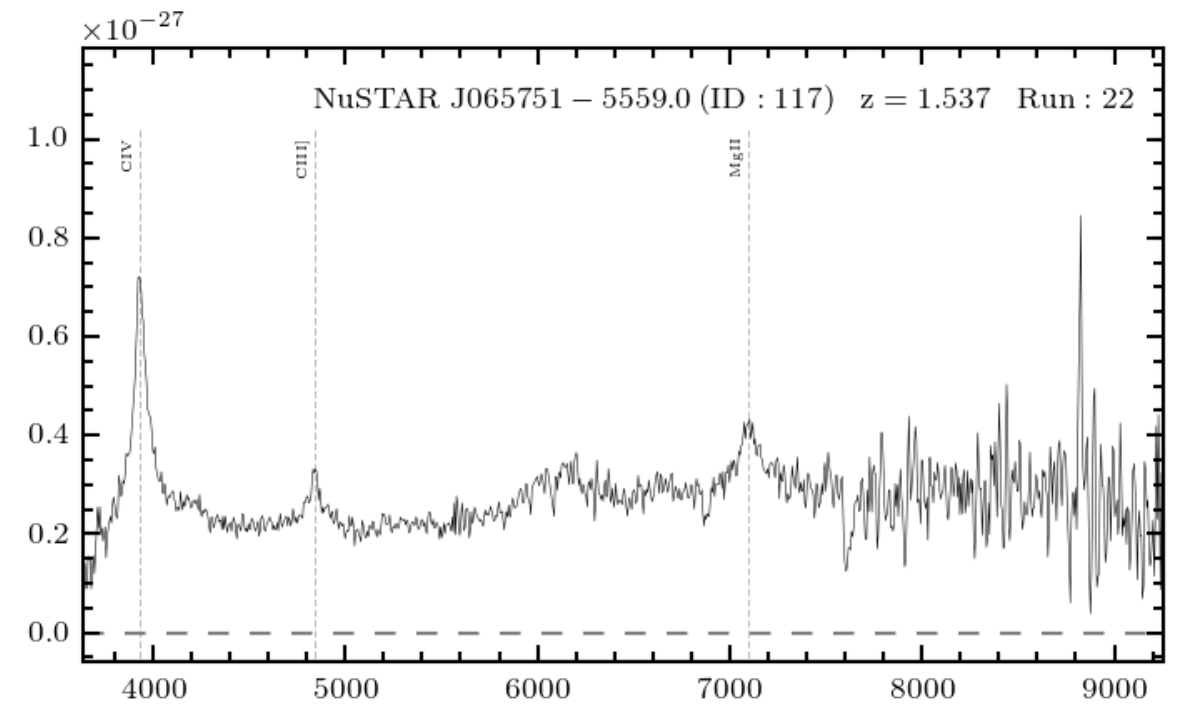}
\end{minipage}
\begin{minipage}[l]{0.325\textwidth}
\includegraphics[width=\textwidth]{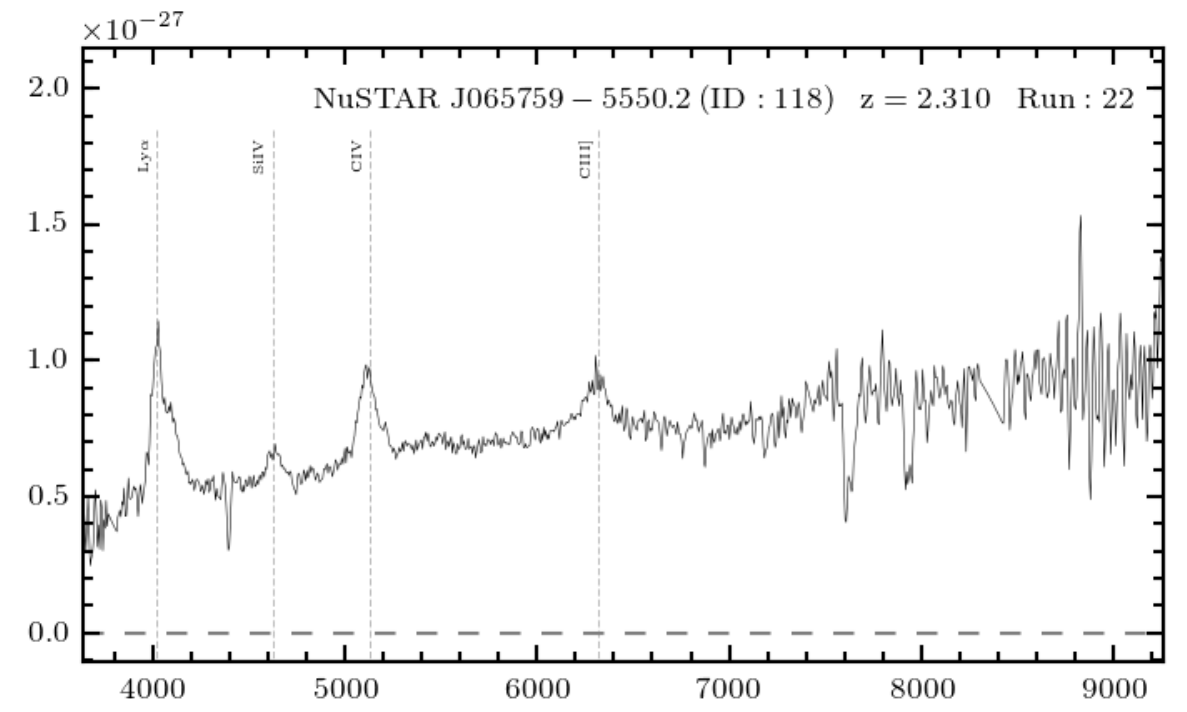}
\end{minipage}
\begin{minipage}[l]{0.325\textwidth}
\includegraphics[width=\textwidth]{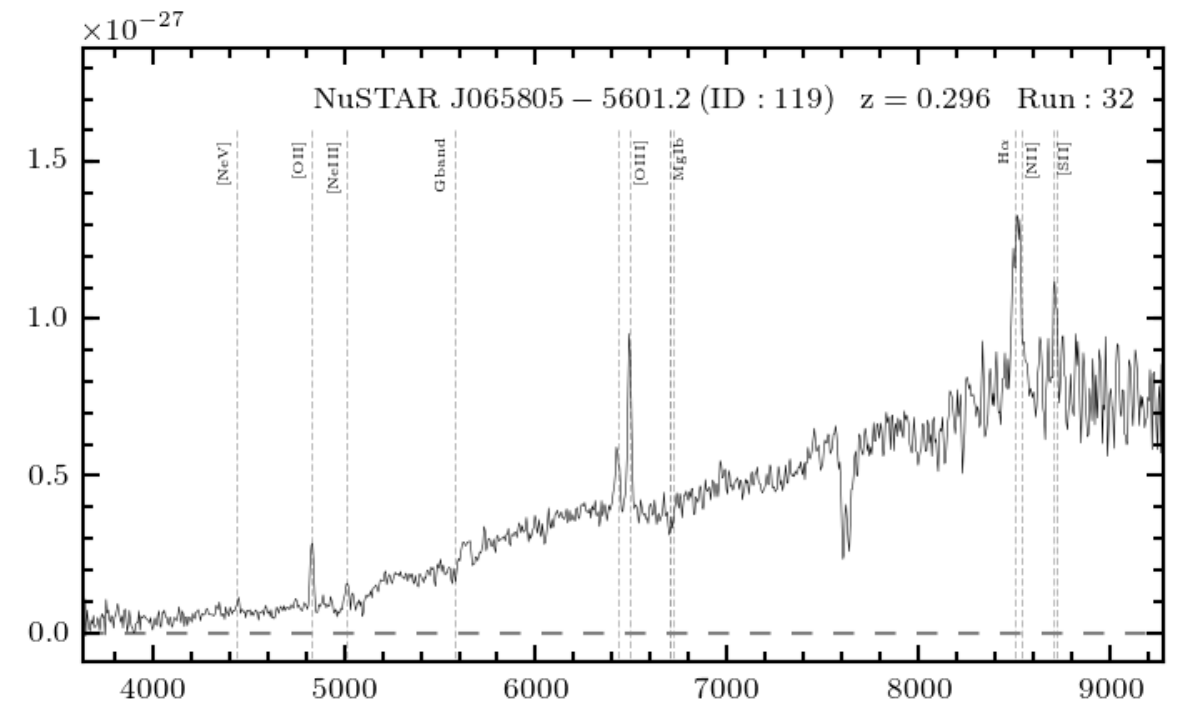}
\end{minipage}
\begin{minipage}[l]{0.325\textwidth}
\includegraphics[width=\textwidth]{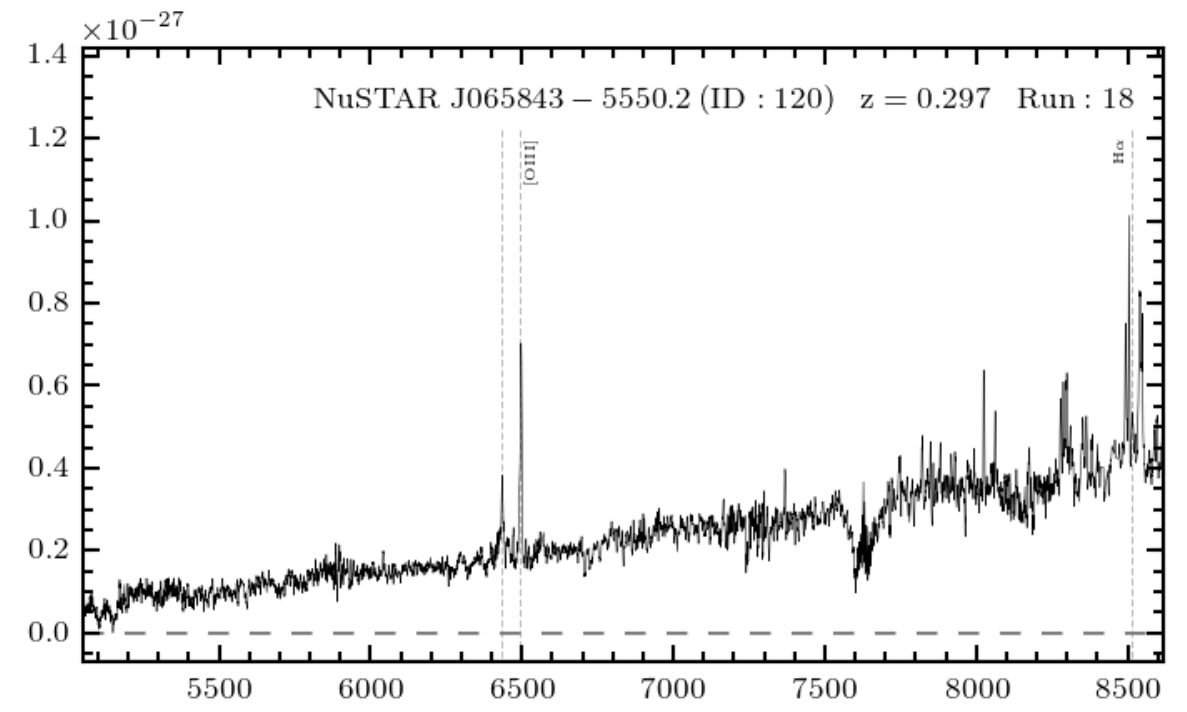}
\end{minipage}
\begin{minipage}[l]{0.325\textwidth}
\includegraphics[width=\textwidth]{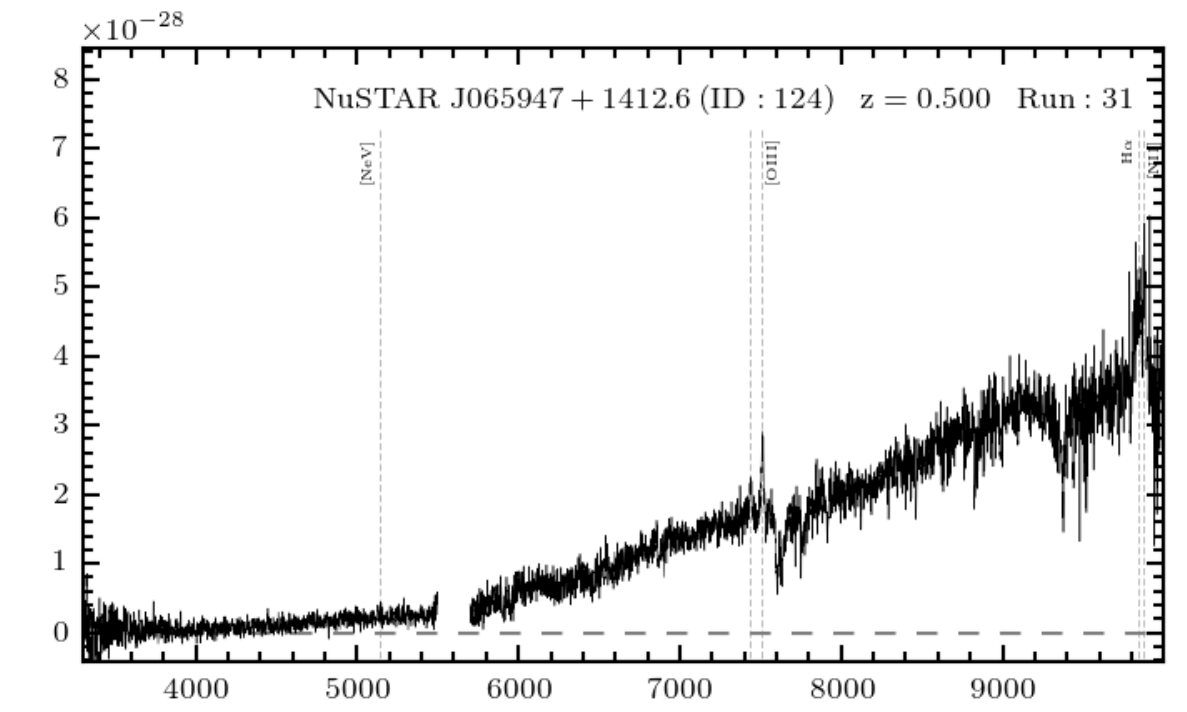}
\end{minipage}
\begin{minipage}[l]{0.325\textwidth}
\includegraphics[width=\textwidth]{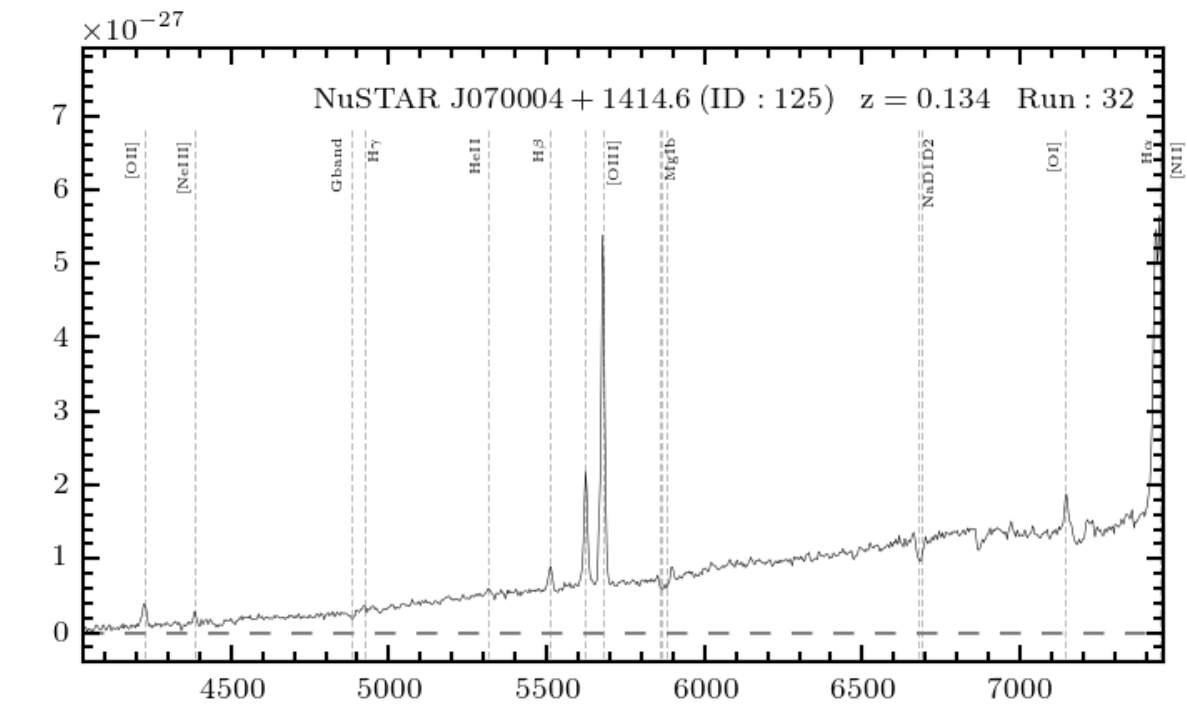}
\end{minipage}
\begin{minipage}[l]{0.325\textwidth}
\includegraphics[width=\textwidth]{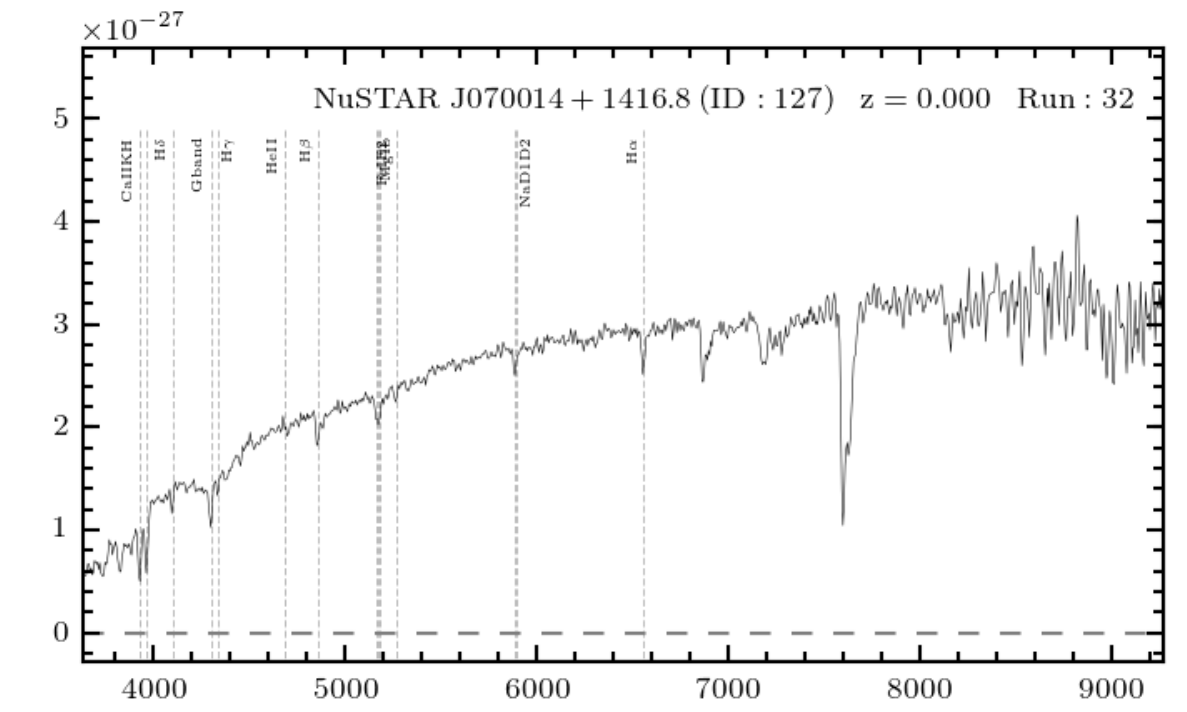}
\end{minipage}
\begin{minipage}[l]{0.325\textwidth}
\includegraphics[width=\textwidth]{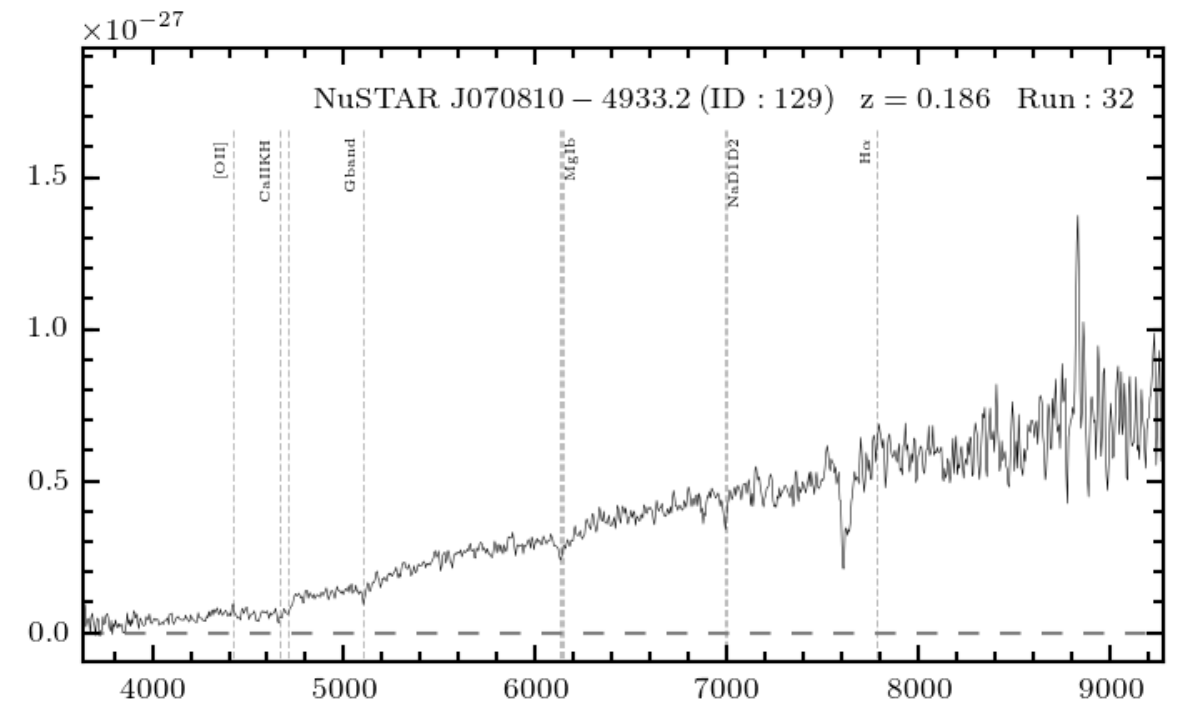}
\end{minipage}
\begin{minipage}[l]{0.325\textwidth}
\includegraphics[width=\textwidth]{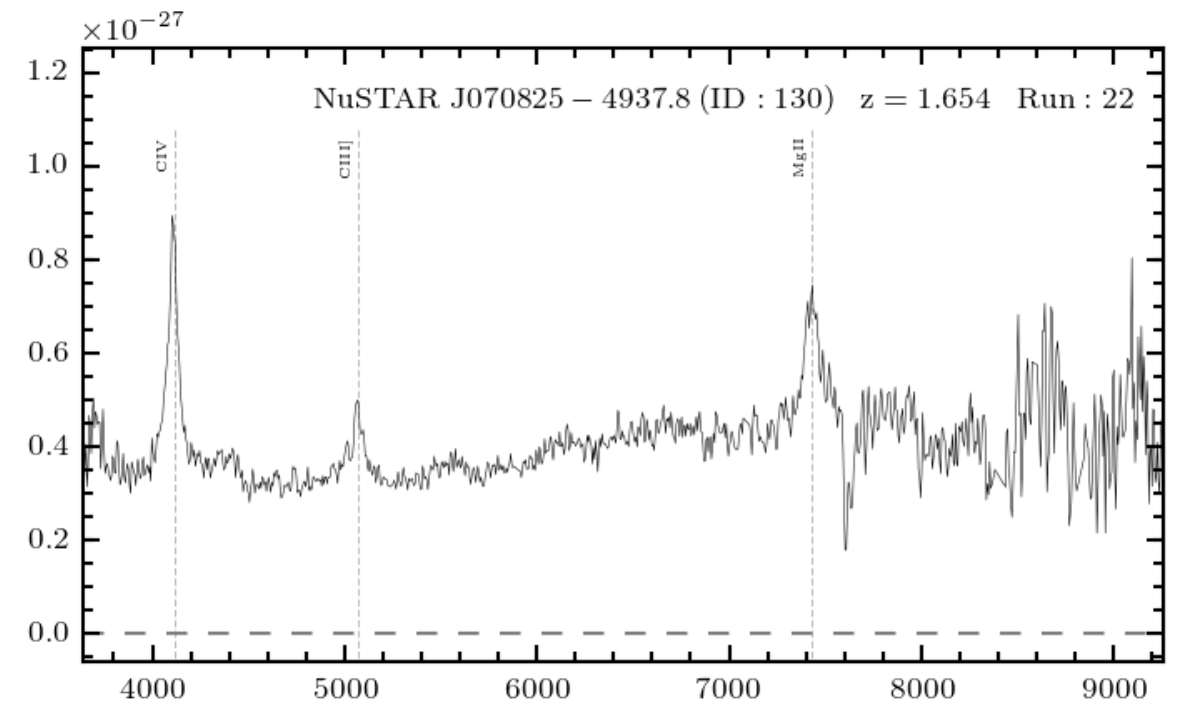}
\end{minipage}
\begin{minipage}[l]{0.325\textwidth}
\includegraphics[width=\textwidth]{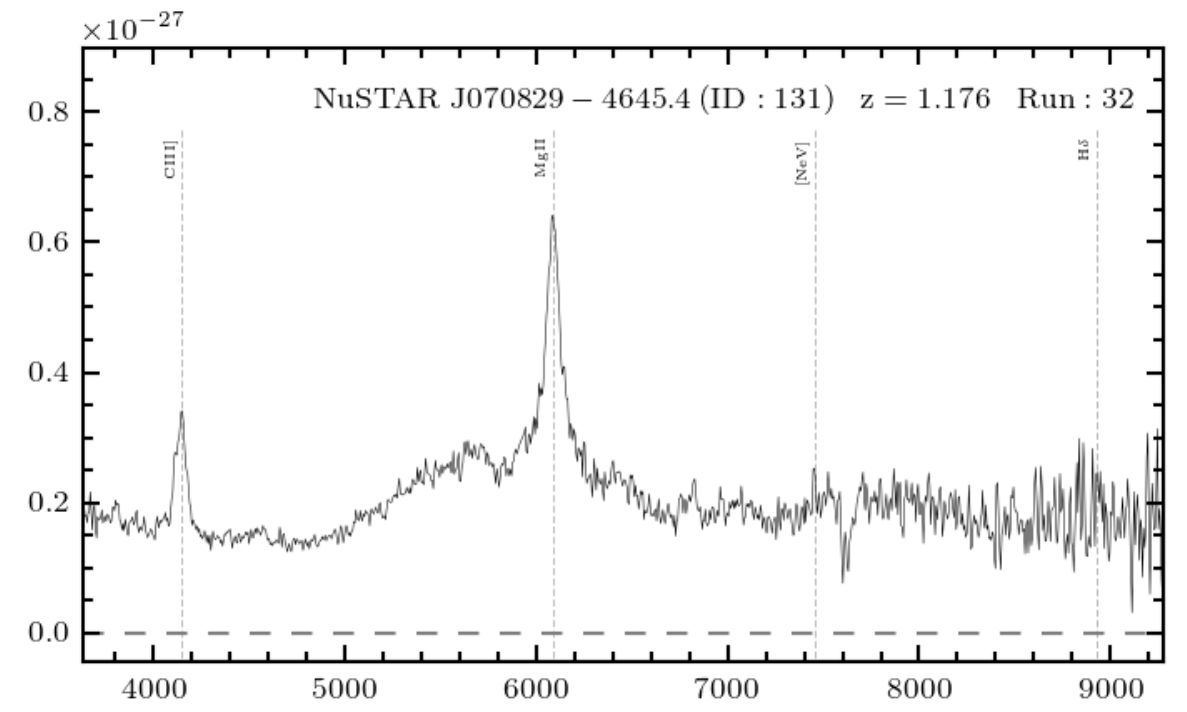}
\end{minipage}
\begin{minipage}[l]{0.325\textwidth}
\includegraphics[width=\textwidth]{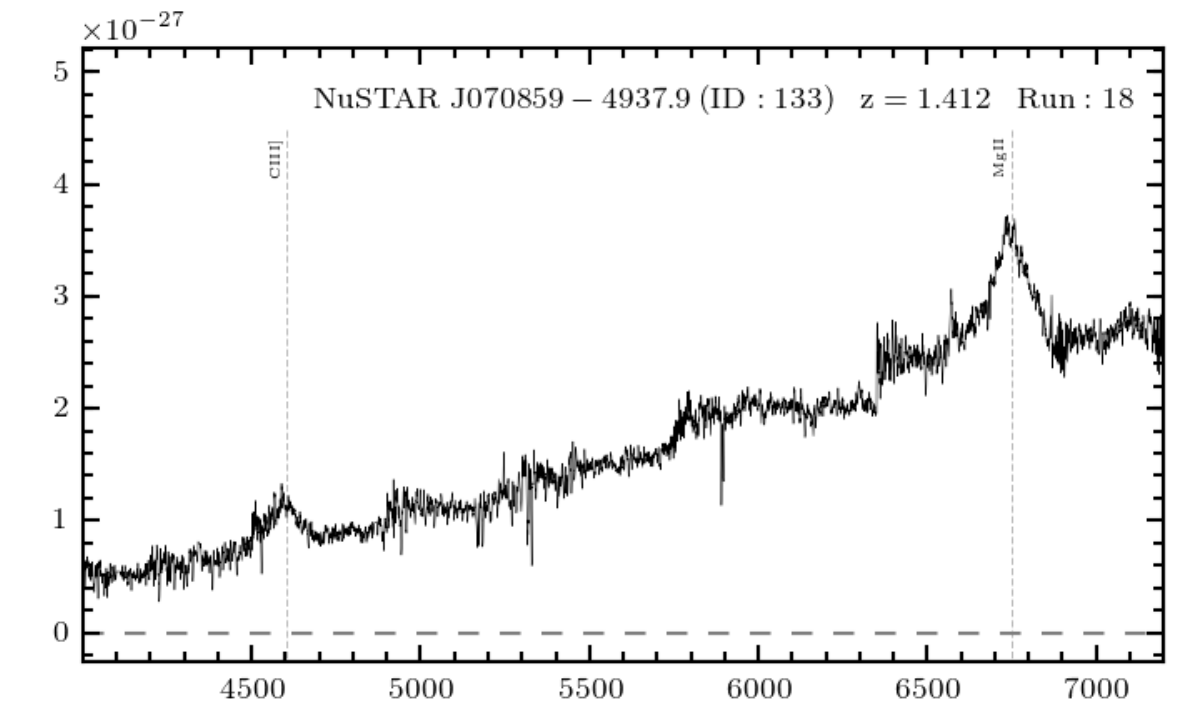}
\end{minipage}
\begin{minipage}[l]{0.325\textwidth}
\includegraphics[width=\textwidth]{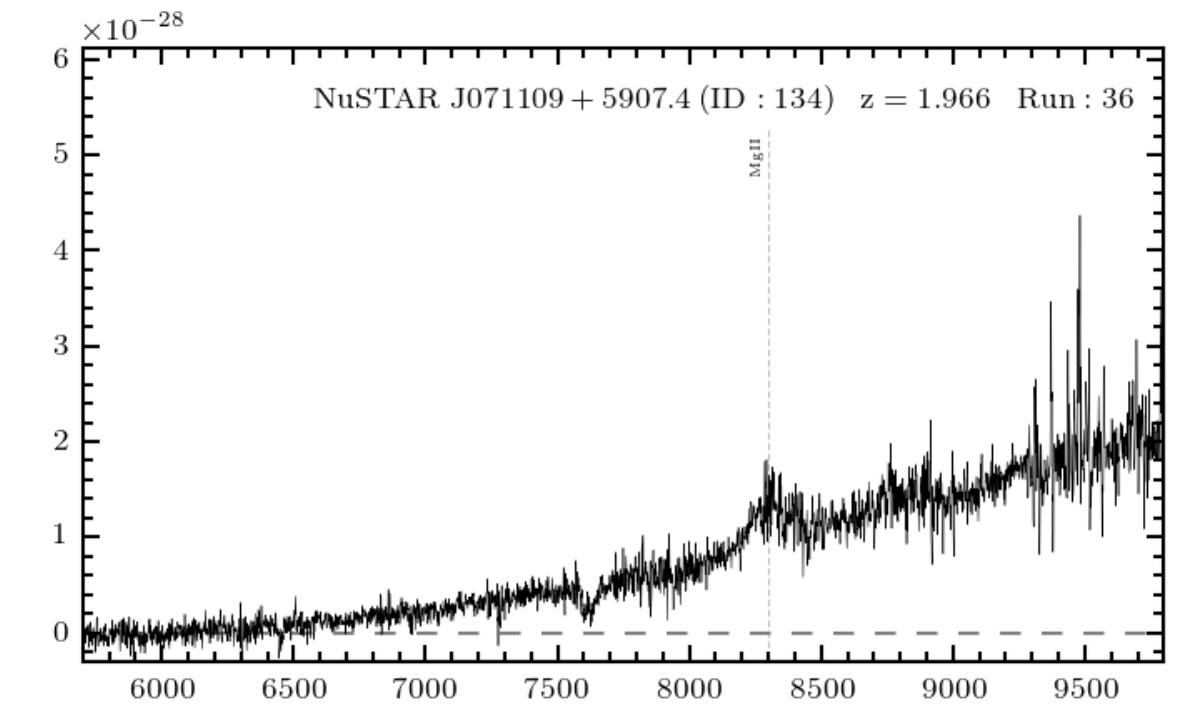}
\end{minipage}
\caption{Continued.}
\end{figure*}
\addtocounter{figure}{-1}
\begin{figure*}
\centering
\begin{minipage}[l]{0.325\textwidth}
\includegraphics[width=\textwidth]{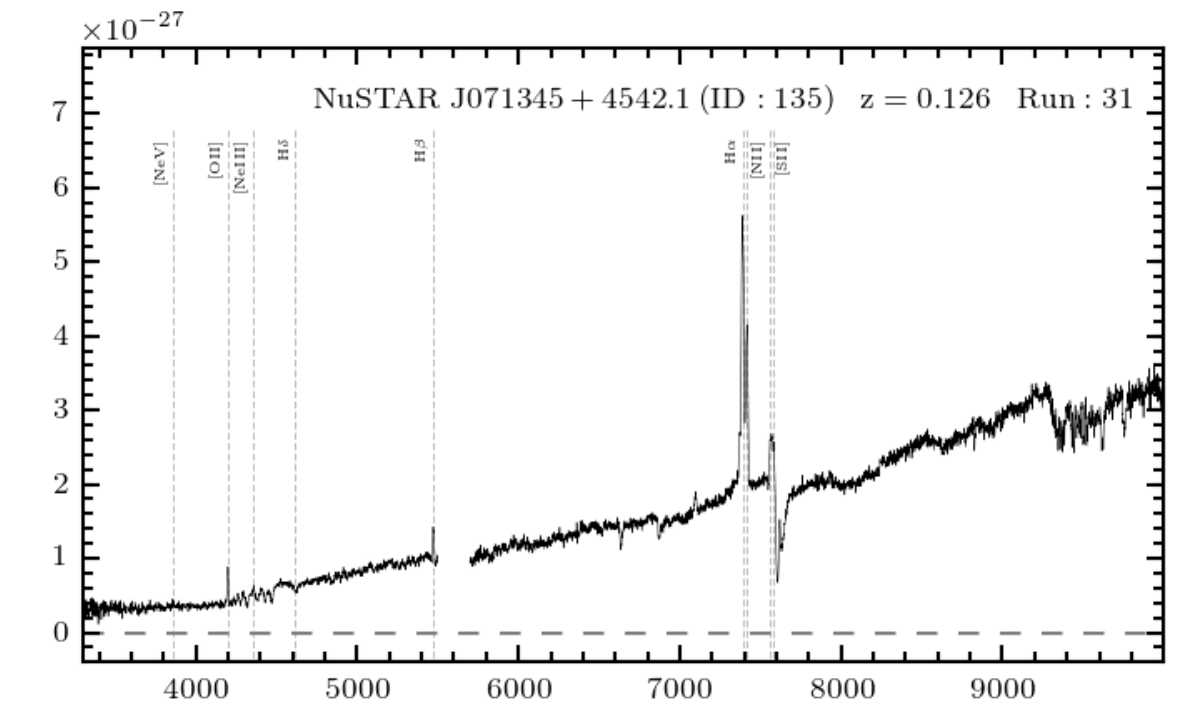}
\end{minipage}
\begin{minipage}[l]{0.325\textwidth}
\includegraphics[width=\textwidth]{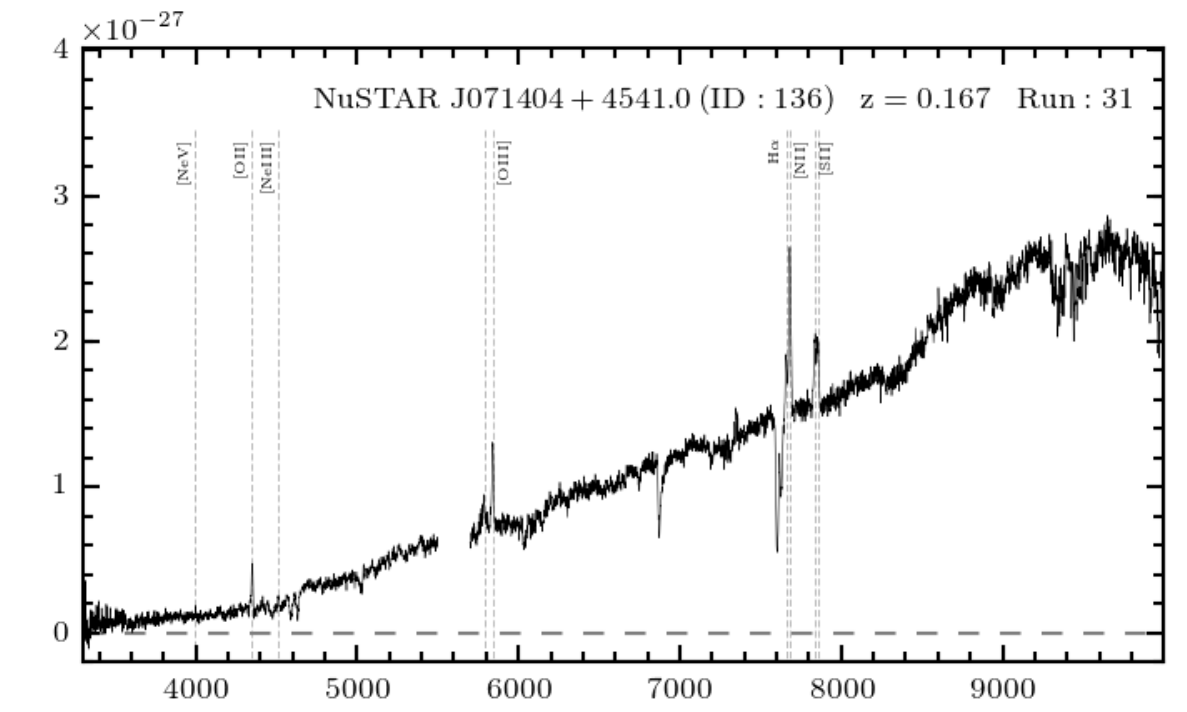}
\end{minipage}
\begin{minipage}[l]{0.325\textwidth}
\includegraphics[width=\textwidth]{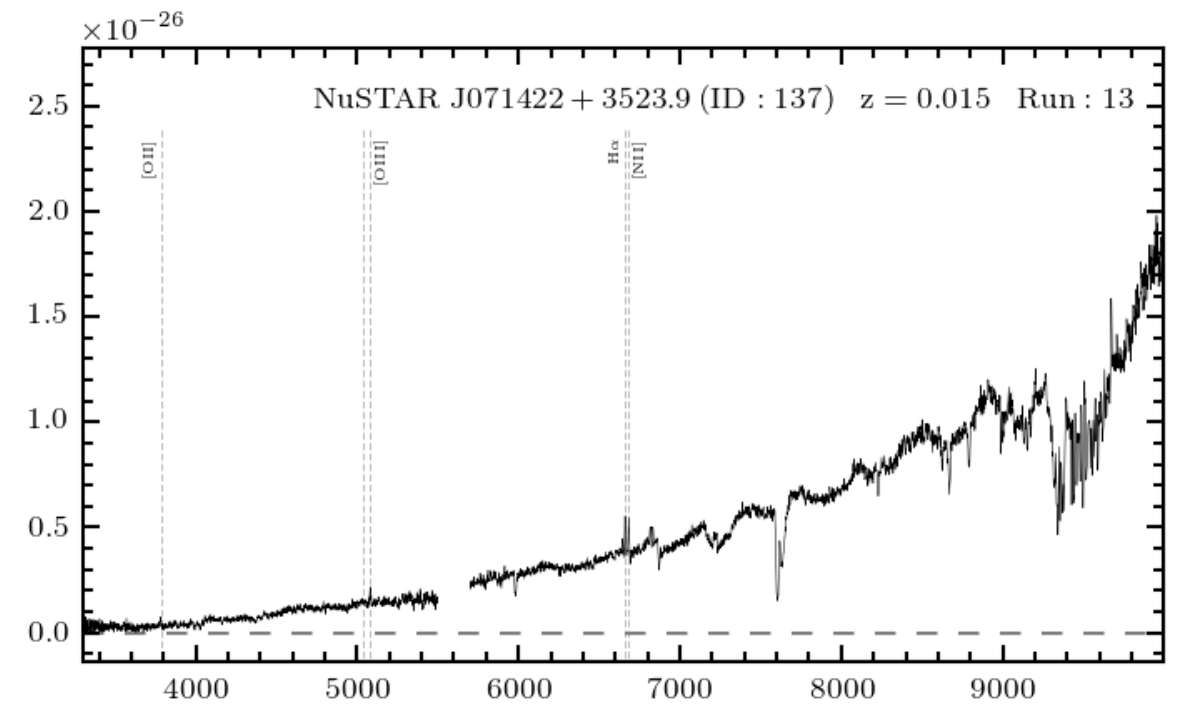}
\end{minipage}
\begin{minipage}[l]{0.325\textwidth}
\includegraphics[width=\textwidth]{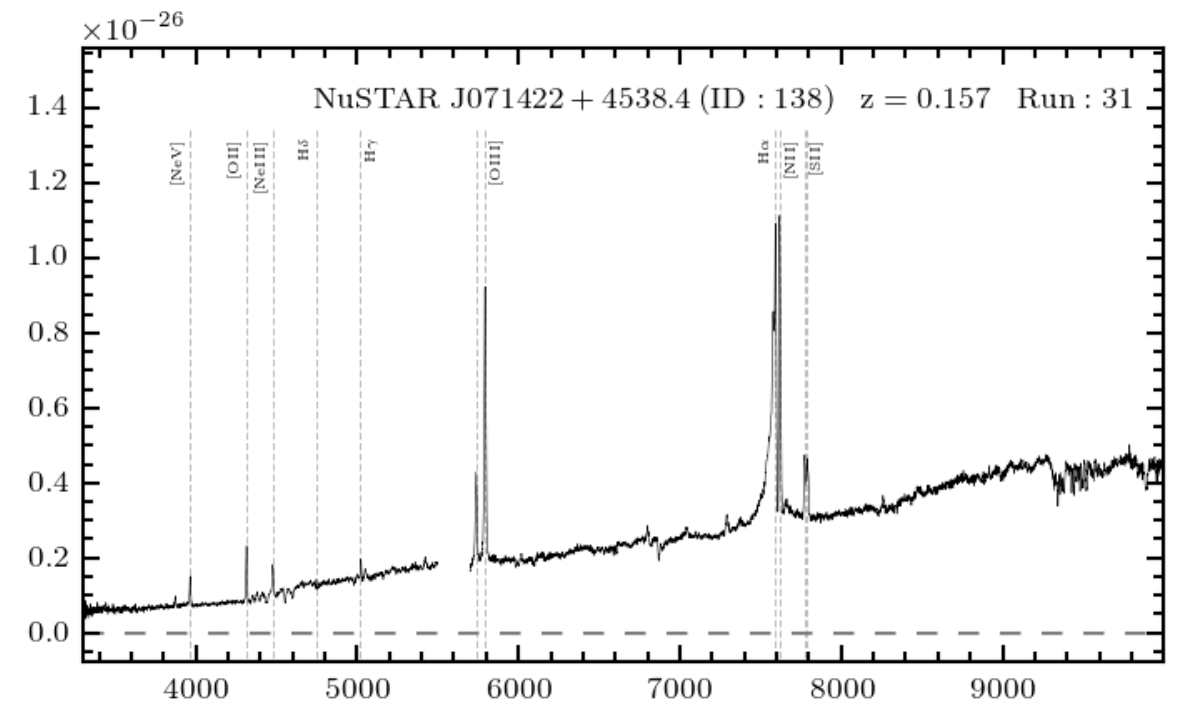}
\end{minipage}
\begin{minipage}[l]{0.325\textwidth}
\includegraphics[width=\textwidth]{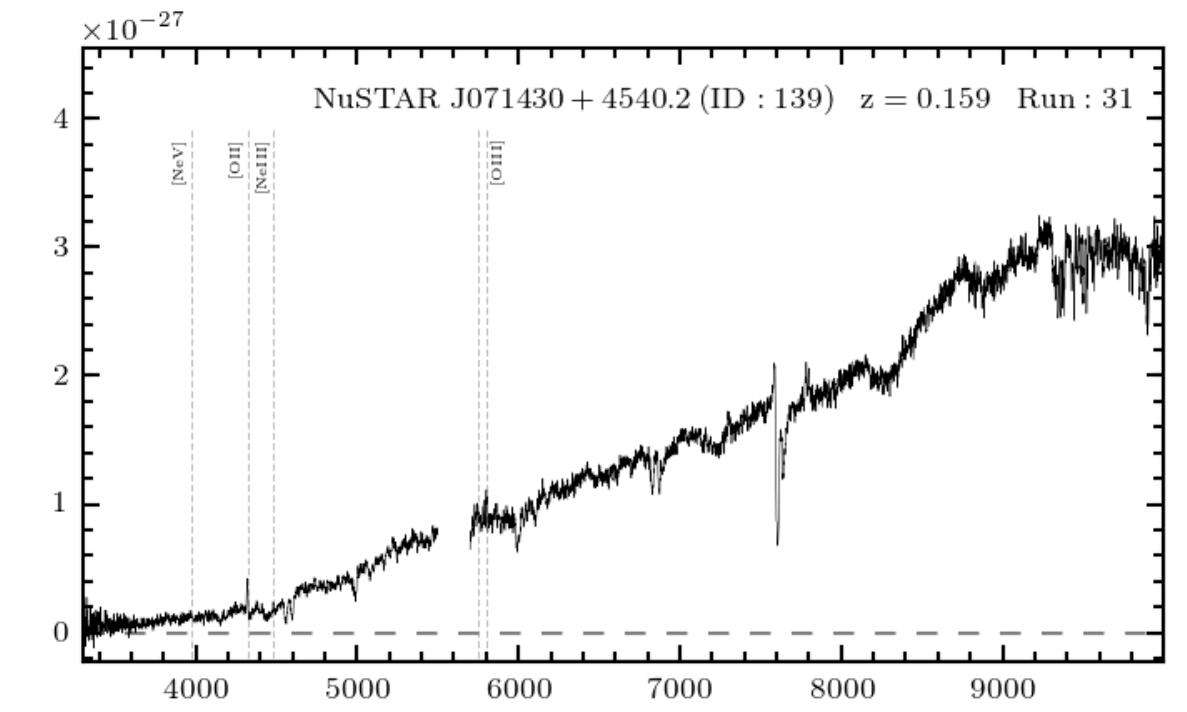}
\end{minipage}
\begin{minipage}[l]{0.325\textwidth}
\includegraphics[width=\textwidth]{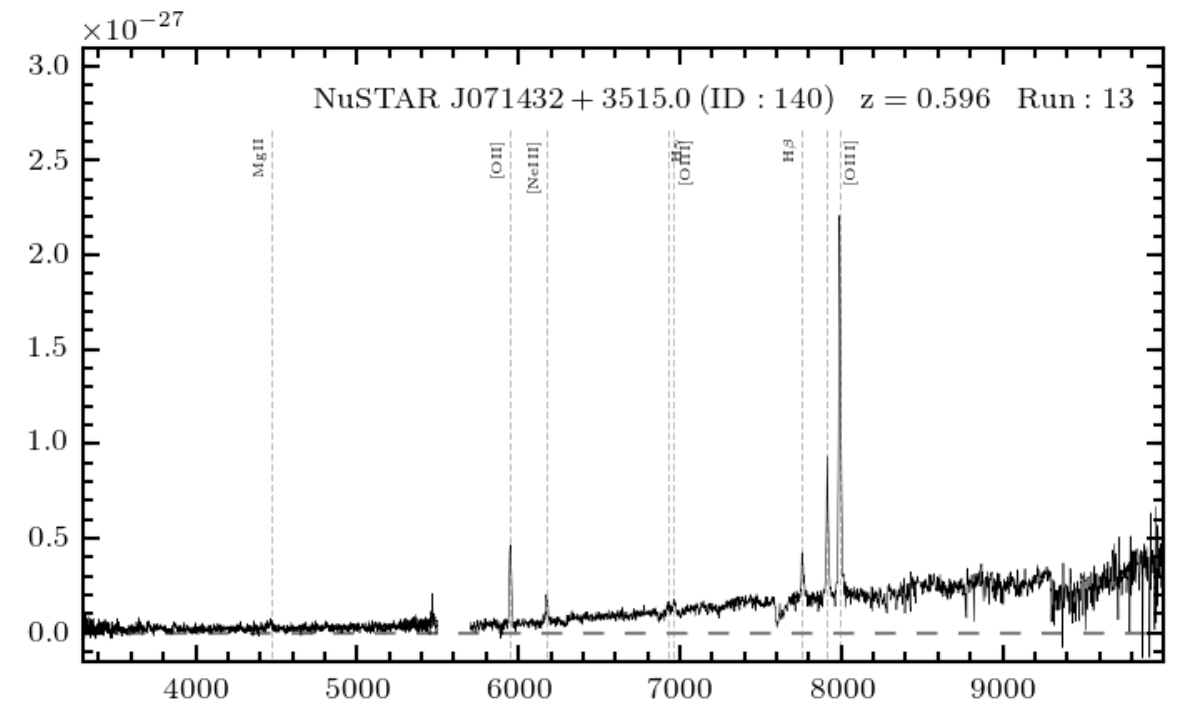}
\end{minipage}
\begin{minipage}[l]{0.325\textwidth}
\includegraphics[width=\textwidth]{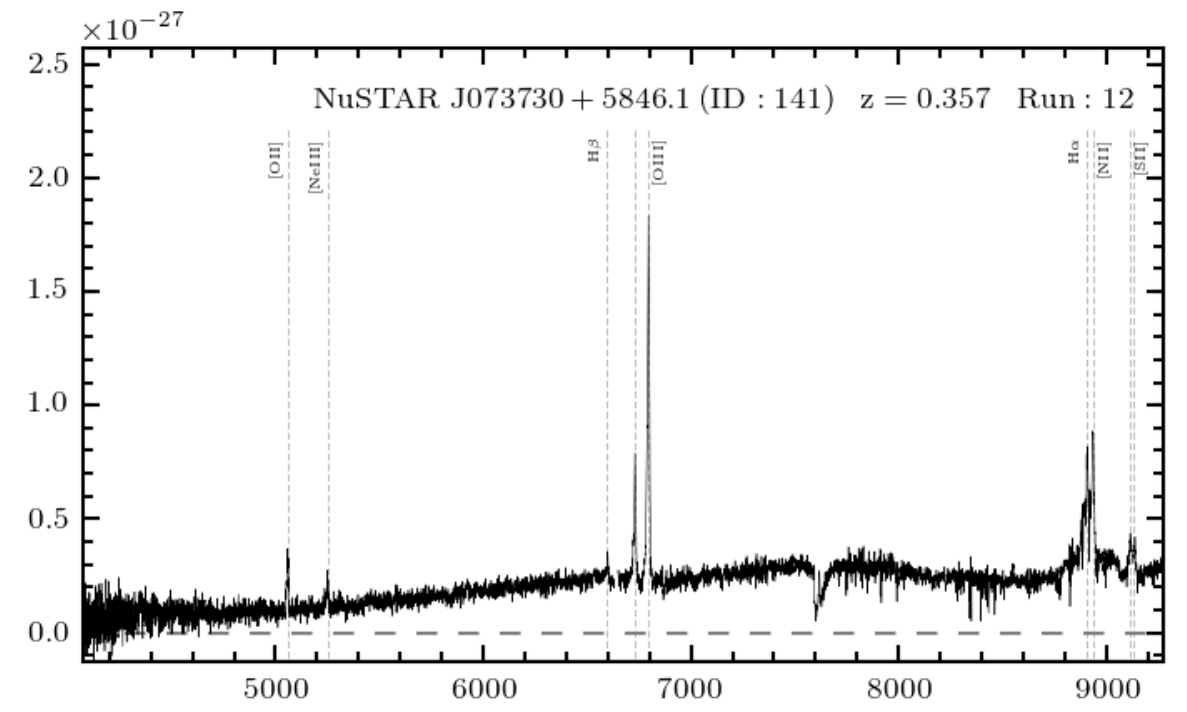}
\end{minipage}
\begin{minipage}[l]{0.325\textwidth}
\includegraphics[width=\textwidth]{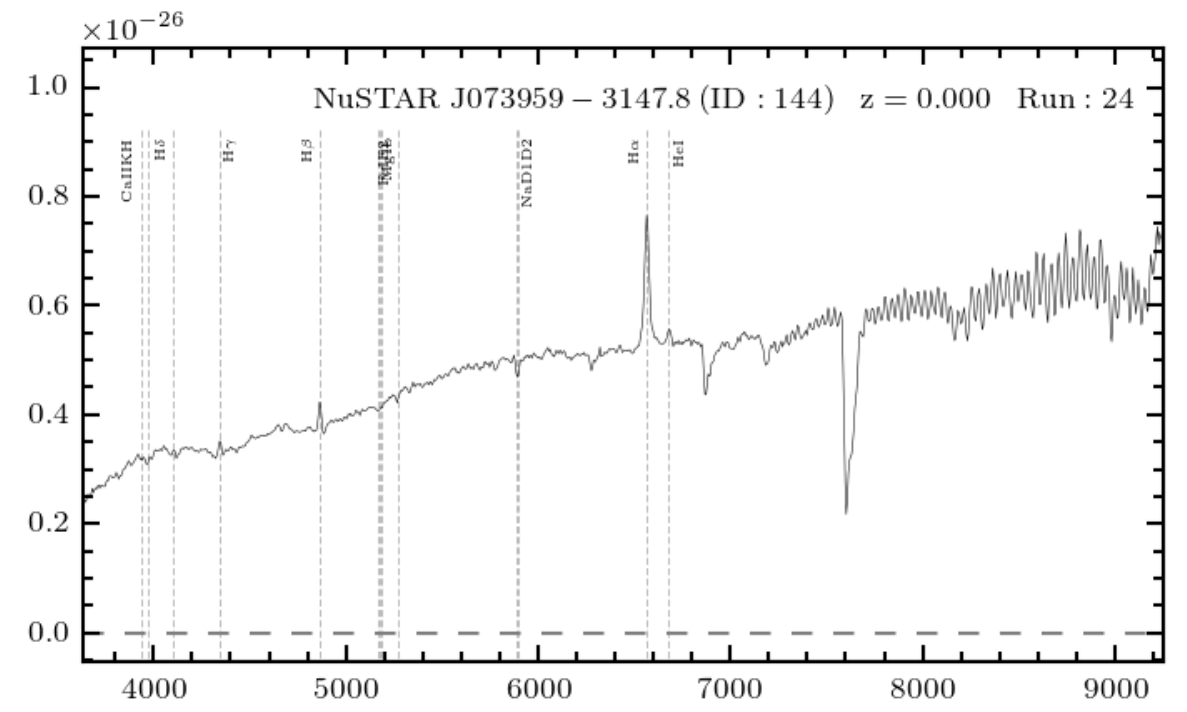}
\end{minipage}
\begin{minipage}[l]{0.325\textwidth}
\includegraphics[width=\textwidth]{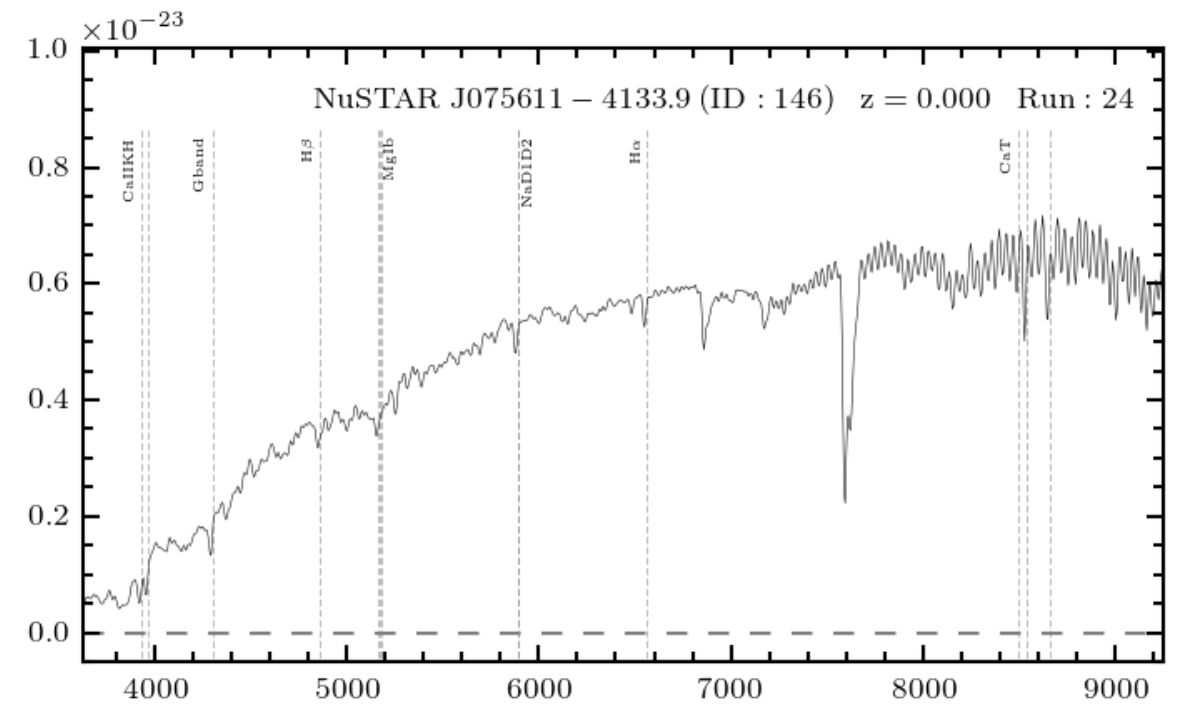}
\end{minipage}
\begin{minipage}[l]{0.325\textwidth}
\includegraphics[width=\textwidth]{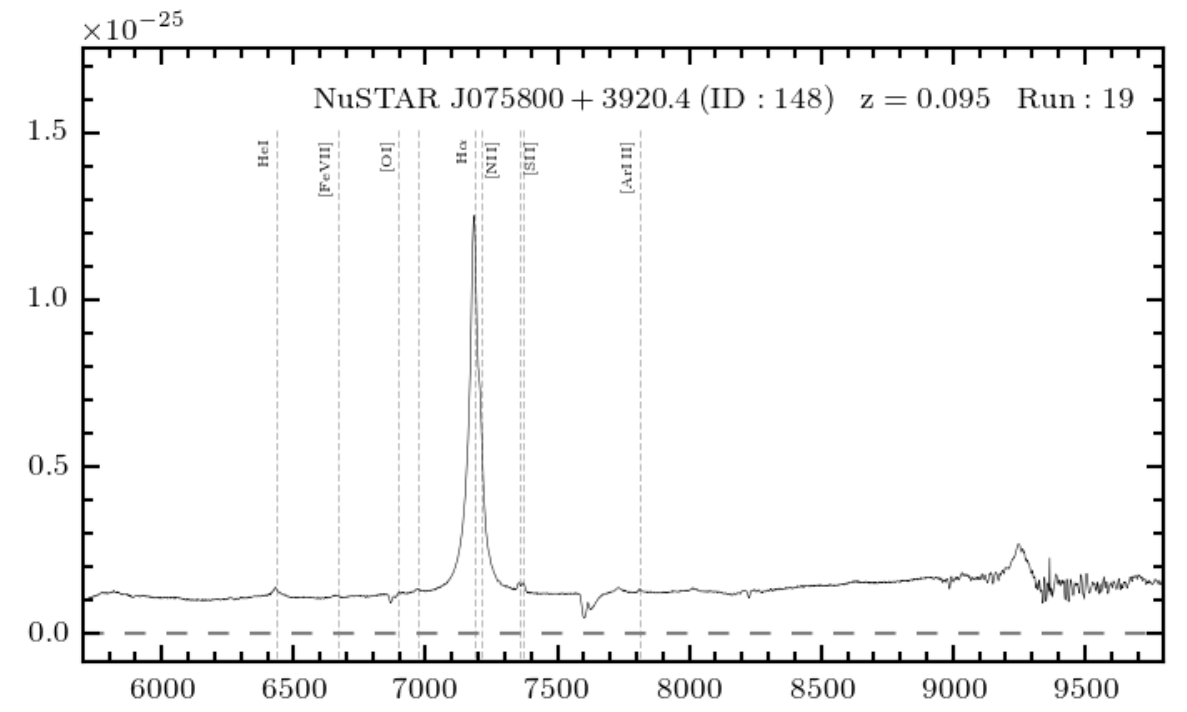}
\end{minipage}
\begin{minipage}[l]{0.325\textwidth}
\includegraphics[width=\textwidth]{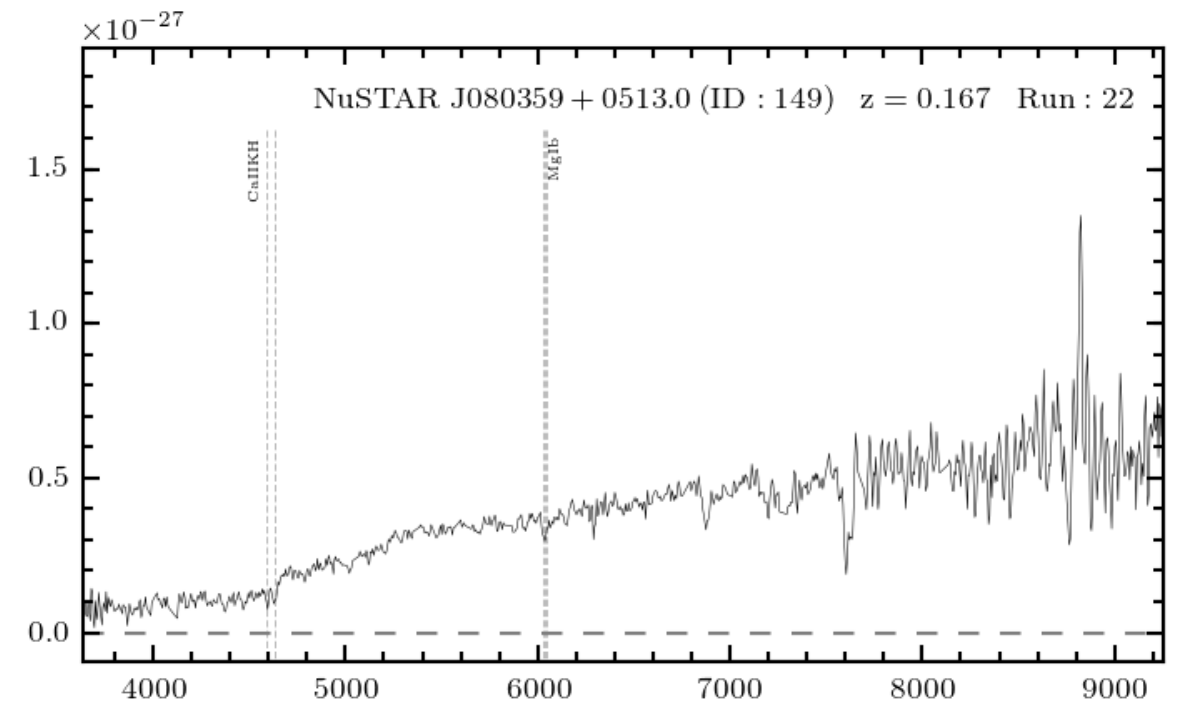}
\end{minipage}
\begin{minipage}[l]{0.325\textwidth}
\includegraphics[width=\textwidth]{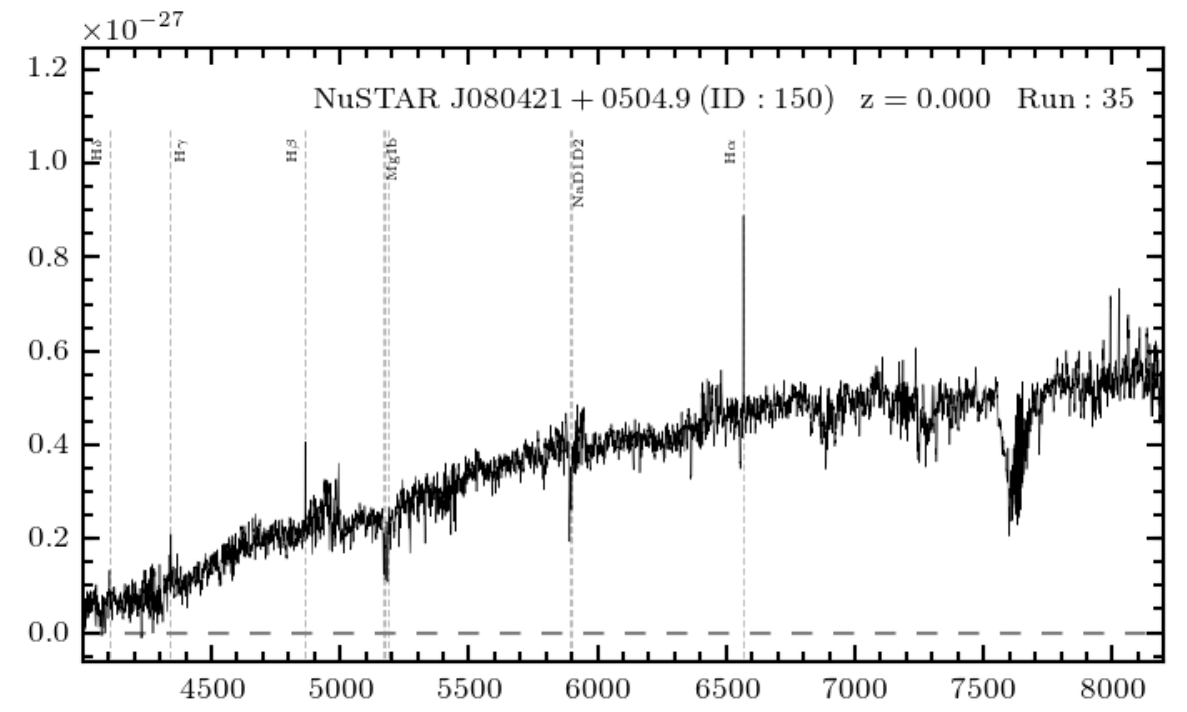}
\end{minipage}
\begin{minipage}[l]{0.325\textwidth}
\includegraphics[width=\textwidth]{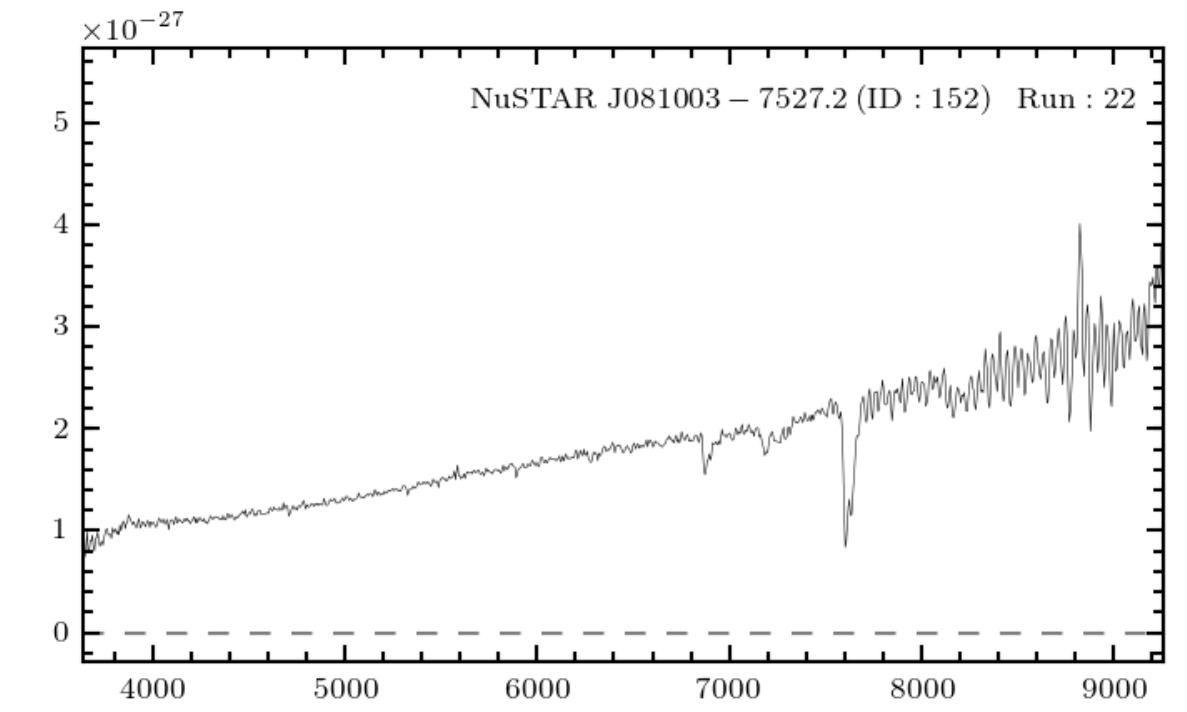}
\end{minipage}
\begin{minipage}[l]{0.325\textwidth}
\includegraphics[width=\textwidth]{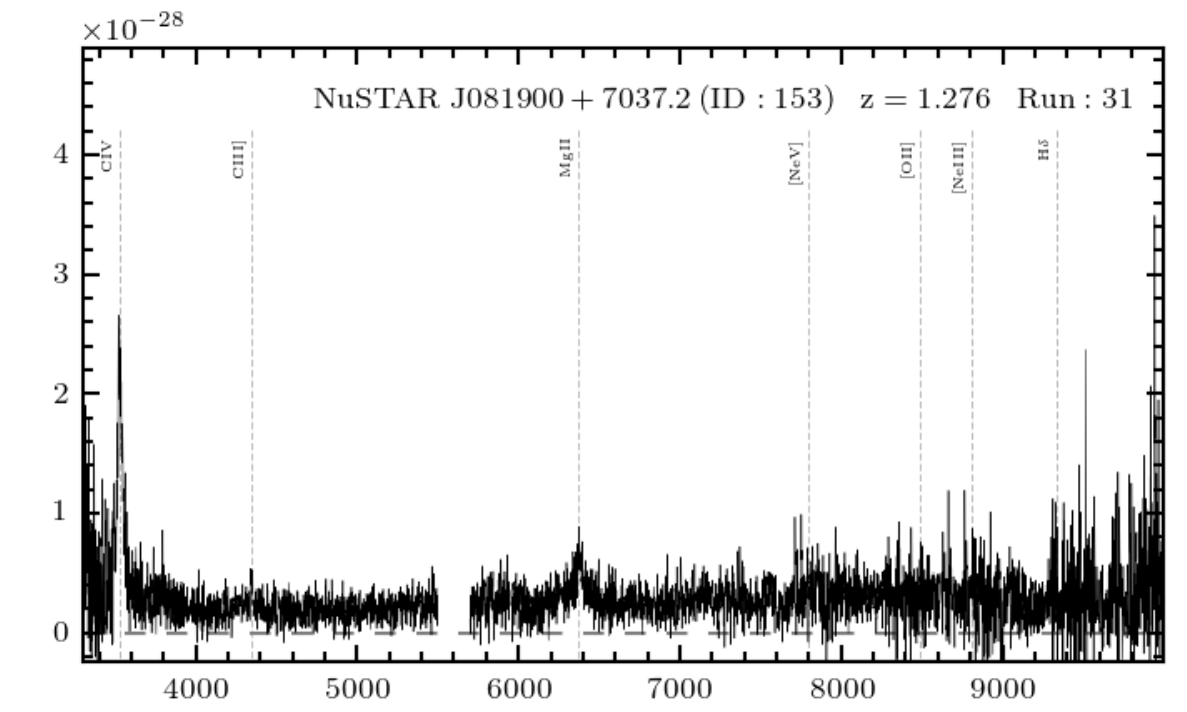}
\end{minipage}
\begin{minipage}[l]{0.325\textwidth}
\includegraphics[width=\textwidth]{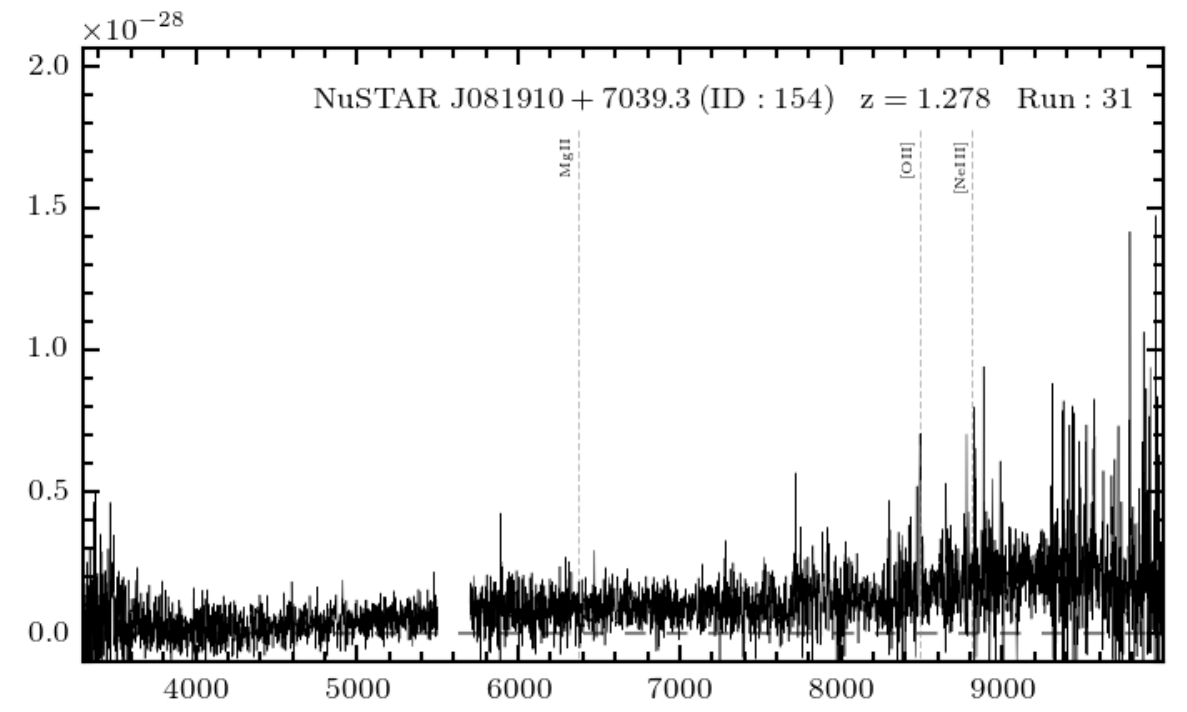}
\end{minipage}
\begin{minipage}[l]{0.325\textwidth}
\includegraphics[width=\textwidth]{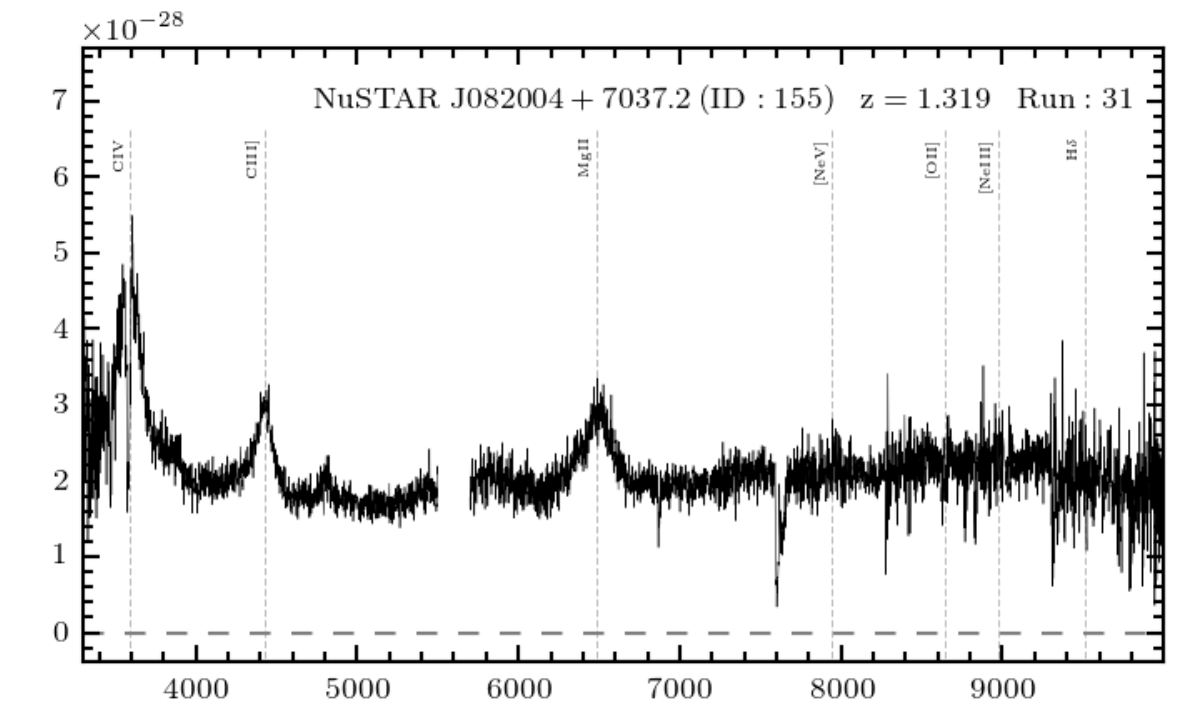}
\end{minipage}
\begin{minipage}[l]{0.325\textwidth}
\includegraphics[width=\textwidth]{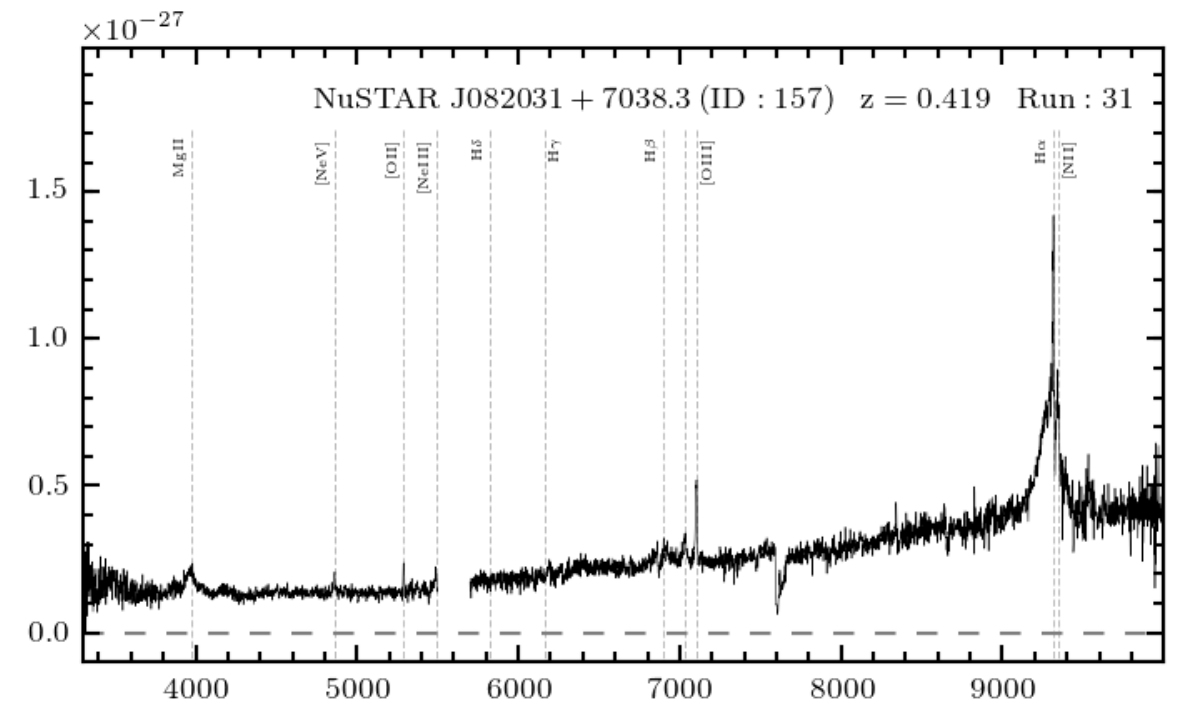}
\end{minipage}
\begin{minipage}[l]{0.325\textwidth}
\includegraphics[width=\textwidth]{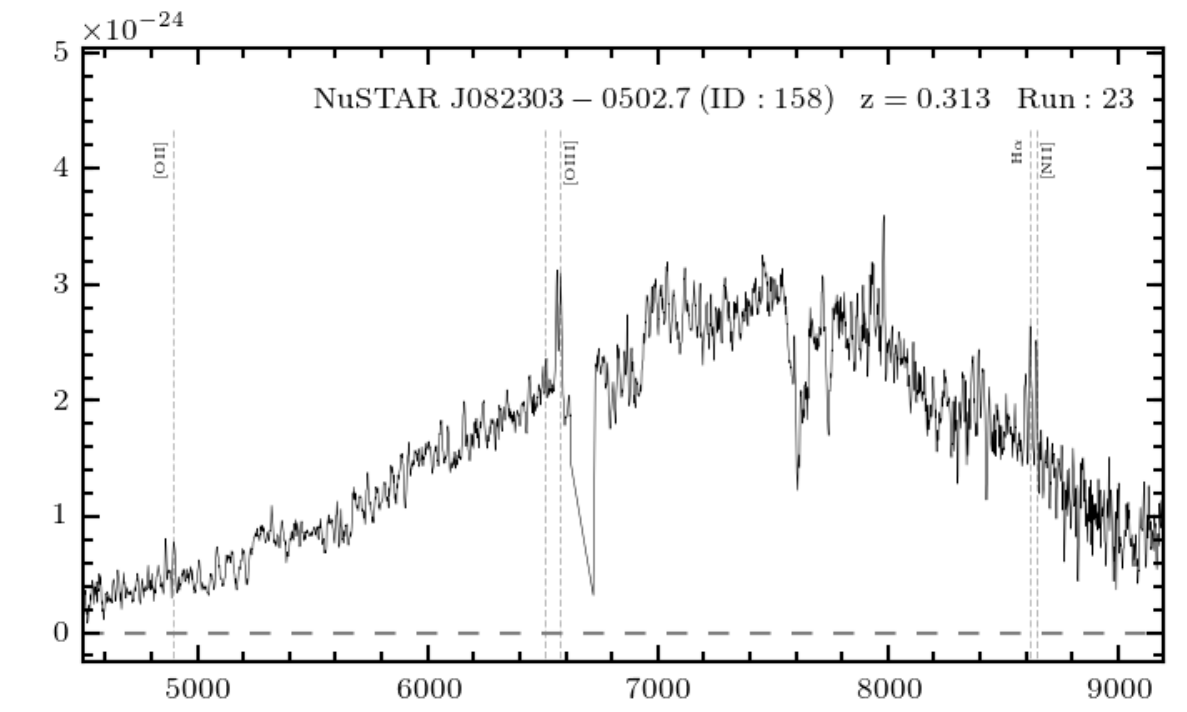}
\end{minipage}
\caption{Continued.}
\end{figure*}
\addtocounter{figure}{-1}
\begin{figure*}
\centering
\begin{minipage}[l]{0.325\textwidth}
\includegraphics[width=\textwidth]{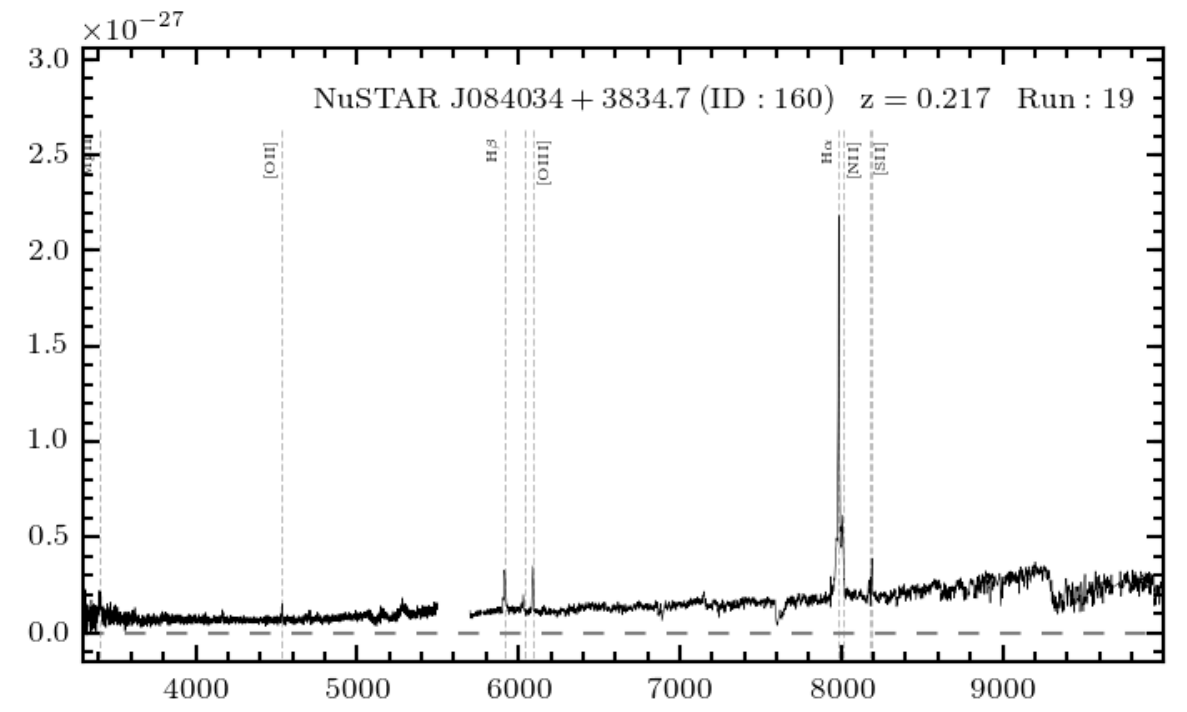}
\end{minipage}
\begin{minipage}[l]{0.325\textwidth}
\includegraphics[width=\textwidth]{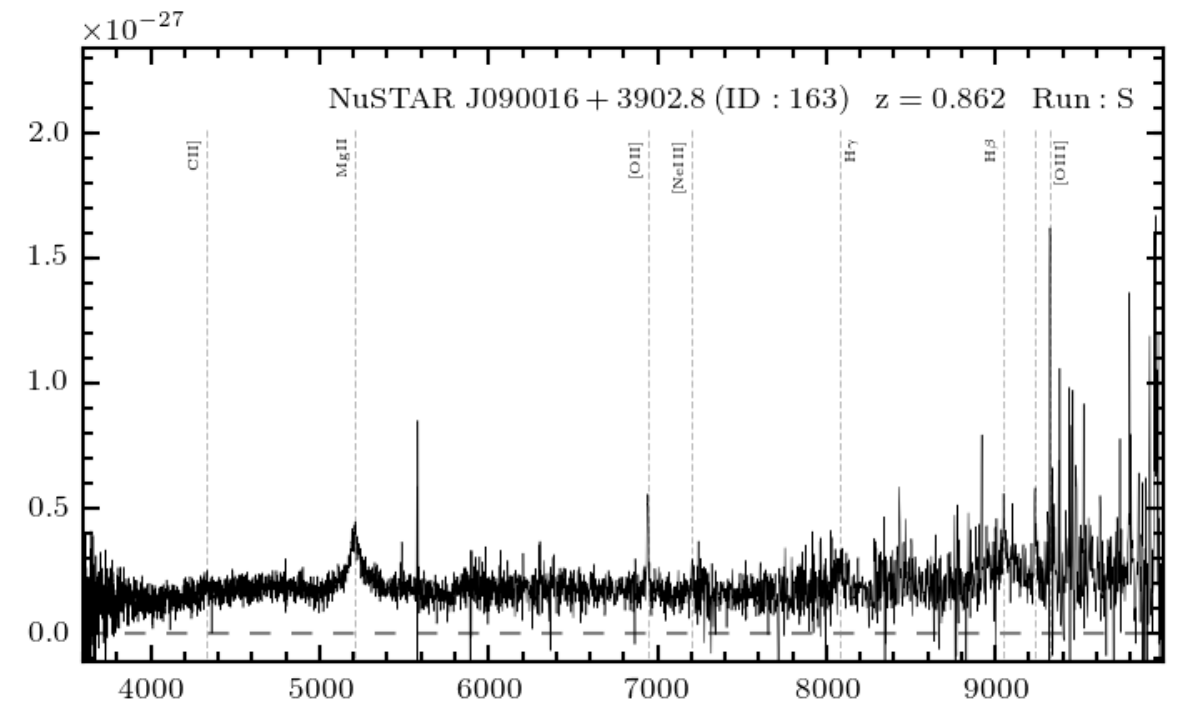}
\end{minipage}
\begin{minipage}[l]{0.325\textwidth}
\includegraphics[width=\textwidth]{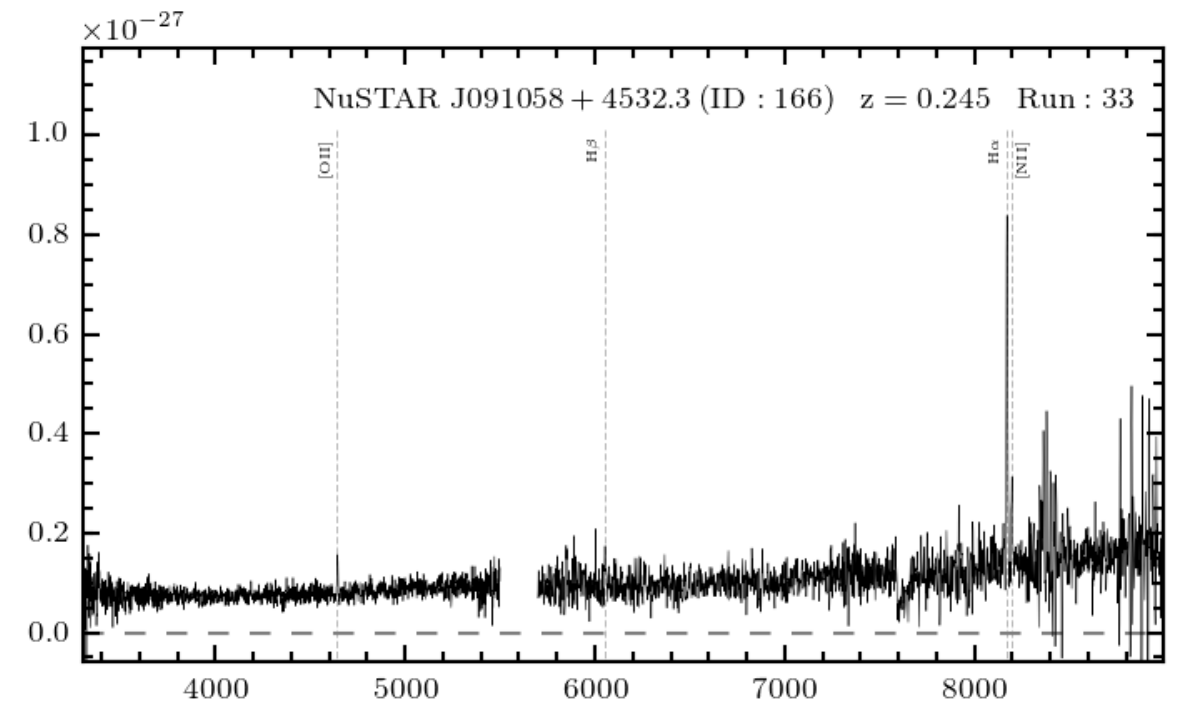}
\end{minipage}
\begin{minipage}[l]{0.325\textwidth}
\includegraphics[width=\textwidth]{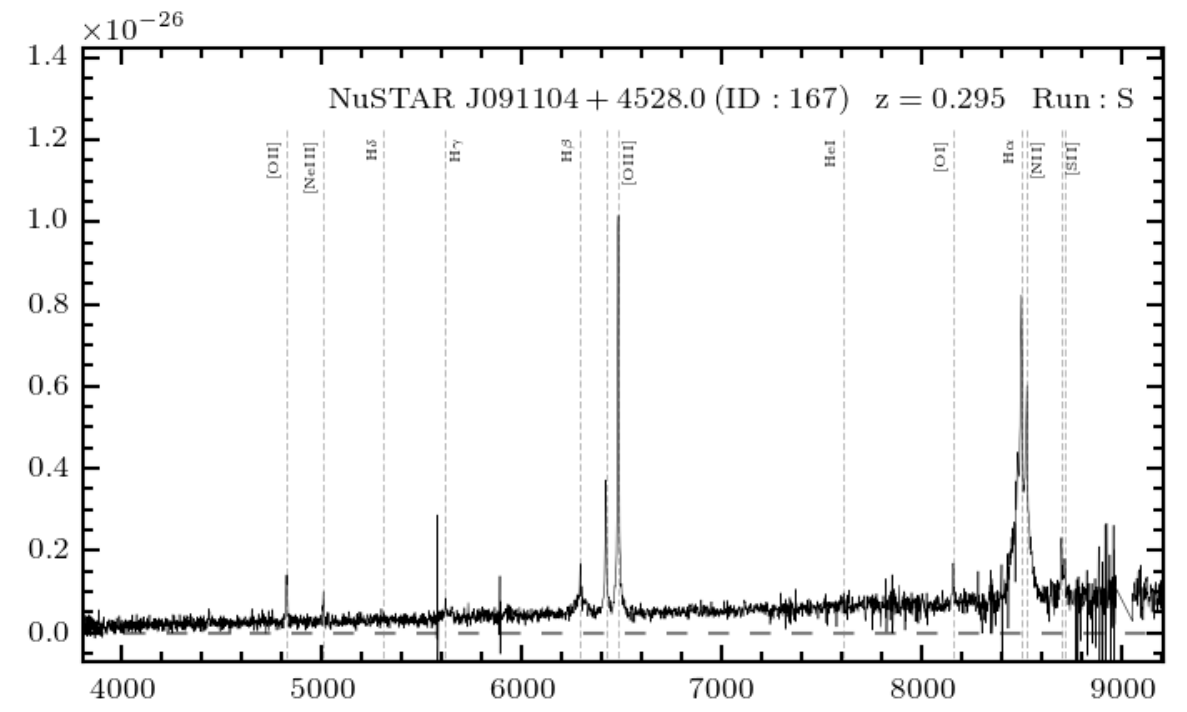}
\end{minipage}
\begin{minipage}[l]{0.325\textwidth}
\includegraphics[width=\textwidth]{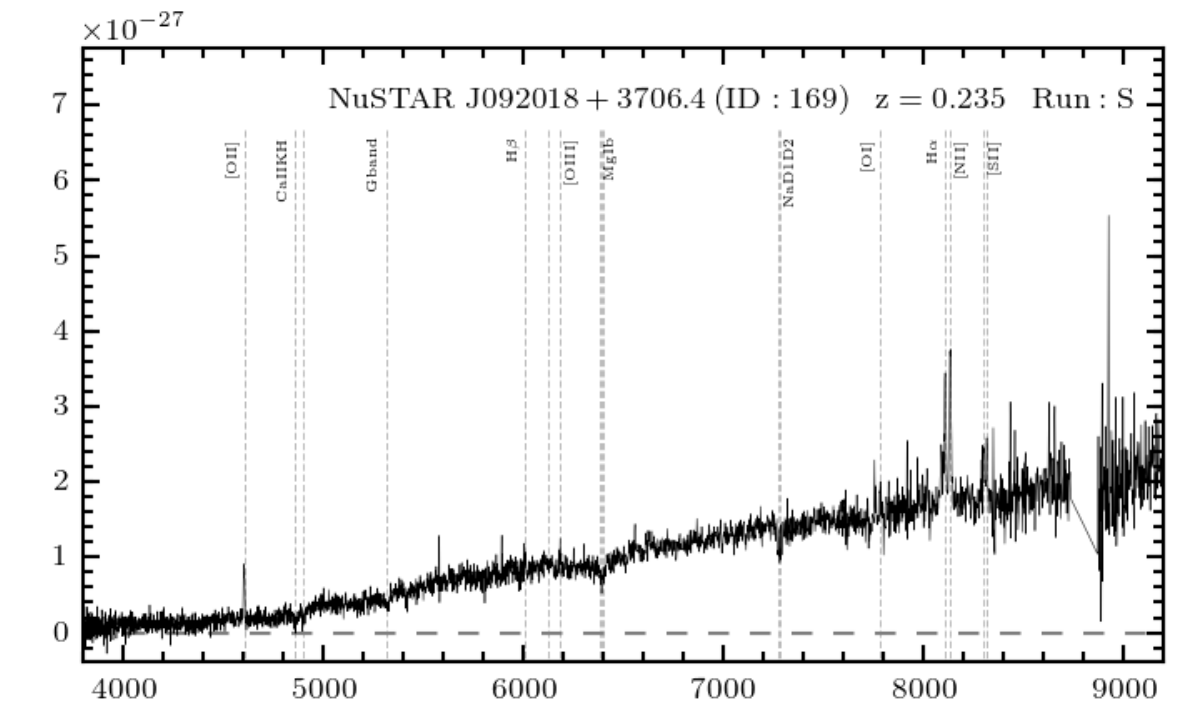}
\end{minipage}
\begin{minipage}[l]{0.325\textwidth}
\includegraphics[width=\textwidth]{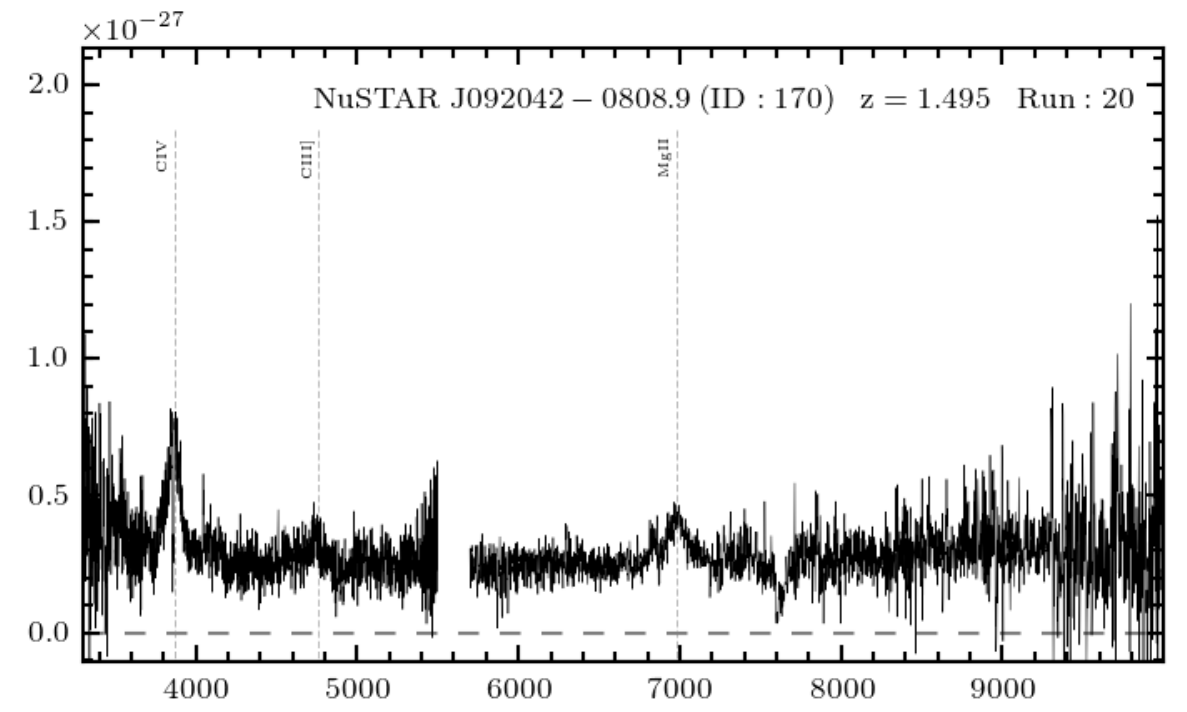}
\end{minipage}
\begin{minipage}[l]{0.325\textwidth}
\includegraphics[width=\textwidth]{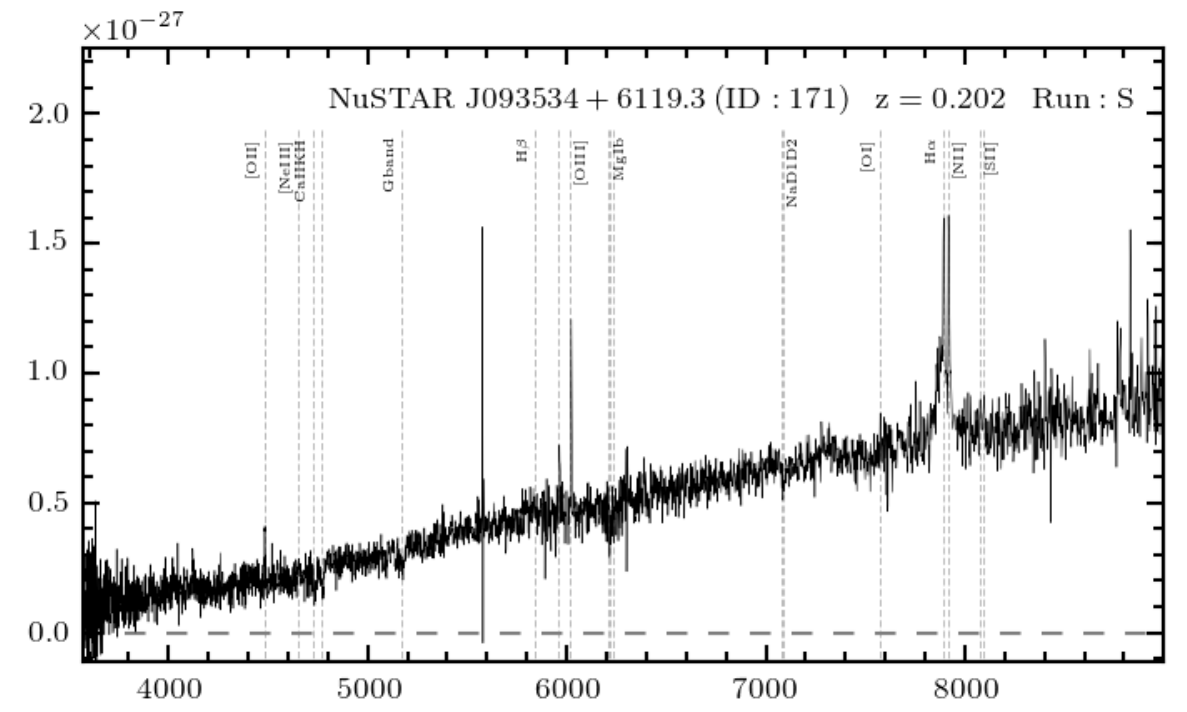}
\end{minipage}
\begin{minipage}[l]{0.325\textwidth}
\includegraphics[width=\textwidth]{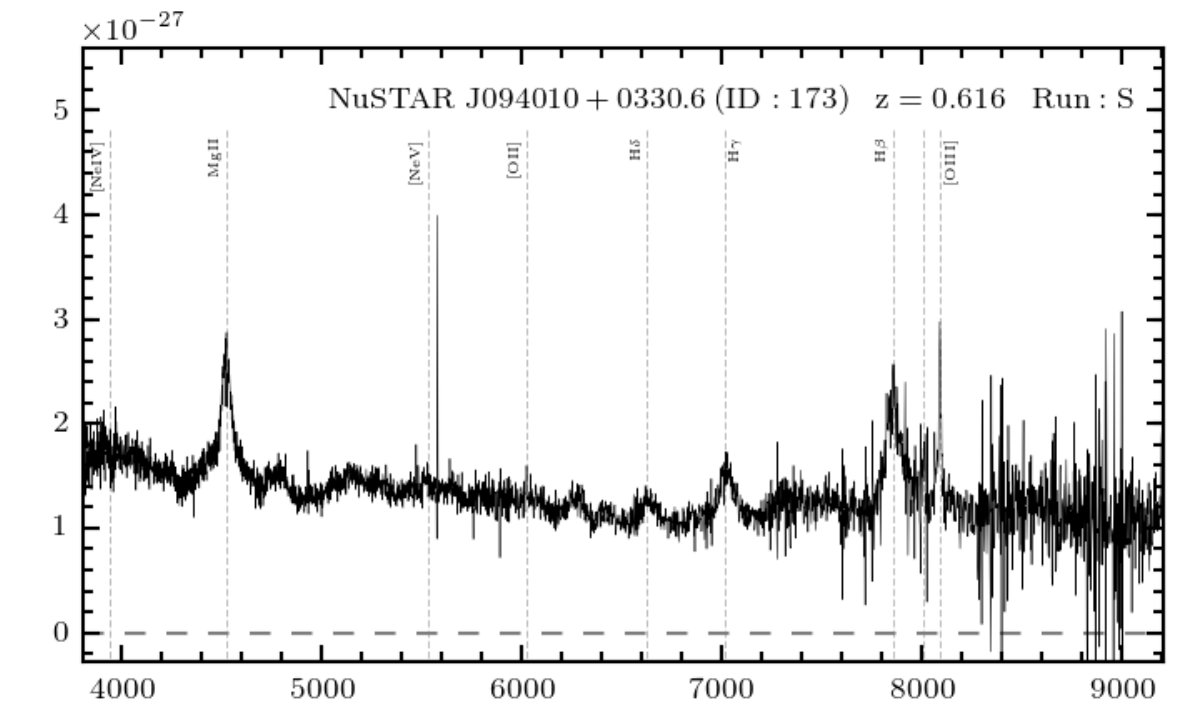}
\end{minipage}
\begin{minipage}[l]{0.325\textwidth}
\includegraphics[width=\textwidth]{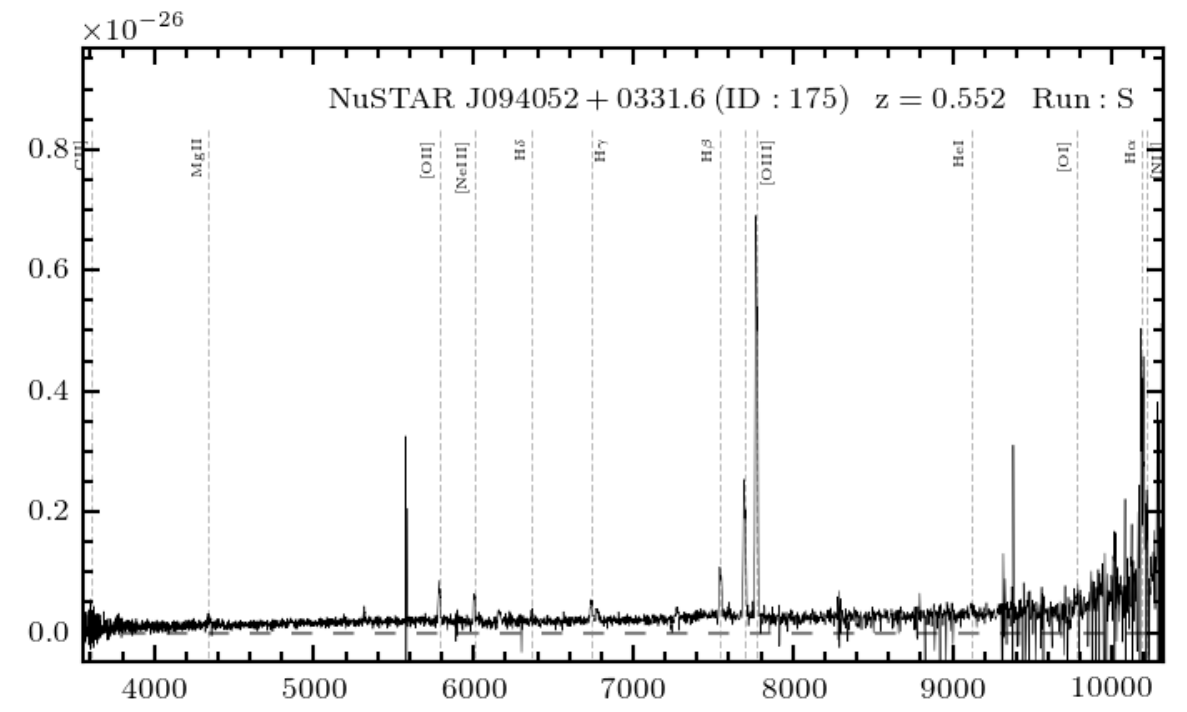}
\end{minipage}
\begin{minipage}[l]{0.325\textwidth}
\includegraphics[width=\textwidth]{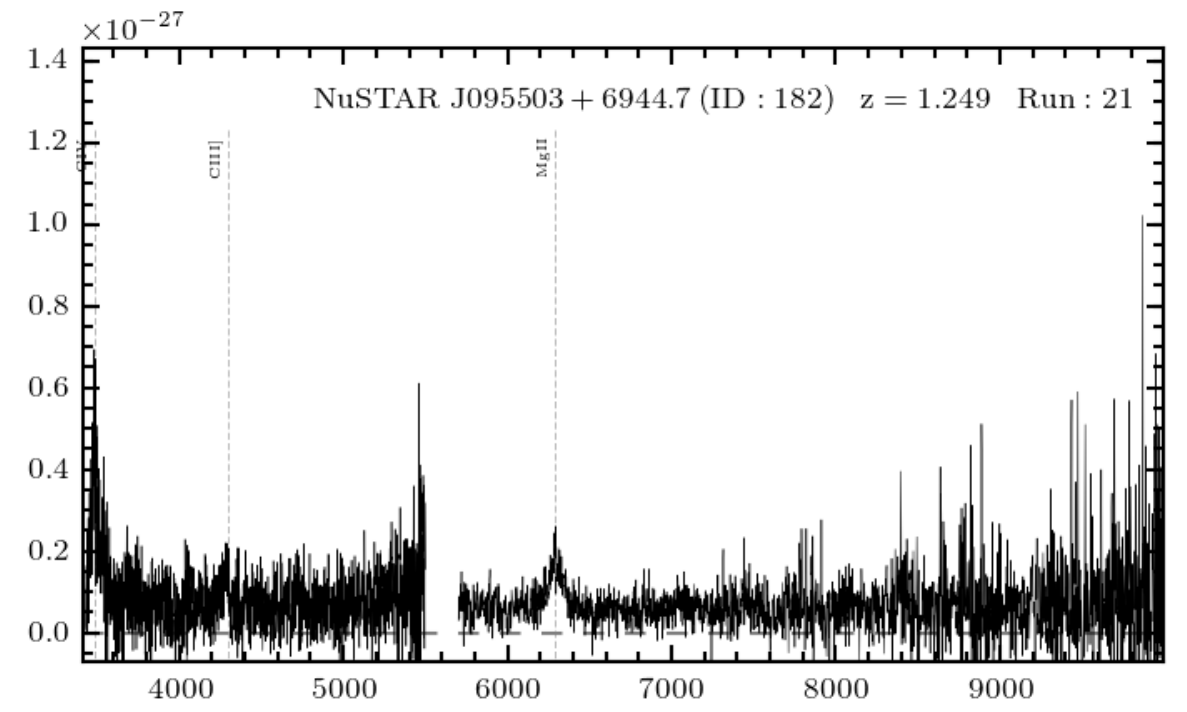}
\end{minipage}
\begin{minipage}[l]{0.325\textwidth}
\includegraphics[width=\textwidth]{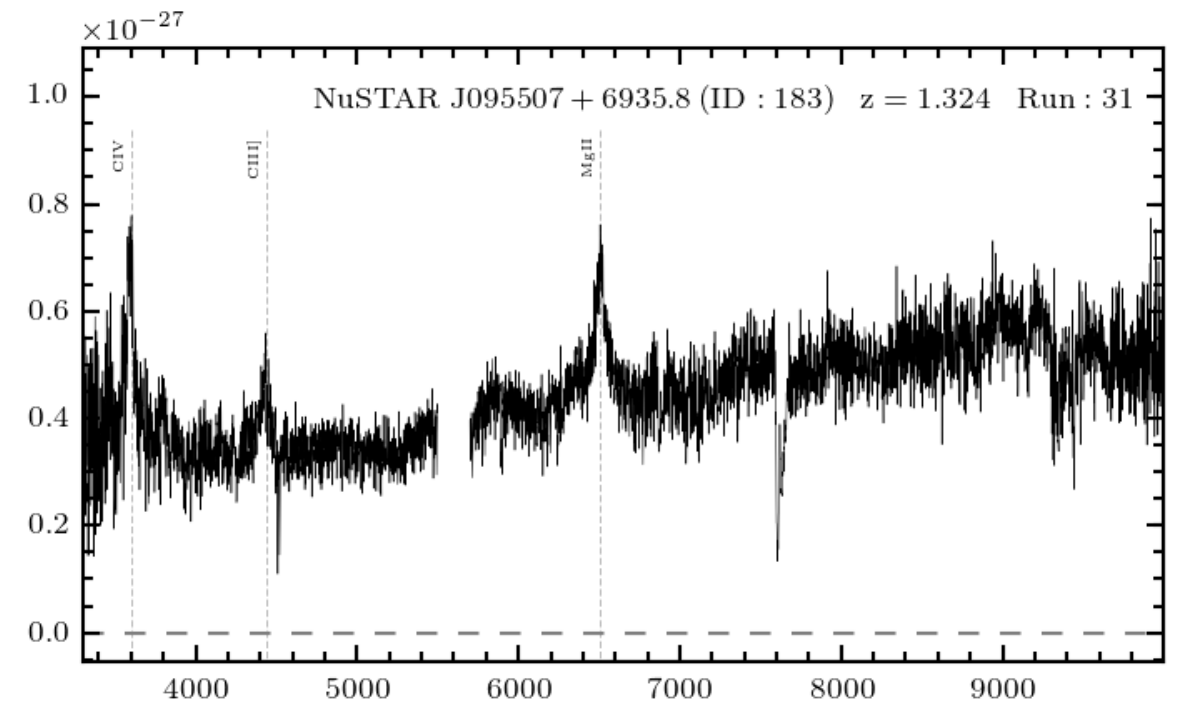}
\end{minipage}
\begin{minipage}[l]{0.325\textwidth}
\includegraphics[width=\textwidth]{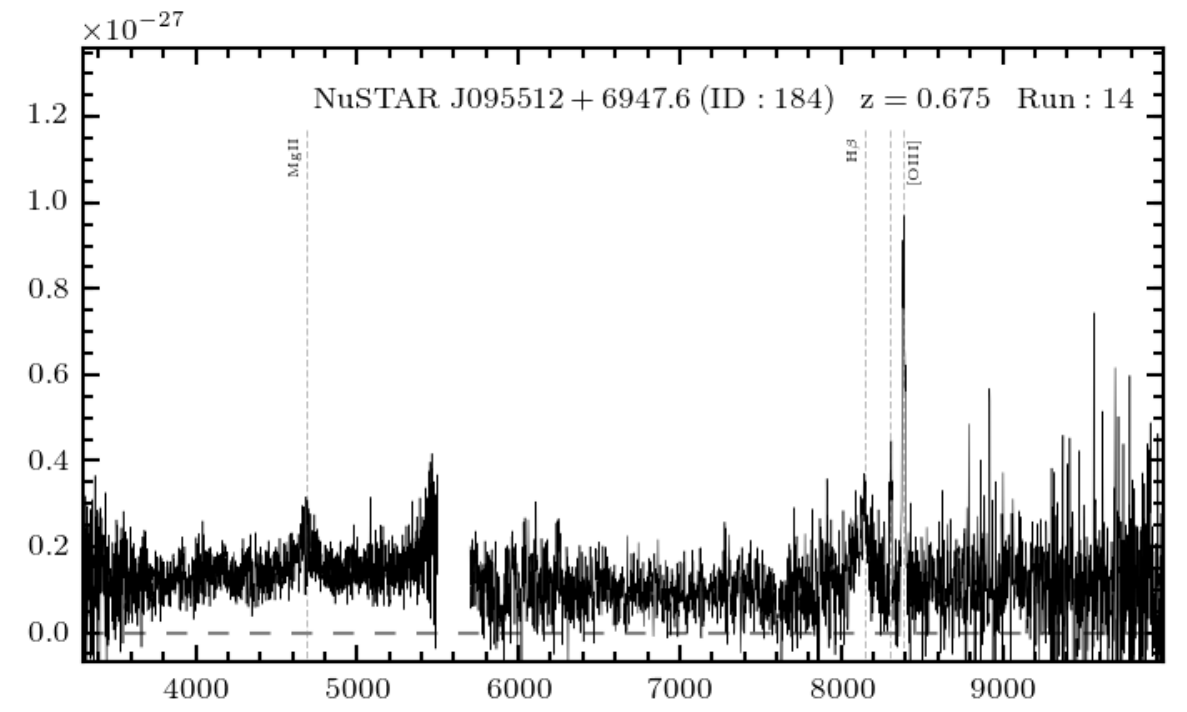}
\end{minipage}
\begin{minipage}[l]{0.325\textwidth}
\includegraphics[width=\textwidth]{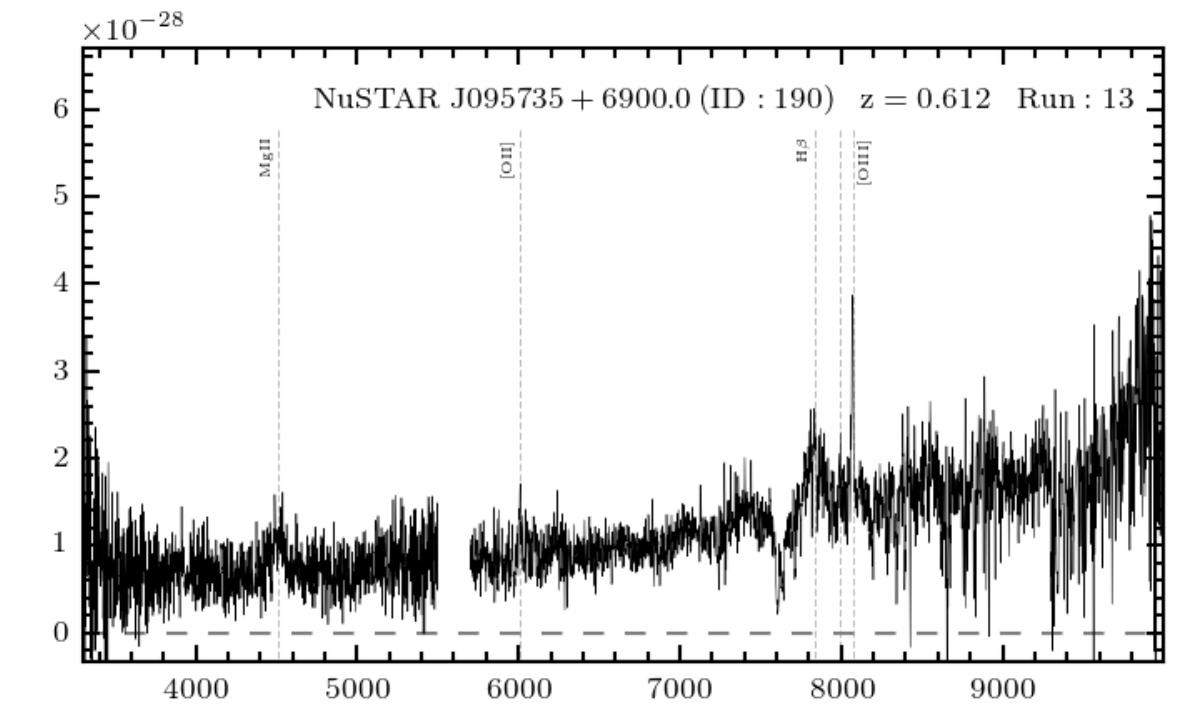}
\end{minipage}
\begin{minipage}[l]{0.325\textwidth}
\includegraphics[width=\textwidth]{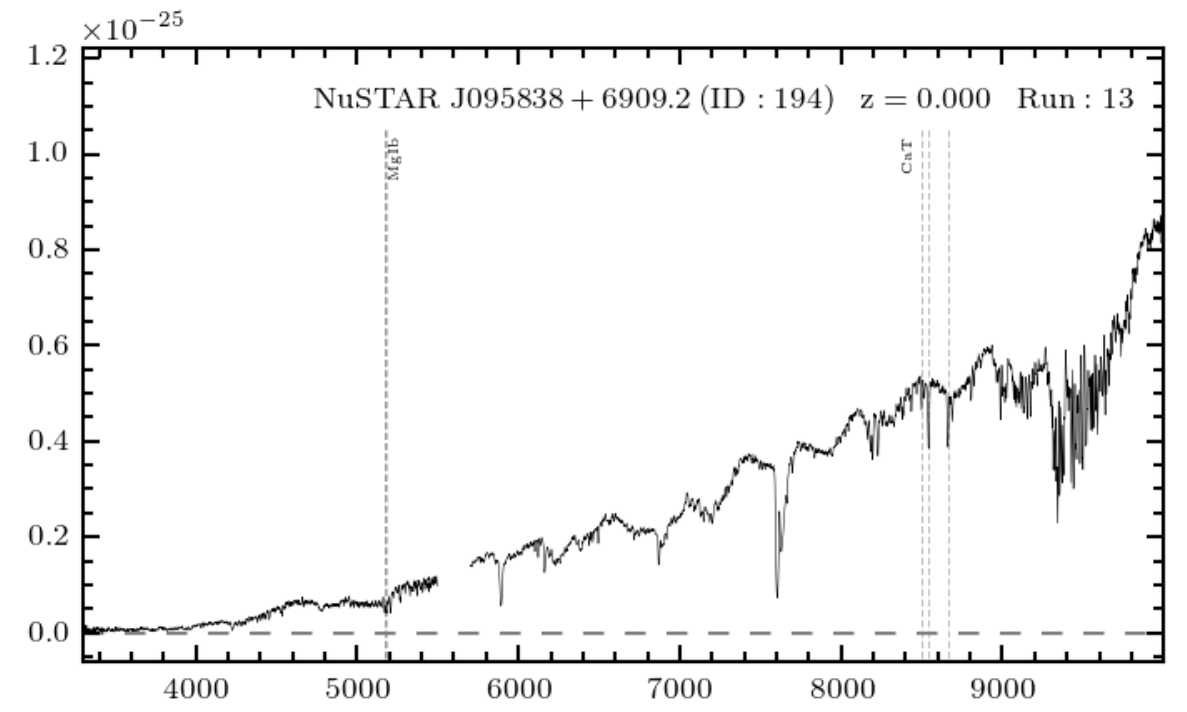}
\end{minipage}
\begin{minipage}[l]{0.325\textwidth}
\includegraphics[width=\textwidth]{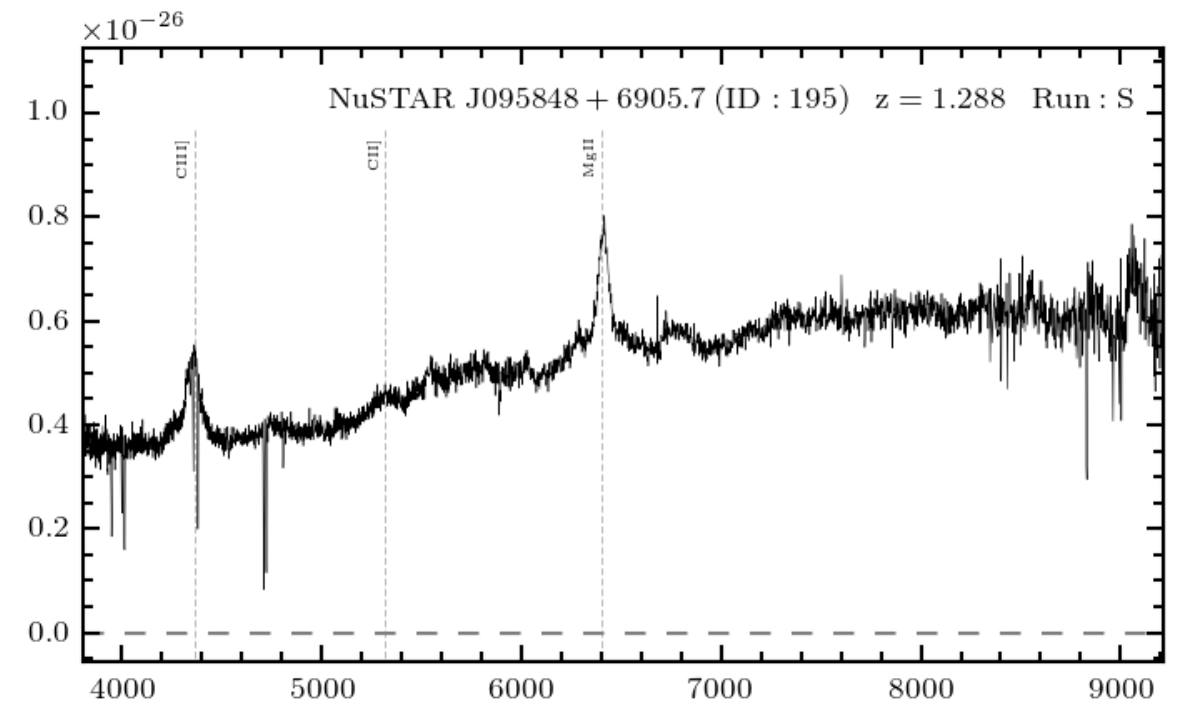}
\end{minipage}
\begin{minipage}[l]{0.325\textwidth}
\includegraphics[width=\textwidth]{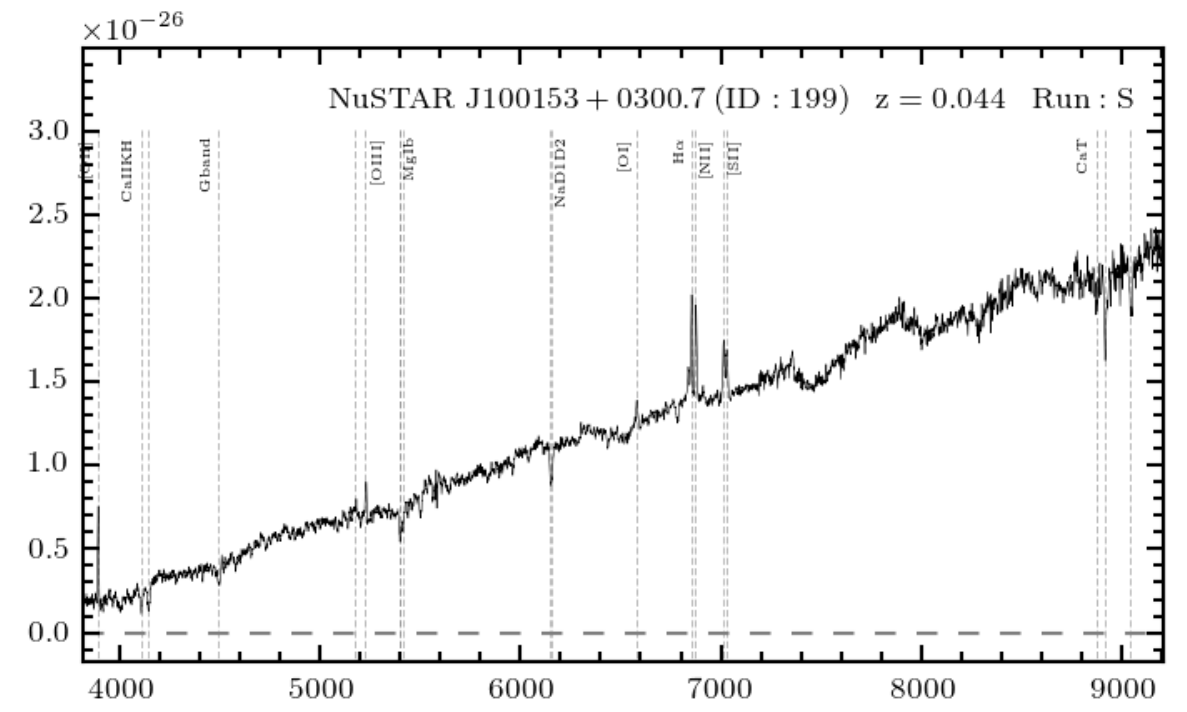}
\end{minipage}
\begin{minipage}[l]{0.325\textwidth}
\includegraphics[width=\textwidth]{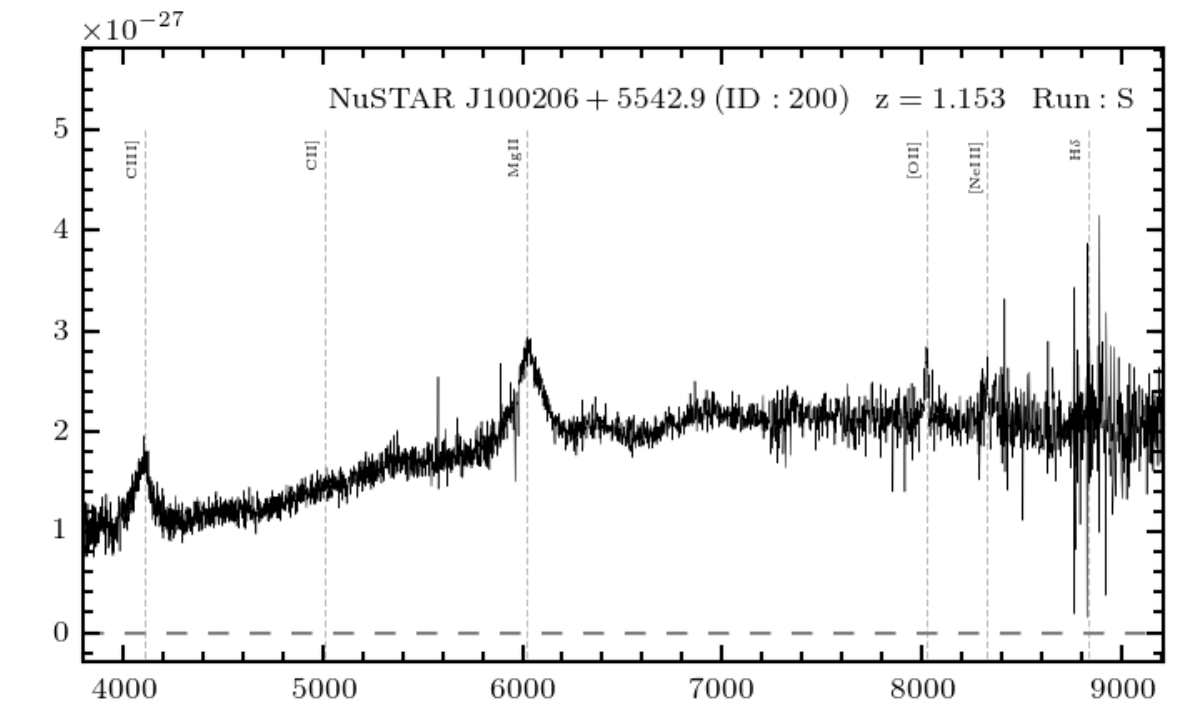}
\end{minipage}
\begin{minipage}[l]{0.325\textwidth}
\includegraphics[width=\textwidth]{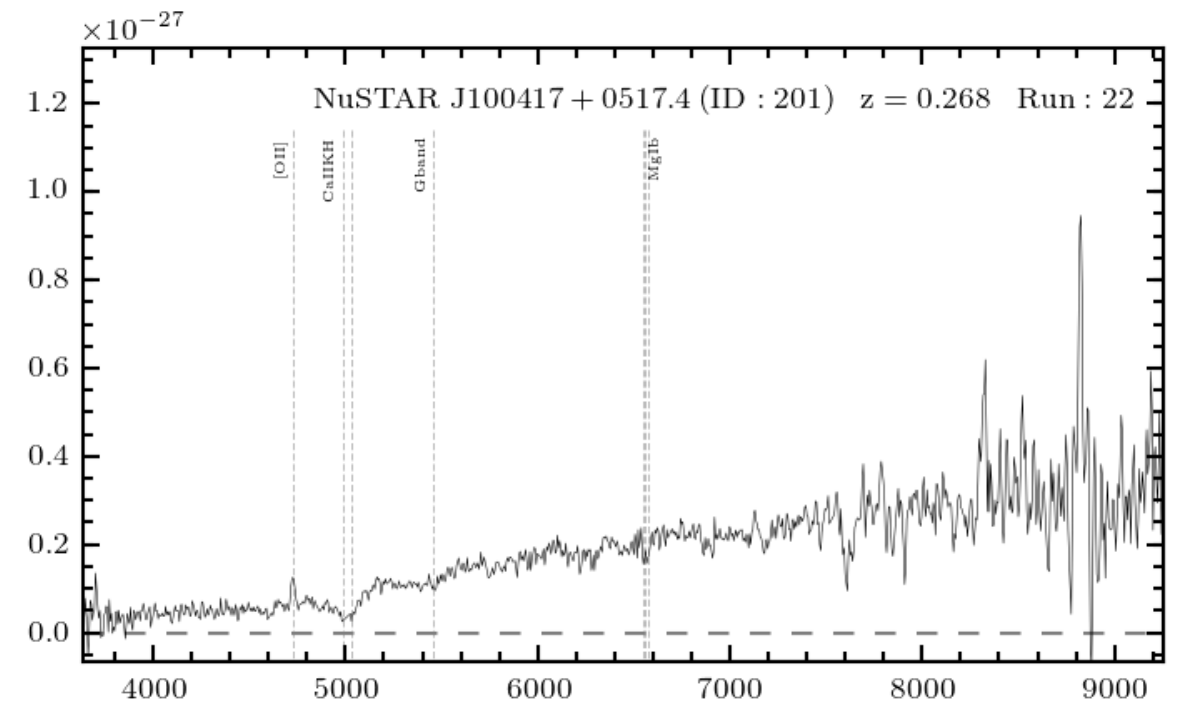}
\end{minipage}
\caption{Continued.}
\end{figure*}
\addtocounter{figure}{-1}
\begin{figure*}
\centering
\begin{minipage}[l]{0.325\textwidth}
\includegraphics[width=\textwidth]{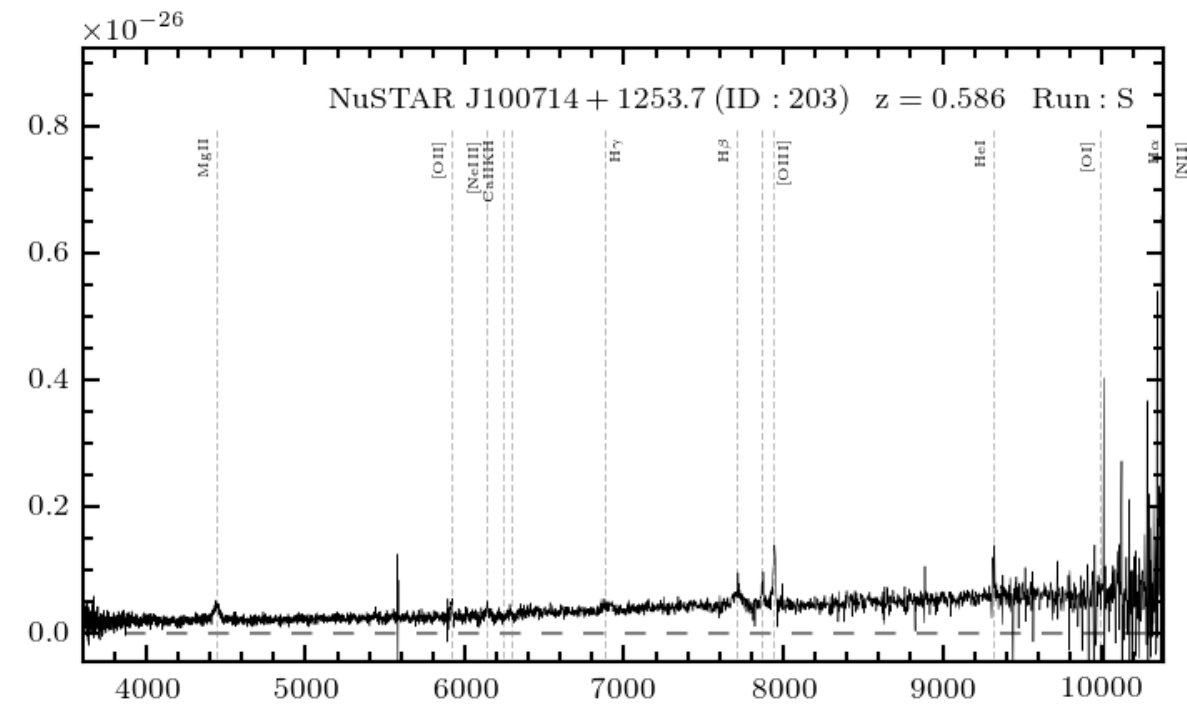}
\end{minipage}
\begin{minipage}[l]{0.325\textwidth}
\includegraphics[width=\textwidth]{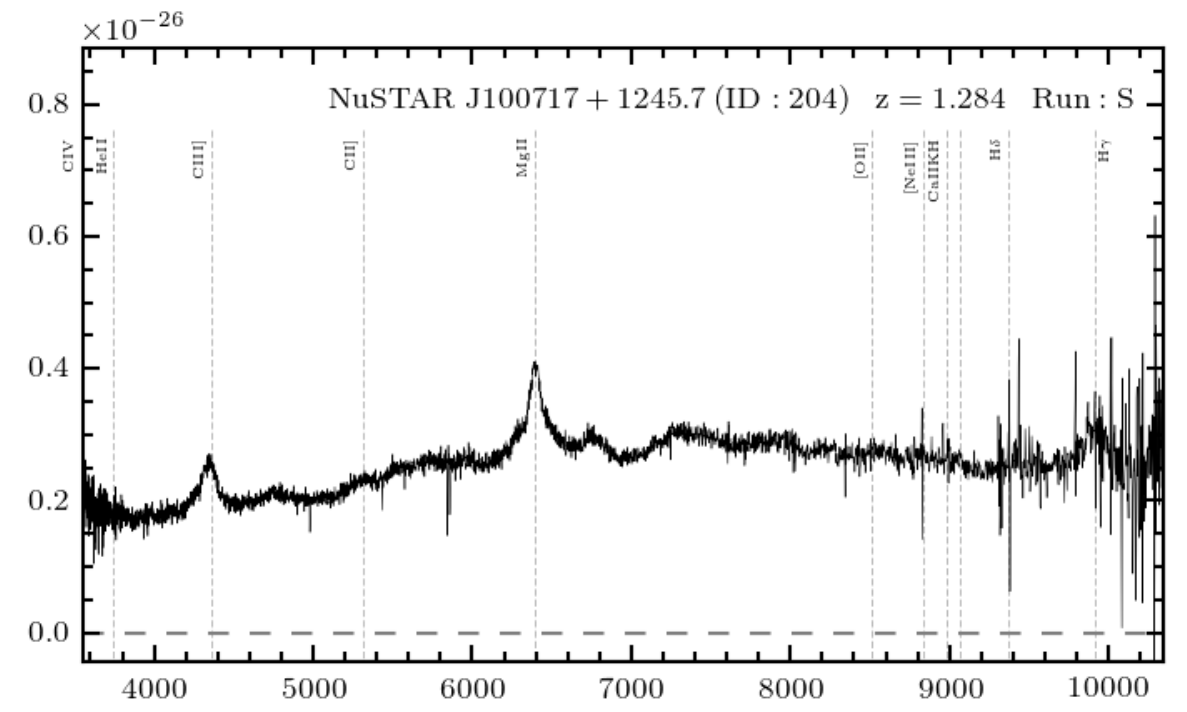}
\end{minipage}
\begin{minipage}[l]{0.325\textwidth}
\includegraphics[width=\textwidth]{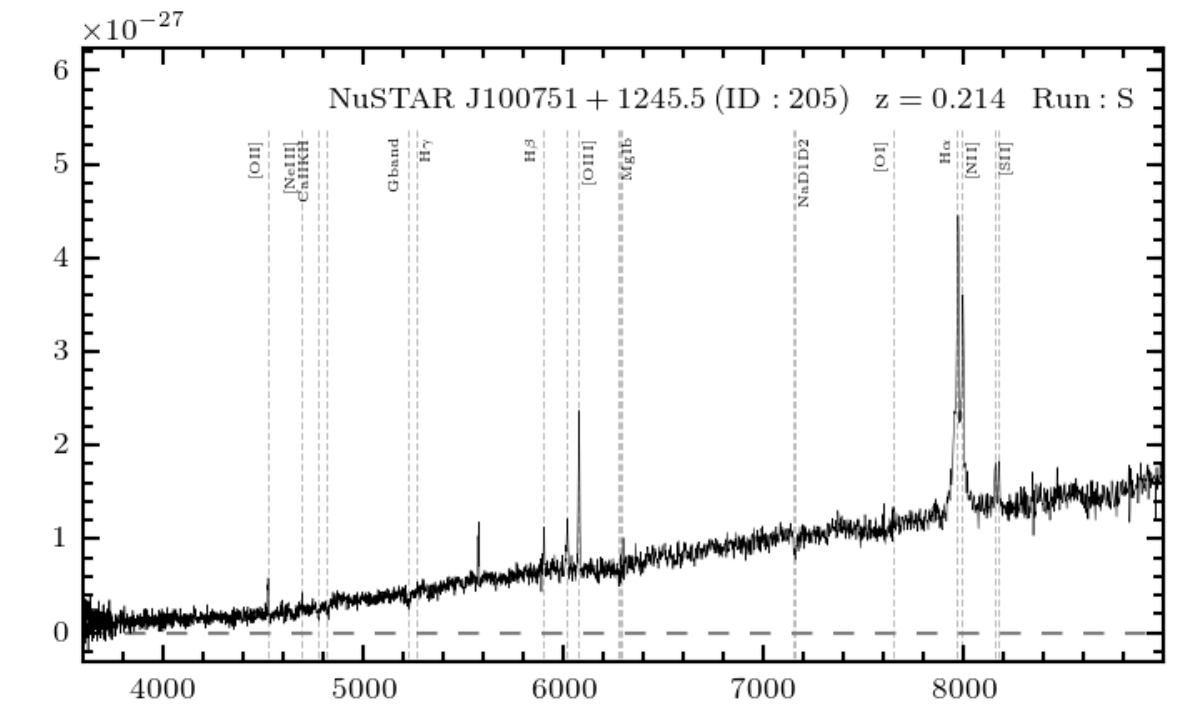}
\end{minipage}
\begin{minipage}[l]{0.325\textwidth}
\includegraphics[width=\textwidth]{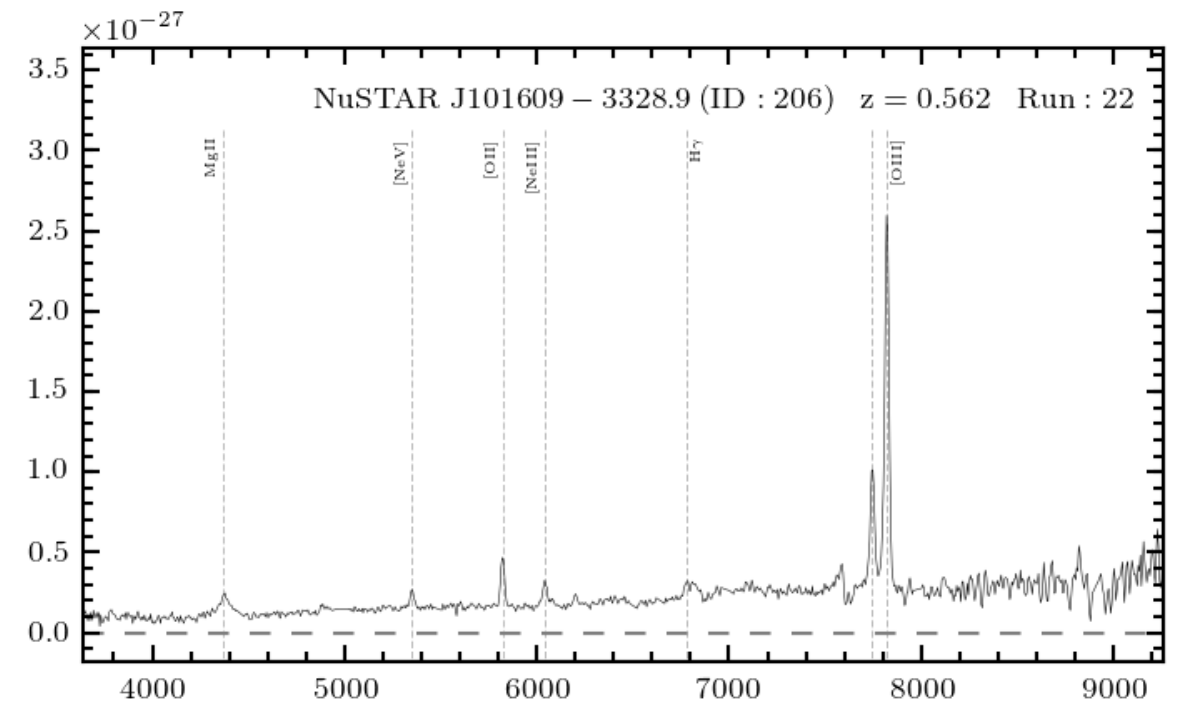}
\end{minipage}
\begin{minipage}[l]{0.325\textwidth}
\includegraphics[width=\textwidth]{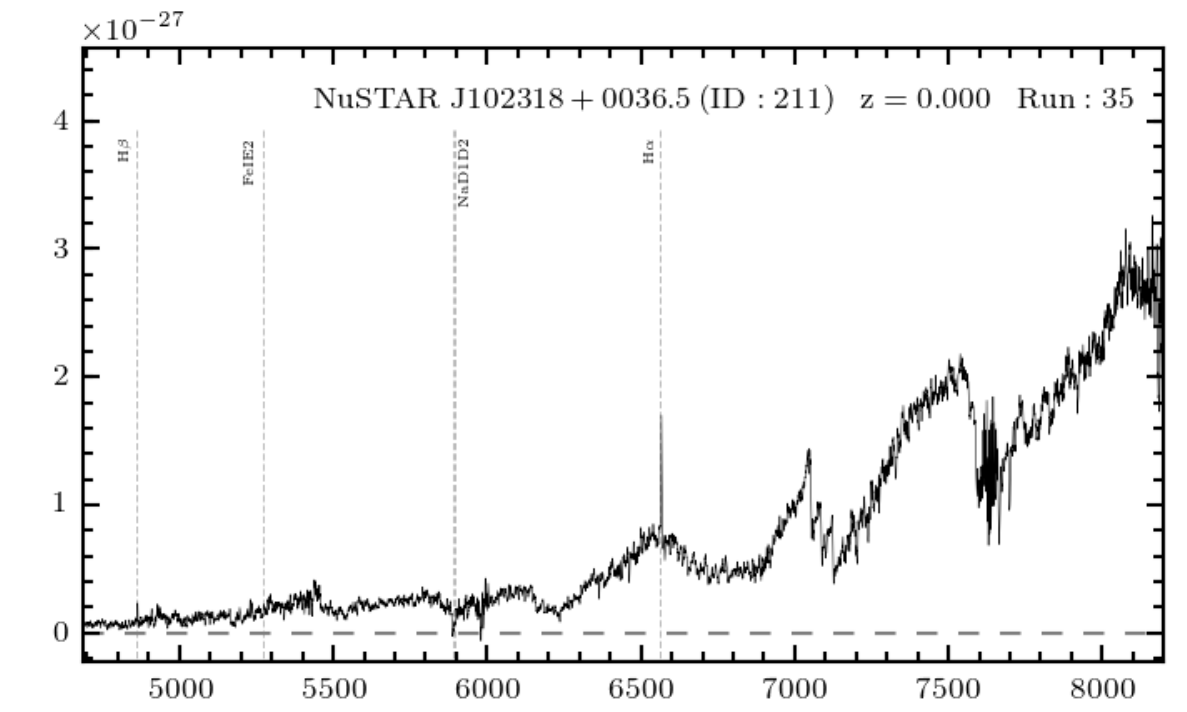}
\end{minipage}
\begin{minipage}[l]{0.325\textwidth}
\includegraphics[width=\textwidth]{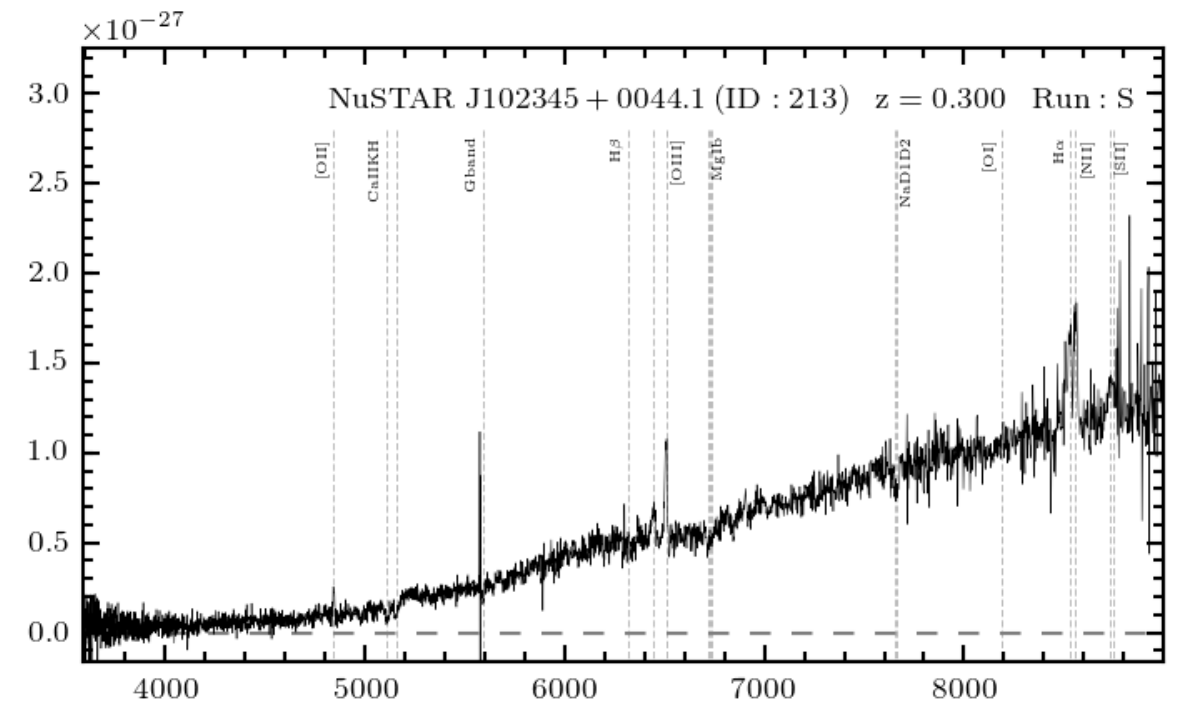}
\end{minipage}
\begin{minipage}[l]{0.325\textwidth}
\includegraphics[width=\textwidth]{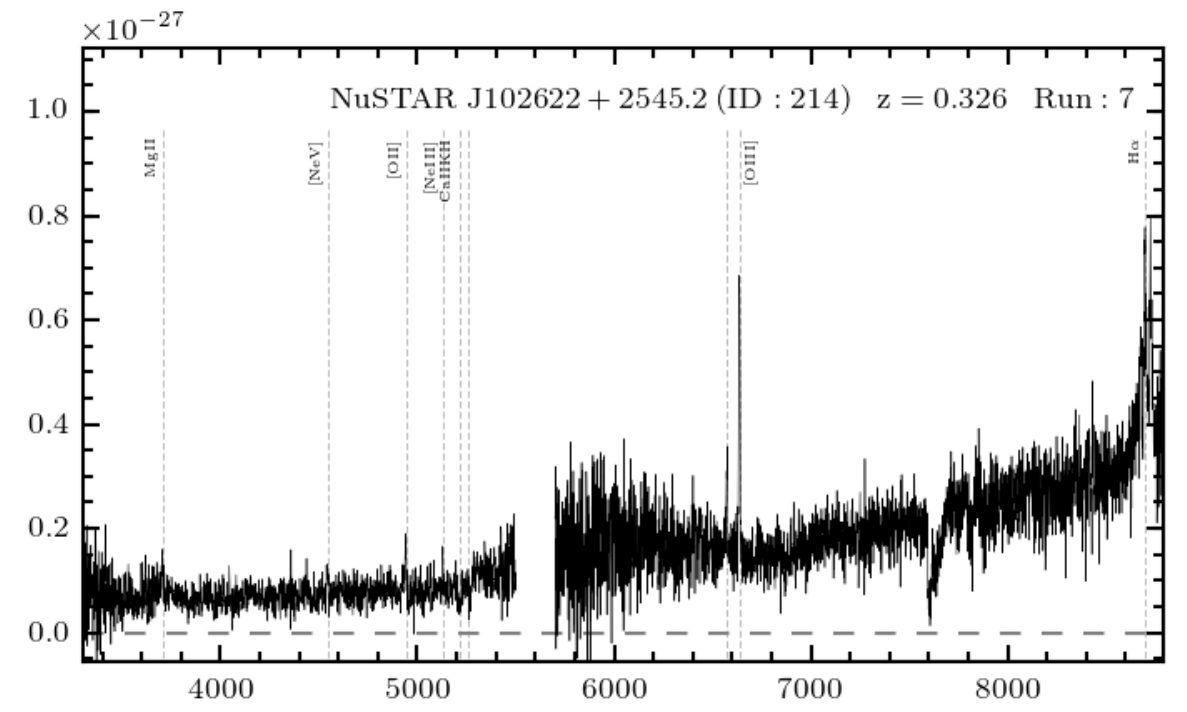}
\end{minipage}
\begin{minipage}[l]{0.325\textwidth}
\includegraphics[width=\textwidth]{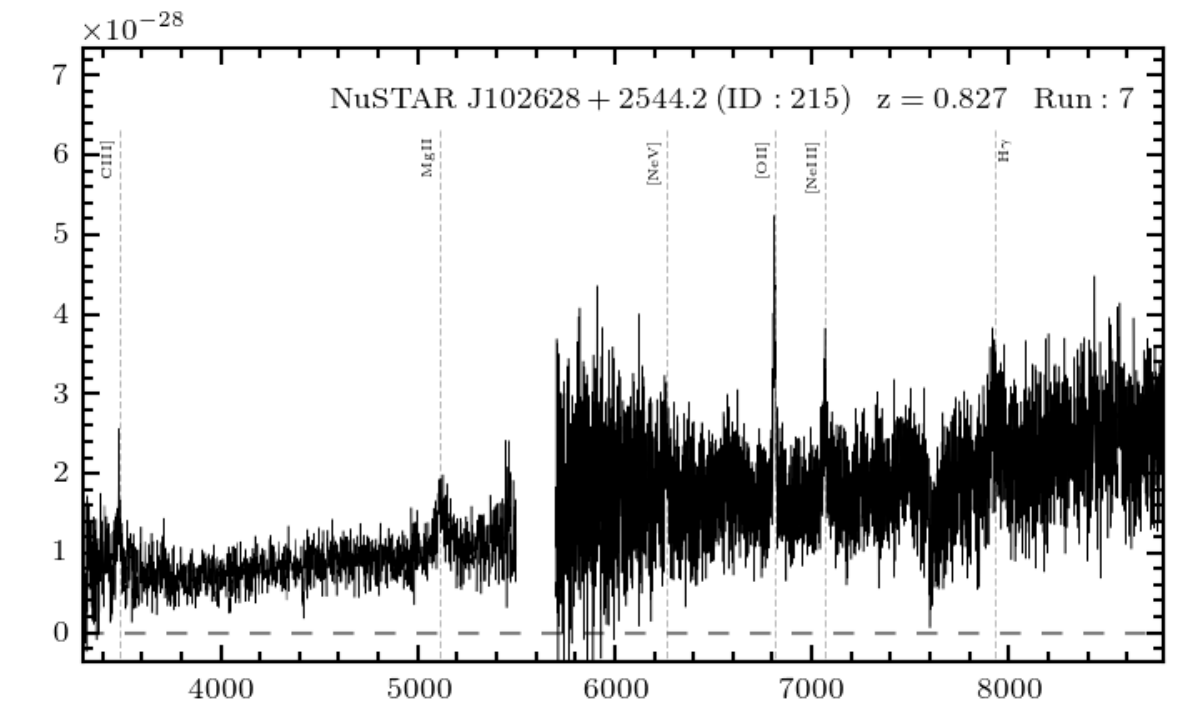}
\end{minipage}
\begin{minipage}[l]{0.325\textwidth}
\includegraphics[width=\textwidth]{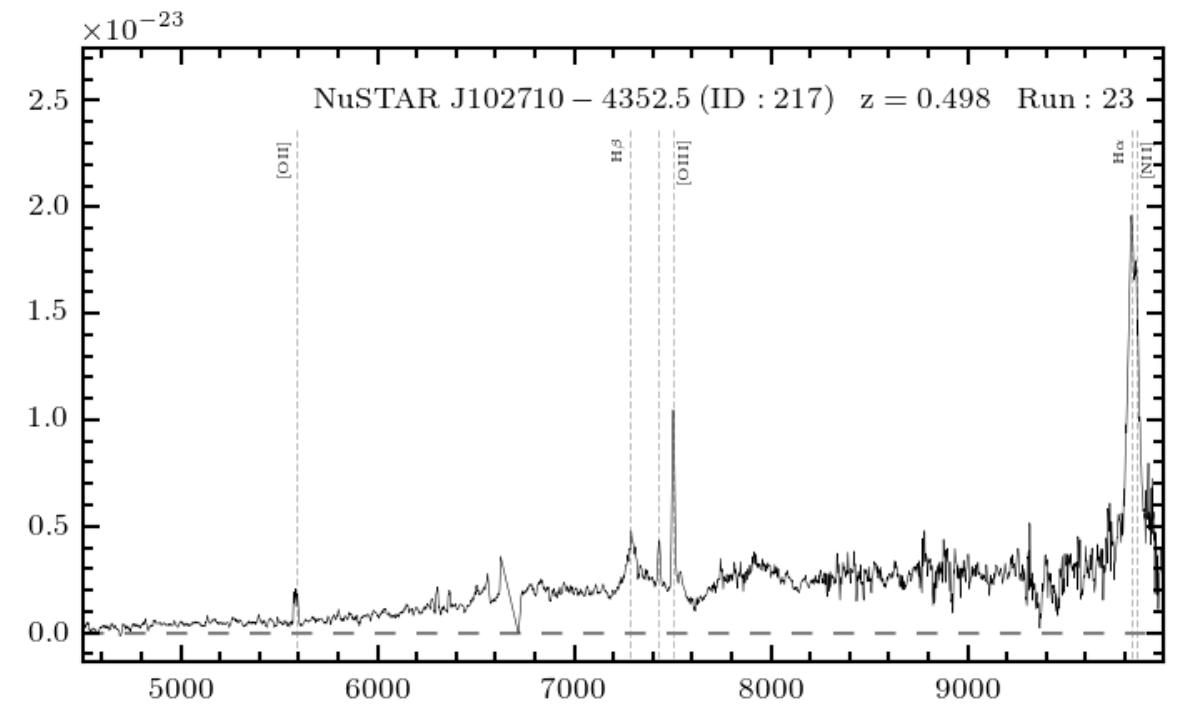}
\end{minipage}
\begin{minipage}[l]{0.325\textwidth}
\includegraphics[width=\textwidth]{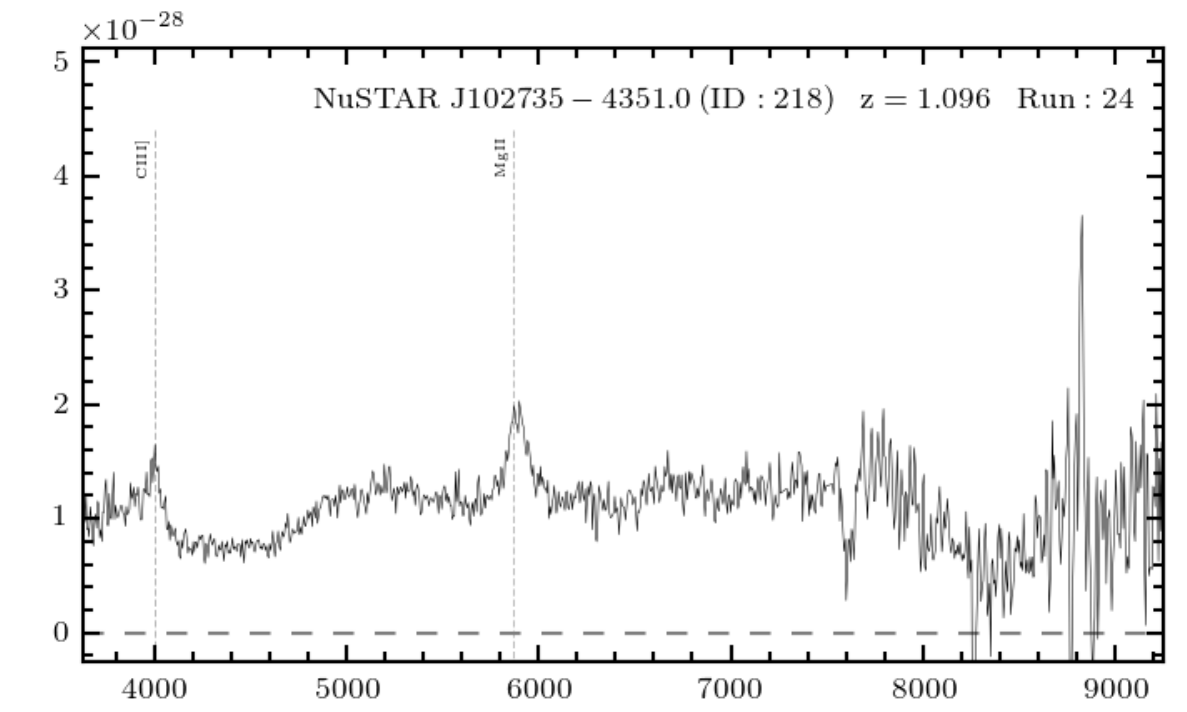}
\end{minipage}
\begin{minipage}[l]{0.325\textwidth}
\includegraphics[width=\textwidth]{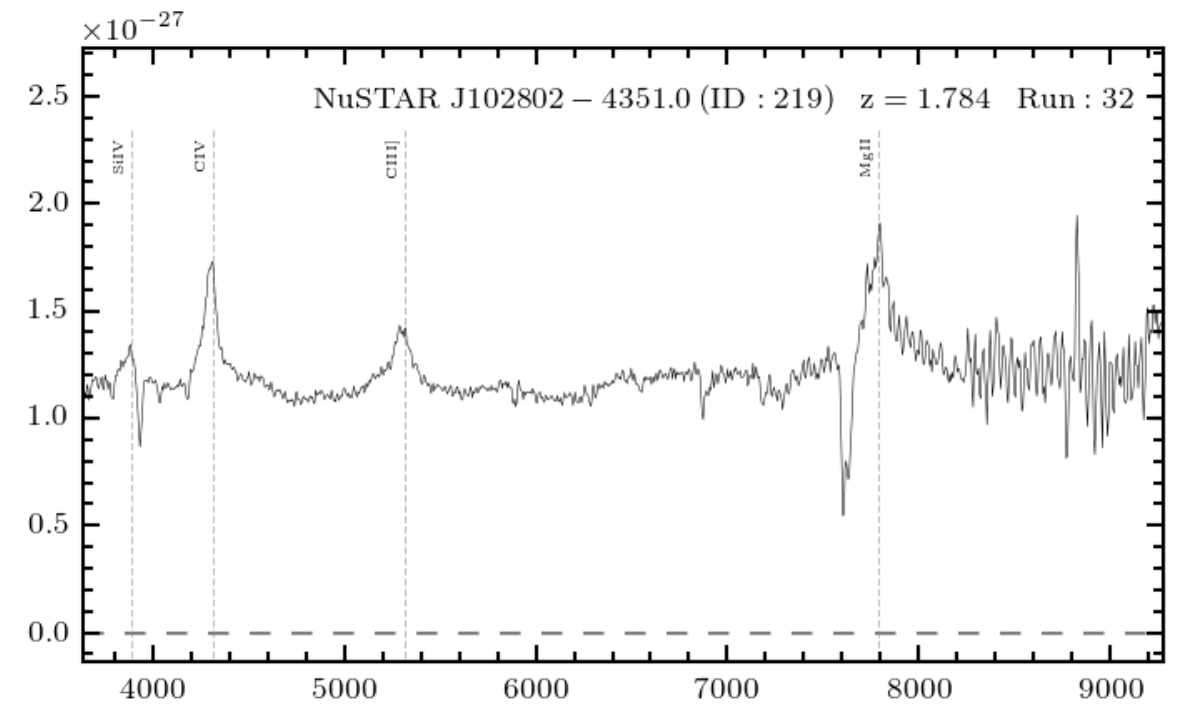}
\end{minipage}
\begin{minipage}[l]{0.325\textwidth}
\includegraphics[width=\textwidth]{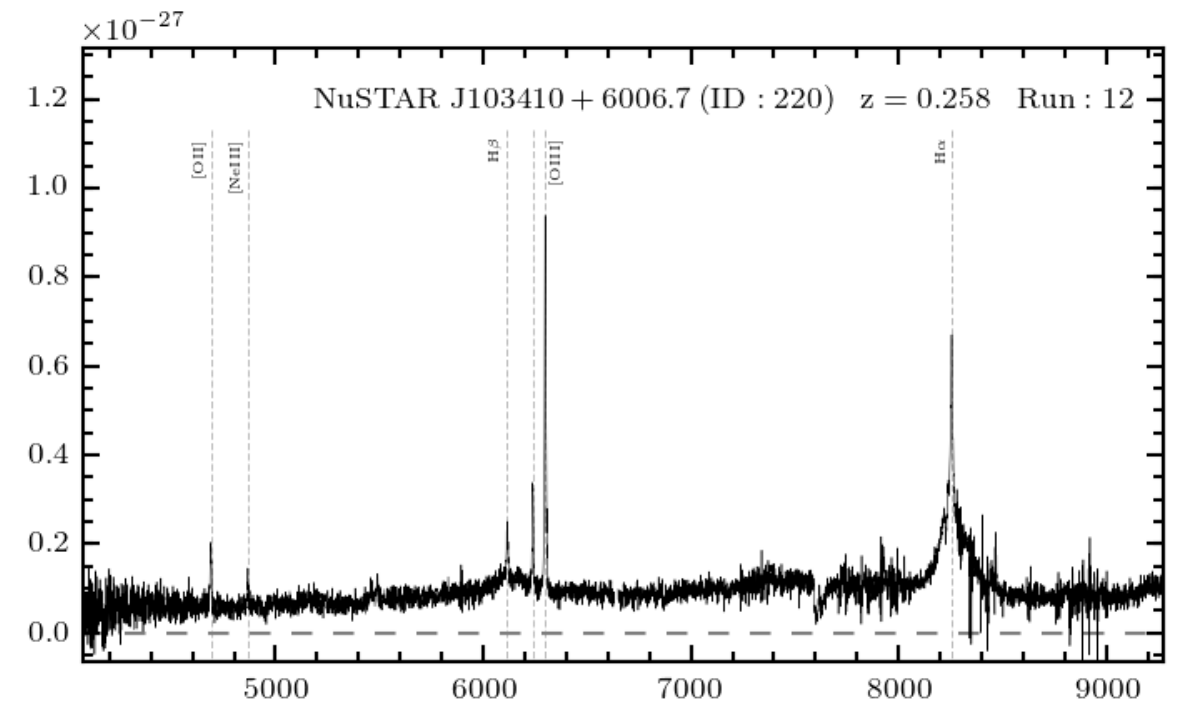}
\end{minipage}
\begin{minipage}[l]{0.325\textwidth}
\includegraphics[width=\textwidth]{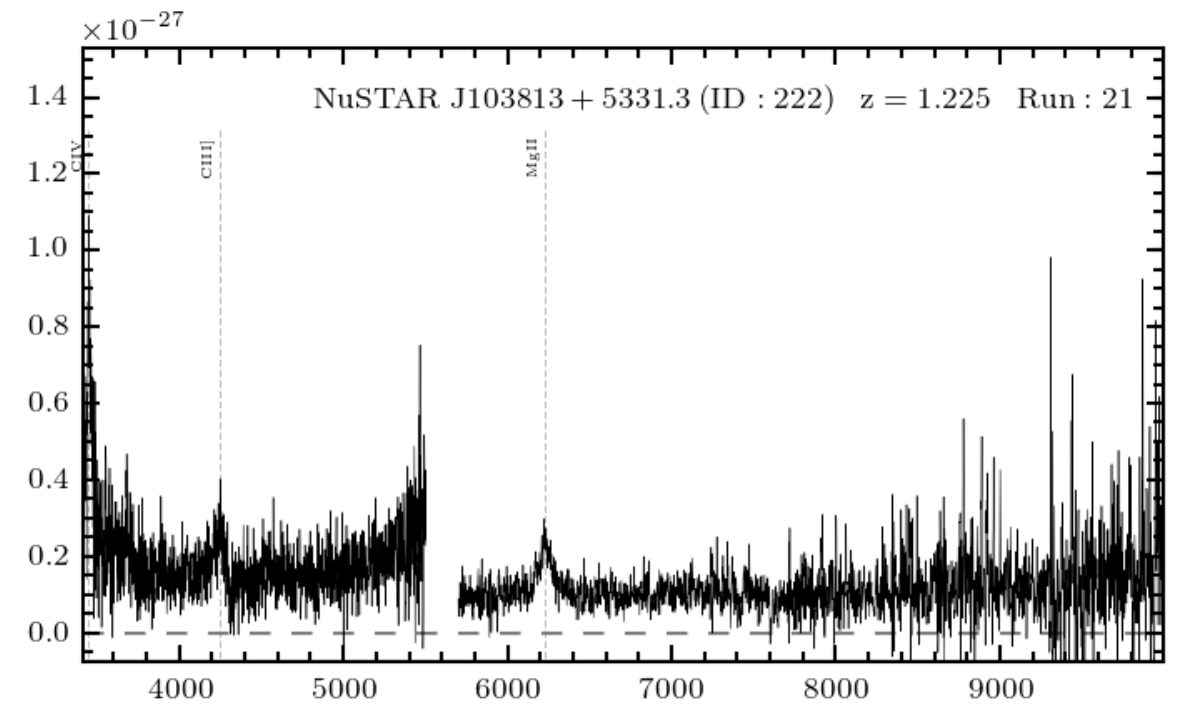}
\end{minipage}
\begin{minipage}[l]{0.325\textwidth}
\includegraphics[width=\textwidth]{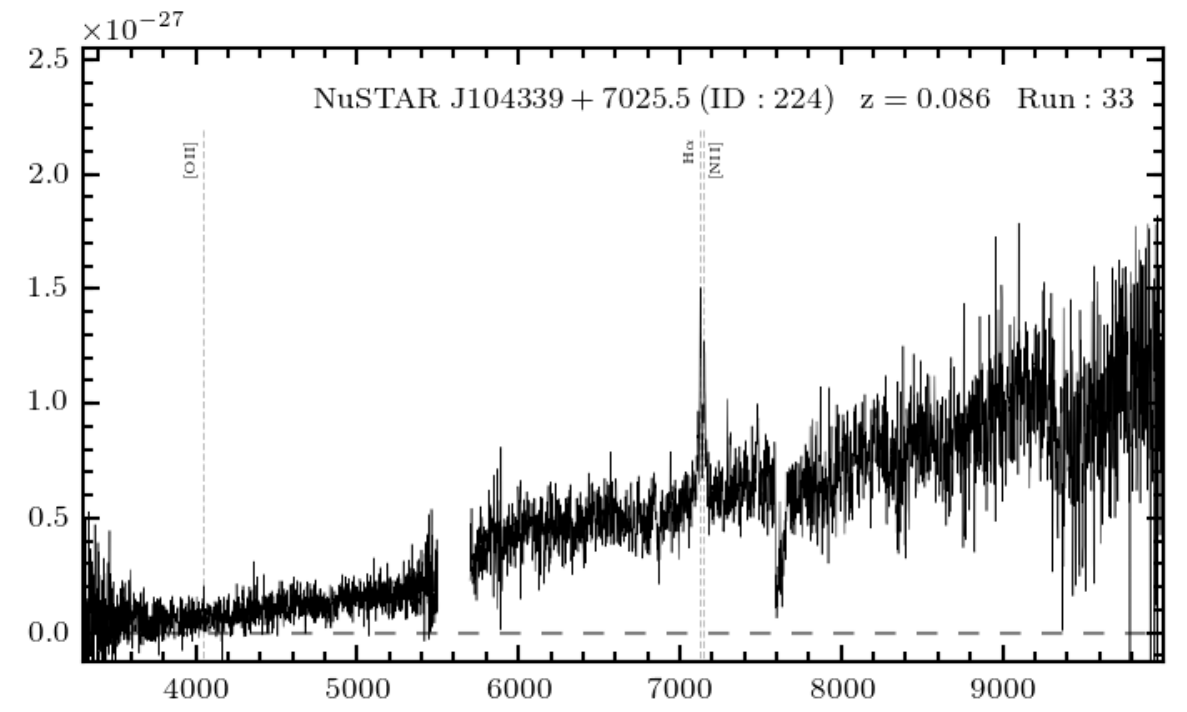}
\end{minipage}
\begin{minipage}[l]{0.325\textwidth}
\includegraphics[width=\textwidth]{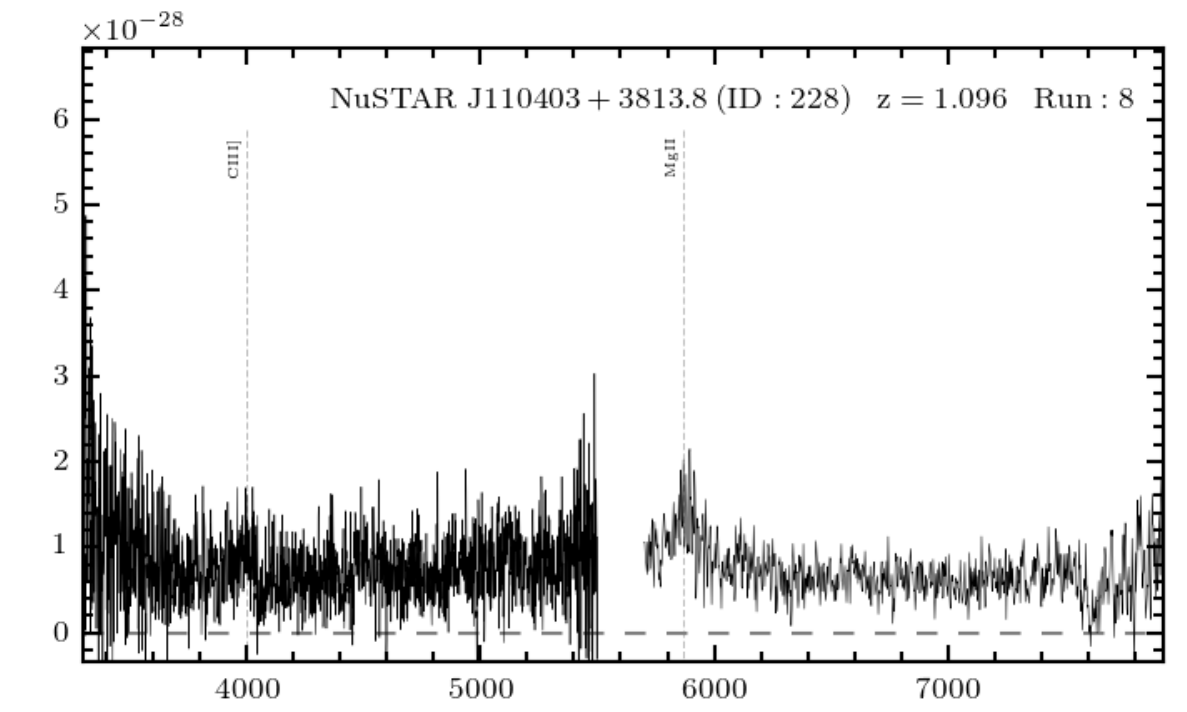}
\end{minipage}
\begin{minipage}[l]{0.325\textwidth}
\includegraphics[width=\textwidth]{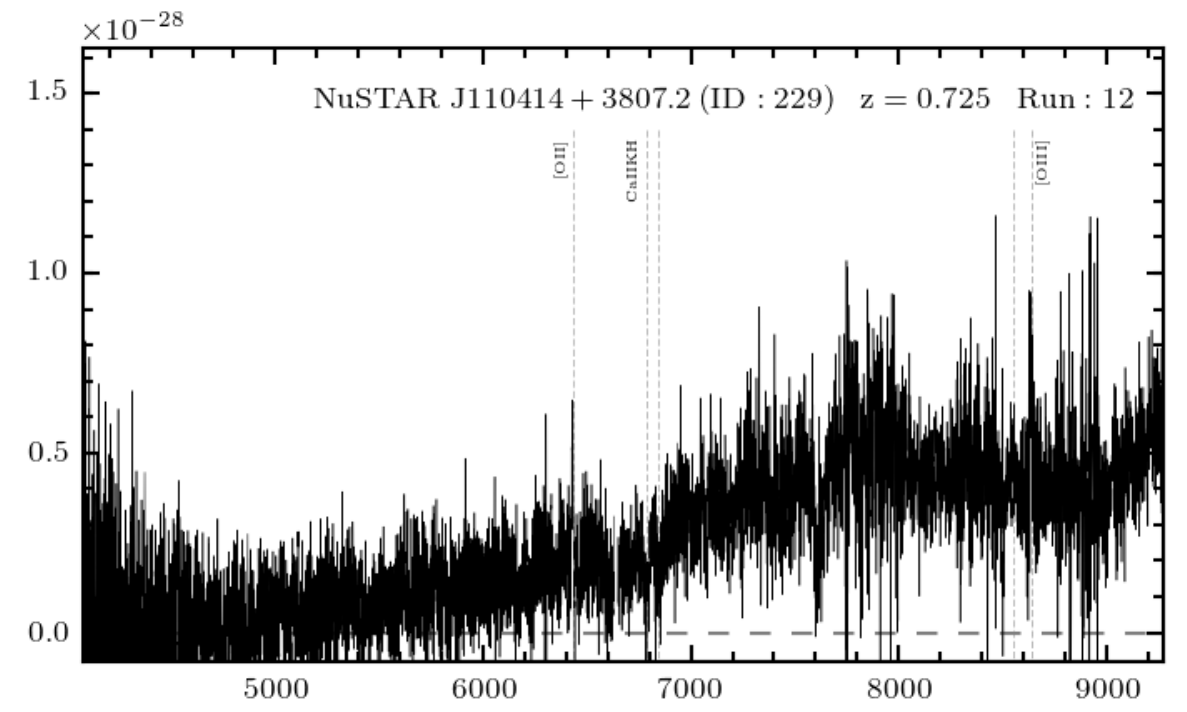}
\end{minipage}
\begin{minipage}[l]{0.325\textwidth}
\includegraphics[width=\textwidth]{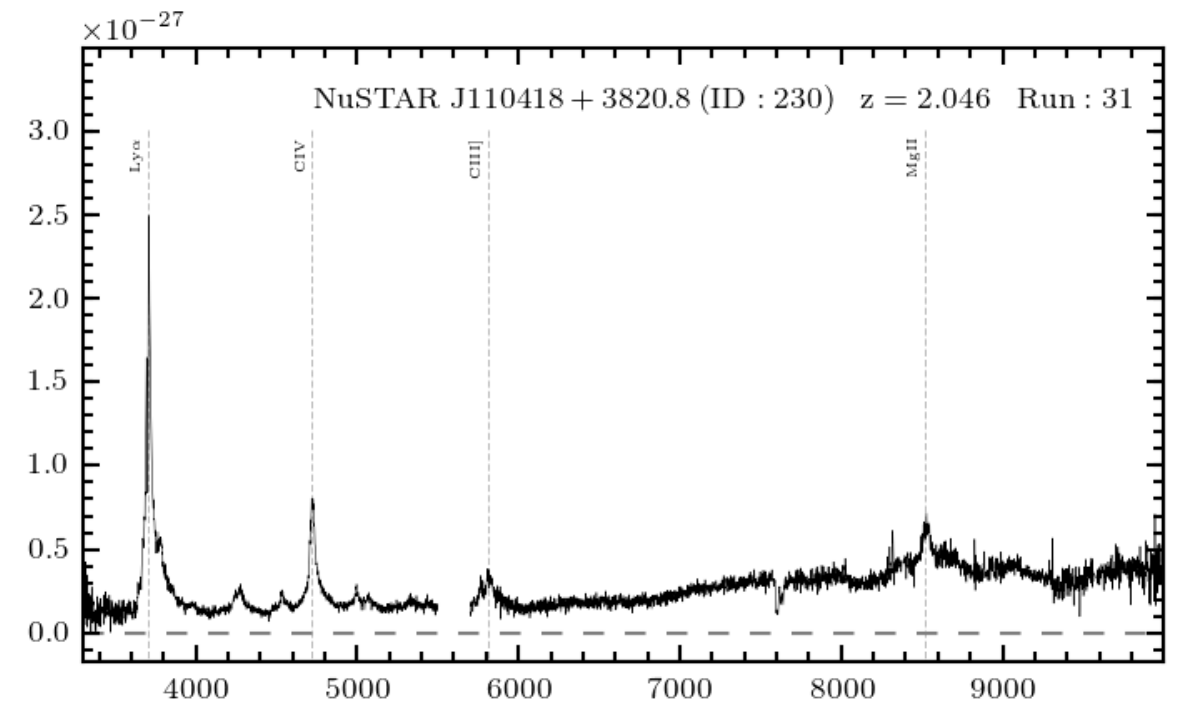}
\end{minipage}
\begin{minipage}[l]{0.325\textwidth}
\includegraphics[width=\textwidth]{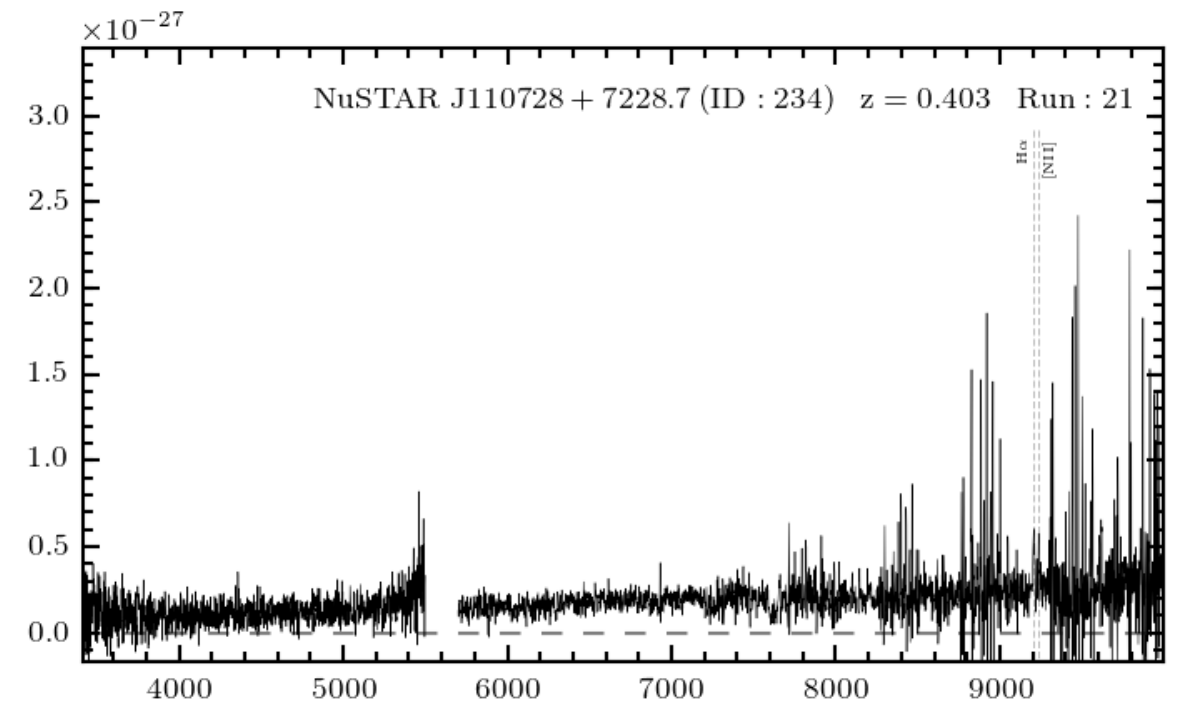}
\end{minipage}
\caption{Continued.}
\end{figure*}
\addtocounter{figure}{-1}
\begin{figure*}
\centering
\begin{minipage}[l]{0.325\textwidth}
\includegraphics[width=\textwidth]{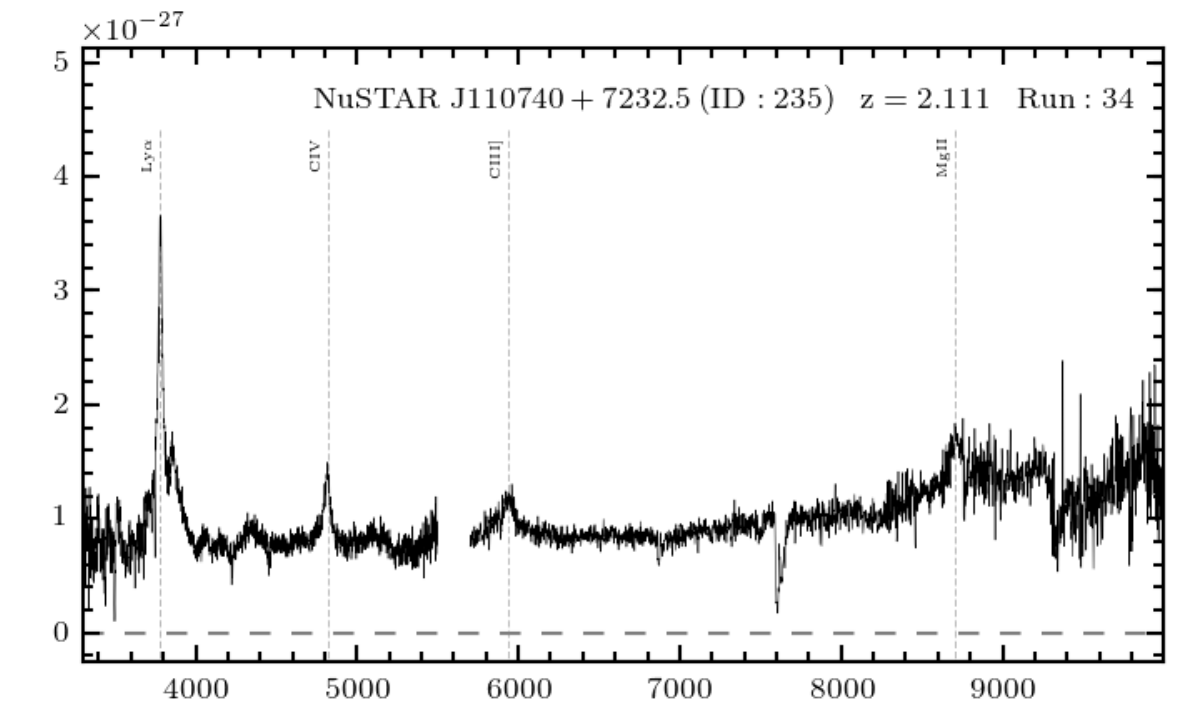}
\end{minipage}
\begin{minipage}[l]{0.325\textwidth}
\includegraphics[width=\textwidth]{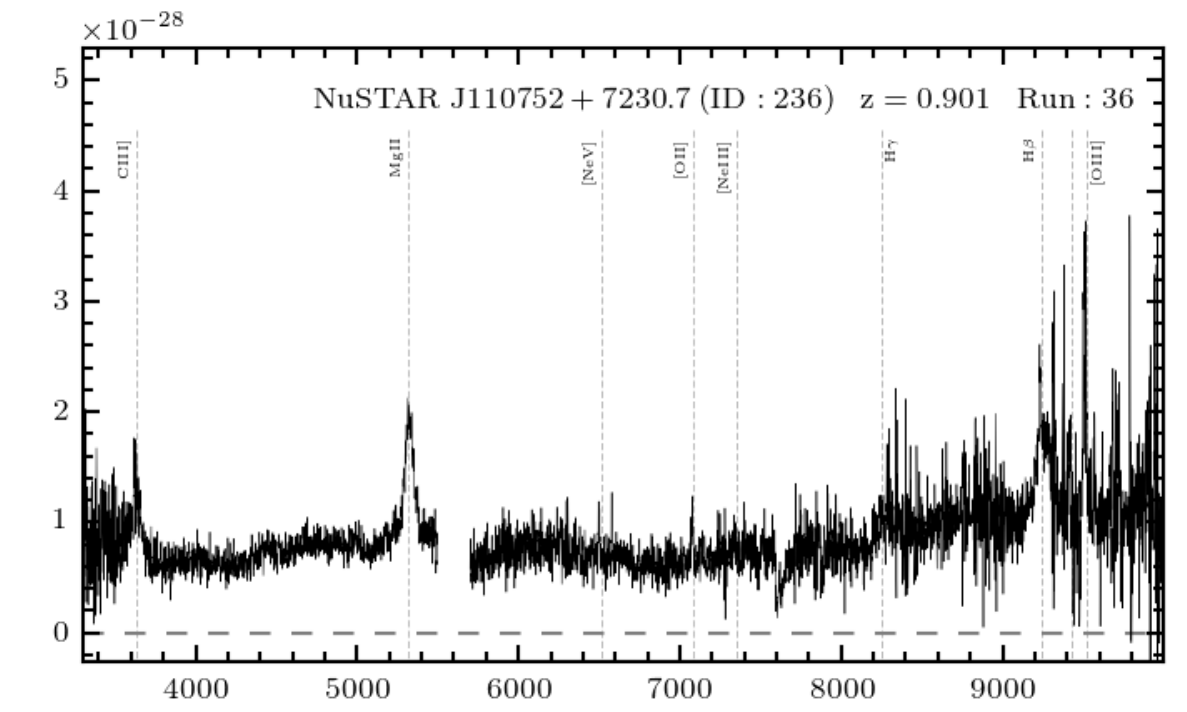}
\end{minipage}
\begin{minipage}[l]{0.325\textwidth}
\includegraphics[width=\textwidth]{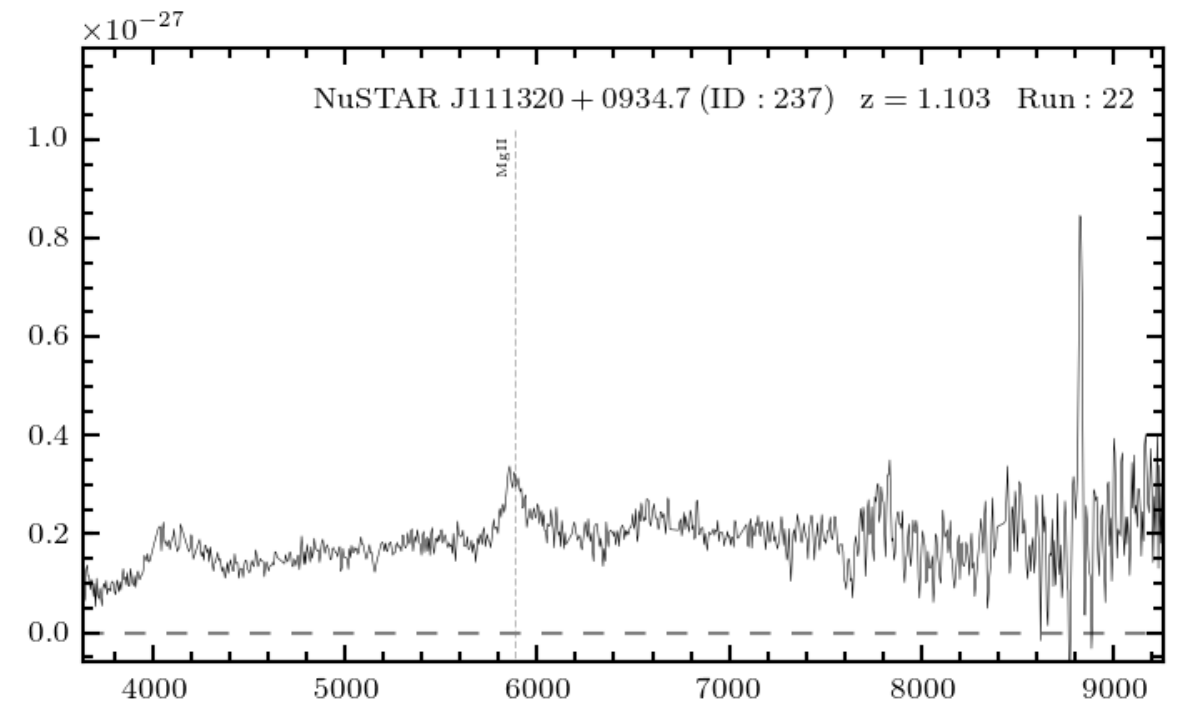}
\end{minipage}
\begin{minipage}[l]{0.325\textwidth}
\includegraphics[width=\textwidth]{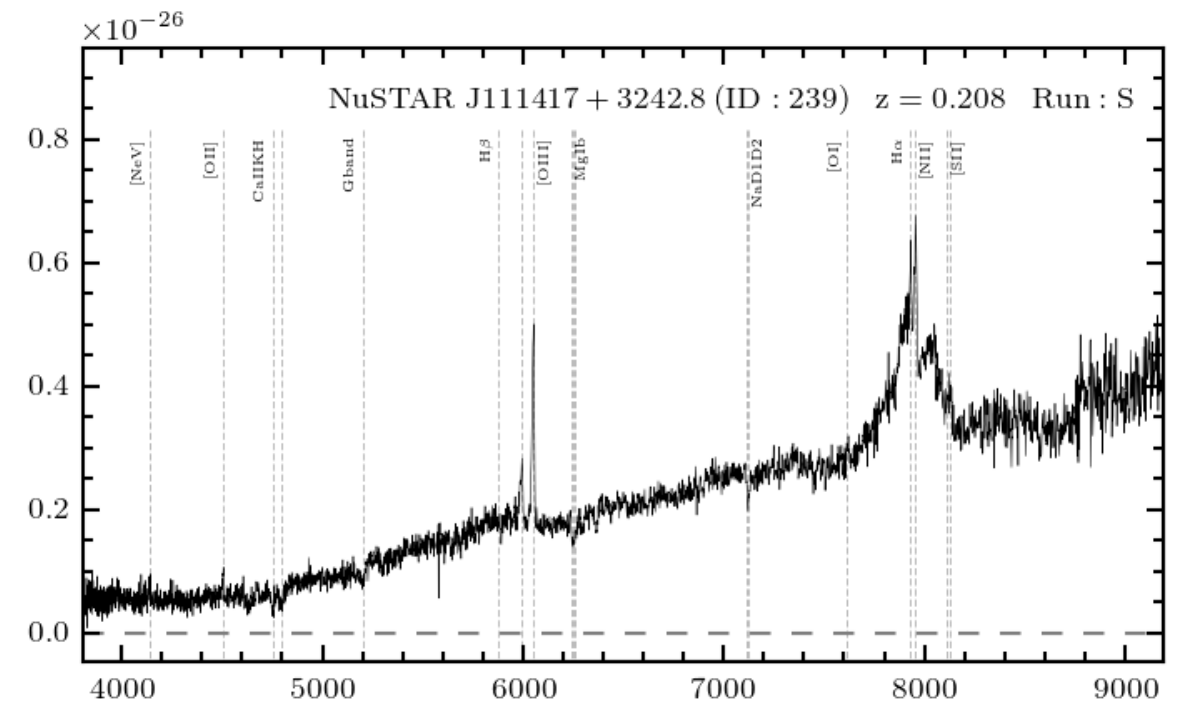}
\end{minipage}
\begin{minipage}[l]{0.325\textwidth}
\includegraphics[width=\textwidth]{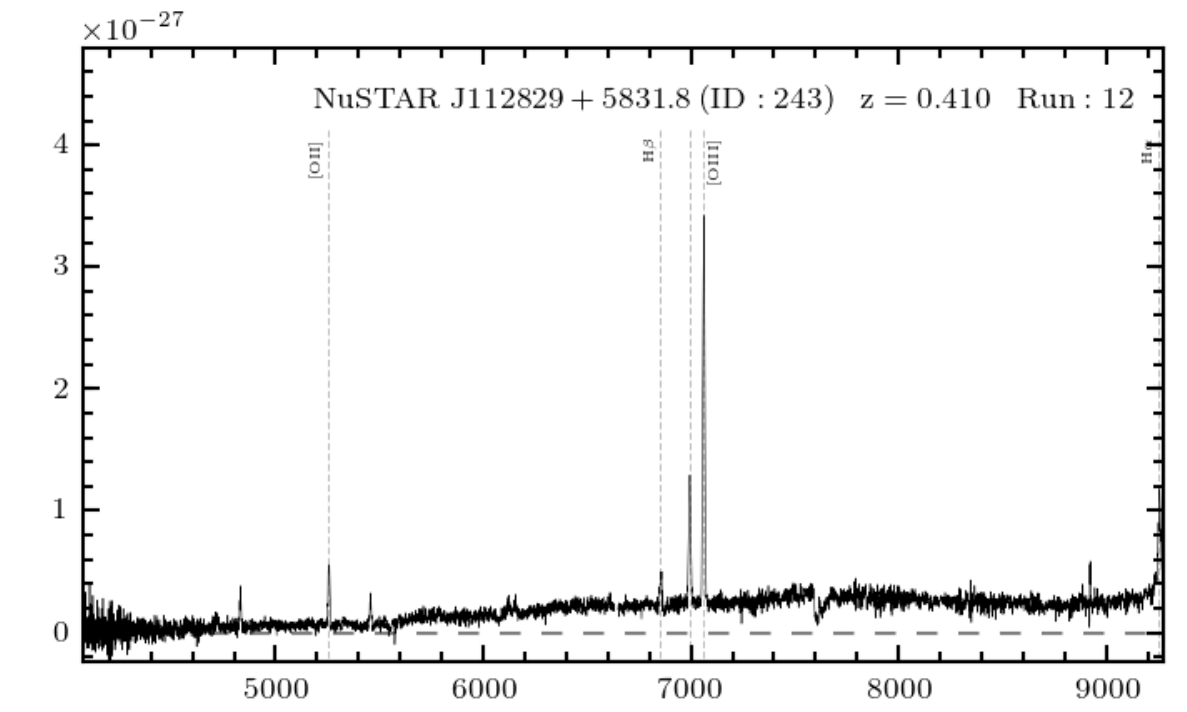}
\end{minipage}
\begin{minipage}[l]{0.325\textwidth}
\includegraphics[width=\textwidth]{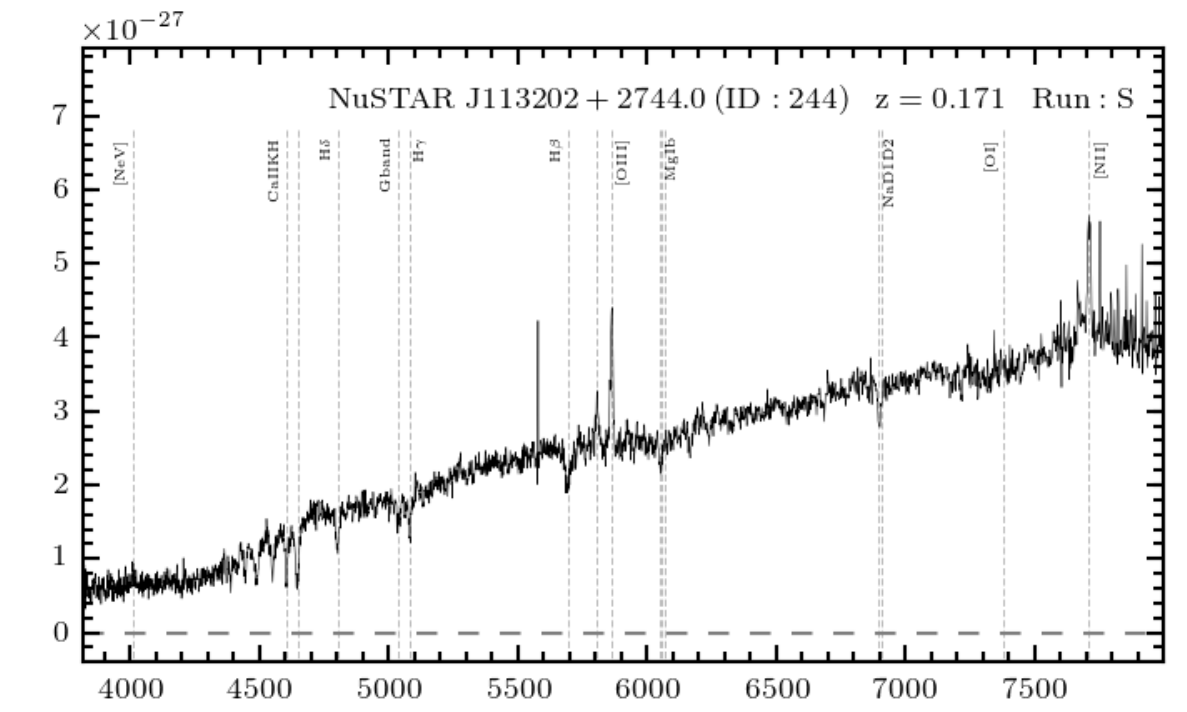}
\end{minipage}
\begin{minipage}[l]{0.325\textwidth}
\includegraphics[width=\textwidth]{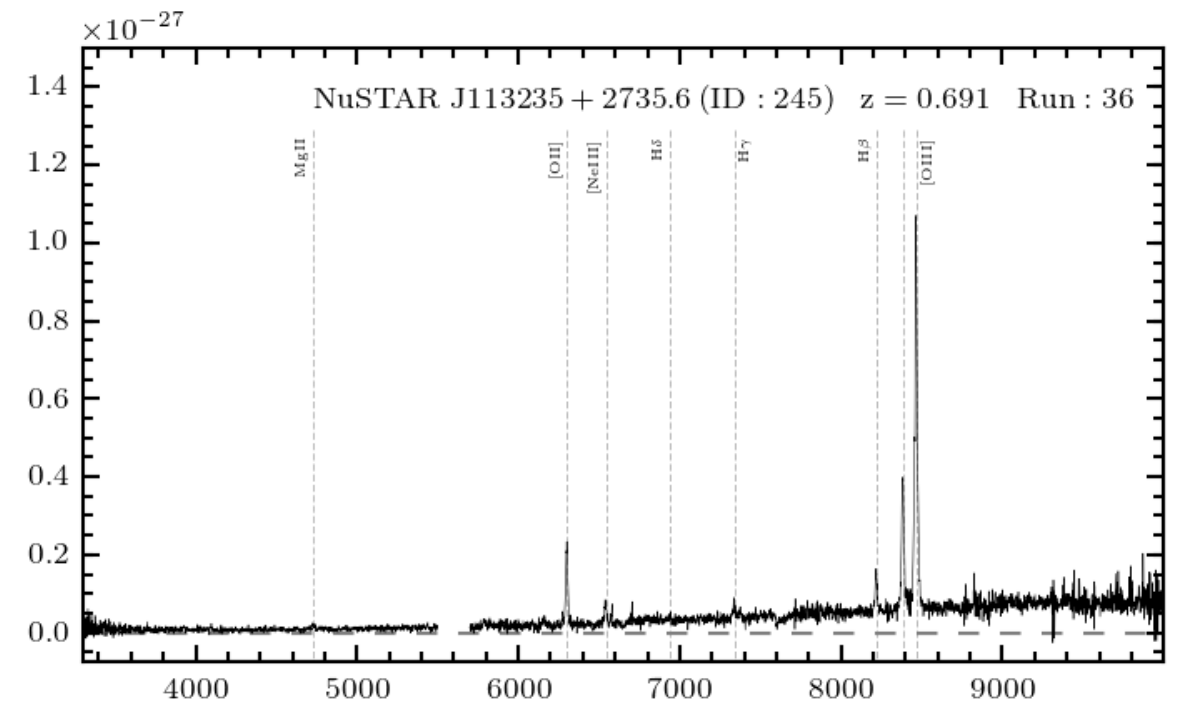}
\end{minipage}
\begin{minipage}[l]{0.325\textwidth}
\includegraphics[width=\textwidth]{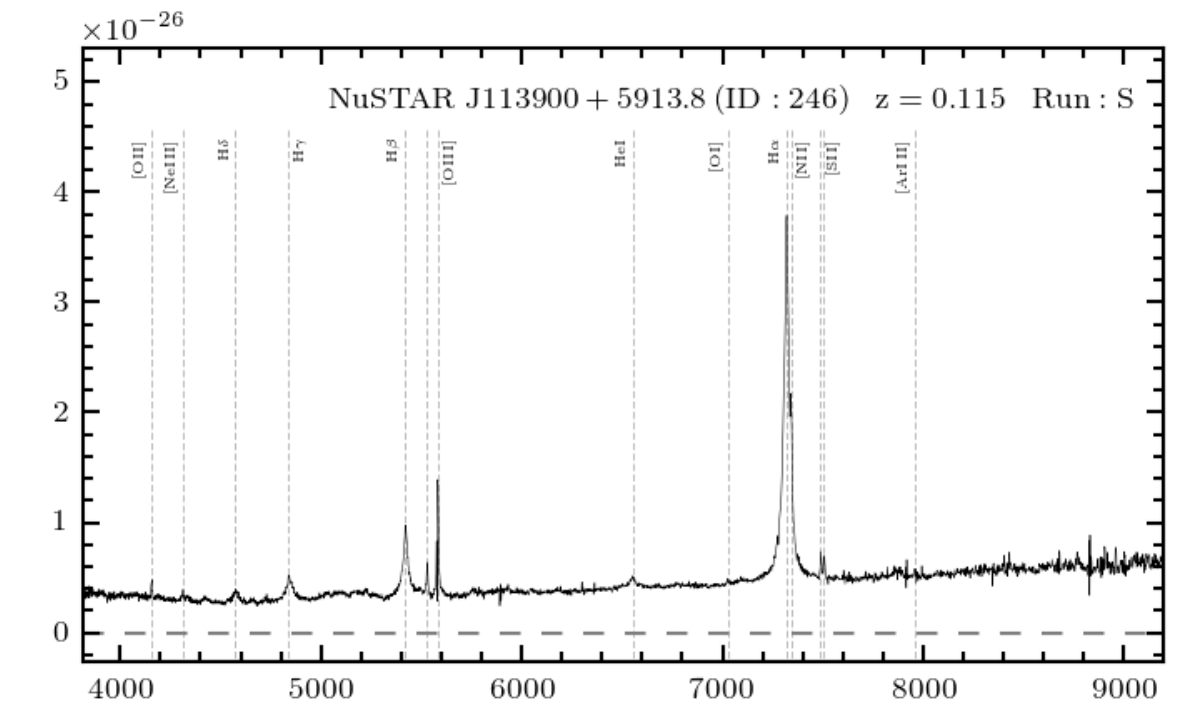}
\end{minipage}
\begin{minipage}[l]{0.325\textwidth}
\includegraphics[width=\textwidth]{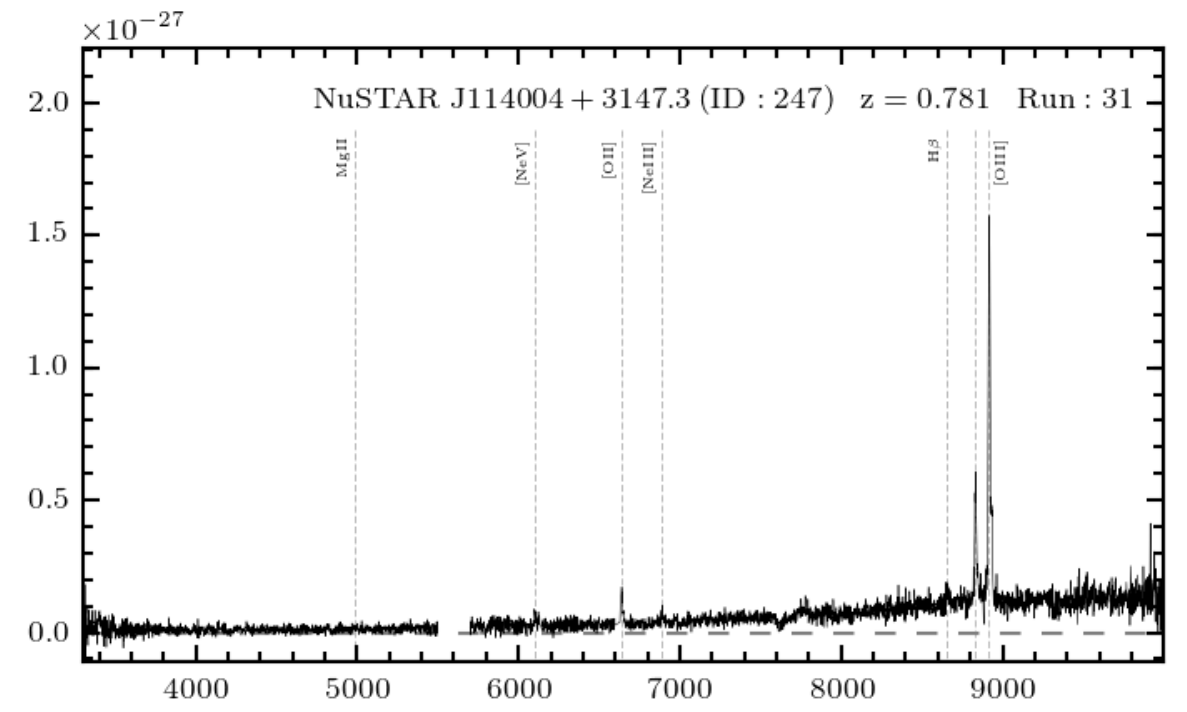}
\end{minipage}
\begin{minipage}[l]{0.325\textwidth}
\includegraphics[width=\textwidth]{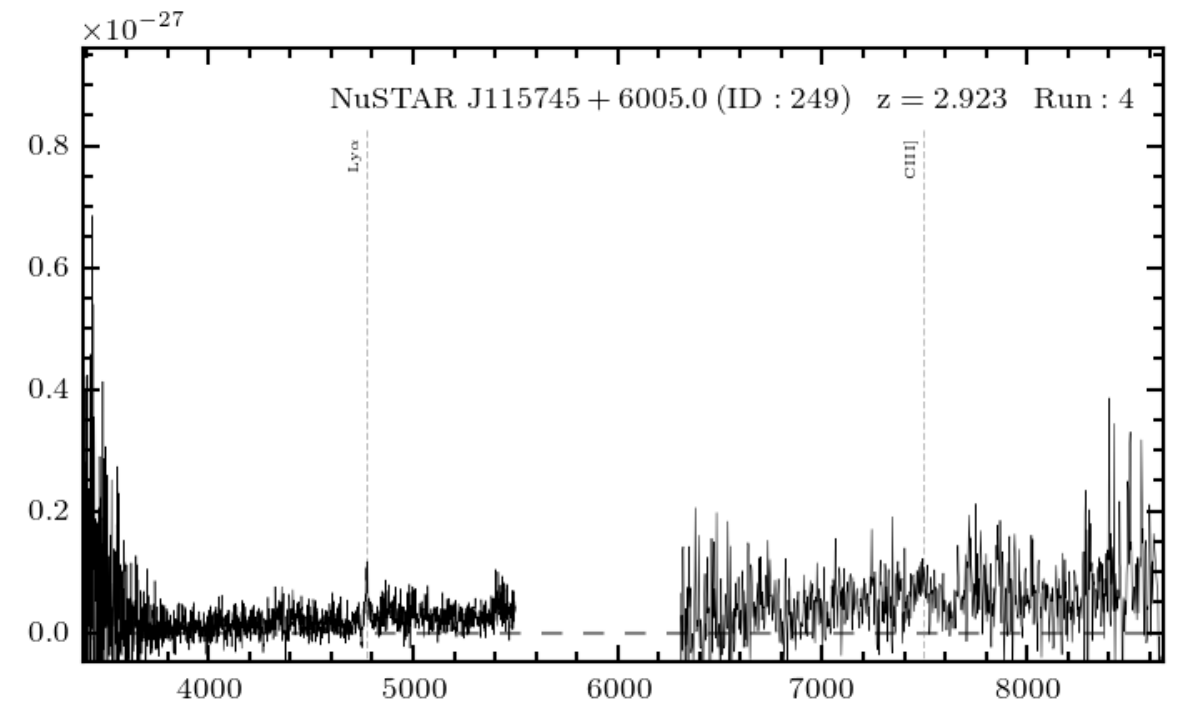}
\end{minipage}
\begin{minipage}[l]{0.325\textwidth}
\includegraphics[width=\textwidth]{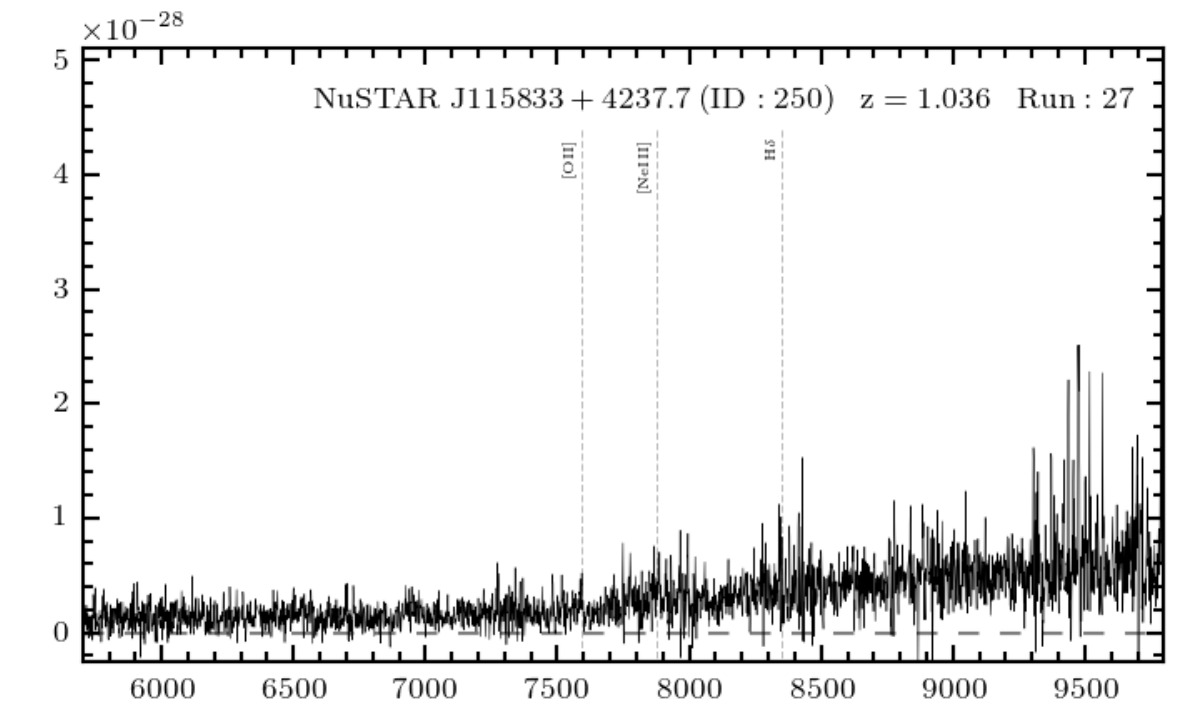}
\end{minipage}
\begin{minipage}[l]{0.325\textwidth}
\includegraphics[width=\textwidth]{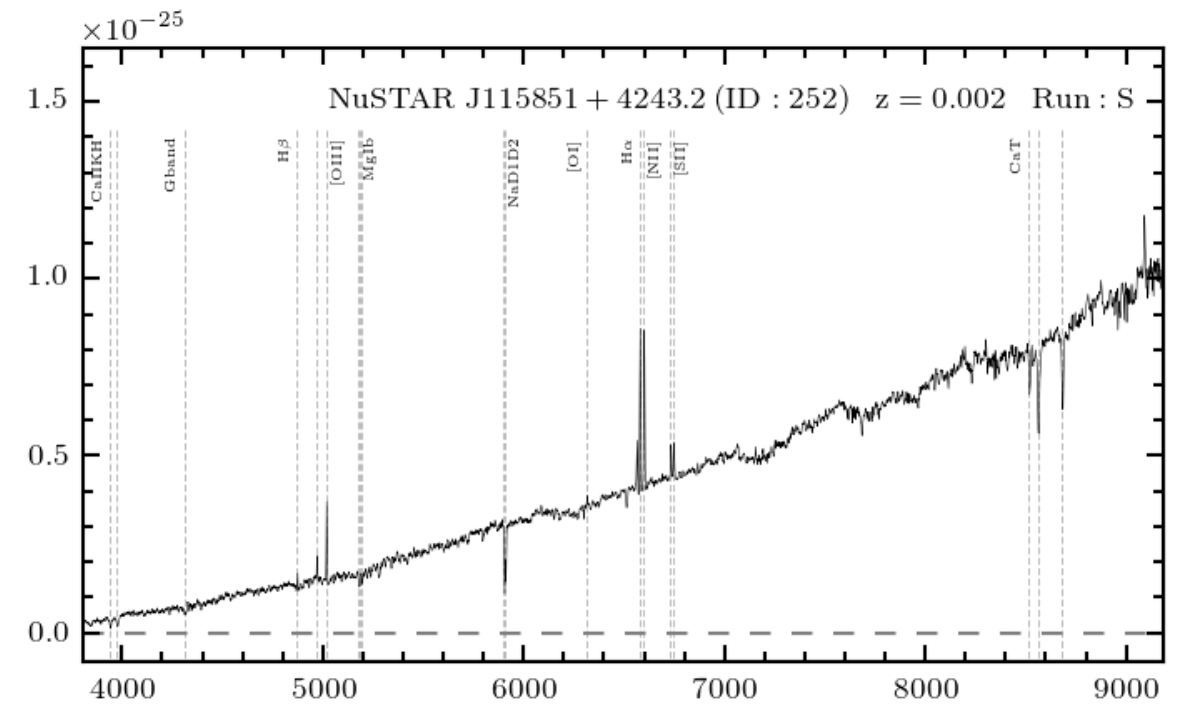}
\end{minipage}
\begin{minipage}[l]{0.325\textwidth}
\includegraphics[width=\textwidth]{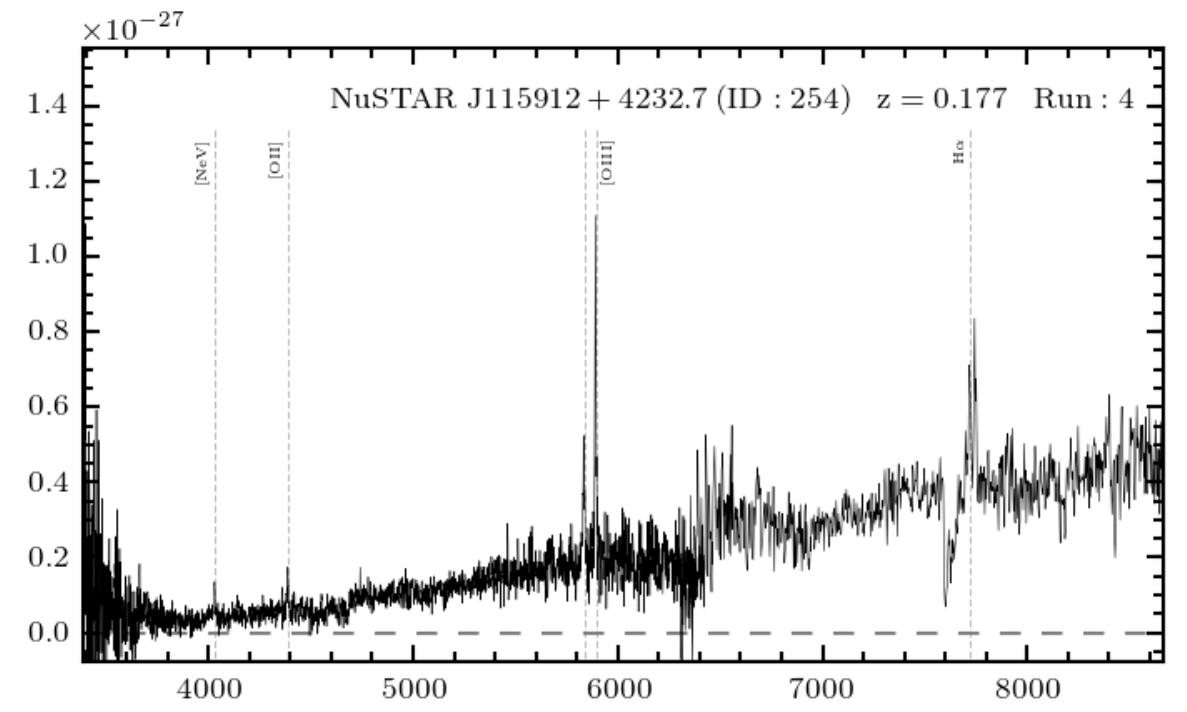}
\end{minipage}
\begin{minipage}[l]{0.325\textwidth}
\includegraphics[width=\textwidth]{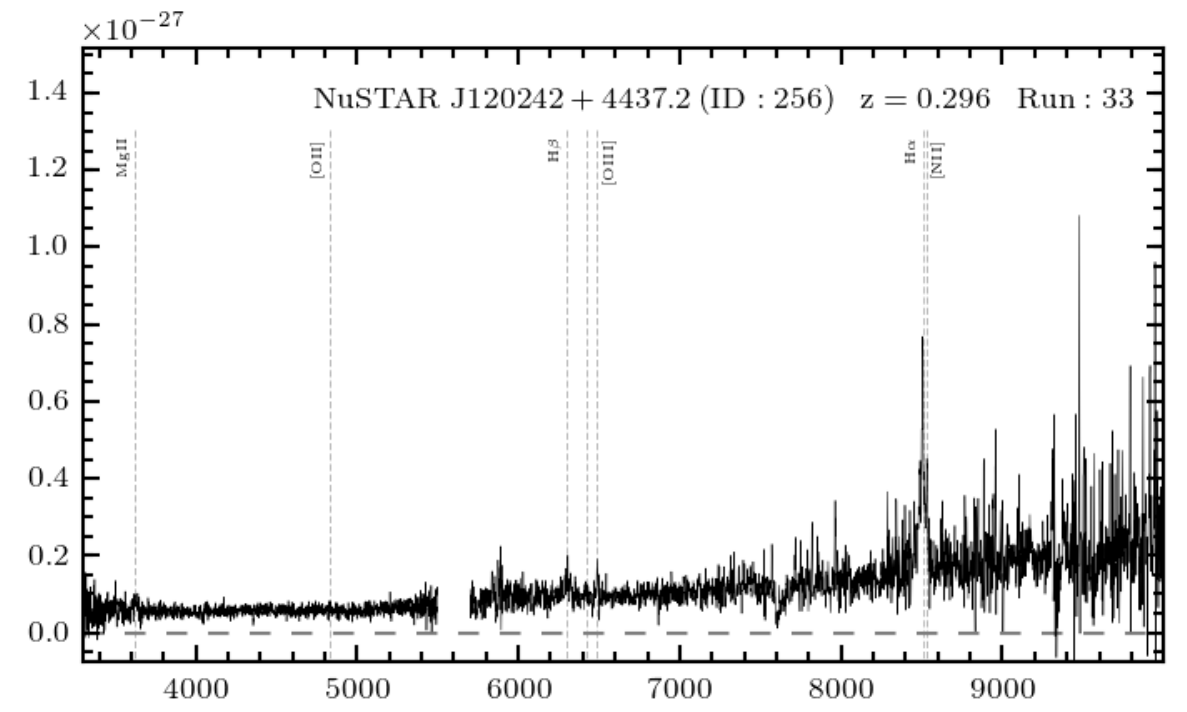}
\end{minipage}
\begin{minipage}[l]{0.325\textwidth}
\includegraphics[width=\textwidth]{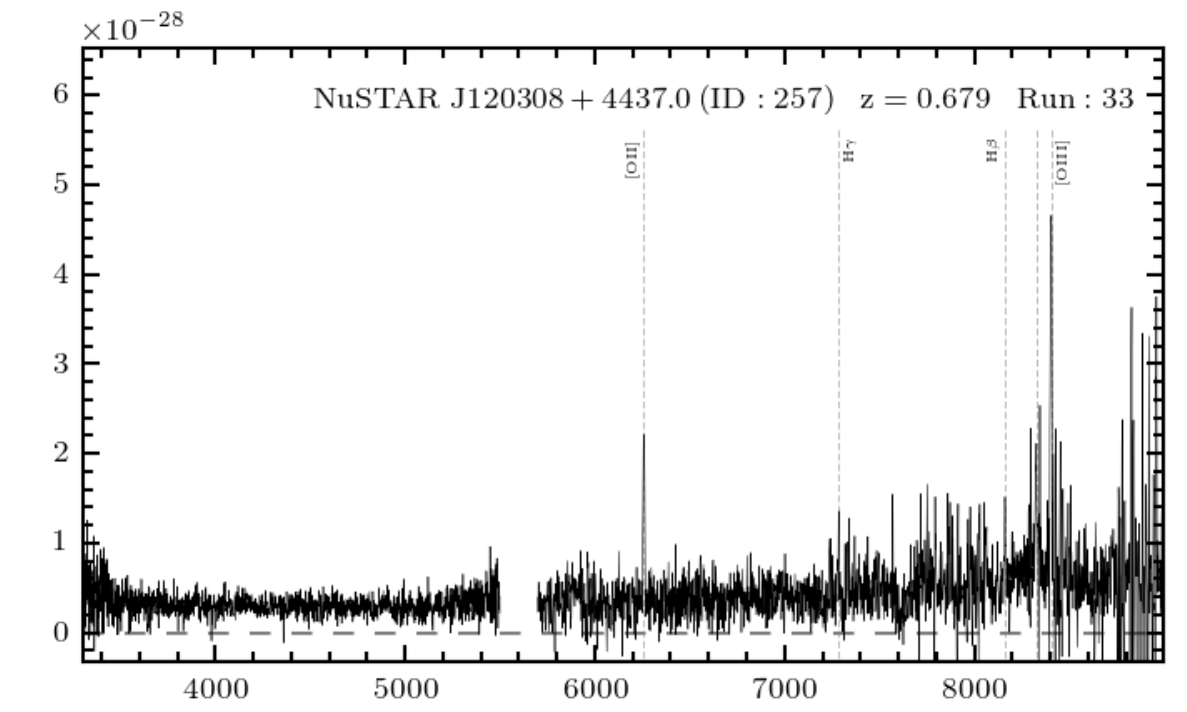}
\end{minipage}
\begin{minipage}[l]{0.325\textwidth}
\includegraphics[width=\textwidth]{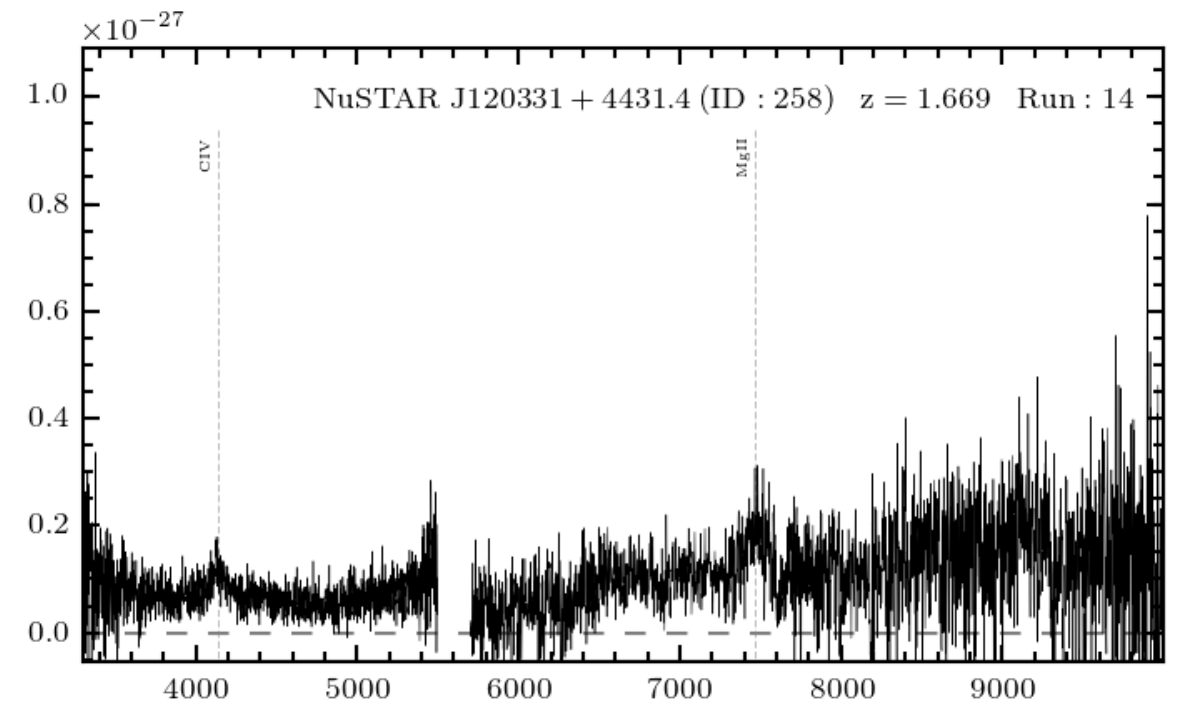}
\end{minipage}
\begin{minipage}[l]{0.325\textwidth}
\includegraphics[width=\textwidth]{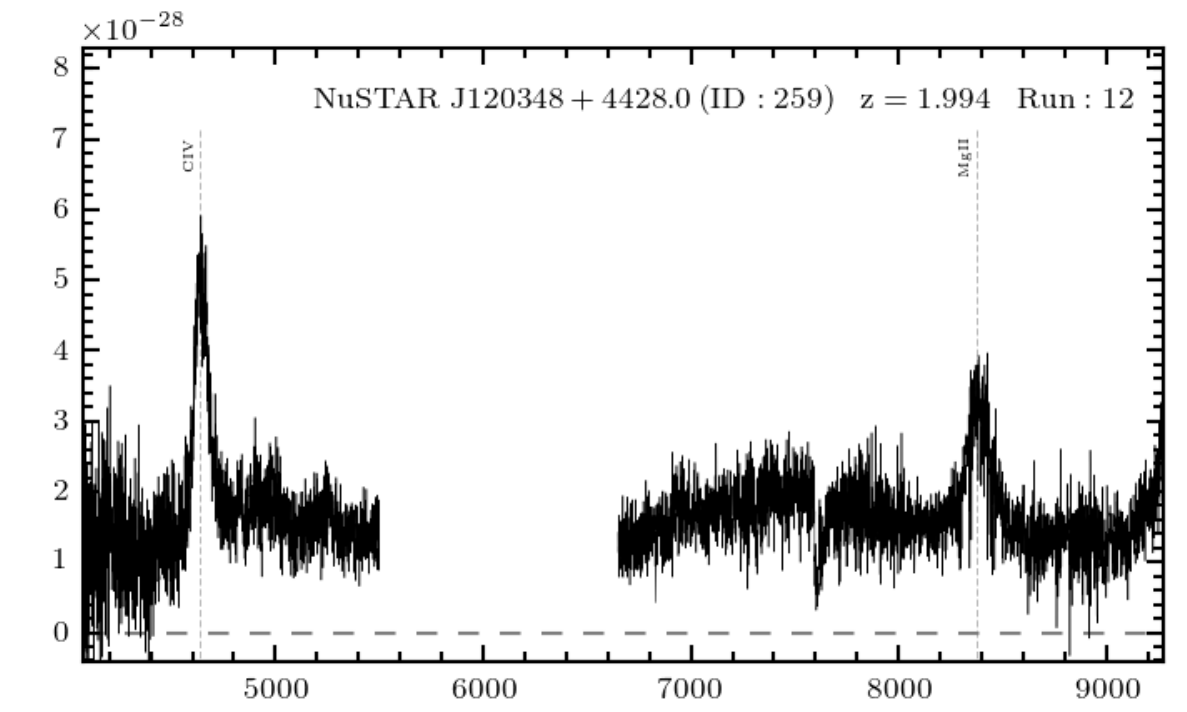}
\end{minipage}
\begin{minipage}[l]{0.325\textwidth}
\includegraphics[width=\textwidth]{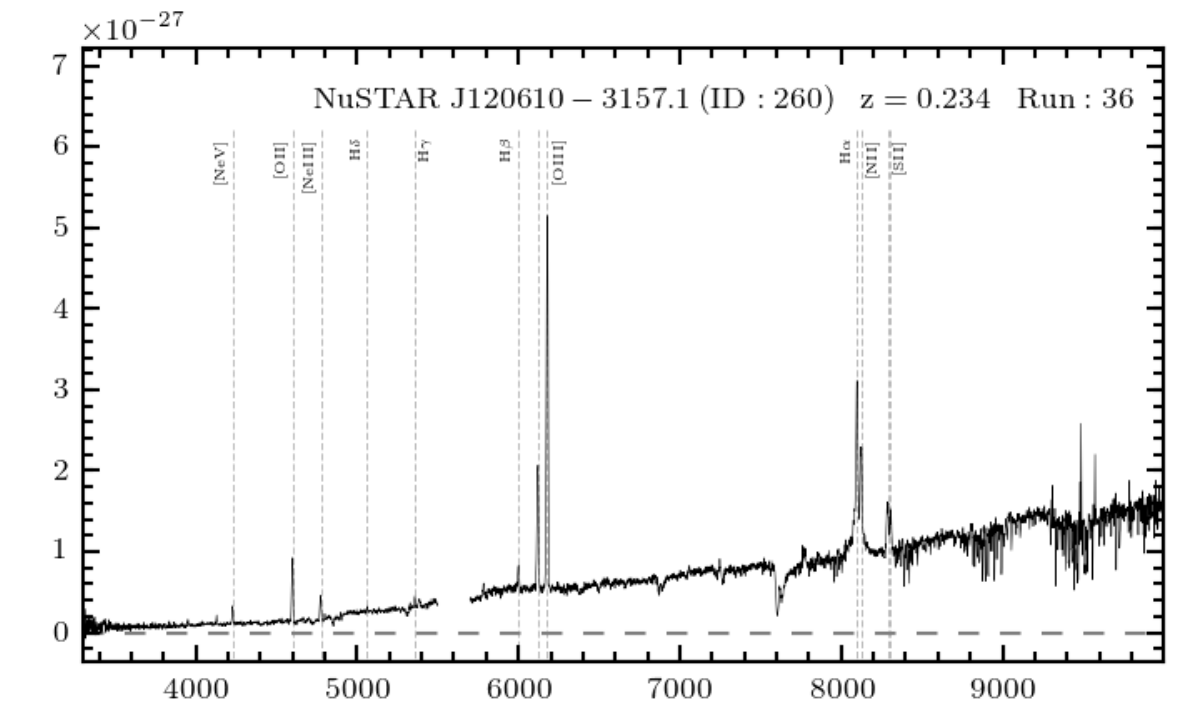}
\end{minipage}
\caption{Continued.}
\end{figure*}
\addtocounter{figure}{-1}
\begin{figure*}
\centering
\begin{minipage}[l]{0.325\textwidth}
\includegraphics[width=\textwidth]{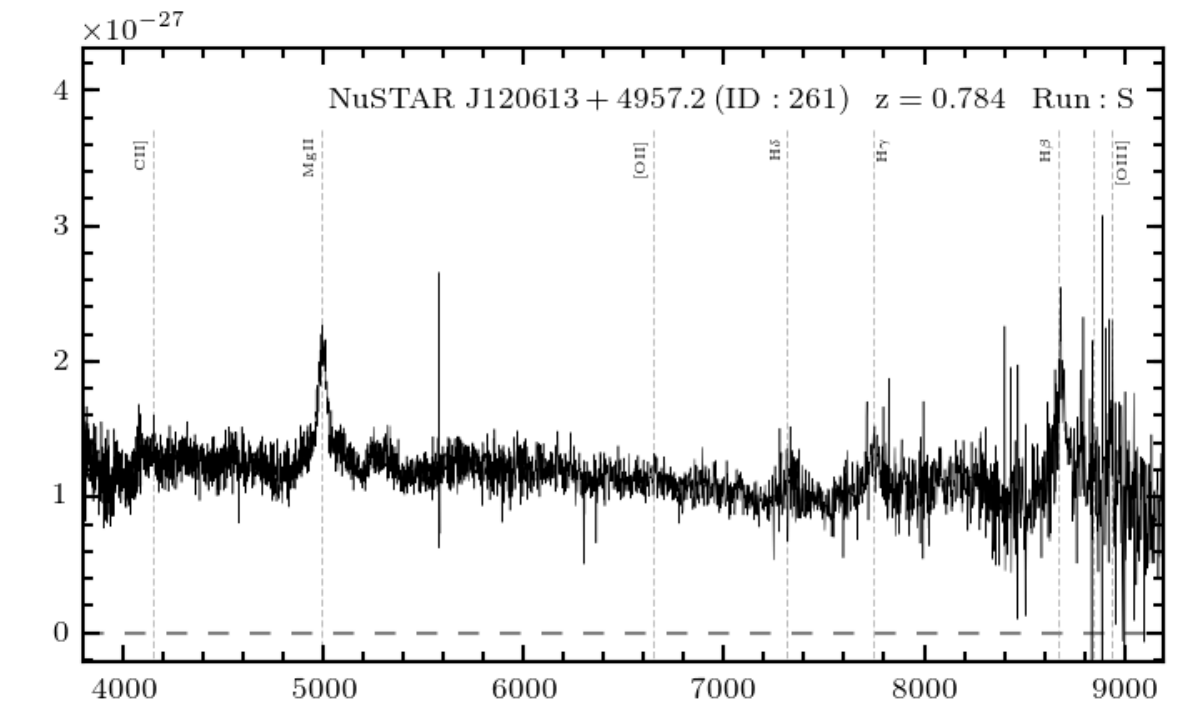}
\end{minipage}
\begin{minipage}[l]{0.325\textwidth}
\includegraphics[width=\textwidth]{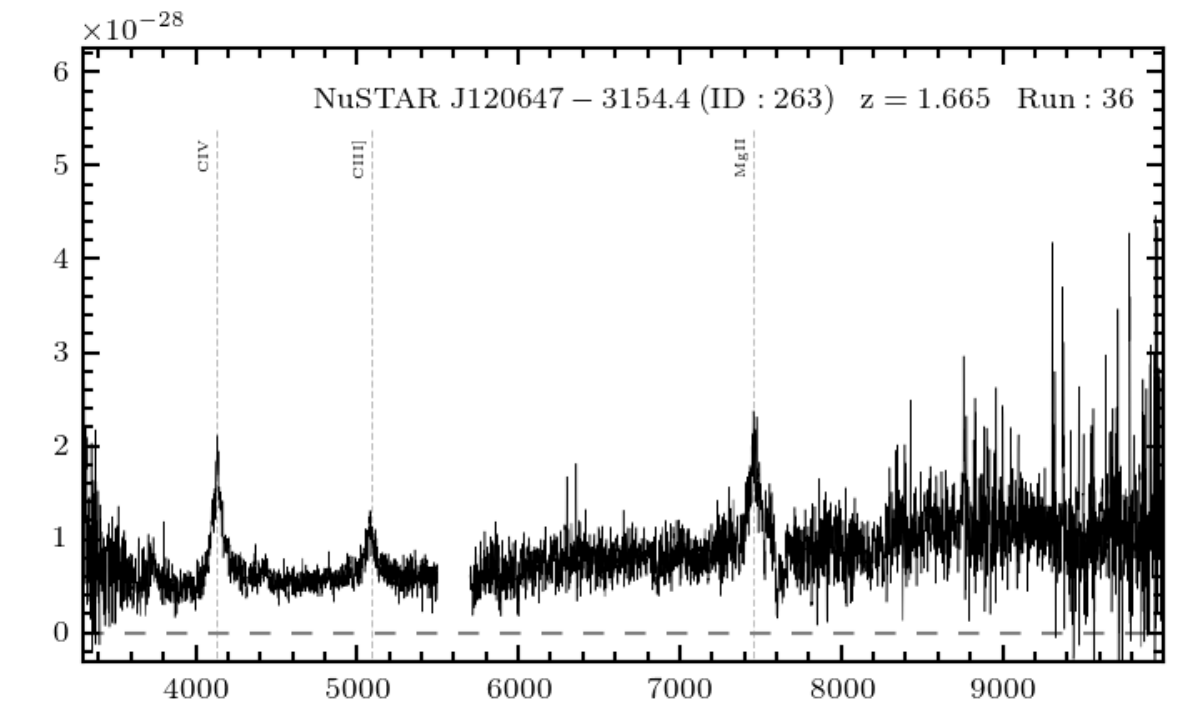}
\end{minipage}
\begin{minipage}[l]{0.325\textwidth}
\includegraphics[width=\textwidth]{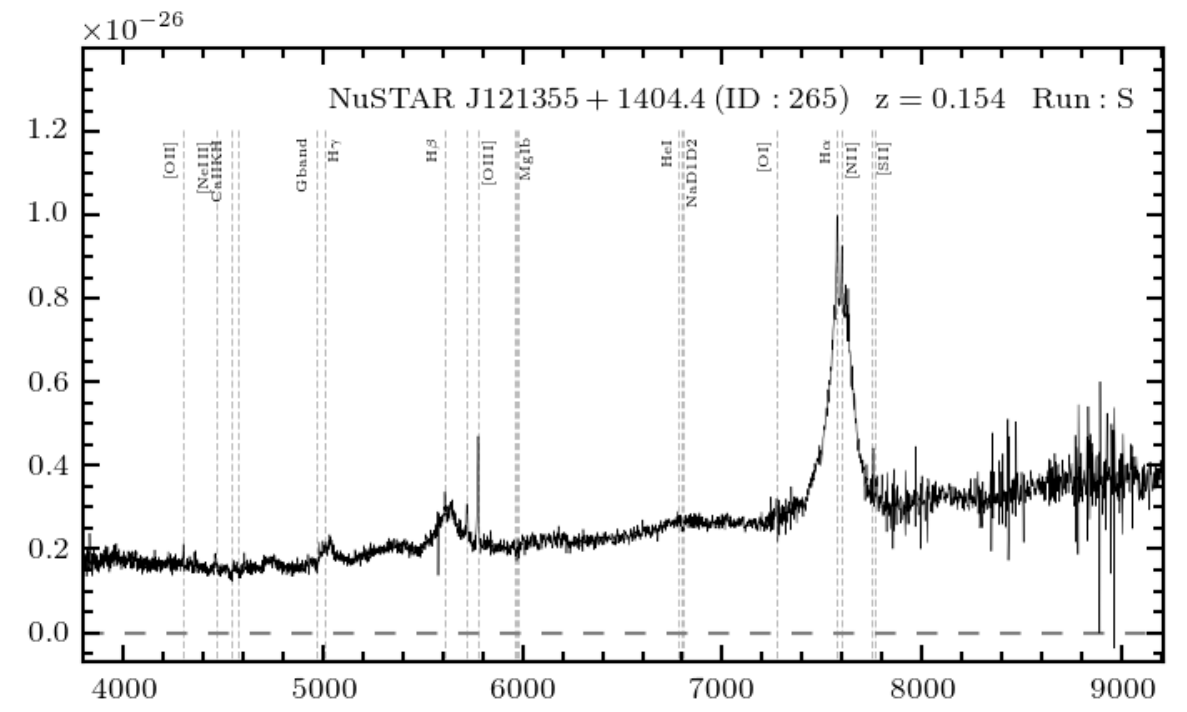}
\end{minipage}
\begin{minipage}[l]{0.325\textwidth}
\includegraphics[width=\textwidth]{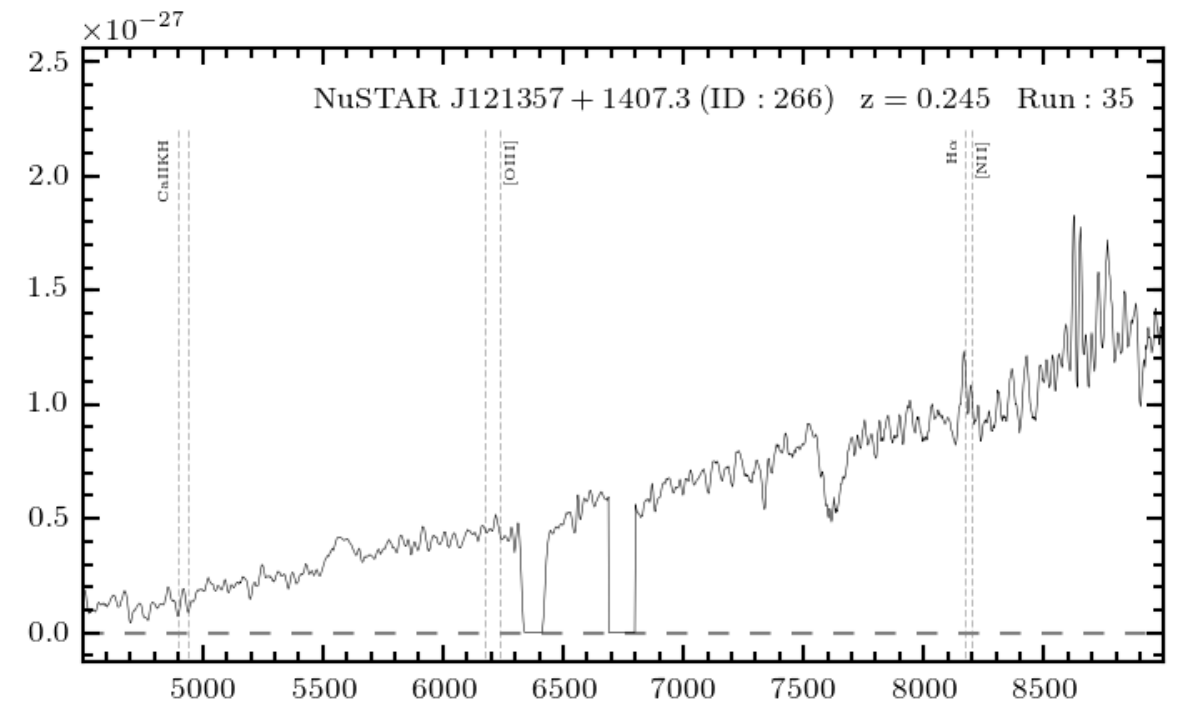}
\end{minipage}
\begin{minipage}[l]{0.325\textwidth}
\includegraphics[width=\textwidth]{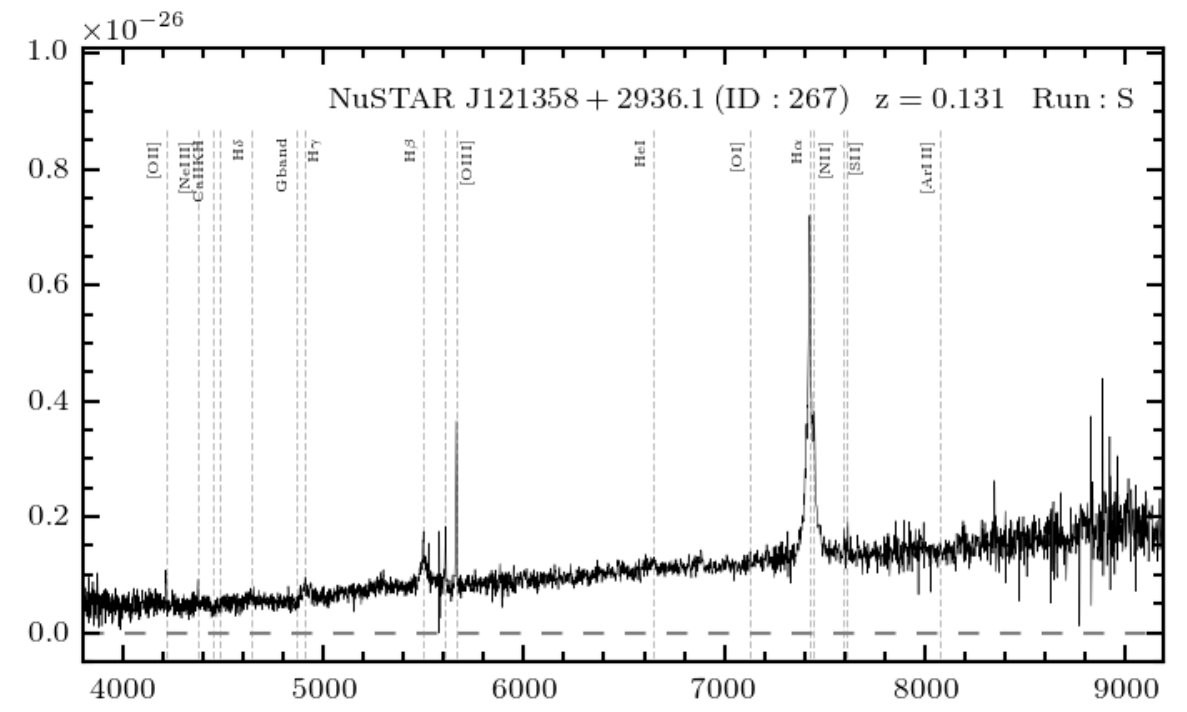}
\end{minipage}
\begin{minipage}[l]{0.325\textwidth}
\includegraphics[width=\textwidth]{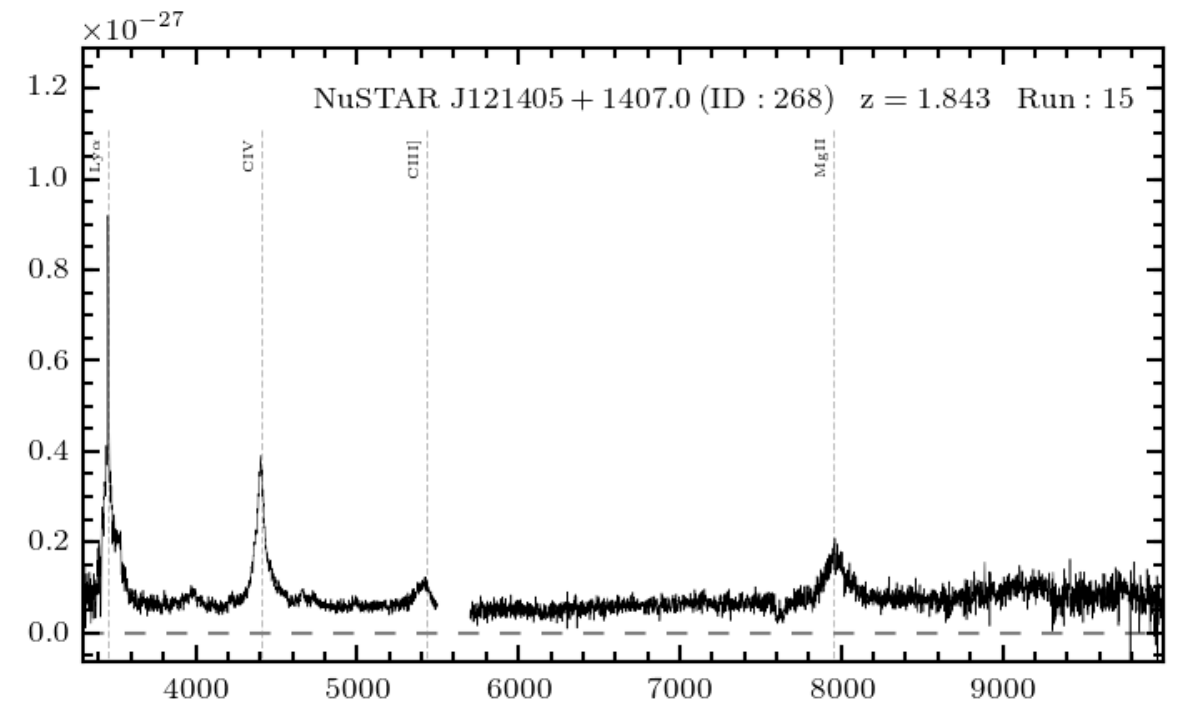}
\end{minipage}
\begin{minipage}[l]{0.325\textwidth}
\includegraphics[width=\textwidth]{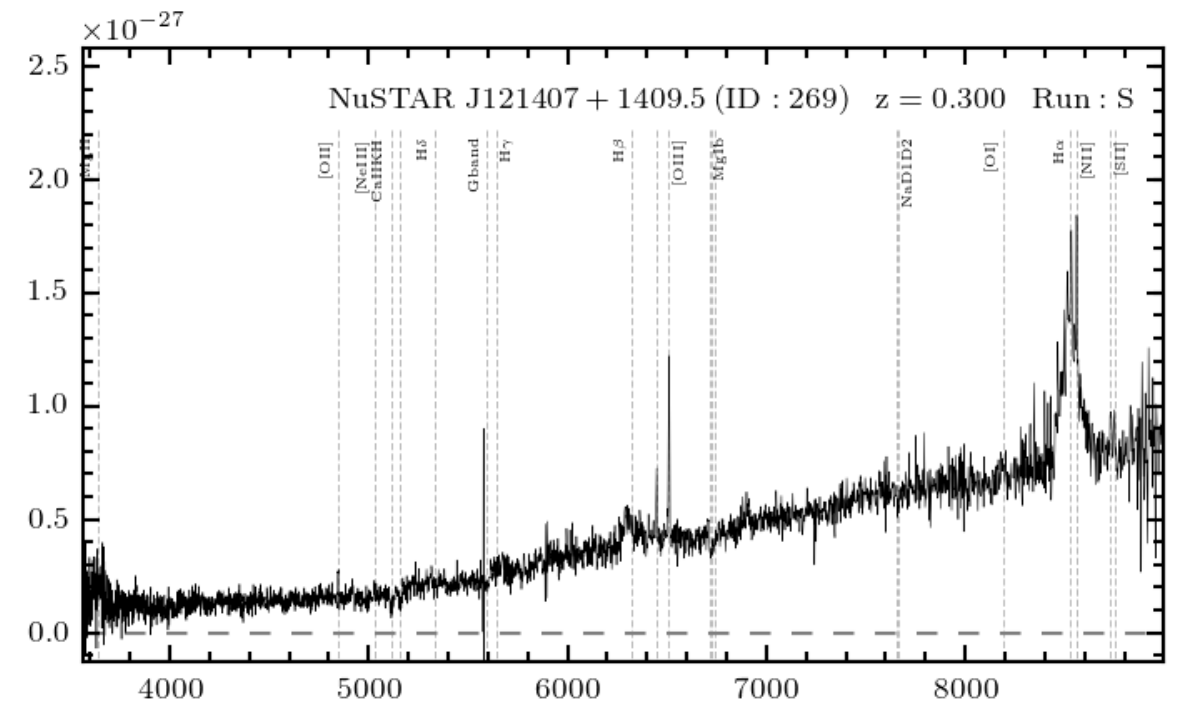}
\end{minipage}
\begin{minipage}[l]{0.325\textwidth}
\includegraphics[width=\textwidth]{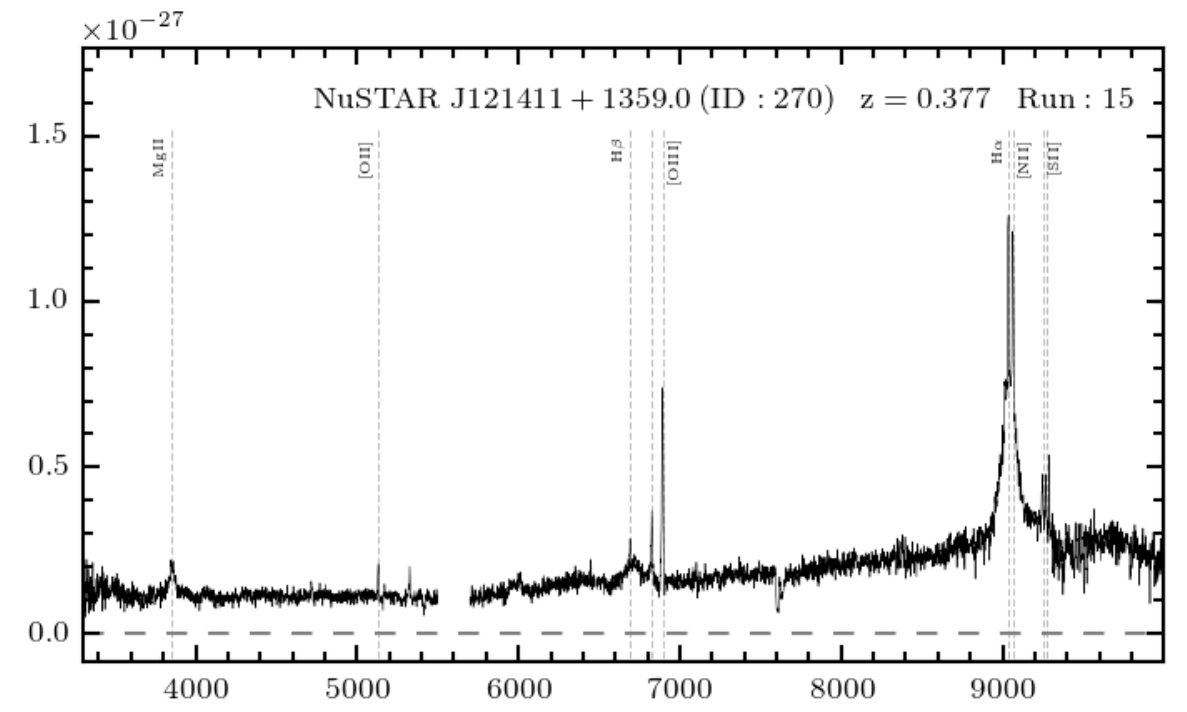}
\end{minipage}
\begin{minipage}[l]{0.325\textwidth}
\includegraphics[width=\textwidth]{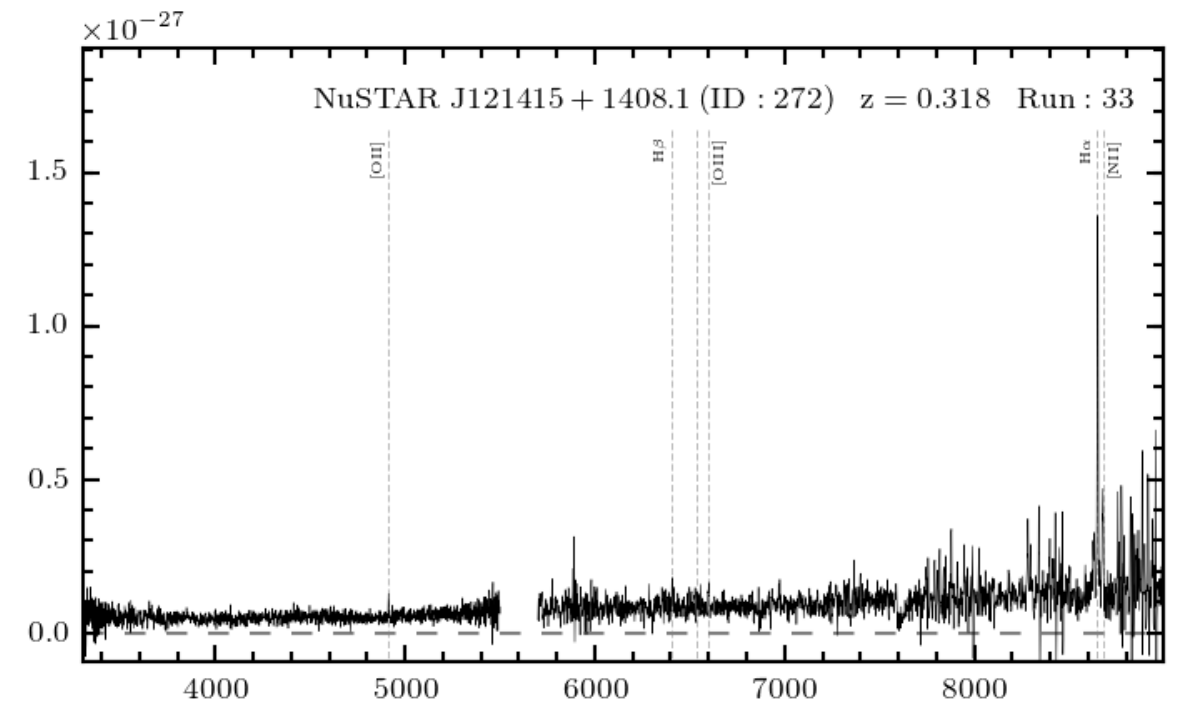}
\end{minipage}
\begin{minipage}[l]{0.325\textwidth}
\includegraphics[width=\textwidth]{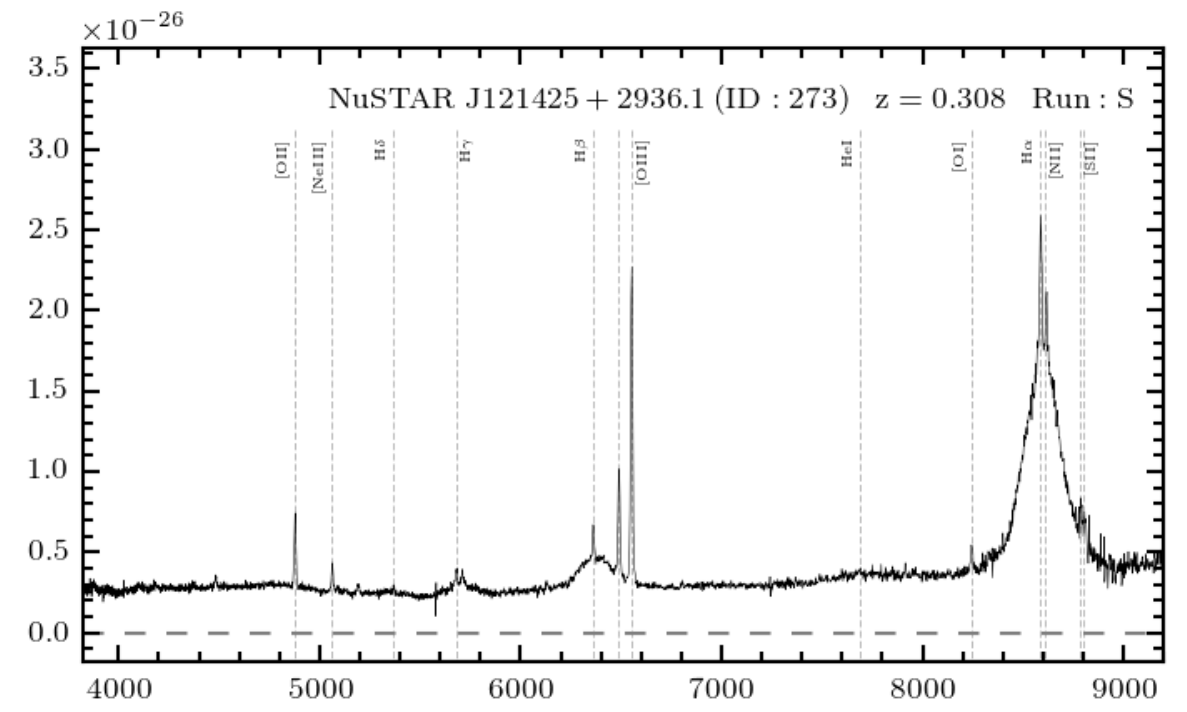}
\end{minipage}
\begin{minipage}[l]{0.325\textwidth}
\includegraphics[width=\textwidth]{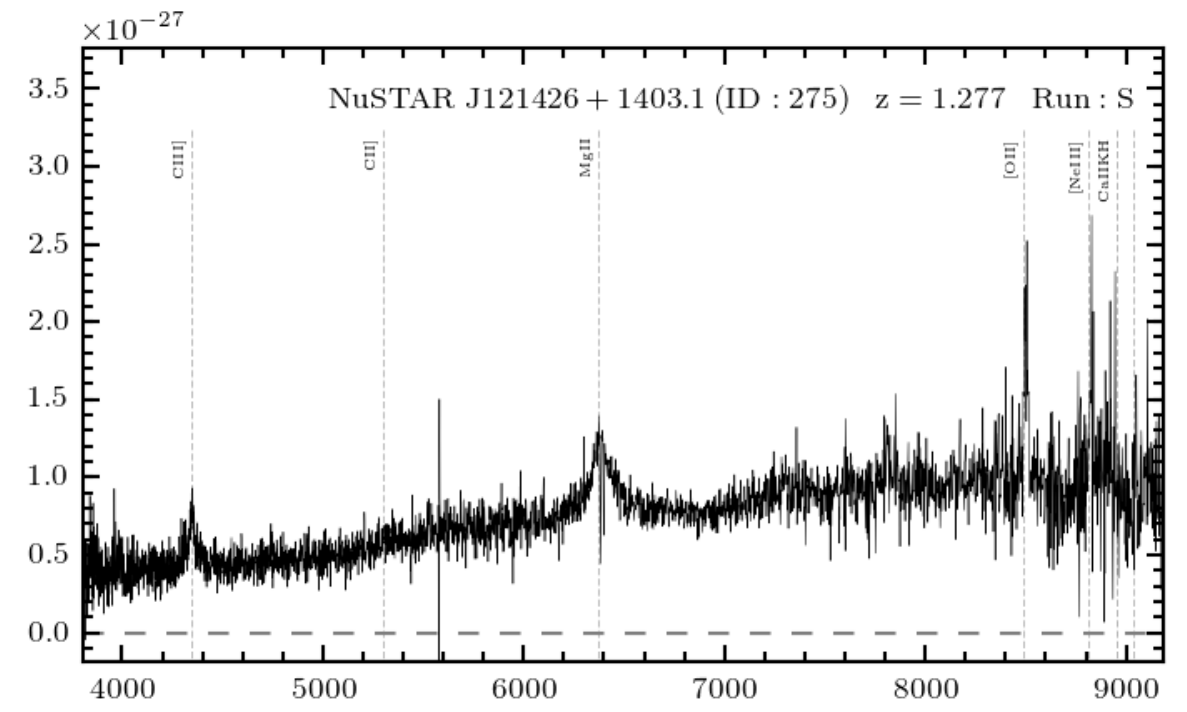}
\end{minipage}
\begin{minipage}[l]{0.325\textwidth}
\includegraphics[width=\textwidth]{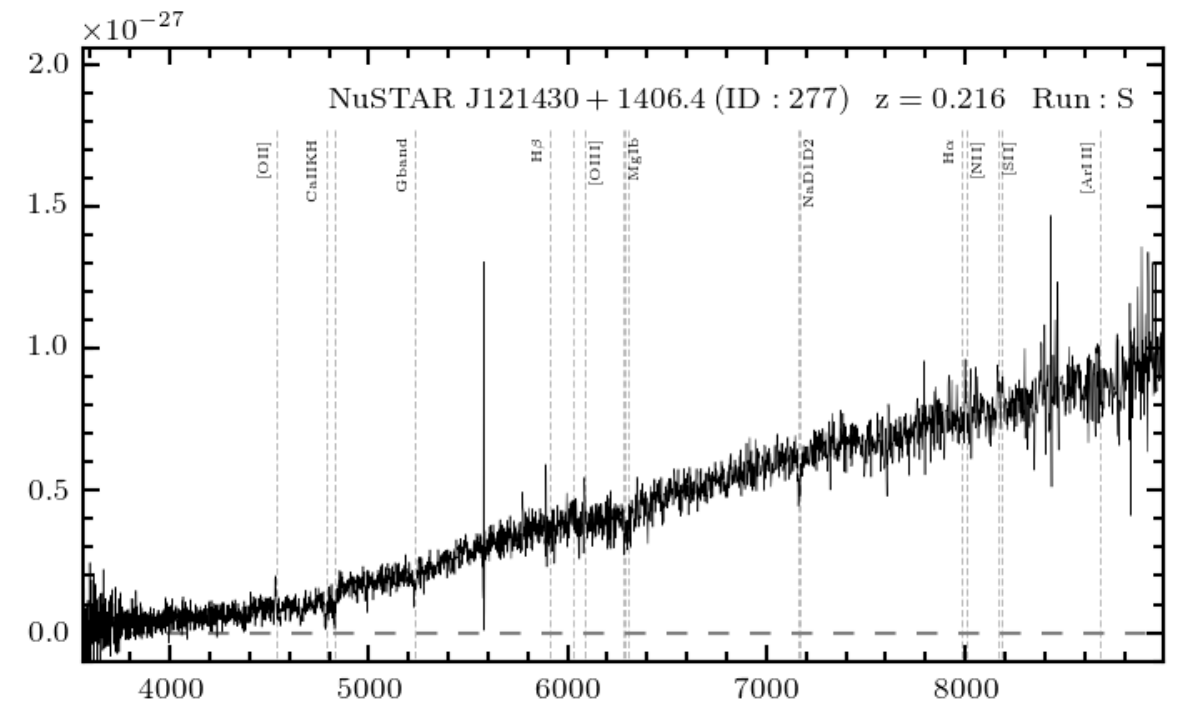}
\end{minipage}
\begin{minipage}[l]{0.325\textwidth}
\includegraphics[width=\textwidth]{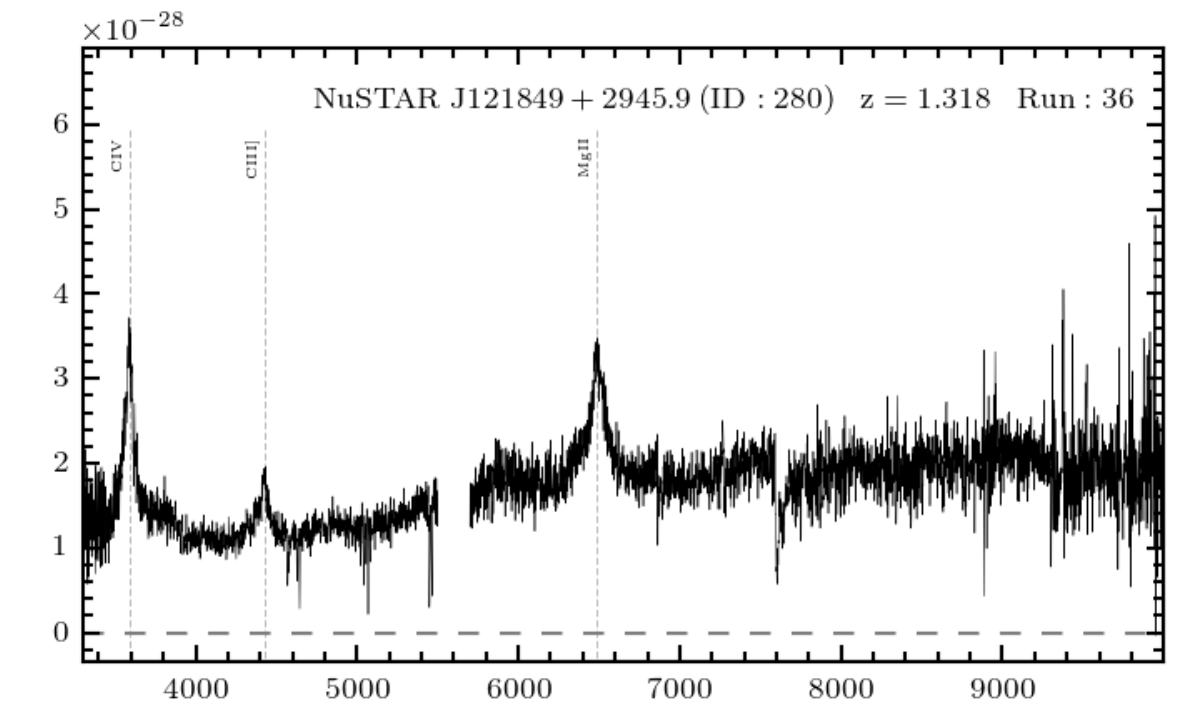}
\end{minipage}
\begin{minipage}[l]{0.325\textwidth}
\includegraphics[width=\textwidth]{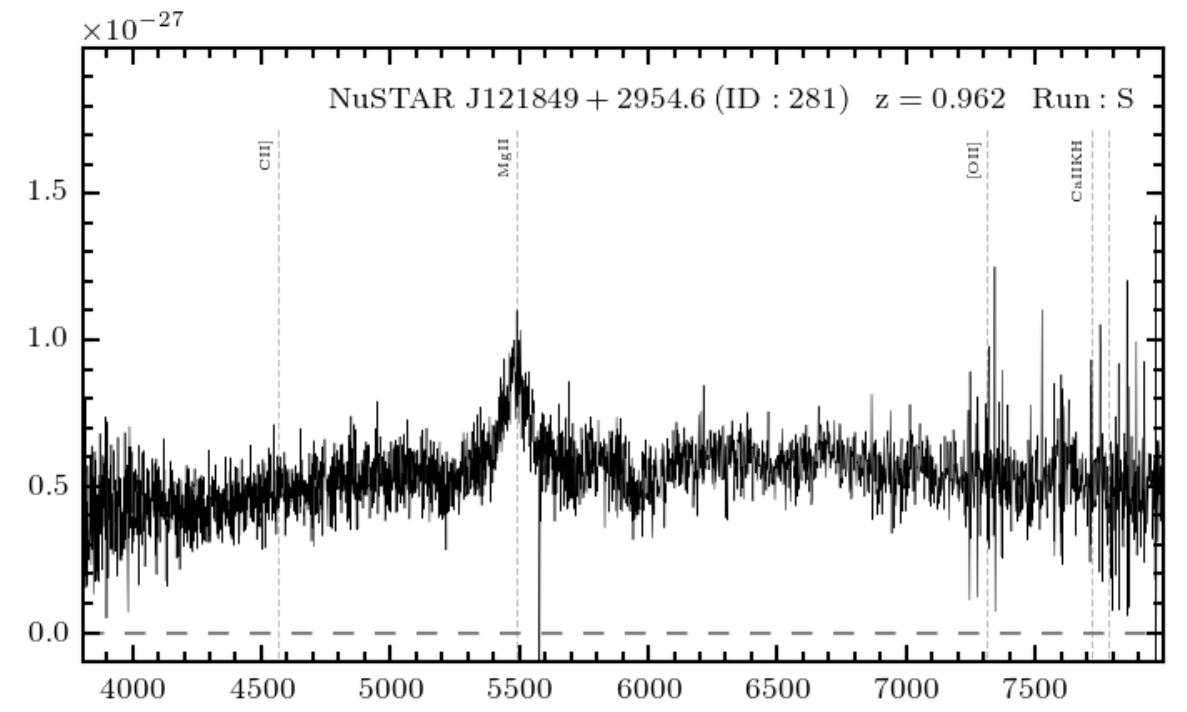}
\end{minipage}
\begin{minipage}[l]{0.325\textwidth}
\includegraphics[width=\textwidth]{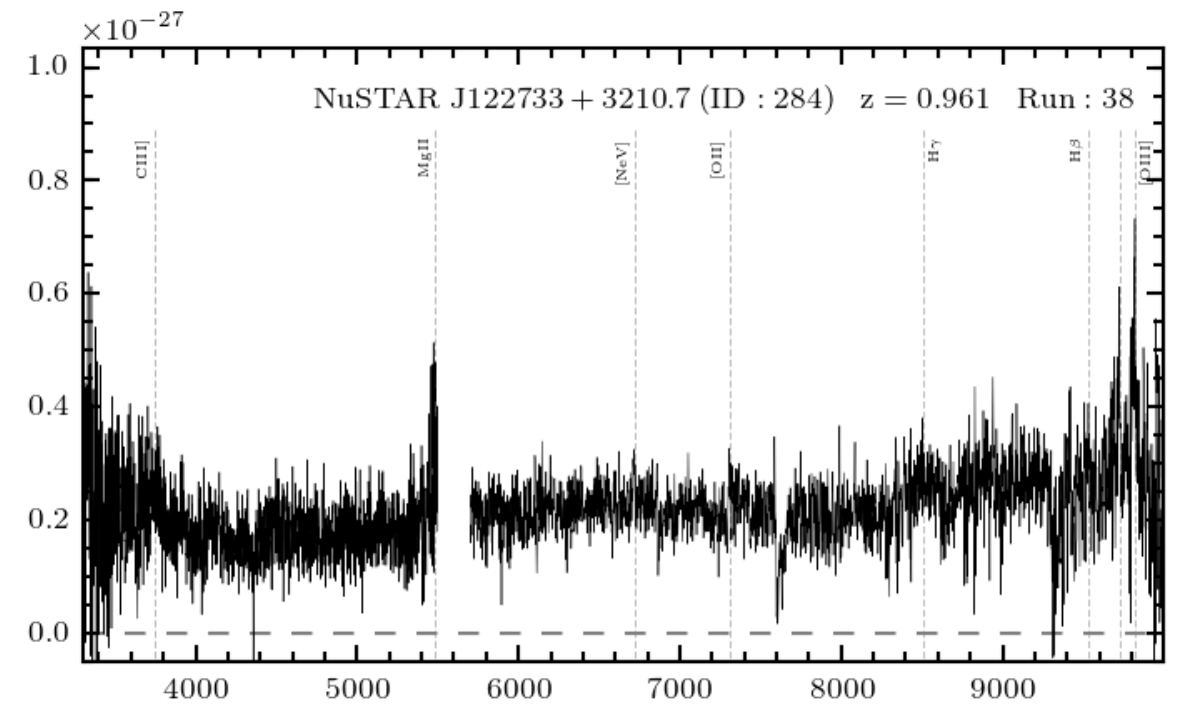}
\end{minipage}
\begin{minipage}[l]{0.325\textwidth}
\includegraphics[width=\textwidth]{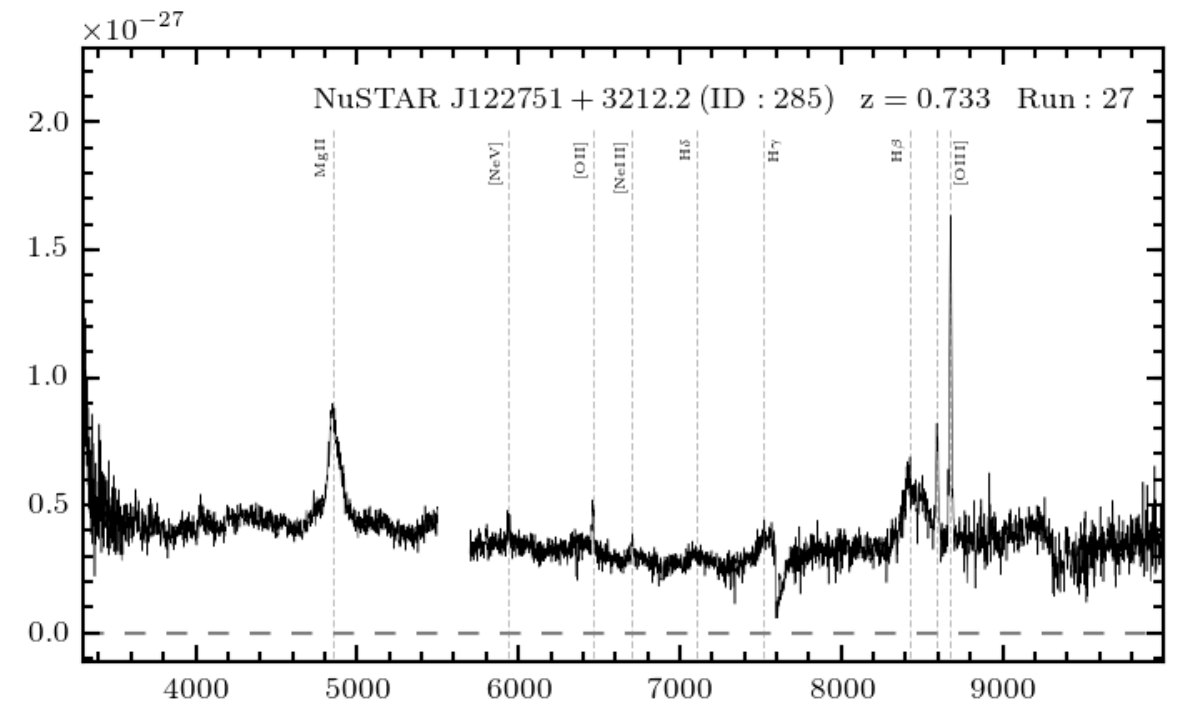}
\end{minipage}
\begin{minipage}[l]{0.325\textwidth}
\includegraphics[width=\textwidth]{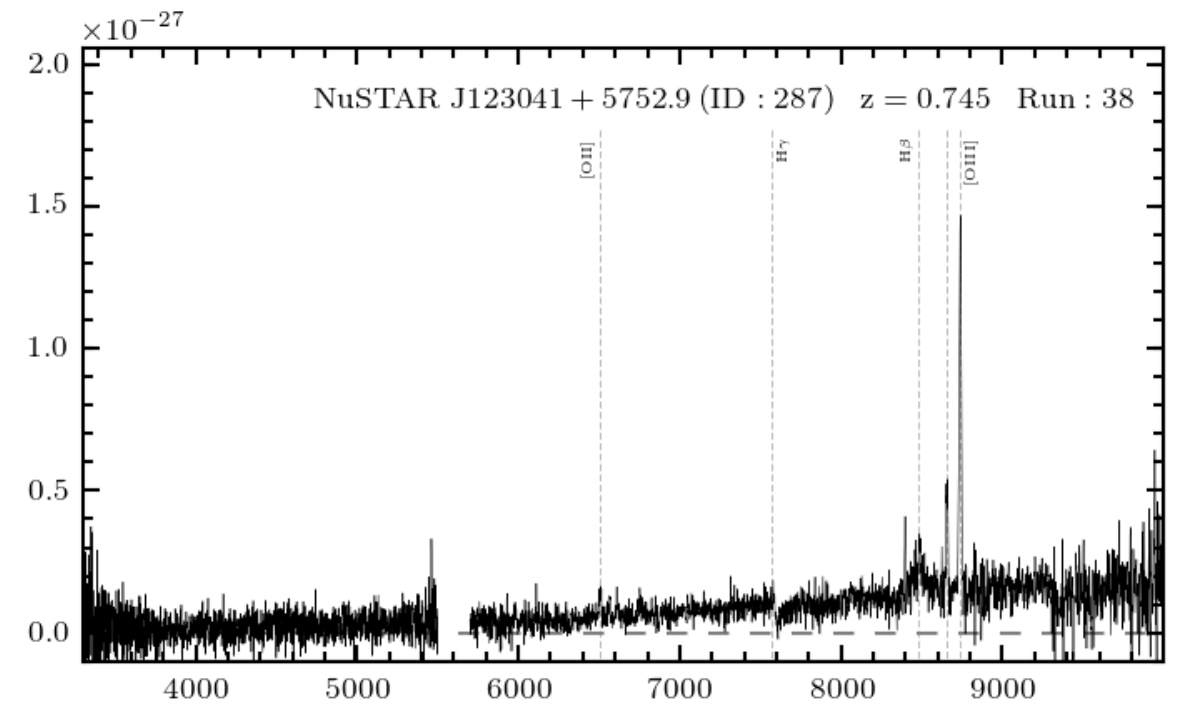}
\end{minipage}
\begin{minipage}[l]{0.325\textwidth}
\includegraphics[width=\textwidth]{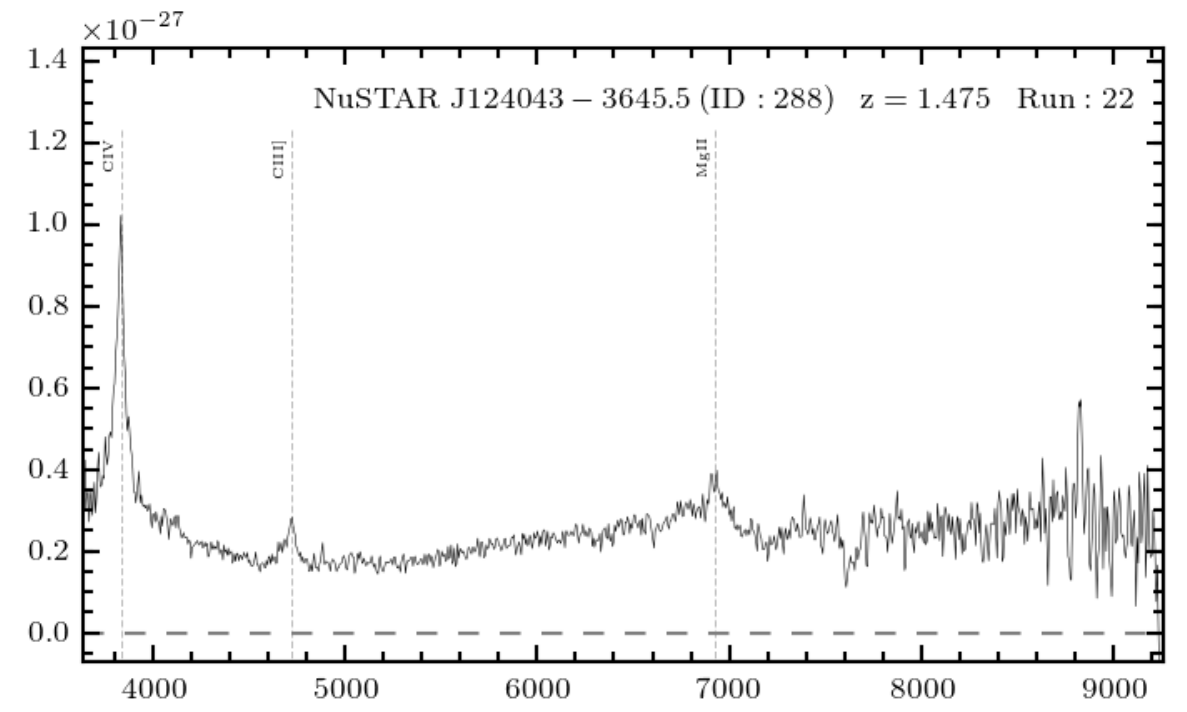}
\end{minipage}
\caption{Continued.}
\end{figure*}
\addtocounter{figure}{-1}
\begin{figure*}
\centering
\begin{minipage}[l]{0.325\textwidth}
\includegraphics[width=\textwidth]{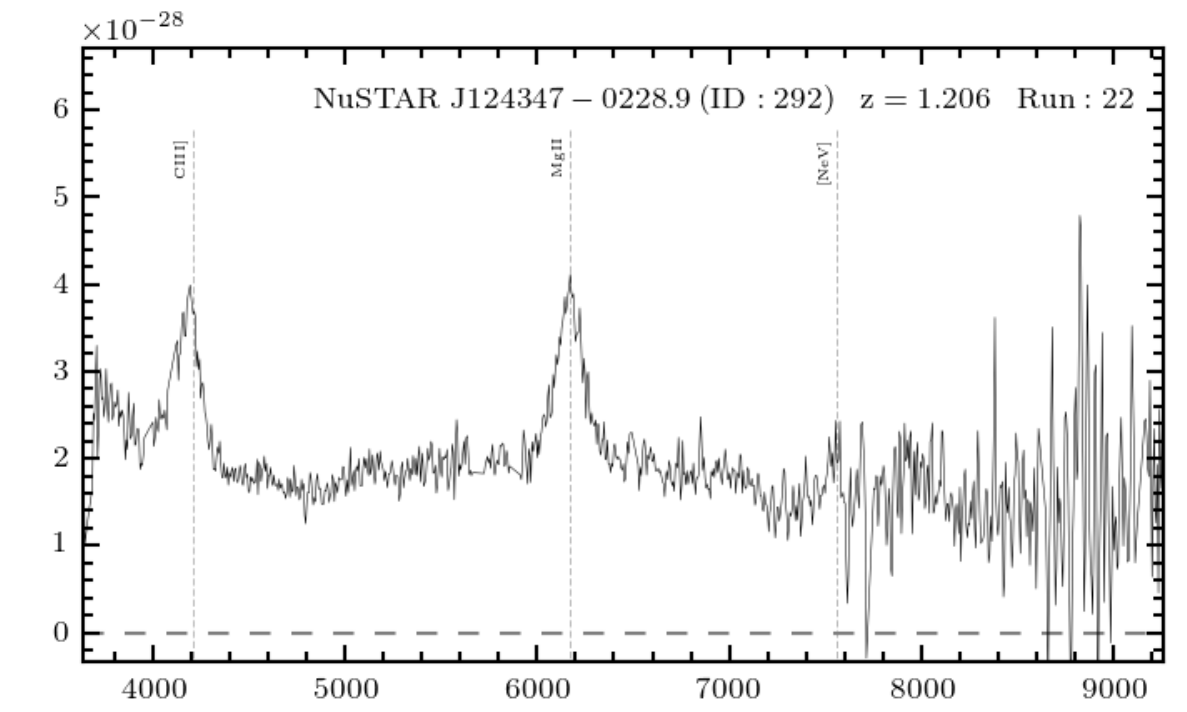}
\end{minipage}
\begin{minipage}[l]{0.325\textwidth}
\includegraphics[width=\textwidth]{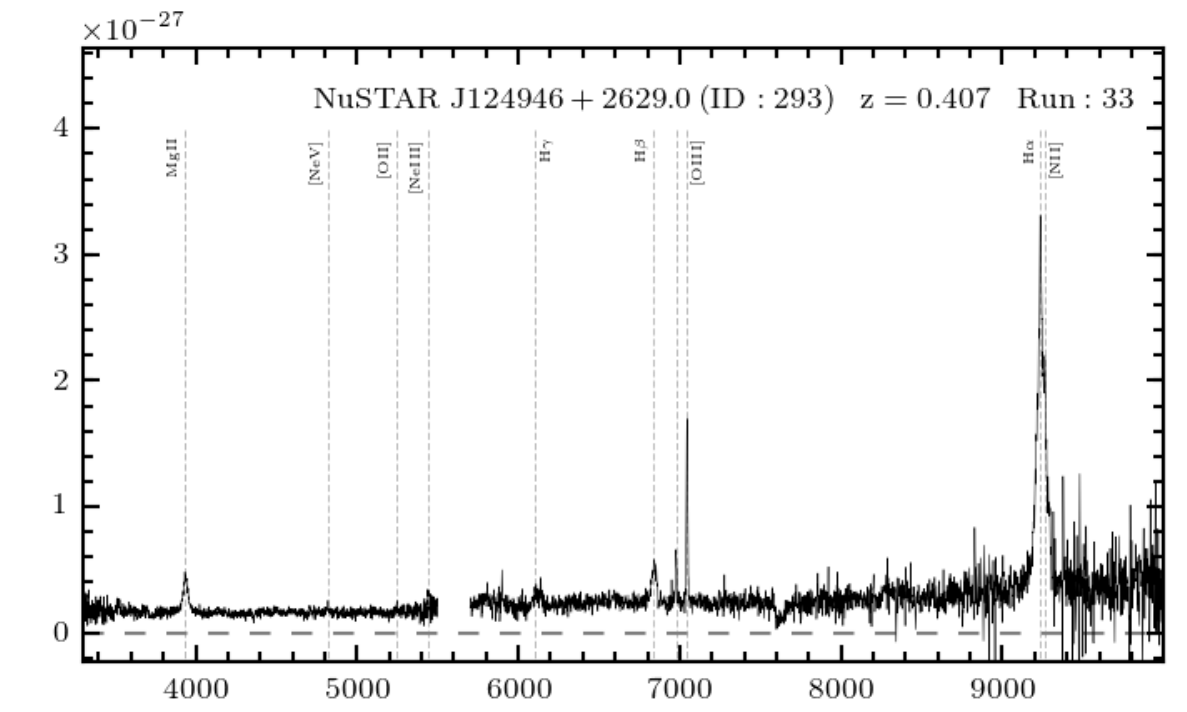}
\end{minipage}
\begin{minipage}[l]{0.325\textwidth}
\includegraphics[width=\textwidth]{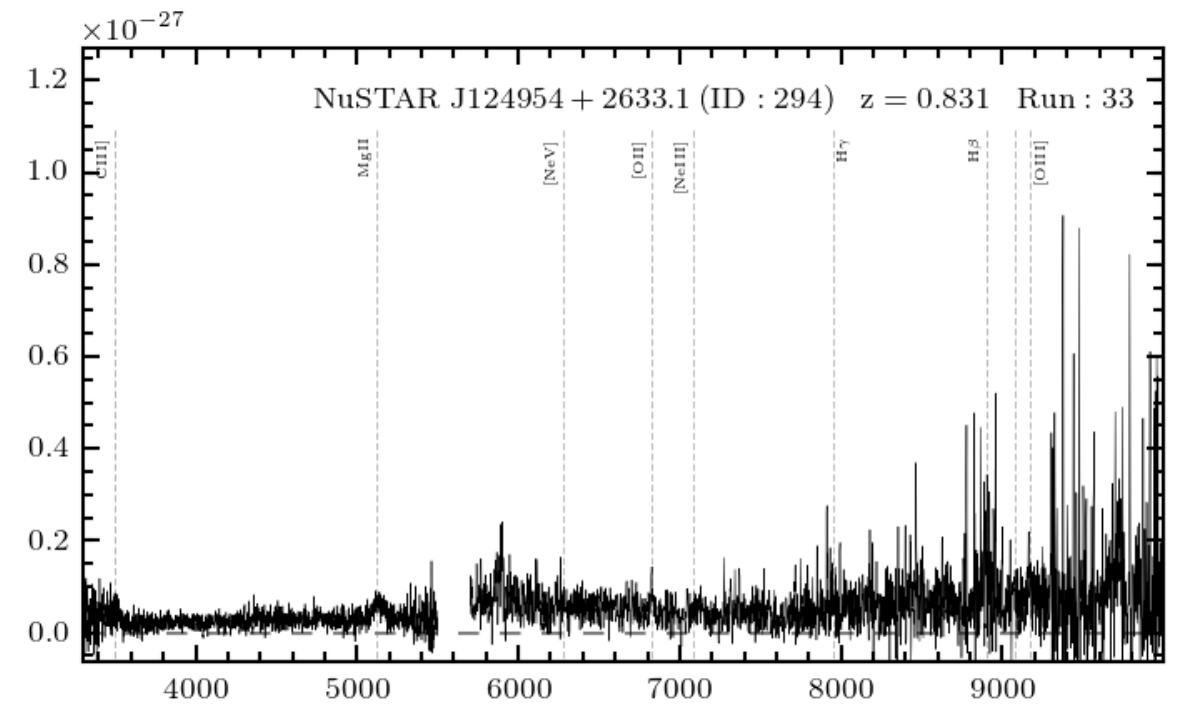}
\end{minipage}
\begin{minipage}[l]{0.325\textwidth}
\includegraphics[width=\textwidth]{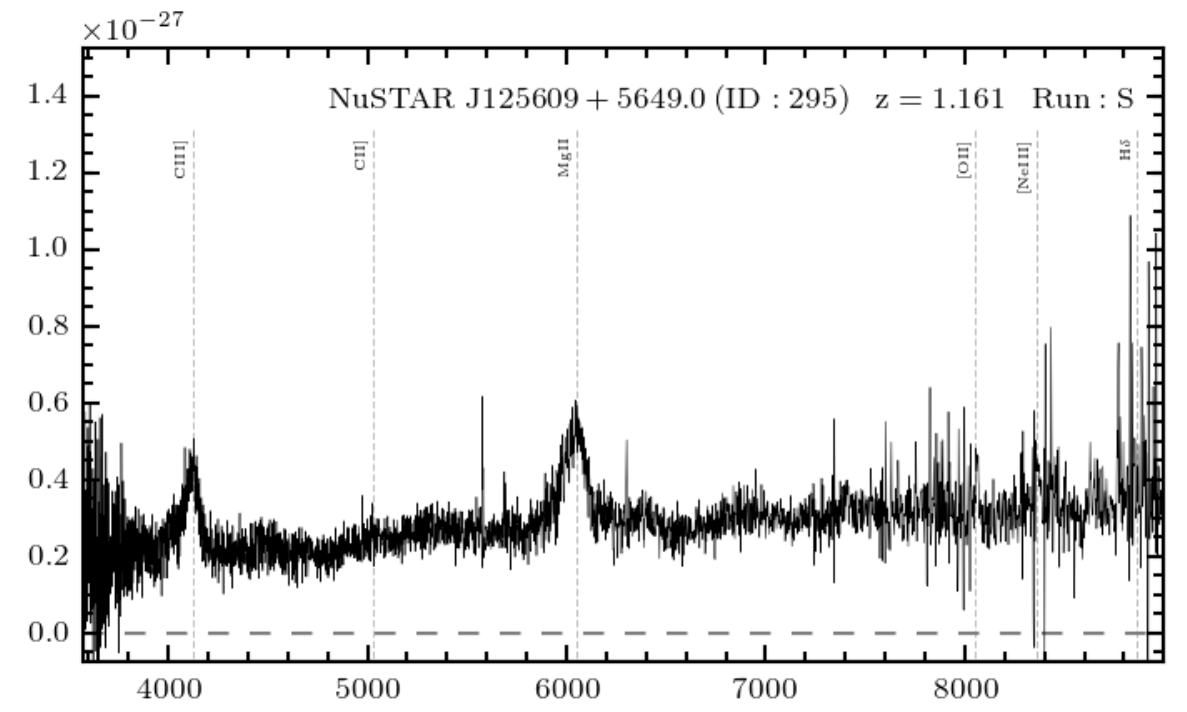}
\end{minipage}
\begin{minipage}[l]{0.325\textwidth}
\includegraphics[width=\textwidth]{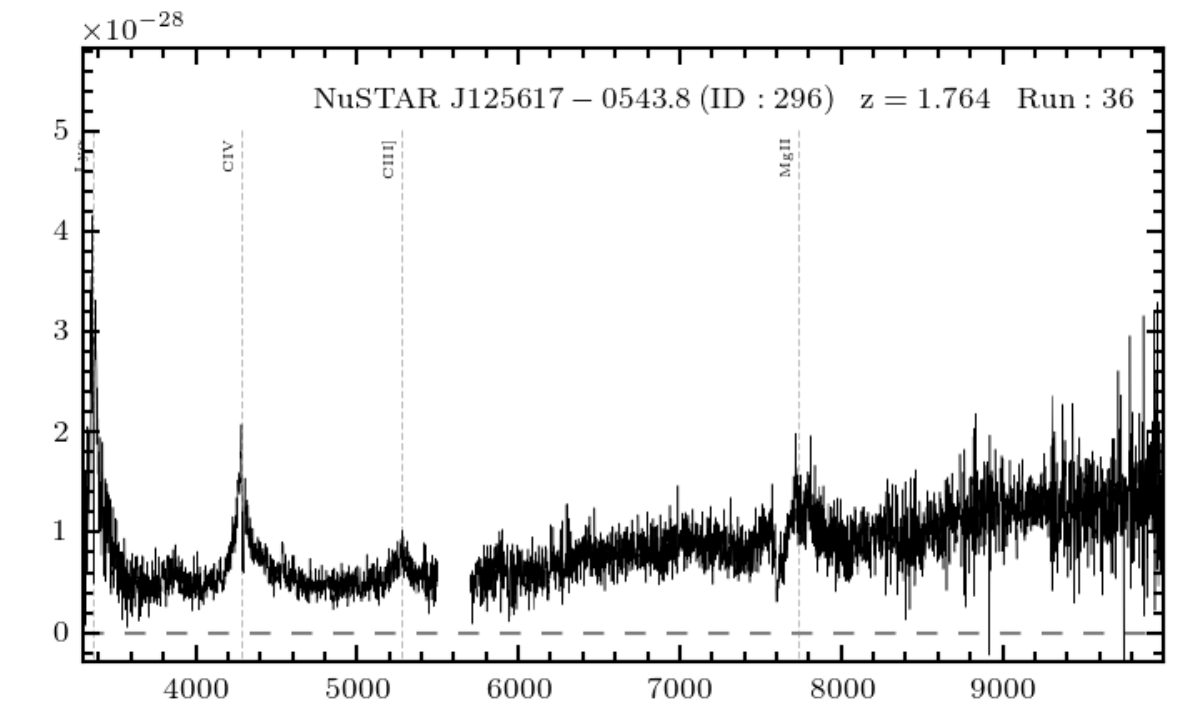}
\end{minipage}
\begin{minipage}[l]{0.325\textwidth}
\includegraphics[width=\textwidth]{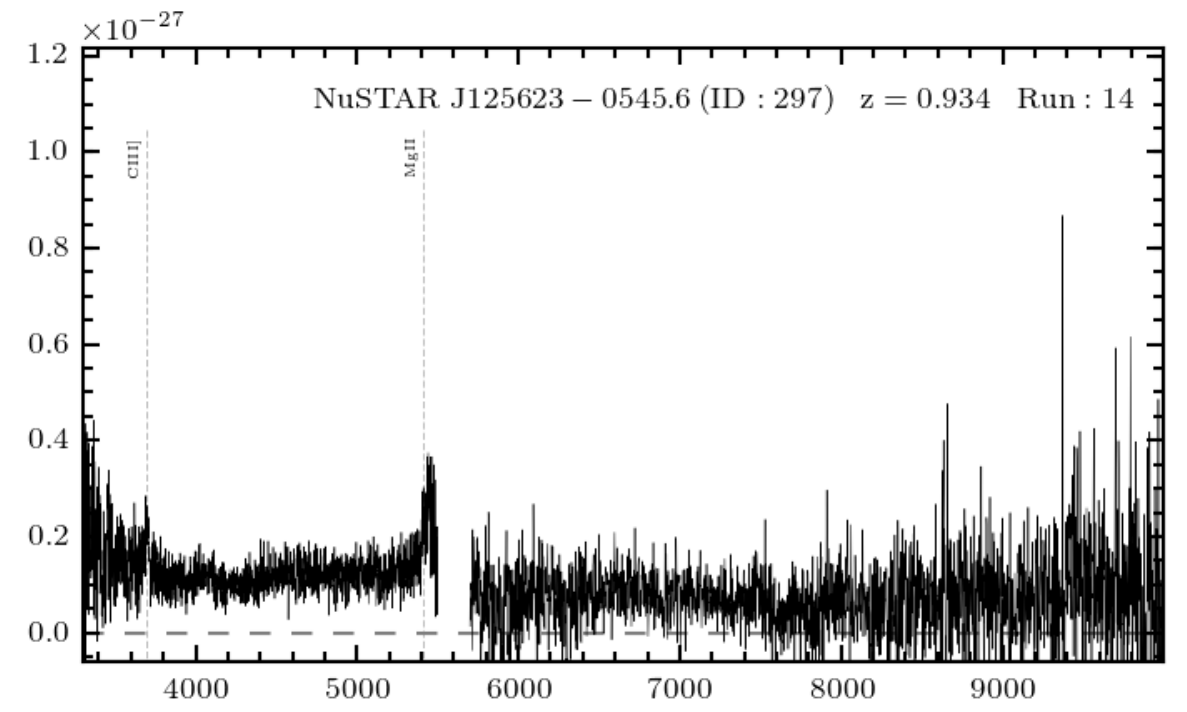}
\end{minipage}
\begin{minipage}[l]{0.325\textwidth}
\includegraphics[width=\textwidth]{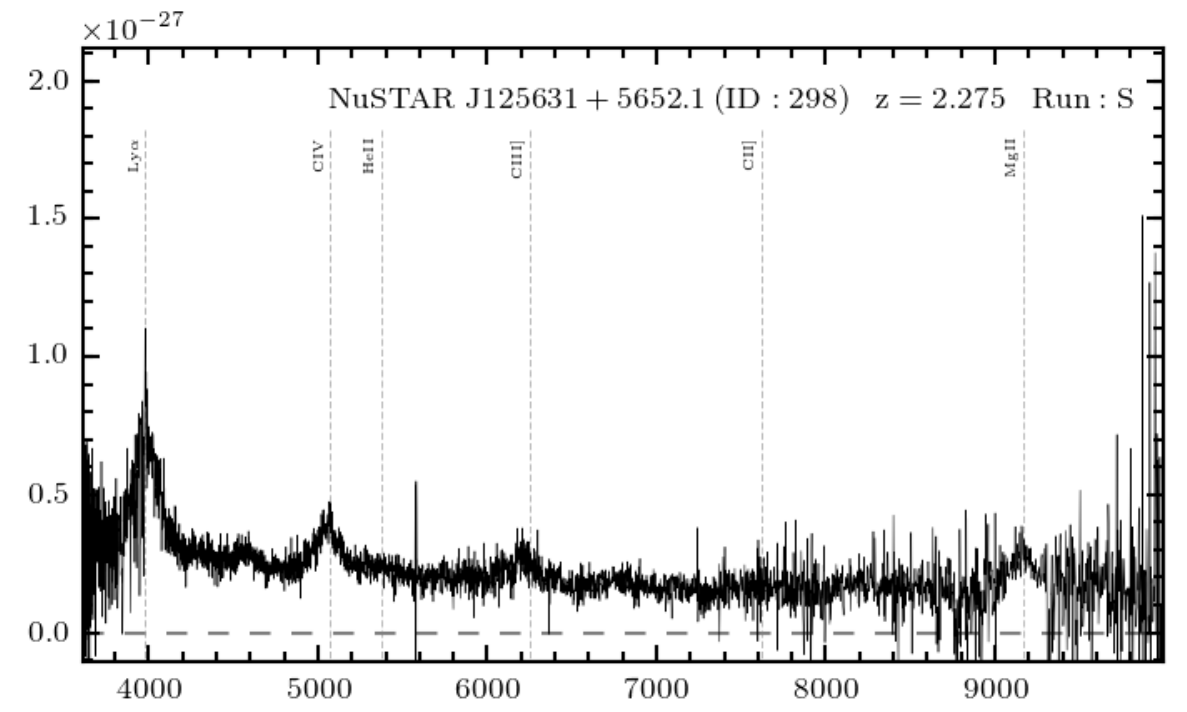}
\end{minipage}
\begin{minipage}[l]{0.325\textwidth}
\includegraphics[width=\textwidth]{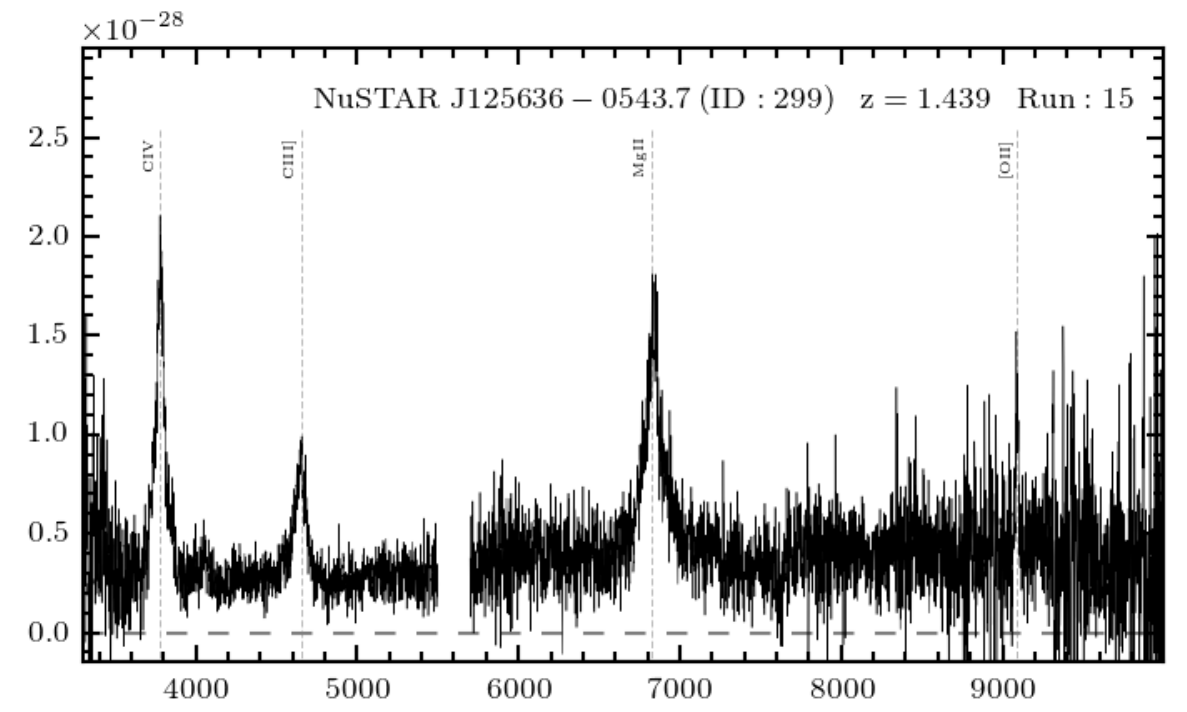}
\end{minipage}
\begin{minipage}[l]{0.325\textwidth}
\includegraphics[width=\textwidth]{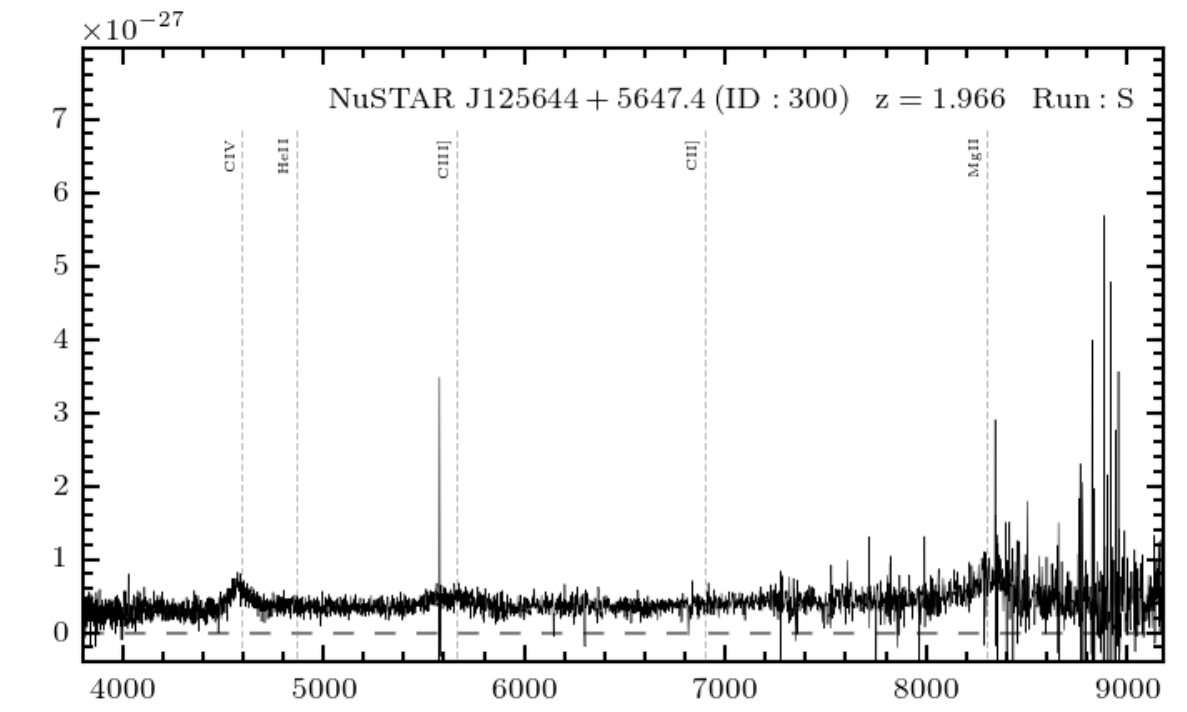}
\end{minipage}
\begin{minipage}[l]{0.325\textwidth}
\includegraphics[width=\textwidth]{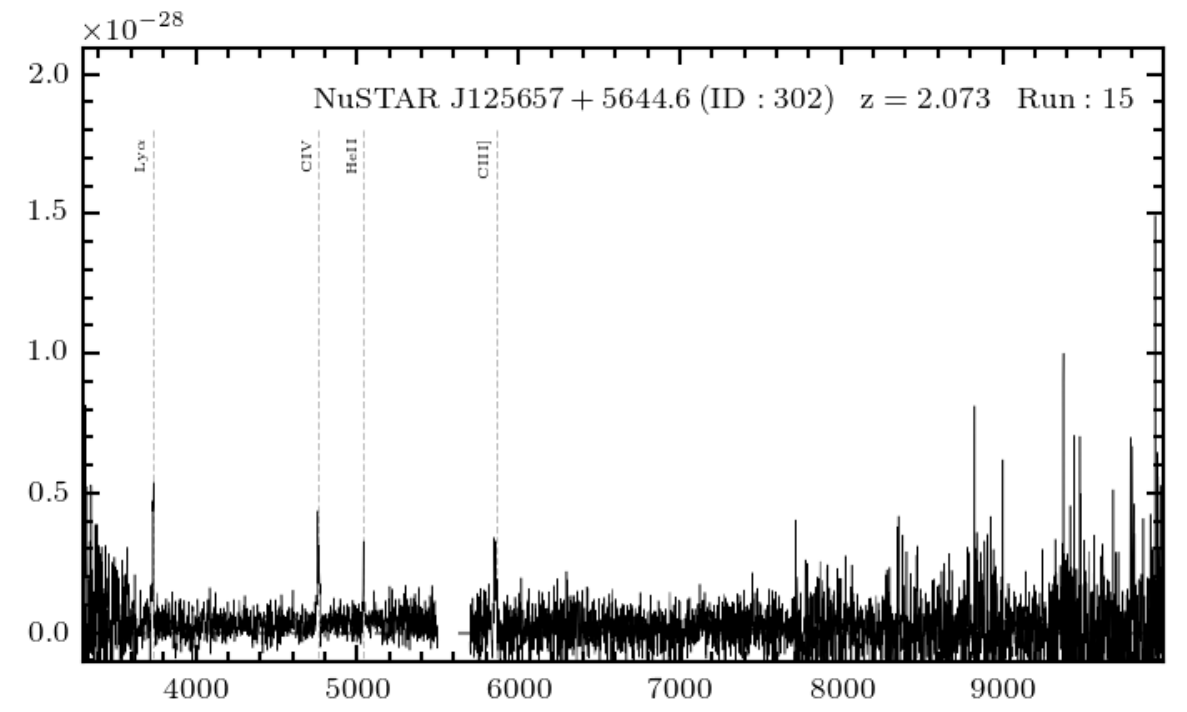}
\end{minipage}
\begin{minipage}[l]{0.325\textwidth}
\includegraphics[width=\textwidth]{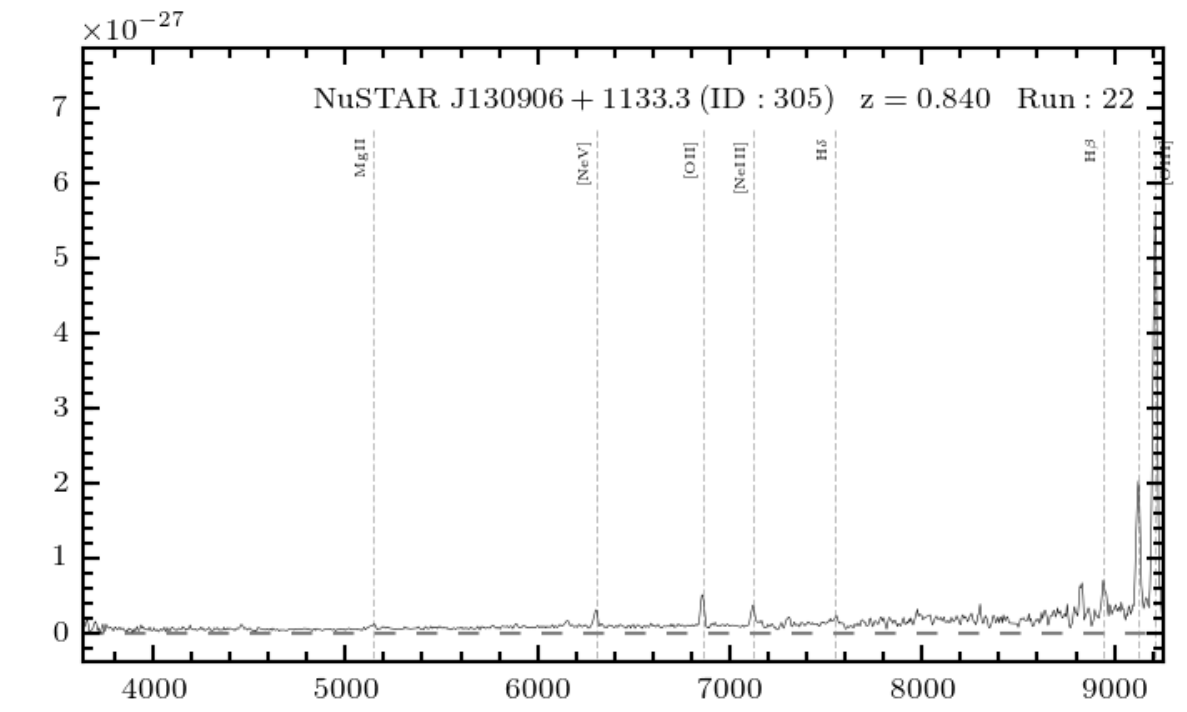}
\end{minipage}
\begin{minipage}[l]{0.325\textwidth}
\includegraphics[width=\textwidth]{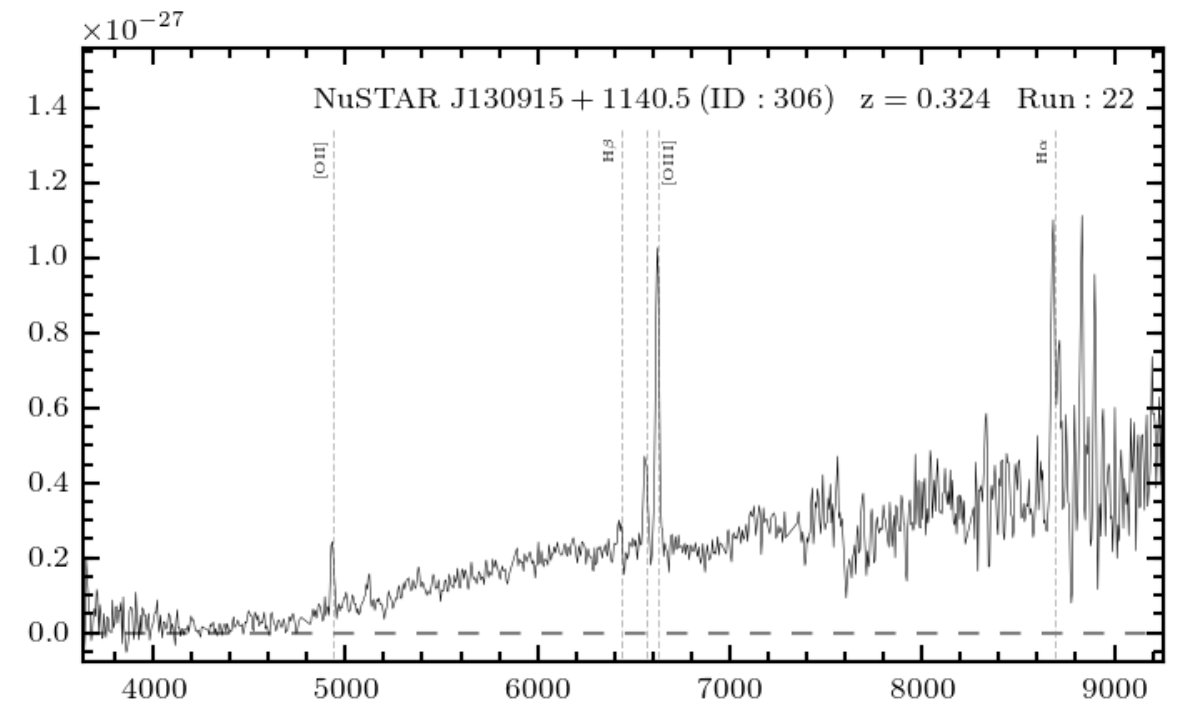}
\end{minipage}
\begin{minipage}[l]{0.325\textwidth}
\includegraphics[width=\textwidth]{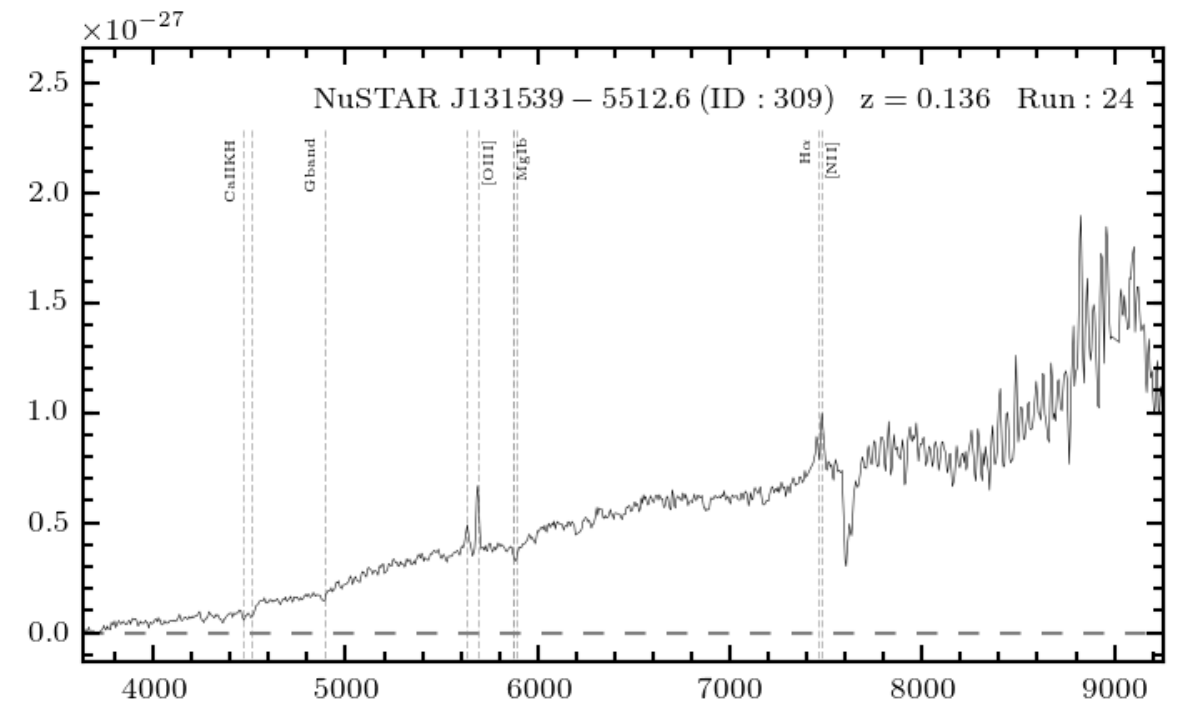}
\end{minipage}
\begin{minipage}[l]{0.325\textwidth}
\includegraphics[width=\textwidth]{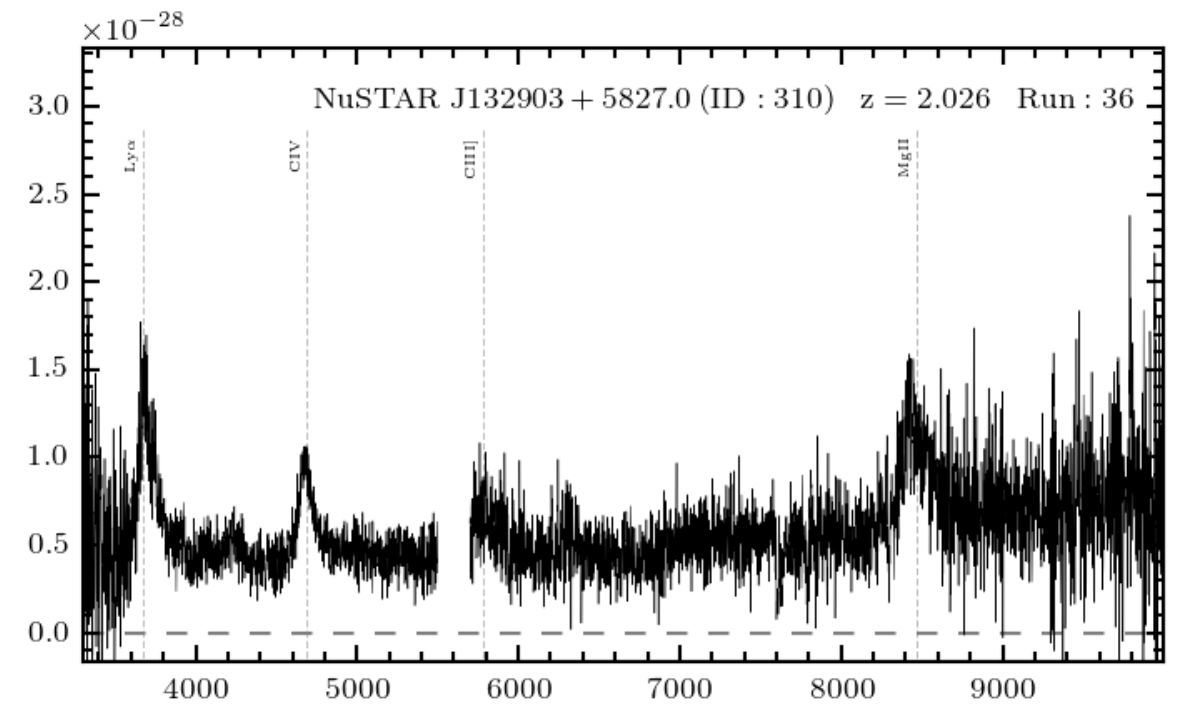}
\end{minipage}
\begin{minipage}[l]{0.325\textwidth}
\includegraphics[width=\textwidth]{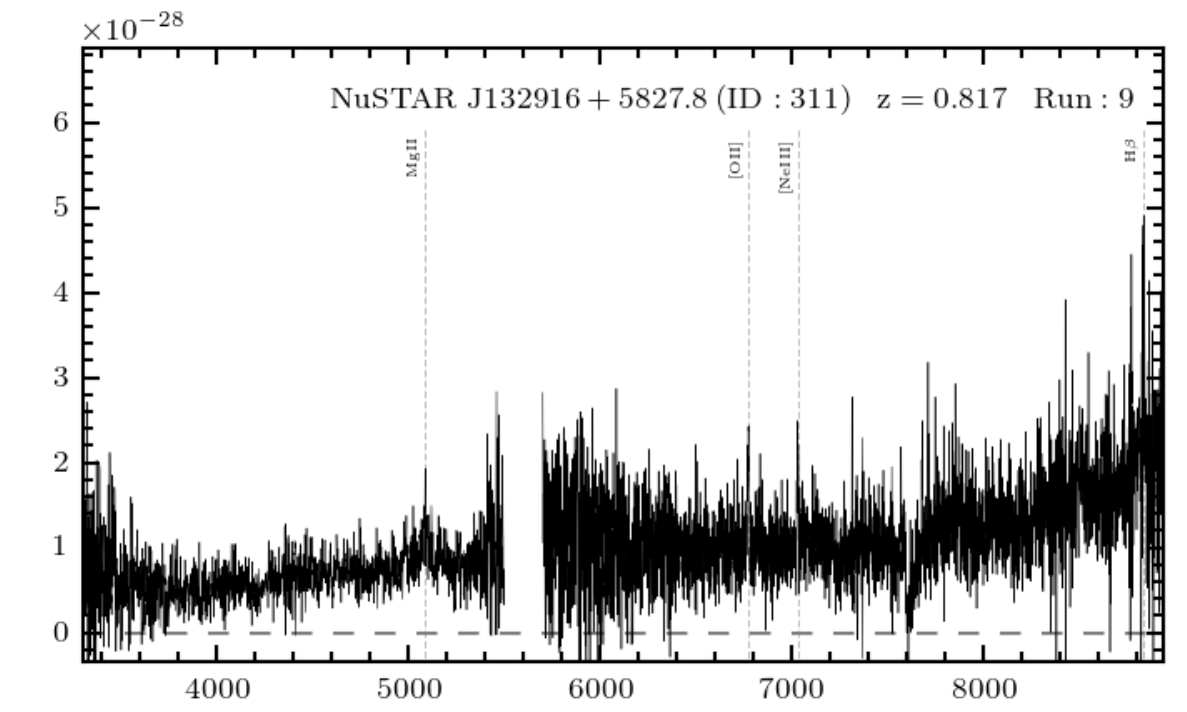}
\end{minipage}
\begin{minipage}[l]{0.325\textwidth}
\includegraphics[width=\textwidth]{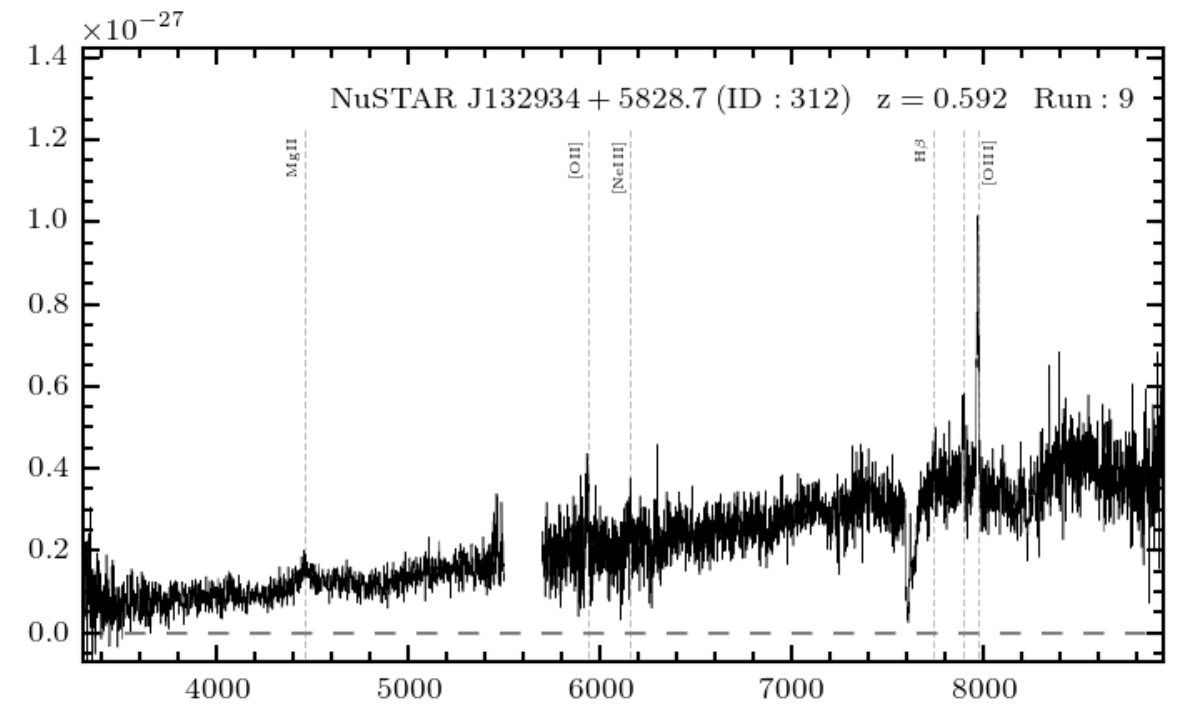}
\end{minipage}
\begin{minipage}[l]{0.325\textwidth}
\includegraphics[width=\textwidth]{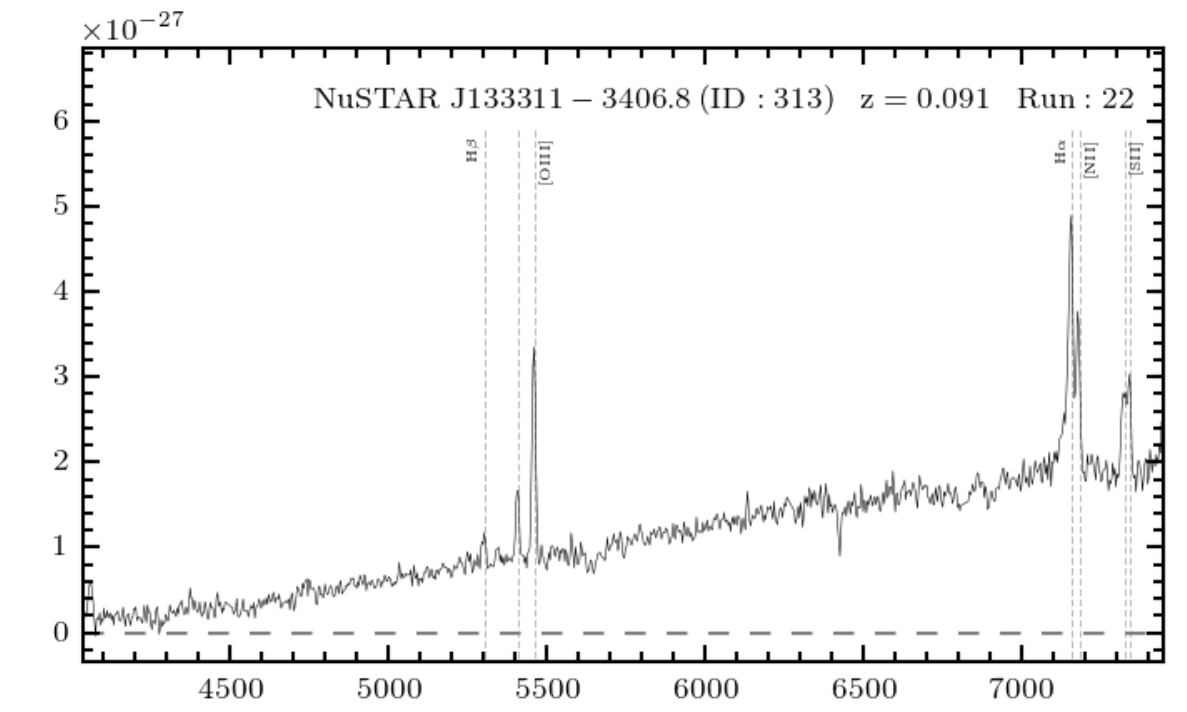}
\end{minipage}
\begin{minipage}[l]{0.325\textwidth}
\includegraphics[width=\textwidth]{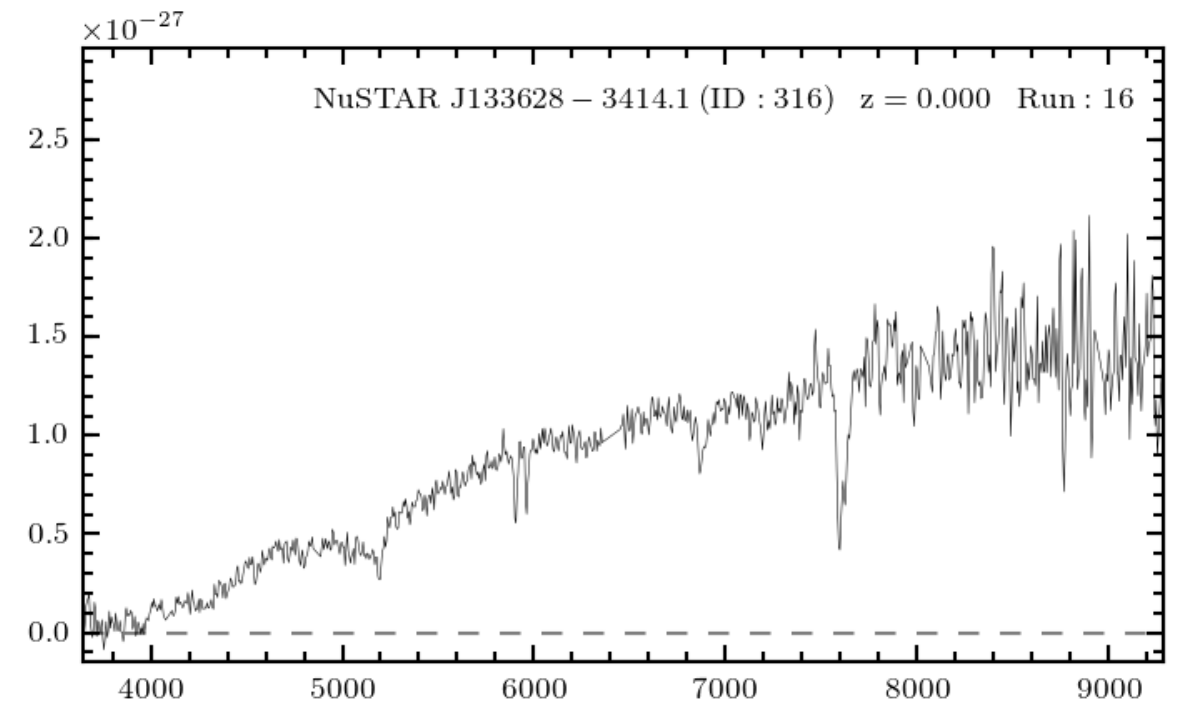}
\end{minipage}
\caption{Continued.}
\end{figure*}
\addtocounter{figure}{-1}
\begin{figure*}
\centering
\begin{minipage}[l]{0.325\textwidth}
\includegraphics[width=\textwidth]{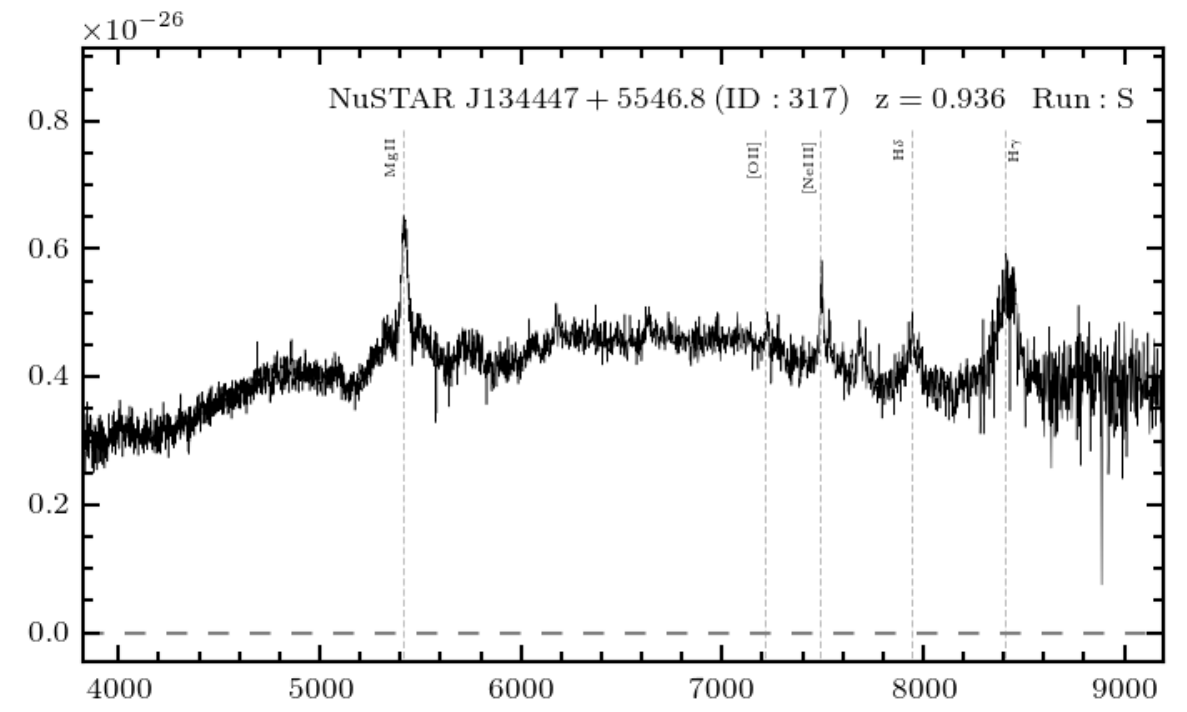}
\end{minipage}
\begin{minipage}[l]{0.325\textwidth}
\includegraphics[width=\textwidth]{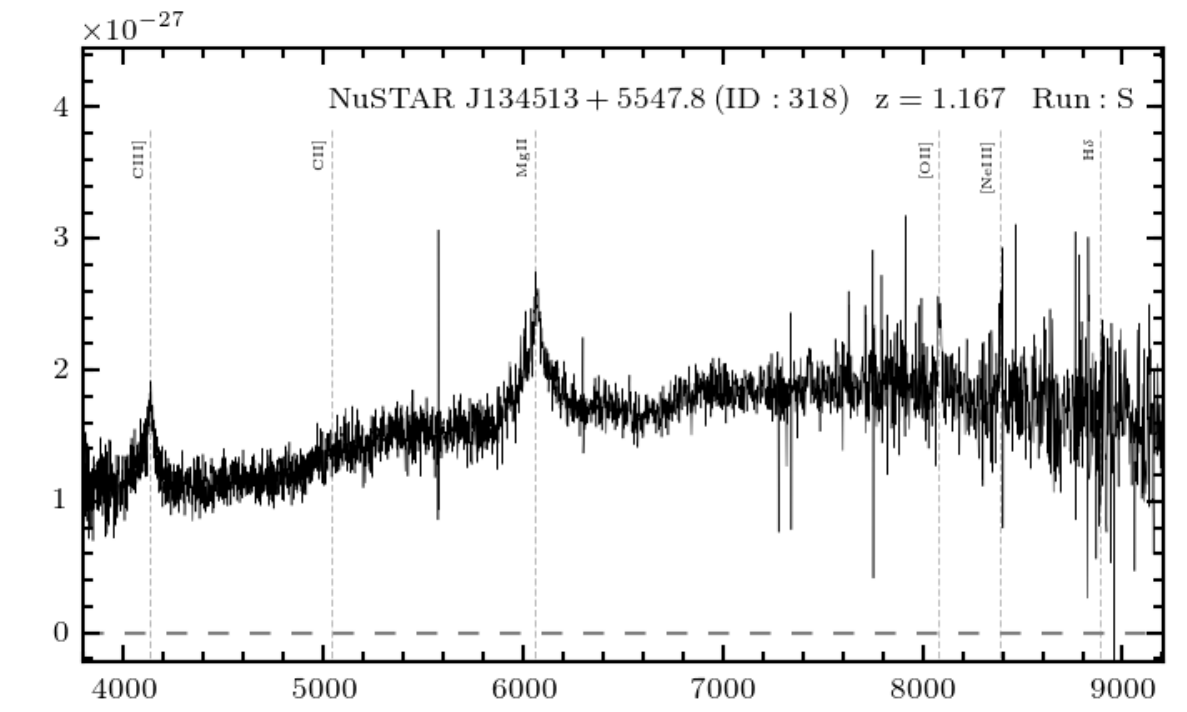}
\end{minipage}
\begin{minipage}[l]{0.325\textwidth}
\includegraphics[width=\textwidth]{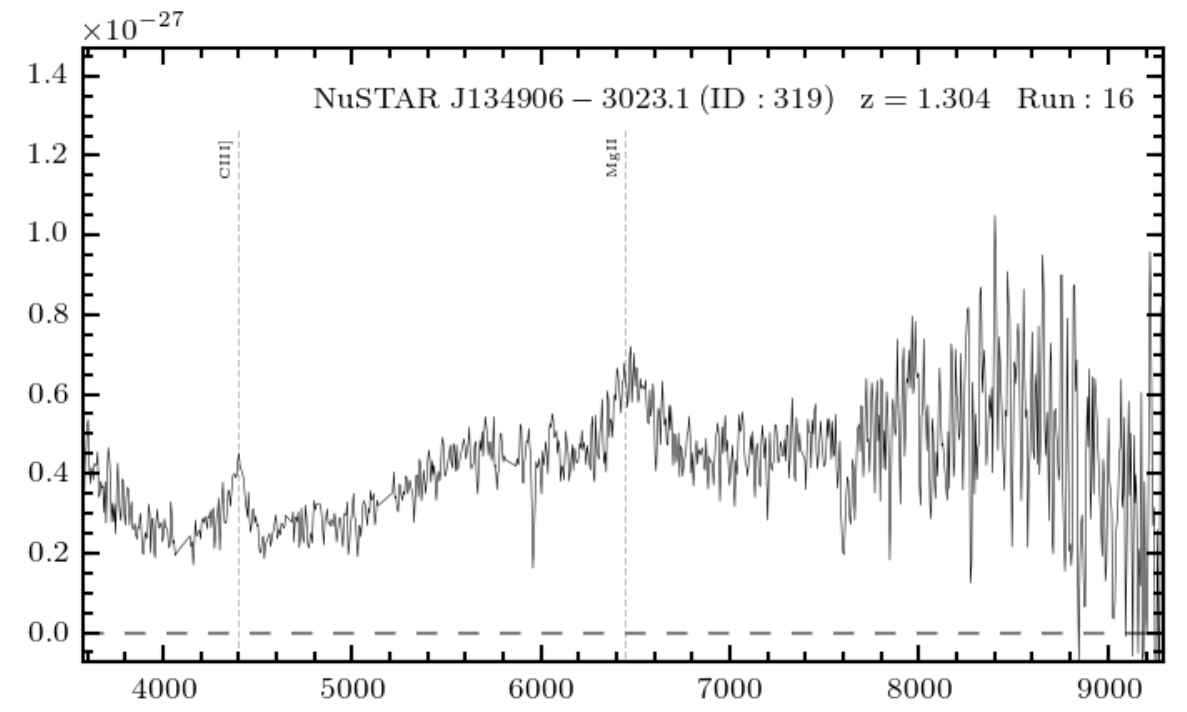}
\end{minipage}
\begin{minipage}[l]{0.325\textwidth}
\includegraphics[width=\textwidth]{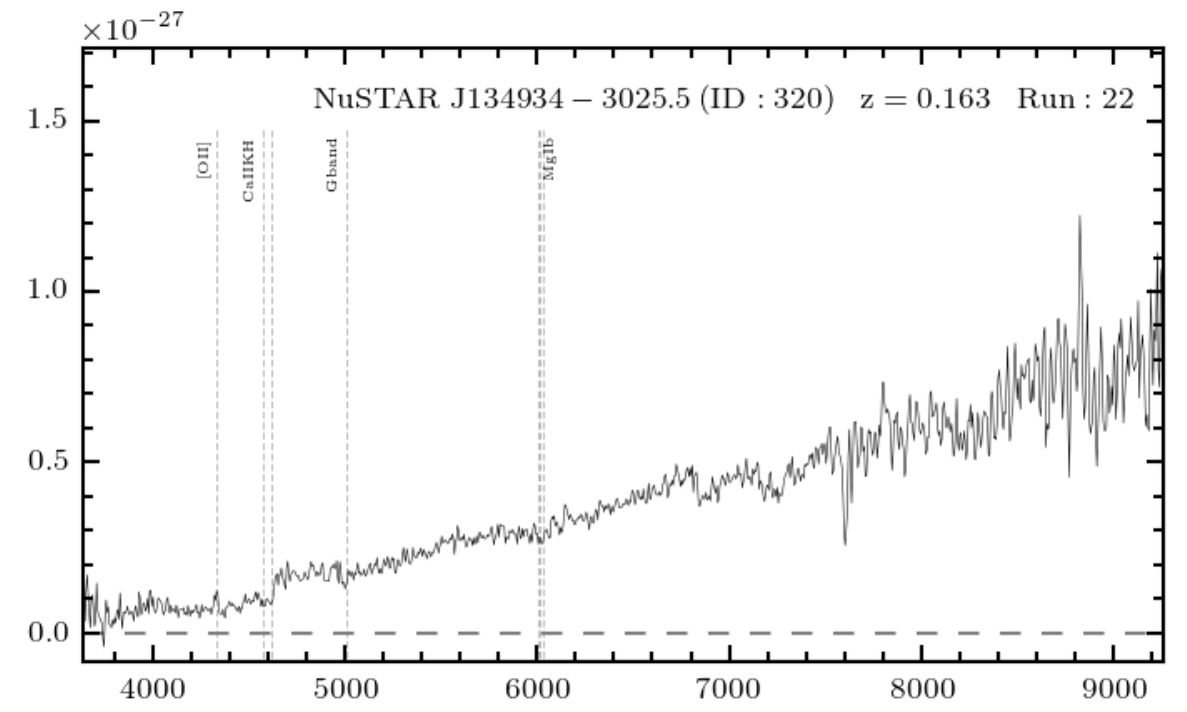}
\end{minipage}
\begin{minipage}[l]{0.325\textwidth}
\includegraphics[width=\textwidth]{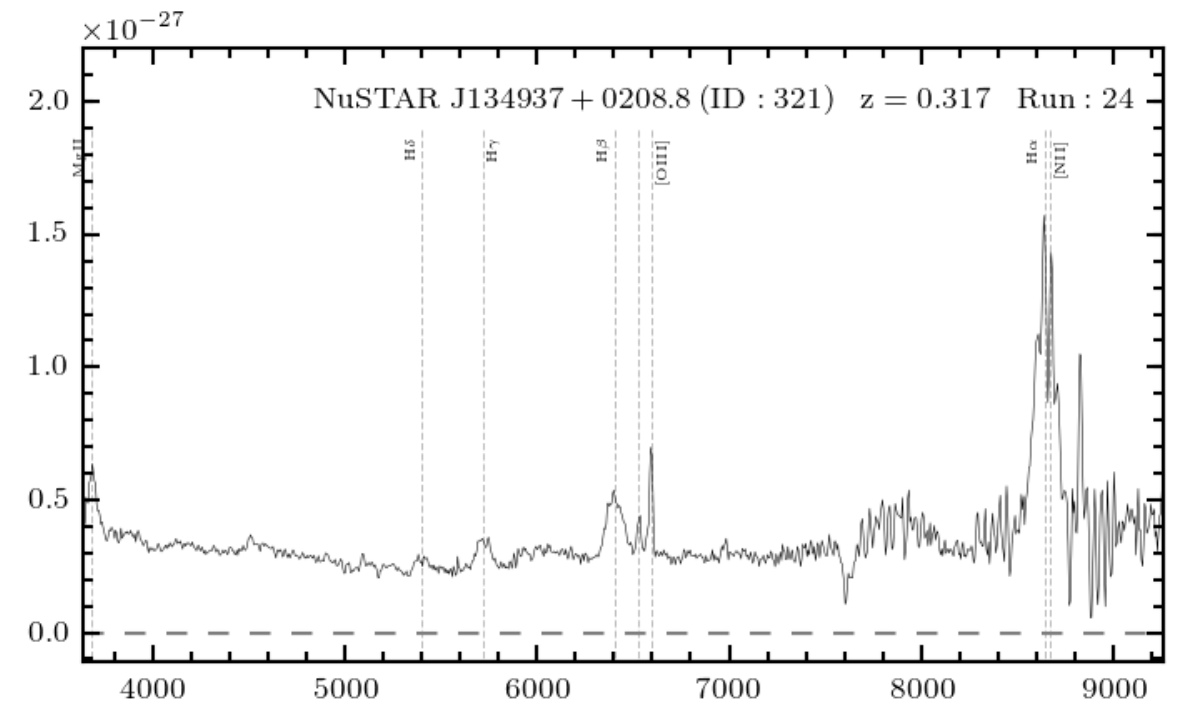}
\end{minipage}
\begin{minipage}[l]{0.325\textwidth}
\includegraphics[width=\textwidth]{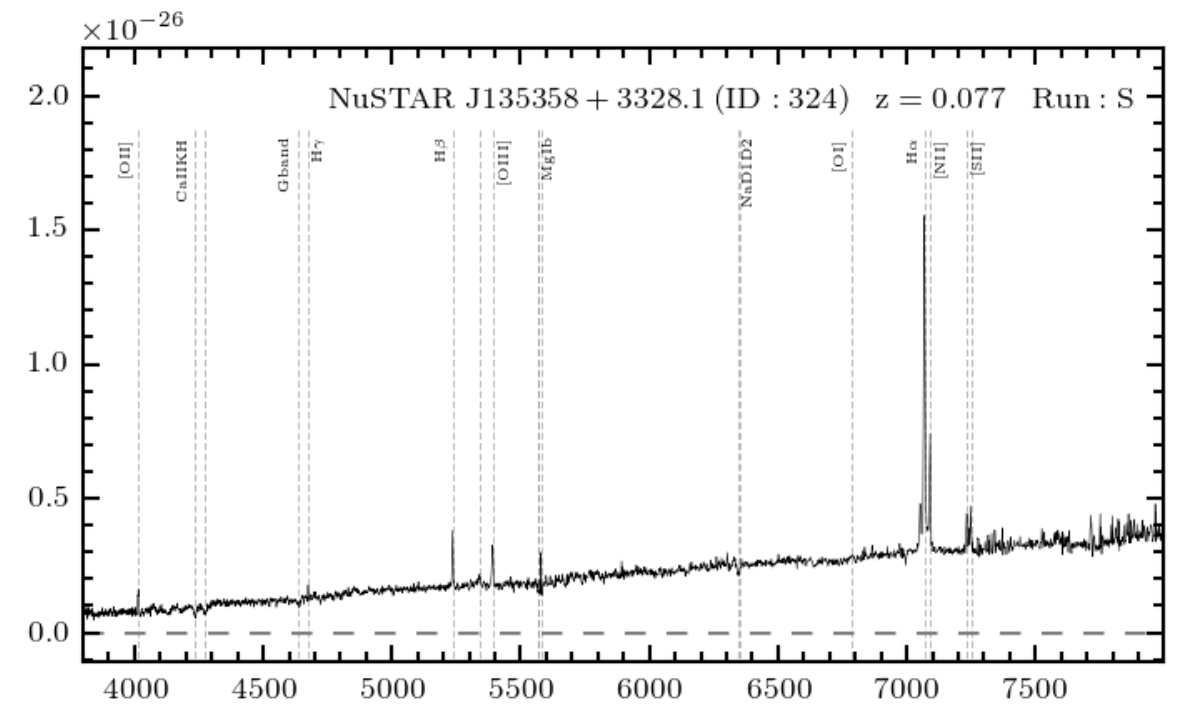}
\end{minipage}
\begin{minipage}[l]{0.325\textwidth}
\includegraphics[width=\textwidth]{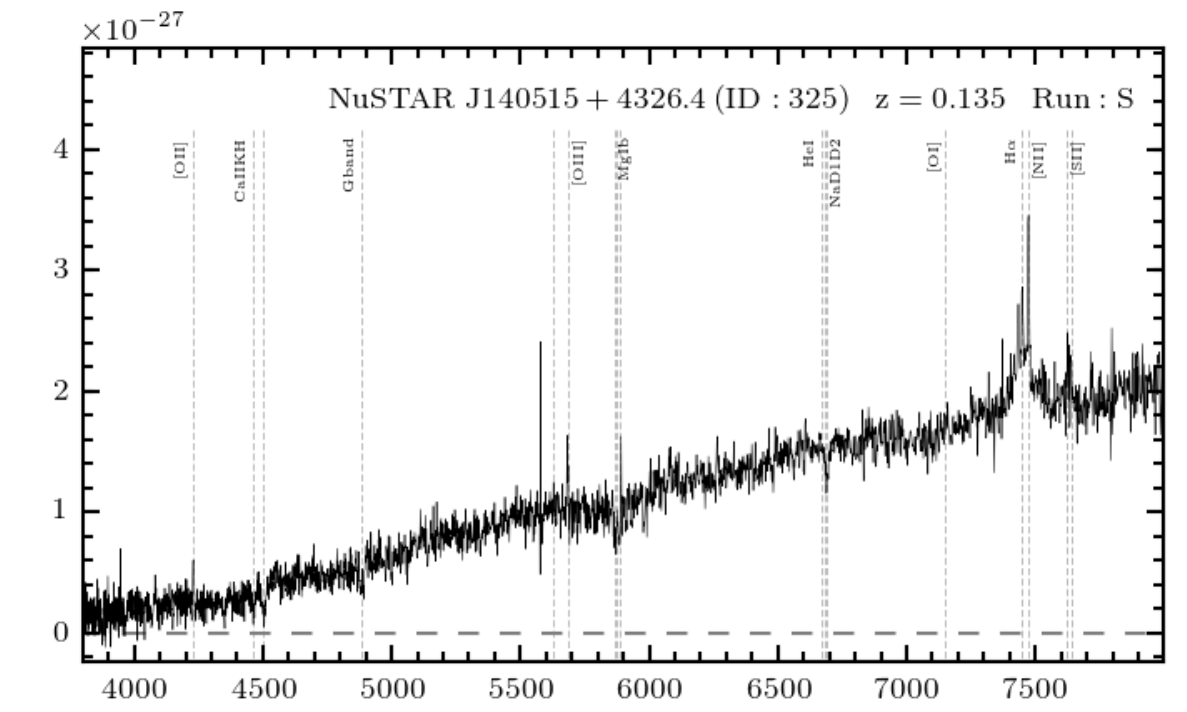}
\end{minipage}
\begin{minipage}[l]{0.325\textwidth}
\includegraphics[width=\textwidth]{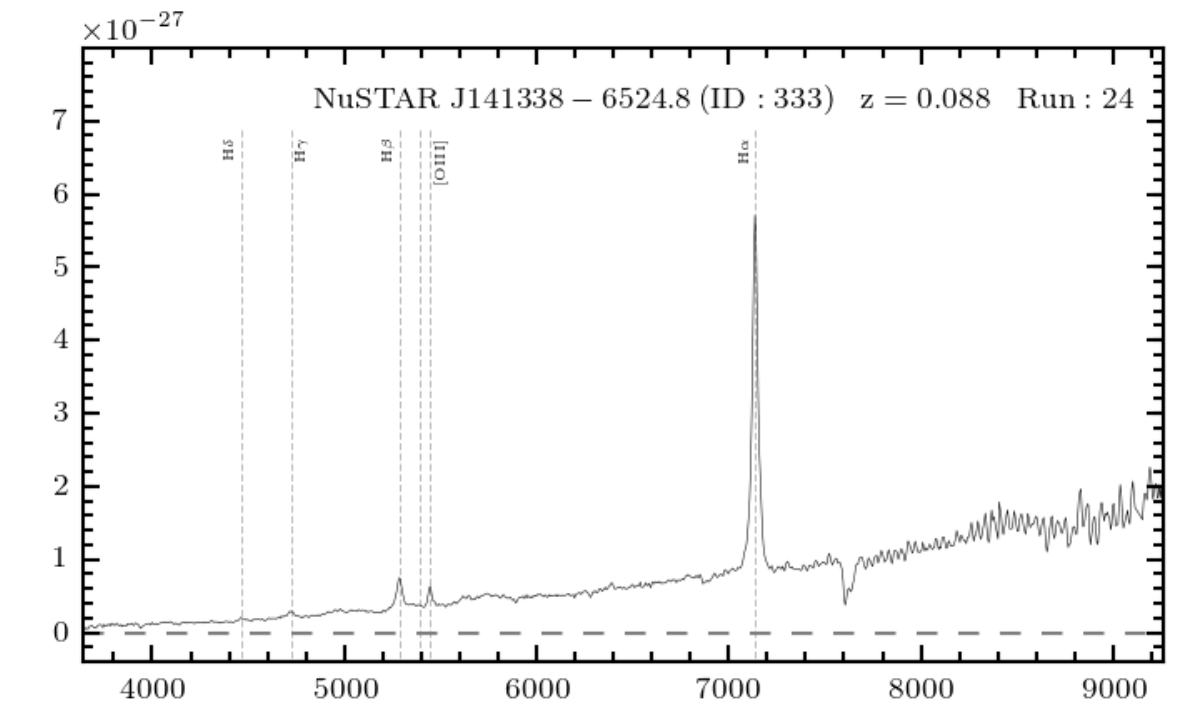}
\end{minipage}
\begin{minipage}[l]{0.325\textwidth}
\includegraphics[width=\textwidth]{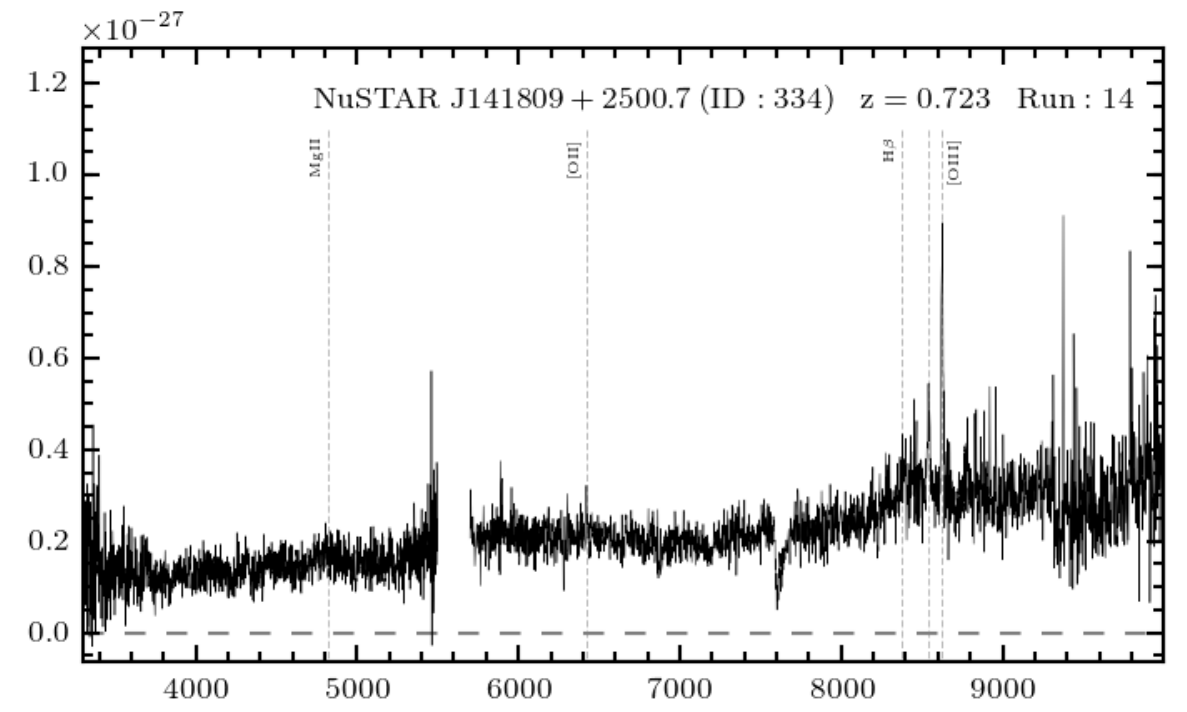}
\end{minipage}
\begin{minipage}[l]{0.325\textwidth}
\includegraphics[width=\textwidth]{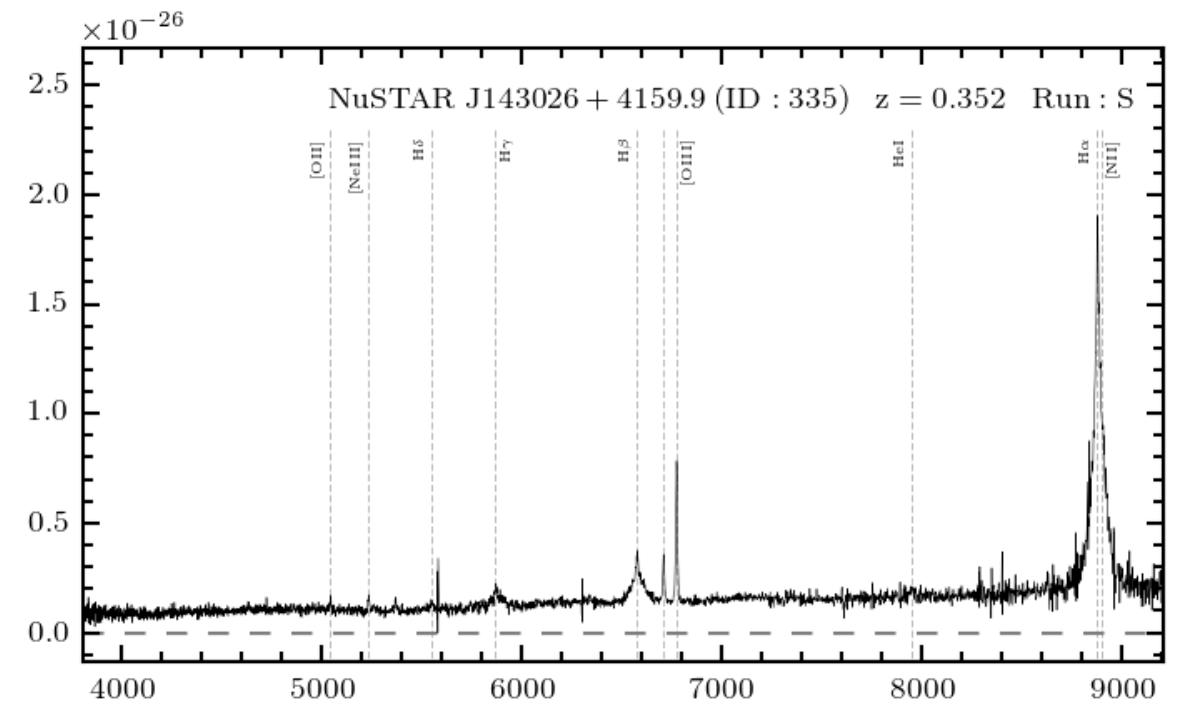}
\end{minipage}
\begin{minipage}[l]{0.325\textwidth}
\includegraphics[width=\textwidth]{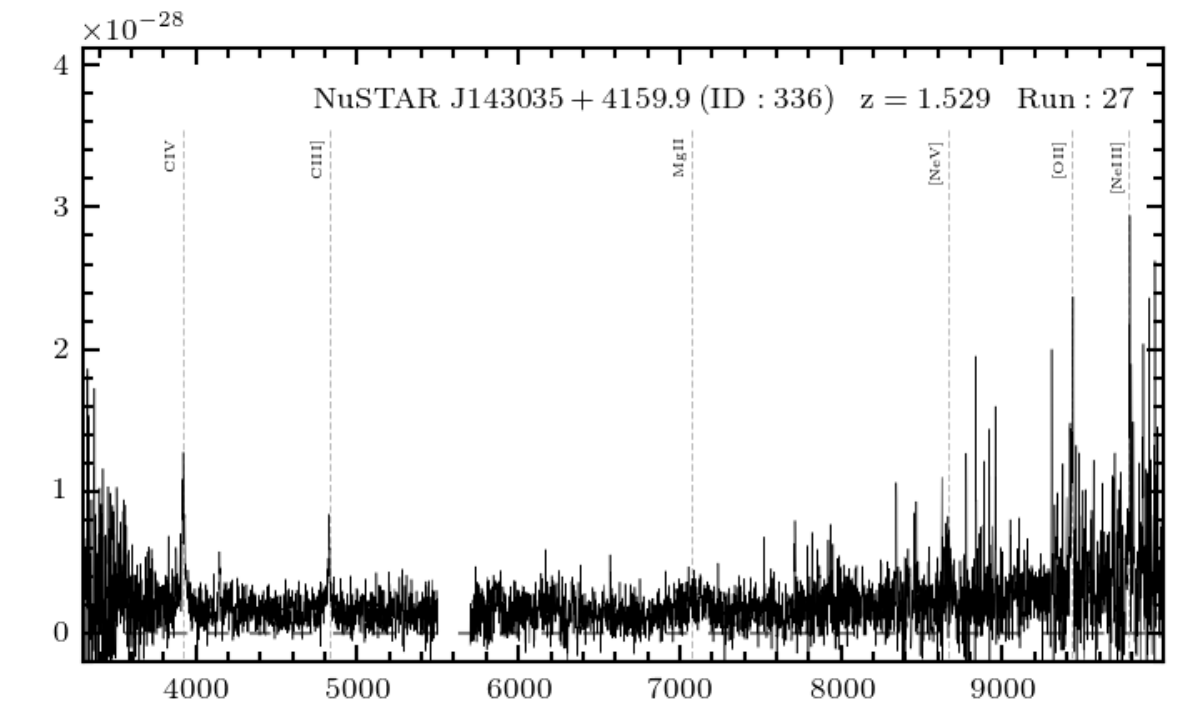}
\end{minipage}
\begin{minipage}[l]{0.325\textwidth}
\includegraphics[width=\textwidth]{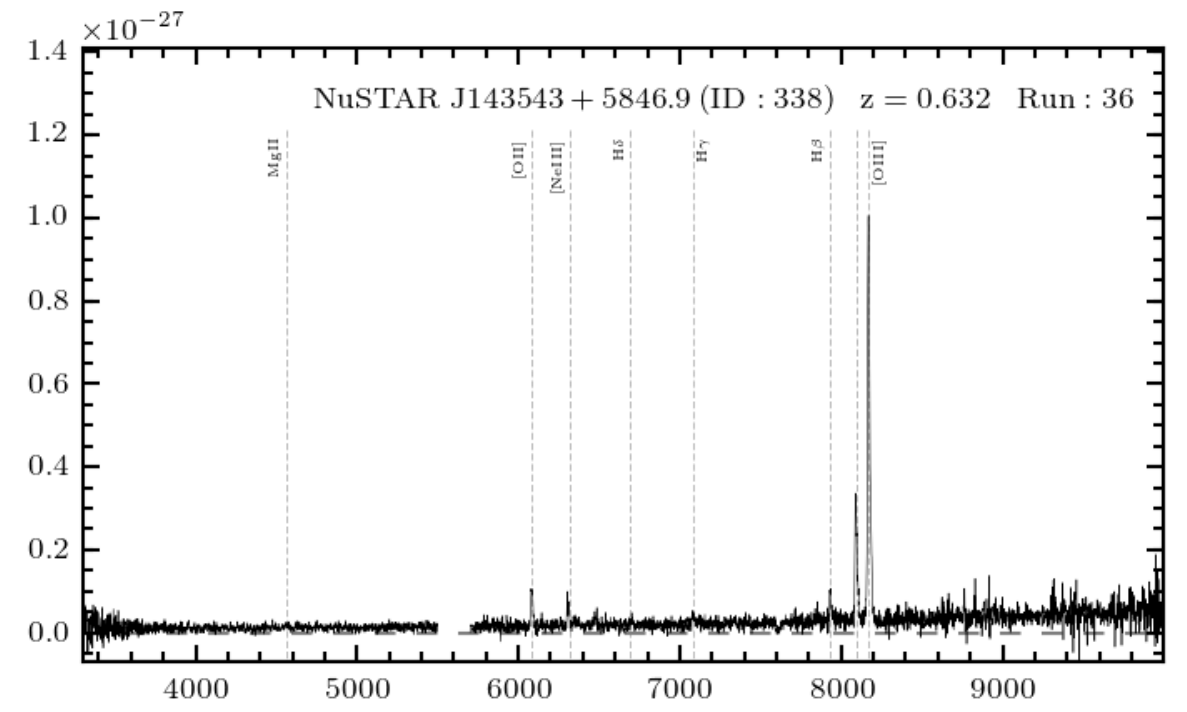}
\end{minipage}
\begin{minipage}[l]{0.325\textwidth}
\includegraphics[width=\textwidth]{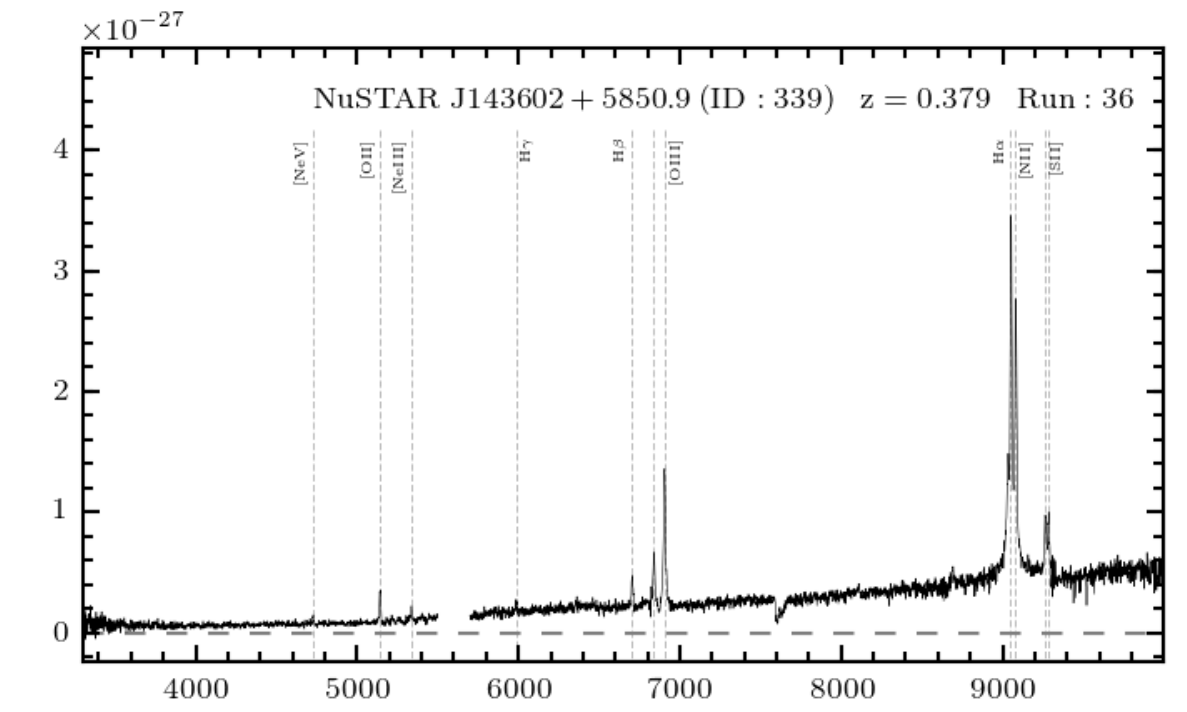}
\end{minipage}
\begin{minipage}[l]{0.325\textwidth}
\includegraphics[width=\textwidth]{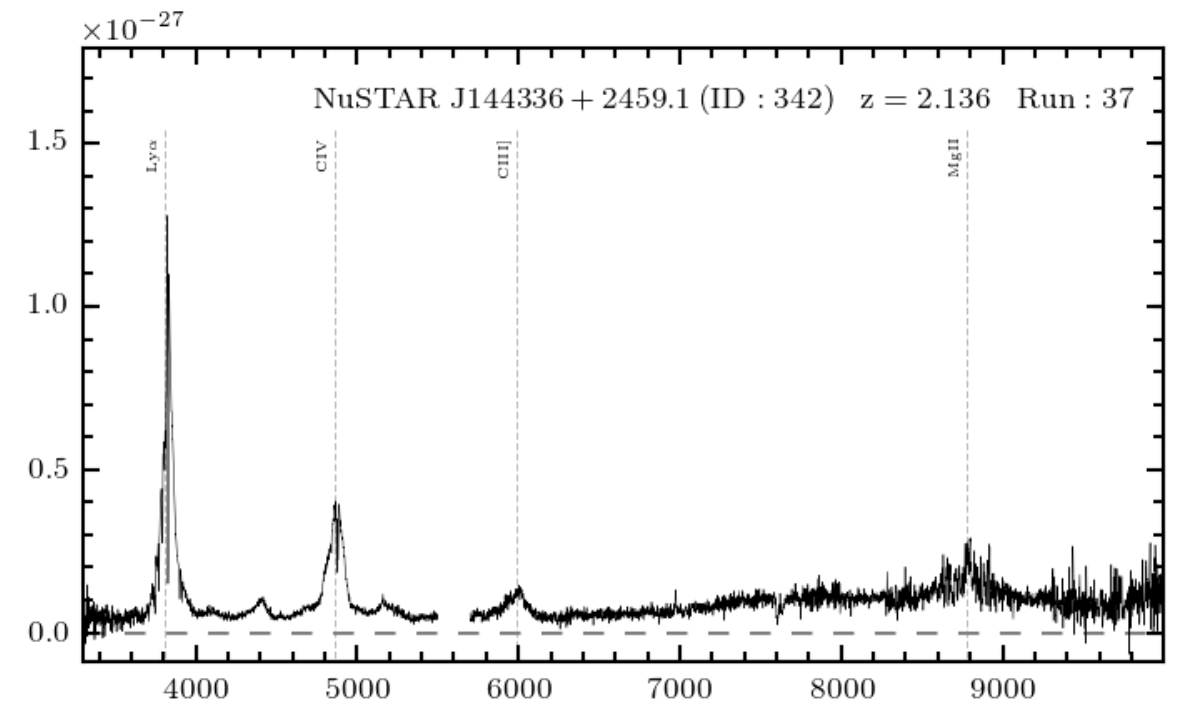}
\end{minipage}
\begin{minipage}[l]{0.325\textwidth}
\includegraphics[width=\textwidth]{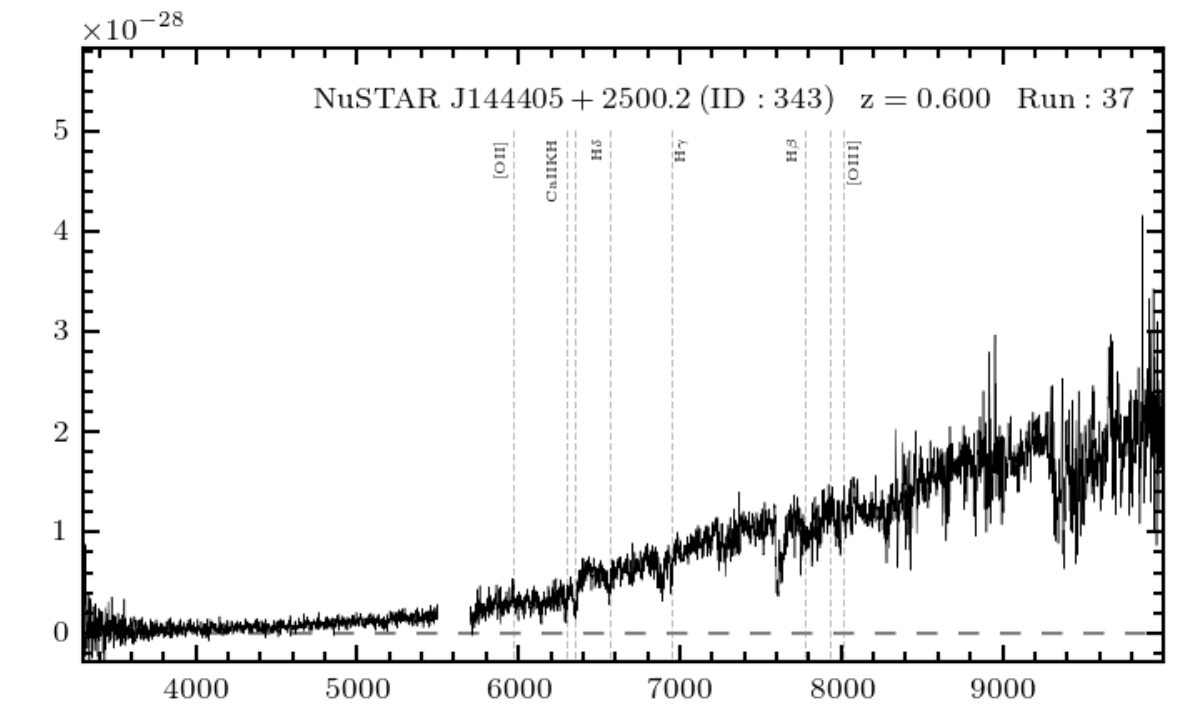}
\end{minipage}
\begin{minipage}[l]{0.325\textwidth}
\includegraphics[width=\textwidth]{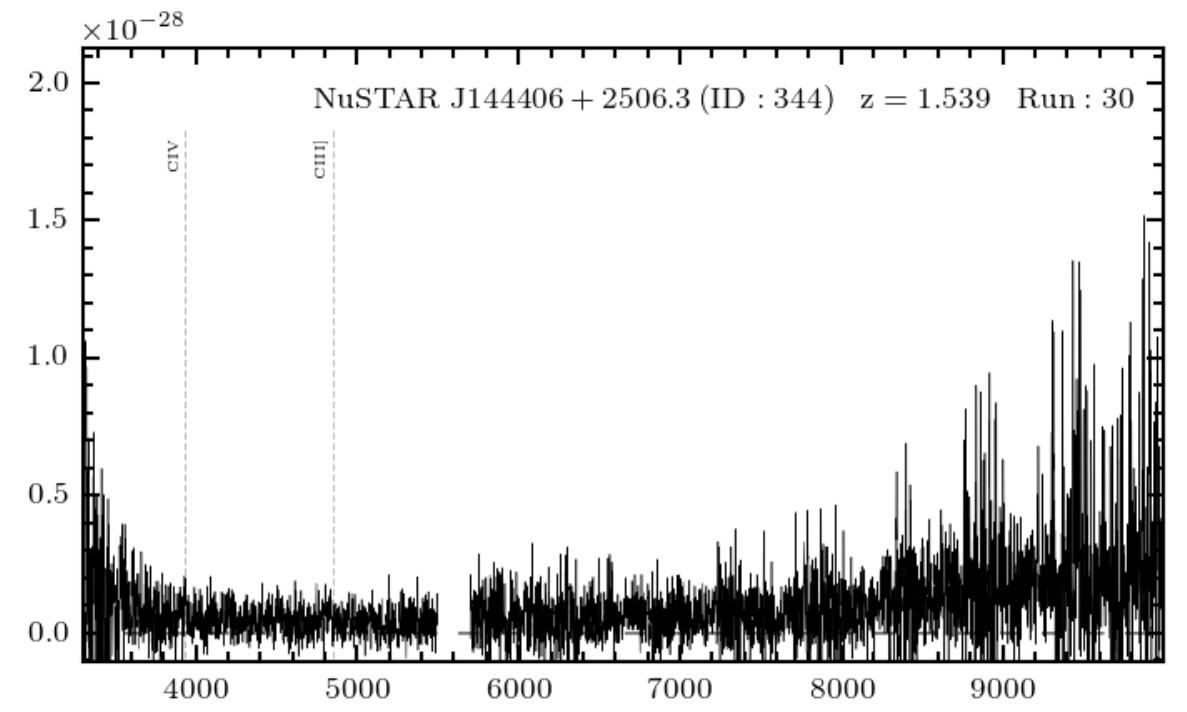}
\end{minipage}
\begin{minipage}[l]{0.325\textwidth}
\includegraphics[width=\textwidth]{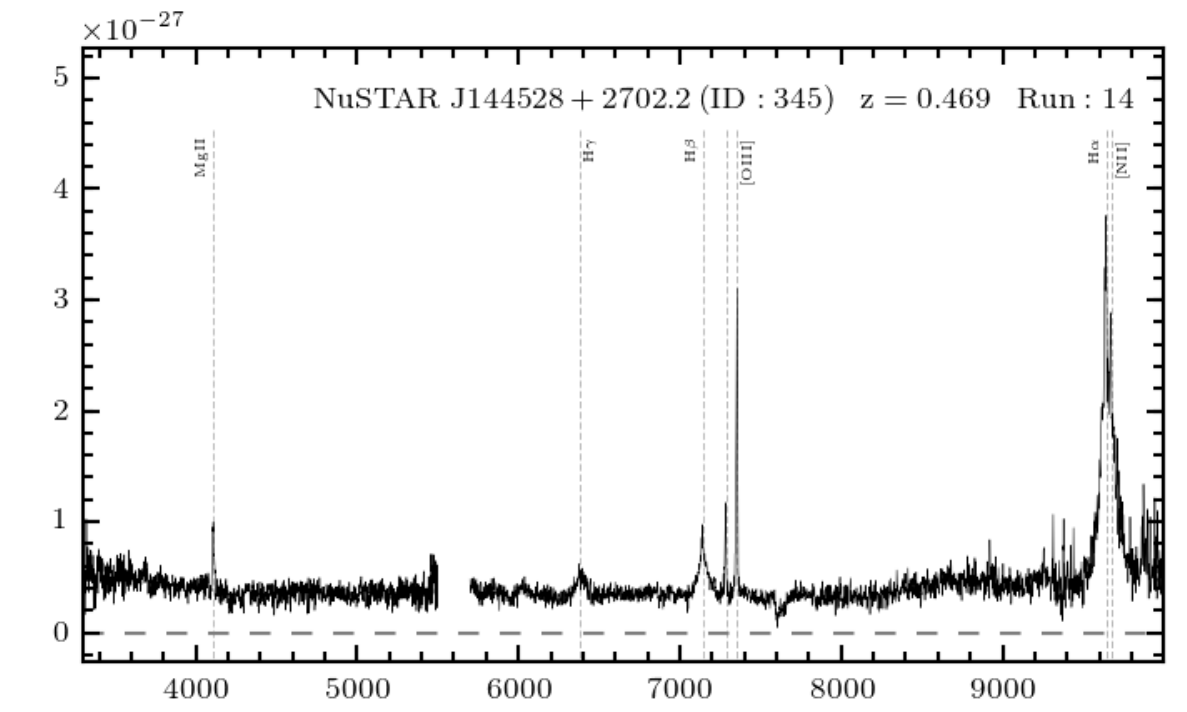}
\end{minipage}
\begin{minipage}[l]{0.325\textwidth}
\includegraphics[width=\textwidth]{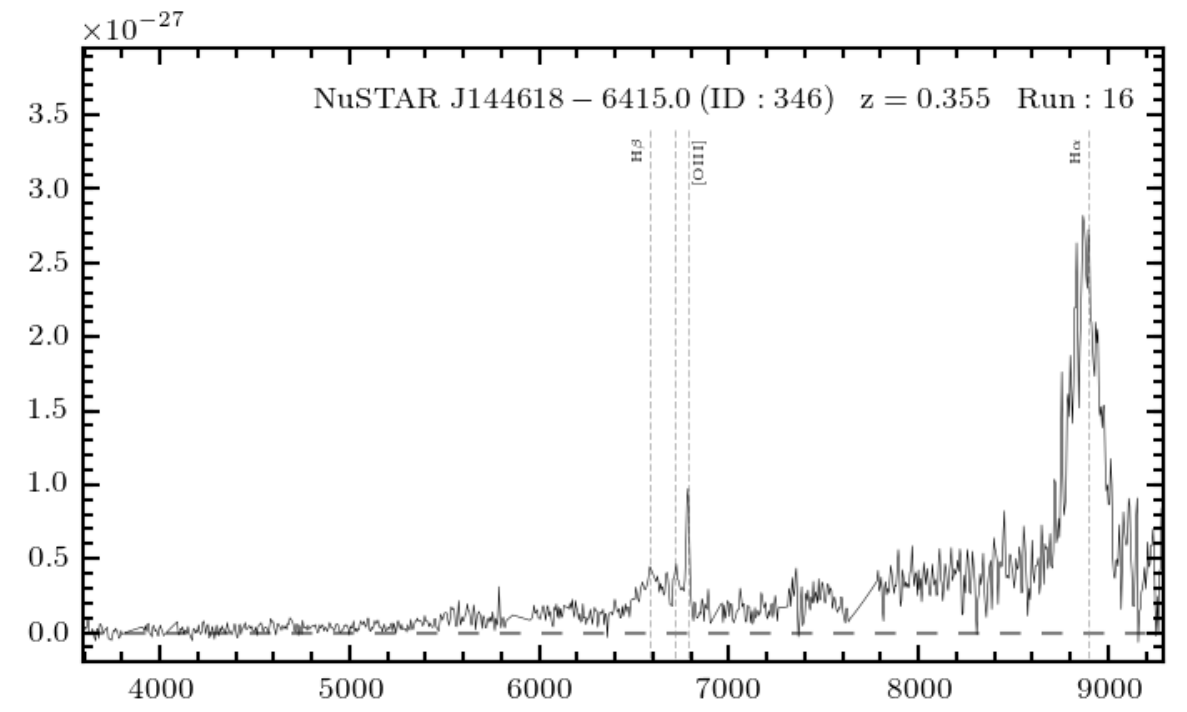}
\end{minipage}
\caption{Continued.}
\end{figure*}
\addtocounter{figure}{-1}
\begin{figure*}
\centering
\begin{minipage}[l]{0.325\textwidth}
\includegraphics[width=\textwidth]{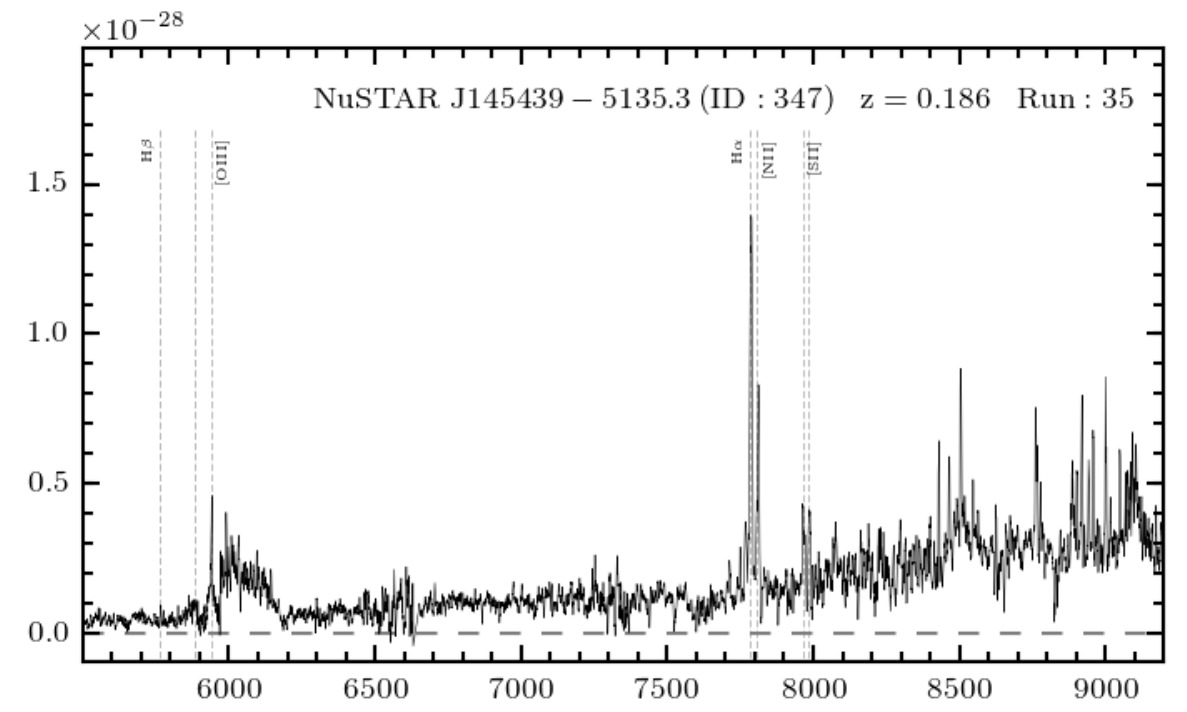}
\end{minipage}
\begin{minipage}[l]{0.325\textwidth}
\includegraphics[width=\textwidth]{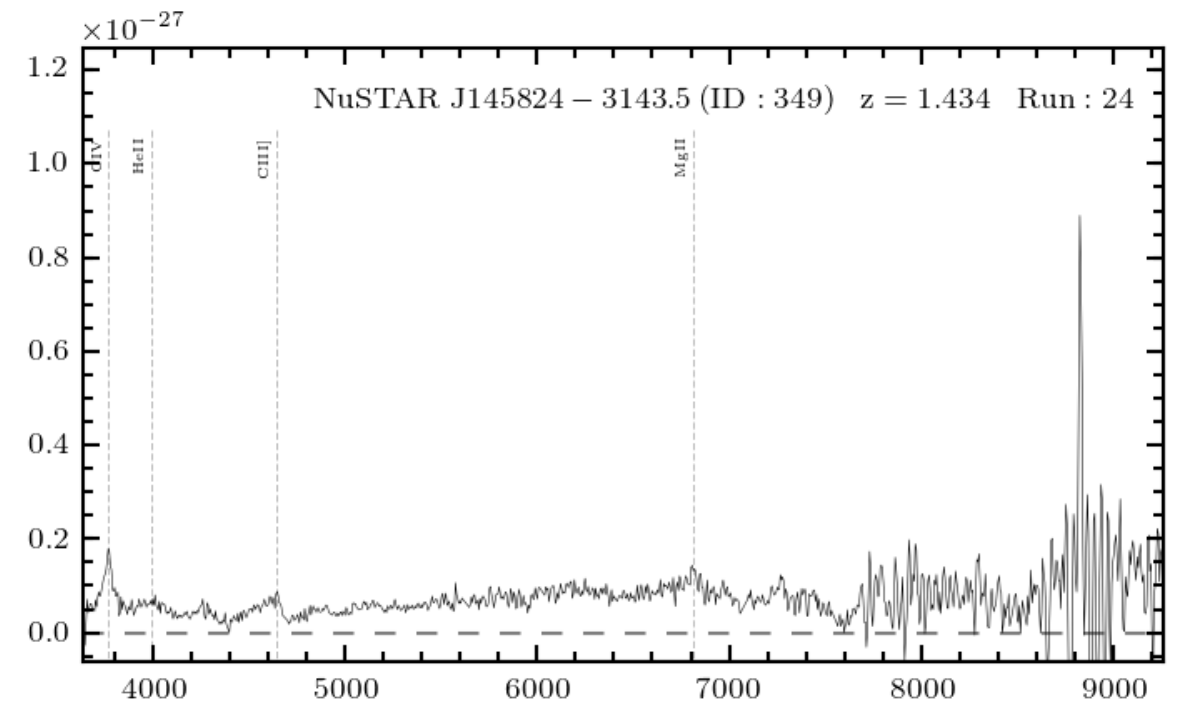}
\end{minipage}
\begin{minipage}[l]{0.325\textwidth}
\includegraphics[width=\textwidth]{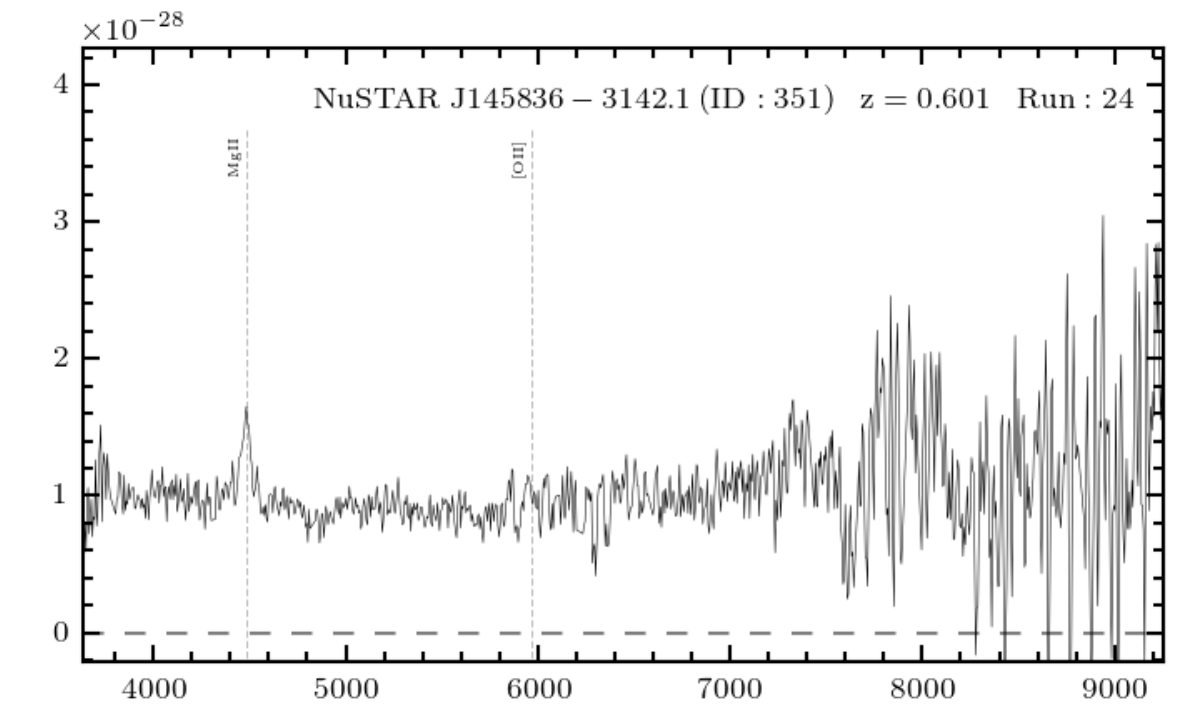}
\end{minipage}
\begin{minipage}[l]{0.325\textwidth}
\includegraphics[width=\textwidth]{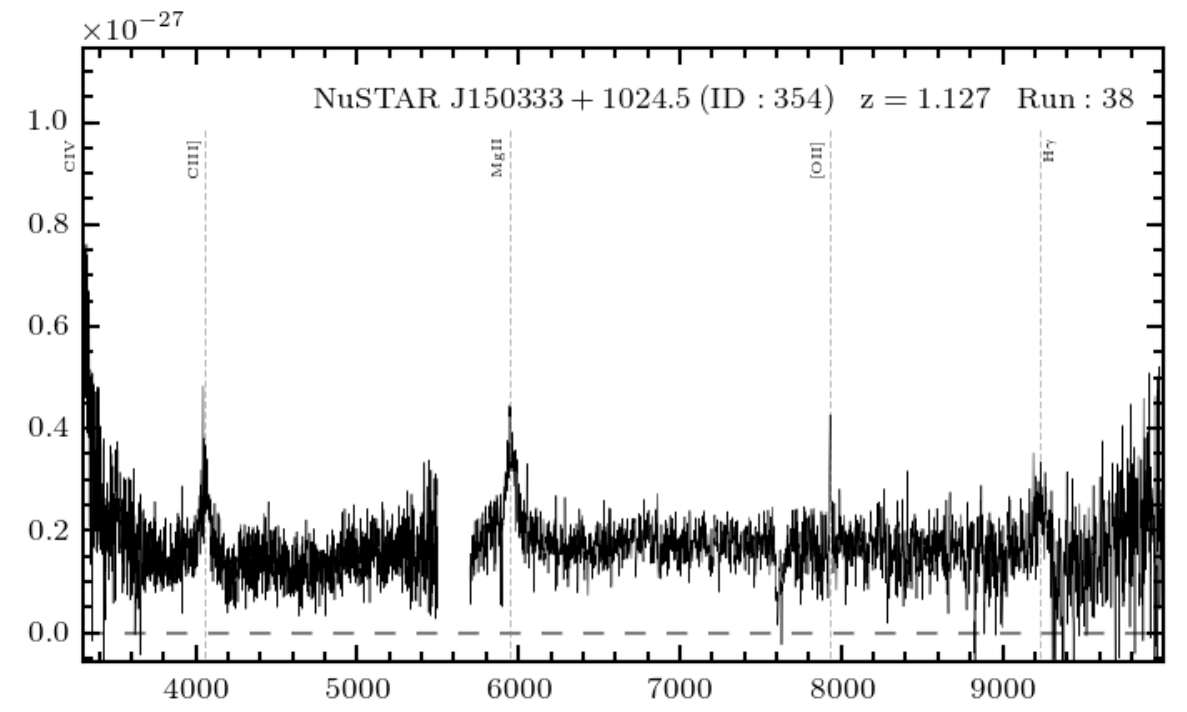}
\end{minipage}
\begin{minipage}[l]{0.325\textwidth}
\includegraphics[width=\textwidth]{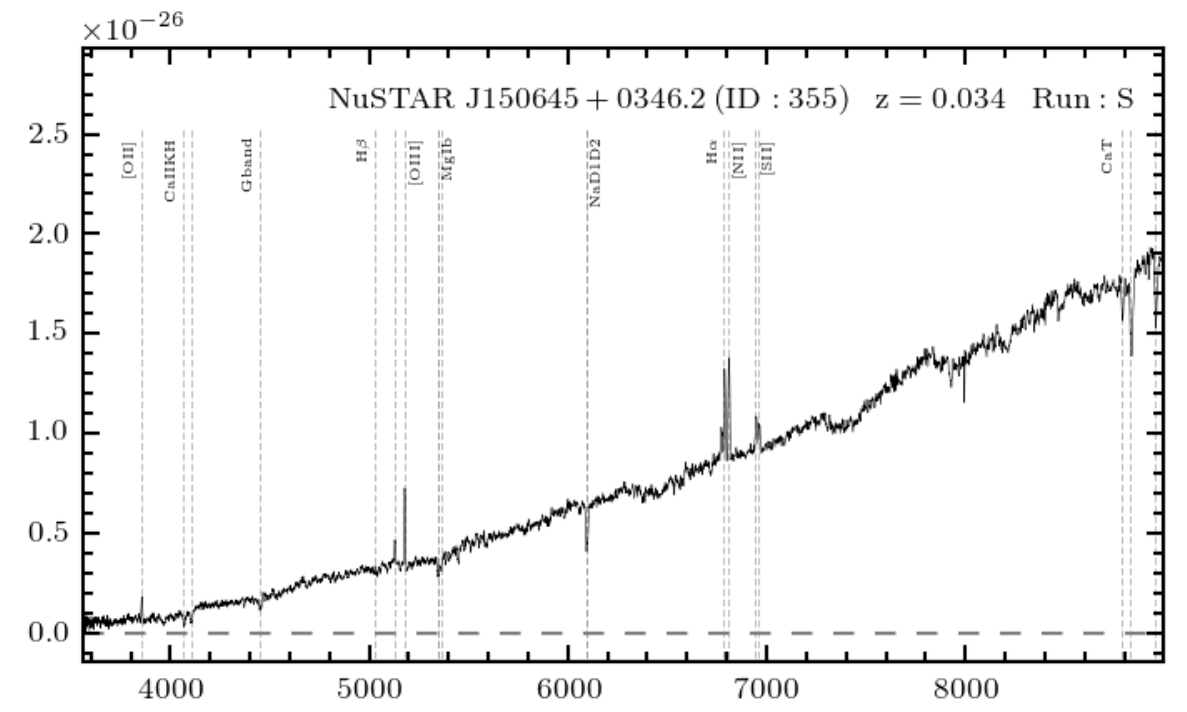}
\end{minipage}
\begin{minipage}[l]{0.325\textwidth}
\includegraphics[width=\textwidth]{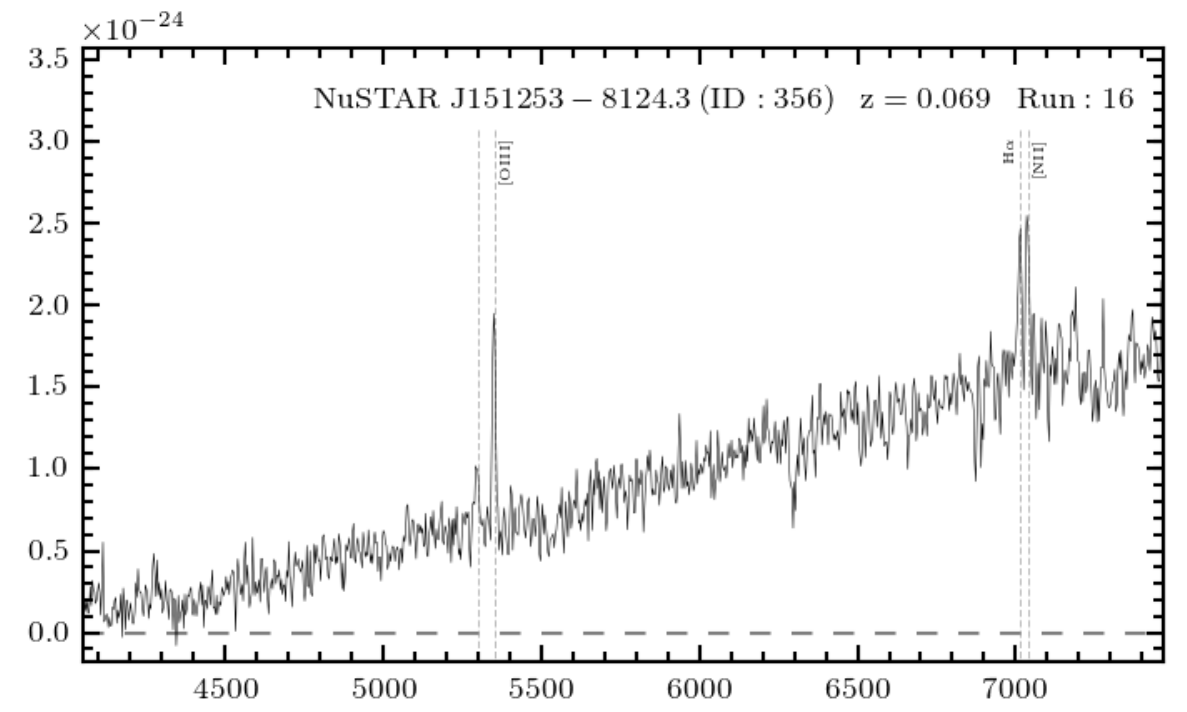}
\end{minipage}
\begin{minipage}[l]{0.325\textwidth}
\includegraphics[width=\textwidth]{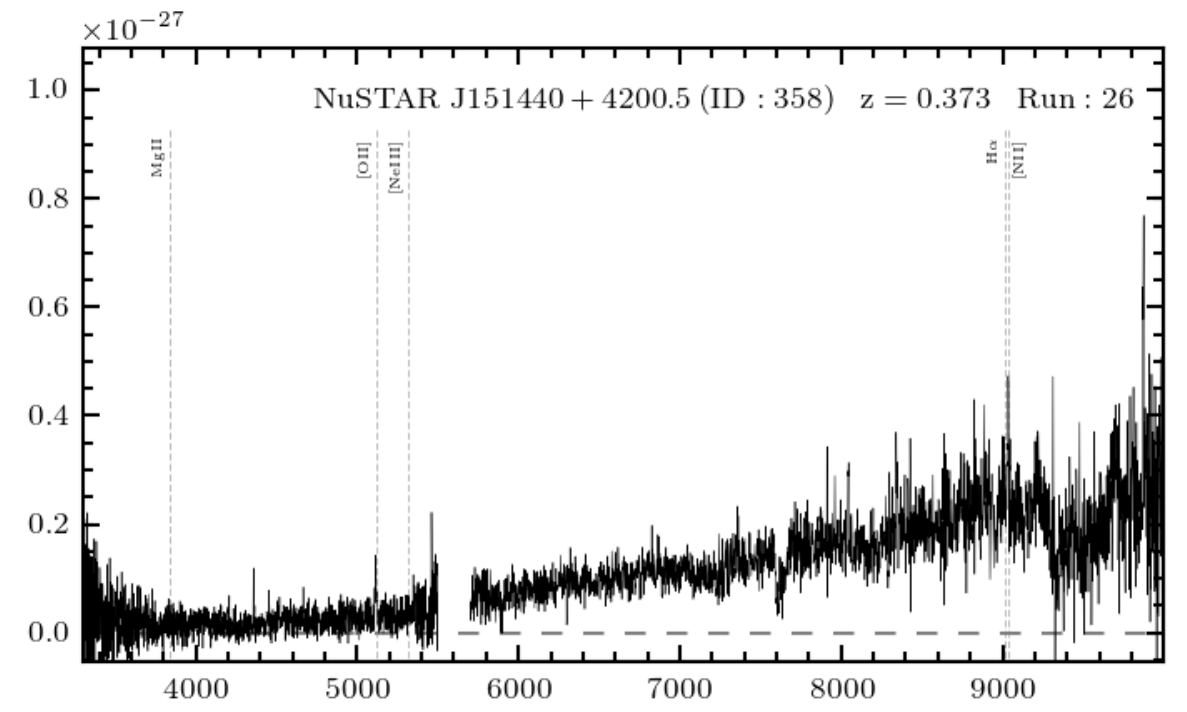}
\end{minipage}
\begin{minipage}[l]{0.325\textwidth}
\includegraphics[width=\textwidth]{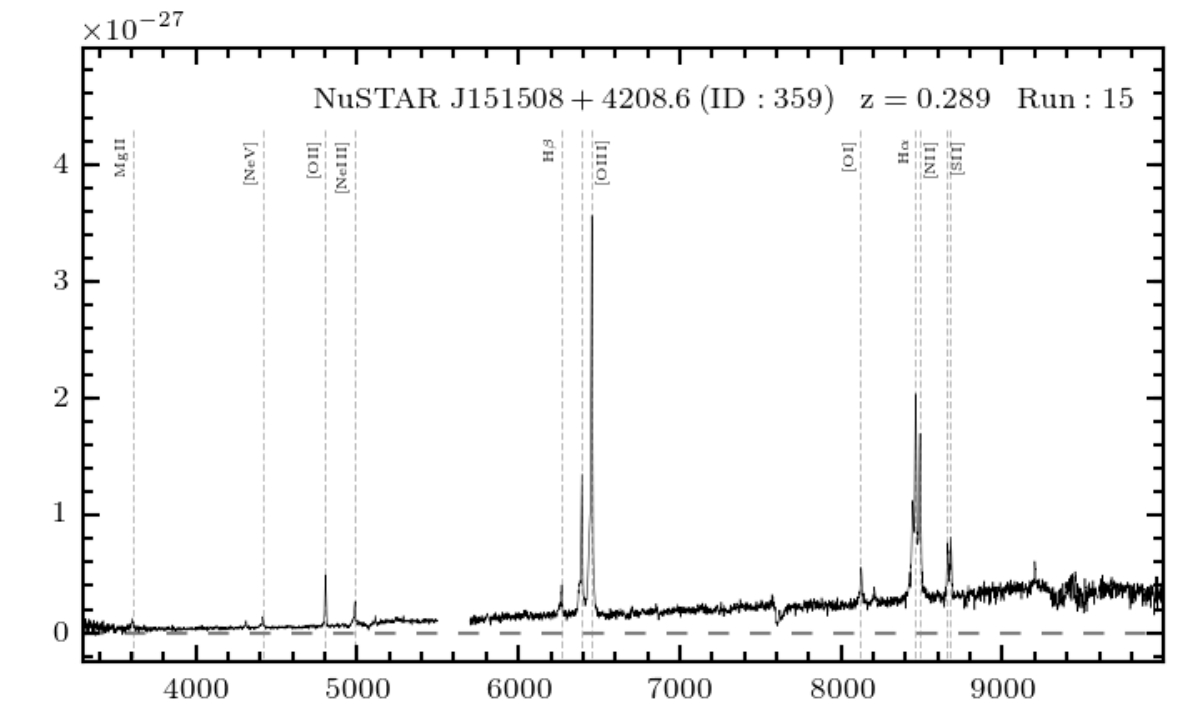}
\end{minipage}
\begin{minipage}[l]{0.325\textwidth}
\includegraphics[width=\textwidth]{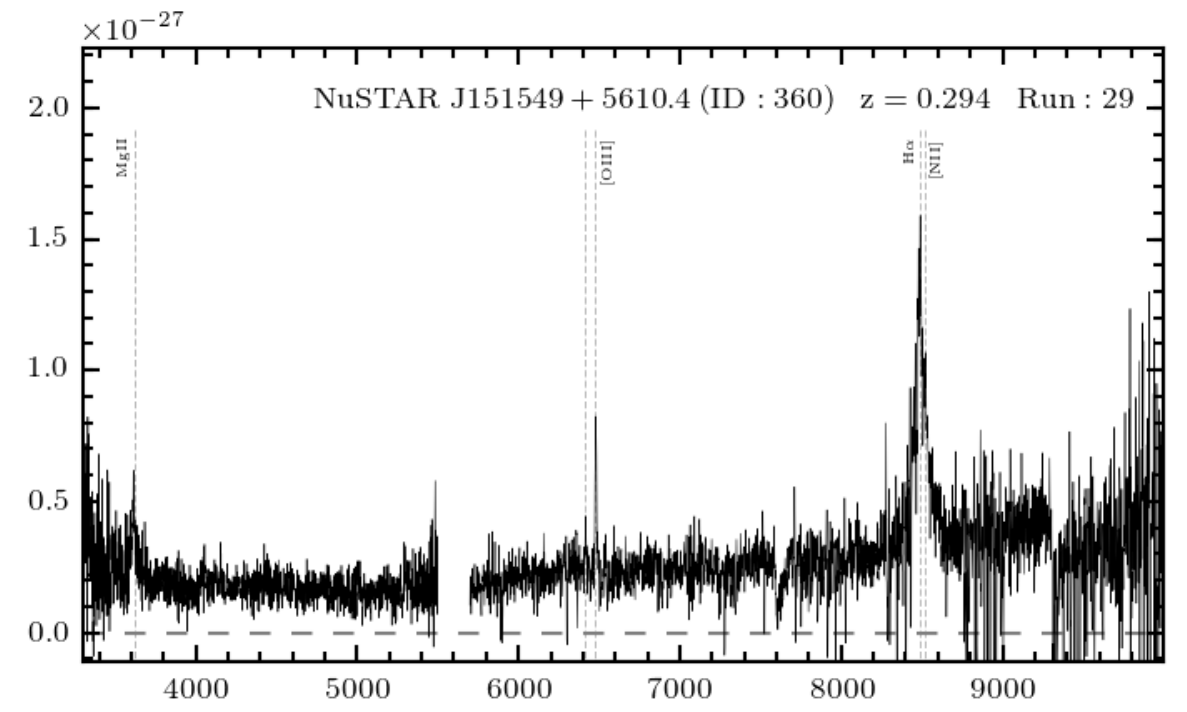}
\end{minipage}
\begin{minipage}[l]{0.325\textwidth}
\includegraphics[width=\textwidth]{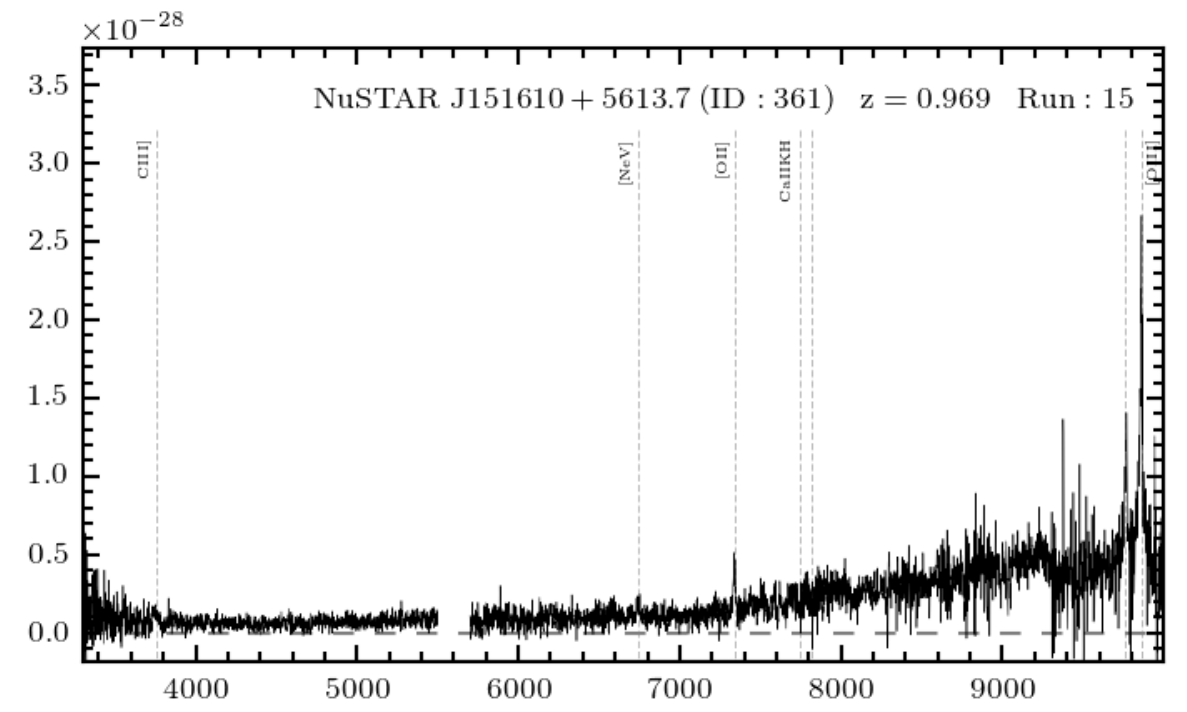}
\end{minipage}
\begin{minipage}[l]{0.325\textwidth}
\includegraphics[width=\textwidth]{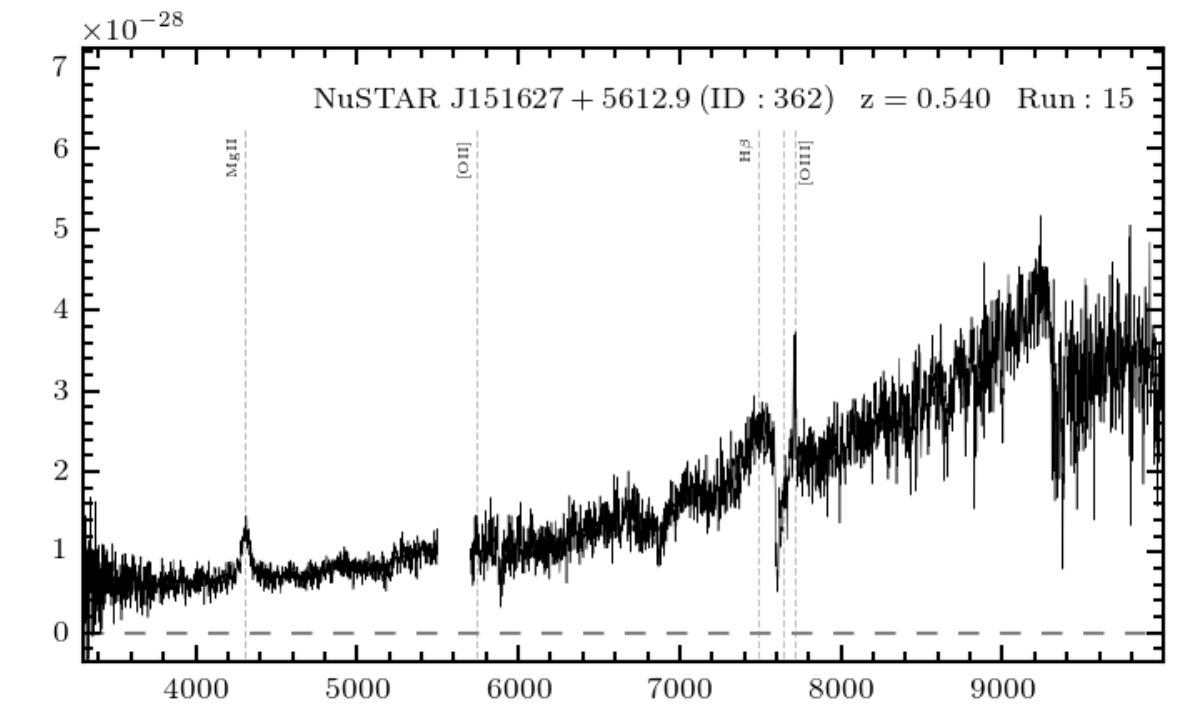}
\end{minipage}
\begin{minipage}[l]{0.325\textwidth}
\includegraphics[width=\textwidth]{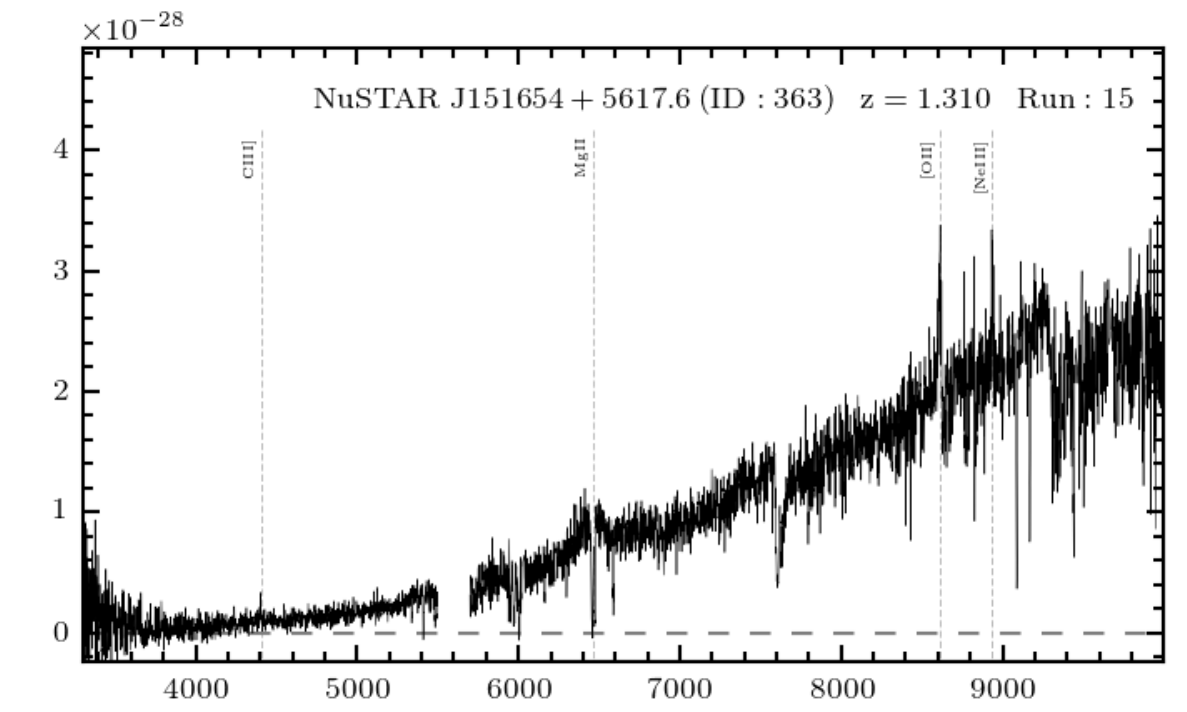}
\end{minipage}
\begin{minipage}[l]{0.325\textwidth}
\includegraphics[width=\textwidth]{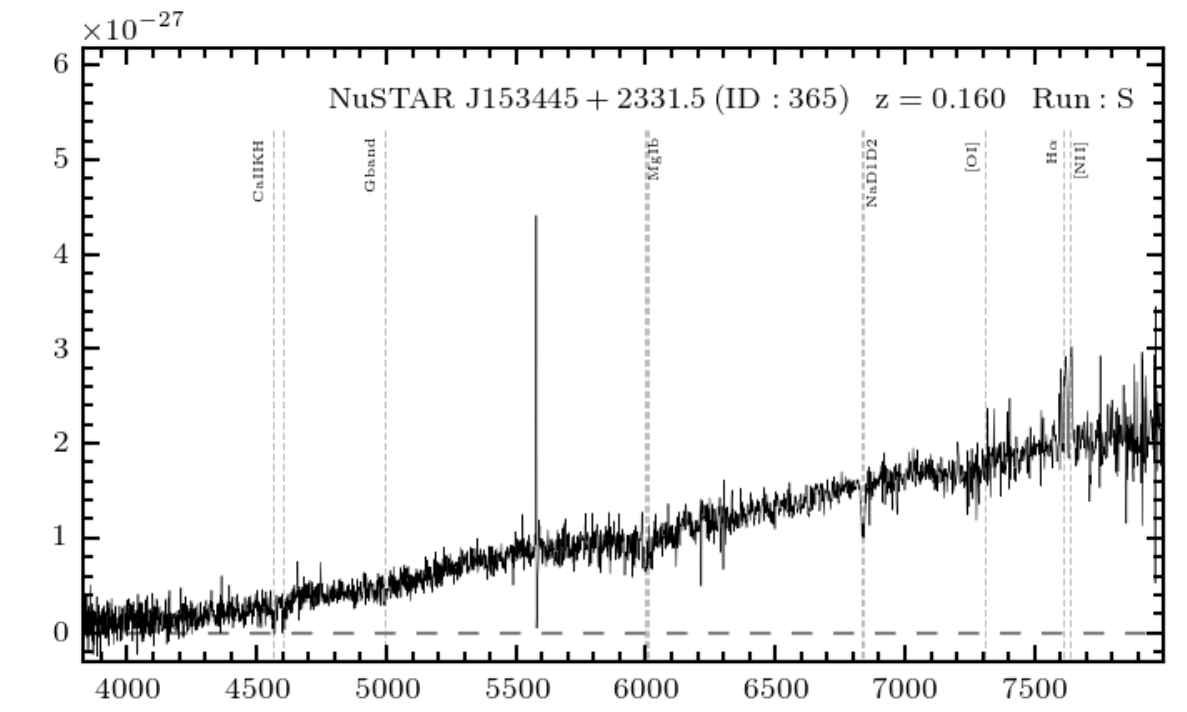}
\end{minipage}
\begin{minipage}[l]{0.325\textwidth}
\includegraphics[width=\textwidth]{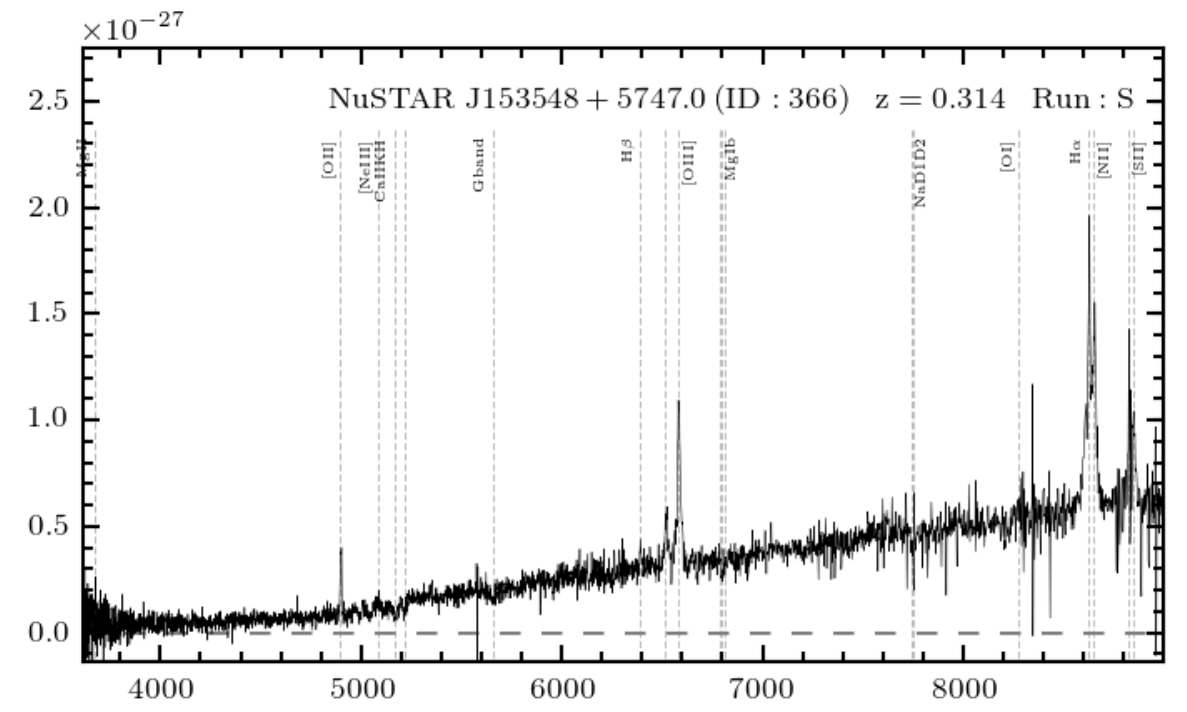}
\end{minipage}
\begin{minipage}[l]{0.325\textwidth}
\includegraphics[width=\textwidth]{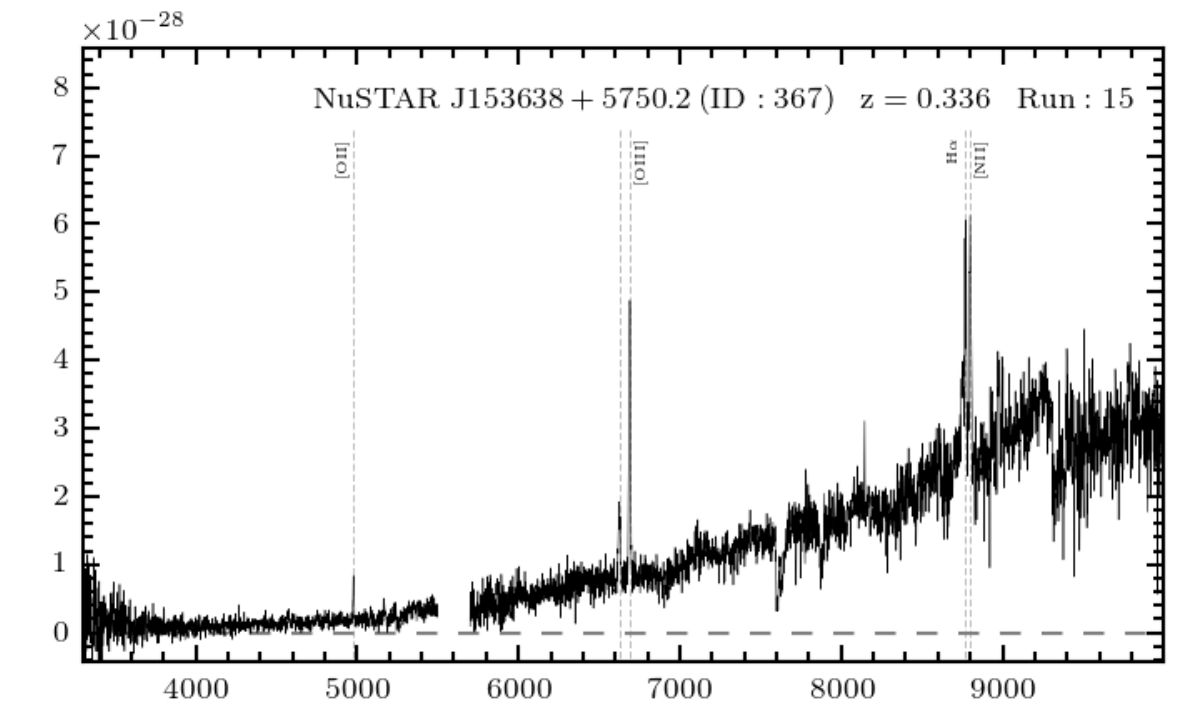}
\end{minipage}
\begin{minipage}[l]{0.325\textwidth}
\includegraphics[width=\textwidth]{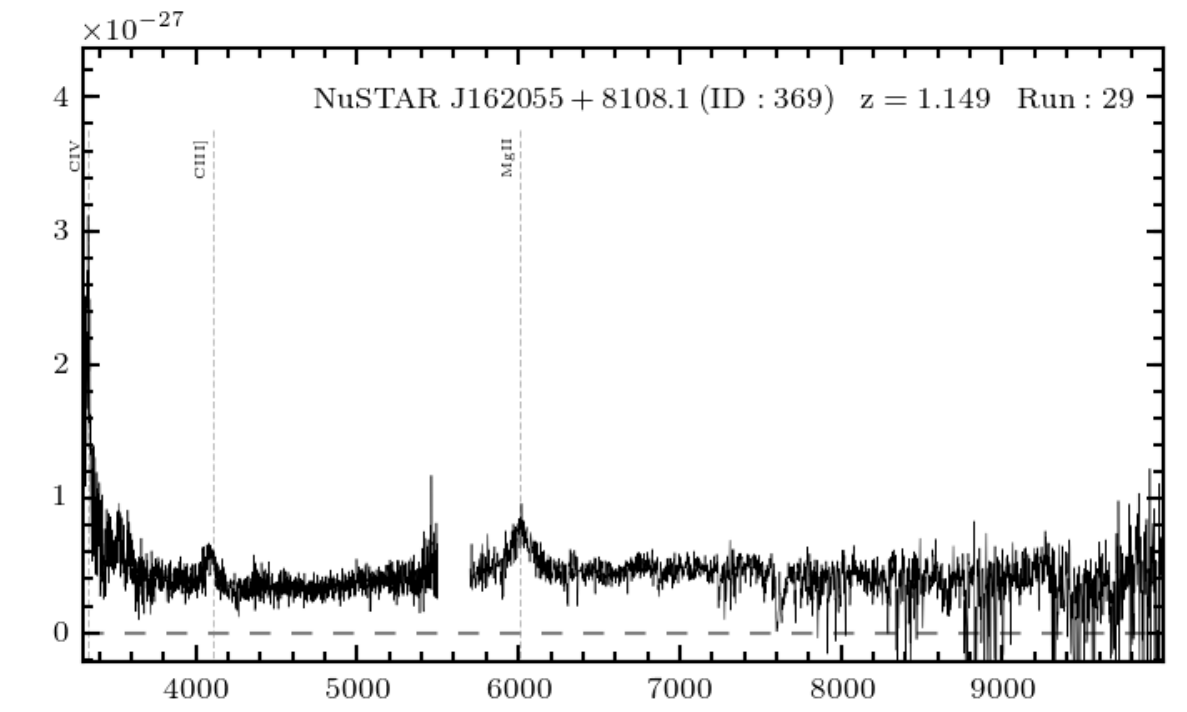}
\end{minipage}
\begin{minipage}[l]{0.325\textwidth}
\includegraphics[width=\textwidth]{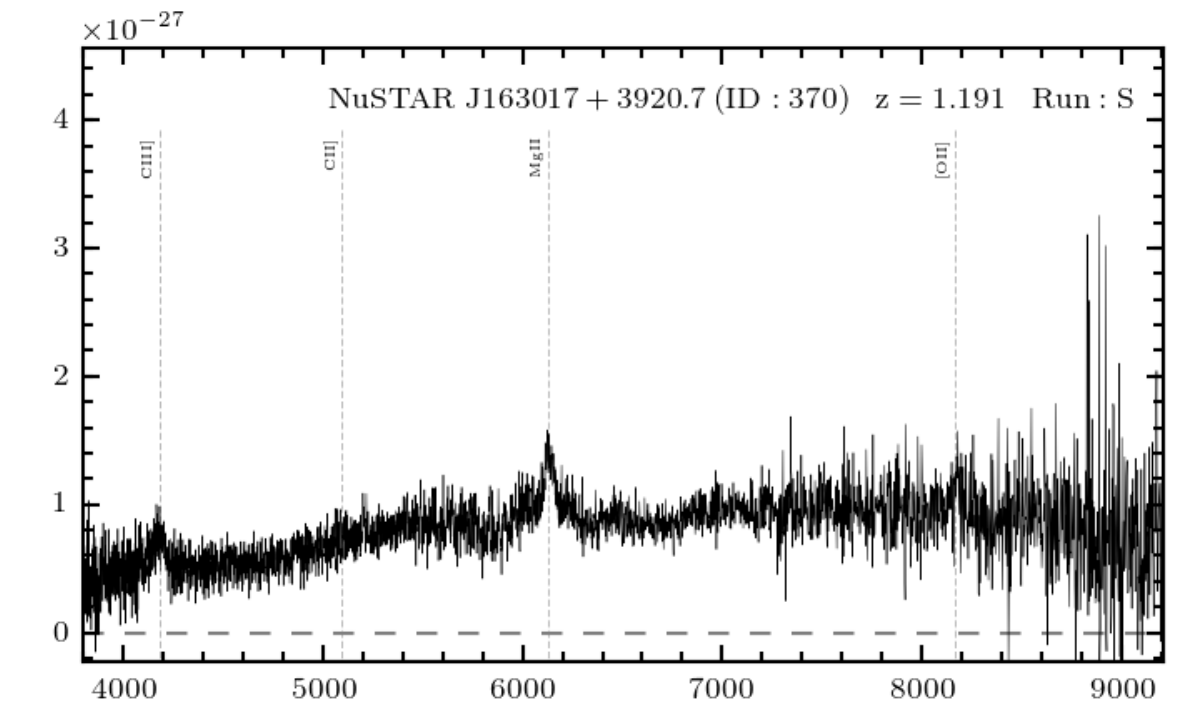}
\end{minipage}
\begin{minipage}[l]{0.325\textwidth}
\includegraphics[width=\textwidth]{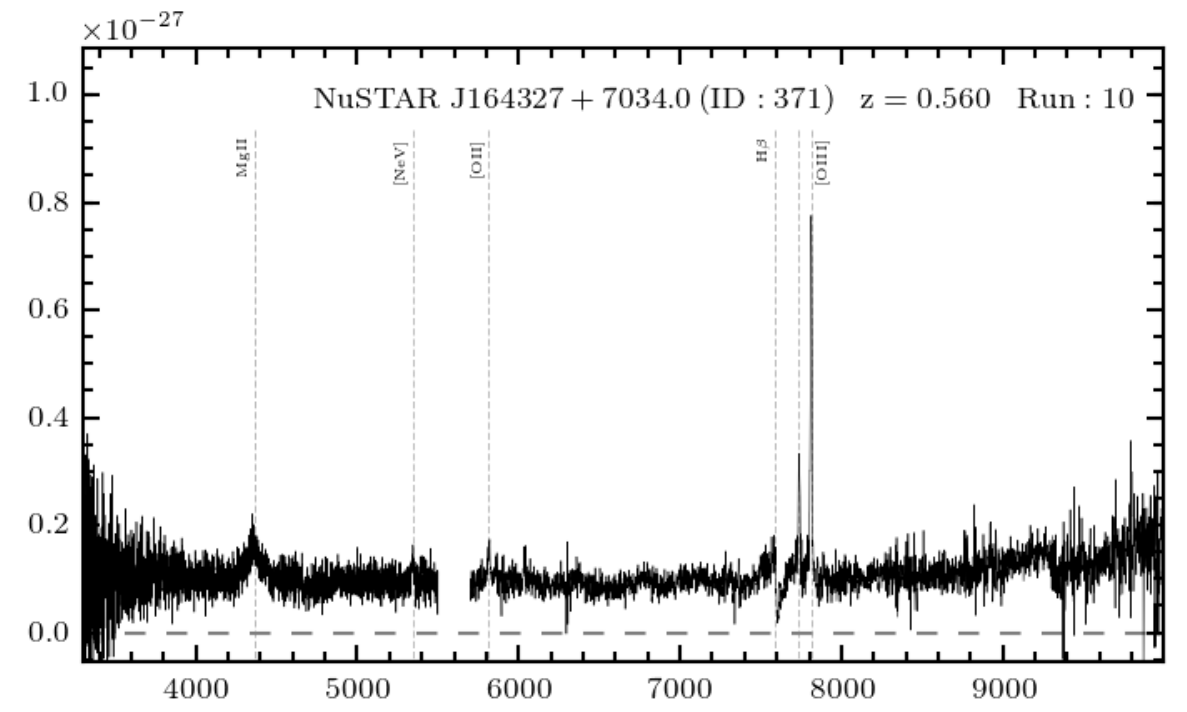}
\end{minipage}
\caption{Continued.}
\end{figure*}
\addtocounter{figure}{-1}
\begin{figure*}
\centering
\begin{minipage}[l]{0.325\textwidth}
\includegraphics[width=\textwidth]{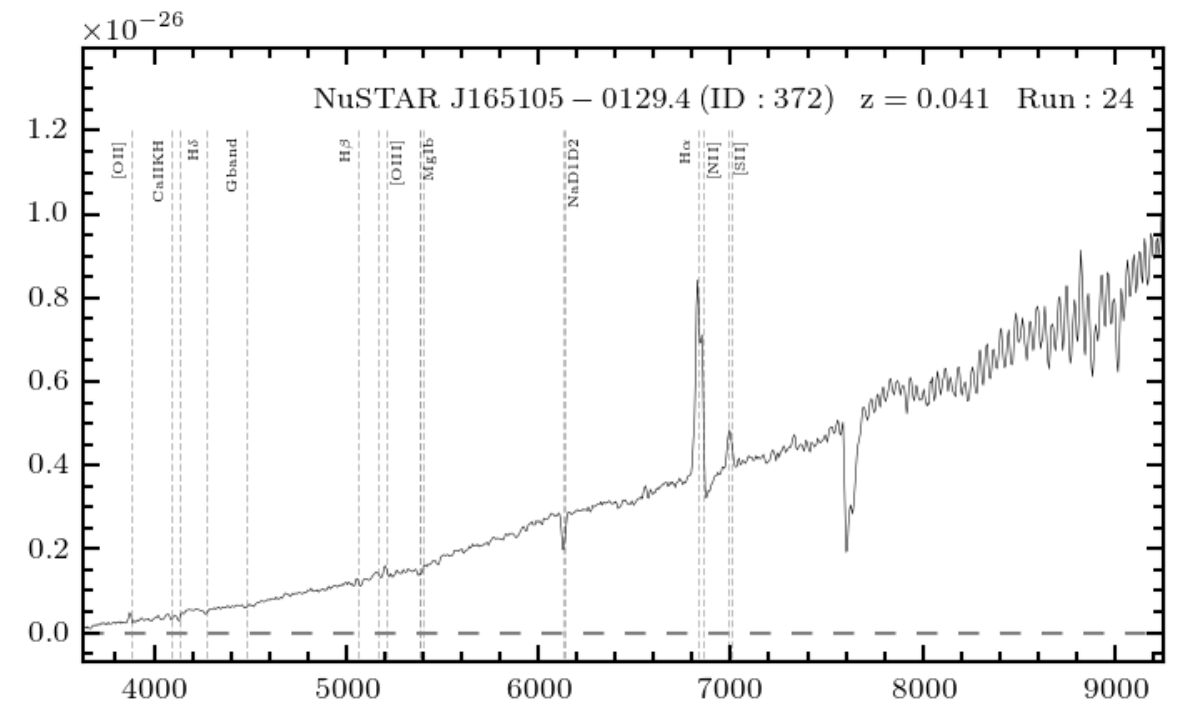}
\end{minipage}
\begin{minipage}[l]{0.325\textwidth}
\includegraphics[width=\textwidth]{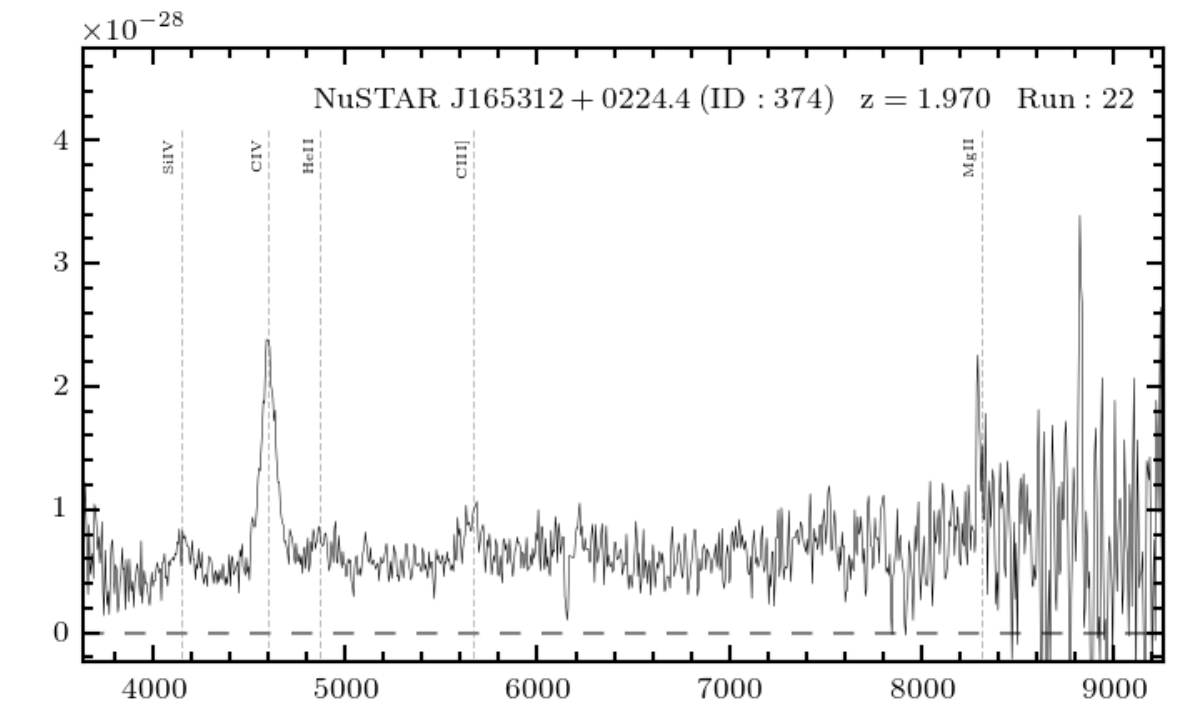}
\end{minipage}
\begin{minipage}[l]{0.325\textwidth}
\includegraphics[width=\textwidth]{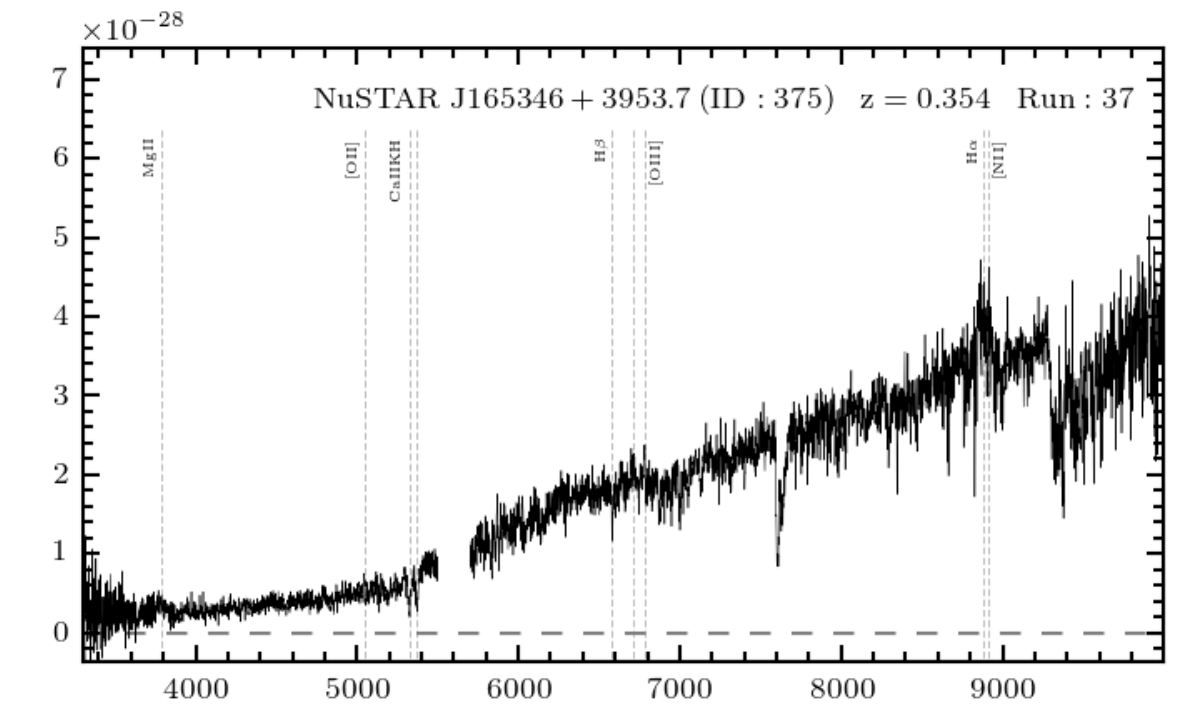}
\end{minipage}
\begin{minipage}[l]{0.325\textwidth}
\includegraphics[width=\textwidth]{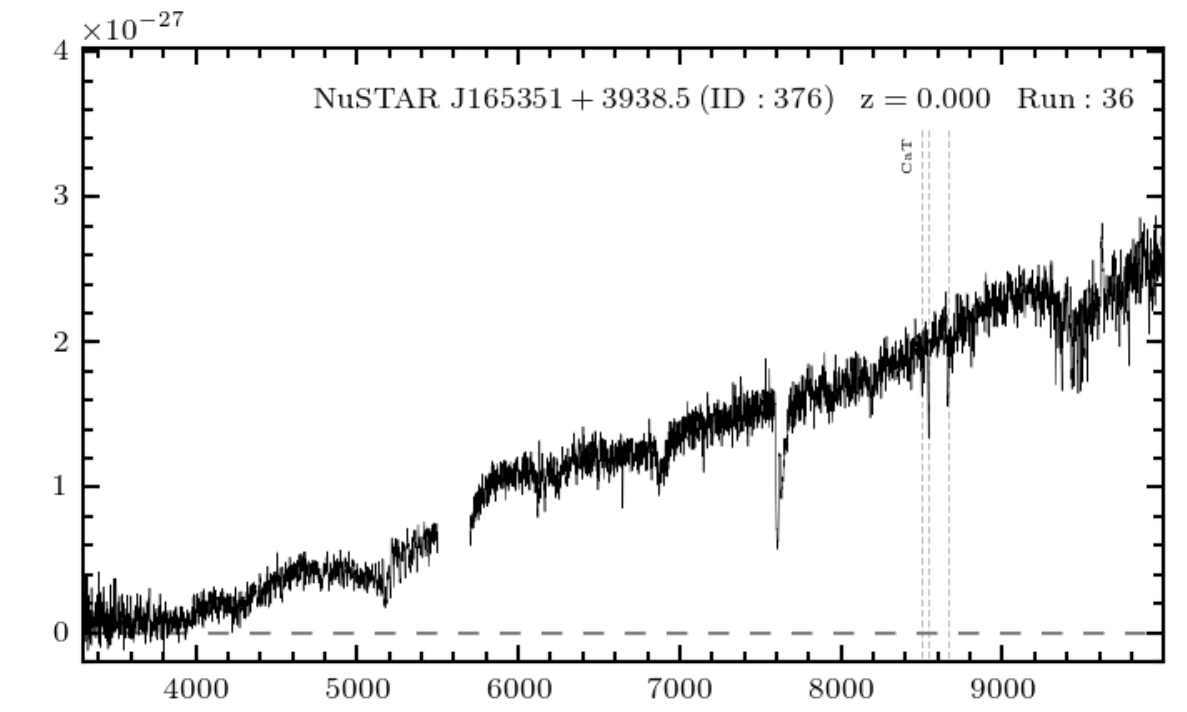}
\end{minipage}
\begin{minipage}[l]{0.325\textwidth}
\includegraphics[width=\textwidth]{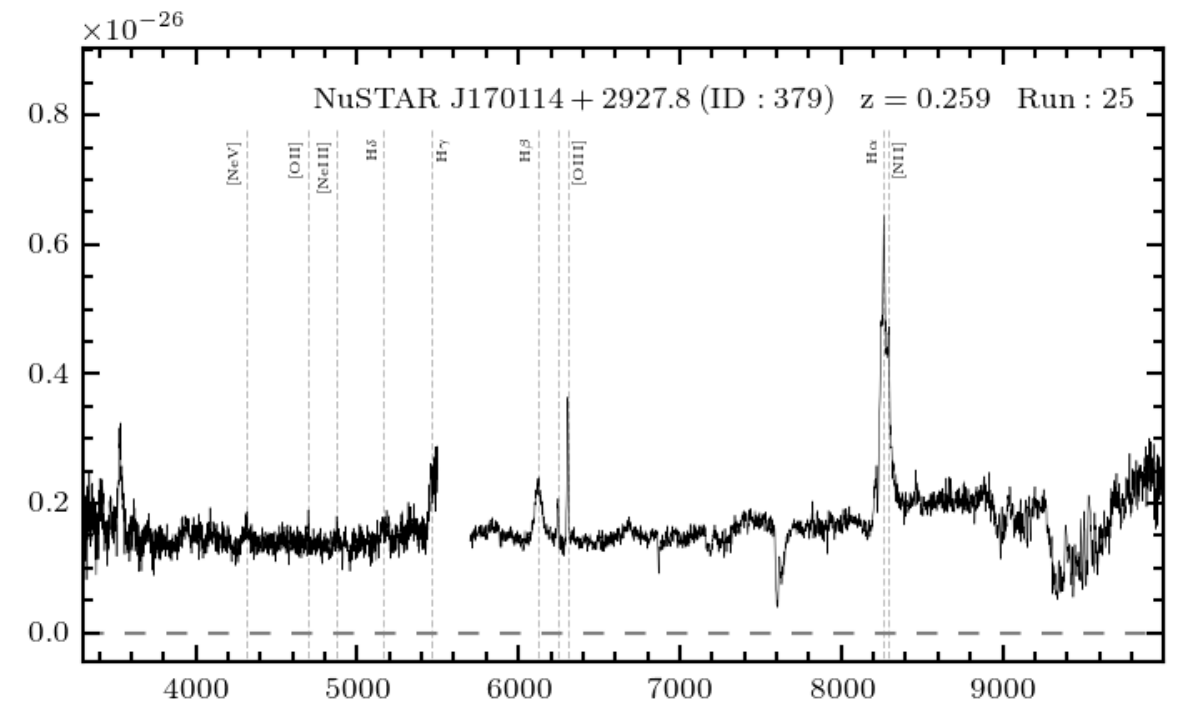}
\end{minipage}
\begin{minipage}[l]{0.325\textwidth}
\includegraphics[width=\textwidth]{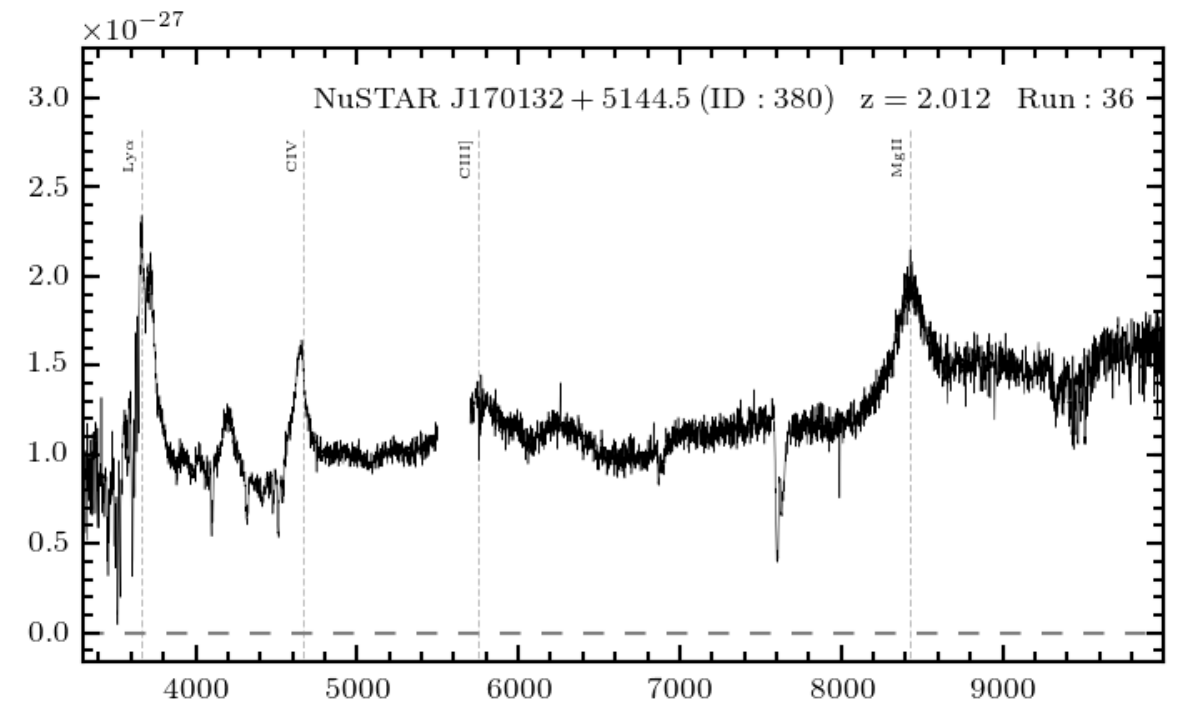}
\end{minipage}
\begin{minipage}[l]{0.325\textwidth}
\includegraphics[width=\textwidth]{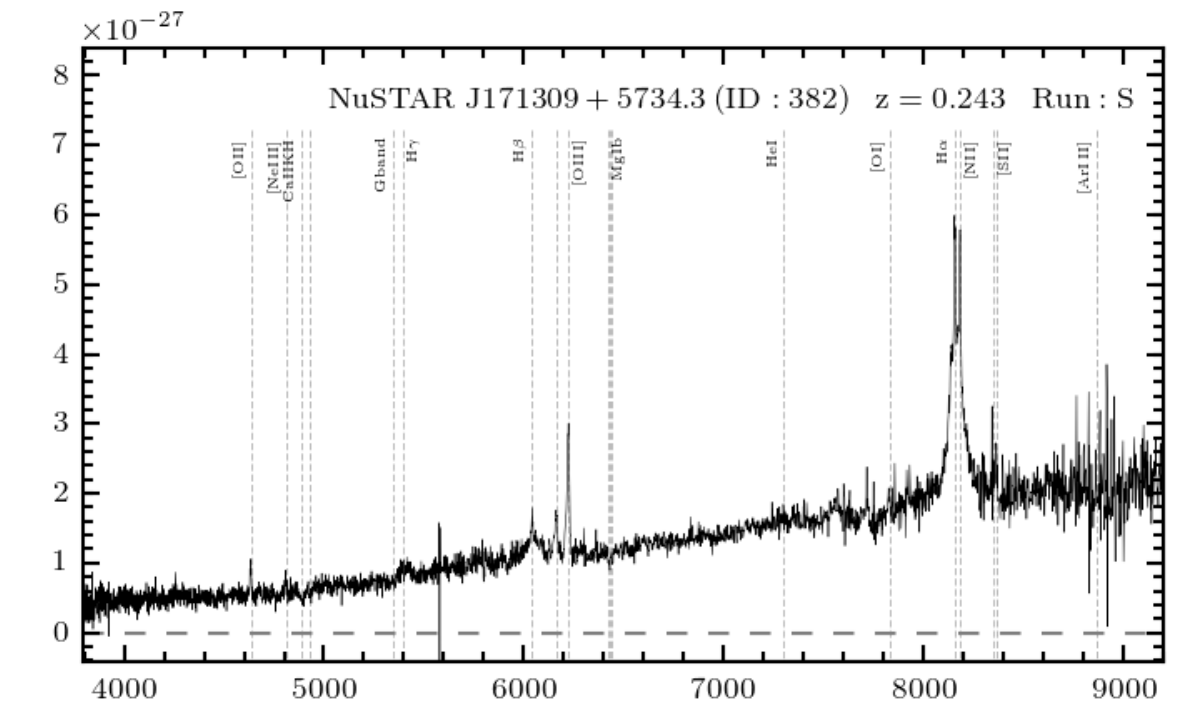}
\end{minipage}
\begin{minipage}[l]{0.325\textwidth}
\includegraphics[width=\textwidth]{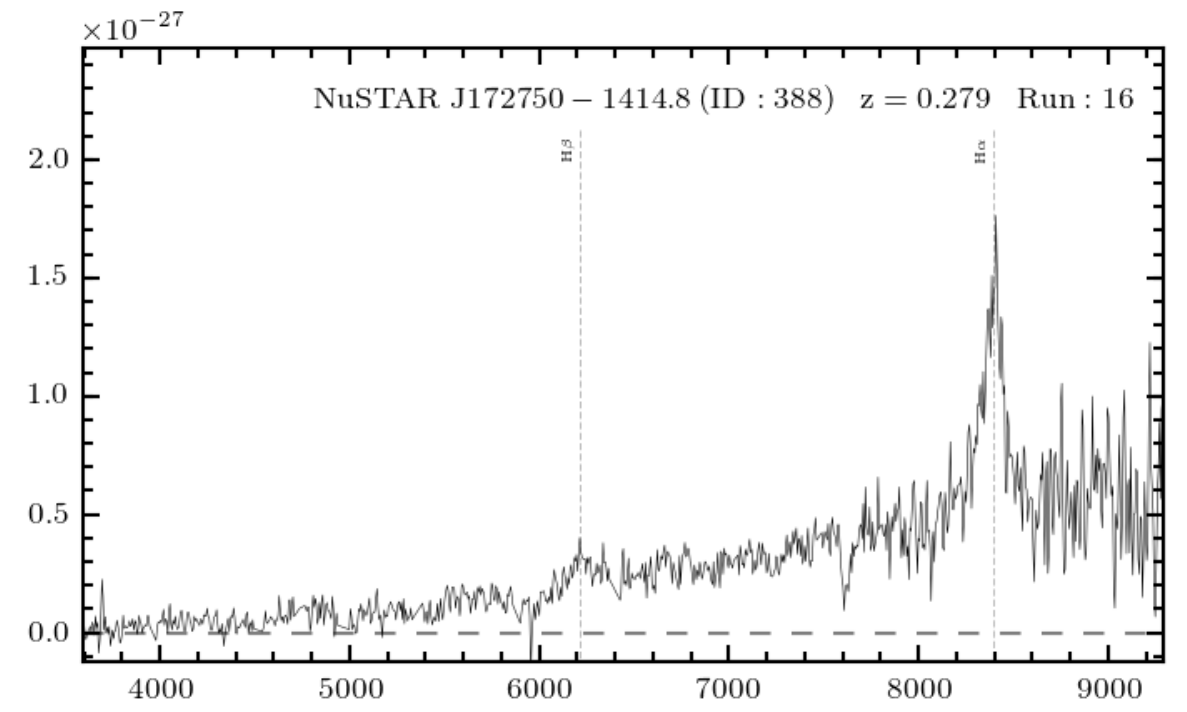}
\end{minipage}
\begin{minipage}[l]{0.325\textwidth}
\includegraphics[width=\textwidth]{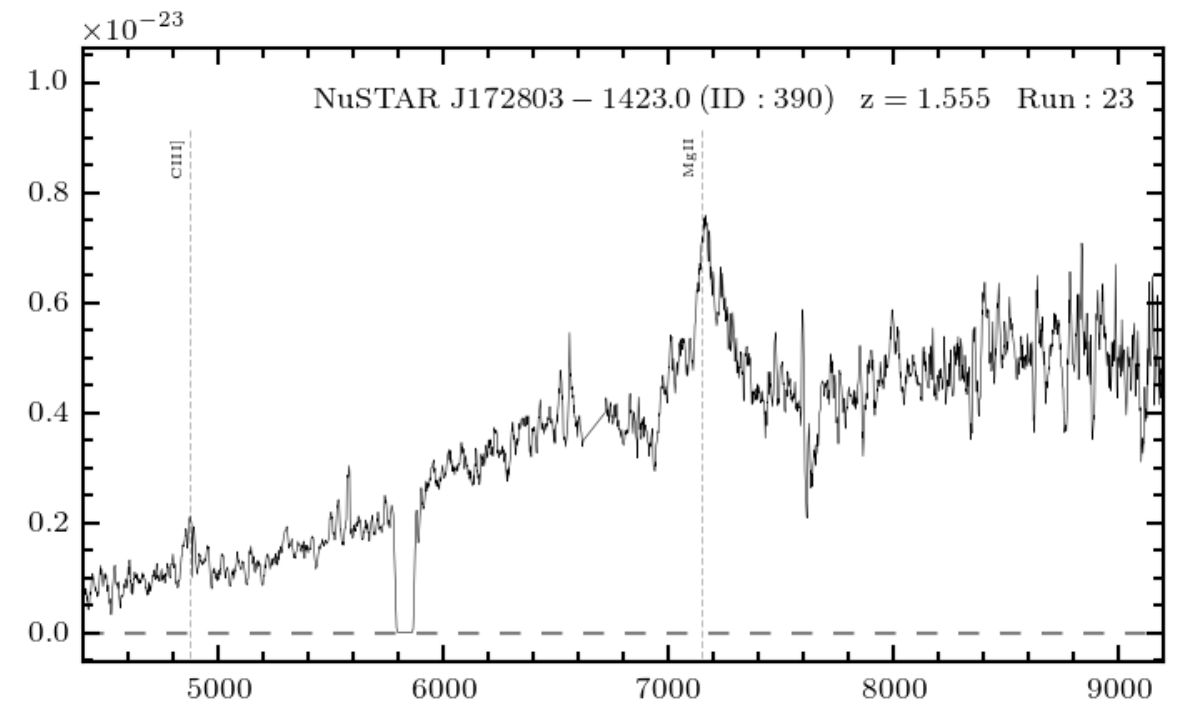}
\end{minipage}
\begin{minipage}[l]{0.325\textwidth}
\includegraphics[width=\textwidth]{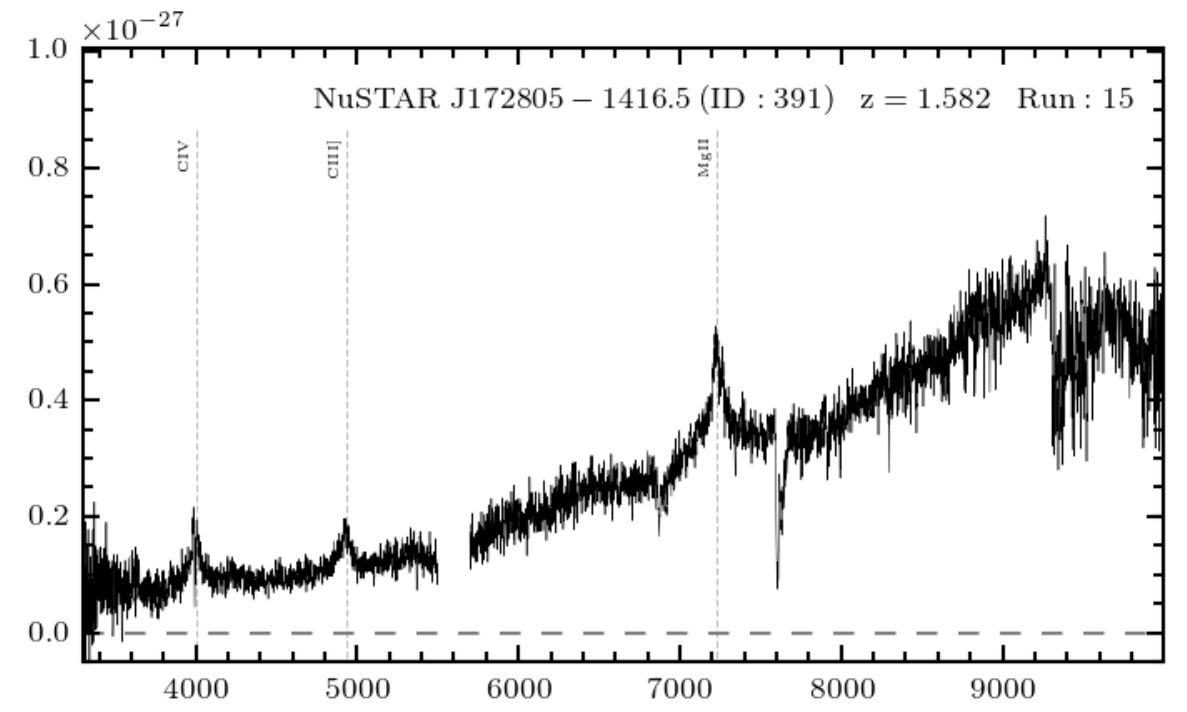}
\end{minipage}
\begin{minipage}[l]{0.325\textwidth}
\includegraphics[width=\textwidth]{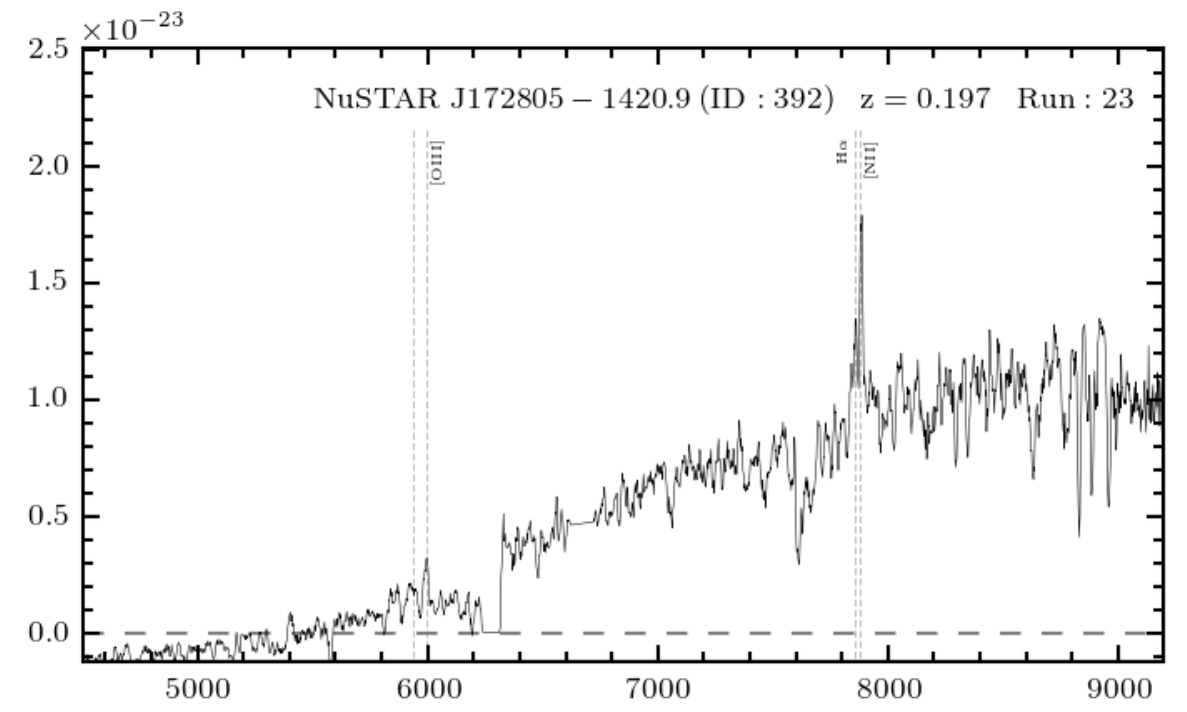}
\end{minipage}
\begin{minipage}[l]{0.325\textwidth}
\includegraphics[width=\textwidth]{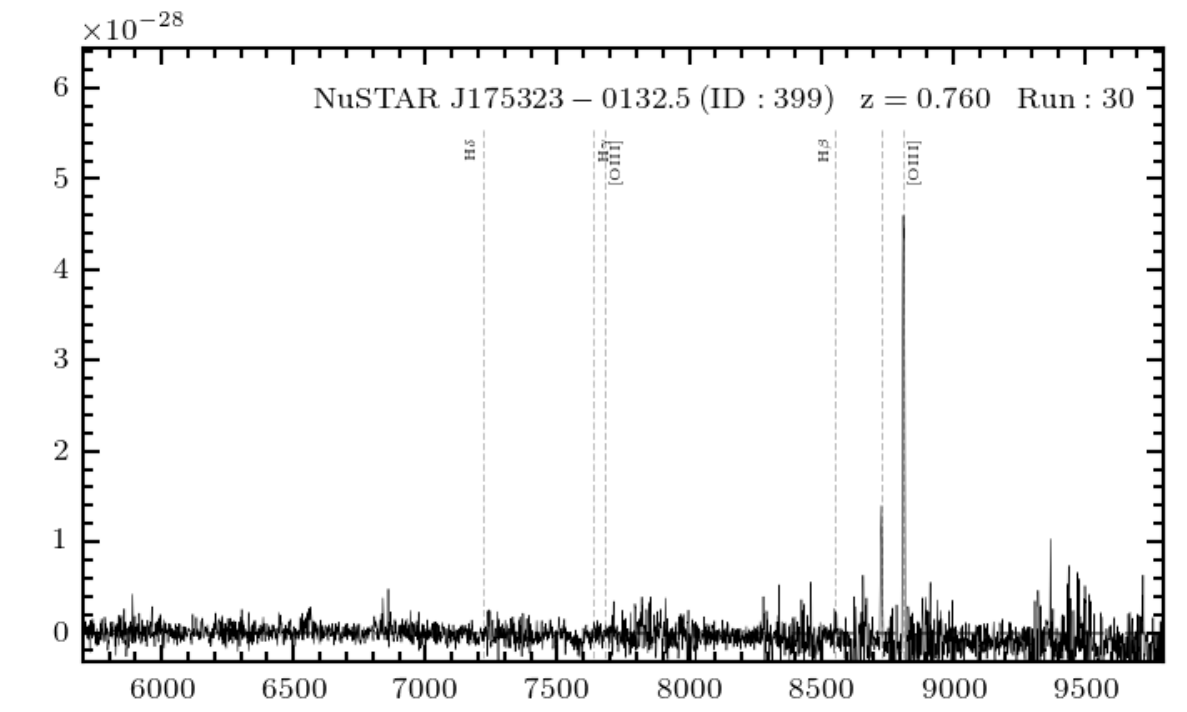}
\end{minipage}
\begin{minipage}[l]{0.325\textwidth}
\includegraphics[width=\textwidth]{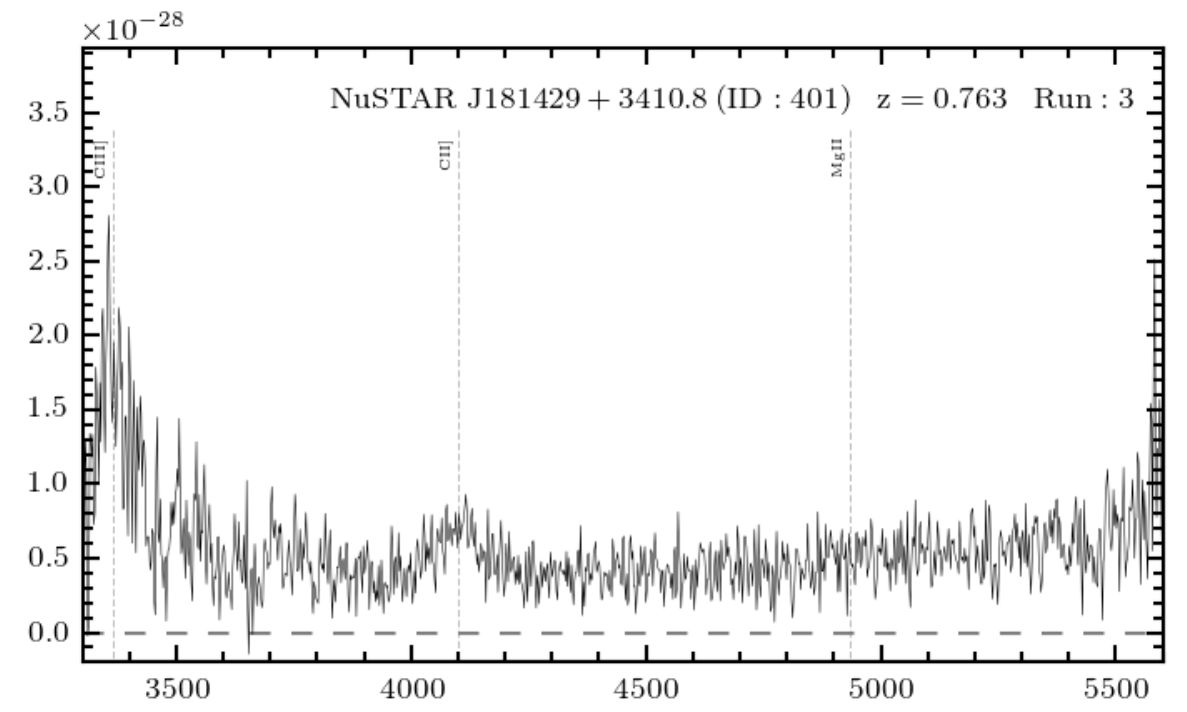}
\end{minipage}
\begin{minipage}[l]{0.325\textwidth}
\includegraphics[width=\textwidth]{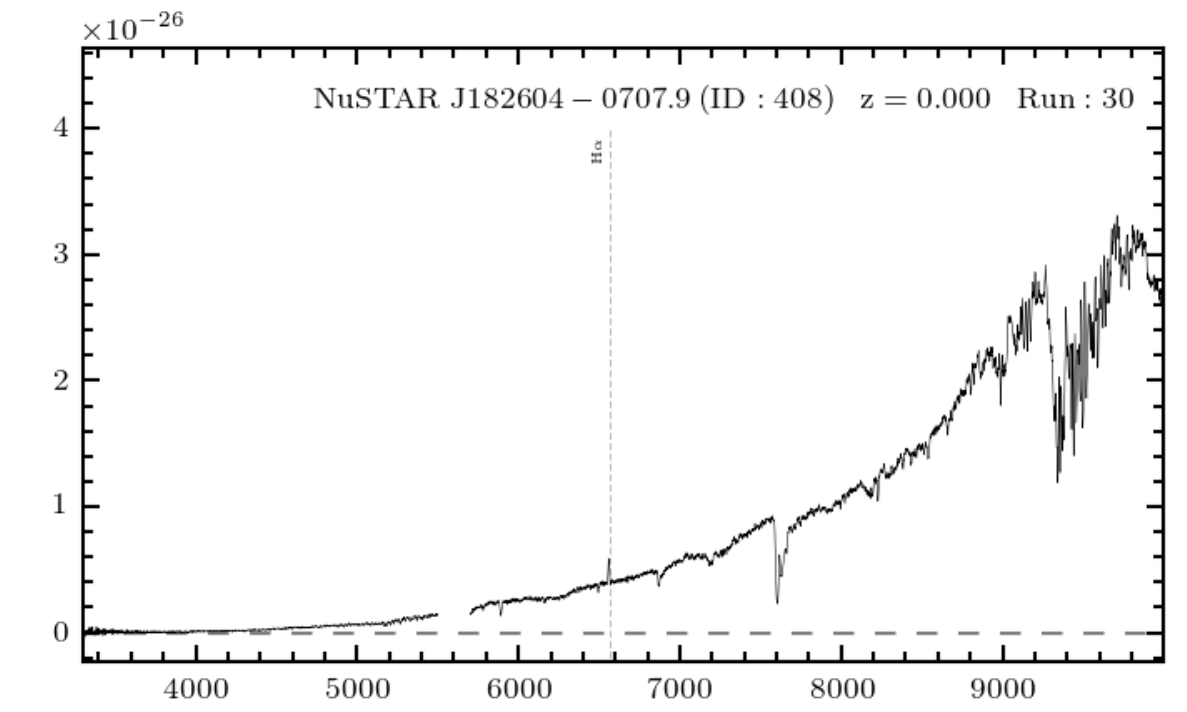}
\end{minipage}
\begin{minipage}[l]{0.325\textwidth}
\includegraphics[width=\textwidth]{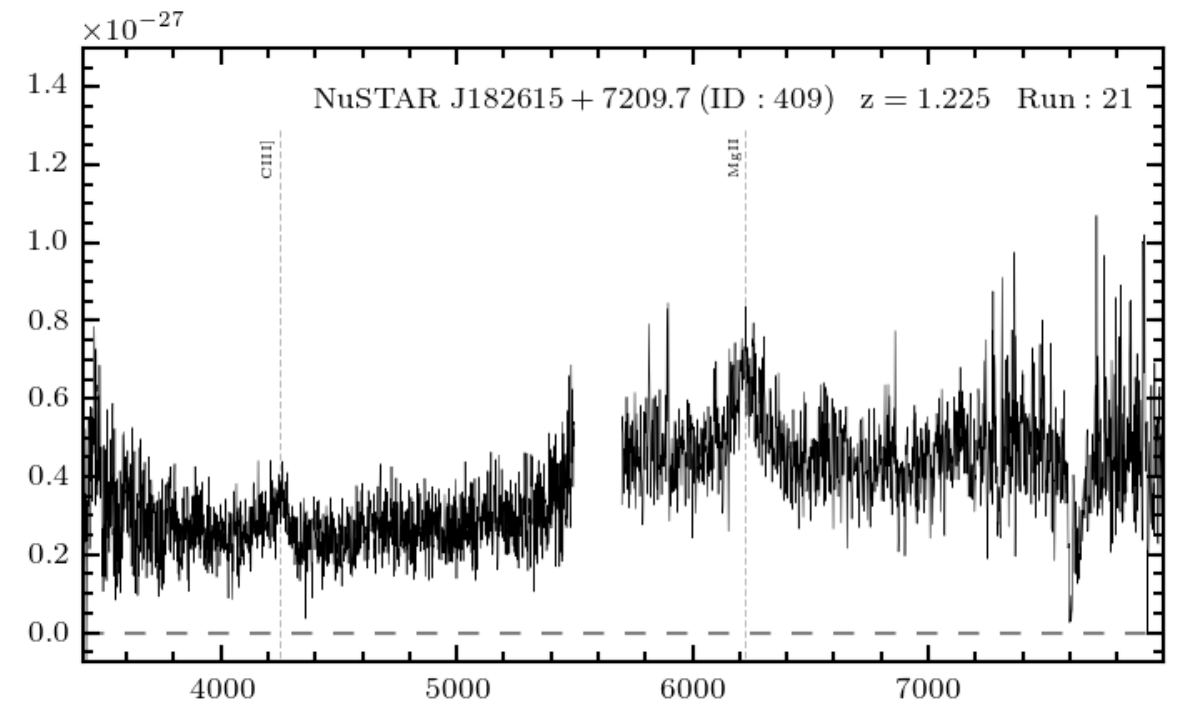}
\end{minipage}
\begin{minipage}[l]{0.325\textwidth}
\includegraphics[width=\textwidth]{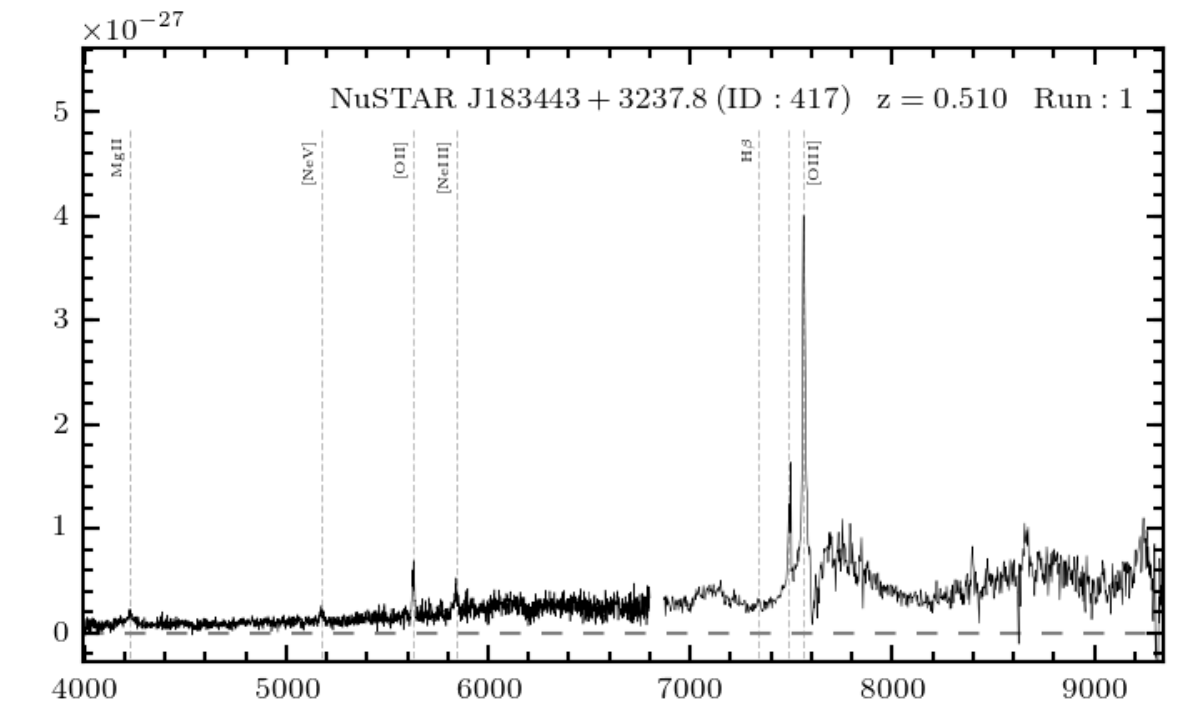}
\end{minipage}
\begin{minipage}[l]{0.325\textwidth}
\includegraphics[width=\textwidth]{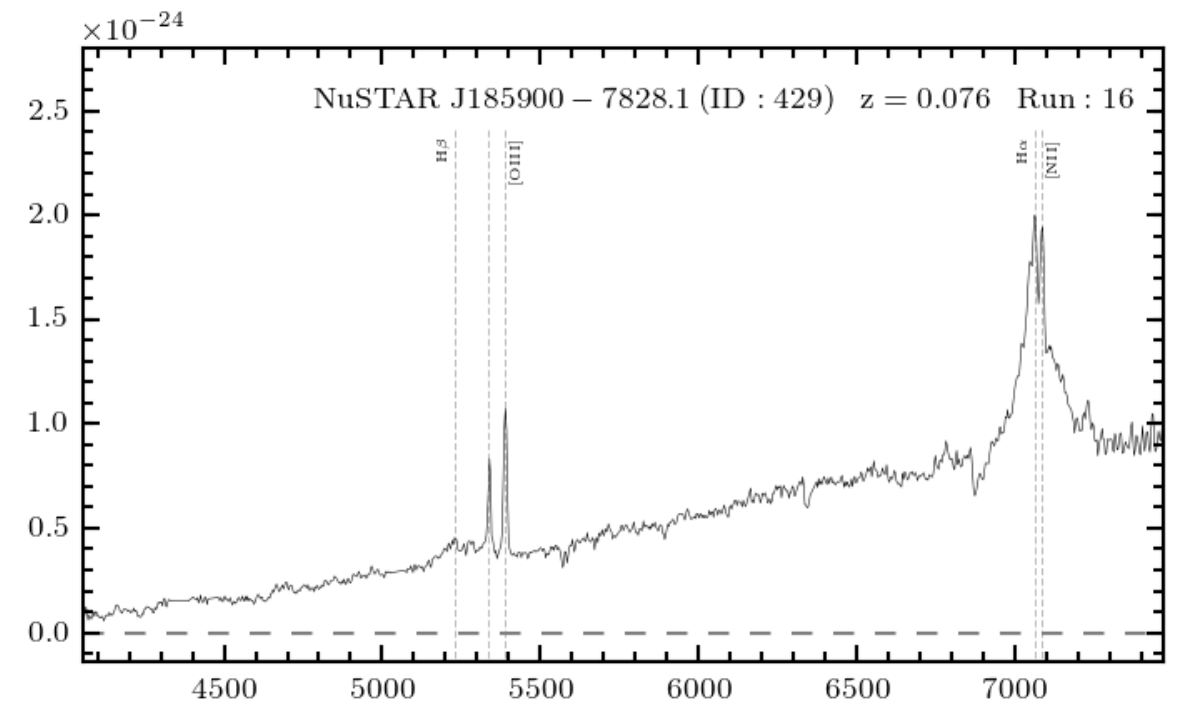}
\end{minipage}
\begin{minipage}[l]{0.325\textwidth}
\includegraphics[width=\textwidth]{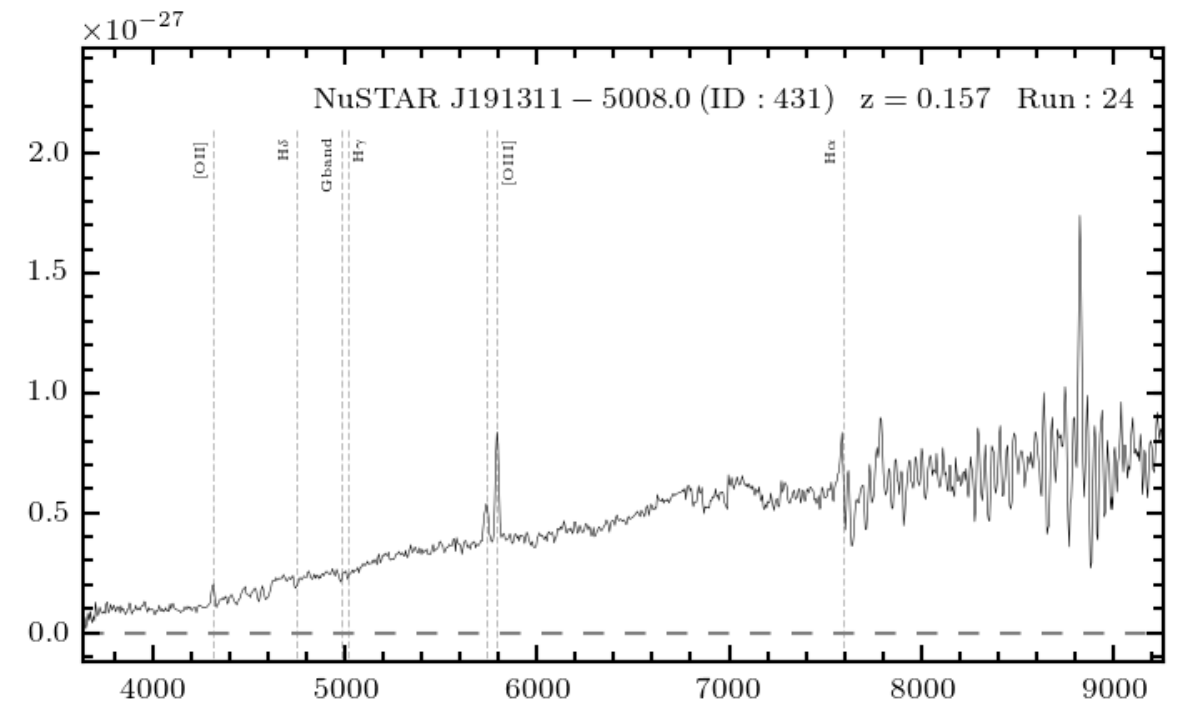}
\end{minipage}
\caption{Continued.}
\end{figure*}
\addtocounter{figure}{-1}
\begin{figure*}
\centering
\begin{minipage}[l]{0.325\textwidth}
\includegraphics[width=\textwidth]{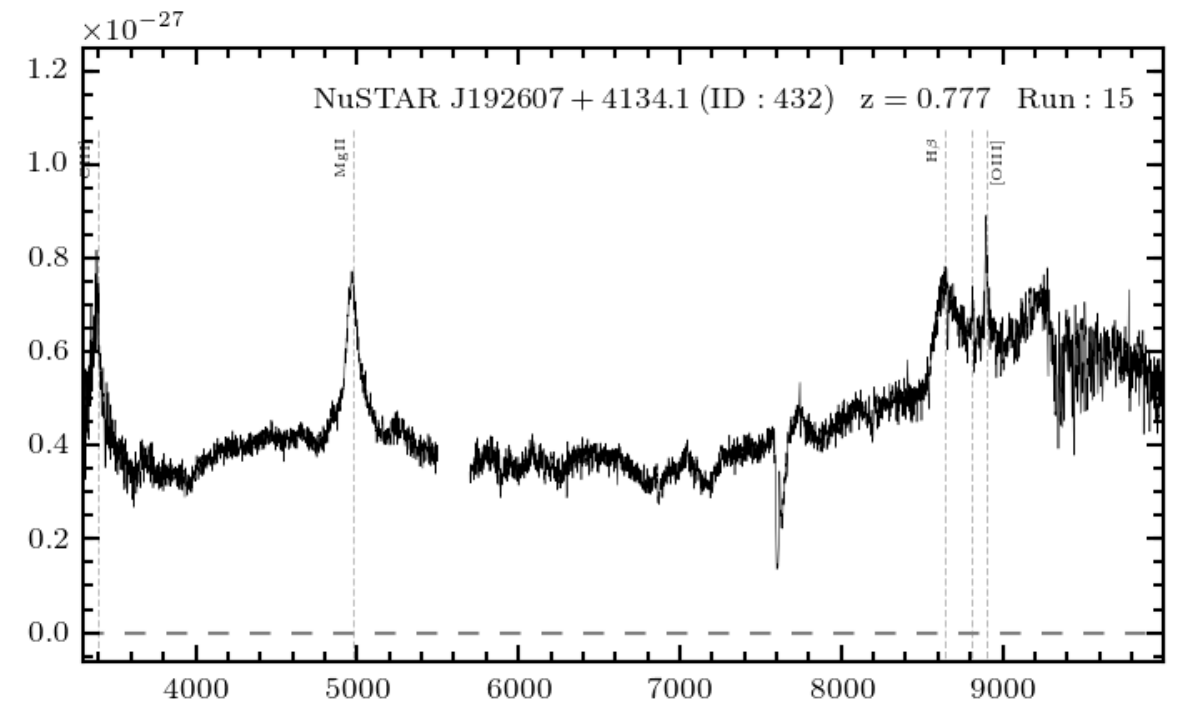}
\end{minipage}
\begin{minipage}[l]{0.325\textwidth}
\includegraphics[width=\textwidth]{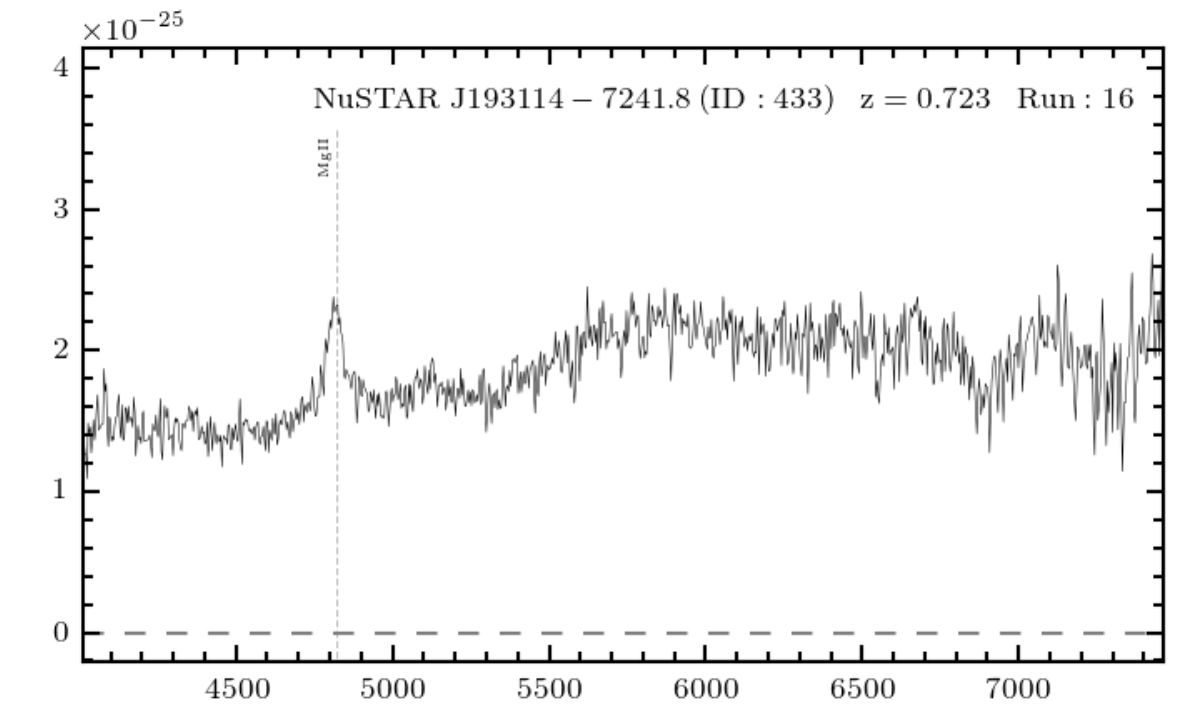}
\end{minipage}
\begin{minipage}[l]{0.325\textwidth}
\includegraphics[width=\textwidth]{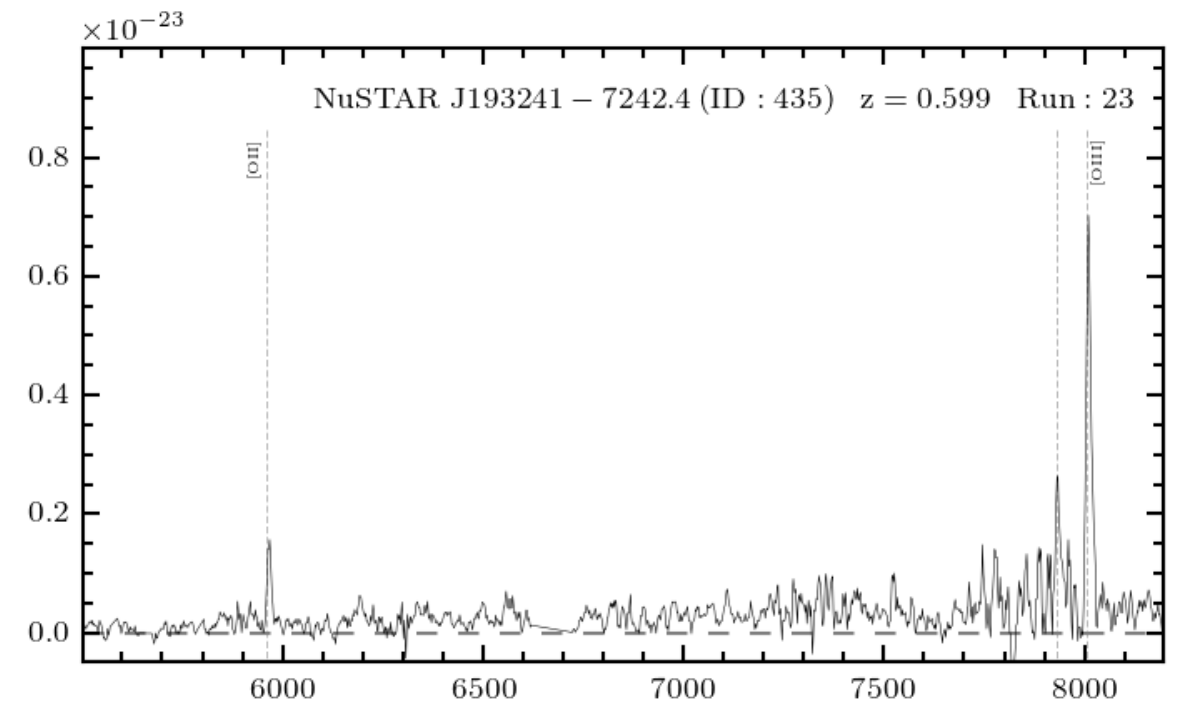}
\end{minipage}
\begin{minipage}[l]{0.325\textwidth}
\includegraphics[width=\textwidth]{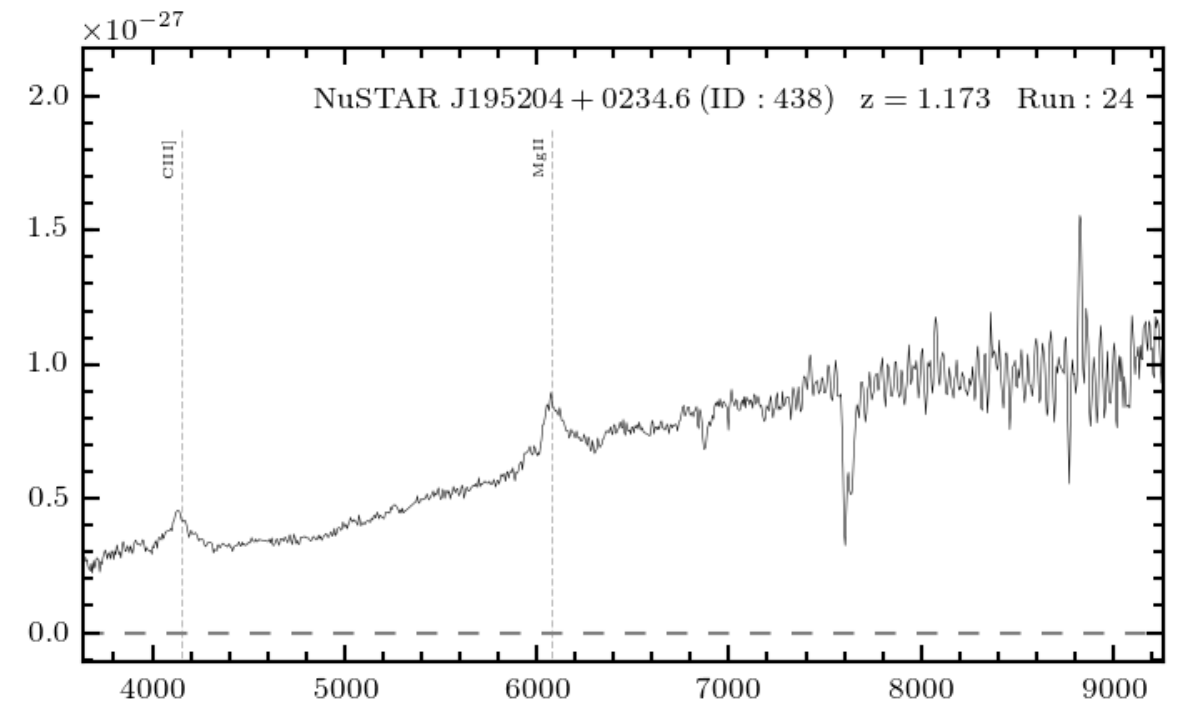}
\end{minipage}
\begin{minipage}[l]{0.325\textwidth}
\includegraphics[width=\textwidth]{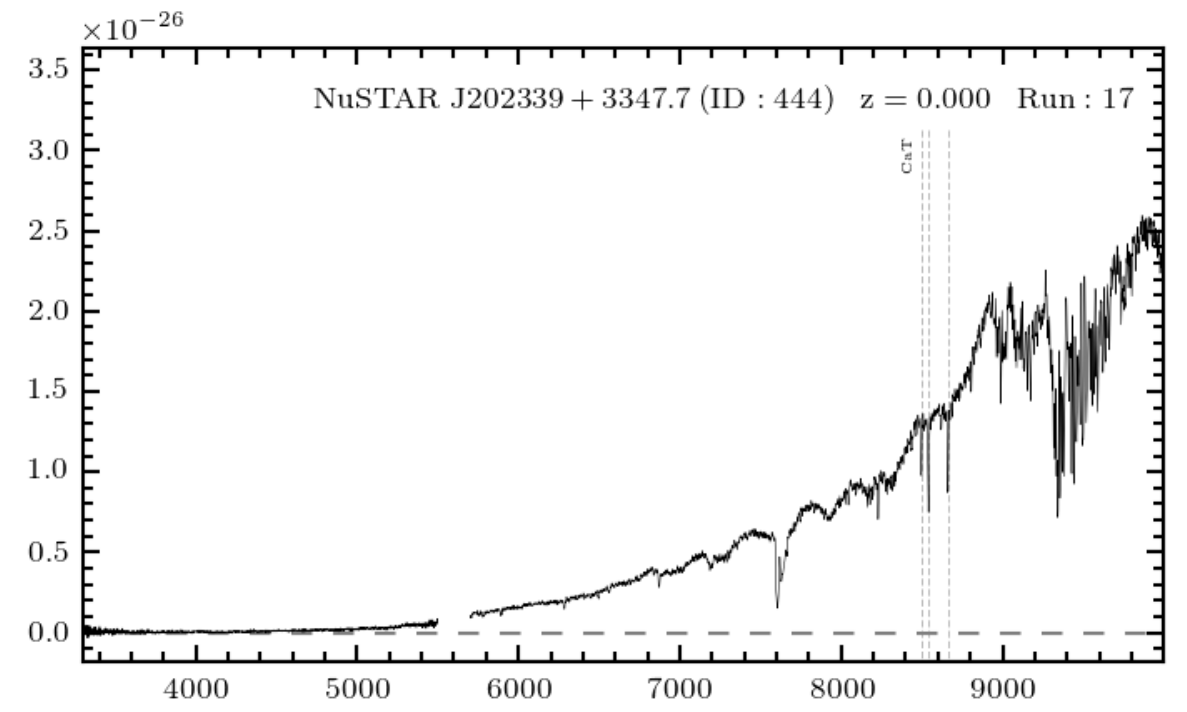}
\end{minipage}
\begin{minipage}[l]{0.325\textwidth}
\includegraphics[width=\textwidth]{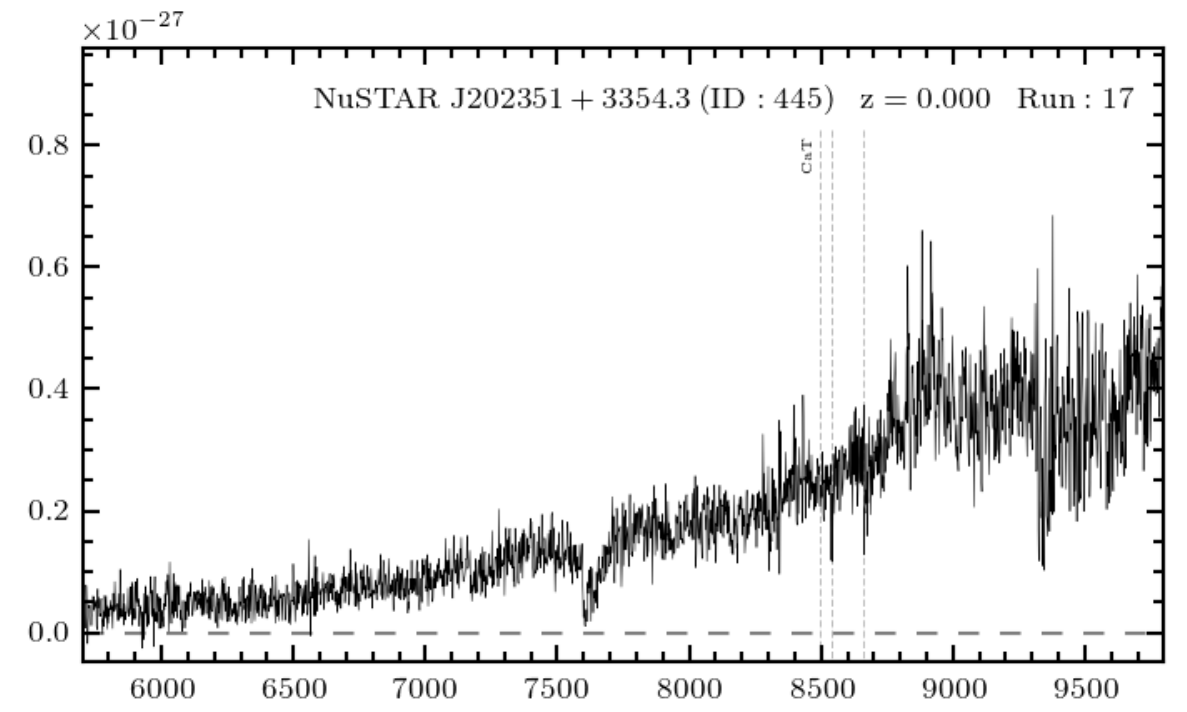}
\end{minipage}
\begin{minipage}[l]{0.325\textwidth}
\includegraphics[width=\textwidth]{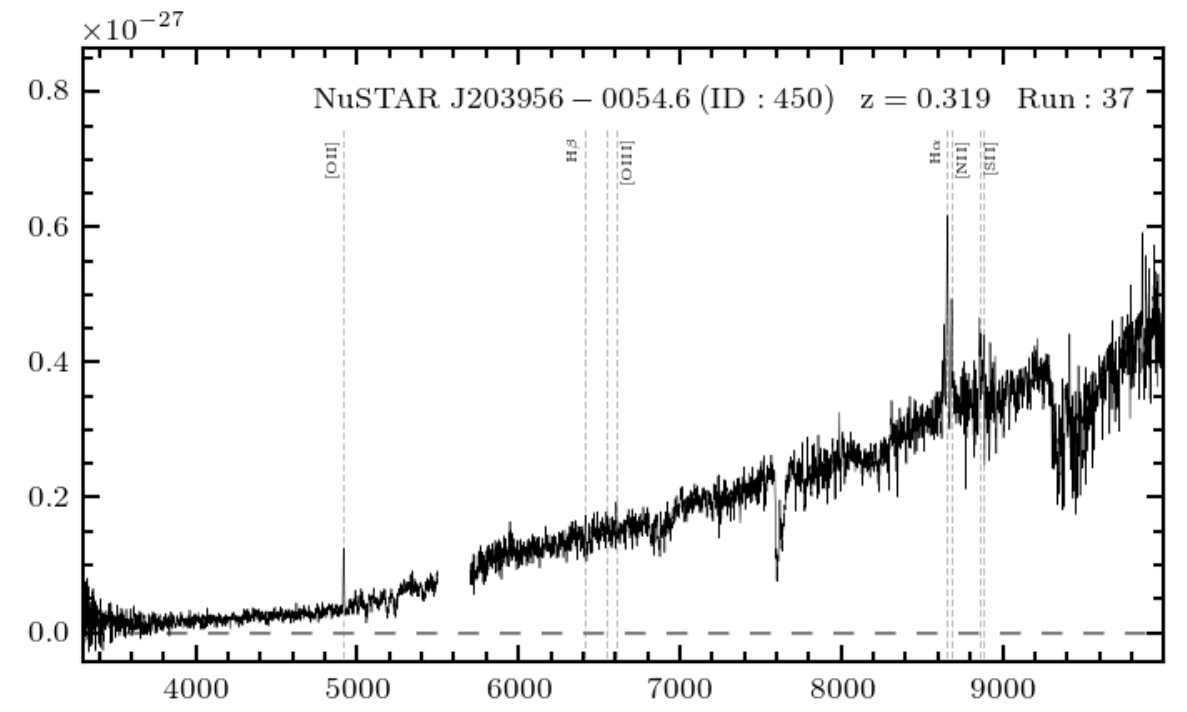}
\end{minipage}
\begin{minipage}[l]{0.325\textwidth}
\includegraphics[width=\textwidth]{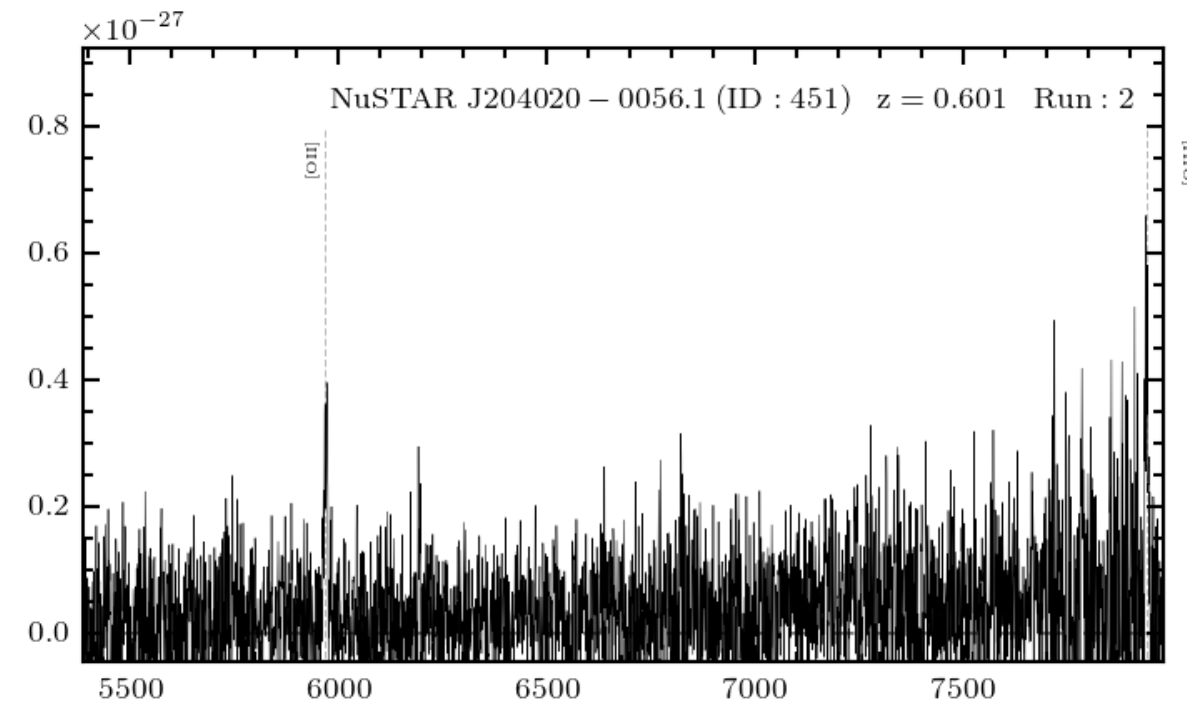}
\end{minipage}
\begin{minipage}[l]{0.325\textwidth}
\includegraphics[width=\textwidth]{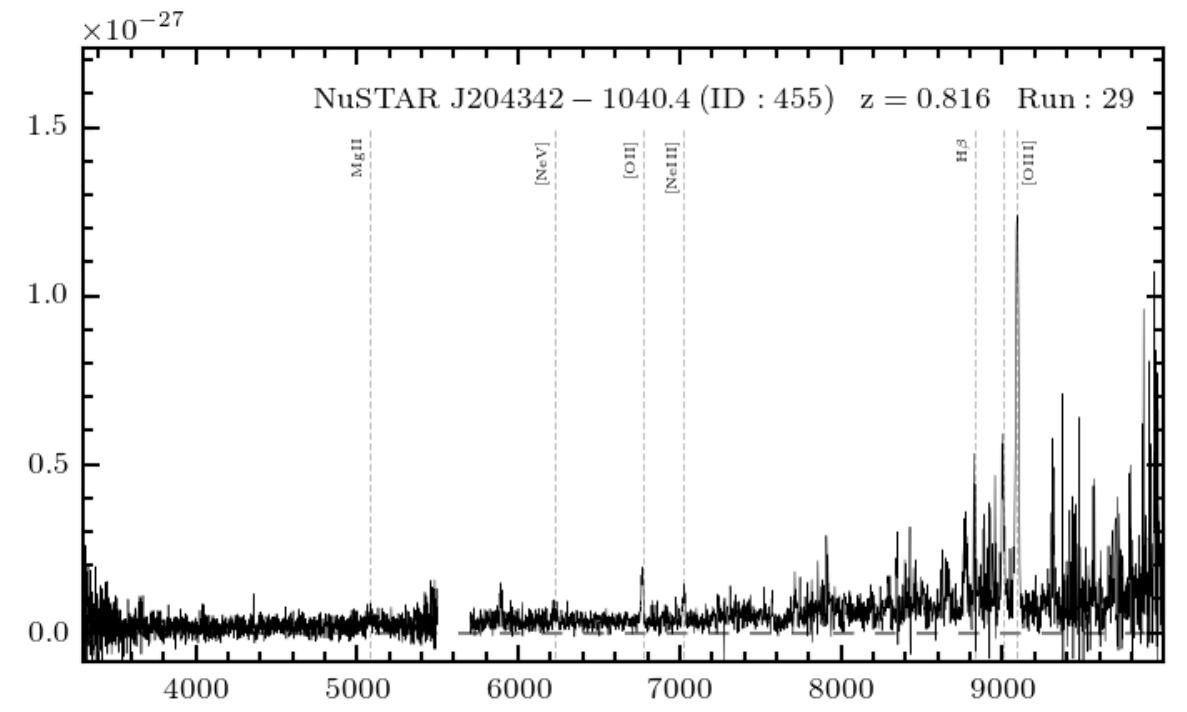}
\end{minipage}
\begin{minipage}[l]{0.325\textwidth}
\includegraphics[width=\textwidth]{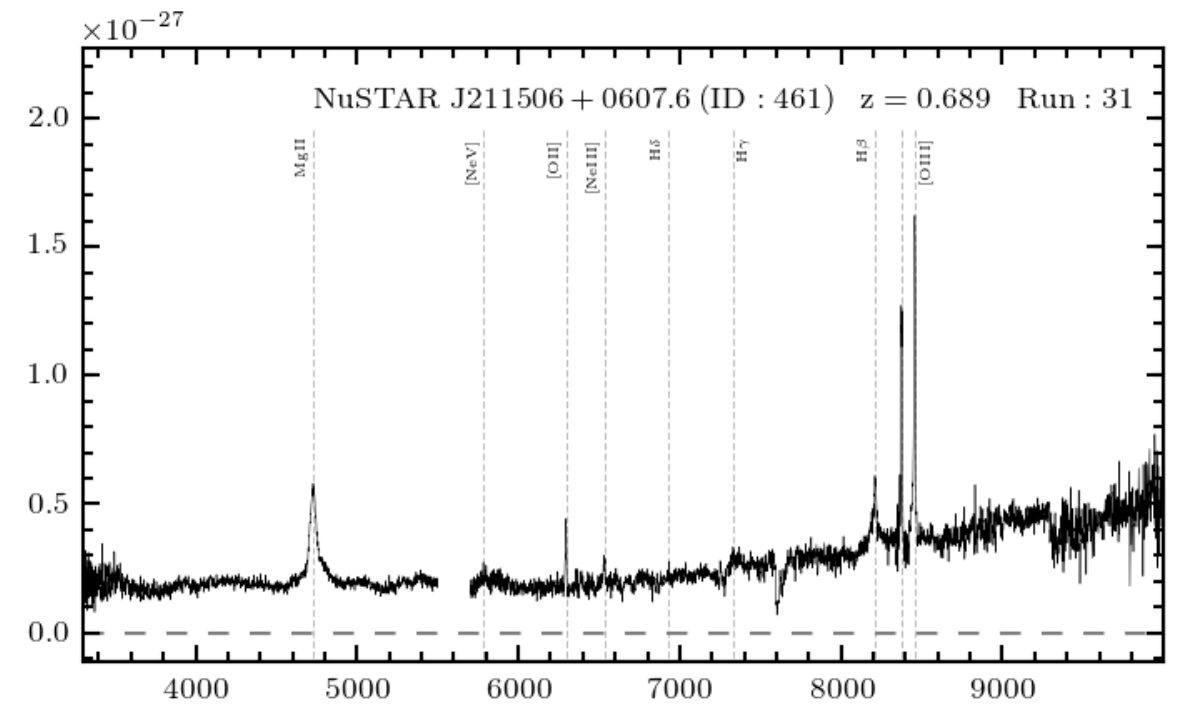}
\end{minipage}
\begin{minipage}[l]{0.325\textwidth}
\includegraphics[width=\textwidth]{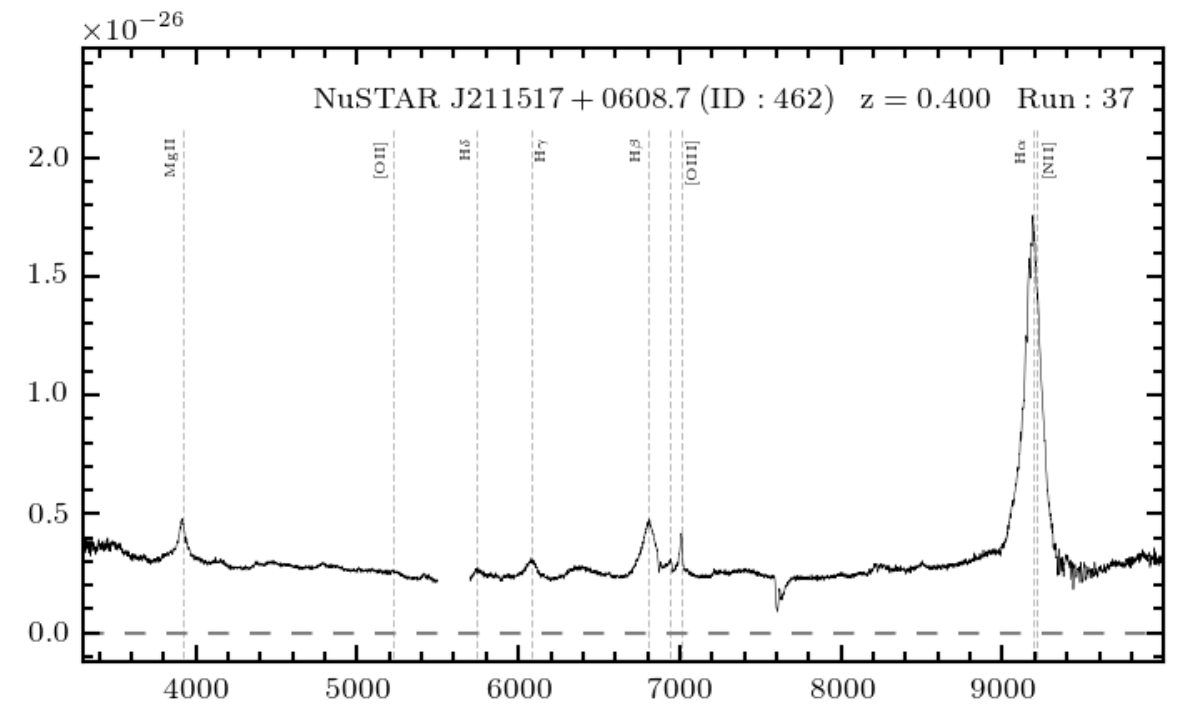}
\end{minipage}
\begin{minipage}[l]{0.325\textwidth}
\includegraphics[width=\textwidth]{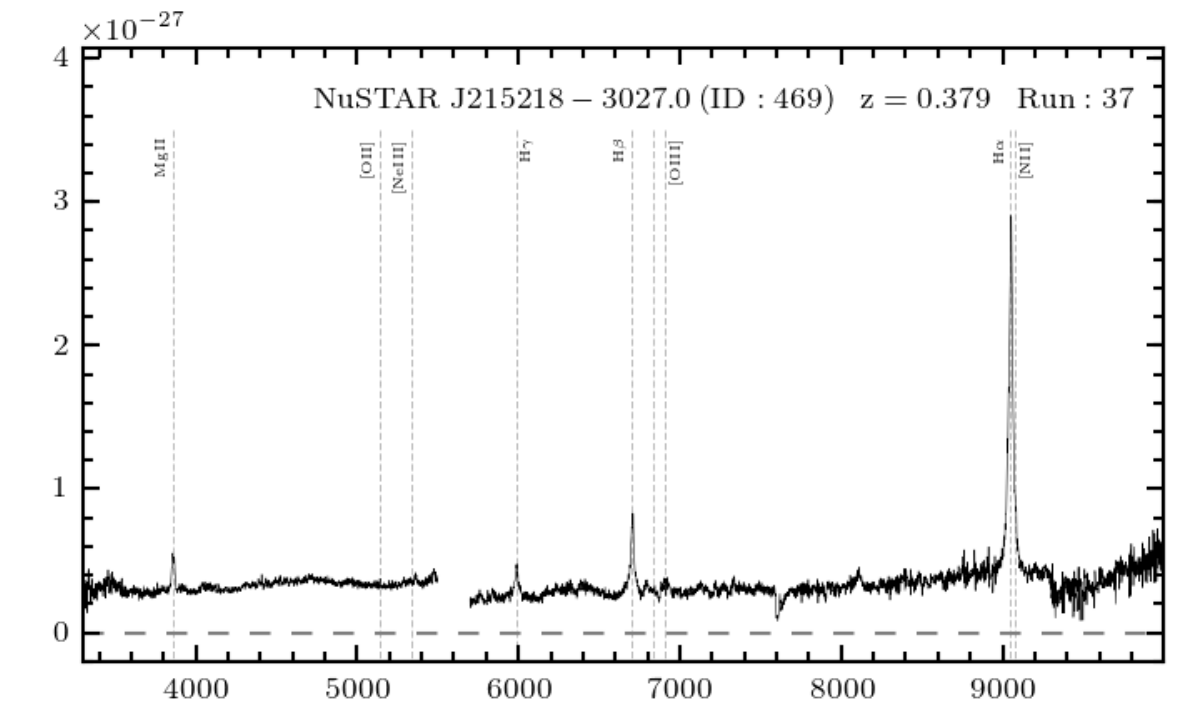}
\end{minipage}
\begin{minipage}[l]{0.325\textwidth}
\includegraphics[width=\textwidth]{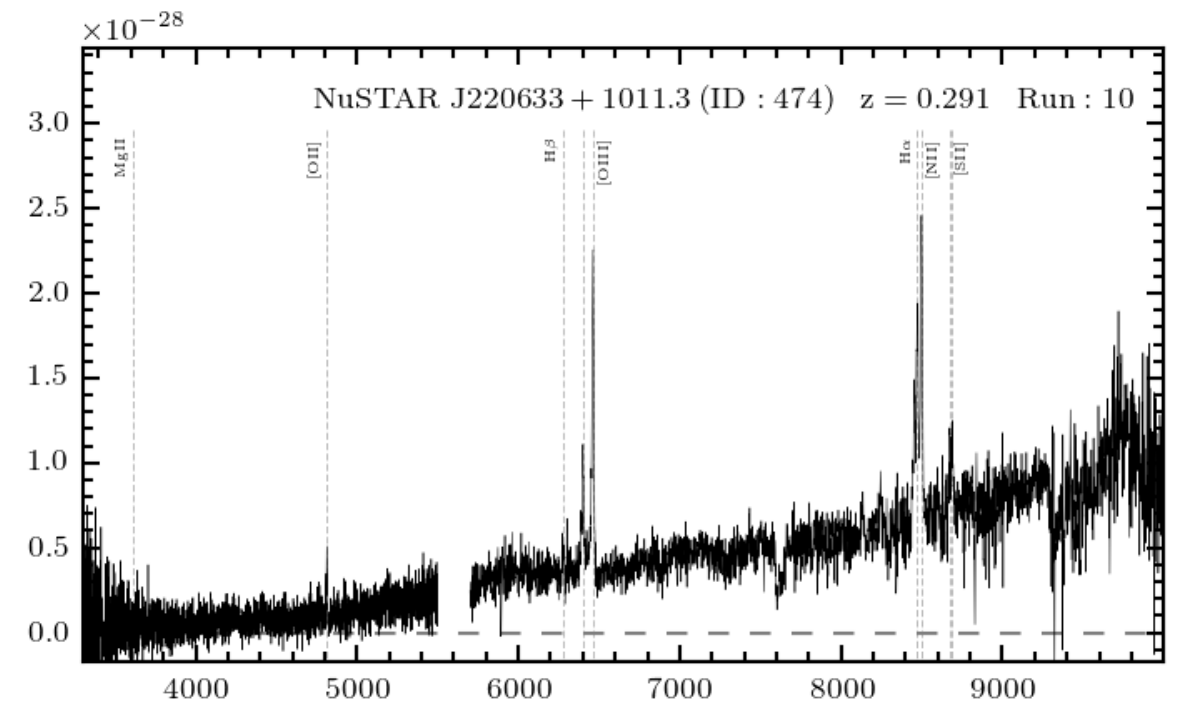}
\end{minipage}
\begin{minipage}[l]{0.325\textwidth}
\includegraphics[width=\textwidth]{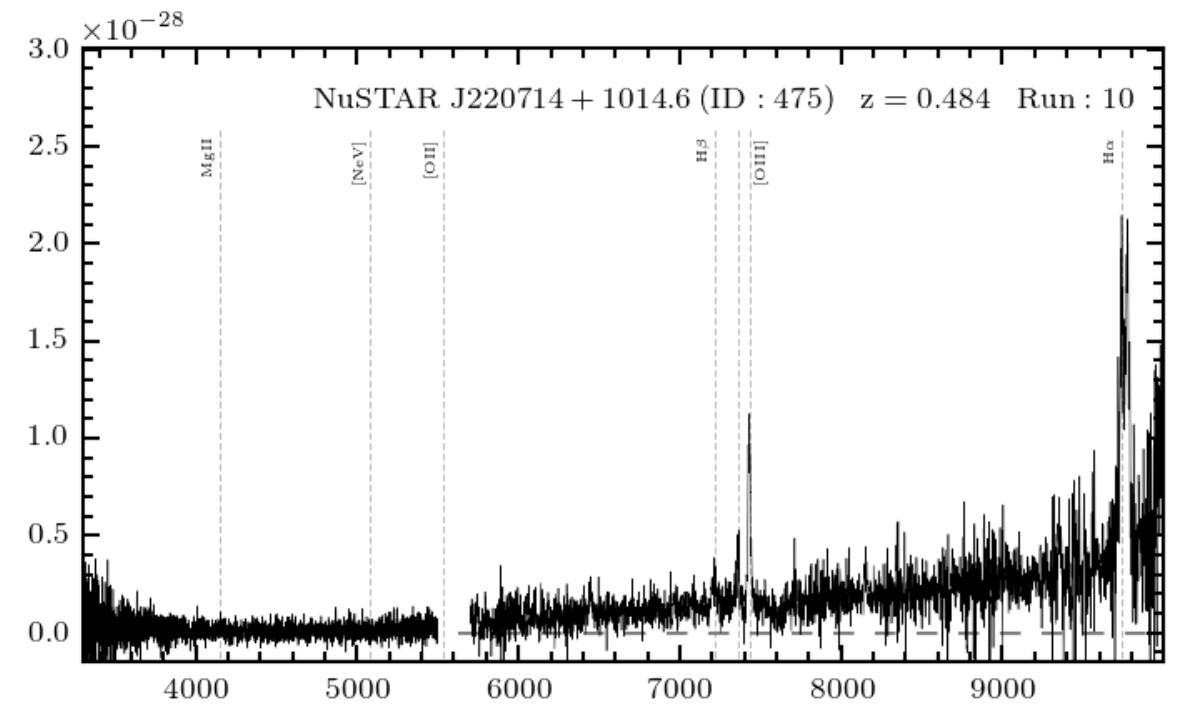}
\end{minipage}
\begin{minipage}[l]{0.325\textwidth}
\includegraphics[width=\textwidth]{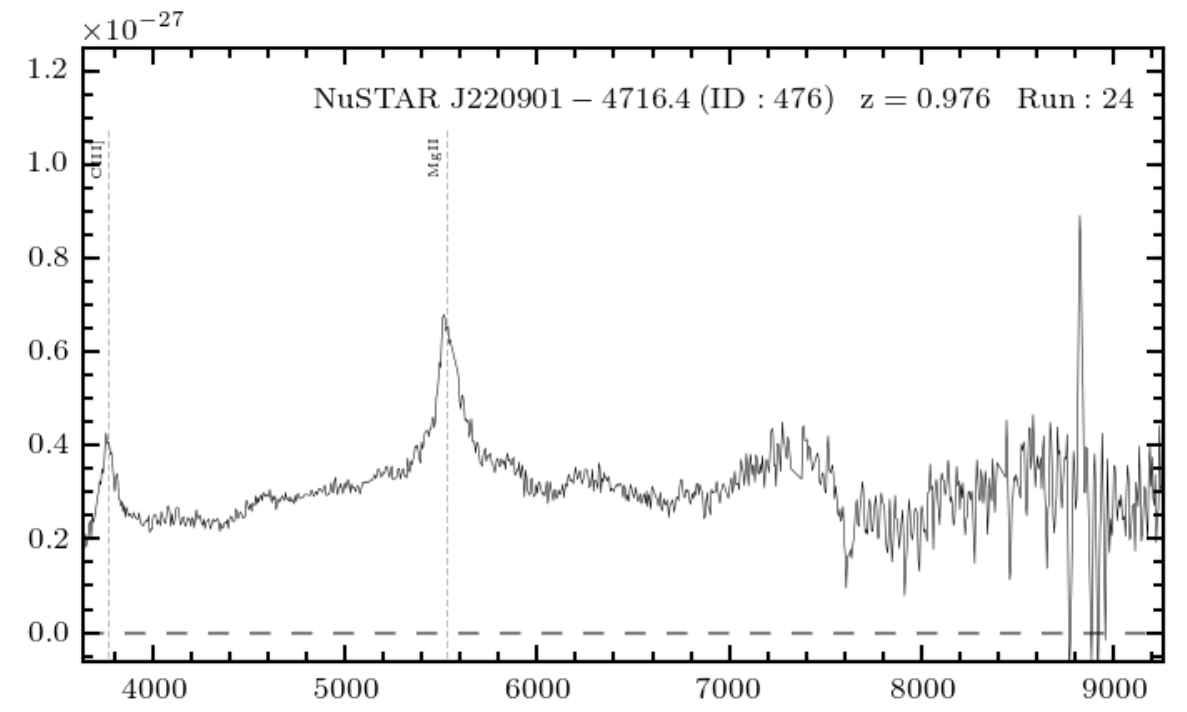}
\end{minipage}
\begin{minipage}[l]{0.325\textwidth}
\includegraphics[width=\textwidth]{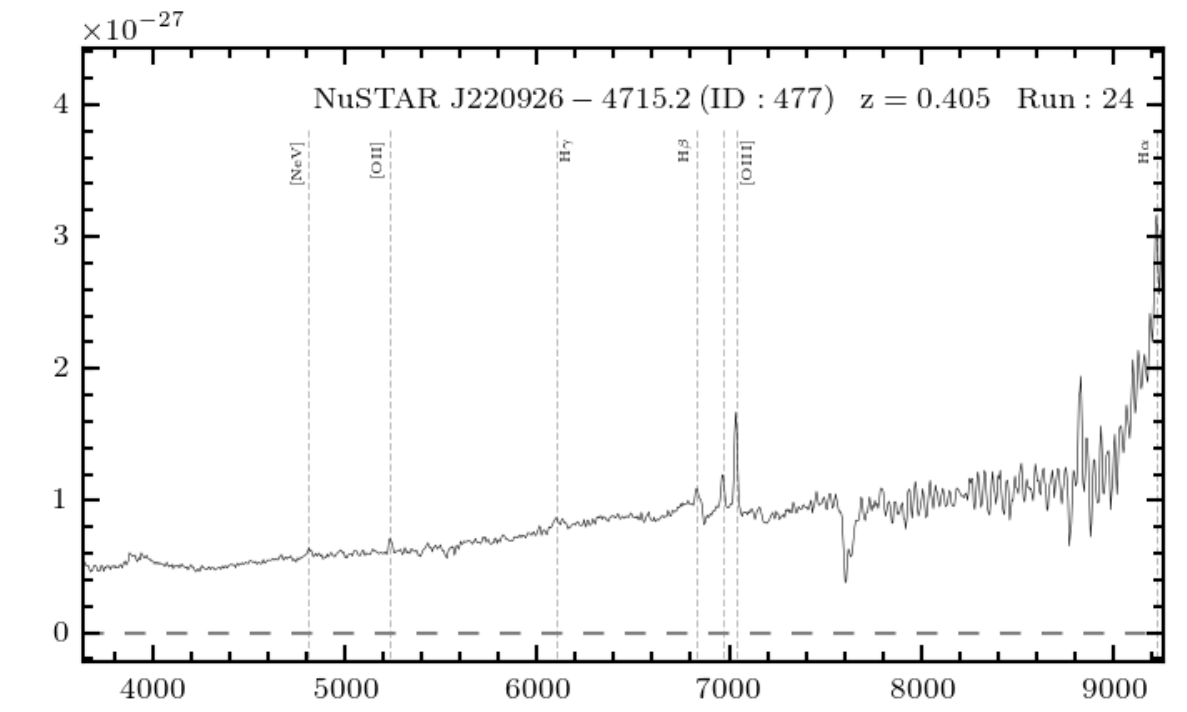}
\end{minipage}
\begin{minipage}[l]{0.325\textwidth}
\includegraphics[width=\textwidth]{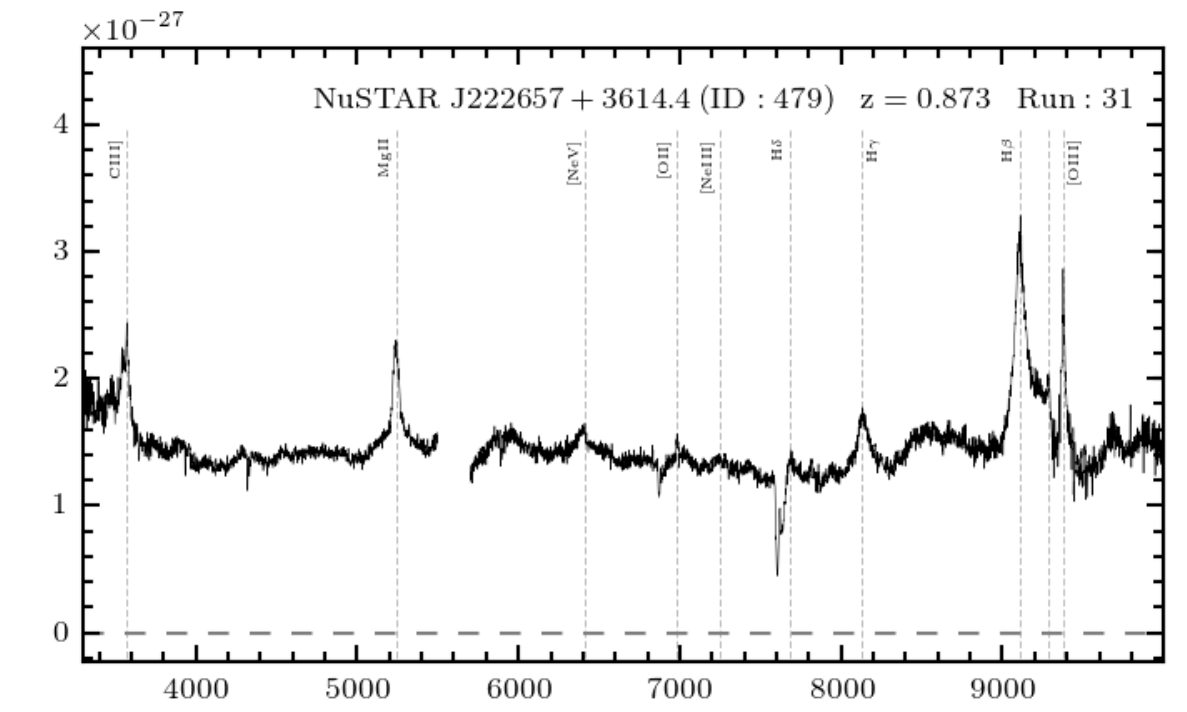}
\end{minipage}
\begin{minipage}[l]{0.325\textwidth}
\includegraphics[width=\textwidth]{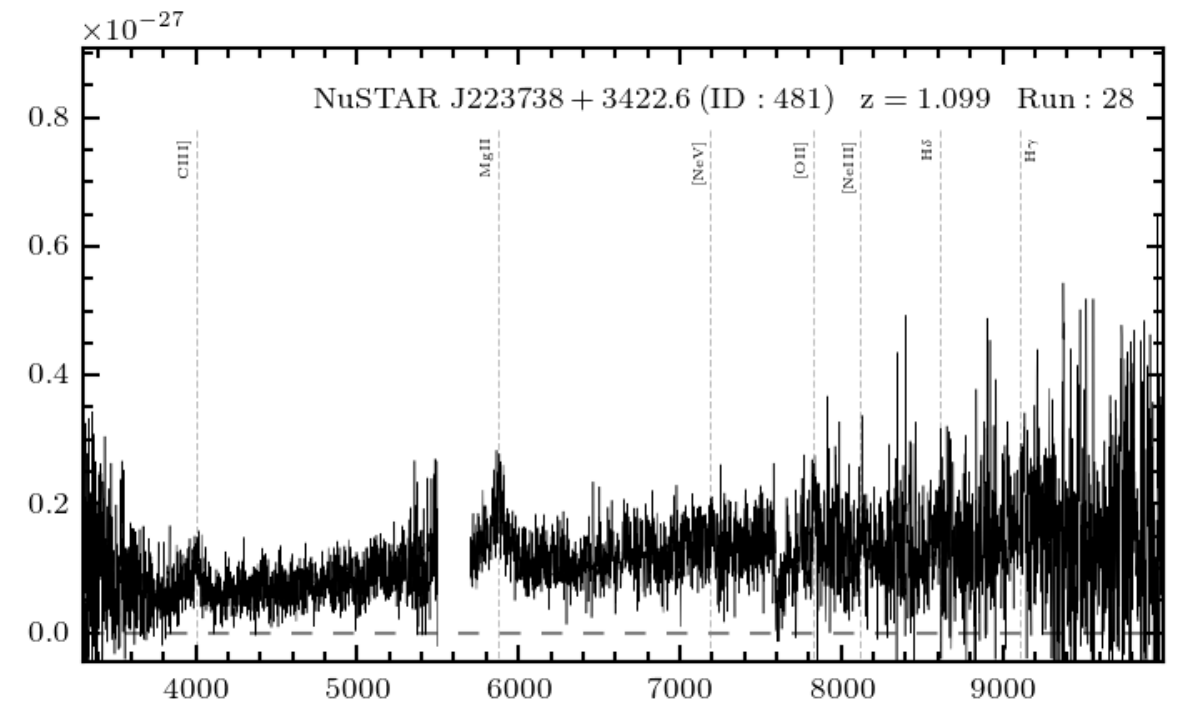}
\end{minipage}
\caption{Continued.}
\end{figure*}
\addtocounter{figure}{-1}
\begin{figure*}
\centering
\begin{minipage}[l]{0.325\textwidth}
\includegraphics[width=\textwidth]{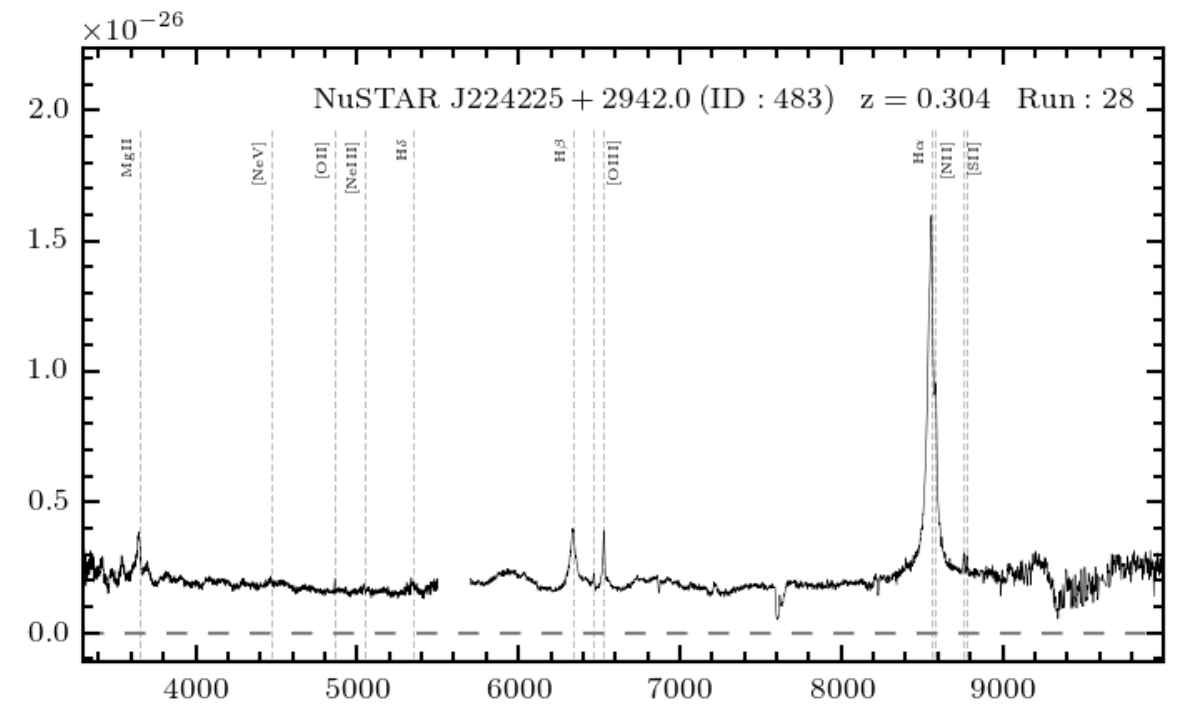}
\end{minipage}
\begin{minipage}[l]{0.325\textwidth}
\includegraphics[width=\textwidth]{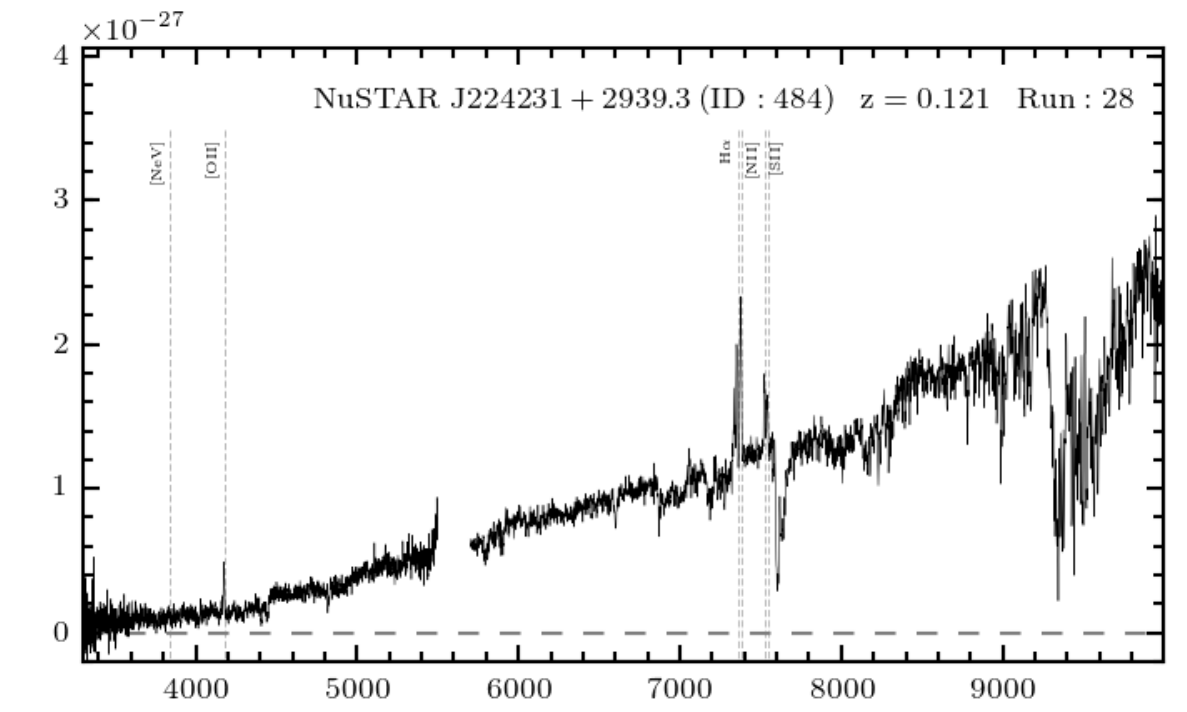}
\end{minipage}
\begin{minipage}[l]{0.325\textwidth}
\includegraphics[width=\textwidth]{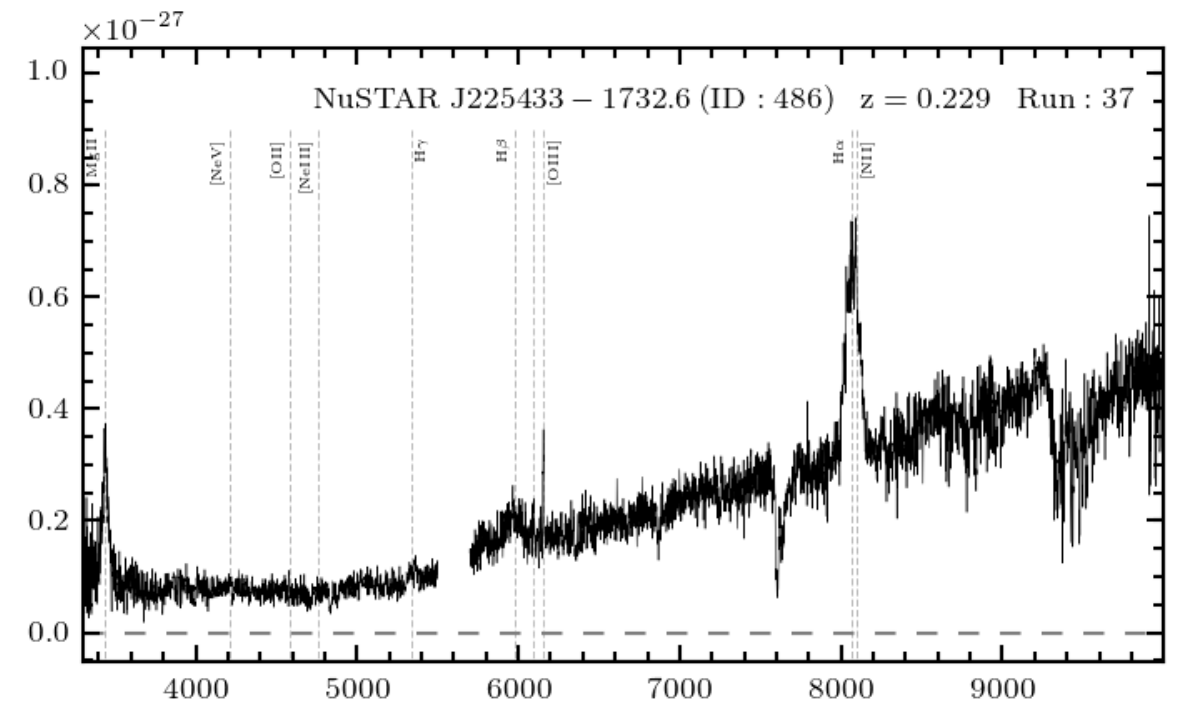}
\end{minipage}
\begin{minipage}[l]{0.325\textwidth}
\includegraphics[width=\textwidth]{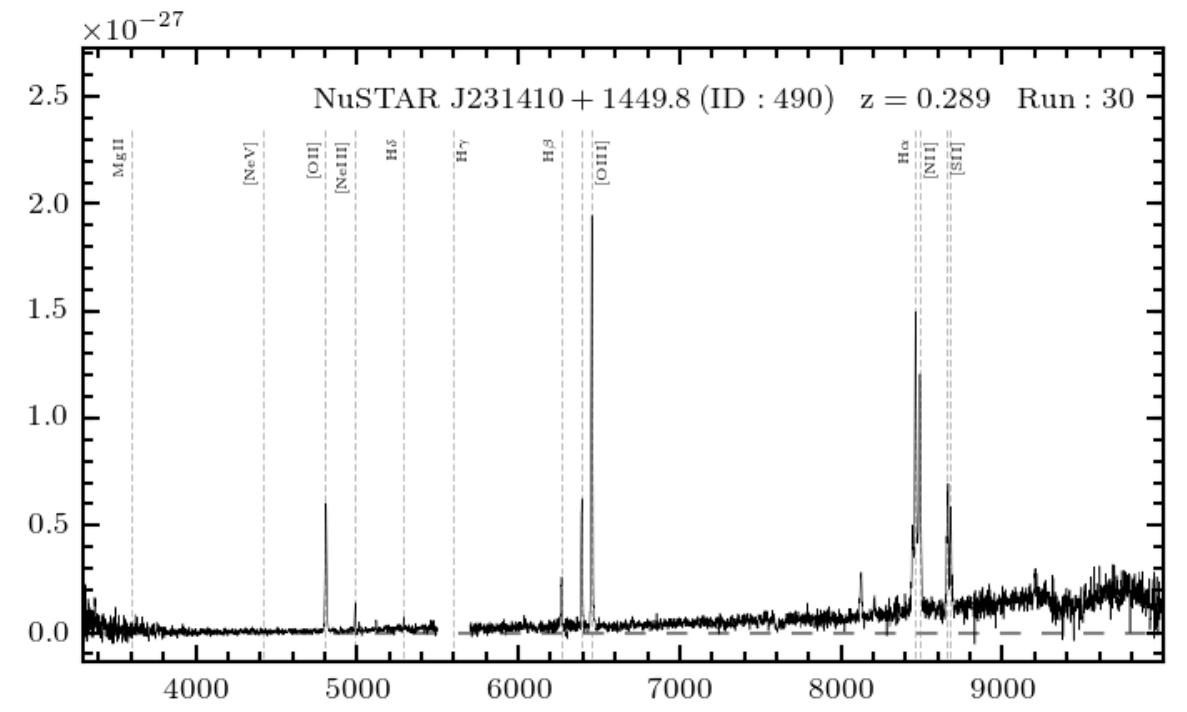}
\end{minipage}
\begin{minipage}[l]{0.325\textwidth}
\includegraphics[width=\textwidth]{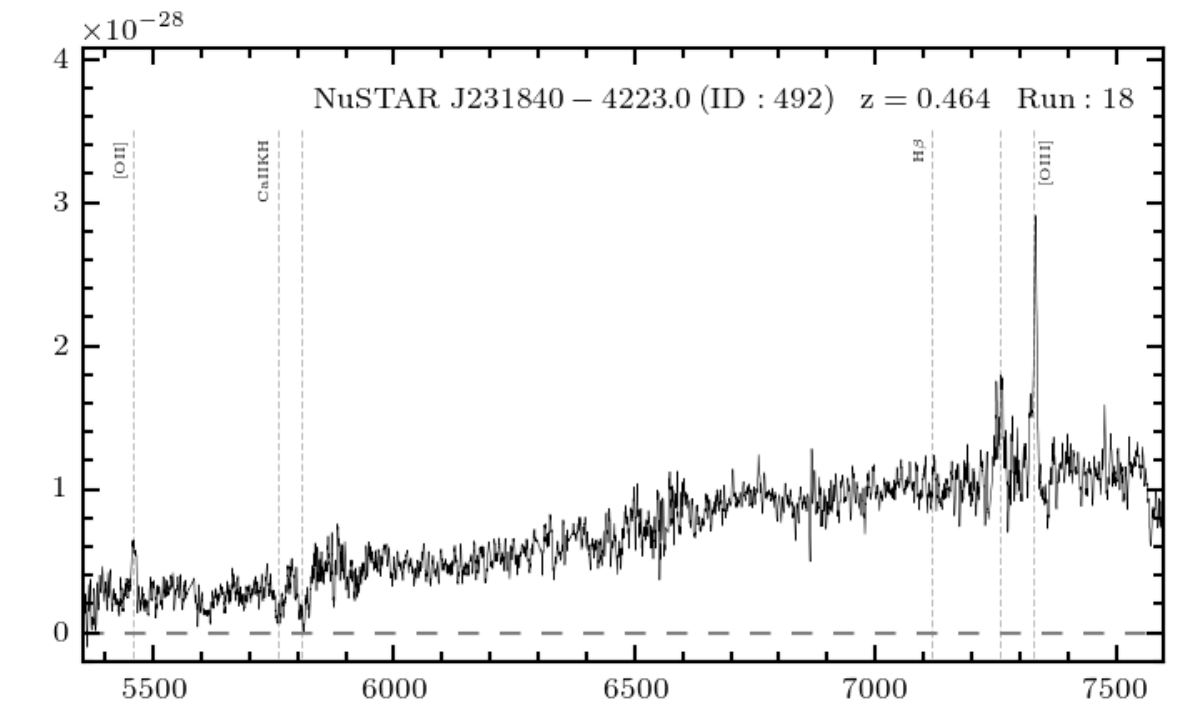}
\end{minipage}
\begin{minipage}[l]{0.325\textwidth}
\includegraphics[width=\textwidth]{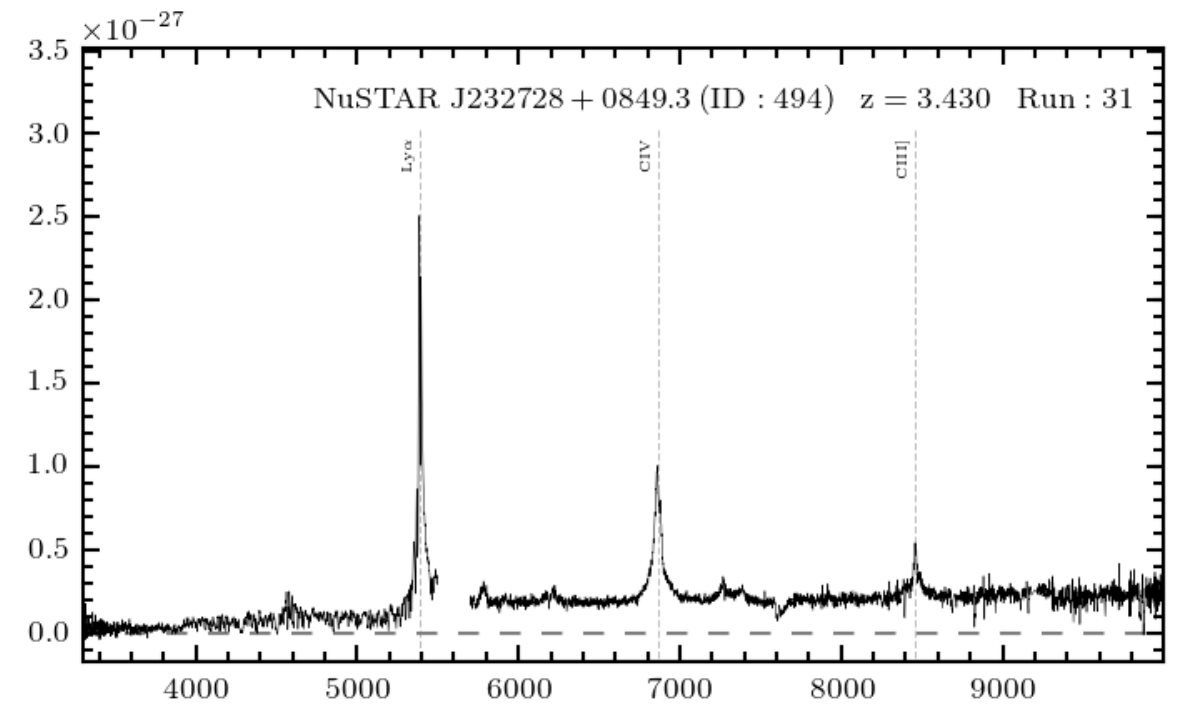}
\end{minipage}
\begin{minipage}[l]{0.325\textwidth}
\includegraphics[width=\textwidth]{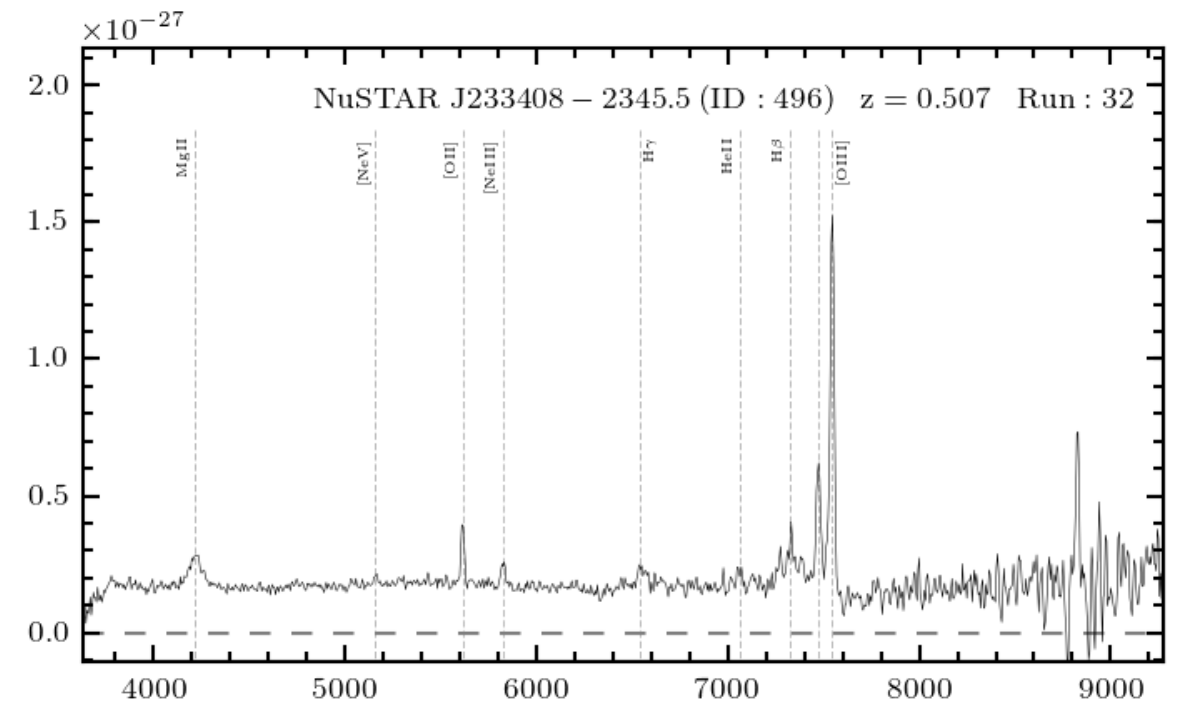}
\end{minipage}
\begin{minipage}[l]{0.325\textwidth}
\includegraphics[width=\textwidth]{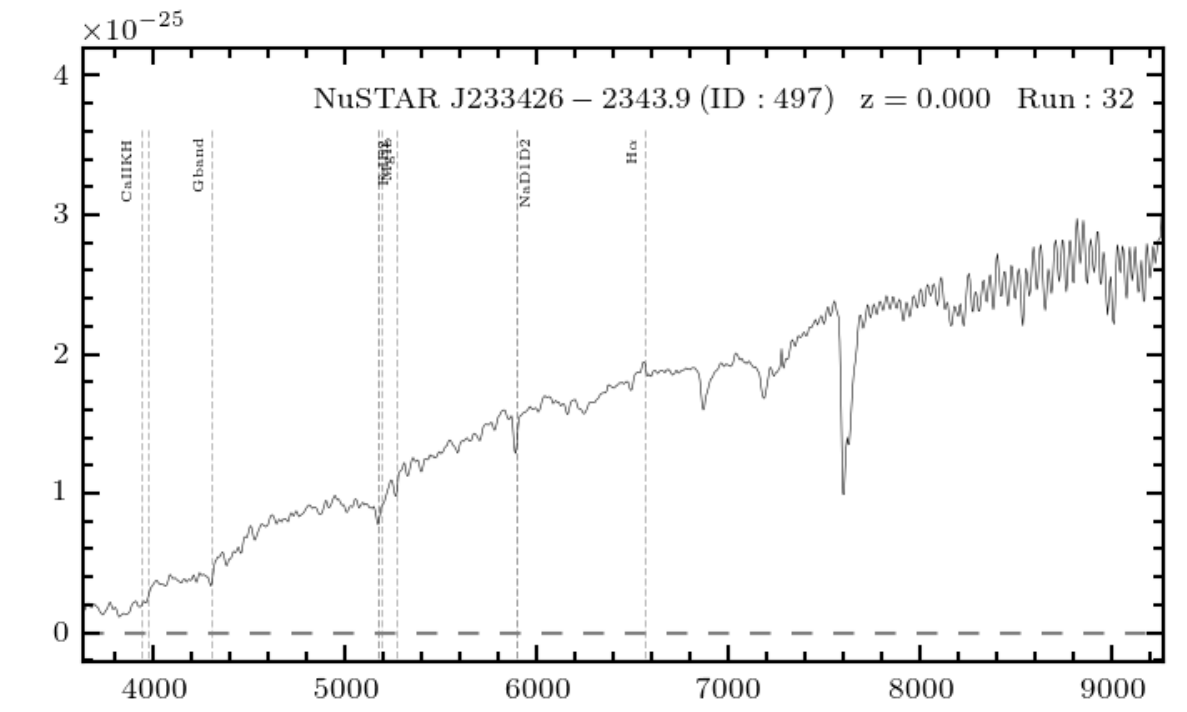}
\end{minipage}
\begin{minipage}[l]{0.325\textwidth}
\includegraphics[width=\textwidth]{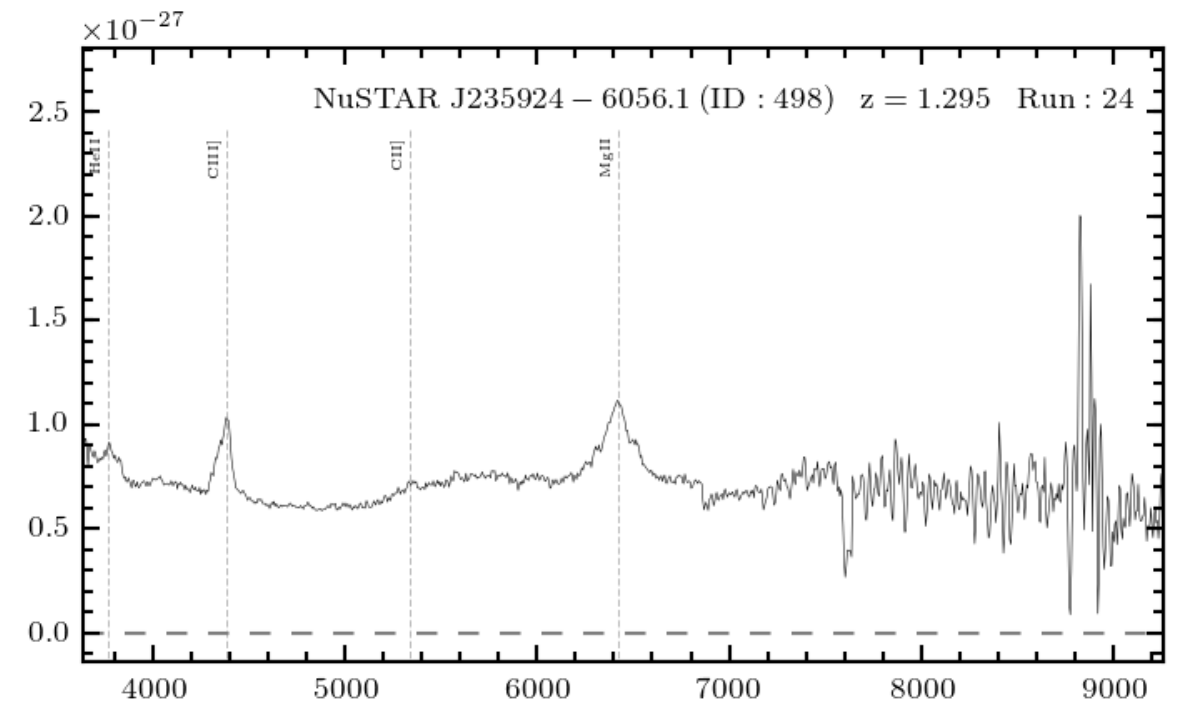}
\end{minipage}
\caption{Continued.}
\label{spec_all}
\end{figure*}

\section{A.3. Assessment of spurious optical and IR counterpart
  matches}

\begin{figure}
\centering
\includegraphics[width=1.0\textwidth]{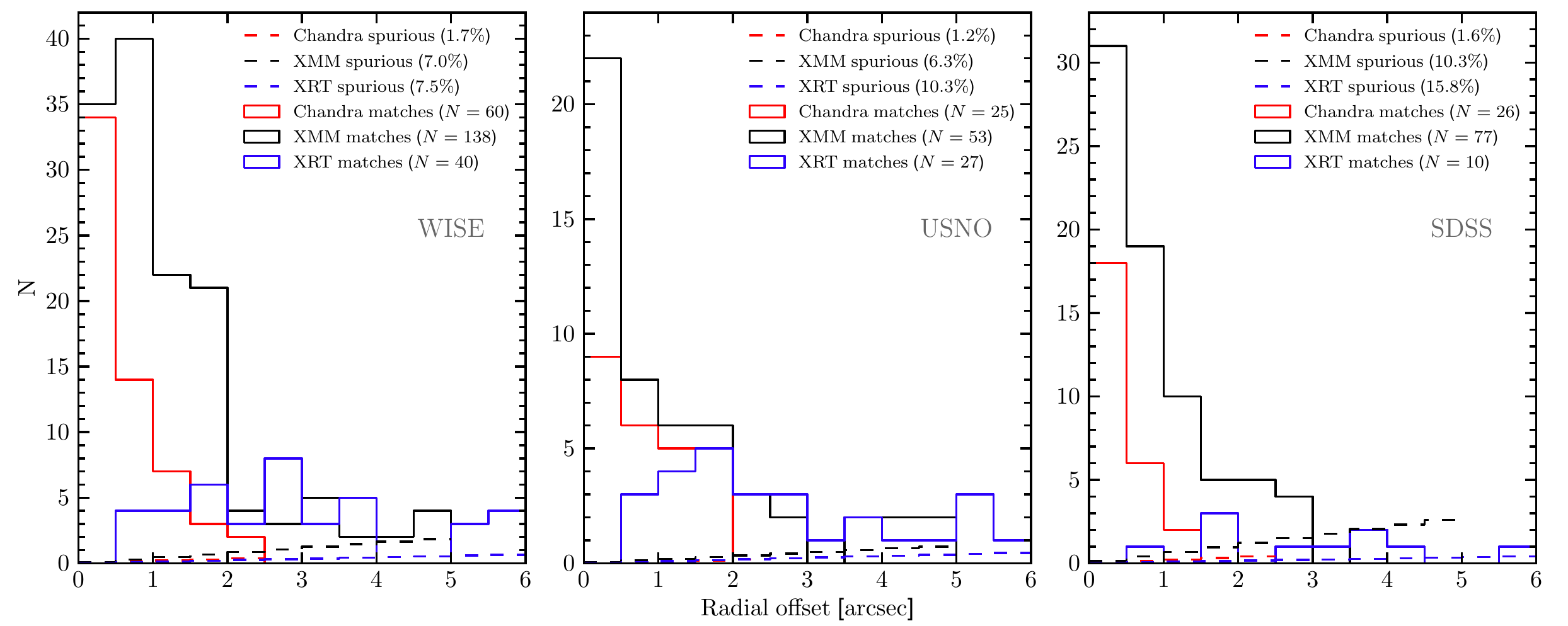}
\caption{Histograms showing the distributions of positional offsets
  between the soft X-ray (\chandra, \xmm, and \swiftxrt; red, black,
  and blue solid lines, respectively) positions and
  the matched \wise (left panel), USNOB1 (middle panel), and SDSS
  (right panel) counterparts, for the \nustar serendipitous survey
  sources with $|b|>10$\degrees. The dashed lines show the distributions
  expected for spurious matches (i.e., assuming there are no true IR or optical
  counterparts to the X-ray sources); these are calculated using the IR and
  optical source densities, taking median values
  across the range of sky positions for the \nustar serendipitous
  survey sources: $3.2$~\persqarcmin for \wise; $3.3$~\persqarcmin for
  USNOB1; and $8.1$~\persqarcmin for SDSS. The bracketed percentages
  show the inferred spurious fraction (i.e., the fraction of the
  X-ray sources with false IR or optical matches) for each subsample.}
\label{offsets_spurious}
\end{figure}

Figure \ref{offsets_spurious} shows histograms of the radial
offsets between soft X-ray counterpart (\chandra, \swiftxrt, and \xmm)
positions and the optical (SDSS and USNOB1) and IR (\wise) matches. We
compare to the radial offset distributions expected for
spurious matches, given the sky density of sources in the IR and optical
surveys, in order to estimate spurious matching fractions. 

\section{A.4. Assessment of spectroscopic completeness for the
  Type 2 fraction subsample}

Here we assess the effective spectroscopic completeness of the subset
of the \nustar serendipitous survey sample used to measure the \typeii
fraction (see Section \ref{type2fraction}). The subset is limited to
hard band ($8$--$24$~keV) selected sources at
redshifts of $0.1<z<0.5$ and luminosities of $2\times10^{43}<L_{\rm
  10-40keV}<2\times10^{44}$~\ergpersec, and includes $30$
spectroscopically identified sources (all NLAGNs or BLAGNs). To assess
the completeness, we must also consider the
unidentified sources which may or may not lie within these redshift
and luminosity ranges (i.e., sources labeled as crosses in
Figure \ref{R_fx}), and their reasons for lacking successful
spectroscopic followup. Since the majority of the spectroscopically identified sources in this
subsample lie at $R<20$, we consider all unidentified sources with
$R<20$ (we conservatively include sources with lower limits in $R$) as potentially lying
within the redshift and luminosity ranges stated above. 
There are $10$ such unidentified sources in total. 
This includes one likely BL~Lac type object which we exclude due to
the possibility of beaming.
A further four of the unidentified sources can be safely
excluded without biasing the \typeii fraction measurement: one
of these has an unambiguous optical counterpart,
and simply has not yet been targetted with ground-based facilities;
two have not been targetted due to the lack of \chandra or \xmm
coverage, required to distinguish between multiple optical
counterparts within the large \nustar positional error circle; and for
one obtaining spectroscopy is problematic due to the close
proximity of brighter optical sources. There are five remaining unidentified sources to consider, where
followup has not been performed due to the lack of a detection in the
available \chandra and/or \xmm coverage (and therefore the lack of an
accurate X-ray position). 
Four of these have high \fprob values at $8$--$24$~keV ($\log P_{\rm False}=-6.5$ to $-6.0$;
i.e., close to our detection threshold of $\log P_{\rm False}=-6.0$)
and comparably deep \chandra or \xmm coverage, which
indicates that they are likely to be spurious sources. 
The remaining single source 
is strongly detected at $8$--$24$~keV ($\log P_{\rm
  False}=-9.6$), and has relatively low quality \chandra and \xmm
coverage, so is consistent with being a genuine astrophysical
source.
We therefore consider the effective spectroscopic
completeness of this subsample to be $97$--$100\%$ (i.e., $30/31$ or
$30/30$, depending on whether the final source lies above or below
$R=20$, since there is only a lower limit in $R$).

\section{A.5. J0650-- A low $L_{\rm X}/L_{\rm MIR}$, likely X-ray weak NLSy1}

Here we consider an outlier in X-ray to MIR luminosity
ratio, NuSTAR~J065003+6046.8 (hereafter J0650; $z=0.319$). 
For this source, the upper limit of $L_{\rm 10-40keV}/L_{\rm 6\mu m}<0.05$ lies
below the CT AGN threshold (as shown in Figure \ref{lx_lmir}). In
other words, the hard X-ray luminosity is very weak
compared to that expected based on the MIR luminosity ($L_{\rm 6\mu
  m}=3.7\times 10^{44}$~\ergpersec). The source is not detected in
the full and hard \nustar bands, but is weakly detected in the soft
band ($\log P_{\rm False}=-6.9$; $\approx 25$ net source counts, for an
effective exposure time of $16$~ks), suggesting a relatively
steep spectral slope. The
properties of the counterparts at X-ray, IR, and optical wavelengths
(see below)
add confidence that the \nustar detection is not spurious. 
J0650 has a strongly detected \xmm counterpart, the $0.5$--$10$~keV
spectrum of which has $107$ net source counts (for an $8$~ks
exposure). A power law fit provides a statistically acceptable fit to
the \xmm spectrum 
($C/n=139/159$), and
the photon index is constrained to be $\Gamma_{\rm eff}=3.1\pm 0.6$, which is
very steep and above the typical range observed for AGNs.
For the $3$--$10$~keV band, where \nustar and \xmm overlap in
sensitivity, the source is undetected with \xmm, with $<19.5$ EPIC counts
overall and a $3$--$8$~keV flux upper limit of $<1.5\times
10^{-14}$~\fluxunit ($99\%$ CL). 
This is significantly lower than our photometric \nustar flux of $4.8 \pm
1.6 \times 10^{-14}$~\fluxunit in the $3$--$8$~keV band. 
The disagreement could in part result from X-ray
variability (between the 2003 \xmm observation and the 2014 \nustar
observation), which is especially likely in this case given the NLSy1
optical classification (see below). The \nustar flux is also likely 
boosted by the Eddington bias, which we have established to be
significant at this low $3$--$8$~keV flux level (see Figure \ref{fnu_fsoft}, and C15).

The X-ray luminosities measured at high and low energies are $L_{\mathrm{10-40\
    keV}}^{\rm obs}<2.0\times 10^{43}$~\ergpersec (from \nustar
photometry) 
 and $L_{\mathrm{2-10\ keV}}^{\rm obs}=2.6\times 10^{42}$~\ergpersec (from the \xmm
spectrum), 
respectively. 
Given the AGN \sixum luminosity measured from \wise photometry
($L_{\rm 6\mu m}=3.7\times 10^{44}$~\ergpersec; the source is
AGN-dominated at this wavelength based on the \wise colors),
these suggest comparatively low X-ray to MIR luminosity ratios, with 
the $2$--$10$~keV and $10$--$40$~keV X-ray luminosities potentially
suppressed by factors of $\approx 50$
and $\gtrsim 7$, respectively, with respect to the
intrinsic relations for AGNs (see Figure \ref{lx_lmir}). 
In the case of the $2$--$10$~keV luminosity, the low value is in part due to
the relatively steep soft X-ray spectral slope. 
If the apparent X-ray suppression were
due to AGN absorption we would expect a
flat X-ray spectral slope ($\Gamma_{\rm eff}<1$), but the observed spectral
slope is comparatively steep ($\Gamma_{\rm eff}\approx 3$). 
One possibility is that the source is an intrinsically X-ray weak,
unobscured AGN. As described below, the source shows the properties of
a NLSy1 in the optical, and intrinsic X-ray weakness has been
identified for some objects in this class (e.g., \citealt{Miniutti12,
  Leighly07a, Leighly07b}). 

Further evidence for the presence of
an AGN in J0650 is given by the \wise colors, which place it
firmly within the MIR AGN selection regions ($W{\rm 1}$--$W{\rm
  2}=1.2$; $W{\rm 2}$--$W{\rm 3}=3.2$). The source is also
comparatively bright in the longer wavelength \wise bands ($W{\rm
  3}=9.50\pm 0.03$ and $W{\rm 4}=7.24\pm 0.07$). On the basis of our
results for \wise colors as a function of X-ray luminosity (Section
\ref{wise_colors}) J0650 is statistically highly likely to have an
intrinsic X-ray luminosity of $L_{\mathrm{10-40\ keV}}^{\rm
  int}>10^{44}$~\ergpersec. The fact that the observed X-ray
luminosity is so much lower may be explained by a combination of
intrinsic X-ray weakness and the steep spectral slope at low
energies, the latter of which may result in a relative increase in
the dust-heating photons which are reprocessed into the MIR
waveband.

Key information for this object is provided by our Keck
spectrum, which reveals a likely NLSy1 AGN.
We detect multiple strong emission lines, from \mgii
at the blue end to \halpha and \nii at the red end.
The source satisfies the conventional NLSy1 definition, with a
relatively narrow \hbeta line ($\mathrm{FWHM}\approx
1710$~km~s$^{-1}$), 
and a low \oiii~$\lambda$5007$/$\hbeta flux ratio
\citep[e.g.,][]{Goodrich89}. 
There are also multiple strong \feii emission lines, another
characteristic feature of NLSy1s \citep[e.g.,][]{Zhou06}. Notably, the
\oiii~$\lambda$5007 line is contaminated by 
strong Fe emission. 
NLSy1s are associated with low black hole masses and high
accretion rates \citep[e.g.,][]{Pounds95,Boller96,Mathur00}, and typically have
significantly steeper X-ray spectral slopes than normal unobscured AGNs
\citep[e.g.,][]{Boller96,Brandt97a}. The latter property is congruous with our
measurement of an extremely steep X-ray photon index for J0650
($\Gamma_{\rm eff}\approx 3$).

\section{A.6. Description of the Secondary Source Catalog}

Here we provide a secondary catalog of $64$ \nustar sources identified
using an independent source detection approach. This independent
(or ``secondary'') method uses \wavdetect to search for significant
emission peaks in the FPMA and FPMB data separately (see Section 2.1.1
of \citealt{Alexander13}) and in the combined A+B data. The method
was developed alongside the primary one (Section \ref{nu_srcdet} of
this paper) in order to investigate the
optimum source detection methodologies for \nustar, and to identify
sources in regions of the \nustar coverage which are automatically
excluded in the primary source detection. 
We emphasise that these secondary sources are
not used in any of the science analyses presented in this paper. The
results in this work therefore correspond to a single, well-defined sample.
Nevertheless, these secondary sources are robust \nustar detections,
some of which will be incorporated in future \nustar studies
(e.g., Chen et al., submitted; Tomsick et al., in prep.), and many for
which ($35$ out of the $43$ sources with spectroscopic
identifications) 
we have obtained new spectroscopic redshifts and
classifications through our followup program. 

The columns for the secondary source catalog are
summarized in Table \ref{secondary_columns_table}. The \nustar columns
are equivalent to the primary catalog columns described in Section
A.1, with the exception that the count rates (columns 20--25) are
aperture-corrected values. The photometric columns are blank where
the A+B data prohibit reliable photometric constraints.
The final column assigns a character to each source, indicating the 
reason for not being included in the primary catalog. These are
categorised into four groups: 
(E) the source is within or very close to the peripheral region
of the \nustar mosaic, which is excluded from the primary source
detection ($33\%$ of cases); (T) the source is narrowly offset from
the central science target position for the
\nustar observation (and thus automatically excluded; see Section \ref{nu_srcdet}), or from another bright source in the field
($11\%$); (X) the source lies in a region which is masked out, or
in a \nustar field which is excluded, from the primary source detection
($44\%$; e.g., due to highly contaminating stray light or a bright science
target); or (L) the source has a comparatively low detection significance
($12\%$).

\begin{table}
\centering
\caption{Column Descriptions for the Secondary {\it NuSTAR} Serendipitous Source Catalog}
\begin{tabular}{ll} \hline\hline \noalign{\smallskip}
Column number & Description \\
\noalign{\smallskip} \hline \noalign{\smallskip}
1 & Unique source identification number (ID). \\
2 & Unique \nustar source name. \\
3, 4 & Right ascension (R.A.) and declination (decl.). \\
5--16 & Total, background, and net source
        counts for the three standard
        energy bands, and associated errors. \\
17--19 & Net vignetting-corrected exposure times at the source
         position, for the combined A+B data. \\
20--25 & Net source
        count rates for the three standard
        energy bands, and associated errors. \\
26--31 & Fluxes in the three standard bands and associated
         errors. \\
32--34 & Observatory name and coordinates (R.A. and decl.) for the lower-energy X-ray counterpart. \\
35--37 & Reference and coordinates (R.A. and decl.) for the optical or \wise counterpart. \\
38 & Spectroscopic redshift. \\
39 & Non-absorption-corrected, rest-frame $10$--$40$~keV luminosity. \\
40 & Character indicating the reason for not appearing in the primary catalog. \\
\noalign{\smallskip} \hline \noalign{\smallskip}
\end{tabular}
\begin{minipage}[l]{0.75\textwidth}
\footnotesize
\textbf{Notes.} The full catalog is available as a machine readable
electronic table.
\end{minipage}
\label{secondary_columns_table}
\end{table}

In Table \ref{sec_specTable} we tabulate details of the optical
spectroscopic properties of individual sources from the secondary
catalog with spectroscopic coverage. The columns are equivalent to those in Table
\ref{specTable}, with the exception of an additional observing run
label (``G'') to mark sources followed up with Gemini-South in January
and February of 2016.
For $78\%$ 
of the sources in Table \ref{sec_specTable} the spectroscopic constraints are from our dedicated followup program
(with Keck, Palomar, NTT, Magellan, and Gemini), and for $22\%$ they are from the
SDSS or the literature. 
Individual source spectra ($F_{\mathrm{\nu}}$ versus $\lambda$) are shown in Figure
\ref{sec_spec_all}. 

\begin{table}
\centering
\caption{Summary of the optical
  spectroscopy for the secondary catalog sources.}
\label{sec_specTable}
\renewcommand*{\arraystretch}{1.1}
\begin{tabular}{llccllc} 
\hline \hline \noalign{\smallskip} 
\multicolumn{1}{c}{ID} & \multicolumn{1}{c}{\nustar Name} & \multicolumn{1}{c}{$z$} &
\multicolumn{1}{c}{Type} & \multicolumn{1}{c}{Lines} &
\multicolumn{1}{c}{Notes} & \multicolumn{1}{c}{Run} \\
\multicolumn{1}{c}{(1)} & \multicolumn{1}{c}{(2)} & \multicolumn{1}{c}{(3)} &
\multicolumn{1}{c}{(4)} & \multicolumn{1}{c}{(5)} &
\multicolumn{1}{c}{(6)} & \multicolumn{1}{c}{(7)} \\
\noalign{\smallskip} \hline \noalign{\smallskip}
1 & NuSTARJ001639+8139.8 & 0.000 & Gal & $\cdot$$\cdot$$\cdot$ & \parbox[t]{4.4cm}{Bright star} & 28 \\
3 & NuSTARJ001844-1022.6 & 0.076 & NL & \parbox[t]{6cm}{\mbox{[\ion{O}{2}]} Ca$_{\mathrm{H,K}}^{\mathrm{\dagger}}$ G-band$^{\mathrm{\dagger}}$ H$\mathrm{\beta}^{\mathrm{\dagger}}$ \mbox{\ion{Mg}{1}{b}}$^{\mathrm{\dagger}}$ Na$_{\mathrm{D1,D2}}^{\mathrm{\dagger}}$ H$\mathrm{\alpha}$ \mbox{[\ion{N}{2}]} \mbox{[\ion{S}{2}]}} & $\cdot$$\cdot$$\cdot$ & S \\
4 & NuSTARJ011018-4612.1 & $\cdot$$\cdot$$\cdot$ & $\cdot$$\cdot$$\cdot$ & $\cdot$$\cdot$$\cdot$ & \parbox[t]{4.4cm}{Continuum detected} & 11 \\
5 & NuSTARJ011053-4602.6 & 0.626 & NL & \parbox[t]{6cm}{\mbox{[\ion{O}{2}]} H$\mathrm{\gamma}$ H$\mathrm{\beta}$ \mbox{[\ion{O}{3}]}} & $\cdot$$\cdot$$\cdot$ & 18 \\
6 & NuSTARJ022801+3115.9 & 1.857 & BL & \parbox[t]{6cm}{Ly$\mathrm{\alpha}$ \mbox{\ion{C}{4}} \mbox{\ion{C}{3}]} \mbox{\ion{Mg}{2}}} & $\cdot$$\cdot$$\cdot$ & 31 \\
7 & NuSTARJ023448-2936.6 & 0.313 & NL & \parbox[t]{6cm}{H$\mathrm{\beta}$ \mbox{[\ion{O}{3}]} H$\mathrm{\alpha}$ \mbox{[\ion{N}{2}]}} & $\cdot$$\cdot$$\cdot$ & G \\
8 & NuSTARJ023459-2944.6 & 0.446 & BL & $\cdot$$\cdot$$\cdot$ & \parbox[t]{4.4cm}{\citet{Caccianiga08}} & L \\
9 & NuSTARJ031602-0221.0 & 0.821 & NL & \parbox[t]{6cm}{\mbox{[\ion{O}{2}]} H$\mathrm{\beta}$ \mbox{[\ion{O}{3}]}} & $\cdot$$\cdot$$\cdot$ & 31 \\
10 & NuSTARJ033313-3612.0 & $\cdot$$\cdot$$\cdot$ & $\cdot$$\cdot$$\cdot$ & $\cdot$$\cdot$$\cdot$ & \parbox[t]{4.4cm}{Two possible counterparts at $z\approx 1$; \citet{Wong08}} & L \\
11 & NuSTARJ033342-3613.9 & 0.559 & NL & \parbox[t]{6cm}{\mbox{[\ion{O}{2}]} H$\mathrm{\beta}$ \mbox{[\ion{O}{3}]}} & $\cdot$$\cdot$$\cdot$ & 11 \\
12 & NuSTARJ033406-3603.9 & 0.910 & BL & \parbox[t]{6cm}{\mbox{\ion{Mg}{2}}} & $\cdot$$\cdot$$\cdot$ & 11 \\
13 & NuSTARJ034403-4441.0 & 0.275 & NL? & \parbox[t]{6cm}{Ca$_{\mathrm{H,K}}^{\mathrm{\dagger}}$ G-band$^{\mathrm{\dagger}}$ \mbox{[\ion{O}{3}]} \mbox{\ion{Mg}{1}{b}}$^{\mathrm{\dagger}}$ Na$_{\mathrm{D1,D2}}^{\mathrm{\dagger}}$ \mbox{[\ion{O}{1}]} H$\mathrm{\alpha}$ \mbox{[\ion{N}{2}]} \mbox{[\ion{S}{2}]}} & $\cdot$$\cdot$$\cdot$ & 32 \\
15 & NuSTARJ043754-4716.0 & 0.337 & NL & \parbox[t]{6cm}{\mbox{[\ion{O}{2}]} H$\mathrm{\beta}$ H$\mathrm{\alpha}$ \mbox{[\ion{N}{2}]}} & $\cdot$$\cdot$$\cdot$ & 23 \\
17 & NuSTARJ081911+7046.6 & 0.720 & NL & \parbox[t]{6cm}{\mbox{[\ion{O}{2}]} H$\mathrm{\beta}$ \mbox{[\ion{O}{3}]}} & \parbox[t]{4.4cm}{\oiii asymmetry} & 13 \\
18 & NuSTARJ090223-4039.4 & 0.087 & BL & \parbox[t]{6cm}{H$\mathrm{\beta}$ \mbox{[\ion{O}{3}]} \mbox{[\ion{O}{1}]} H$\mathrm{\alpha}$ \mbox{[\ion{N}{2}]}} & $\cdot$$\cdot$$\cdot$ & 11 \\
20 & NuSTARJ092418-3142.2 & 0.000 & Gal & $\cdot$$\cdot$$\cdot$ & \parbox[t]{4.4cm}{Continuum detected} & 22 \\
21 & NuSTARJ094734-3104.2 & $\cdot$$\cdot$$\cdot$ & $\cdot$$\cdot$$\cdot$ & $\cdot$$\cdot$$\cdot$ & \parbox[t]{4.4cm}{Continuum detected} & 24 \\
22 & NuSTARJ101600-3329.6 & 0.231 & NL & \parbox[t]{6cm}{\mbox{[\ion{O}{2}]} \mbox{[\ion{O}{3}]} H$\mathrm{\alpha}$ \mbox{[\ion{N}{2}]}} & $\cdot$$\cdot$$\cdot$ & 22 \\
27 & NuSTARJ105008-5958.8 & 0.000 & Gal & \parbox[t]{6cm}{H$\mathrm{\delta}$ H$\mathrm{\gamma}$ H$\mathrm{\beta}$ Na$_{\mathrm{D1,D2}}^{\mathrm{\dagger}}$ H$\mathrm{\alpha}$} & $\cdot$$\cdot$$\cdot$ & 11 \\
28 & NuSTARJ110445+3811.1 & 0.144 & NL & \parbox[t]{6cm}{\mbox{[\ion{O}{2}]} \mbox{[\ion{Ne}{3}]} Ca$_{\mathrm{H,K}}^{\mathrm{\dagger}}$ H$\mathrm{\delta}^{\mathrm{\dagger}}$ H$\mathrm{\beta}$ \mbox{[\ion{O}{3}]} \mbox{\ion{Mg}{1}{b}}$^{\mathrm{\dagger}}$ Na$_{\mathrm{D1,D2}}^{\mathrm{\dagger}}$ \mbox{[\ion{O}{1}]} H$\mathrm{\alpha}$ \mbox{[\ion{N}{2}]} \mbox{[\ion{S}{2}]}} & $\cdot$$\cdot$$\cdot$ & S \\
30 & NuSTARJ120259+4430.4 & 0.465 & NL & \parbox[t]{6cm}{H$\mathrm{\beta}$ \mbox{[\ion{O}{3}]}} & $\cdot$$\cdot$$\cdot$ & 14 \\
31 & NuSTARJ120711+3348.4 & 0.135 & BL & \parbox[t]{6cm}{\mbox{[\ion{O}{2}]} \mbox{[\ion{Ne}{3}]} Ca$_{\mathrm{H,K}}^{\mathrm{\dagger}}$ H$\mathrm{\gamma}$ H$\mathrm{\beta}$ \mbox{[\ion{O}{3}]} H$\mathrm{\alpha}$ \mbox{[\ion{N}{2}]}} & $\cdot$$\cdot$$\cdot$ & S \\
32 & NuSTARJ120930-0500.1 & 0.391 & NL & \parbox[t]{6cm}{H$\mathrm{\delta}^{\mathrm{\dagger}}$ H$\mathrm{\beta}$ \mbox{[\ion{O}{3}]} \mbox{[\ion{O}{1}]}} & $\cdot$$\cdot$$\cdot$ & G \\
33 & NuSTARJ121027+3929.1 & 0.615 & BL\ Lac & \parbox[t]{6cm}{Ca$_{\mathrm{H,K}}^{\mathrm{\dagger}}$ H$\mathrm{\delta}^{\mathrm{\dagger}}$ G-band$^{\mathrm{\dagger}}$ H$\mathrm{\gamma}^{\mathrm{\dagger}}$ H$\mathrm{\beta}^{\mathrm{\dagger}}$ \mbox{[\ion{O}{3}]} \mbox{\ion{Mg}{1}{b}}$^{\mathrm{\dagger}}$} & \parbox[t]{4.4cm}{\citet{Morris91}} & S \\
34 & NuSTARJ121038+3930.7 & 1.033 & NL & \parbox[t]{6cm}{\mbox{\ion{C}{3}]} \mbox{\ion{C}{2}]} \mbox{[\ion{Ne}{4}]} \mbox{\ion{Mg}{2}} \mbox{[\ion{Ne}{5}]} \mbox{[\ion{O}{2}]} \mbox{[\ion{Ne}{3}]} H$\mathrm{\beta}$ \mbox{[\ion{O}{3}]}} & $\cdot$$\cdot$$\cdot$ & 15 \\
35 & NuSTARJ121049+3928.5 & 0.023 & NL & \parbox[t]{6cm}{\mbox{[\ion{O}{2}]} Ca$_{\mathrm{H,K}}^{\mathrm{\dagger}}$ G-band$^{\mathrm{\dagger}}$ H$\mathrm{\gamma}$ H$\mathrm{\beta}$ \mbox{[\ion{O}{3}]} \mbox{\ion{Mg}{1}{b}}$^{\mathrm{\dagger}}$ Na$_{\mathrm{D1,D2}}^{\mathrm{\dagger}}$ \mbox{[\ion{O}{1}]} H$\mathrm{\alpha}$ \mbox{[\ion{N}{2}]} \mbox{[\ion{S}{2}]} CaT$^{\mathrm{\dagger}}$} & $\cdot$$\cdot$$\cdot$ & S \\
36 & NuSTARJ121854+2958.0 & 0.175 & NL? & \parbox[t]{6cm}{\mbox{[\ion{O}{2}]} \mbox{[\ion{Ne}{3}]} H$\mathrm{\delta}$ H$\mathrm{\gamma}$ H$\mathrm{\beta}$ \mbox{[\ion{O}{3}]} \mbox{[\ion{O}{1}]} H$\mathrm{\alpha}$ \mbox{[\ion{N}{2}]} \mbox{[\ion{S}{2}]} \mbox{[\ion{Ar}{3}]}} & $\cdot$$\cdot$$\cdot$ & S \\
37 & NuSTARJ123559-3951.9 & 0.000 & Gal & $\cdot$$\cdot$$\cdot$ & \parbox[t]{4.4cm}{Star} & L \\
38 & NuSTARJ125021+2635.9 & 0.751 & BL & \parbox[t]{6cm}{\mbox{\ion{Mg}{2}} \mbox{[\ion{Ne}{5}]} \mbox{[\ion{O}{2}]} \mbox{[\ion{Ne}{3}]} H$\mathrm{\beta}$ \mbox{[\ion{O}{3}]}} & $\cdot$$\cdot$$\cdot$ & 27 \\
39 & NuSTARJ125605+5643.8 & 0.984 & NL & \parbox[t]{6cm}{\mbox{\ion{Mg}{2}} \mbox{[\ion{O}{2}]} \mbox{[\ion{Ne}{3}]}} & $\cdot$$\cdot$$\cdot$ & 12 \\
40 & NuSTARJ125711+2748.1 & 0.306 & NL & \parbox[t]{6cm}{\mbox{[\ion{O}{2}]} Ca$_{\mathrm{H,K}}^{\mathrm{\dagger}}$ \mbox{[\ion{O}{3}]} H$\mathrm{\alpha}$ \mbox{[\ion{N}{2}]}} & $\cdot$$\cdot$$\cdot$ & 15 \\
41 & NuSTARJ125715+2746.6 & $\cdot$$\cdot$$\cdot$ & $\cdot$$\cdot$$\cdot$ & \parbox[t]{6cm}{CaT$^{\mathrm{\dagger}}$} & \parbox[t]{4.4cm}{Counterpart uncertainty} & 15 \\
42 & NuSTARJ125744+2751.2 & 0.325 & NL & \parbox[t]{6cm}{\mbox{[\ion{O}{2}]} Ca$_{\mathrm{H,K}}^{\mathrm{\dagger}}$ \mbox{[\ion{O}{3}]} H$\mathrm{\alpha}$ \mbox{[\ion{N}{2}]}} & $\cdot$$\cdot$$\cdot$ & 15 \\
43 & NuSTARJ130157-6358.1 & 0.000 & Gal & $\cdot$$\cdot$$\cdot$ & \parbox[t]{4.4cm}{\citet{Krivonos15}} & L \\
45 & NuSTARJ130616-4930.8 & 0.284 & BL? & \parbox[t]{6cm}{H$\mathrm{\beta}$ \mbox{[\ion{O}{3}]} H$\mathrm{\alpha}$ \mbox{[\ion{N}{2}]}} & $\cdot$$\cdot$$\cdot$ & 16 \\
46 & NuSTARJ134447+5554.0 & 0.458 & BL? & \parbox[t]{6cm}{\mbox{\ion{Mg}{2}} H$\mathrm{\beta}$ \mbox{[\ion{O}{3}]}} & $\cdot$$\cdot$$\cdot$ & 14 \\
48 & NuSTARJ143256-4419.3 & 0.119 & NL & \parbox[t]{6cm}{Ca$_{\mathrm{H,K}}^{\mathrm{\dagger}}$ G-band$^{\mathrm{\dagger}}$ H$\mathrm{\beta}^{\mathrm{\dagger}}$ \mbox{\ion{Mg}{1}{b}}$^{\mathrm{\dagger}}$ Na$_{\mathrm{D1,D2}}^{\mathrm{\dagger}}$} & \parbox[t]{4.4cm}{Cluster ABELL 3602} & 24 \\
49 & NuSTARJ155520-3315.1 & 0.551 & NL & \parbox[t]{6cm}{\mbox{[\ion{O}{2}]} H$\mathrm{\beta}$ \mbox{[\ion{O}{3}]}} & $\cdot$$\cdot$$\cdot$ & 23 \\
50 & NuSTARJ165050-0126.6 & 0.791 & BL & \parbox[t]{6cm}{\mbox{\ion{Mg}{2}} H$\mathrm{\beta}$} & $\cdot$$\cdot$$\cdot$ & 24 \\
51 & NuSTARJ165104-0127.2 & 0.852 & NL & \parbox[t]{6cm}{\mbox{[\ion{Ne}{5}]} \mbox{[\ion{O}{2}]} \mbox{[\ion{Ne}{3}]} \mbox{[\ion{O}{3}]}} & $\cdot$$\cdot$$\cdot$ & 30 \\
52 & NuSTARJ170016+7840.7 & 0.778 & BL & \parbox[t]{6cm}{\mbox{\ion{Mg}{2}} \mbox{[\ion{O}{2}]} \mbox{[\ion{O}{3}]}} & $\cdot$$\cdot$$\cdot$ & 9 \\
53 & NuSTARJ172822-1421.4 & 0.688 & NL & \parbox[t]{6cm}{\mbox{[\ion{O}{2}]} H$\mathrm{\gamma}$ H$\mathrm{\beta}$ \mbox{[\ion{O}{3}]}} & $\cdot$$\cdot$$\cdot$ & 23 \\
55 & NuSTARJ175307-0123.7 & 0.451 & NL? & \parbox[t]{6cm}{\mbox{[\ion{O}{2}]} \mbox{[\ion{Ne}{3}]} H$\mathrm{\beta}$ \mbox{[\ion{O}{3}]}} & $\cdot$$\cdot$$\cdot$ & 24 \\
56 & NuSTARJ181417+3411.6 & 0.714 & BL & \parbox[t]{6cm}{\mbox{\ion{Mg}{2}} \mbox{[\ion{Ne}{5}]} \mbox{[\ion{O}{2}]} H$\mathrm{\beta}$ \mbox{[\ion{O}{3}]}} & $\cdot$$\cdot$$\cdot$ & 30 \\
63 & NuSTARJ223654+3423.4 & 0.148 & NL & \parbox[t]{6cm}{\mbox{[\ion{O}{2}]} \mbox{[\ion{O}{3}]} H$\mathrm{\alpha}$ \mbox{[\ion{N}{2}]}} & $\cdot$$\cdot$$\cdot$ & 31 \\
64 & NuSTARJ224037+0802.6 & 1.418 & BL & \parbox[t]{6cm}{\mbox{\ion{C}{4}} \mbox{\ion{C}{3}]} \mbox{\ion{Mg}{2}} \mbox{[\ion{O}{2}]}} & $\cdot$$\cdot$$\cdot$ & 30 \\
\noalign{\smallskip} \hline \hline \noalign{\smallskip}
\end{tabular}
\end{table}

\begin{figure*}
\centering
\begin{minipage}[l]{0.325\textwidth}
\includegraphics[width=\textwidth]{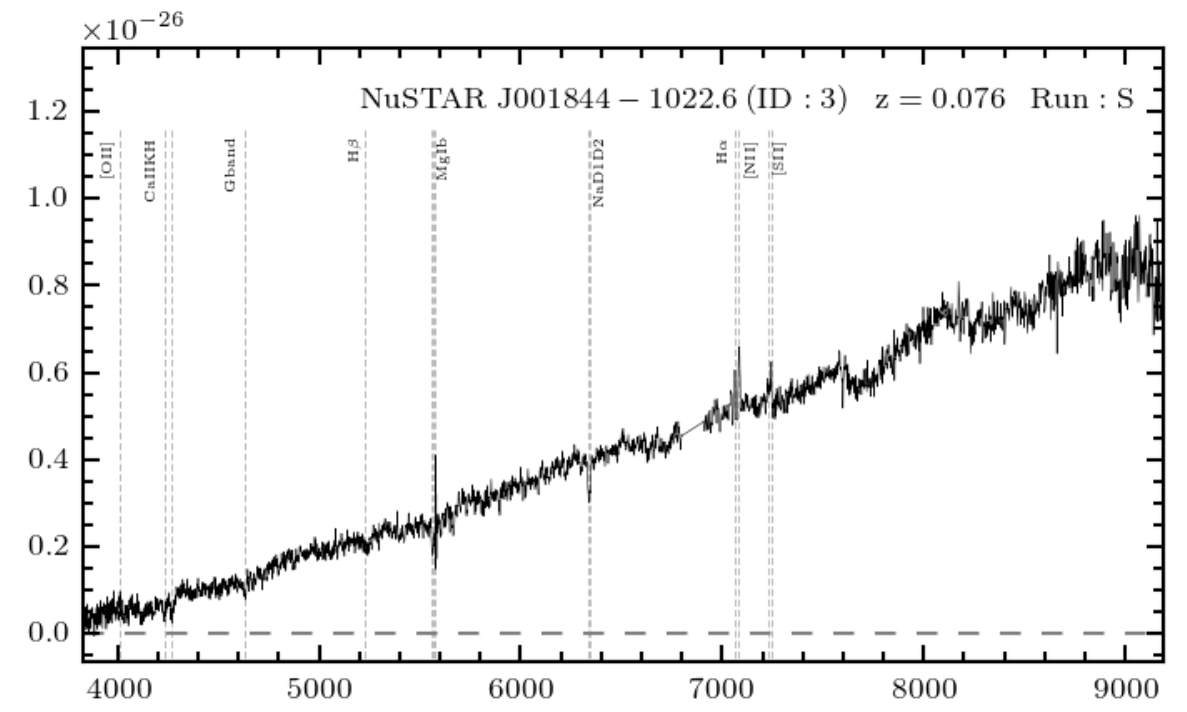}
\end{minipage}
\begin{minipage}[l]{0.325\textwidth}
\includegraphics[width=\textwidth]{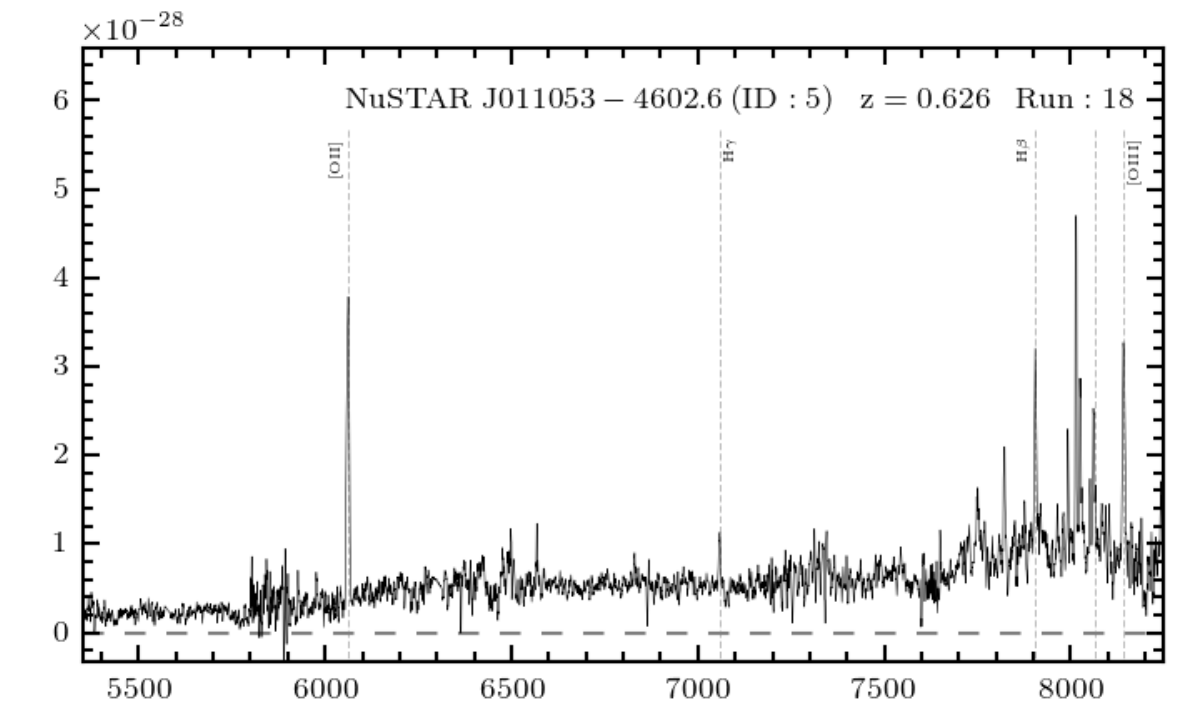}
\end{minipage}
\begin{minipage}[l]{0.325\textwidth}
\includegraphics[width=\textwidth]{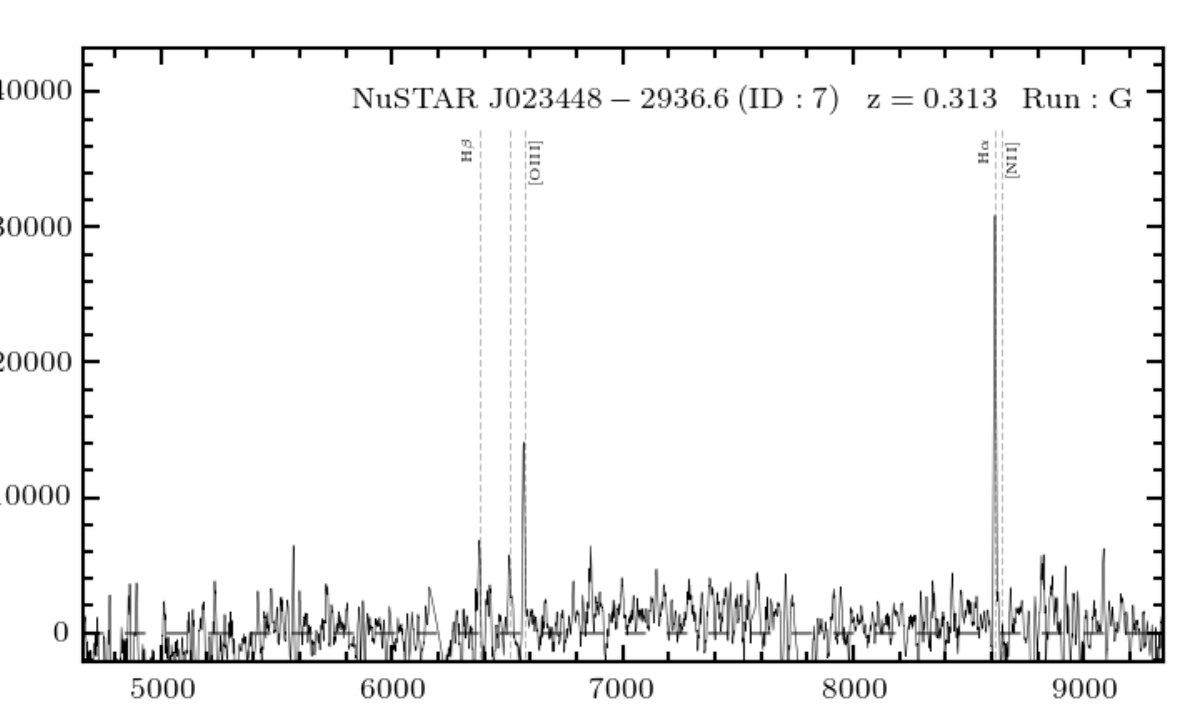}
\end{minipage}
\begin{minipage}[l]{0.325\textwidth}
\includegraphics[width=\textwidth]{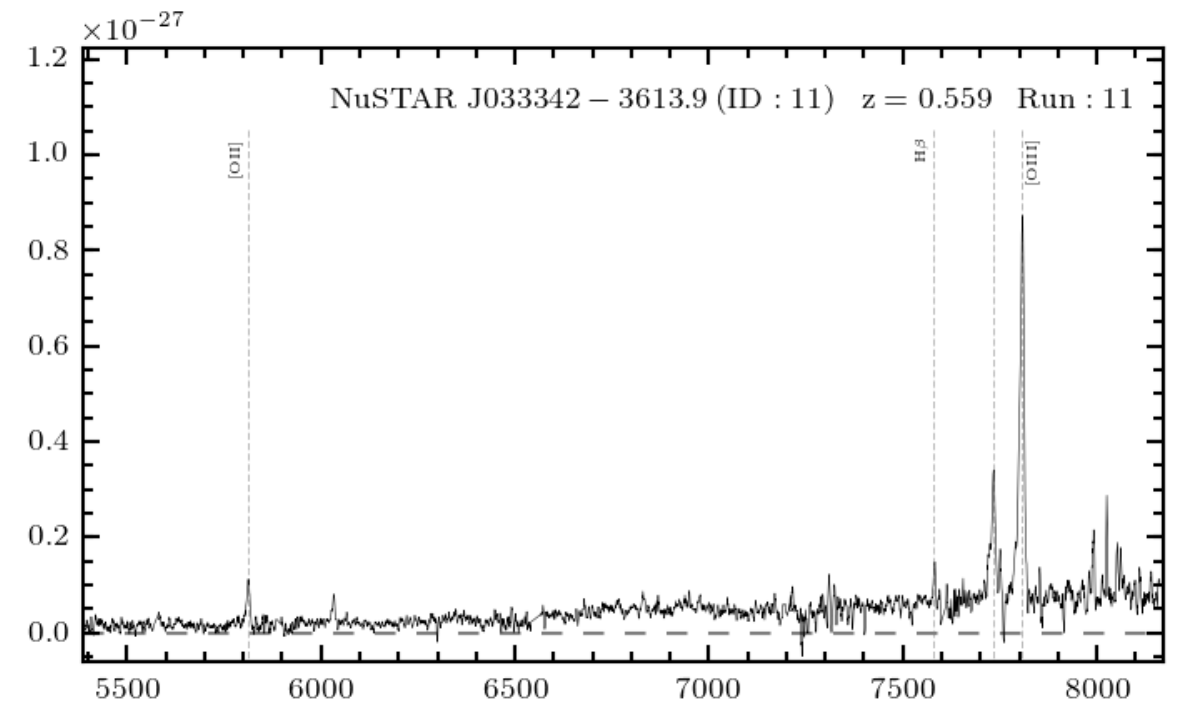}
\end{minipage}
\begin{minipage}[l]{0.325\textwidth}
\includegraphics[width=\textwidth]{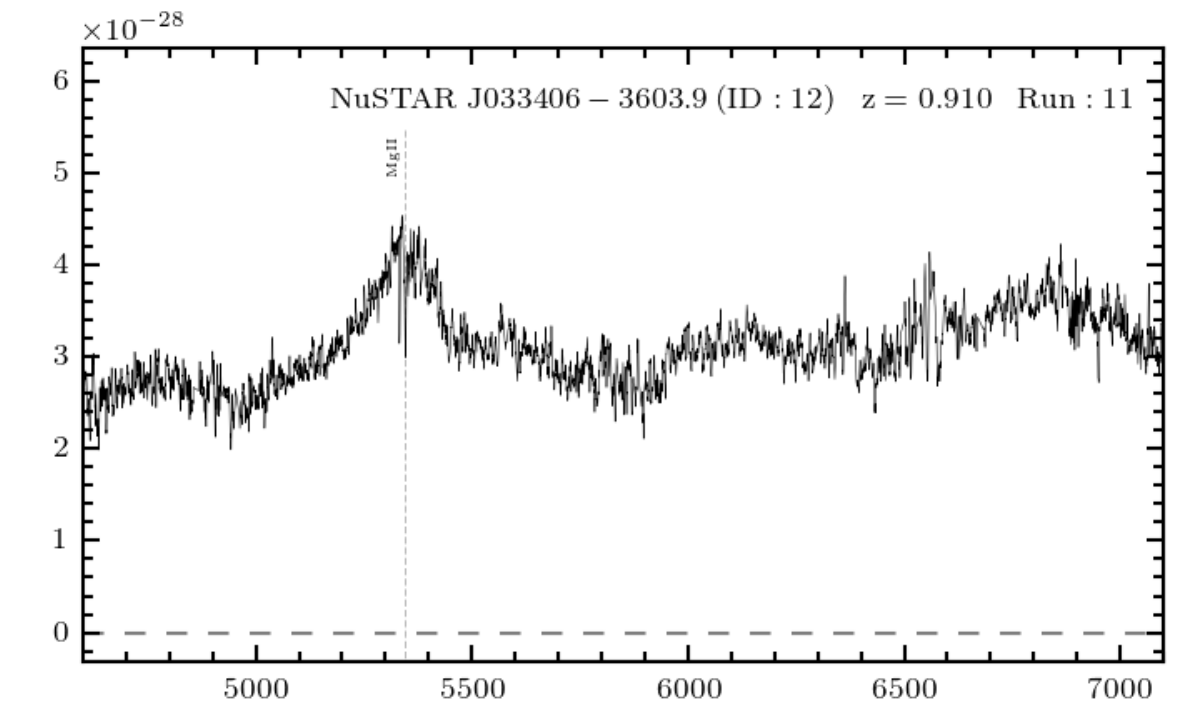}
\end{minipage}
\begin{minipage}[l]{0.325\textwidth}
\includegraphics[width=\textwidth]{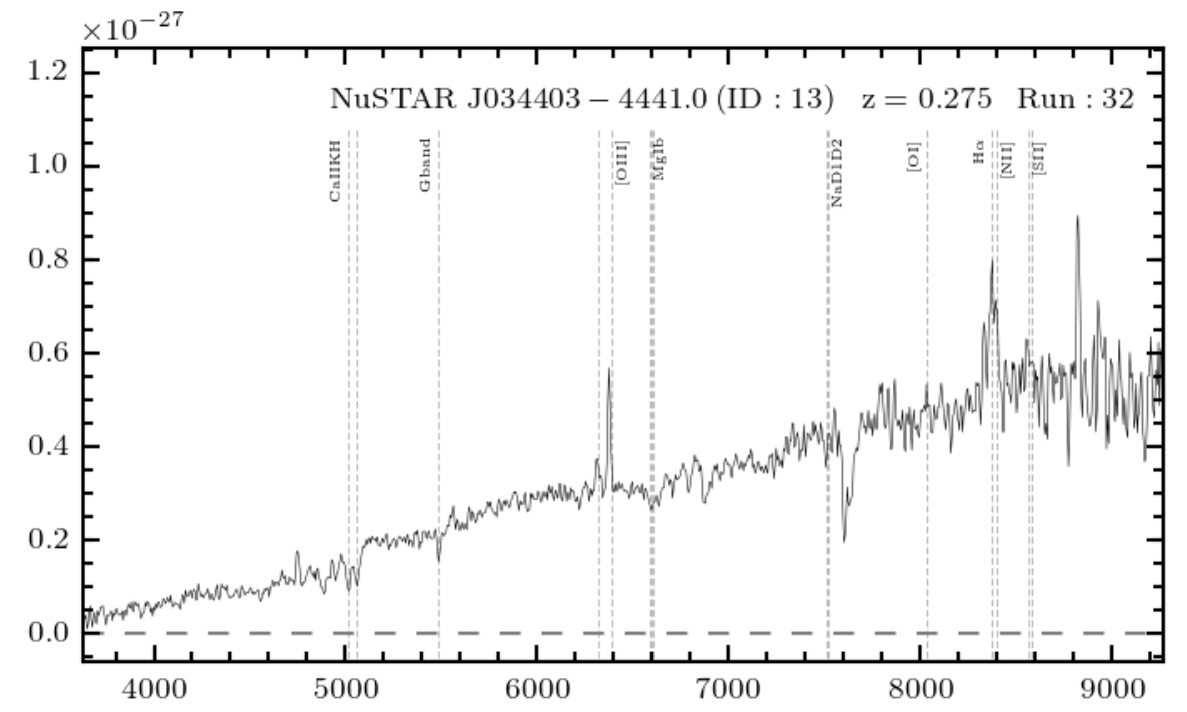}
\end{minipage}
\begin{minipage}[l]{0.325\textwidth}
\includegraphics[width=\textwidth]{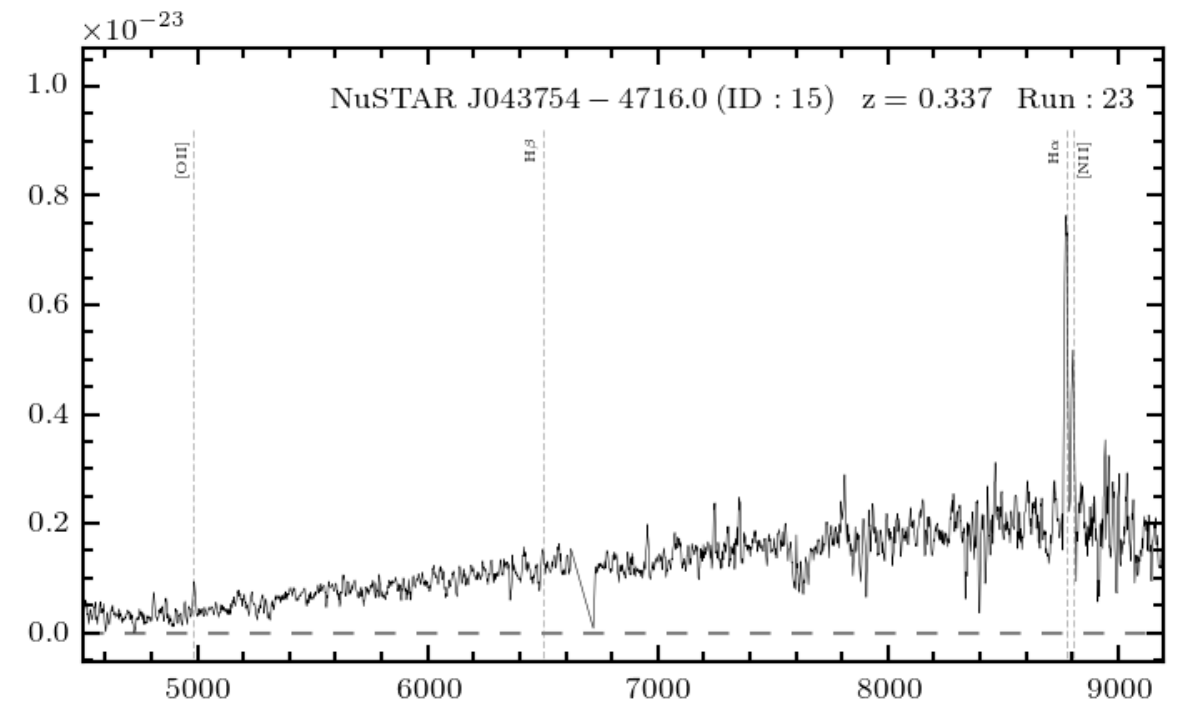}
\end{minipage}
\begin{minipage}[l]{0.325\textwidth}
\includegraphics[width=\textwidth]{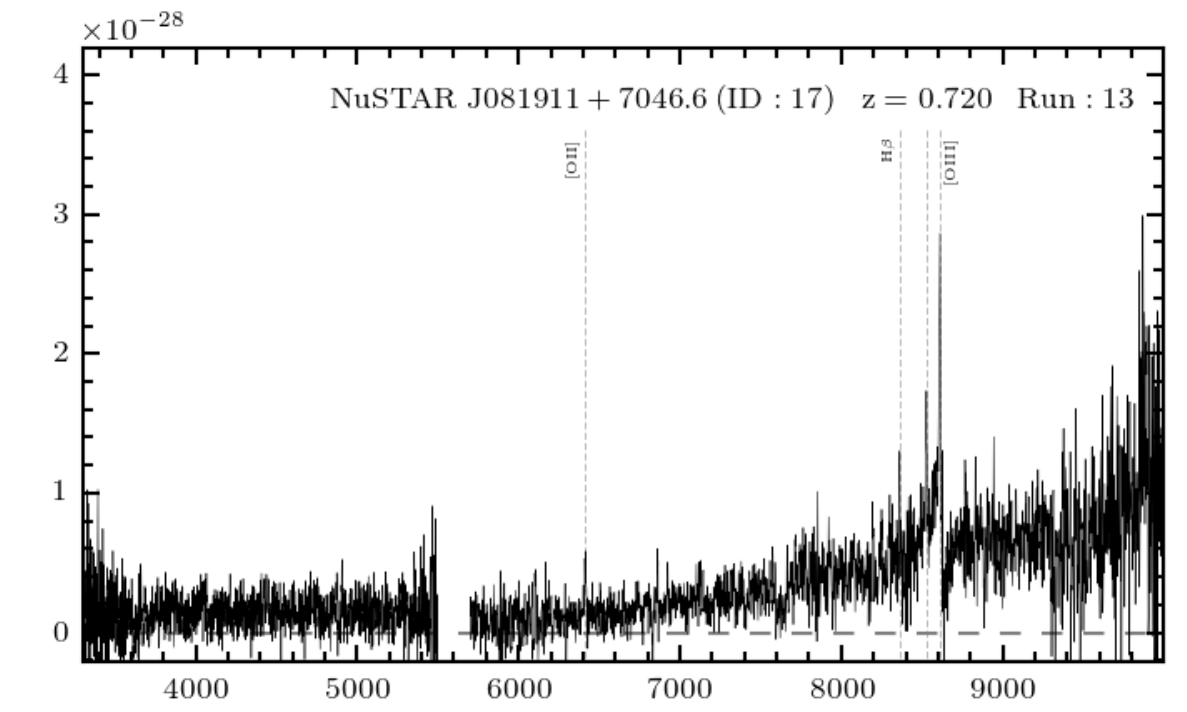}
\end{minipage}
\begin{minipage}[l]{0.325\textwidth}
\includegraphics[width=\textwidth]{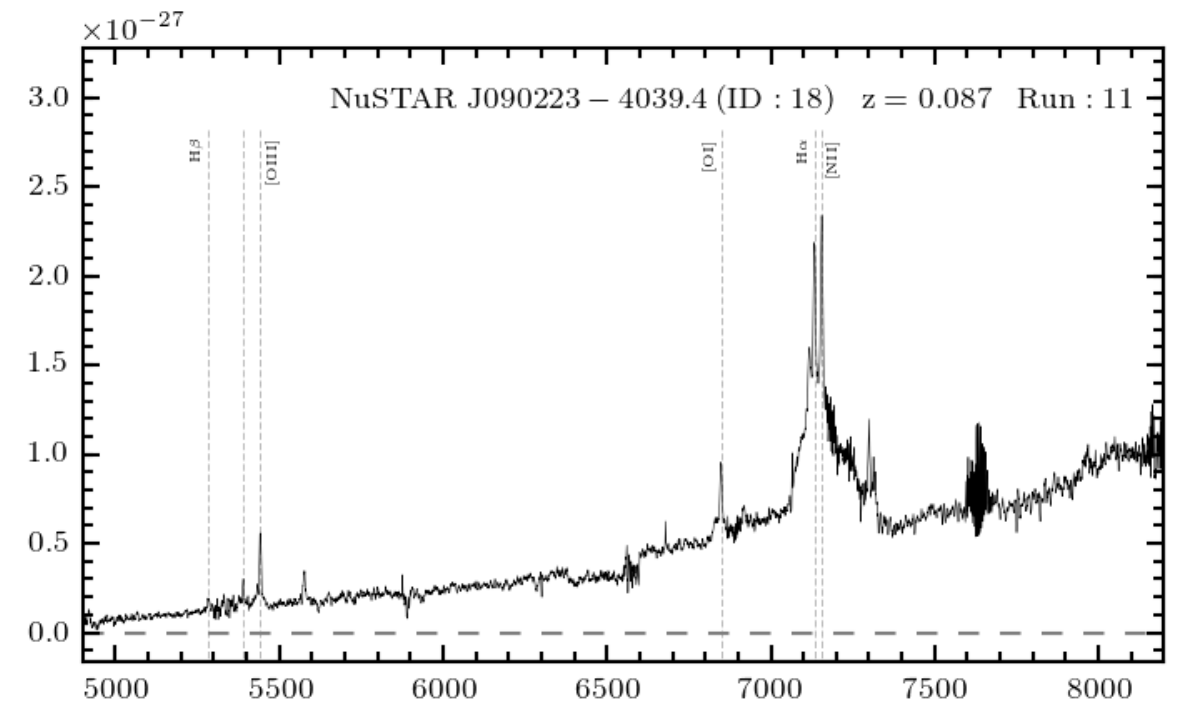}
\end{minipage}
\begin{minipage}[l]{0.325\textwidth}
\includegraphics[width=\textwidth]{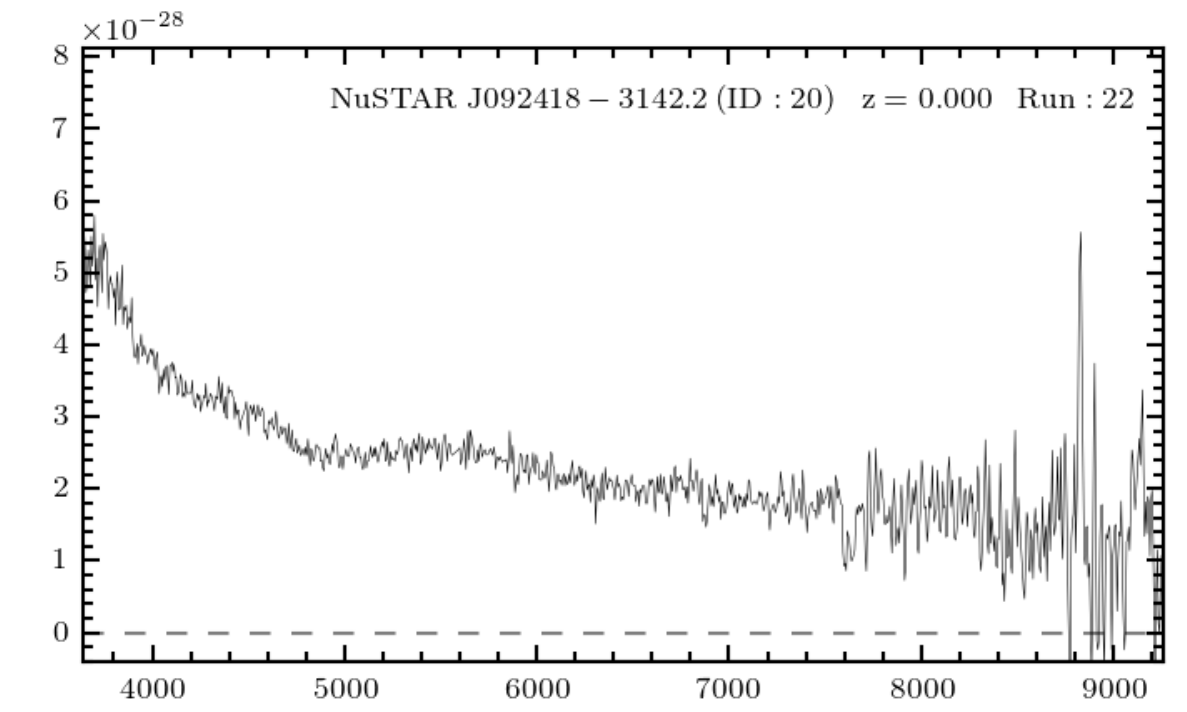}
\end{minipage}
\begin{minipage}[l]{0.325\textwidth}
\includegraphics[width=\textwidth]{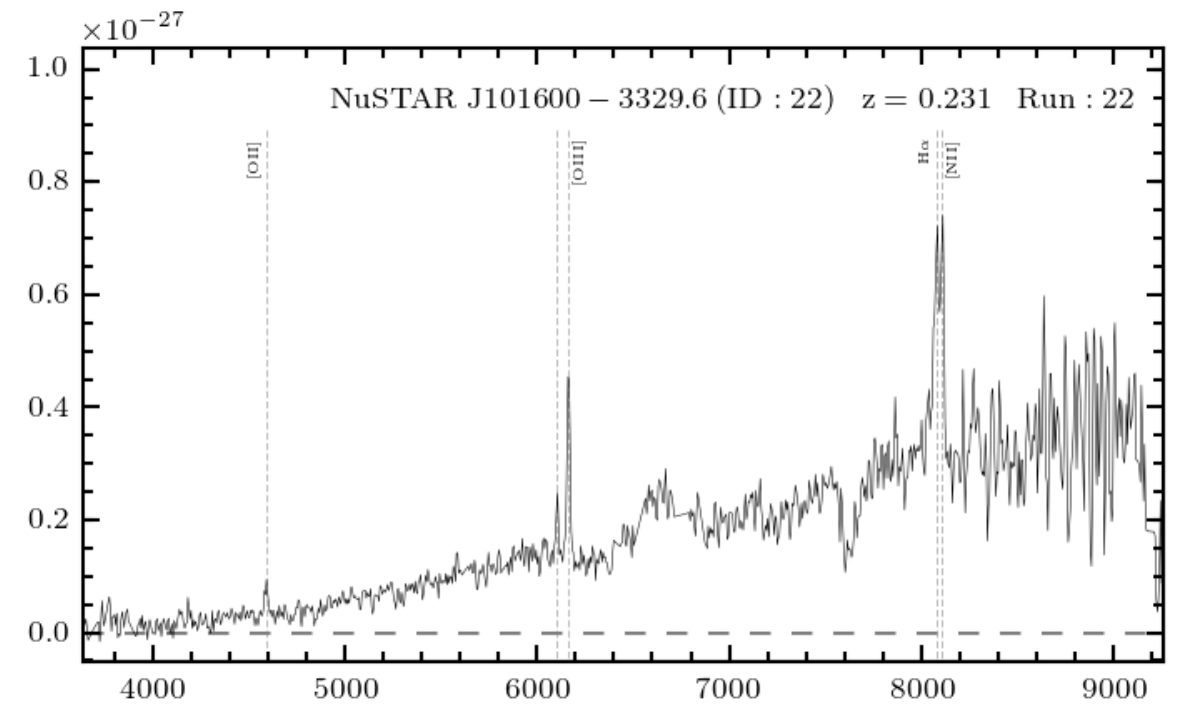}
\end{minipage}
\begin{minipage}[l]{0.325\textwidth}
\includegraphics[width=\textwidth]{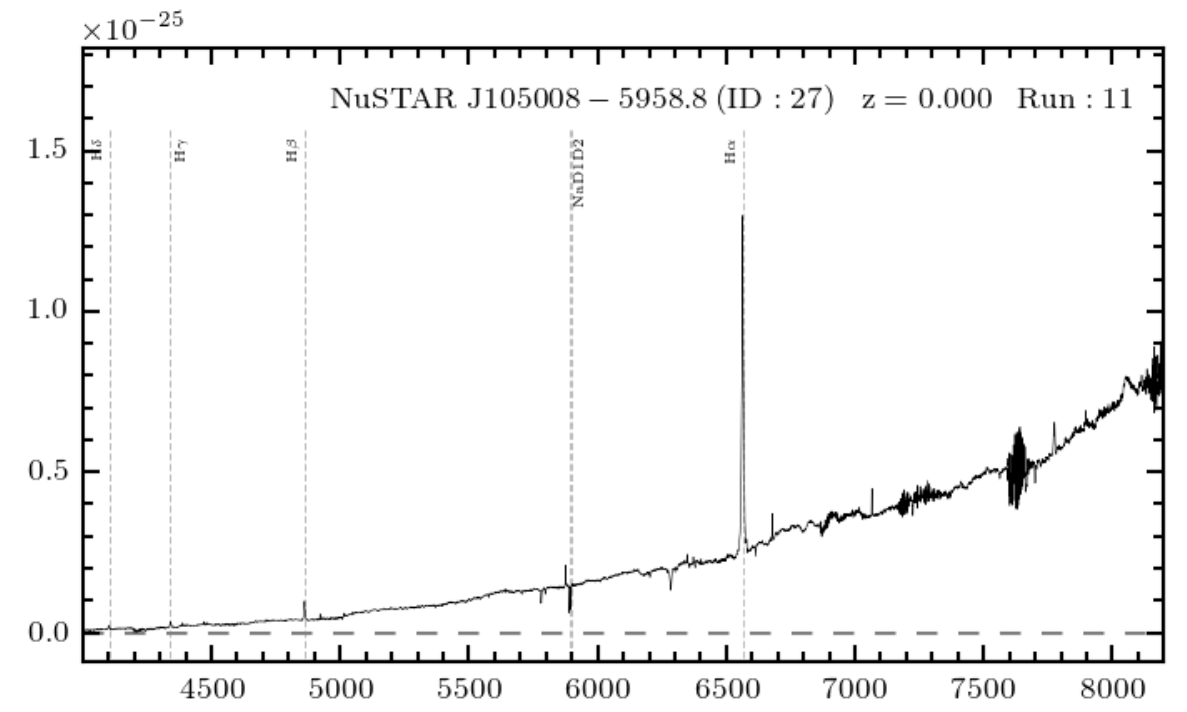}
\end{minipage}
\begin{minipage}[l]{0.325\textwidth}
\includegraphics[width=\textwidth]{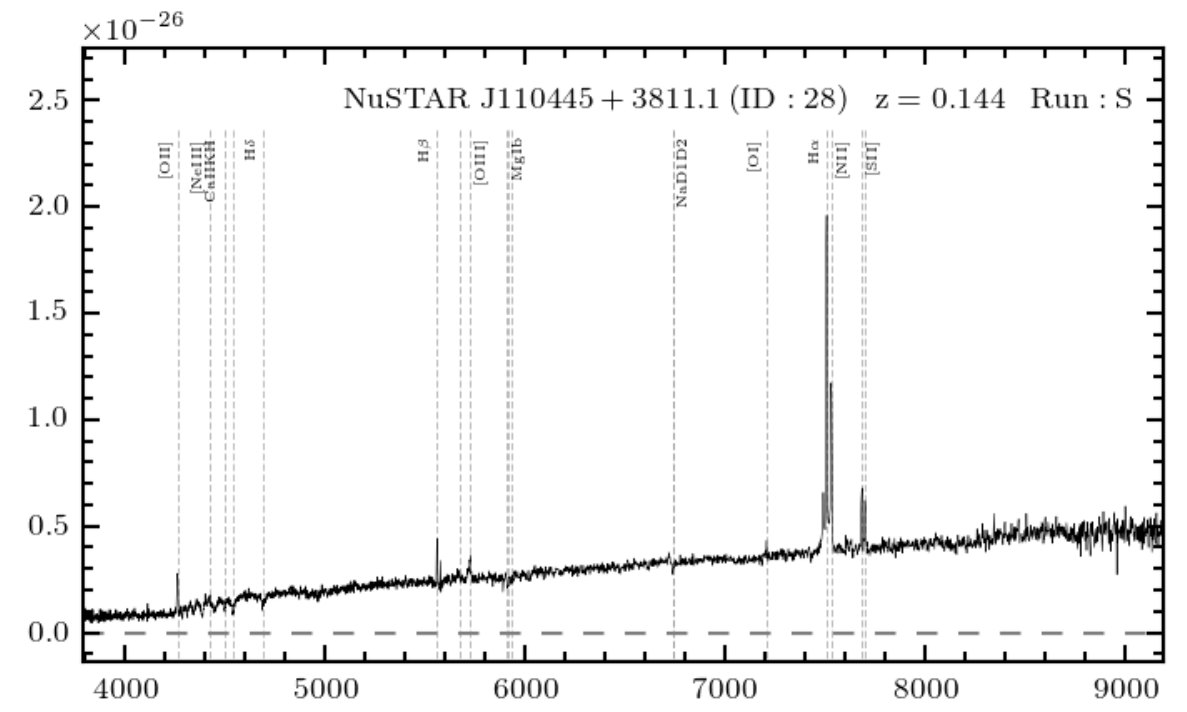}
\end{minipage}
\begin{minipage}[l]{0.325\textwidth}
\includegraphics[width=\textwidth]{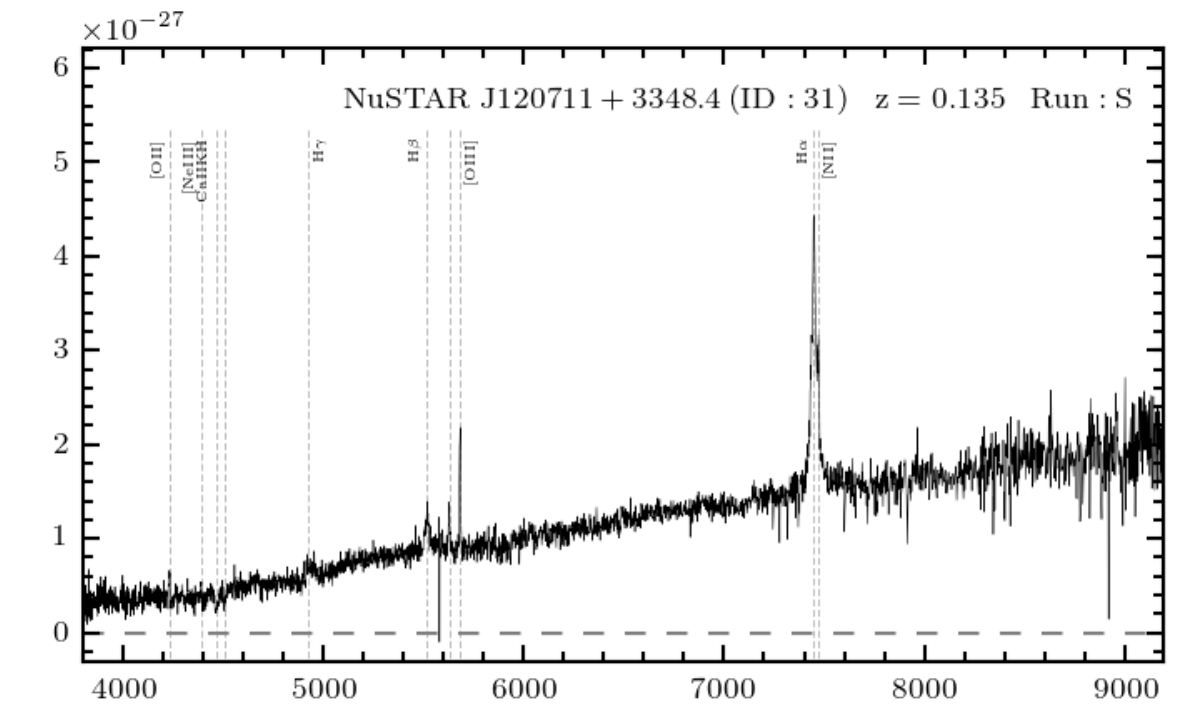}
\end{minipage}
\begin{minipage}[l]{0.325\textwidth}
\includegraphics[width=\textwidth]{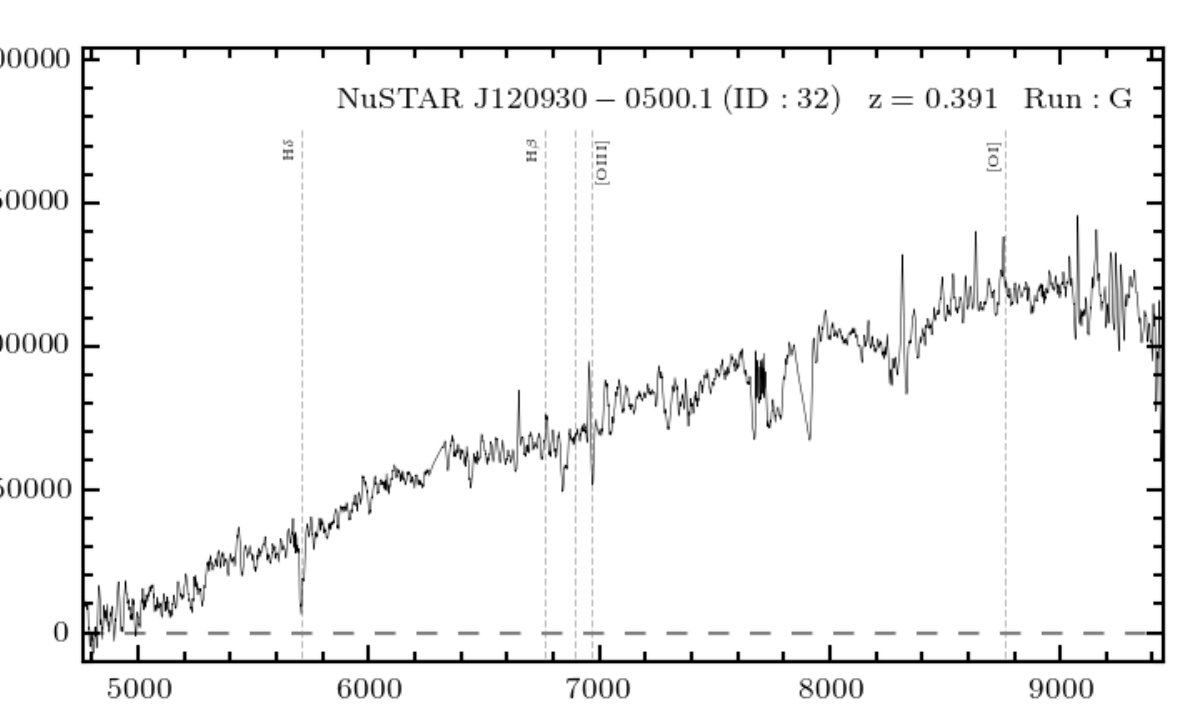}
\end{minipage}
\begin{minipage}[l]{0.325\textwidth}
\includegraphics[width=\textwidth]{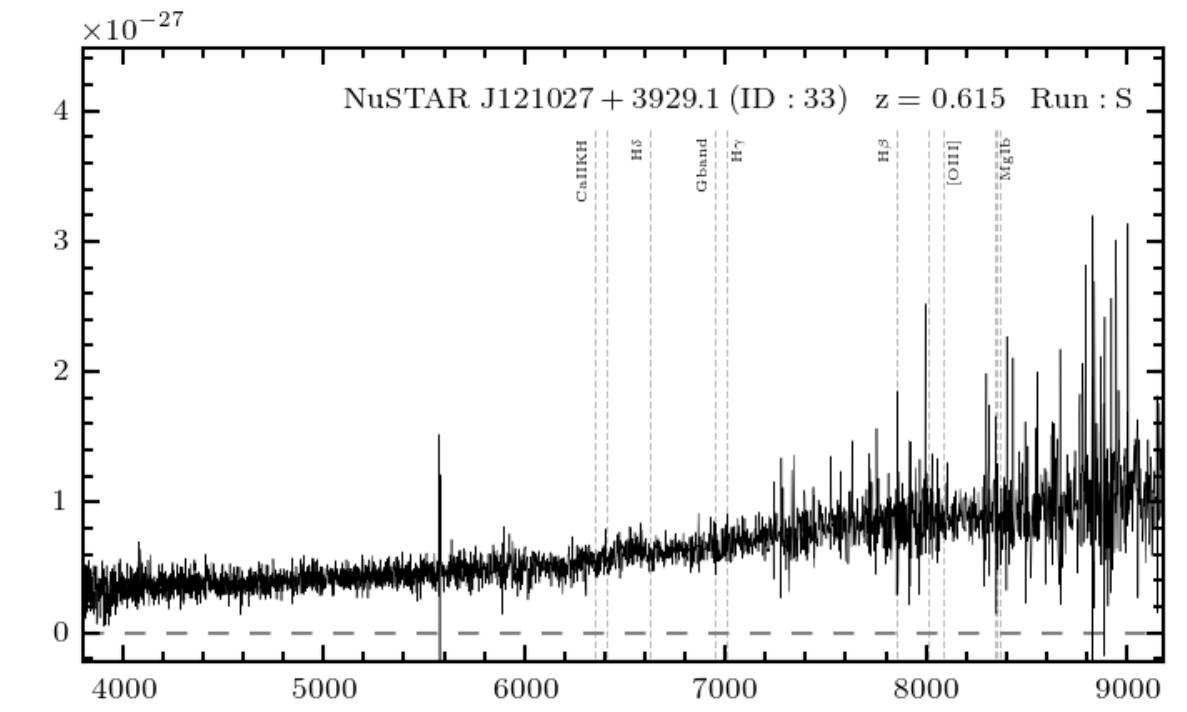}
\end{minipage}
\begin{minipage}[l]{0.325\textwidth}
\includegraphics[width=\textwidth]{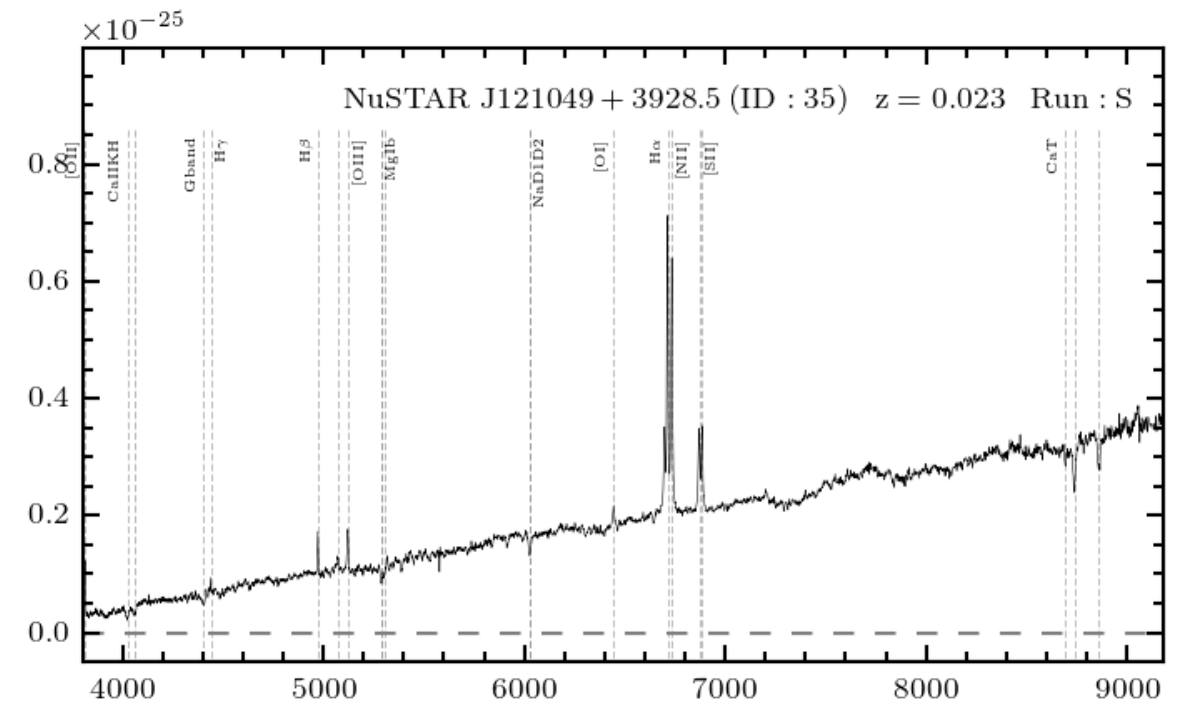}
\end{minipage}
\begin{minipage}[l]{0.325\textwidth}
\includegraphics[width=\textwidth]{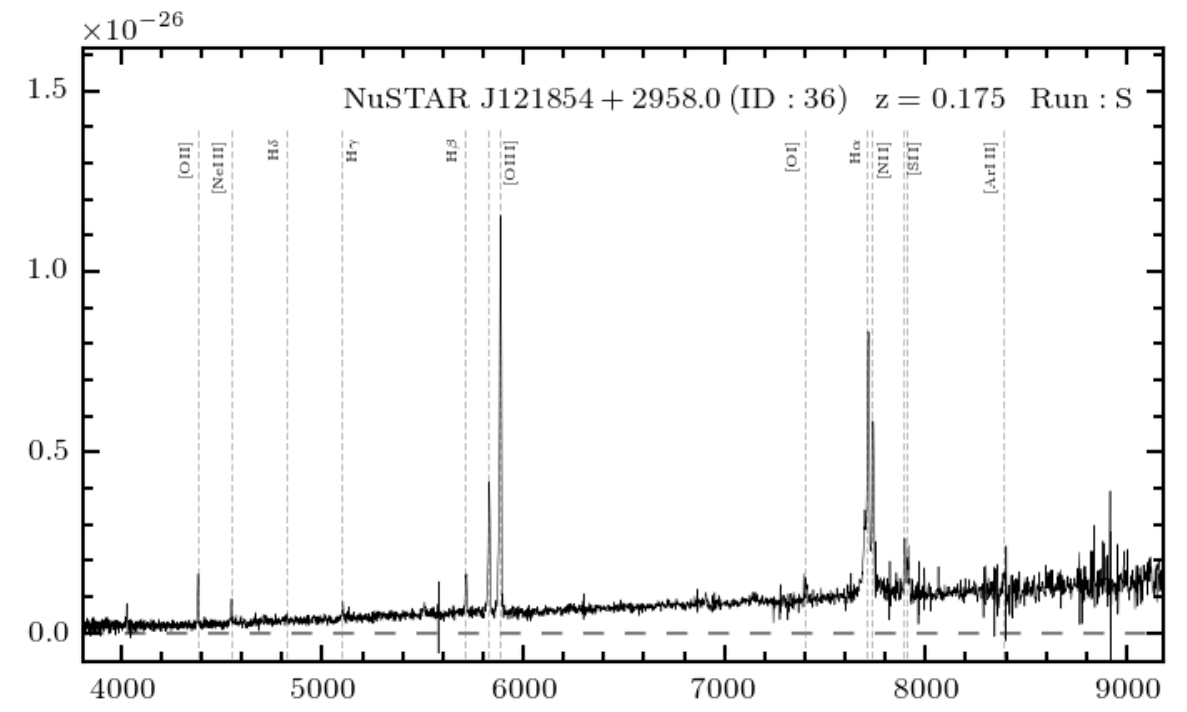}
\end{minipage}
\caption{Optical spectra for the secondary catalog sources (continued
  on the following pages). The axes and labelling are the same as for
  Figure \ref{spec_all}. Full
  resolution figures are available online.}
\end{figure*}
\addtocounter{figure}{-1}
\begin{figure*}
\centering
\begin{minipage}[l]{0.325\textwidth}
\includegraphics[width=\textwidth]{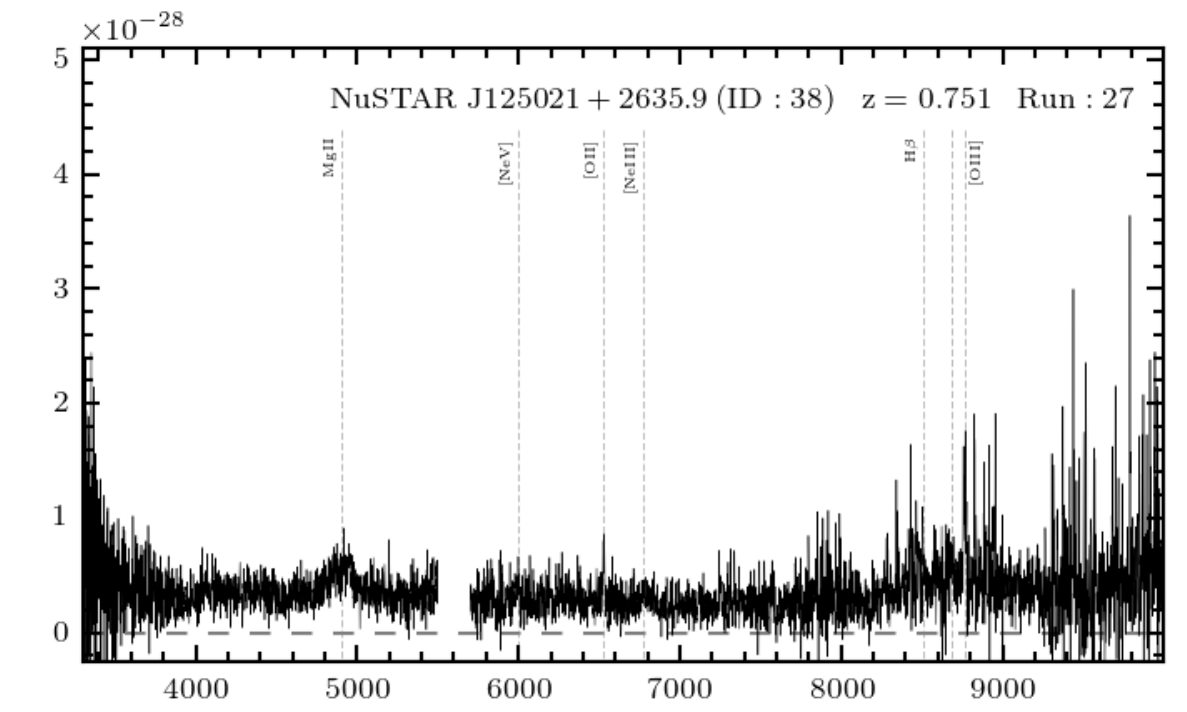}
\end{minipage}
\begin{minipage}[l]{0.325\textwidth}
\includegraphics[width=\textwidth]{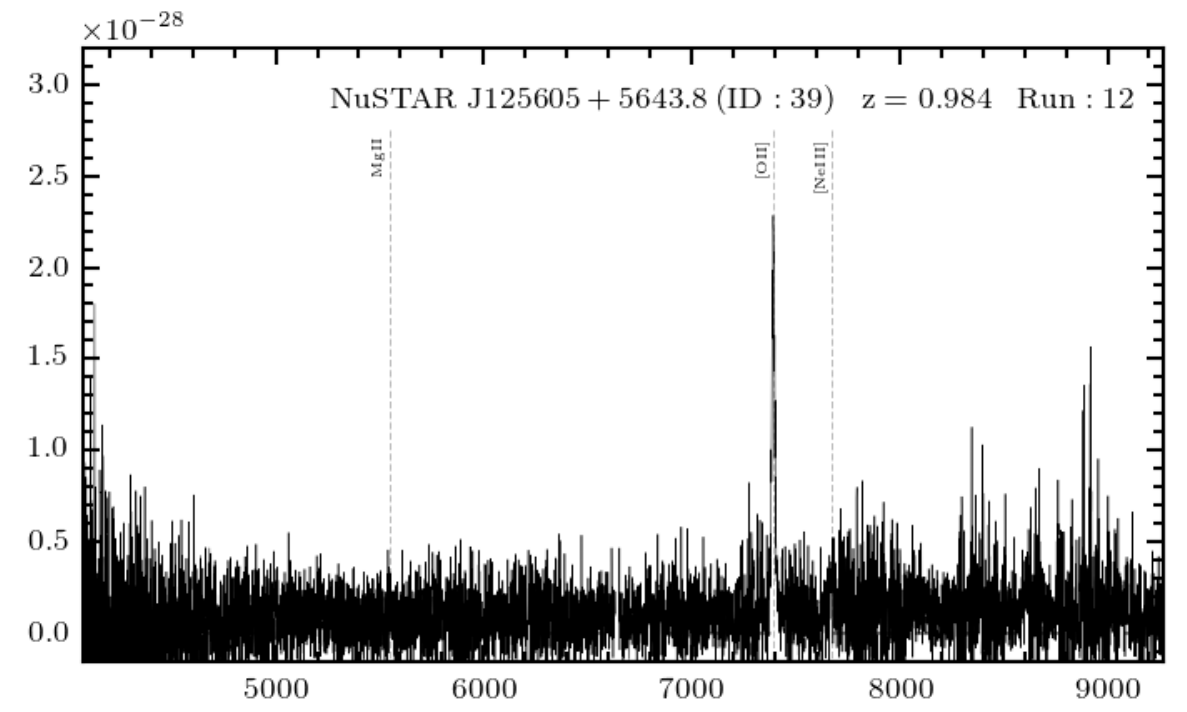}
\end{minipage}
\begin{minipage}[l]{0.325\textwidth}
\includegraphics[width=\textwidth]{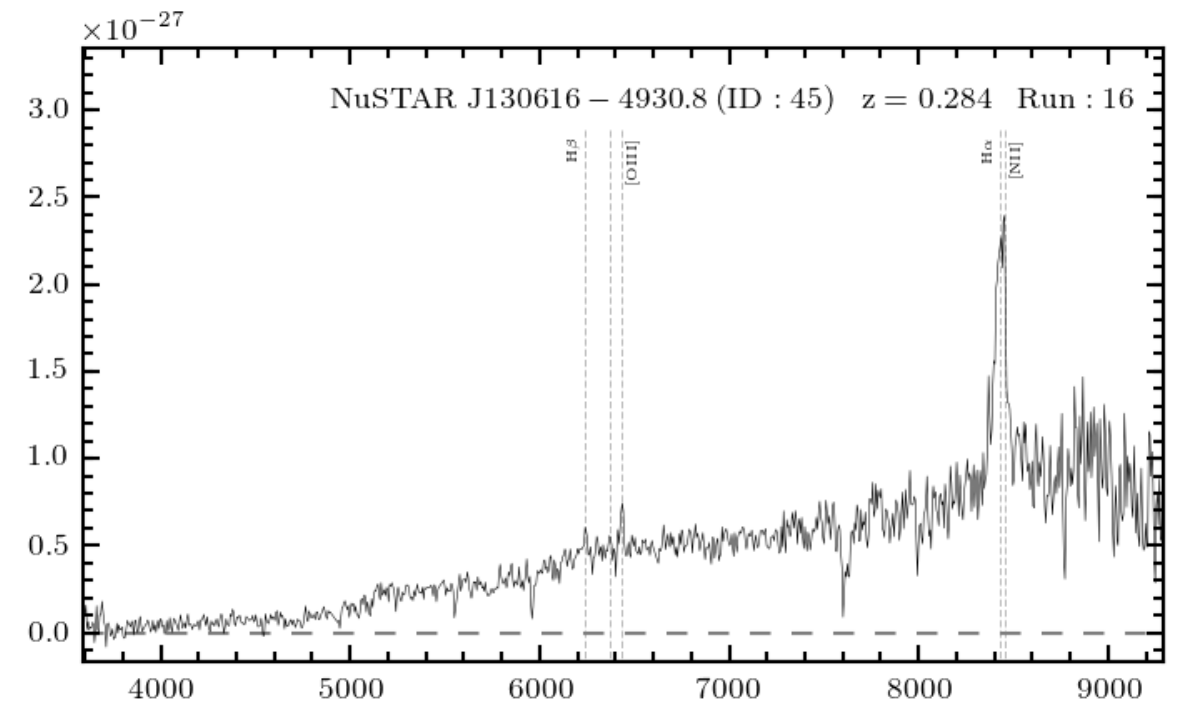}
\end{minipage}
\begin{minipage}[l]{0.325\textwidth}
\includegraphics[width=\textwidth]{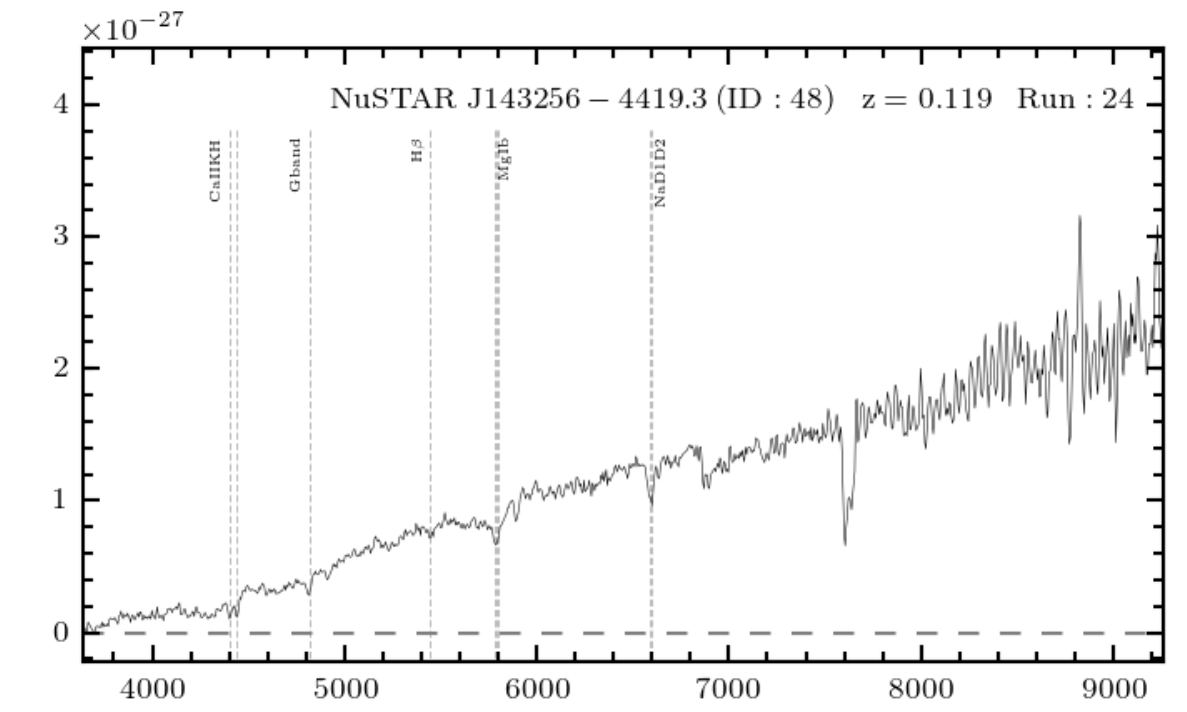}
\end{minipage}
\begin{minipage}[l]{0.325\textwidth}
\includegraphics[width=\textwidth]{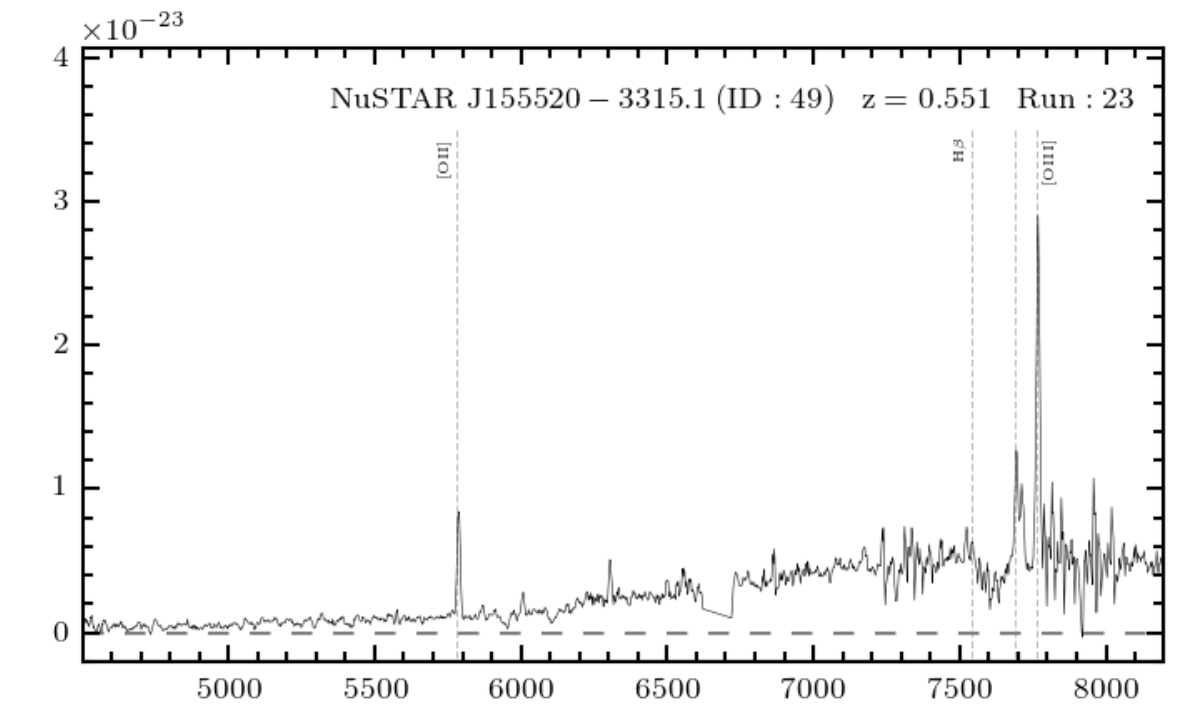}
\end{minipage}
\begin{minipage}[l]{0.325\textwidth}
\includegraphics[width=\textwidth]{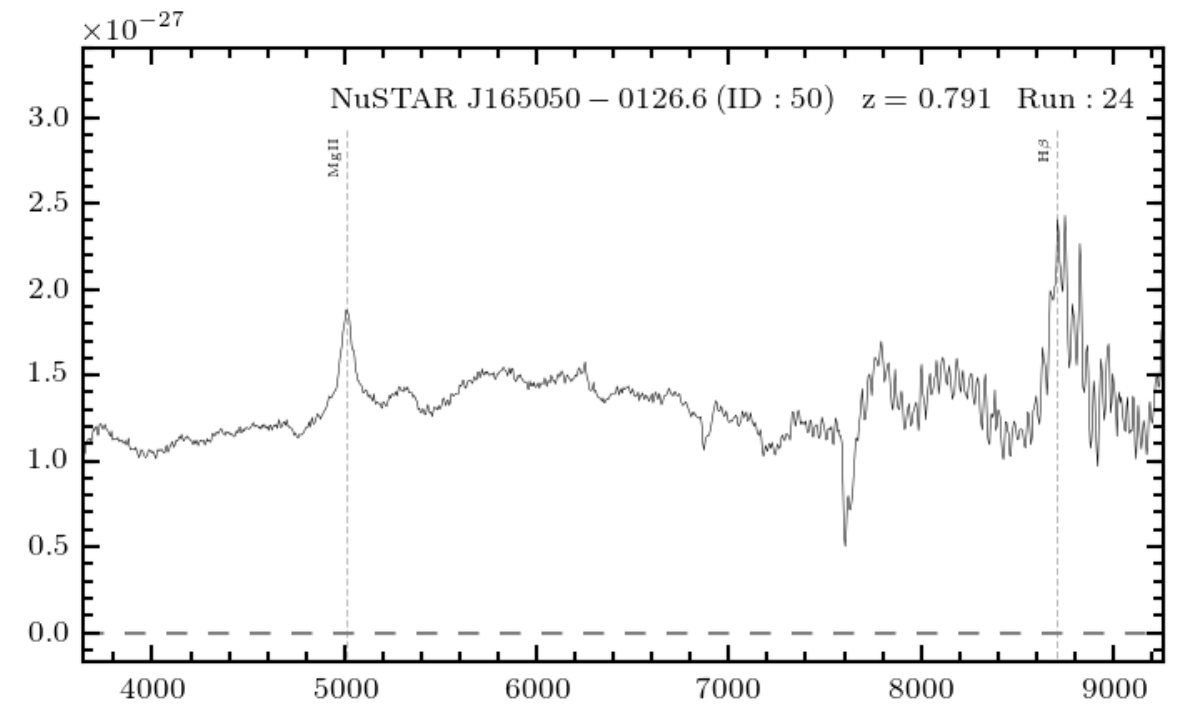}
\end{minipage}
\begin{minipage}[l]{0.325\textwidth}
\includegraphics[width=\textwidth]{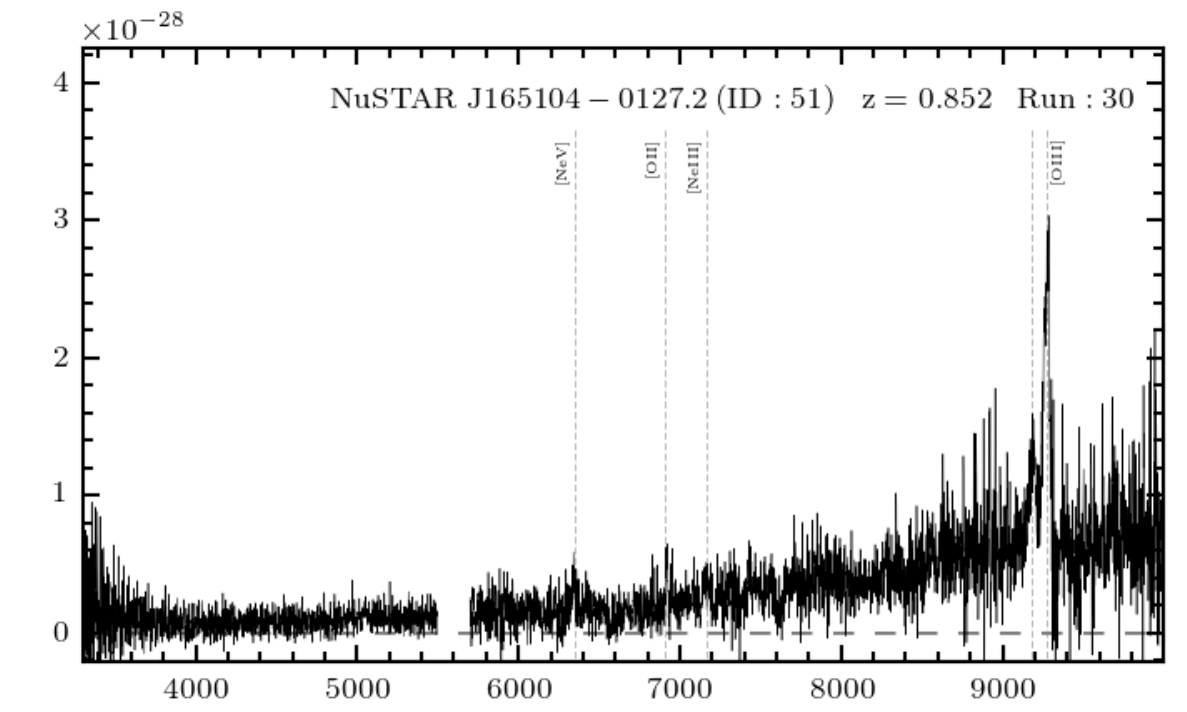}
\end{minipage}
\begin{minipage}[l]{0.325\textwidth}
\includegraphics[width=\textwidth]{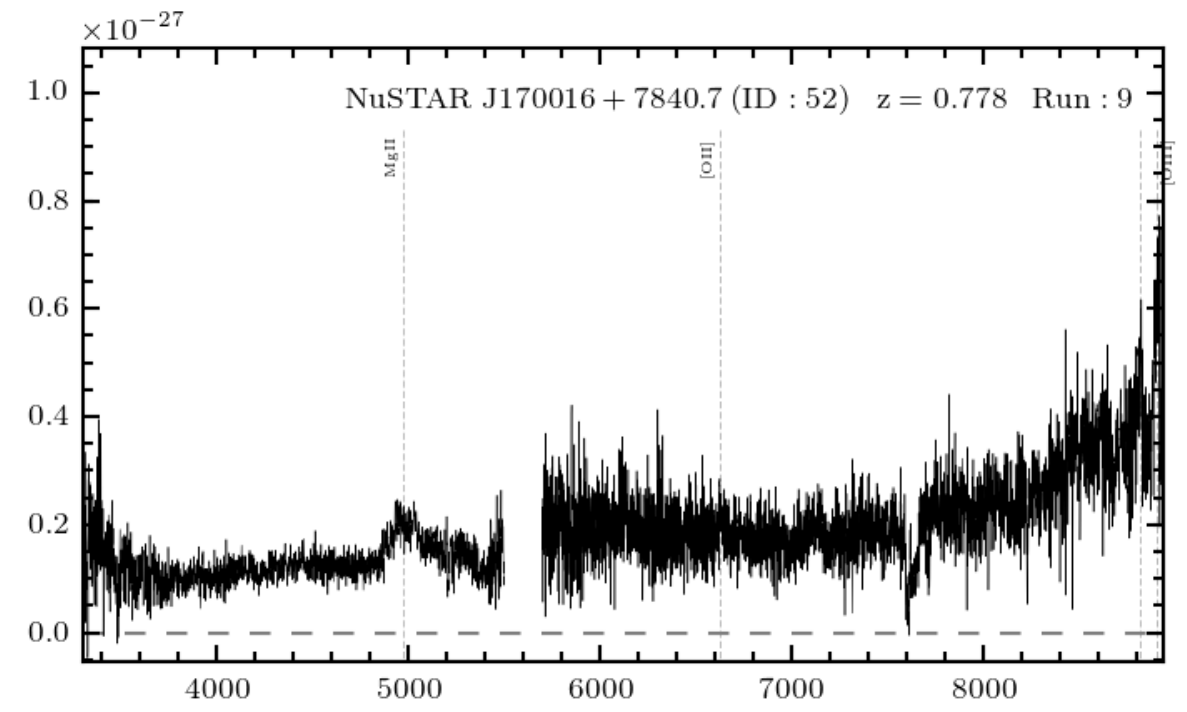}
\end{minipage}
\begin{minipage}[l]{0.325\textwidth}
\includegraphics[width=\textwidth]{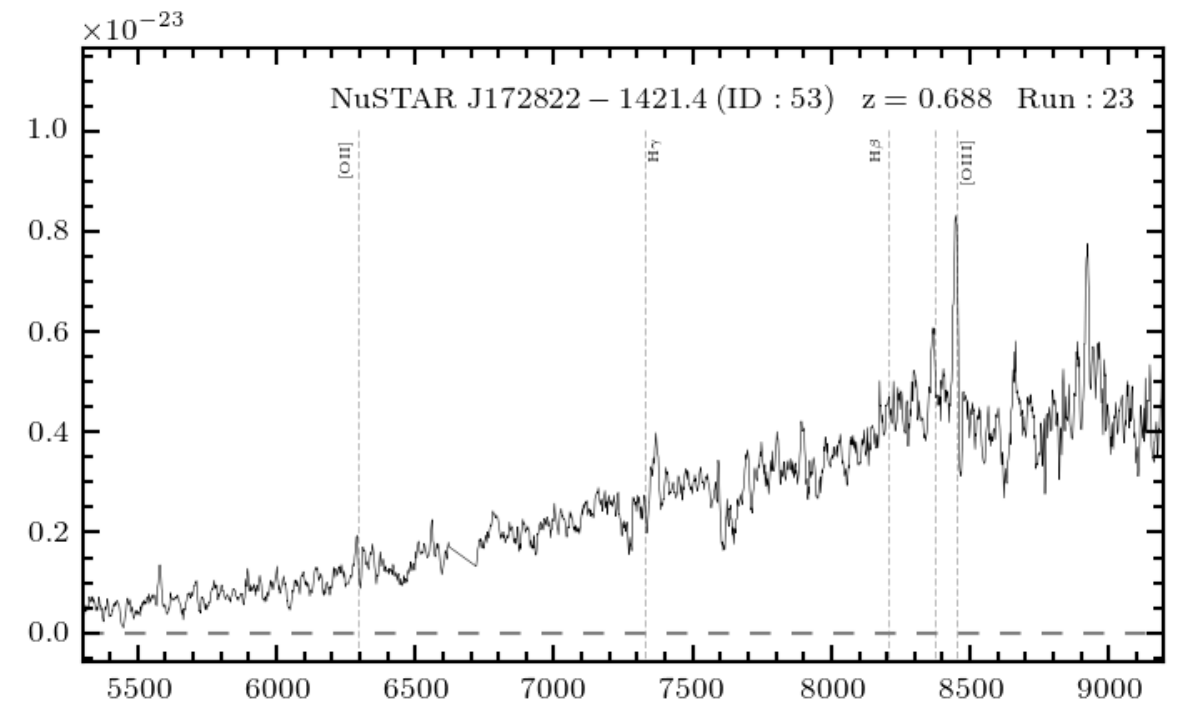}
\end{minipage}
\begin{minipage}[l]{0.325\textwidth}
\includegraphics[width=\textwidth]{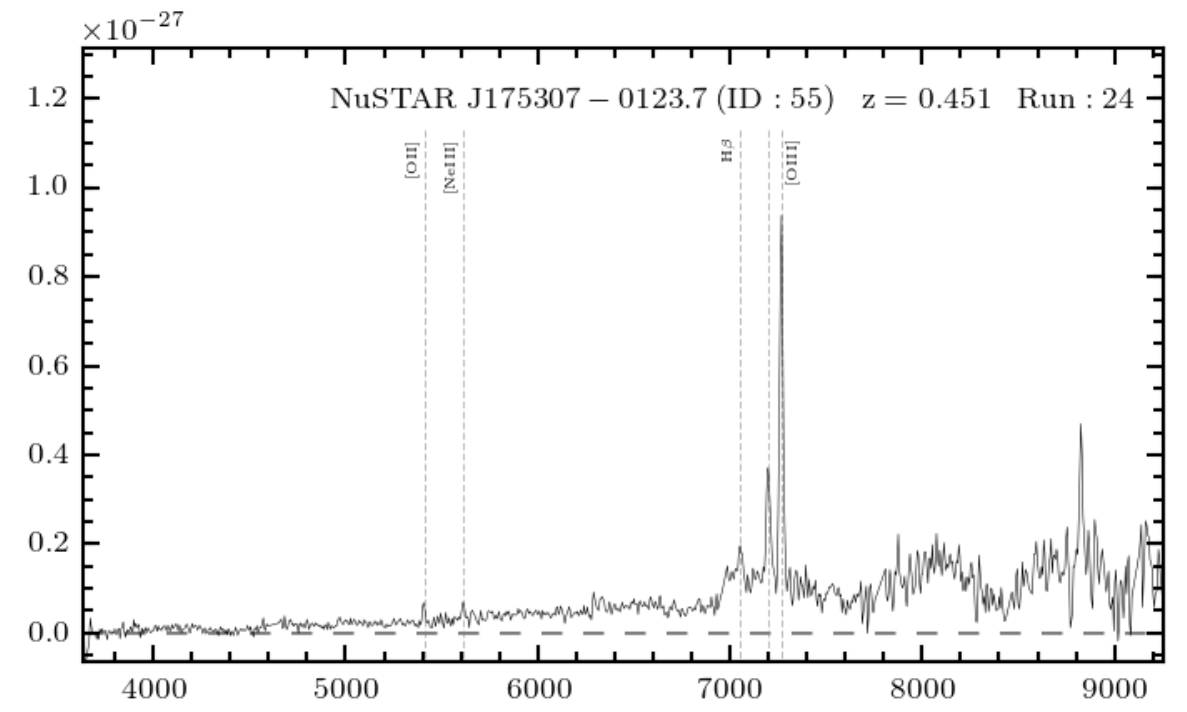}
\end{minipage}
\begin{minipage}[l]{0.325\textwidth}
\includegraphics[width=\textwidth]{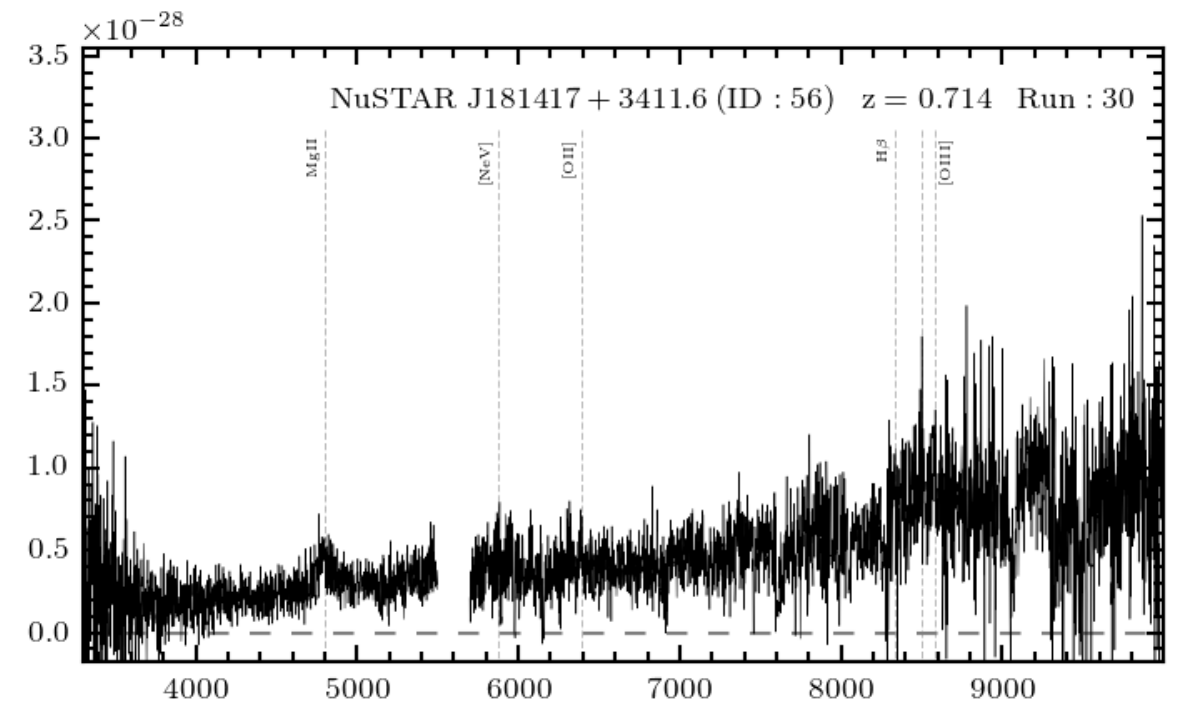}
\end{minipage}
\begin{minipage}[l]{0.325\textwidth}
\includegraphics[width=\textwidth]{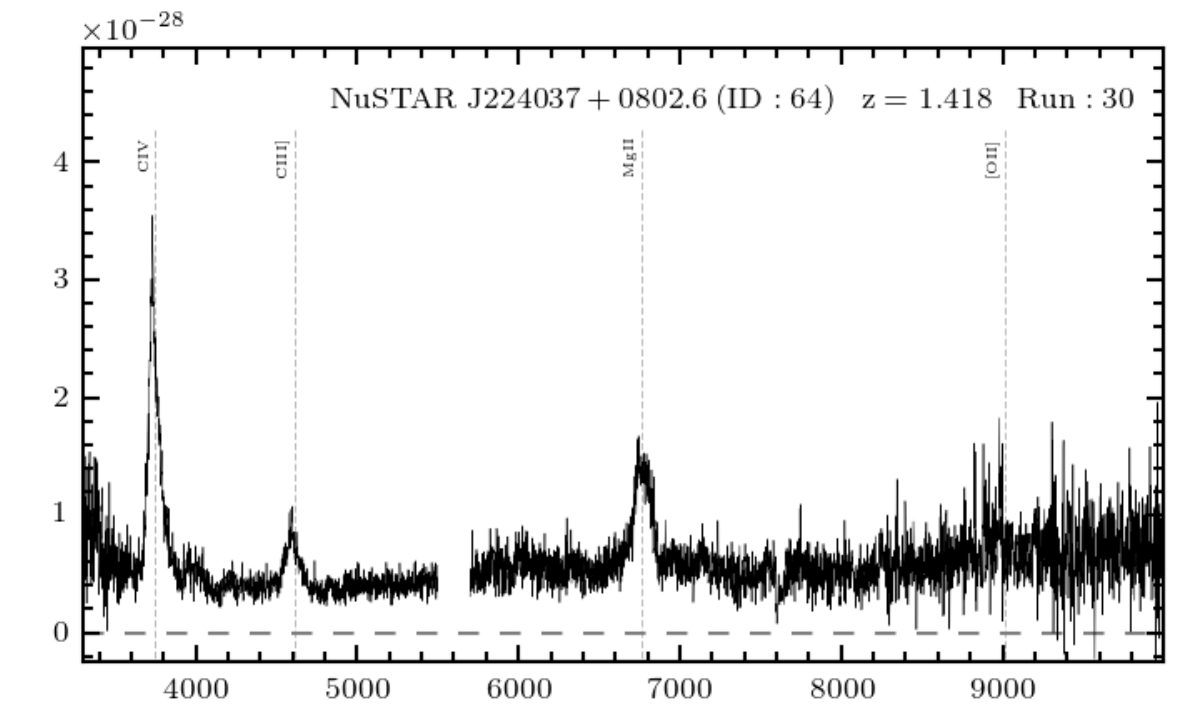}
\end{minipage}
\caption{Continued.}
\label{sec_spec_all}
\end{figure*}


\end{document}